\documentclass[letterpaper, 10pt, twoside]{LTJournalArticle}
\usepackage{graphicx}
\graphicspath{ {./images/} }
\usepackage{wrapfig}
\usepackage{enumitem}
\usepackage{amsmath,stackengine}
\stackMath
\usepackage{amssymb}
\usepackage{marginnote}
\usepackage{comment}
\usepackage{hyperref}
\usepackage{url}
\usepackage{esint}
\usepackage{multirow}
\usepackage{comment}
\usepackage{caption}
\usepackage{subcaption}
\usepackage{tikz}
\usetikzlibrary{arrows.meta}
\newcommand{\transpose}{^\intercal}
\newcommand{\e}[1]{^{[#1]}}

\DeclareMathOperator{\tr}{tr}
\DeclareMathOperator{\Arg}{Arg}
\usepackage{mathtools}
\usepackage{listings}
\usepackage{xcolor}
\usepackage{siunitx}
\usepackage{pgfplots}
\pgfplotsset{compat=1.9, , every axis/.append style={font=\scriptsize}}
\numberwithin{equation}{section}

\usepackage[nottoc]{tocbibind}
\usepackage[square,numbers]{natbib}
\definecolor{codegray}{rgb}{0.5,0.5,0.5}
\definecolor{codered}{rgb}{0.75,0,0}
\definecolor{backcolour}{rgb}{0.95,0.95,0.92}
\lstdefinestyle{mystyle}{
    backgroundcolor=\color{backcolour},
    keywordstyle=\color{blue},
    numberstyle=\tiny\color{codegray},
    stringstyle=\color{codered},
    basicstyle=\ttfamily\footnotesize,
    breakatwhitespace=false,         
    captionpos=b,                    
    keepspaces=true,                 
    numbers=left,                    
    numbersep=5pt,                  
    showspaces=false,                
    showstringspaces=false,
    showtabs=false,                  
    tabsize=2
}
\runninghead{Geometrical Properties of Chaotic Neuron Systems}
\footertext{}
\title{Exploring Geometrical Properties of Chaotic Systems Through an Analysis of the Rulkov Neuron Maps}
\author{Brandon B. Le\thanks{Author for correspondence: \href{mailto:brandon.bd.le@gmail.com}{brandon.bd.le@gmail.com}} , Nivika A. Gandhi}
\date{\footnotesize Quantum Lab, Thomas Jefferson High School for Science and Technology, Alexandria, VA \\[4px] 2024}

\renewcommand{\maketitlehookd}{
    \begin{abstract}
        \noindent Dynamical systems theory is a branch of mathematical physics with countless applications to numerous fields. Some dynamical systems exhibit chaotic behavior, characterized by a sensitive dependence on initial conditions commonly known as the ``butterfly effect.'' While extensive research has been conducted on chaos emerging from a dynamical system's temporal dynamics, our research examines extreme sensitivity to initial conditions in discrete-time dynamical systems from a geometrical perspective. Specifically, we develop methods of detecting, classifying, and quantifying geometric structures that lead to chaotic behavior in maps, including certain bifurcations, fractal geometry, strange attractors, multistability, fractal basin boundaries, and Wada basins of attraction. We also develop slow-fast dynamical systems theory for discrete-time systems, with a specific application to modeling the spiking and bursting behavior emerging from the electrophysiology of biological neurons. Our research mainly focuses on two simple low-dimensional slow-fast Rulkov maps, which model both non-chaotic and chaotic spiking-bursting neuronal behavior. We begin by exploring the maps' individual dynamics and parameter spaces, performing bifurcation analyses, describing and quantifying their chaotic dynamics, and modeling an injection of current into them. Then, by putting these neurons into different physical arrangements and coupling them with a flow of current, we find that complex dynamics and geometries emerge from the existence of multistability and sensitivity to initial conditions in higher-dimensional state space. We then analyze the complexity and fractalization of these coupled neuron systems' attractors and basin boundaries using our mathematical and computational methods. This paper begins with a conversational introduction to the geometry of chaos, then integrates mathematics, physics, neurobiology, computational modeling, and electrochemistry to present original research that provides a novel perspective on how types of geometrical sensitivity to initial conditions appear in discrete-time neuron systems.
    \end{abstract}
}
 
\begin{document}
 
\maketitle

\tableofcontents

\newpage
\clearpage

\section{Introduction}

In daily life, the word ``chaos'' often brings to mind a state of complete disorder and randomness. Edward Lorenz, the father of chaos theory, famously summarized chaos as ``when the present determines the future but the approximate present does not approximately determine the future.'' These two ideas almost entirely capture the meaning of chaos in dynamical systems theory: chaos is characterized by a sensitive dependence on initial conditions, often leading to behavior that appears random. One of the ways in which chaos arises in a system is through sensitive dependence on initial conditions emerging from its temporal dynamics, commonly referred to as the ``butterfly effect.'' This is the idea that one small disturbance can result in large implications later on, or in the words of Lorenz, a butterfly flapping its wings in Brazil can set off a tornado in Texas. In dynamical systems theory, this magnification of an initial perturbation can be quantified using Lyapunov exponents, which describe the rate at which nearby trajectories diverge from each other. This type of ``temporal'' chaos has been well-studied in dynamical systems theory; however, there is another type of chaos that has been less studied in previous research. Commonly referred to as unpredictability, this second type of chaos emerges from a geometrical property or structure of a dynamical system, specifically in systems that exhibit a property known as multistability, which is when a system can achieve multiple different stable states depending on its initial conditions. It follows that this geometrical sensitivity occurs when close initial conditions can end up in completely different final states. Interestingly, both these ``temporal'' and ``geometrical'' forms of chaos are related to each other and associated with the presence of fractal geometry. A major focus of this paper is on the types of geometric structures that lead to or result from sensitivity to initial conditions. Specifically, we explore methods of detecting, classifying, and quantifying these structures and their relation to chaos.

Another major focus of this paper is the application of dynamical systems to neurobiology and neuronal modeling. Specifically, we research two systems established by Nikolai F. Rulkov in the early 2000s. These systems are simple models of biological neurons capable of modeling a variety of neuronal behaviors, including silence, spiking, bursts of spikes, and chaotic spiking-bursting. The simplicity of the Rulkov neuron maps allows for experimentation that would not be possible with more computationally intensive models. For example, we explore arranging neurons into complex physical structures and coupling them with a flow of current. 

In the past, researchers have mainly focused on the Rulkov maps' biological properties, with minimal research being done on the chaotic dynamics and possibility of multistability and complex geometric structures that could lead to sensitivity to initial conditions in these maps. Therefore, the aims of this project and the main source of original research presented in this paper are as follows:
\begin{enumerate}
    \item to analyze and quantify the chaotic dynamics of uncoupled Rulkov neurons,
    \item to model the behavior and analyze the dynamics of complex Rulkov neuron systems,
    \item to explore the possibility of the existence of multistability and fractal geometry in the Rulkov maps,
    \item and if this multistability and geometry exists, to detect, classify, and quantify the geometrical properties associated with chaos and unpredictability.
\end{enumerate}
We believe that this research will provide a novel perspective on how these geometrical properties emerging from and resulting in sensitivity to initial conditions might appear in mathematical neuron systems and biological neurons in general. Another major aim of this paper was to curate the language and methods to be approachable and accessible for a more general audience than the target audience of most papers in chaos theory. For this reason, much of this paper is written in a conversational manner, and concepts are presented in an intuitive way. A reader with only a background in vector algebra and elementary calculus should be able to understand this paper, with most of the higher-level mathematics being delegated to the appendices.

This paper is organized as follows. In Section \ref{background}, we cover important background information from the theory of dynamical systems. In Section \ref{geometry_dynamics_chaos}, we detail much of the past and current literature, as well as our own additions and modifications, relating to the geometrical properties and structures that emerge in dynamical systems, with a focus on those that result from or lead to chaos and unpredictability. In Section \ref{slow-fast-systems-and-dynamics}, we discuss the concept of a slow-fast system and develop slow-fast dynamical systems theory for discrete-time systems. To motivate the Rulkov maps, we also discuss neurons and neuronal modeling from a biological perspective. In Section \ref{rulkov-maps}, we introduce the Rulkov maps, performing a thorough analysis of their non-chaotic and chaotic dynamics, bifurcations, and regimes of neuronal behavior. In Section \ref{injection-of-current}, we model the injection of direct current into the Rulkov maps, using the example of a pulse of electrical current. In Section \ref{coupling-of-rulkov-neurons}, we establish the method of coupling Rulkov neurons with a flow of current and perform an in-depth analysis of the dynamics of multiple coupled neuron systems. In Section \ref{geometrical-analysis-of-rulkov-neuron-systems}, we combine all of our research, applying the mathematical and computational theory from Section \ref{geometry_dynamics_chaos} to the Rulkov map systems established in Sections \ref{rulkov-maps}, \ref{injection-of-current}, and \ref{coupling-of-rulkov-neurons} to analyze three systems for geometrical properties associated with sensitivity to initial conditions.

\section{Background}
\label{background}

A dynamical system can be thought of as being a point that evolves, or moves around, according to a specific set of rules \cite{meiss}. The location of this point is known as the dynamical system's state, which is dependent on and evolves according to a time variable denoted as $t$ \cite[p. 1]{arrowsmith}. All of the possible states of a dynamical system live in the system's state space, and each state in state space corresponds to a unique point. A simple example of a dynamical system is a ball rolling across a table, which evolves in accordance with Newton's laws of motion. This system's state at a given time $t$ is the ball's location at that time $t$, and the system's state space is the entire surface of the table, which contains all the possible locations of the ball.

In this paper, we will concern ourselves with deterministic dynamical systems, where every state can be determined uniquely from past states, in contrast with stochastic dynamical systems, where randomness is involved \cite[p. 2]{alligood}. For this reason, we can calculate all of a deterministic dynamical system's future states if we know its initial state \cite{meiss}. It is important to note that the initial state of a system is not necessarily the state of the system when it was created, but rather, the state at the beginning of any relevant stretch of time \cite[p. 9]{lorenz}.

There are two main types of dynamical systems: continuous-time and discrete-time. In continuous-time dynamical systems, the time variable is continuous ($t\in\mathbb{R}$).\footnote{The symbol $\in$ means ``is an element of.''} Consider a system in real $n$-dimensional state space, or $\mathbb{R}^n$, with the state vector $\mathbf{x}$: 
\begin{equation}
    \mathbf{x} = \begin{pmatrix}
        x\e{1} \\
        x\e{2} \\
        \vdots \\
        x\e{n}
    \end{pmatrix}
    = \Big\langle x\e{1},\,x\e{2},\,\hdots\,,\,x\e{n}\Big\rangle
    \label{eq:statevector}
\end{equation}
where $x\e{m}$ is the component of $\mathbf{x}$ in the $m$th dimension of space.\footnote{$[m]$ is not an exponent.} We can describe the dynamics of a continuous-time dynamical system using a set of differential equations \cite[p. 1]{arrowsmith}:
\begin{equation}
    \frac{d\mathbf{x}}{dt} = \Dot{\mathbf{x}} = \mathbf{g}(\mathbf{x})
    \label{eq:diffeq-general}
\end{equation}
For discrete-time dynamical systems, the time variable is discrete ($t\in\mathbb{N}$), and the dynamics are governed by a mapping or iteration function \cite[p. 9]{martelli}:
\begin{equation}
    \mathbf{x}_{k+1} = \mathbf{f}(\mathbf{x}_k)
    \label{eq:iterationcond}
\end{equation}
where $\mathbf{x}_k$ is the state of the system at the time $t=k$. In this paper, we will mainly focus on discrete-time dynamical systems.

The forward orbit or trajectory $O^+(\mathbf{x}_0)$ of a discrete-time dynamical system is the set of all its iterates, generated by iterating an initial state $\mathbf{x}_0$ using the iteration function $\mathbf{f}$ an infinite number of times \cite[p. 411]{layek}:
\begin{align}
    \begin{split}
        O^+(\mathbf{x}_0) &= \{\mathbf{f}^t(\mathbf{x}_0)\}_{t=0}^{\infty} \\
        &= \{\mathbf{x}_0,\, \mathbf{f}(\mathbf{x}_0),\, \mathbf{f}^2(\mathbf{x}_0),\, \mathbf{f}^3(\mathbf{x}_0),\, \hdots\} \\
        &= \{\mathbf{x}_0,\, \mathbf{x}_1,\, \mathbf{x}_2,\, \mathbf{x}_3,\, \hdots\}
    \end{split}
\end{align}
where 
\begin{equation}
    \mathbf{f}^t(\mathbf{x}) = (\mathbf{f} \circ \mathbf{f} \circ \hdots \circ \mathbf{f})(\mathbf{x}) = \mathbf{f}(\mathbf{f}(\hdots \mathbf{f}(\mathbf{x})\hdots)
\end{equation}
$\mathbf{f}$ being repeated $t$ times. A general orbit $O(\mathbf{x})$ of length $j$ is simply a subset of $O^+(\mathbf{x}_0)$ that starts at $\mathbf{x}$ and ends at $\mathbf{f}^j(\mathbf{x})$:
\begin{equation}
    \begin{split}
        O(\mathbf{x}) &= \{\mathbf{f}^t(\mathbf{x})\}_{t=0}^j \\
        &= \{\mathbf{x},\, \mathbf{f}(\mathbf{x}),\, \mathbf{f}^2(\mathbf{x}),\,\hdots\,,\, \mathbf{f}^j(\mathbf{x})\} \\
    \end{split}
\end{equation}
An orbit can then be thought of as a subset of state space connected through time evolution.

To establish some basic definitions, a state $\mathbf{x}_s$ is stationary if it satisfies
\begin{equation}
    \mathbf{f}(\mathbf{x}_s) = \mathbf{x}_s
    \label{eq:stationary}
\end{equation}
If this holds, the point $\mathbf{x}_s$ is called a fixed point of $\mathbf{f}$ \cite[p. 16]{martelli}. Similarly, a state $\mathbf{x}_p$ is a periodic point of $\mathbf{f}$ if 
\begin{equation}
    \mathbf{f}^q(\mathbf{x}_p) = \mathbf{x}_p
    \label{eq:period}
\end{equation}
where $q\geq 1$. The periodic orbit or $q$-cycle of $\mathbf{x}_p$ is
\begin{equation}
    O^q(\mathbf{x}_p) = \{\mathbf{x}_p,\, \mathbf{f}(\mathbf{x}_p),\, \mathbf{f}^2(\mathbf{x}_p),\, \hdots \,,\, \mathbf{f}^{q-1}(\mathbf{x}_p)\}
\end{equation}
where the smallest $q$ satisfying Equation \ref{eq:period} is the period of the orbit \cite[p. 5]{arrowsmith}.

\subsection{Stability}

In simple terms, a given orbit $O(\mathbf{x})$ is said to be ``Lyapunov stable'' if nearby orbits will remain in a neighborhood of $O(\mathbf{x})$ \cite{holmes}.\footnote{We define a neighborhood of a given orbit $O(\mathbf{x})$ to be an open set (a set that does not contain its boundary) containing $O(\mathbf{x})$.} To formalize this for a fixed point of a discrete-time dynamical system, we will use the definition outlined by Bof, Carli, and Schenato \cite{bof}: a stationary state $\mathbf{x}_s$ is Lyapunov stable if, for every $\epsilon>0$, there exists a $\delta>0$ such that for all $t>0$,
\begin{equation}
    |\mathbf{x}_0-\mathbf{x}_s|<\delta \implies |\mathbf{f}^t(\mathbf{x}_0)-\mathbf{x}_s|<\epsilon
    \label{eq:stable}
\end{equation}
where 
\begin{equation}
    |\mathbf{x}| = \sqrt{\left(x\e{1}\right)^2+\left(x\e{2}\right)^2+\hdots+\left(x\e{n}\right)^2}
    \label{eq:magnitude}
\end{equation}

According to the definition by Bof, Carli, and Schenato \cite{bof}, a Lyapunov stable stationary state $\mathbf{x}_s$ is stable, or asymptotically stable, if $\delta$ can be chosen such that 
\begin{equation}
    |\mathbf{x}_0-\mathbf{x}_s|<\delta \implies \lim_{t\to\infty}\mathbf{f}^t(\mathbf{x}_0) =\mathbf{x}_s
\end{equation}
In other words, $\mathbf{x}_s$ is stable if and only if there is a neighborhood of $\mathbf{x}_s$ where all initial states will asymptotically approach $\mathbf{x}_s$ \cite[p. 5]{arrowsmith}. If a stationary state is not stable, then we call it unstable.

We will say that $O^q(\mathbf{x}_p)$, a periodic orbit of period $q$, is Lyapunov stable if each point $\mathbf{x}\in O^q(\mathbf{x}_p)$ is a Lyapunov stable point of $\mathbf{f}^q$ \cite[p. 23]{martelli}. A simple map with Lyapunov stable periodic orbits is the one-dimensional map with iteration function $f(x)=1-x$. In this map, every point except $x=0.5$ is in a Lyapunov stable 2-cycle. For example, $f(0)=1$ and $f(1)=0$, so $\{0,\,1\}$ is a periodic orbit with $q=2$. It is Lyapunov stable because $f^q(x) = f^2(x) = f(f(x)) = x$, which is obviously Lyapunov stable for all $x$.

\subsection{Quantification of Chaos} \label{quantification}

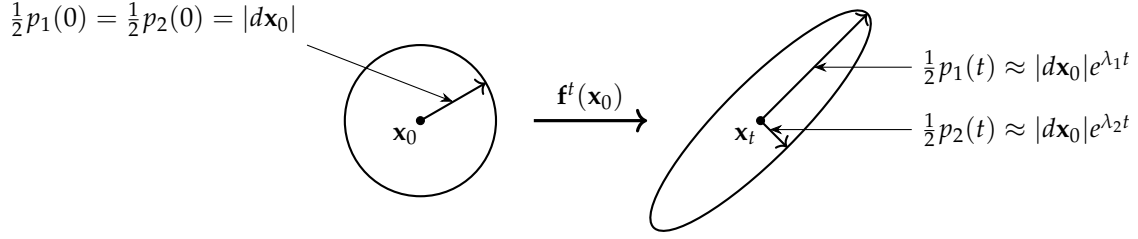
\begin{figure*}
    \centering
    \begin{tikzpicture}
        \draw[thick] (0, 0) circle [radius=1];
        \filldraw (0, 0) circle [radius=1.5pt];
        \draw[->, thick] (0,0) -> (0.866,0.5);
        \node at (-0.2, -0.2) {$\mathbf{x}_0$};
        \draw[-Stealth] (-1.5, 1) -> (0.433, 0.25);
        \node[above left] at (-1.5, 1) {$\frac{1}{2}p_1(0) = \frac{1}{2}p_2(0) = |d\mathbf{x}_0|$};
        \draw[->, very thick] (1.5, 0) -> (3, 0);
        \node[above] at (2.25, 0) {$\mathbf{f}^t(\mathbf{x}_0)$};
        \draw[rotate around={45:(4.5,0)}, thick] (4.5,0) ellipse (2 and 0.5);
        \filldraw (4.5, 0) circle [radius=1.5pt];
        \node at (4.3, -0.2) {$\mathbf{x}_t$};
        \draw[->, thick] (4.5, 0) -> (5.914, 1.414);
        \draw[->, thick] (4.5, 0) -> (4.854, -0.354);
        \draw[-Stealth] (6.5, 0.707) -> (5.207, 0.707);
        \draw[-Stealth] (6.5, -0.118) -> (4.618, -0.118);
        \node[right] at (6.5, 0.707) {$\frac{1}{2}p_1(t) \approx |d\mathbf{x}_0|e^{\lambda_1t}$};
        \node[right] at (6.5, -0.118) {$\frac{1}{2}p_2(t) \approx |d\mathbf{x}_0|e^{\lambda_2t}$};
    \end{tikzpicture}
    \vspace{4px}
    \caption{Evolution of an initial infinitesimal one-dimensional sphere of perturbations from $\mathbf{x}_0$ by $d\mathbf{x}_0$ after $t$ iterations, with the semi-principal axes of the one-dimensional ellipsoid approximated by a modification of Equation \ref{eq:lyap-prin-axes}}
    \label{fig:perturbation-evolution}
\end{figure*}

According to Lorenz \cite[p. 8]{lorenz}, the father of chaos theory, a dynamical system exhibits chaos if it is ``sensitively dependent on initial conditions.'' To get a better idea of what this means, we will turn to a quantitative definition of chaos provided by Alligood, Sauer, and Yorke \cite[p. 106]{alligood}: chaos is defined by ``a Lyapunov exponent greater than zero.'' The Lyapunov exponent is a quantity that characterizes the separation rate of close trajectories \cite{brandon}. To define this precisely, let us first consider a simple one-dimensional discrete-time dynamical system with an initial state $x_0$ and an iteration function $x_{k+1}=f(x_k)$. If we perturb this initial state by some small amount $\delta x_0$, we can expect the separation of the initial and the perturbed states after some time $t$, denoted by $\delta x_t$, to be approximated by
\begin{equation}
    |\delta x_t| = |f^t(x_0+\delta x_0)-f^t(x_0)| \approx |\delta x_0|e^{\lambda t}
    \label{eq:lyapunovapprox}
\end{equation}
where $\lambda$ is the Lyapunov exponent. Here, we can see that if $\lambda>0$, any difference in initial conditions will be magnified as time goes on; even if the initial perturbation is infinitesimal, $\delta x_t$ will become significant as $t$ goes to infinity. This is how we quantify the chaos of a dynamical system: the more positive $\lambda$ is, the faster close trajectories will diverge from each other, so the more chaotic a system is. Similarly, the more negative $\lambda$ is, the faster close trajectories will converge on each other, so the more non-chaotic a system is \cite{wolf}.

The Lyapunov exponent $\lambda$ is calculated from Equation \ref{eq:lyapunovapprox} as $t$ goes to infinity and $\delta x_0$ approaches 0:
\begin{equation}
    \begin{split}
        \lambda &= \lim_{t\to\infty}\lim_{\delta x_0\to 0}\frac{1}{t}\ln\left|\frac{\delta x_t}{\delta x_0}\right| \\
        &= \lim_{t\to\infty}\lim_{\delta x_0\to 0}\frac{1}{t}\ln\left|\frac{f^t(x_0+\delta x_0)-f^t(x_0)}{\delta x_0}\right|
        \label{eq:lyapunovunsimp}
    \end{split}
\end{equation}
It is shown by Le \cite{brandon} that this can be simplified to\footnote{See Appendix \ref{lyap1d-deriv} for a full derivation.}
\begin{equation}
    \lambda = \lim_{t\to\infty}\frac{1}{t}\sum_{i=0}^{t-1}\ln |f'(x_i)|
    \label{eq:1dlyap}
\end{equation}
which is simply the average of the logs of the derivatives taken at all the points visited by $O^+(x_0)$.

Now, let us consider an $n$-dimensional system with an initial state $\mathbf{x}_0$ and an iteration function $\mathbf{x}_{k+1} = \mathbf{f}(\mathbf{x}_k)$. Given some initial perturbation in a general direction 
\begin{equation}
    \delta \mathbf{x}_0 = \begin{pmatrix}
        \delta x\e{1}_0 \\[4pt]
        \delta x\e{2}_0 \\[1pt]
        \vdots \\[1pt]
        \delta x\e{n}_0
    \end{pmatrix}
    \label{eq:perturbation-gendirec}
\end{equation}
we can rewrite Equation \ref{eq:lyapunovunsimp} for our $n$-dimensional system as
\begin{equation}
    \begin{split}
        \lambda &= \lim_{t\to\infty}\lim_{\delta x\e{1}_0,\, \delta x\e{2}_0,\, \hdots\,,\, \delta x\e{n}_0\to0}\frac{1}{t}\ln\frac{|\delta \mathbf{x}_t|}{|\delta \mathbf{x}_0|} \\
        &= \lim_{t\to\infty}\frac{1}{t}\ln\frac{|d\mathbf{x}_t|}{|d\mathbf{x}_0|}
        \label{eq:lyap-nd-unsimp}
    \end{split}
\end{equation}
where $d\mathbf{x}_0$ is an infinitesimal initial perturbation and $d\mathbf{x}_t$ is the evolution of that perturbation after $t$ steps. 

If we write Equation \ref{eq:iterationcond} fully as
\begin{align}
    \begin{split}
        \begin{pmatrix}
        x\e{1}_{k+1} \\[4pt]
        x\e{2}_{k+1} \\[1pt]
        \vdots \\[1pt]
        x\e{n}_{k+1}
    \end{pmatrix} =
    \begin{pmatrix}
        f\e{1}(x\e{1}_k,\, x\e{2}_k,\,\hdots\,,\, x\e{n}_k) \\[4pt]
        f\e{2}(x\e{1}_k,\, x\e{2}_k,\,\hdots\,,\, x\e{n}_k) \\[1pt]
        \vdots \\[1pt]
        f\e{n}(x\e{1}_k,\, x\e{2}_k,\,\hdots\,,\, x\e{n}_k)
    \end{pmatrix}
    \end{split}
    \label{eq:expandediteration}
\end{align}
where $f\e{m}$ is the iteration function corresponding to the $m$th dimension of $\mathbf{f}$, then we can write the Jacobian matrix $J(\mathbf{x})$ of the system as 
\begin{equation}
    J(\mathbf{x}) = \begin{pmatrix}
        \frac{\partial f\e{1}}{\partial x\e{1}} & \frac{\partial f\e{1}}{\partial x\e{2}} & \hdots & \frac{\partial f\e{1}}{\partial x\e{n}} \\[4pt]
        \frac{\partial f\e{2}}{\partial x\e{1}} & \frac{\partial f\e{2}}{\partial x\e{2}} & \hdots & \frac{\partial f\e{2}}{\partial x\e{n}} \\[1pt]
        \vdots & \vdots & \ddots & \vdots \\[1pt]
        \frac{\partial f\e{n}}{\partial x\e{1}} & \frac{\partial f\e{n}}{\partial x\e{2}} & \hdots & \frac{\partial f\e{n}}{\partial x\e{n}}
    \end{pmatrix}
    \label{eq:jacobian}
\end{equation}
It is shown by Le \cite{brandon} that an infinitesimal perturbation can be iterated by the Jacobian matrix:\footnote{See Appendix \ref{lyapspec-deriv} for a full derivation.}
\begin{equation}
    d\mathbf{x}_{k+1} = J(\mathbf{x}_k)d\mathbf{x}_k
    \label{eq:jacobian-iteration}
\end{equation}
It follows that
\begin{equation}
    d\mathbf{x}_{t} = J(\mathbf{x}_{t-1})J(\mathbf{x}_{t-2})\hdots J(\mathbf{x}_0)d\mathbf{x}_0
    = J^td\mathbf{x}_0
    \label{eq:jacobian-alltheway}
\end{equation}
Rewriting Equation \ref{eq:lyap-nd-unsimp} by substituting Equation \ref{eq:jacobian-alltheway}, we get that
\begin{equation}
    \begin{split}
        \lambda &= \lim_{t\to\infty}\frac{1}{t}\ln\frac{|J^td\mathbf{x}_0|}{|d\mathbf{x}_0|} \\
        &= \lim_{t\to\infty}\frac{1}{t}\ln\left|J^t\mathbf{u}_0\right|
    \end{split}
    \label{eq:lyapunitvector}
\end{equation}
where $\mathbf{u}_0$ is a unit vector in the direction of $d\mathbf{x}_0$.

In an $n$-dimensional system, there is a spectrum of Lyapunov exponents $\lambda = \{\lambda_1,\, \lambda_2,\,\hdots\,,\, \lambda_n\}$ ordered from largest to smallest. In the spectrum, each Lyapunov exponent corresponds to the rate of separation of trajectories for a specific starting direction of $\mathbf{u}_0$ \cite{wolf}. Geometrically, we can imagine all possible $d\mathbf{x}_0$ forming an $(n-1)$-dimensional sphere\footnote{In this paper, we say that a one-dimensional sphere is a circle, a two-dimensional sphere is a standard sphere, and so on. Topologically, this is because a circle ``looks like'' a one-dimensional line up close, a sphere ``looks like'' a two-dimensional plane up close, and so on.} with infinitesimal radius centered around $\mathbf{x}_0$. As time evolves, this $(n-1)$-dimensional sphere will be deformed into an $(n-1)$-dimensional ellipsoid with $n$ principal axes (see Figure \ref{fig:perturbation-evolution}). If we say $p_i(t)$ represents the length of the $i$th principal axis\footnote{The 1st principal axis is the longest, the 2nd principal is the second longest, and so on.} at time $t$, then the Lyapunov exponent $\lambda_i$ corresponds to the growth of $p_i(t)$. Then, from Equation \ref{eq:lyap-nd-unsimp},
\begin{equation}
    \lambda_i = \lim_{t\to\infty}\frac{1}{t}\ln\frac{p_i(t)}{p_i(0)} = \lim_{t\to\infty}\frac{1}{t}\ln\frac{p_i(t)}{2|d\mathbf{x}_0|}
    \label{eq:lyap-prin-axes}
\end{equation}
because the radius of a circle is half the length of its principal axis. For an arbitrarily chosen direction of $\mathbf{u}_0$, the perturbation will grow in magnitude according to the maximal Lyapunov exponent $\lambda_1$ because of the exponential nature of Lyapunov exponents \cite{brandon}. Geometrically, this is because, after a long time, the 1st principal axis of the ellipse will dwarf all of the other principal axes in comparison (see Figure \ref{fig:perturbation-evolution}). Any $d\mathbf{x}_0$ that has a non-zero component in the direction of the 1st principal axis will eventually be overwhelmed by the growth of that component, resulting in $d\mathbf{x}_t$ approaching the direction of the 1st principal axis and growing according to $\lambda_1$.

While Equation \ref{eq:lyap-prin-axes} provides good intuition for what is going on, it isn't practical to use it to calculate the Lyapunov spectrum. Instead, Le \cite{brandon} shows that the Lyapunov spectrum can be manipulated from Equation \ref{eq:lyapunitvector} into the form\footnote{See Appendix \ref{lyapspec-deriv} for a full derivation.}
\begin{equation}
    \lambda_i = \lim_{t\to\infty}\frac{1}{2t}\ln\mu_i
    \label{eq:lyapunov-eigenvalue}
\end{equation}
where $\mu_i$ is an eigenvalue\footnote{For the unfamiliar reader, if $M\mathbf{v} = m\mathbf{v}$ for some matrix $M$, some vector $\mathbf{v}$, and some scalar $m$, $m$ is an eigenvalue of $M$ and $\mathbf{v}$ is its associated eigenvector.  Lay, Lay, and McDonald \cite[pp. 268-273]{linear} provide an excellent in-depth explanation of this concept.} of the matrix $J^{t\intercal}J^t$.\footnote{$M^{\intercal}$ is the transpose of $M$.} These eigenvalues $\mu_i$ are labelled with their subscript $i$ so that the Lyapunov exponents satisfy $\lambda_1\geq\lambda_2\geq\hdots\geq\lambda_n$. Mirroring the aforementioned definition from Alligood, Sauer, and Yorke \cite[p. 106]{alligood}, at least one Lyapunov exponent being greater than 0 indicates chaos \cite{wolf}. 

\section{The Geometry of Dynamical and Chaotic Systems}
\label{geometry_dynamics_chaos}

As defined in Section \ref{quantification}, a chaotic system is a dynamical system exhibiting sensitive dependence on initial conditions. We have already discussed the type of sensitivity to initial conditions known as ``temporal'' chaos, which emerges from a system's dynamics and can be quantified using Lyapunov exponents. However, another type of sensitivity to initial conditions, a ``geometrical'' form of chaos, arises in some systems exhibiting a phenomenon known as multistability.

In general, dynamical systems can have many interesting geometrical properties, including attractors, basins of attraction, and fractal basin boundaries. We are interested in these geometric structures because some of them are strongly connected with these ``temporal'' and ``geometrical'' forms of chaos. In this section, we will explore some of the geometries of dynamical and chaotic systems and methods of detecting, classifying, and quantifying them.

\subsection{Non-Chaotic Attractors}
\label{nonchaoticattractors}

An attractor, as defined by Strogatz \cite[p. 332]{strogatz}, is ``a set to which all neighboring trajectories converge.'' In more specific terms, an attractor can be defined as a set of points $A$ in state space that satisfies the following three properties:
\begin{enumerate}
    \item Trajectories that start in $A$ will stay in $A$. Specifically, any forward orbit $O^+(\mathbf{x}_0)$ with $\mathbf{x}_0\in A$ will satisfy $\mathbf{x}_t\in A$ for all $t>0$.
    \item $A$ will attract an open set of initial conditions. Specifically, say that the distance from an arbitrary state $\mathbf{x}$ to $A$ is the smallest $|\mathbf{x} - \mathbf{a}|$ such that $\mathbf{a}\in A$. Then, there is some open set $U$ that $A$ is a subset of, or $A\subseteq U$,\footnote{The symbol $\subseteq$ means ``is a subset of.''} such that if some $\mathbf{x}_0\in U$, then the distance from $\mathbf{x}_t$ to $A$ approaches 0 as $t\to\infty$. Essentially, all forward orbits that start sufficiently close to $A$ will be attracted to $A$.
    \item $A$ is as small as possible. Specifically, this means there is no set $A'$ satisfying $A'\subset A$\footnote{The symbol $\subset$ means ``is a proper subset of.'' $X\subset Y$ if and only if $X\subseteq Y$ but $X\neq Y$.} that also satisfies the previous conditions.
\end{enumerate} 

Attractors are key geometrical features of dynamical systems because, by definition, they represent the typical behavior of a set of a dynamical system's state space. We will begin our discussion of attractors with non-chaotic attractors, that is, attractors that don't exhibit sensitivity to initial conditions.

\begin{figure}[b]
    \centering
    \begin{subfigure}[b]{0.2\textwidth}
        \centering
        \begin{tikzpicture}
            \draw[thick, domain=-1:1]plot(\x,{1-(\x)^2});
            \draw[fill](0, 1.05) circle [radius=0.05];
            \node[above] at (0, 1.05) {$\mathbf{x}_s$};
        \end{tikzpicture}
        \caption{Ball at the top of a hill, an example of a fixed point repeller}
        \label{fig:fixedpointrepeller}
    \end{subfigure}
    \hfill
    \begin{subfigure}[b]{0.2\textwidth}
        \centering
        \begin{tikzpicture}
            \draw[thick, domain=-1:1]plot(\x,{(\x)^2});
            \draw[fill](0, 0.05) circle [radius=0.05];
            \node[above] at (0, 0.05) {$\mathbf{x}_s$};
        \end{tikzpicture}
        \caption{Ball at the bottom of a valley, an example of a fixed point attractor}
        \label{fig:fixedpointattractor}
    \end{subfigure}
    \caption{Two systems with fixed points}
    \label{fig:twofixedpoints}
    \vspace{10px}
\end{figure}
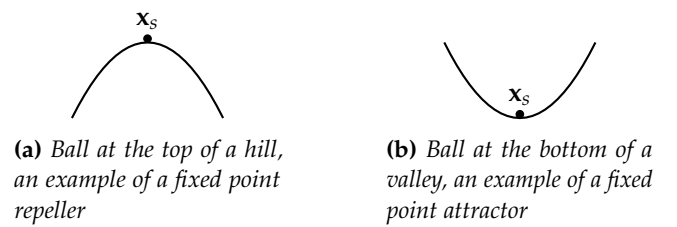

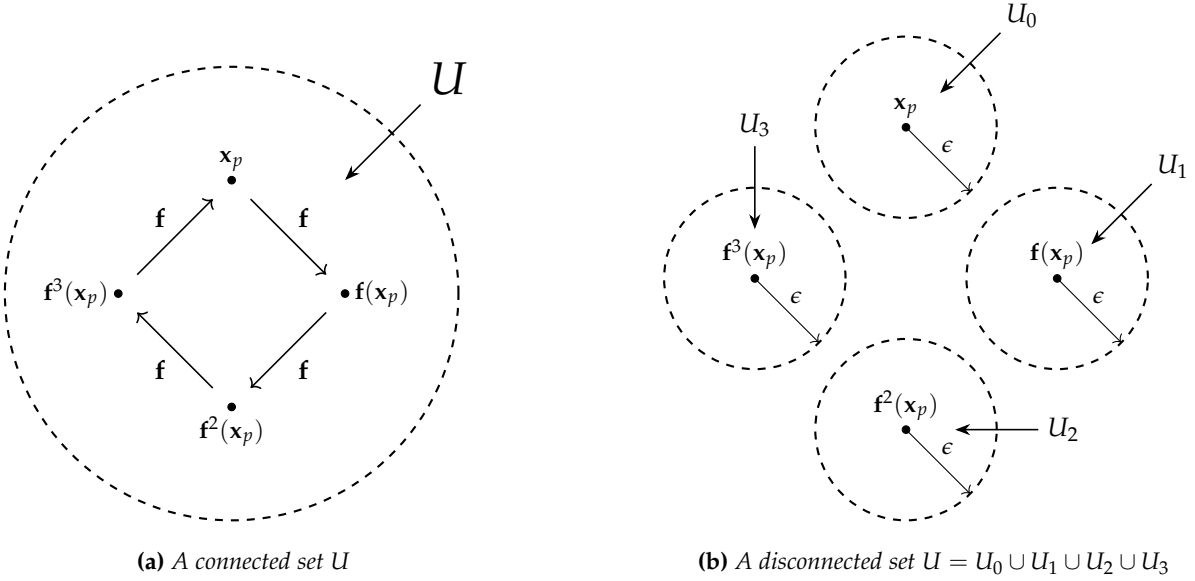
\begin{figure*}
    \centering
    \begin{subfigure}[b]{0.475\textwidth}
        \centering
        \begin{tikzpicture}
            \draw[dashed, thick] (0,0) circle [radius=3];
            \node[above right] at (2.5,2.5) {\LARGE $U$};
            \draw[-Stealth, semithick](2.5,2.5)--(1.5,1.5);
            \draw[fill] (0,1.5) circle [radius=0.05];
            \node[above] at (0, 1.5) {\small $\mathbf{x}_p$};
            \draw[->, semithick](0.25,1.25)--(1.25,0.25);
            \node[above right] at (0.75,0.75) {$\mathbf{f}$};
            \draw[fill] (1.5,0) circle [radius=0.05];
            \node[right] at (1.5,0) {\small $\mathbf{f}(\mathbf{x}_p)$};
            \draw[<-, semithick](0.25,-1.25)--(1.25,-0.25);
            \node[below right] at (0.75,-0.75) {$\mathbf{f}$};
            \draw[fill] (0,-1.5) circle [radius=0.05];
            \node[below] at (0, -1.5) {\small $\mathbf{f}^2(\mathbf{x}_p)$};
            \draw[->, semithick](-0.25,-1.25)--(-1.25,-0.25);
            \node[below left] at (-0.75,-0.75) {$\mathbf{f}$};
            \draw[fill] (-1.5,0) circle [radius=0.05];
            \node[left] at (-1.5,0) {\small $\mathbf{f}^3(\mathbf{x}_p)$};
            \draw[->, semithick] (-1.25,0.25)--(-0.25,1.25);
            \node[above left] at (-0.75,0.75) {$\mathbf{f}$};
        \end{tikzpicture}
        \vspace{4px}
        \caption{A connected set $U$}
        \label{fig:periodicorbitsa}
    \end{subfigure}
    \hfill
    \begin{subfigure}[b]{0.475\textwidth}
        \centering
        \begin{tikzpicture}
            \draw[fill] (0,2) circle [radius=0.05];
            \node[above] at (0, 2) {\small $\mathbf{x}_p$};
            \draw[dashed, thick] (0,2) circle [radius=1.2];
            \draw[-Stealth, semithick](1.25,3.25)--(0.45,2.45);
            \node[above right] at (1.2,3.2) {$U_0$};
            \draw[->, thin](0,2)--(0.849,1.151);
            \node[above right] at (0.35, 1.55) {\small $\epsilon$};
            
            \draw[fill] (2,0) circle [radius=0.05];
            \node[above] at (2,0) {\small $\mathbf{f}(\mathbf{x}_p)$};
            \draw[dashed, thick] (2,0) circle [radius=1.2];
            \draw[-Stealth, semithick](3.25,1.25)--(2.45,0.45);
            \node[above right] at (3.2,1.2) {$U_1$};
            \draw[->, thin] (2,0)--(2.849,-0.849);
            \node[above right] at (2.35, -0.45) {\small $\epsilon$};
            
            \draw[fill] (0,-2) circle [radius=0.05];
            \node[above] at (0, -2) {\small $\mathbf{f}^2(\mathbf{x}_p)$};
            \draw[dashed, thick] (0,-2) circle [radius=1.2];
            \draw[-Stealth, semithick](1.75, -2)--(0.65,-2);
            \node[right] at (1.75,-2) {$U_2$};
            \draw[->, thin] (0,-2)--(0.849,-2.849);
            \node[above right] at (0.35, -2.45) {\small $\epsilon$};
            
            \draw[fill] (-2,0) circle [radius=0.05];
            \node[above] at (-2,0) {\small $\mathbf{f}^3(\mathbf{x}_p)$};
            \draw[dashed, thick] (-2,0) circle [radius=1.2];
            \draw[-Stealth, semithick](-2, 1.75)--(-2,0.65);
            \node[above] at (-2,1.75) {$U_3$};
            \draw[->, thin] (-2,0)--(-1.151,-0.849);
            \node[above right] at (-1.65,-0.45) {\small $\epsilon$};
        \end{tikzpicture}
        \vspace{4px}
        \caption{A disconnected set $U = U_0\cup U_1\cup U_2\cup U_3$}
        \label{fig:periodicorbitsb}
    \end{subfigure}
    \vspace{2px}
    \caption{Two possible open sets of initial conditions $U$ that 4-cycle attractor $A = \{\mathbf{x}_p,\, \mathbf{f}(\mathbf{x}_p),\, \mathbf{f}^2(\mathbf{x}_p),\, \mathbf{f}^3(\mathbf{x}_p)\}$ attracts}
    \label{fig:periodicorbits}
\end{figure*}

\subsubsection{Fixed Points}

As previously defined in Section \ref{background} (Equation \ref{eq:stationary}), a state $\mathbf{x}_s$ is a fixed point of $\mathbf{f}$ if it satisfies $\mathbf{f}(\mathbf{x}_s) = \mathbf{x}_s$. A fixed point $\mathbf{x}_s$ is also an attractor if it satisfies all of the previously listed properties.

To get an intuitive understanding of what makes a given fixed point an attractor or not, we will consider two example systems. First, let us consider a ball on top of a hill (see Figure \ref{fig:fixedpointrepeller}). The point at the top of this hill $\mathbf{x}_s$ is our potential attractor. This is clearly a fixed point by Equation \ref{eq:stationary}: as time evolves, the state of the system won't change. For the same reason, Property 1 of attractors is satisfied. Now, let us consider an open set that contains $\mathbf{x}_s$. If a ball starts in a neighborhood of $\mathbf{x}_s$, it will roll down the hill and be repelled from $\mathbf{x}_s$ no matter how small that neighborhood is. This means that there is no open set that satisfies the requirements of Property 2, so $\mathbf{x}_s$ is not an attractor. In fact, because states that start in a neighborhood of $\mathbf{x}_s$ are repelled from it, we call $\mathbf{x}_s$ a repeller \cite[p. 17]{strogatz}.

Our second system is a ball at the bottom of a valley (see Figure \ref{fig:fixedpointattractor}). Again, this point $\mathbf{x}_s$ is clearly a fixed point and satisfies Property 1. However, if we now consider open sets that contain our fixed point, we can see that a ball that starts in a neighborhood of $\mathbf{x}_s$ will oscillate around $\mathbf{x}_s$ before eventually settling at it. Therefore, $\mathbf{x}_s$ satisfies Property 2: there exists a neighborhood of the point such that initial states in that neighborhood are attracted to it. Finally, notice that our attractor $A$ contains only one point: $A = \{\mathbf{x}_s\}$. There exists no proper subset of $A$ besides the empty set, which is obviously not an attractor, so Property 3 is automatically satisfied. Therefore, $\mathbf{x}_s$ is an attractor.

We prove in Appendix \ref{fixedpointattractor-criteria} that there is a simple set of criteria to determine whether a fixed point is an attractor, a repeller, or a saddle point.\footnote{Saddle points attract some states in a small neighborhood of it and repel others in that same neighborhood. They only exist in systems higher than one dimension.} For a one-dimensional system $x_{k+1} = f(x_k)$ with a fixed point $x_s$, the criteria are as follows:
\begin{enumerate}
    \item $x_s$ is an attractor if $|f'(x_s)|<1$.
    \item $x_s$ is a repeller if $|f'(x_s)|>1$.
\end{enumerate}
For an $n$-dimensional system $\mathbf{x}_{k+1} = \mathbf{f}(\mathbf{x}_k)$ with a fixed point $\mathbf{x}_s$,
\begin{enumerate}
    \item $\mathbf{x}_s$ is an attractor if the absolute value of each eigenvalue of the Jacobian matrix at $\mathbf{x}_s$ is less than 1. That is, $|\nu_i|<1$ for $i = 1,\,2,\,\hdots\,,\,n$.
    \item $\mathbf{x}_s$ is a repeller if $|\nu_i|>1$ for $i = 1,\,2,\,\hdots\,,\,n$.
    \item $\mathbf{x}_s$ is a saddle point if at least one $|\nu_i|>1$ and at least one $|\nu_i|<1$ for $i = 1,\,2,\,\hdots\,,\,n$.
\end{enumerate}

\subsubsection{Periodic Orbits}

Consider a periodic orbit of the periodic point $\mathbf{x}_p$ that has a period $q$: 
\begin{equation}
    O^q(\mathbf{x}_p) = \{\mathbf{x}_p,\, \mathbf{f}(\mathbf{x}_p),\, \mathbf{f}^2(\mathbf{x}_p),\, \hdots \,,\, \mathbf{f}^{q-1}(\mathbf{x}_p)\}
\end{equation}
Let us consider what has to be true for this periodic orbit to be attractor $A = O^q(\mathbf{x}_p)$. Of course, Property 1 is immediately satisfied by the definition of a periodic orbit established in Section \ref{background}. By Property 2, in order for $A$ to be an attractor, there must be an open set such that any forward orbit $O^+(\mathbf{x}_0)$ that starts in this set contains an orbit $O(\mathbf{x})$ that is arbitrarily close to $O^q(\mathbf{x}_p)$. In other words, a state that starts in the open set will eventually fall into a periodic orbit that approaches $O^q(\mathbf{x}_p)$. A simple possibility for an open set of initial conditions that $A$ attracts is a connected set\footnote{A connected set is a set where every point in the set can be reached by a continuous path from every other point in the set without leaving it.} that contains all the points in the periodic orbit (see Figure \ref{fig:periodicorbitsa}). Another possibility is a disconnected set composed of neighborhoods of all the points visited by $O^q(\mathbf{x}_p)$ (see Figure \ref{fig:periodicorbitsb}). For example, if we define these neighborhoods as open sets $U_i$ that are the sets of states $\mathbf{x}$ that satisfy $|\mathbf{x} - \mathbf{f}^i(\mathbf{x}_p)| < \epsilon$, where $i = 0,\, 1,\, 2,\,\hdots\,,\,q-1$, then our open set of initial conditions is $U = U_0\cup U_1\cup U_2\cup\hdots\cup U_{q-1}$.\footnote{$S_1\cup S_2$ is the union of $S_1$ and $S_2$, which is the set of all the elements contained in $S_1$, $S_2$, or both $S_1$ and $S_2$.} Finally, because we defined the period $q$ of our periodic orbit $O^q(\mathbf{x}_p)$ in Section \ref{background} to be the smallest $q$ satisfying $\mathbf{f}^q(\mathbf{x}_p) = \mathbf{x}_p$ (Equation \ref{eq:period}), any $A'$ will be missing at least one point in $O^q(\mathbf{x}_p)$. A trajectory that starts in $A'$ will leave $A'$ when it reaches one of these missing points, so $A'$ fails Property 1. Therefore, a periodic orbit $O^q(\mathbf{x}_p)$ that satisfies Property 2 is an attractor. We will delve deeper into and see examples of periodic orbit attractors in Section \ref{bifurcations}.

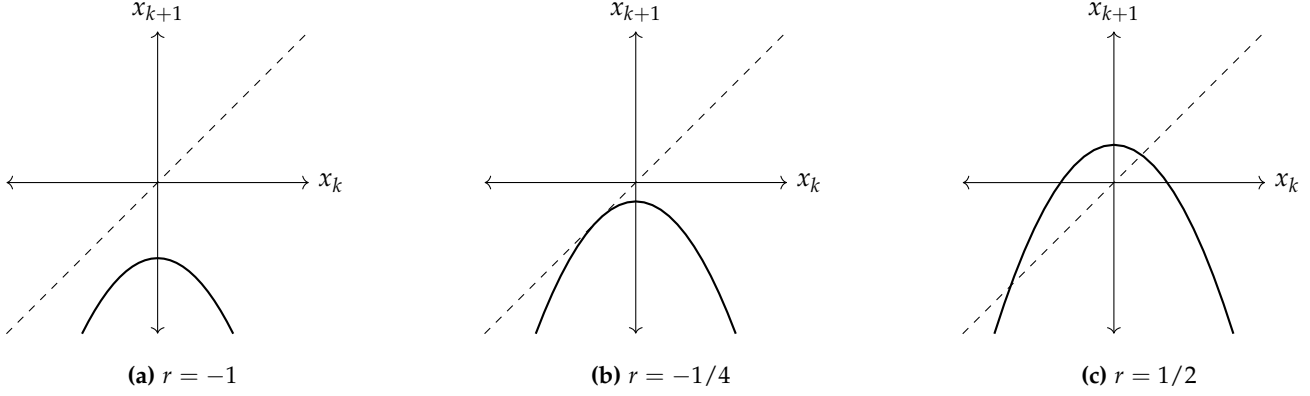
\begin{figure*}
    \centering
    \begin{subfigure}[b]{0.275\textwidth}
        \centering
        \begin{tikzpicture}
            \draw[<->] (0,2)--(0,-2);
            \draw[<->] (2,0)--(-2,0);
            \draw[thick, domain=-1:1]plot(\x,{-1-(\x)^2});
            \draw[dashed, domain=-2:2]plot(\x,{\x});
            \node[above] at (0,2) {$x_{k+1}$};
            \node[right] at (2,0) {$x_{k}$};
        \end{tikzpicture}
        \vspace{4px}
        \caption{$r=-1$}
        \label{fig:fixedpoint-graph1}
    \end{subfigure}
    \hfill
    \begin{subfigure}[b]{0.275\textwidth}
        \centering
        \begin{tikzpicture}
            \draw[<->] (0,2)--(0,-2);
            \draw[<->] (2,0)--(-2,0);
            \draw[thick, domain=-1.323:1.323]plot(\x,{-1/4-(\x)^2});
            \draw[dashed, domain=-2:2]plot(\x,{\x});
            \node[above] at (0,2) {$x_{k+1}$};
            \node[right] at (2,0) {$x_{k}$};
        \end{tikzpicture}
        \vspace{4px}
        \caption{$r=-1/4$}
        \label{fig:fixedpoint-graph2}
    \end{subfigure}
    \hfill
    \begin{subfigure}[b]{0.275\textwidth}
        \centering
        \begin{tikzpicture}
            \draw[<->] (0,2)--(0,-2);
            \draw[< ->] (2,0)--(-2,0);
            \draw[thick, domain=-1.581:1.581]plot(\x,{1/2-(\x)^2});
            \draw[dashed, domain=-2:2]plot(\x,{\x});
            \node[above] at (0,2) {$x_{k+1}$};
            \node[right] at (2,0) {$x_{k}$};
        \end{tikzpicture}
        \vspace{4px}
        \caption{$r=1/2$}
        \label{fig:fixedpoint-graph3}
    \end{subfigure}
    \vspace{2px}
    \caption{Graphs of the quadratic function $x_{k+1} = r-x_k^2$ for values of $r$ before, at, and after a saddle-node bifurcation, with fixed points at the intersection of the function with the line $x_{k+1} = x_k$}
    \label{fig:fixedpoint-graphs}
\end{figure*}

\subsection{Bifurcations}
\label{bifurcations}

One of the interesting properties of some dynamical systems is a dependence on parameter. Namely, say a map $\mathbf{x}_{k+1} = \mathbf{f}(\mathbf{x}_k;\,r)$ is dependent on some parameter $r$, where we use the semicolon to separate changing variables from fixed parameters. For most values of $r$, varying $r$ slightly will cause the system's dynamics to change quantitatively, such as a fixed point changing its location slightly. Sometimes, however, the qualitative geometry of a system can change as $r$ is varied. Specifically, there can be an appearance or disappearance of stable orbits, a change in stability, or the emergence of complex geometric structures like strange attractors \cite{crawford}.\footnote{See Section \ref{strangeattractors} for details on strange attractors.} Bifurcations are defined as these qualitative changes in a system's dynamics, and bifurcation points are the values of the system's parameters at which bifurcations occur \cite[p. 45]{strogatz}. Bifurcation theory is a very rich and complex field,\footnote{For the interested reader, Crawford \cite{crawford} provides a far more comprehensive introduction to bifurcation theory than we give in this paper.} and there are many different types of bifurcations, but we will focus on two simple ones in this section: saddle-node bifurcations and period-doubling bifurcations.

\subsubsection{Saddle-Node Bifurcations}

As an example of a system that exhibits bifurcations, let us explore the one-dimensional $r$-dependent quadratic map with iteration function 
\begin{equation}
    f(x;\,r) = r-x^2
    \label{eq:quadratic-map}
\end{equation}
To analyze the qualitative changes in the dynamics of this system, let us first examine the fixed points of this map and their stability. By Equation \ref{eq:stationary}, the fixed points of this map satisfy
\begin{equation}
    x_s = f(x_s;\,r) = r-x_s^2
\end{equation}
Rearranging and using the quadratic formula, we get that
\begin{equation}
    x_s(r) = \frac{-1\pm\sqrt{1+4r}}{2}
    \label{eq:fixedpoint-quadratic-eq}
\end{equation}
where we write $x_s$ as a function of $r$ to emphasize its $r$-dependence. By analysis of the discriminant, we can see that there are no fixed points for $r<-1/4$, one fixed point for $r=-1/4$, and two fixed points for $r>-1/4$. Here, as we continuously increase the parameter $r$, two fixed points appear at the bifurcation point $r=-1/4$ in a region where there were no fixed points before. This type of bifurcation is known as a saddle-node bifurcation, named after the appearance of this bifurcation in higher-dimensional systems \cite[p. 48]{strogatz}. 

We can see how these fixed points appear graphically by putting $x_k$ on the horizontal axis and $x_{k+1}$ on the vertical axis. Then, the intersections between the parabola $x_{k+1} = f(x_k;\,r)$ and the line $x_{k+1} = x_k$ are where the fixed points are located. This is shown in Figure \ref{fig:fixedpoint-graphs}, where $x_{k+1} = f(x_k;\,r)$ is plotted for three values of $r$. In Figure \ref{fig:fixedpoint-graph1}, $x_{k+1} = f(x_k;\,r)$ doesn't intersect $x_{k+1} = x_k$ anywhere, so there are no fixed points for $r=-1<-1/4$. In Figure \ref{fig:fixedpoint-graph2}, $x_{k+1} = f(x_k;\,r)$ intersects $x_{k+1} = x_k$ at one point, the point of tangency, and we can see from Equation \ref{eq:fixedpoint-quadratic-eq} that this fixed point is at
\begin{equation}
    x_s\left(-\frac{1}{4}\right) = \frac{-1\pm\sqrt{1+4\left(-\frac{1}{4}\right)}}{2} = -\frac{1}{2}
    \label{eq:quadratic-saddlenode-bifurcationpoint}
\end{equation}
Finally, in Figure \ref{fig:fixedpoint-graph3}, $x_{k+1} = f(x_k;\,r)$ intersects $x_{k+1} = x_k$ at two points, so from Equation \ref{eq:fixedpoint-quadratic-eq}, the two fixed points are located at
\begin{equation}
    \begin{split}
        x_{s,\,1}\left(\frac{1}{2}\right) &= \frac{-1+\sqrt{1+4\left(\frac{1}{2}\right)}}{2} = \frac{-1+\sqrt{3}}{2}\\
        x_{s,\,2}\left(\frac{1}{2}\right) &= \frac{-1-\sqrt{1+4\left(\frac{1}{2}\right)}}{2} = \frac{-1-\sqrt{3}}{2}
    \end{split}
\end{equation}
where we define the functions
\begin{align}
    x_{s,\,1}(r) &= \frac{-1+\sqrt{1+4r}}{2} \label{eq:quadratic-fixed-point-function1} \\
    x_{s,\,2}(r) &= \frac{-1-\sqrt{1+4r}}{2} \label{eq:quadratic-fixed-point-function2}
\end{align}
as the locations of the two fixed points that exist for $r>-1/4$.

Bifurcations are also concerned with the stability of these fixed points, namely, whether they are attractors or repellers. Let us first consider $x_{s,\,2}(r)$. By the criteria for the attractiveness of fixed points in one dimension established in Section \ref{nonchaoticattractors} and Appendix \ref{fixedpointattractor-criteria}, the attractiveness of $x_{s,\,2}(r)$ is dependent on $|f'(x_{s,\,2};\,r)|$. Since $f'(x;\,r) = -2x$,
\begin{equation}
    \begin{split}
        |f'(x_{s,\,2};\,r)| &= \left|-2\left(\frac{-1-\sqrt{1+4r}}{2}\right)\right| \\
        &= 1+\sqrt{1+4r}
    \end{split}
    \label{eq:quadratic-fixedpoint2}
\end{equation}
This is obviously greater than 1 for all $r>-1/4$, so on this interval, $x_{s,\,2}$ is a repeller. 

Now, let us consider $x_{s,\,1}$. By the criteria for the attractiveness of fixed points in one dimension, $x_{s,\,1}$ is an attractor when $|f'(x_{s,\,1};\,r)|<1$ and a repeller when $|f'(x_{s,\,1};\,r)|>1$. Let us first consider for what $r$ values $x_{s,\,1}$ is an attractor:
\begin{equation}
    \begin{gathered}
        |f'(x_{s,\,1};\,r)|<1 \\
        -1 < f'(x_{s,\,1};\,r) < 1 \\
        -1 < -2\left(\frac{-1+\sqrt{1+4r}}{2}\right) < 1 \\
        -\frac{1}{4} < r < \frac{3}{4} \\
    \end{gathered}
    \label{eq:quadratic-fixedpoint1-attractor}
\end{equation}
Similarly, the values of $r$ where $x_{s,\,1}$ is a repeller are
\begin{equation}
    \begin{gathered}
        |f'(x_{s,\,1};\,r)|>1 \\
        \begin{aligned}
            f'(x_{s,\,1};\,r) &< -1 \hspace{20px}& f'(x_{s,\,1};\,r) &> 1 \\
            1-\sqrt{1+4r} &< -1 & 1-\sqrt{1+4r} &> 1 \\
            \sqrt{1+4r} &> 2 & \sqrt{1+4r} &< 0 \\
            r &> \frac{3}{4} 
        \end{aligned} 
    \end{gathered}
    \label{eq:quadratic-fixedpoint1-repeller}
\end{equation}
In summary, $x_{s,\,2}$ is a repeller for $r > -1/4$, while $x_{s,\,1}$ starts as an attractor for $-1/4 < r < 3/4$ then turns into a repeller for $r > 3/4$. 

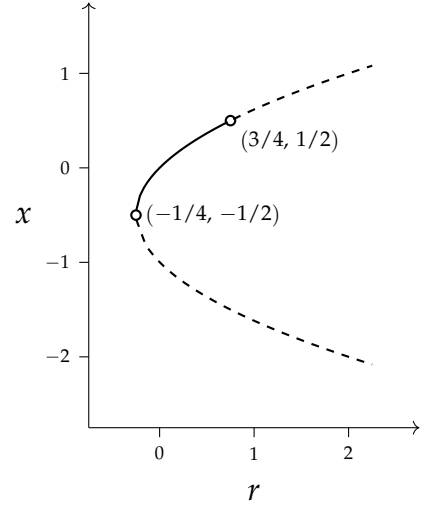
\begin{figure}
    \centering
    \begin{tikzpicture}[scale=1.25]
        \draw[->] (-0.75,-2.75)--(2.75,-2.75);
        \draw[->] (-0.75,-2.75)--(-0.75,1.75);
        \draw[thick, domain=-0.248:0.72]plot(\x,{(-1+sqrt(1+4*\x))/2});
        \draw[thick, dashed, domain=0.78:2.25]plot(\x,{(-1+sqrt(1+4*\x))/2});
        \draw[thick, dashed, domain=-0.248:2.25]plot(\x,{(-1-sqrt(1+4*\x))/2});
        \draw[thick] (-1/4,-1/2) circle [radius=0.05];
        \draw[thick] (3/4,1/2) circle [radius=0.05];
        \node[right] at (-1/4,-1/2) {\footnotesize $\left(-1/4,\, -1/2\right)$};
        \node[below right] at (3/4,1/2) {\footnotesize $\left(3/4,\, 1/2\right)$};
        \node[below] at (1,-3.25) {\large $r$};
        \node[left] at (-1.25,-0.5) {\large $x$};
        \draw (0, -2.75)--(0, -2.85); 
        \draw (1, -2.75)--(1, -2.85); 
        \draw (2, -2.75)--(2, -2.85); 
        \node[below] at (0, -2.85) {\scriptsize $0$};
        \node[below] at (1, -2.85) {\scriptsize $1$};
        \node[below] at (2, -2.85) {\scriptsize $2$};
        \draw (-0.75, -2)--(-0.85, -2);
        \draw (-0.75, -1)--(-0.85, -1);
        \draw (-0.75, 0)--(-0.85, 0);
        \draw (-0.75, 1)--(-0.85, 1);
        \node[left] at (-0.85, -2) {\scriptsize $-2$};
        \node[left] at (-0.85, -1) {\scriptsize $-1$};
        \node[left] at (-0.85, 0) {\scriptsize $0$};
        \node[left] at (-0.85, 1) {\scriptsize $1$};
    \end{tikzpicture}
    \vspace{2px}
    \caption{Bifurcation diagram showing the fixed point attractors (solid curve) and repellers (dashed curves) for the quadratic map $f(x;\,r)=r-x^2$, with a saddle-node bifurcation at $(-1/4,\,-1/2)$}
    \label{fig:quadratic-bifurcation-diagram1}
\end{figure}

In Figure \ref{fig:quadratic-bifurcation-diagram1}, we plot a bifurcation diagram of this system, which shows the locations of the attractors and repellers of the system as a function of $r$. Specifically, the solid curves represent fixed point attractors, while the dashed curves represent fixed point repellers. We make the bifurcation diagram by graphing Equation \ref{eq:quadratic-fixed-point-function1} (the top half of the parabola) and Equation \ref{eq:quadratic-fixed-point-function2} (the bottom half of the parabola). We know the saddle-node bifurcation is at $(-1/4,\,-1/2)$ by Equation \ref{eq:quadratic-saddlenode-bifurcationpoint}, and we know where the fixed points are attracting and repelling by Equations \ref{eq:quadratic-fixedpoint2}, \ref{eq:quadratic-fixedpoint1-attractor}, and \ref{eq:quadratic-fixedpoint1-repeller}. 

\begin{figure*}
    \centering
    \begin{subfigure}[b]{0.275\textwidth}
        \centering
        \begin{tikzpicture}
            \draw[<->] (0,2)--(0,-2);
            \draw[<->] (2,0)--(-2,0);
            \draw[thick, domain=-1.443:1.443, samples=100]plot(\x,{-(\x)^4+2*(1/2)*(\x)^2-(1/2)^2+1/2});
            \draw[dashed, domain=-2:2]plot(\x,{\x});
            \node[above] at (0,2) {$x_{k+2}$};
            \node[right] at (2,0) {$x_{k}$};
        \end{tikzpicture}
        \vspace{4px}
        \caption{$r=1/2$}
        \label{fig:periodicpoint-graph1}
    \end{subfigure}
    \hfill
    \begin{subfigure}[b]{0.275\textwidth}
        \centering
        \begin{tikzpicture}
            \draw[<->] (0,2)--(0,-2);
            \draw[<->] (2,0)--(-2,0);
            \draw[thick, domain=-1.552:1.552, samples=100]plot(\x,{-(\x)^4+2*(3/4)*(\x)^2-(3/4)^2+3/4});
            \draw[dashed, domain=-2:2]plot(\x,{\x});
            \node[above] at (0,2) {$x_{k+2}$};
            \node[right] at (2,0) {$x_{k}$};
        \end{tikzpicture}
        \vspace{4px}
        \caption{$r=3/4$}
        \label{fig:periodicpoint-graph2}
    \end{subfigure}
    \hfill
    \begin{subfigure}[b]{0.275\textwidth}
        \centering
        \begin{tikzpicture}
            \draw[<->] (0,2)--(0,-2);
            \draw[< ->] (2,0)--(-2,0);
            \draw[thick, domain=-1.653:1.653, samples=100]plot(\x,{-(\x)^4+2*(1)*(\x)^2-(1)^2+1});
            \draw[dashed, domain=-2:2]plot(\x,{\x});
            \node[above] at (0,2) {$x_{k+2}$};
            \node[right] at (2,0) {$x_{k}$};
        \end{tikzpicture}
        \vspace{4px}
        \caption{$r=1$}
        \label{fig:periodicpoint-graph3}
    \end{subfigure}
    \vspace{2px}
    \caption{Graphs of the second iterate of the quadratic map $x_{k+1} = r-x_k^2$, namely, the quartic function $x_{k+2} = -x_k^4+2rx_k^2-r^2+r$, for values of $r$ before, at, and after a period-doubling bifurcation, with fixed points and periodic points of period 2 at the intersection of the function with the line $x_{k+2} = x_k$}
    \label{fig:periodicpoint-graphs}
\end{figure*}
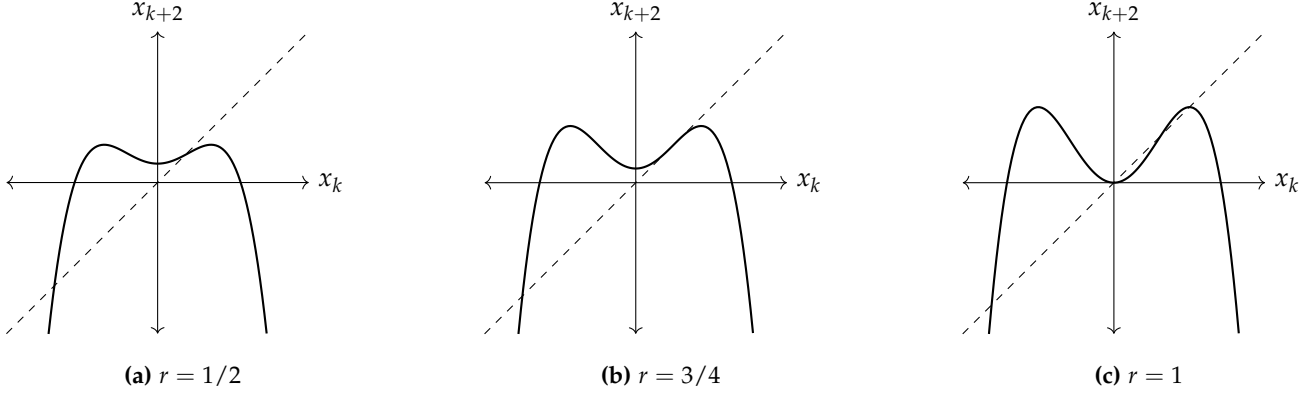

\subsubsection{Period-Doubling Bifurcations}

We can see from Figure \ref{fig:quadratic-bifurcation-diagram1} that $x_{s,\,1}$ changes from an attractor to a repeller at $r=3/4$ since $f'(x_{s,\,1}(3/4);\,3/4) = -1$. This is clearly a qualitative change in the system's dynamics, so it must be another kind of bifurcation. To see what this bifurcation is, we need to explore the attractiveness of 2-cycles in this system. To do this, we consider the mapping function that iterates $f(x;\,r)$ twice:
\begin{equation}
    \begin{split}
        f^2(x;\,r) &= r-\left(r-x^2\right)^2 \\
        &= -x^4+2rx^2-r^2+r
    \end{split}
    \label{eq:iterate-quadratic-twice}
\end{equation}
By Equation \ref{eq:period}, periodic points of 2-cycles will follow
\begin{equation}
    x_p = f^2(x_p;\,r) = -x_p^4+2rx_p^2-r^2+r
\end{equation}
Numerically, we can determine that the four solutions to this quartic equation are
\begin{align}
    x_{p,\,{1}}(r) &= \frac{1+\sqrt{-3+4r}}{2} \label{eq:quadratic-periodicpoint-function1} \\
    x_{p,\,{2}}(r) &= \frac{1-\sqrt{-3+4r}}{2} \label{eq:quadratic-periodicpoint-function2} \\
    x_{p,\,{3}}(r) &= \frac{-1+\sqrt{1+4r}}{2} \label{eq:quadratic-periodicpoint-function3} \\
    x_{p,\,{4}}(r) &= \frac{-1-\sqrt{1+4r}}{2} \label{eq:quadratic-periodicpoint-function4}
\end{align}
where $x_{p,\,1}(r)$ and $x_{p,\,2}(r)$ are two periodic points in a 2-cycle and $x_{p,\,3}(r)$ and $x_{p,\,4}(r)$ are stationary points of $f(x;\,r)$ since, by Equations \ref{eq:quadratic-fixed-point-function1} and \ref{eq:quadratic-fixed-point-function2}, $x_{p,\,{3}}(r) = x_{s,\,{1}}(r)$ and $x_{p,\,{4}}(r) = x_{s,\,{2}}(r)$. This makes sense because a stationary point of $f(x;\,r)$ is also a stationary point of $f^2(x;\,r)$. 

These four periodic points can be seen graphically in a very similar manner to Figure \ref{fig:fixedpoint-graphs}: by graphing $x_{k+2} = f^2(x_k;\,r)$ and looking at where it intersects with $x_{k+2}=x_k$, the line of all periodic points of period 2. We show this in Figure \ref{fig:periodicpoint-graphs} by plotting the second iterate of the quadratic map for three values of $r$. In Figure \ref{fig:periodicpoint-graph1}, we plot before the second bifurcation shown in Figure \ref{fig:quadratic-bifurcation-diagram1}. The two points of intersection are the fixed points $x_{p,\,{3}}(1/2)$ and $x_{p,\,{4}}(1/2)$, which match up graphically with $x_{s,\,{1}}(1/2)$ and $x_{s,\,{2}}(1/2)$ in Figure \ref{fig:fixedpoint-graph3}. In Figure \ref{fig:periodicpoint-graph2}, we plot at the bifurcation, and we can see that there is a qualitative change in $f^2(x_k;\,r)$ at the new point of tangency. Finally, after the bifurcation (Figure \ref{fig:periodicpoint-graph3}), we can see there are now four points of intersection, two of which are $x_{p,\,{1}}(1)$ and $x_{p,\,{2}}(1)$.

Using Equations \ref{eq:quadratic-periodicpoint-function1} and \ref{eq:quadratic-periodicpoint-function2}, we can determine that these points $x_{p,\,{1}}(1)$ and $x_{p,\,{2}}(1)$ are
\begin{equation}
    \begin{split}
        x_{p,\,1}(1) &= \frac{1+\sqrt{-3+4(1)}}{2} = 1 \\
        x_{p,\,2}(1) &= \frac{1-\sqrt{-3+4(1)}}{2} = 0
    \end{split}
\end{equation}
These points clearly make up a 2-cycle: $f\left(x_{p,\,1}(1);\,1\right) = 1-1^2 = 0 = x_{p,\,2}$ and $f\left(x_{p,\,2}(1);\,1\right) = 1-0^2 = 1 = x_{p,\,1}$. To determine whether or not this 2-cycle is an attractor or a repeller, we have to consider $\left|\left(f^2\right)'\left(x_{p,\,{1}}(1);\,1\right)\right|$ and $\left|\left(f^2\right)'\left(x_{p,\,{2}}(1);\,1\right)\right|$. Taking the derivative of Equation \ref{eq:iterate-quadratic-twice}, we get that
\begin{equation}
    \left(f^2\right)'(x;\,r) = -4x^3+4rx
\end{equation}
Substituting,
\begin{equation}
    \begin{split}
        \left(f^2\right)'\left(x_{p,\,1}(1);\,1\right) &= -4(1)^3+4(1)(1) = 0 \\
        \left(f^2\right)'\left(x_{p,\,2}(1);\,1\right) &= -4(0)^3+4(1)(0) = 0 \\
    \end{split}
\end{equation}
which is also clear graphically in Figure \ref{fig:periodicpoint-graph3}. Then, by the criteria for fixed point attractiveness established in Section \ref{nonchaoticattractors},\footnote{We can use the criteria for fixed point attractiveness here because these periodic points are fixed points in $f^2(x;\,r)$.} $x_{p,\,{1}}(1)$ and $x_{p,\,{2}}(1)$ are both attractors in $f^2(x;\,r)$, meaning the 2-cycle $\{x_{p,\,1}(1),\,x_{p,\,2}(1)\}$ is an attractor.

Now, we would like to establish whether the 2-cycles for all values of $r$ are attractors or repellers. To make this easier, we realize that since $\left(f^2\right)'(x;\,r)$ is a polynomial, it changes continuously as $r$ changes. Therefore, instead of going through the inequality work we did for the map's fixed points, we can establish where the derivative's absolute value equals $1$, then check points on either side of these $r$ values for attractiveness. Considering $x_{p,\,1}(r)$ first, we can solve $\left|\left(f^2\right)'\left(x_{p,\,{1}}(r);\,r\right)\right| = 1$ by substituting in Equation \ref{eq:quadratic-periodicpoint-function1} and solving for $r$, which yields $r=3/4,\,5/4$. The same values of $r$ satisfy $\left|\left(f^2\right)'\left(x_{p,\,{2}}(r);\,r\right)\right| = 1$, which makes sense because a 2-cycle won't have one attracting fixed point and one repelling fixed point. Because the 2-cycle changes stability at $r=3/4$ and $r=5/4$, we want to consider their attractiveness on the intervals $r<3/4$, $3/4<r<5/4$, and $r>5/4$. First, we can see from Equations \ref{eq:quadratic-periodicpoint-function1} and \ref{eq:quadratic-periodicpoint-function2} that there are no 2-cycles when $r<3/4$. We have already done a test for a 2-cycle with an $r$ between $3/4$ and $5/4$, namely $r=1$, which we determined was an attractor. This indicates that 2-cycles for $3/4<r<5/4$ are attractors. Finally, since $\left|\left(f^2\right)'\left(x_{p,\,{1,\,2}}(r);\,r\right)\right|$ crosses $1$ at $r=5/4$, 2-cycles for $r>5/4$ must be repellers, which we can easily verify using any test value of $r>5/4$.

\begin{figure}
    \centering
    \begin{tikzpicture}[scale=1.25]
        \draw[->] (-0.75,-2.75)--(2.75,-2.75);
        \draw[->] (-0.75,-2.75)--(-0.75,1.75);
        \draw[thick, domain=-0.248:0.72, samples=100]plot(\x,{(-1+sqrt(1+4*\x))/2});
        \draw[thick, dashed, domain=0.78:2, samples=100]plot(\x,{(-1+sqrt(1+4*\x))/2});
        \draw[thick, dashed, domain=-0.248:2, samples=100]plot(\x,{(-1-sqrt(1+4*\x))/2});
        \draw[thick, domain=0.753:1.2, samples=100]plot(\x,{(1+sqrt(-3+4*\x))/2});
        \draw[thick, domain=0.753:1.2, samples=100]plot(\x,{(1-sqrt(-3+4*\x))/2});
        \draw[thick, dashed, domain=1.3:2.25, samples=100]plot(\x,{(1+sqrt(-3+4*\x))/2});
        \draw[thick, dashed, domain=1.3:2.25, samples=100]plot(\x,{(1-sqrt(-3+4*\x))/2});
        \draw[thick] (-1/4,-1/2) circle [radius=0.05];
        \draw[thick] (3/4,1/2) circle [radius=0.05];
        \draw[thick] (5/4,1.207) circle [radius=0.05];
        \draw[thick] (5/4,-0.207) circle [radius=0.05];
        \node[right] at (-1/4,-1/2) {\scriptsize $\left(-1/4,\, -1/2\right)$};
        \node[right] at (3/4,0.35) {\scriptsize $\left(3/4,\, 1/2\right)$};
        \node[right] at (5/4,1.15) {\scriptsize $(5/4,\, (1+\sqrt{2})/2)$};
        \node[right] at (5/4,-0.15) {\scriptsize $(5/4,\, (1-\sqrt{2})/2)$};
        \node[below] at (1,-3.25) {\large $r$};
        \node[left] at (-1.25,-0.5) {\large $x$};
        \draw (0, -2.75)--(0, -2.85); 
        \draw (1, -2.75)--(1, -2.85); 
        \draw (2, -2.75)--(2, -2.85); 
        \node[below] at (0, -2.85) {\scriptsize $0$};
        \node[below] at (1, -2.85) {\scriptsize $1$};
        \node[below] at (2, -2.85) {\scriptsize $2$};
        \draw (-0.75, -2)--(-0.85, -2);
        \draw (-0.75, -1)--(-0.85, -1);
        \draw (-0.75, 0)--(-0.85, 0);
        \draw (-0.75, 1)--(-0.85, 1);
        \node[left] at (-0.85, -2) {\scriptsize $-2$};
        \node[left] at (-0.85, -1) {\scriptsize $-1$};
        \node[left] at (-0.85, 0) {\scriptsize $0$};
        \node[left] at (-0.85, 1) {\scriptsize $1$};
    \end{tikzpicture}
    \vspace{2px}
    \caption{Bifurcation diagram showing the fixed point and 2-cycle attractors (solid curves) and repellers (dashed curves) for the quadratic map $f(x;\,r)=r-x^2$, with a saddle-node bifurcation at $(-1/4,\,-1/2)$ and period-doubling bifurcations at $(3/4,\, 1/2)$, $(5/4,\, (1+\sqrt{2})/2)$, and $(5/4,\, (1-\sqrt{2})/2)$}
    \label{fig:quadratic-bifurcation-diagram2}
    \vspace{-0.5cm}
\end{figure}
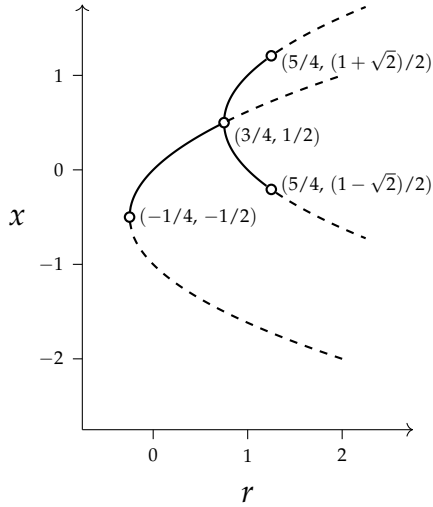
    
Using this information, we can now extend our bifurcation diagram from Figure \ref{fig:quadratic-bifurcation-diagram1} to Figure \ref{fig:quadratic-bifurcation-diagram2}, which includes 2-cycles by also graphing the two functions in Equations \ref{eq:quadratic-periodicpoint-function1} and \ref{eq:quadratic-periodicpoint-function2} and indicating their stability. We can see now why the bifurcation point at $r=3/4$ with $f'(x_{s,\,1}(3/4);\,3/4) = -1$ is known as a period-doubling bifurcation: at $r=3/4$, the attractor changes from a fixed point attractor at $x=1/2$ to a periodic orbit with a point right above and right below $x=1/2$. Namely, the attractor's period doubles from 1 to 2. Similarly, at $r=5/4$, we have $\left(f^2\right)'\left(x_{p,\,{2}}(r);\,r\right)=-1$, which implies that the bifurcation of the 2-cycle from stable to unstable at $r=5/4$ is also a period-doubling bifurcation (from 2 to 4). We could prove this by continuing the algebra for $f^4(x;\,r)$, but instead, we will demonstrate a method to find these bifurcations numerically using a different, well-studied map. 

\subsubsection{The Logistic Map}

The logistic map is another simple quadratic map defined by the following iteration function \cite{may}:
\begin{equation}
    f(x;\,r) = rx(1-x)
\end{equation}
It is probably the most famous example of complex dynamics emerging from a simple model. Despite its mathematical simplicity, rather than analytically analyzing the dynamics of the map as we did for the previous quadratic map, we will do it numerically. 

\begin{figure*}
    \centering
    \includegraphics[scale=0.225]{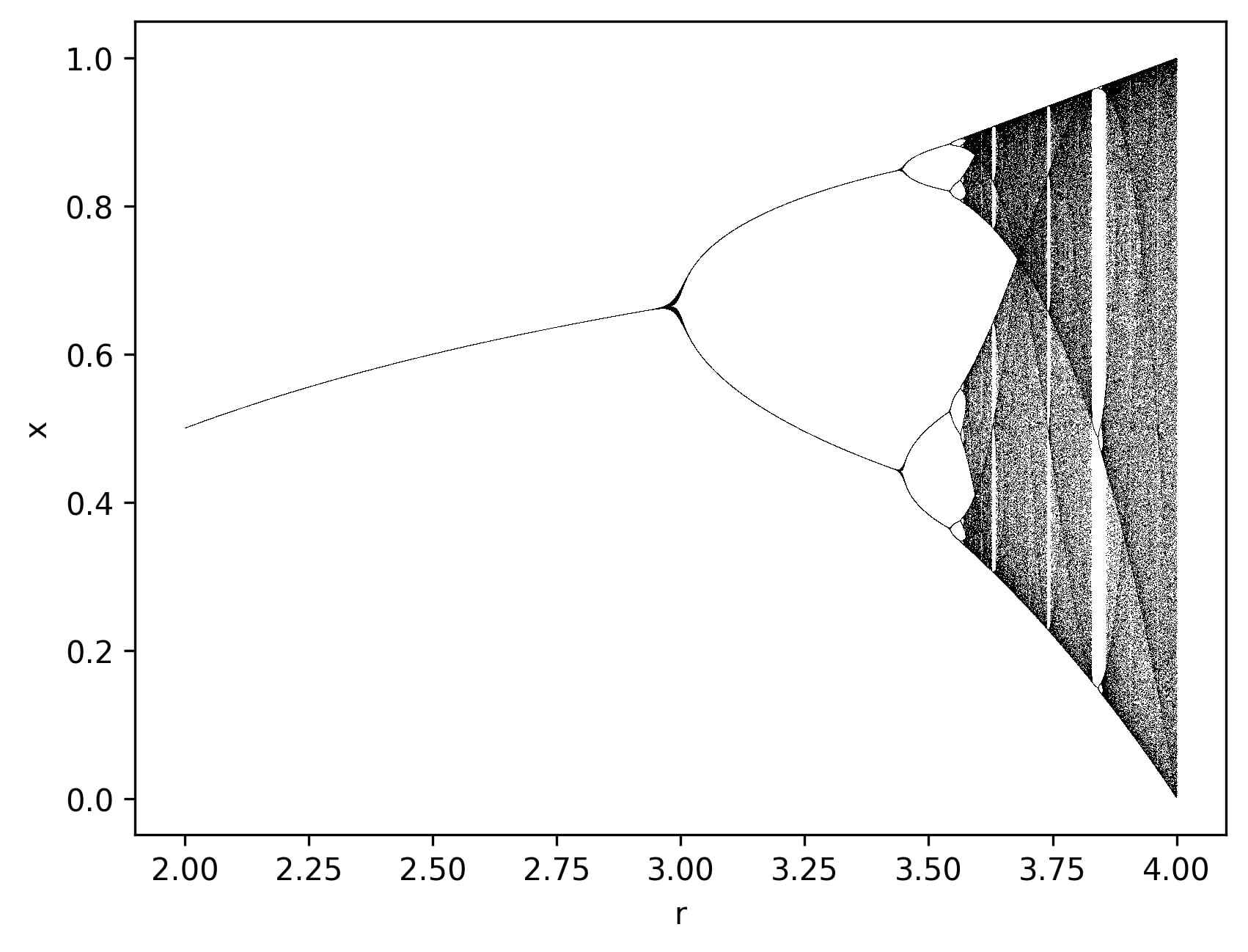}
    \caption{Bifurcation diagram of the logistic map for $2\leq r\leq 4$, graphed with the code in Appendix \ref{bifurcation_diagram_logistic_code}}
    \label{fig:logistic_bifurcation}
\end{figure*}

To numerically plot a bifurcation diagram, we will use the method described by Strogatz \cite[p. 363]{strogatz} that plots the attractors of a system for many different values of $r$ and displays any saddle-node or period-doubling bifurcations.\footnote{These diagrams are more accurately called orbit diagrams \cite[p. 368]{strogatz} because a true bifurcation diagram plots both the attractors and repellers of a system. However, since repellers are not important for this discussion, we will also refer to these diagrams that plot only attractors as bifurcation diagrams.} The method goes as follows: for a given value of $r$, start from an initial state $x_0$ and iterate some number of transients, say $100$ times, which allows the system to settle down to its attractor. Then, generate an orbit starting from $x_{100}$ for some number of iterations, say $100$ times again. This leaves us with the orbit $\{x_{100},\, x_{101},\,\hdots\,,\,x_{200}\}$. For our purposes, this orbit contains enough points in the attractor $A$ for a given value of $r$. Calculating these attractor orbits for many values of $r$, we get our bifurcation diagram by plotting these orbits on a graph of $x$ vs. $r$. The Python code in Appendix \ref{bifurcation_diagram_logistic_code} accomplishes this for $2\leq r\leq 4$, and the resulting graph is shown in Figure \ref{fig:logistic_bifurcation}.

In this bifurcation diagram, we can see that the system starts with a fixed point attractor, but at around $r=3$, there is a period-doubling bifurcation, and the attractor turns into a 2-cycle. Near $r=3.45$, there is another period-doubling bifurcation to a 4-cycle, then another to an 8-cycle, and so on, the successive period-doubling bifurcations becoming faster and faster.\footnote{The factor by which the distance between successive bifurcations shrinks is a universal constant for systems approaching chaos by period-doubling known as the Feigenbaum constant \cite[p. 523]{layek}. An interested reader is recommended to see the original paper by Feigenbaum \cite{feigenbaum} for details.}

\begin{figure*}
    \centering
    \begin{subfigure}[b]{0.45\textwidth}
        \centering
        \includegraphics[scale = 0.125]{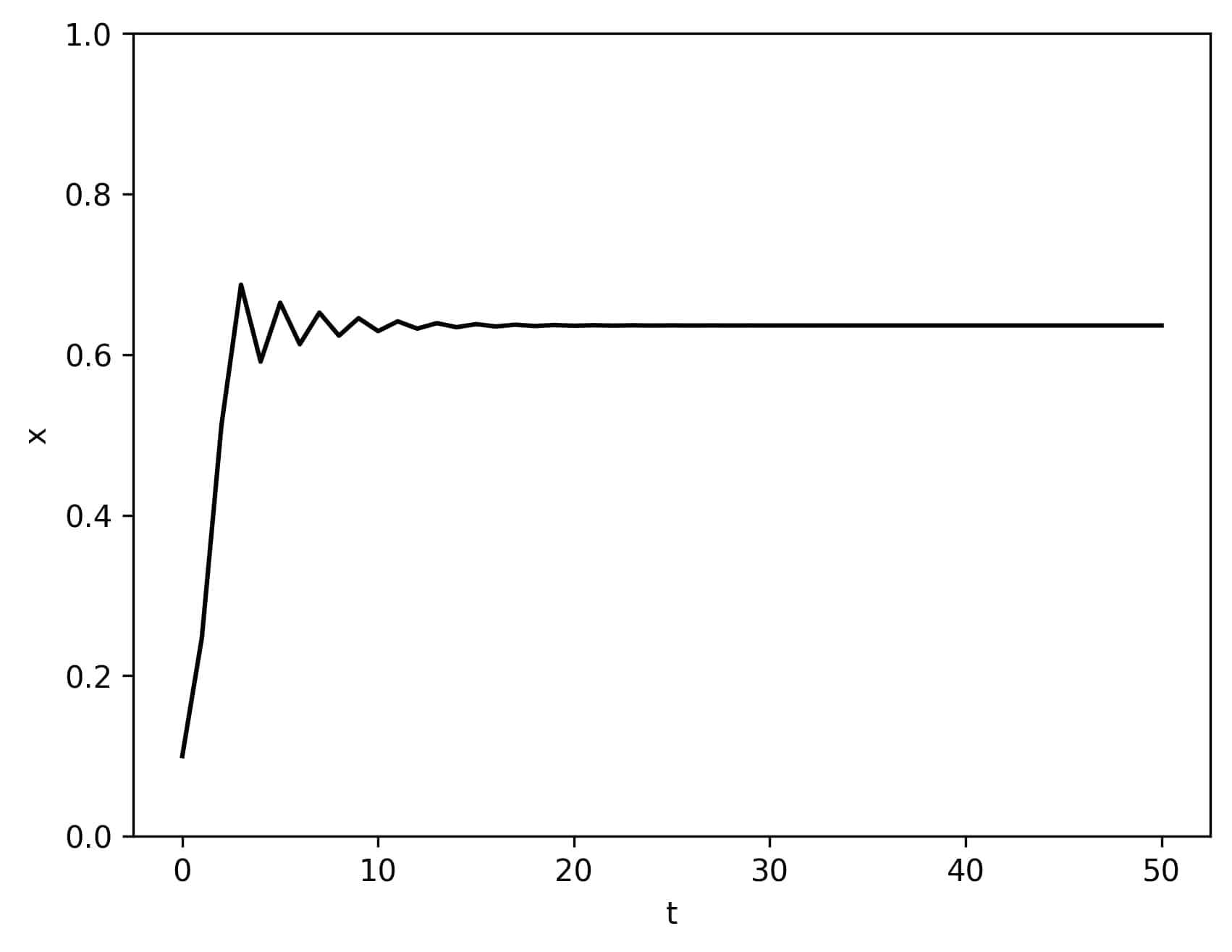}
        \vspace{4px}
        \caption{$r=2.75$}
        \label{fig:logistic_2.75}
    \end{subfigure}
    \hfill
    \begin{subfigure}[b]{0.45\textwidth}
        \centering
        \includegraphics[scale = 0.125]{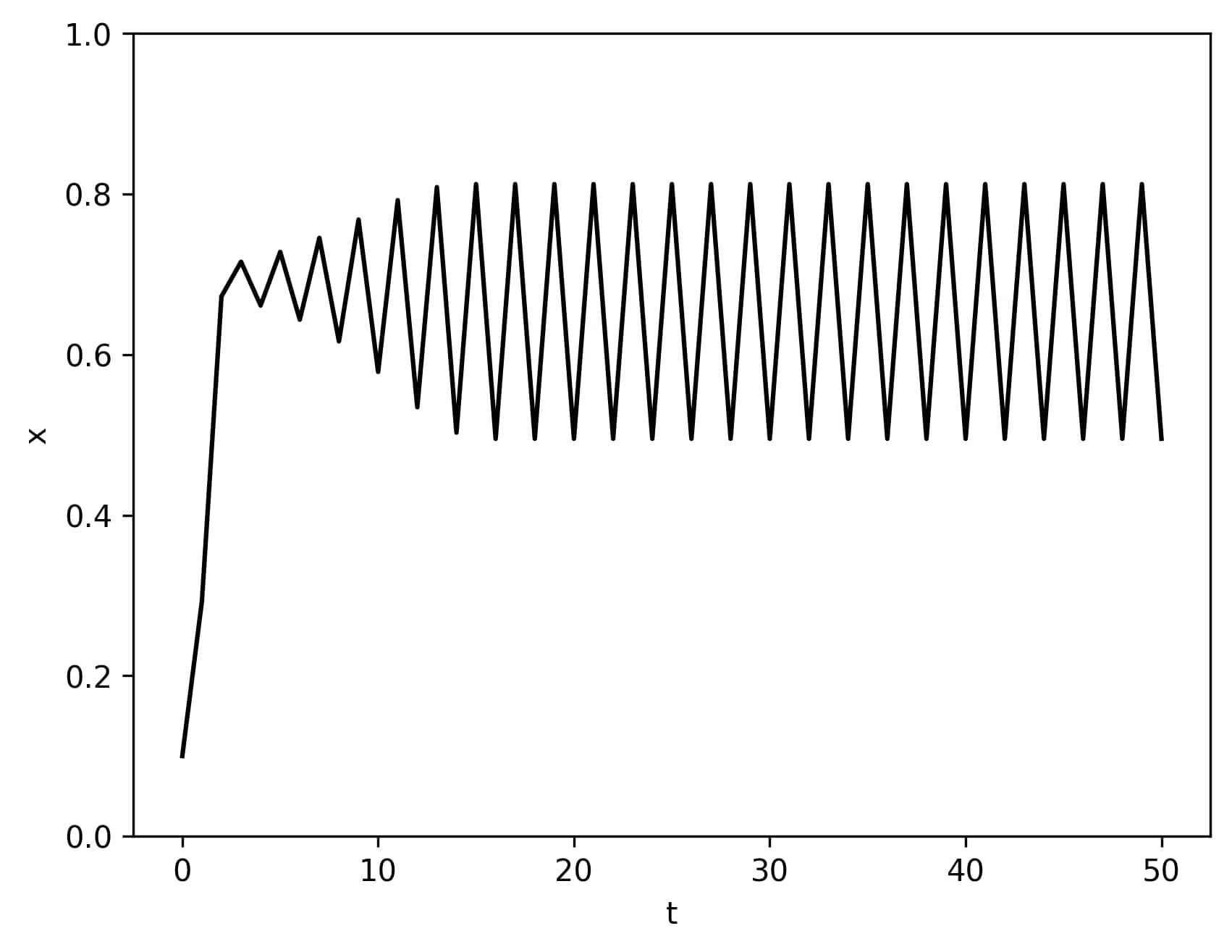}
        \vspace{4px}
        \caption{$r=3.25$}
        \label{fig:logistic_3.25}
    \end{subfigure}
    \vspace{1cm} \\
    \begin{subfigure}[b]{0.45\textwidth}
        \centering
        \includegraphics[scale = 0.125]{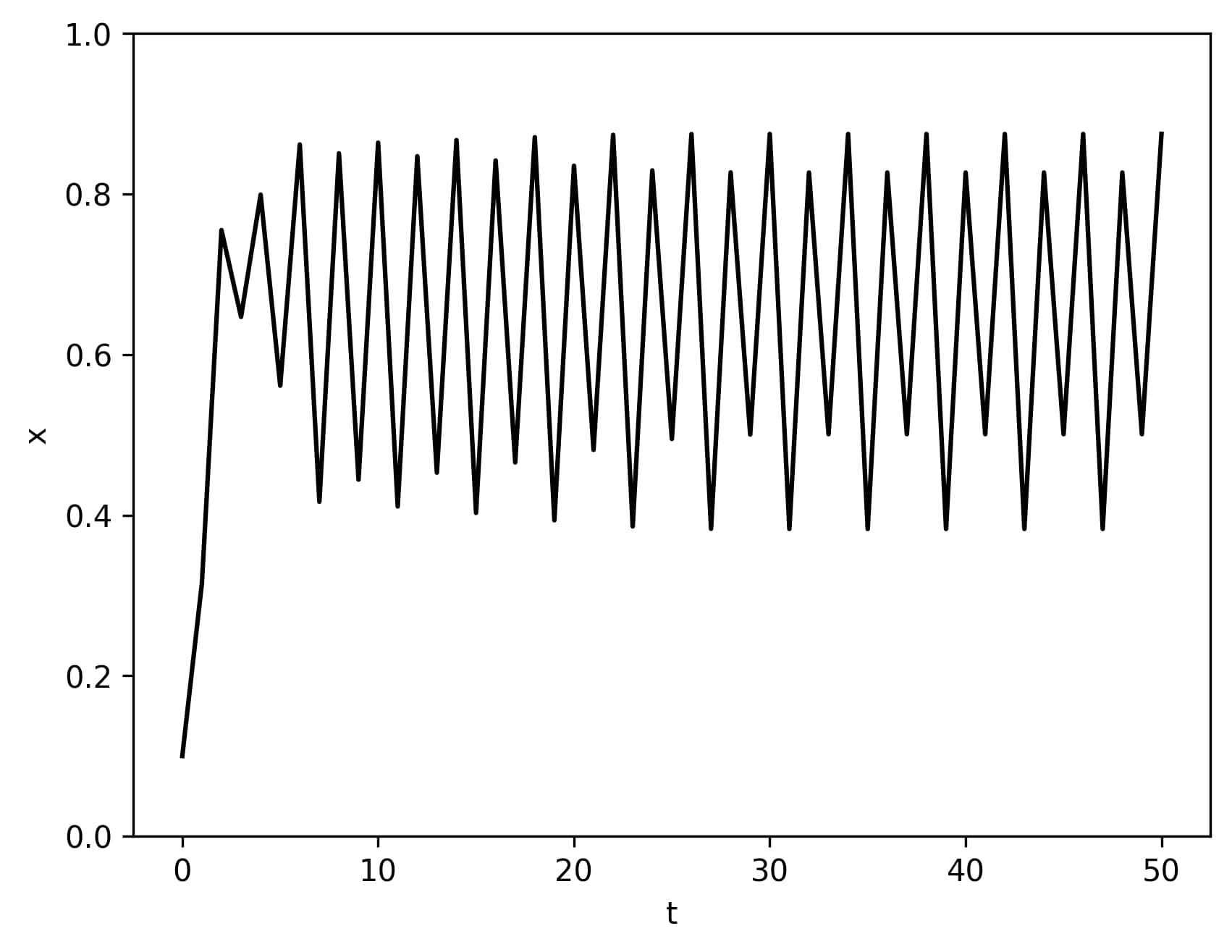}
        \vspace{4px}
        \caption{$r=3.5$}
        \label{fig:logistic_3.5}
    \end{subfigure}
    \hfill
    \begin{subfigure}[b]{0.45\textwidth}
        \centering
        \includegraphics[scale = 0.125]{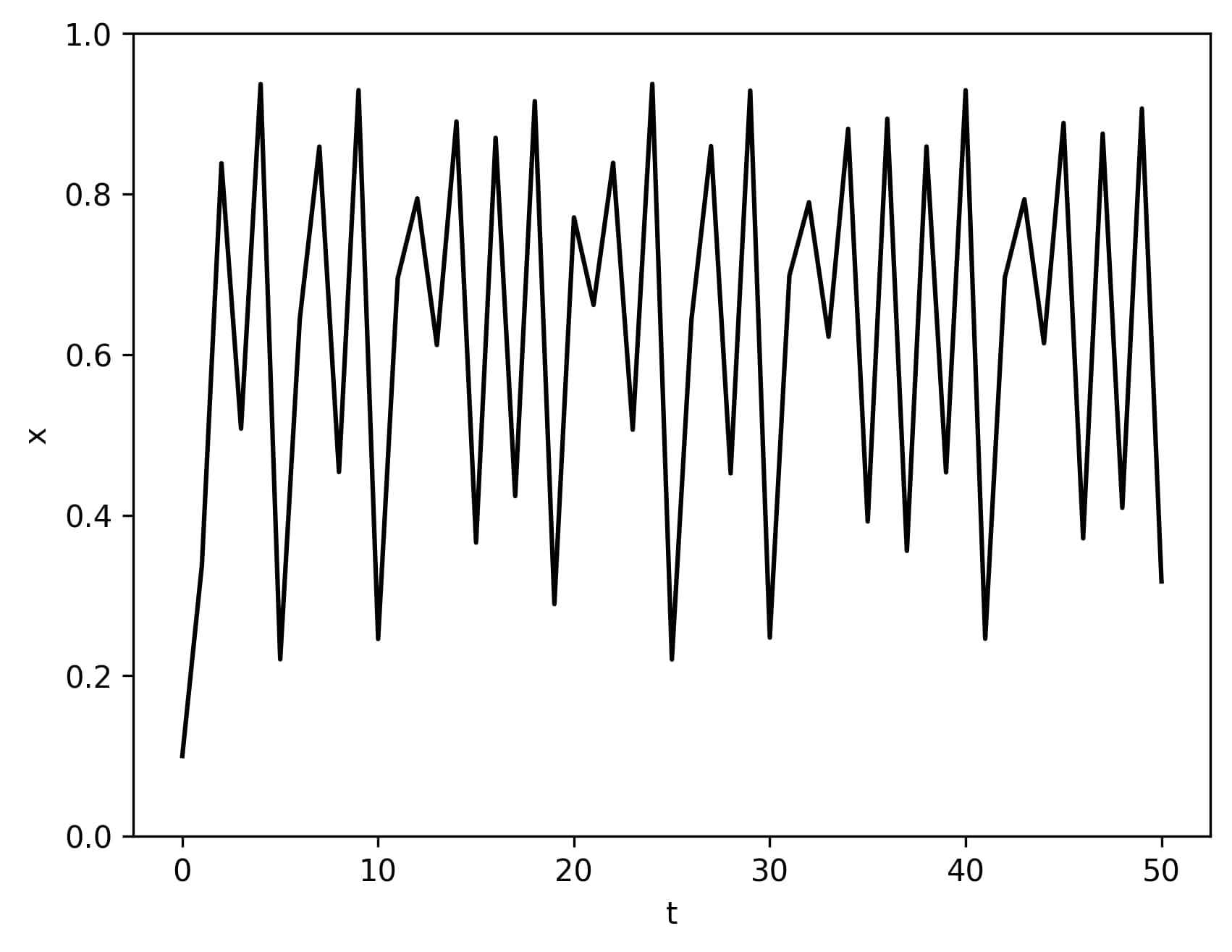}
        \vspace{4px}
        \caption{$r=3.75$}
        \label{fig:logistic_3.75}
    \end{subfigure}
    \vspace{0.5cm}
    \caption{Orbits of the logistic map for $0\leq t\leq 50$, graphed with the code in Appendix \ref{orbits_logistic_code}}
    \label{fig:logistic_orbits}
\end{figure*}

\begin{figure*}
    \centering
    \includegraphics[scale=0.225]{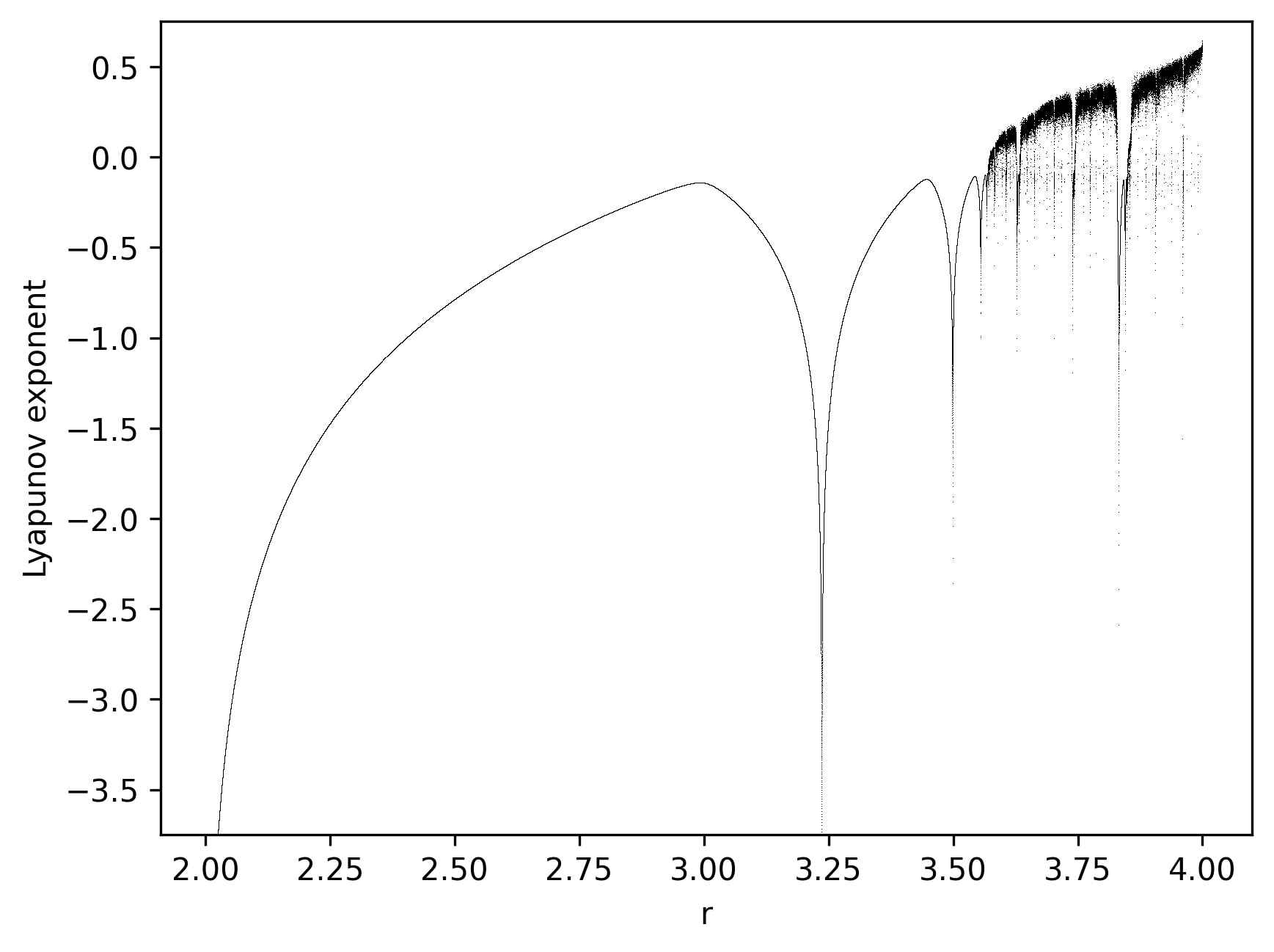}
    \caption{Lyapunov exponents of the logistic map for $2\leq r\leq 4$, graphed with the code in Appendix \ref{logistic_lyapunov_code}}
    \label{fig:logistic_lyapunov}
\end{figure*}

To visualize these different periodic orbit attractors for different values of $r$, we can plot some points of the forward orbit $O^+(x_0)$ against time $t$ at different values of $r$. We use the code in Appendix \ref{orbits_logistic_code} to accomplish this for $x_0=0.1$ and the orbit $\{x_0,\,x_1,\,\hdots\,,\,x_{50}\}$, which is shown in Figure \ref{fig:logistic_orbits}. In the figure, we draw lines between consecutive points $(k,\, x_k)$ and $(k+1,\, x_{k+1})$ in the orbit to make seeing the dynamics easier, but it is important to note that only the corners of these curves matter. In Figure \ref{fig:logistic_2.75}, we plot at $r=2.75$, which is before the first period-doubling bifurcation. Therefore, the orbit settles down at a fixed point around $x=0.64$, corresponding to the attractor value shown in the bifurcation diagram (Figure \ref{fig:logistic_bifurcation}) at $r=2.75$. After the first period-doubling bifurcation, we plot at $r=3.25$ (Figure \ref{fig:logistic_3.25}), then after the second, we plot at $r=3.5$ (Figure \ref{fig:logistic_3.5}). Comparing the 2-cycle and 4-cycle shown in Figures \ref{fig:logistic_3.25} and \ref{fig:logistic_3.5} to our bifurcation diagram in Figure \ref{fig:logistic_bifurcation}, the periodic orbits correspond to their respective attractor sets.

As we can see in the bifurcation diagram in Figure \ref{fig:logistic_bifurcation}, at around $r=3.57$, the successive bifurcations reach their limit, and the map approaches an attractor with essentially an infinite number of points. In other words, a forward orbit $O^+(x_0)$ for most values of $r>3.57$ will never settle down to a fixed point or periodic orbit; it will exhibit the aperiodic long-term behavior characteristic of chaotic systems. We can definitively show for which $r$ values the logistic map is chaotic by examining its sensitivity to initial conditions. Namely, we can calculate the Lyapunov exponent of the map for different values of $r$, which we accomplish using the code in Appendix \ref{logistic_lyapunov_code} and show a graph of in Figure \ref{fig:logistic_lyapunov}. In this graph, we can see that the Lyapunov exponents stay less than or equal to 0 for all $r<3.57$. We should expect this because orbits approach a fixed point or periodic orbit attractor for these parameter values, meaning the system is not chaotic. However, when $r>3.57$, the Lyapunov exponents remain mainly positive, indicating that the system is chaotic. This matches with the chaotic-looking bifurcation diagram above it. Interestingly, there are some values of $r$ in the field of chaos above $r=3.57$ where the Lyapunov exponent is less than 0, most prominently where $r\approx3.83$. Matching this with the bifurcation diagram, we can see that when $r\approx3.83$, the logistic map briefly has a stable 3-cycle before dissolving into chaos again. These intervals of stability between regions of chaos displayed in Figure \ref{fig:logistic_bifurcation} are known as periodic windows \cite[p. 363]{strogatz} or islands of stability \cite{brandon}.

Another interesting observation when comparing Figures \ref{fig:logistic_bifurcation} and \ref{fig:logistic_lyapunov} is that the period-doubling bifurcations always seem to occur when $\lambda = 0$. We can show this is the case by recalling from earlier in this section that a period-doubling bifurcation occurs when 
\begin{equation}
    \left(f^q\right)'(x_{p};\,r) = -1
    \label{eq:period-doubling-bifurcation-requirement}
\end{equation}
where $x_{p}$ is a periodic point in an attracting $q$-cycle. Then, from Equation \ref{eq:1dlyap}\footnote{We manipulate Equation \ref{eq:1dlyap} by changing the chain rule step in Appendix \ref{lyap1d-deriv} (Equation \ref{eq:chain_rule_step_1dlyap_deriv}) so that the derivatives are split by $q$ iterations rather than 1. In order for this to work, $t$ must be chosen so that $(t-1)/q\in\mathbb{N}$.} and choosing $x_0=x_p$,\footnote{If we didn't make this choice, the result would still be the same because $x_{qt}\to x_{p}$ for large $t$, and since $t\to\infty$, the factor $1/t$ will take care of all the non-zero terms. However, making this choice makes the math cleaner.}
\begin{equation}
    \begin{split}
        \lambda &= \lim_{t\to\infty}\frac{1}{t}\sum_{i=0}^{(t-1)/q}\ln\left|(f^q)'(x_{qi};\,r)\right| \\
        &= \lim_{t\to\infty}\frac{1}{t}\sum_{i=0}^{(t-1)/q}\ln\left|(f^q)'(x_{p};\,r)\right| \\
        &= \lim_{t\to\infty}\frac{1}{t}\sum_{i=0}^{(t-1)/q}\ln1 \\
        &= 0
    \end{split}
\end{equation}
We have now demonstrated a few connections between bifurcations and Lyapunov exponents using the logistic map example, which will be useful in our later analysis.

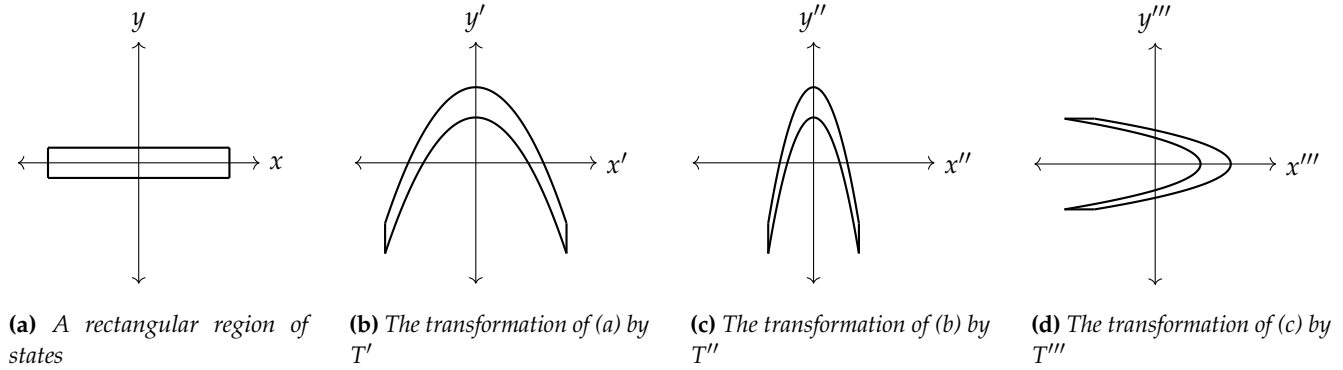
\begin{figure*}
    \centering
    \begin{subfigure}[b]{0.225\textwidth}
        \centering
        \begin{tikzpicture}[scale=0.8]
            \draw[<->] (0,2)--(0,-2);
            \draw[<->] (2,0)--(-2,0);
            \node[above] at (0,2) {$y$};
            \node[right] at (2,0) {$x$};
            \draw[thick, domain=-1.5:1.5, samples=100]plot(\x,0.25);
            \draw[thick, domain=-1.5:1.5, samples=100]plot(\x,-0.25);
            \draw[thick, domain=-0.25:0.25, variable=\y]  plot ({-1.5}, {\y});
            \draw[thick, domain=-0.25:0.25, variable=\y]  plot ({1.5}, {\y});
        \end{tikzpicture}
        \vspace{4px}
        \caption{A rectangular region of states}
        \label{fig:henon-transformation-0}
    \end{subfigure}
    \hfill
    \begin{subfigure}[b]{0.225\textwidth}
        \centering
        \begin{tikzpicture}[scale=0.8]
            \draw[<->] (0,2)--(0,-2);
            \draw[<->] (2,0)--(-2,0);
            \node[above] at (0,2) {$y'$};
            \node[right] at (2,0) {$x'$};
            \draw[thick, domain=-1.5:1.5, samples=100]plot(\x,{0.25+1-\x*\x});
            \draw[thick, domain=-1.5:1.5, samples=100]plot(\x,{-0.25+1-\x*\x});
            \draw[thick, domain=-0.25:0.25, variable=\y]  plot ({-1.5}, {\y+1-2.25});
            \draw[thick, domain=-0.25:0.25, variable=\y]  plot ({1.5}, {\y+1-2.25});
        \end{tikzpicture}
        \vspace{4px}
        \caption{The transformation of (a) by $T'$}
        \label{fig:henon-transformation-1}
    \end{subfigure}
    \hfill
    \begin{subfigure}[b]{0.225\textwidth}
        \centering
        \begin{tikzpicture}[scale=0.8]
            \draw[<->] (0,2)--(0,-2);
            \draw[<->] (2,0)--(-2,0);
            \node[above] at (0,2) {$y''$};
            \node[right] at (2,0) {$x''$};
            \draw[thick, domain=-1.5:1.5, samples=100]plot(0.5*\x,{0.25+1-\x*\x});
            \draw[thick, domain=-1.5:1.5, samples=100]plot(0.5*\x,{-0.25+1-\x*\x});
            \draw[thick, domain=-0.25:0.25, variable=\y]  plot ({0.5*-1.5}, {\y+1-2.25});
            \draw[thick, domain=-0.25:0.25, variable=\y]  plot ({0.5*1.5}, {\y+1-2.25});
        \end{tikzpicture}
        \vspace{4px}
        \caption{The transformation of (b) by $T''$}
        \label{fig:henon-transformation-2}
    \end{subfigure}
    \hfill
    \begin{subfigure}[b]{0.225\textwidth}
        \centering
        \begin{tikzpicture}[scale=0.8]
            \draw[<->] (0,2)--(0,-2);
            \draw[<->] (2,0)--(-2,0);
            \node[above] at (0,2) {$y'''$};
            \node[right] at (2,0) {$x'''$};
            \draw[thick, domain=-1.5:1.5, samples=100]plot({0.25+1-\x*\x},0.5*\x);
            \draw[thick, domain=-1.5:1.5, samples=100]plot({-0.25+1-\x*\x},0.5*\x);
            \draw[thick, domain=-0.25:0.25, variable=\y]  plot ({\y+1-2.25}, {0.5*-1.5});
            \draw[thick, domain=-0.25:0.25, variable=\y]  plot ({\y+1-2.25}, {0.5*1.5});
        \end{tikzpicture}
        \vspace{-9px}
        \caption{The transformation of (c) by $T'''$}
        \label{fig:henon-transformation-3}
    \end{subfigure}
    \vspace{2px}
    \caption{A chain of transformations that results in the Hénon map}
    \label{fig:henon-transformations}
\end{figure*}

\subsection{Strange Attractors and Fractal Geometry}
\label{strangeattractors}

So far, our discussion of attractors has been limited to fixed points and periodic orbits, which are not chaotic. In this section, however, we will discuss attractors that do exhibit chaotic dynamics. According to Ruelle \cite{ruelle}, who coined the term, an attractor set $A$ is called strange if there are points in the attractor $\mathbf{x},\,\mathbf{a}\in A$ such that the distance between $\mathbf{f}^t(\mathbf{x})$ and $\mathbf{f}^t(\mathbf{a})$ grows exponentially with $t$ until it is bounded by the size of the attractor. In other words, strange attractors are chaotic, exhibiting sensitive dependence on initial conditions.\footnote{More generally, the term strange attractor can be used to describe any attractor with a fractal structure and not necessarily chaotic ones \cite{grebogi}, but for the purposes of this paper, we will say that a strange attractor is one that is both chaotic and fractal.}

At first, the idea of an attractor being chaotic can seem counterintuitive. How can nearby trajectories get exponentially farther apart from each other while staying bounded in the attractor? Mathematically, this can be answered by looking at the Lyapunov exponents of the dynamics on a strange attractor. We show in Appendix \ref{fixedpointattractor-criteria} that all the Lyapunov exponents of a fixed point attractor are negative because all nearby trajectories are attracted towards it. However, strange attractors have Lyapunov exponents that are both positive and negative; for example, a three-dimensional strange attractor has a Lyapunov spectrum of signs $\{+,\,0,\,-\}$ \cite{wolf}. This is because a strange attractor is globally stable, but locally unstable, exhibiting both chaotic and attractive behavior. Infinitesimally close initial states grow further apart exponentially according to the $\lambda_1$ until the edge of the attractor prevents them from getting any further apart.

We can get an even better understanding of the concept of a strange attractor by looking at it geometrically. Strange attractors are generated through a ``stretching and folding process'' \cite{grassberger}. As an intuitive example of how this works, we can imagine the stretching and folding of pastry dough. If we put a drop of food coloring in our pastry dough (representing a set of close initial states), then stretch and fold it a bunch of times, the food coloring will eventually spread throughout the entire thing. A map with a strange attractor will mathematically stretch and fold the attractor as it iterates, so a set of close initial conditions will eventually spread throughout the entire attractor. This gives us a good understanding of the seemingly counterintuitive nature of strange attractors: the stretching and folding property magnifies any initial perturbation in initial conditions, but by the nature of the attractor itself, trajectories stay bounded. 

\begin{figure*}
    \centering
    \includegraphics[scale=0.06]{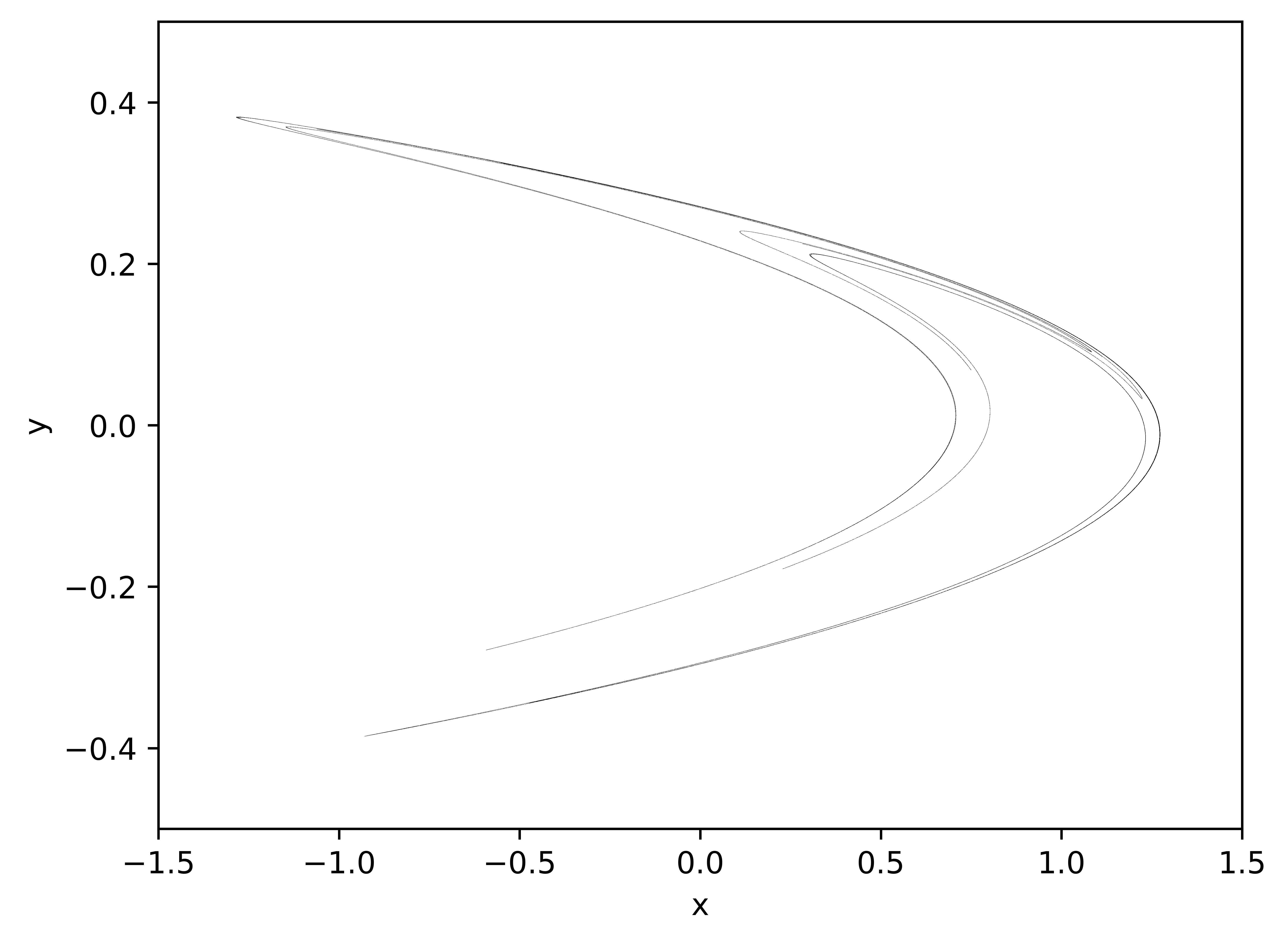}
    \caption{The Hénon attractor, graphed with the code in Appendix \ref{henon_attractor_code}}
    \label{fig:henon-attractor}
\end{figure*}

\subsubsection{The Hénon Map}

To illustrate an example of a strange attractor, we will use a simple, well-studied map discovered by Hénon \cite{henon}, who simulated the mathematical stretching and folding property of strange attractors by through a series of simple transformations in two-dimensional space. First, let us define the state vector $\mathbf{x}$ of the Hénon map as
\begin{equation}
    \mathbf{x} = \begin{pmatrix}
        x\e{1} \\
        x\e{2}
    \end{pmatrix}
    =
    \begin{pmatrix}
        x \\
        y
    \end{pmatrix}
\end{equation}
and two Hénon map parameters $a$ and $b$. Starting with a rectangular region of states (Figure \ref{fig:henon-transformation-0}), the first transformation $T'$ transforms $x$ into $x'$ and $y$ into $y'$ by stretching and folding the rectangle into a parabolic shape (Figure \ref{fig:henon-transformation-1}):
\begin{equation}
    T':\begin{pmatrix}
        x' \\
        y'
    \end{pmatrix}
    =
    \begin{pmatrix}
        x \\
        1+y-a x^2
    \end{pmatrix}
\end{equation}
The next transformation $T''$ completes the folding process by contracting along the $x$-axis (Figure \ref{fig:henon-transformation-2}):
\begin{equation}
    T'':\begin{pmatrix}
        x'' \\
        y''
    \end{pmatrix}
    =
    \begin{pmatrix}
        b x' \\
        y'
    \end{pmatrix}
\end{equation}
Finally, we complete the process with the transformation $T'''$ that reflects across the line $y=x$, which brings us back to our initial orientation with a stretched and folded set of states (Figure \ref{fig:henon-transformation-3}): 
\begin{equation}
    T''':\begin{pmatrix}
        x''' \\
        y'''
    \end{pmatrix}
    =
    \begin{pmatrix}
        y'' \\
        x''
    \end{pmatrix}
\end{equation}
The composite transformation $T=T'''\,T''\,T'$ is the Hénon mapping:
\begin{equation}
    T:\begin{pmatrix}
        x''' \\
        y'''
    \end{pmatrix}
    =
    \begin{pmatrix}
        1 + y - a x^2 \\
        b x
    \end{pmatrix}
\end{equation}
Relabelling to our standard notation $\mathbf{x}_{k+1} = \mathbf{f}(\mathbf{x}_k)$,
\begin{equation}
    \begin{pmatrix}
        x_{k+1} \\
        y_{k+1}
    \end{pmatrix}
    =
    \begin{pmatrix}
        1 + y_k - a x_k^2 \\
        b x_k
    \end{pmatrix}
    \label{eq:henon-mapping}
\end{equation}

The Hénon attractor can be visualized computationally by using Equation \ref{eq:henon-mapping} to generate an orbit $O(\mathbf{x}_0)$. We accomplish this using the code in Appendix \ref{henon_attractor_code}, which generates the orbit $O(\mathbf{x}_0) = \{ \mathbf{x}_{0},\, \mathbf{x}_{1},\, \hdots \,,\, \mathbf{x}_{250010} \}$ using $a = 1.4$, $b = 0.3$,\footnote{The choice of values of parameters $a$ and $b$ is the standard choice to ensure the proper amount of stretching and folding in order to generate the strange attractor. See the paper by Hénon \cite{henon} for more insight into how these numbers were chosen.} and $\mathbf{x}_0 = \langle 0,\,0 \rangle$, then plots the orbit $\{ \mathbf{x}_{10},\, \mathbf{x}_{11},\, \hdots \,,\, \mathbf{x}_{250010}\}$. The result is shown in Figure \ref{fig:henon-attractor}.

To examine the strange chaotic nature of the Hénon attractor, specifically its global stability but local instability, let us examine its Lyapunov exponents. Recall from Section \ref{quantification} that the Lyapunov spectrum can be calculated using the eigenvalues of the matrix $J^{t\intercal}J^t$. From Equation \ref{eq:henon-mapping}, we know that the Hénon map is defined by the function
\begin{equation}
    \mathbf{f}(\mathbf{x};\,a,\,b) = 
    \begin{pmatrix}
        f^{[1]}(x,\,y;\,a,\,b) \\
        f^{[2]}(x,\,y;\,a,\,b)
    \end{pmatrix}
    =
    \begin{pmatrix}
        1+y-a x^2 \\
        b x
    \end{pmatrix}
\end{equation}
Therefore, by Equation \ref{eq:jacobian}, the Jacobian matrix $J(\mathbf{x})$ of the Hénon map is
\begin{equation}
    J(\mathbf{x}) = 
    \begin{pmatrix}
        \frac{\partial f^{[1]}}{\partial x} & \frac{\partial f^{[1]}}{\partial y} \\[4px]
        \frac{\partial f^{[2]}}{\partial x} & \frac{\partial f^{[2]}}{\partial y}
    \end{pmatrix}
    =
    \begin{pmatrix}
        -2a x & 1 \\
        b & 0
    \end{pmatrix}
\end{equation}
which depends on $x$. Because we are finding the Lyapunov spectrum on the strange attractor, this $x$ value will keep changing and jumping around the attractor, so finding the matrix $J^{t\intercal}J^t$ and its eigenvalues analytically is impossible. Even numerically, approximating the spectrum using a large value of $t$ will require computationally expensive matrix multiplication, and the matrix product will likely overflow before reaching a sufficiently high value of $t$ for an accurate approximation. For this reason, we will use the QR factorization method for Lyapunov spectrum calculation that we derive in Appendix \ref{qr-meth-lyap-spec-calc}. Specifically, in the code in Appendix \ref{lyap_spec_henon_code}, we implement the algorithm described in Appendix \ref{qr-meth-lyap-spec-calc} and approximate the spectrum using a large value of $t$ in Equation \ref{eq:qr-comp-friendly}:
\begin{equation}
    \lambda_i = \lim_{t\to\infty}\frac{1}{t}\sum_{j=1}^t \ln \left|r^{(j)}_{ii}\right|
\end{equation}
This yields that
\begin{equation}
    \lambda = \{\lambda_1,\,\lambda_2\} \approx \{0.419,\, -1.623\}
    \label{eq:henon_lyap_spec}
\end{equation}
As we can see, the maximal Lyapunov exponent is positive, indicating the chaotic behavior on the strange attractor and its local instability. However, $\lambda_2$ is negative, indicating its global stability.\footnote{The values we calculate in Appendix \ref{lyap_spec_henon_code} and display in Equation \ref{eq:henon_lyap_spec} agree with the values calculated in the thesis by Asbroek \cite{asbroek}, which we recommend an interested reader to see for a more in-depth analysis.}

\begin{figure*}
    \centering
    \hfill
    \begin{subfigure}[b]{0.45\textwidth}
        \centering
        \includegraphics[scale = 0.035]{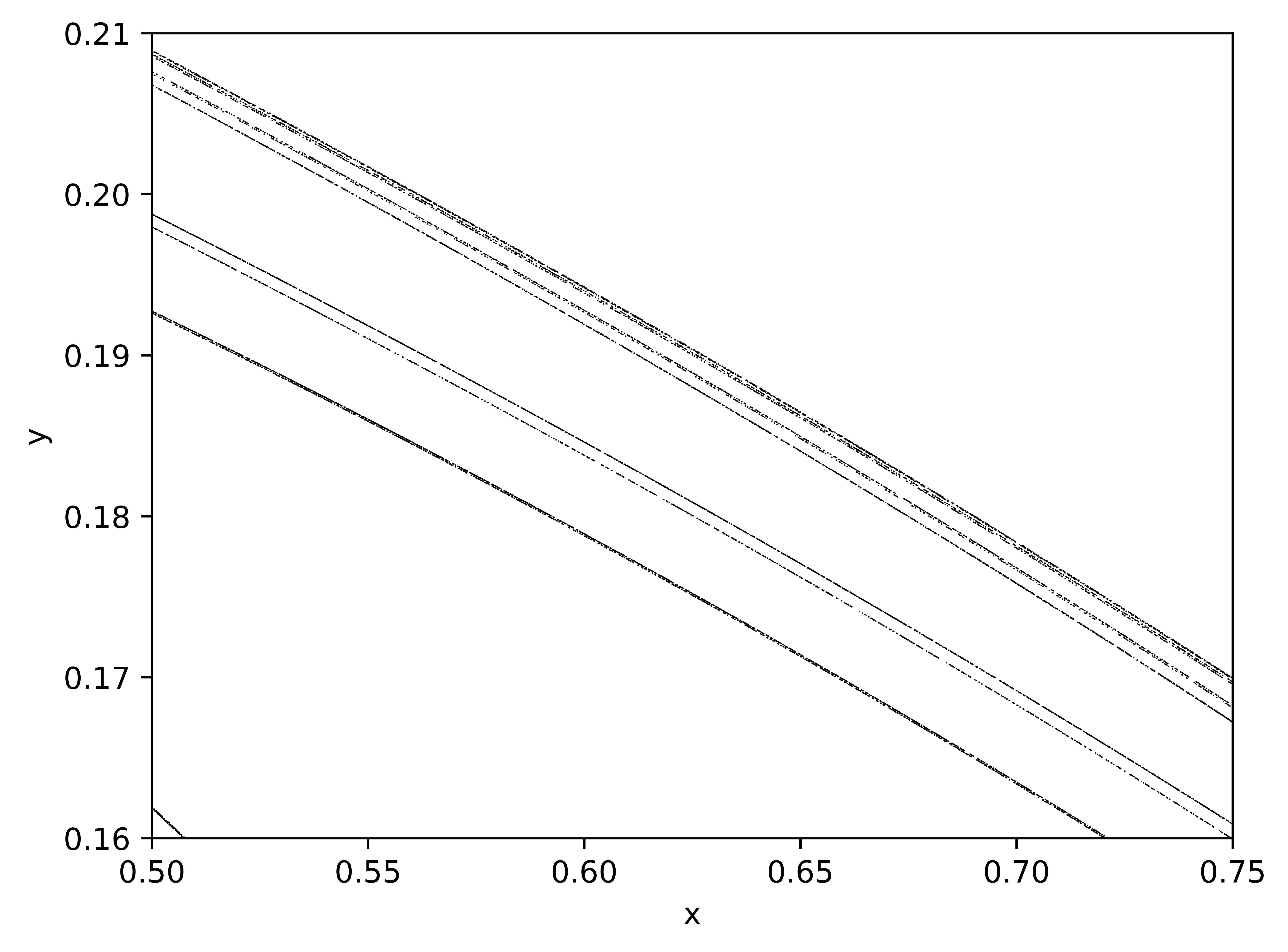}
        \vspace{4px}
        \caption{$0.5\leq x\leq 0.75$, $0.16\leq y\leq 0.21$}
        \label{fig:henon-zoom-1}
    \end{subfigure}
    \hfill
    \begin{subfigure}[b]{0.45\textwidth}
        \centering
        \includegraphics[scale = 0.035]{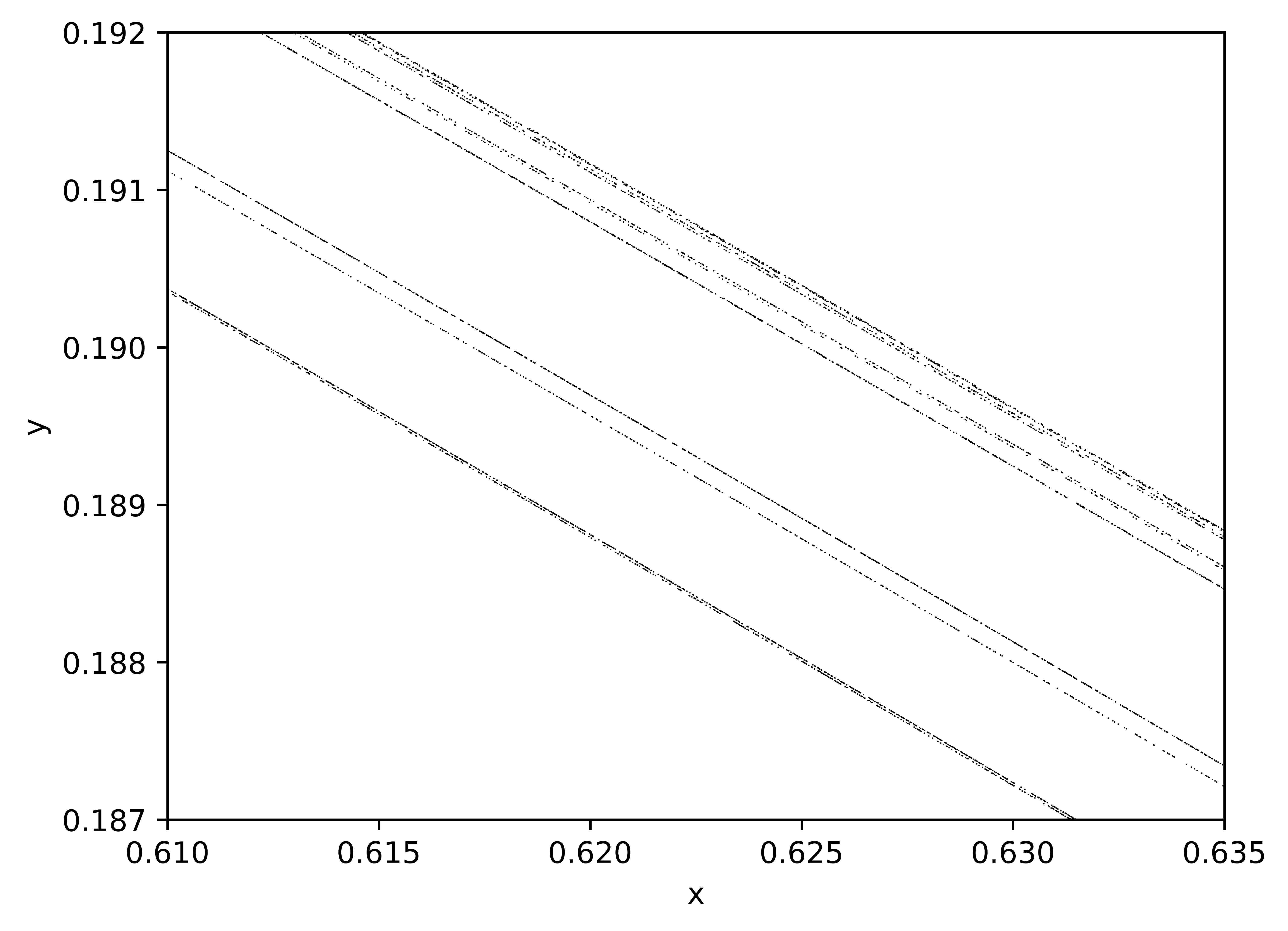}
        \vspace{4px}
        \caption{$0.61\leq x\leq 0.635$, $0.187\leq y\leq 0.192$}
        \label{fig:henon-zoom-2}
    \end{subfigure}
    \hfill
    \vspace{2px}    
    \caption{Zooming in twice on the Hénon attractor, graphed by changing the $x$-limits and $y$-limits of the code in Appendix \ref{henon_attractor_code}}
    \label{fig:henon-zooms}
\end{figure*}

\subsubsection{Fractal Dimension and Measure}

Now that we have quantified the chaotic dynamics on the Hénon attractor, we are ready to explore its geometry. To begin, let us zoom in on the Hénon attractor in Figure \ref{fig:henon-attractor} to examine its fine structure, or detail at a small scale. By altering the $x$ and $y$-limits of the code in Appendix \ref{henon_attractor_code} to $0.5\leq x\leq 0.75$ and $0.16\leq y\leq 0.21$, we get the graph in Figure \ref{fig:henon-zoom-1}. As we can see, there appear to be six parallel curves: three bunched together at the top, two together below that, and one lone curve below that. However, if we zoom in again on the three curves at the top, which we show in Figure \ref{fig:henon-zoom-2}, we can see that it is actually composed of six curves grouped exactly as it was before: three, two, and one. If we were to zoom in more, we would see the same thing over and over no matter how far we zoom in \cite{henon}. This property of the Hénon attractor's fine structure and detail not vanishing, but rather, remaining essentially unchanged as we zoom in to arbitrarily small scales is characteristic of geometric objects known as fractals \cite{mandelbrot1}. In a way, fractals are a rebellion against calculus, which generally assumes that things eventually appear smooth if we look closely enough.

Fractal geometry is important for us to study because it is a defining feature of strange attractors and, more generally, multiple types of sensitivity to initial conditions. To understand why this is the case, recall that a strange attractor is generated through a process of stretching and folding. Once again using the Hénon map as an example, let us look at Figure \ref{fig:henon-transformations} again. We see that starting with a rectangle of initial conditions, iterating the map once gives us two layers. Doing it again would give us four, then eight, and so on. Taking this to the limit, we would have infinitely many infinitely thin layers, all separated by gaps of different sizes. This is essentially what the Hénon attractor is, and it is why strange attractors are fractals.

The geometry of fractal attractors can be complex and difficult to describe because of its departure from calculus and our standard mathematical tools. One way that we can quantitatively characterize fractal geometry is through the idea of a fractal dimension, which generalizes our standard idea of a dimension in a clever way. 

In order to motivate the concept of a fractal dimension, we must first introduce the concept of measure. Measure is a way of generalizing the concepts of length in one dimension, area in two dimensions, and volume in three dimensions to any geometric object \cite[p. 2]{tao}. Let us denote the measure of some set $S$ as $\mu(S)$. If $S$ is $n$-dimensional, meaning we can describe any point in it using $n$ numbers, we can denote its measure as $\mu^n(S)$, but we often omit this superscript when the dimension is clear from context. A valid measure $\mu(S)$ has the following properties \cite[p. 250]{alligood}:
\begin{enumerate}
    \item For any set $S$, 
    \begin{equation}
        \mu(S)\geq 0
    \end{equation}
    This makes sense intuitively because measure can't be negative.
    \item For a union set $S_1\cup S_2\cup\hdots = \bigcup_{i=1}^{\infty} S_i$ where no two sets $S_i$ intersect,
    \begin{equation}
        \mu\left(\bigcup_{i=1}^{\infty} S_i\right) = \sum_{i=1}^{\infty}\mu(S_i)
    \end{equation}
    In other words, the measure of a set is equal to the sum of the measures of all the distinct subsets that the set is made of. This property also works for a finite number of sets. Physically, this means that if we divide a three-dimensional object into a bunch of pieces, adding up the volumes of all the individual pieces will give the total volume of the object.
    \item If $\emptyset$ is the empty set, or the set with nothing in it, it is clear that
    \begin{equation}
        \mu(\emptyset) = 0
    \end{equation}
    \item If $S$ is $n$-dimensional and $m>n$, then
    \begin{equation}
        \mu^{m}(S) = 0
        \label{eq:measure-property-4}
    \end{equation}
    In other words, the length of a point, or the area of a line, or the volume of a sheet are all 0.
    \item For $S\subset\mathbb{R}^n$ and $\mathbf{x}\in\mathbb{R}^n$,
    \begin{equation}
        \mu(S+\mathbf{x}) = \mu(S)
    \end{equation}
    In other words, if we move something around in space, its measure doesn't change.
\end{enumerate}

Now that we have defined measure, we can now begin to talk about dimension. First, let us consider the one-dimensional set $S=(0,\,a)$, where $(0,\,a)$ is the open interval from 0 to $a$. This line segment clearly has a length of $a$, so $\mu^1(S) = a$. Now, if we scale up this segment by a factor of 2, we get the set $2S = (0,\,2a)$. Splitting this up, we can see that $2S$ is the union of two segments with measure $a$, namely, the interval from 0 to $a$ and the interval from $a$ to $2a$ (by Property 5 of measures). Then, by Property 2 of measures, $\mu^1(2S) = 2\mu^1(S) = 2a$, which intuitively makes sense because the line segment is two times longer. In other words, scaling a one-dimensional object by 2 scales its measure by 2.

Let us now consider a two-dimensional set $S = (0,\,a)\times(0,\,a)$,\footnote{$S_1\times S_2$ is the Cartesian product of $S_1$ and $S_2$, which is the set of all vectors $\langle a,\,b \rangle$ such that $a\in S_1$ and $b\in S_2$.} which is a square with side length $a$. This square has an area of $a^2$, so $\mu^2(S) = a^2$. Scaling up this square by a factor of 2, we get $2S = (0,\,2a)\times(0,\,2a)$, which is a square of side length $2a$. This square has area $4a^2$, so $\mu^2(2S) = 4\mu^2(S)$, meaning scaling a two-dimensional object by 2 scales its measure by 4. For a three-dimensional set $S = (0,\,a)\times(0,\,a)\times(0,\,a)$, which is a cube with side length $a$ and volume $a^3$, scaling $S$ up by a factor of 2 will result in a cube with side length $2a$ and volume $8a^3$, so $\mu^3(2S) = 8\mu^3(S)$. In other words, scaling a three-dimensional object by 2 scales its measure by 8.

In summary, we have 
\begin{equation}
    \begin{split}
        \mu^1(2S) &= 2\mu^1(S) \\
        \mu^2(2S) &= 4\mu^2(S) \\
        \mu^3(2S) &= 8\mu^3(S)
    \end{split}
\end{equation}
The pattern here is clear: the dimension $n$ of a given object is the power to which a scaling factor is raised to scale its measure. In other words, if we scale an $n$-dimensional object $S$ by some scaling factor $\sigma$,
\begin{equation}
    \mu(\sigma S) = \sigma^n\mu(S)
    \label{eq:scaling-factor-measure-nonfractal}
\end{equation}
Then, if we have some set $S$ that defines a fractal object and it follows the equation
\begin{equation}
    \mu(\sigma S) = \sigma^d\mu(S)
    \label{eq:scaling-factor-measure}
\end{equation}
for any $\sigma$, we can interpret the exponent $d$ as the fractal dimension of $S$.

\begin{figure}
    \centering
    \includegraphics[scale=0.24]{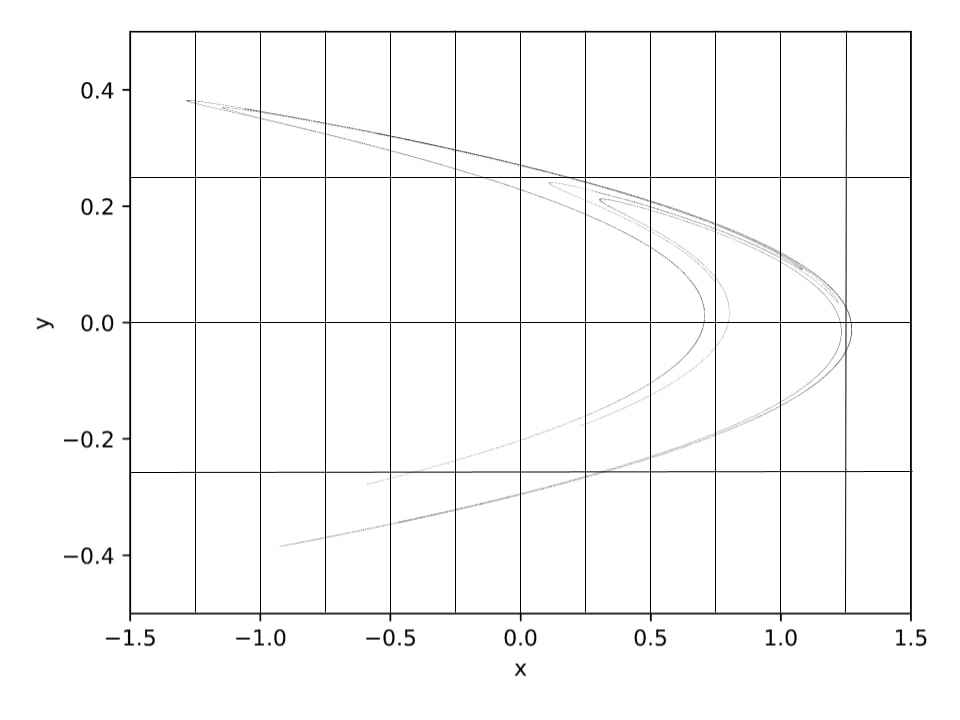}
    \vspace{-8px}
    \caption{Covering the Hénon attractor in Figure \ref{fig:henon-attractor} with two-dimensional boxes of side length $\epsilon=1/4$}
    \label{fig:henon-grid}
\end{figure}

A clever way of practically determining this fractal dimension is through the method of box-counting. The process goes as follows: for a set $S\subset\mathbb{R}^n$, fill state space with $n$-dimensional boxes of side length $\epsilon$.\footnote{A one-dimensional box is an interval of length $\epsilon$, a two-dimensional box is a square of side length $\epsilon$, and so on.} Say that the number of boxes that contain at least one element of $S$ is $N(\epsilon)$, which is approximately proportional to the set's measure $\mu(S)$. In Figure \ref{fig:henon-grid}, we cover the Hénon attractor with boxes of size $\epsilon=1/4$. Out of the 48 boxes shown, the attractor touches 28 of them, so $N(1/4)=28$.

Notice that scaling $S$ by some factor $\sigma$ is equivalent to scaling the box size $\epsilon$ by $1/\sigma$. For example, consider a rectangle in two-dimensional space. If we scale it up by 2, the rectangle will cover 4 times as much area, so $N(\epsilon)$ will scale by 4. However, if we cut $\epsilon$ in half, all the boxes will get 4 times smaller, so the same thing happens. Therefore, we can rewrite Equation \ref{eq:scaling-factor-measure} as
\begin{equation}
    N(\epsilon) \approx \left(\frac{1}{\epsilon^d}\right)c\mu(S)
    \label{eq:box-counting-power-function}
\end{equation}
where $c$ is the constant of proportionality. Taking logs on both sides,\footnote{We use the natural logarithm, but of course, this will work for any convenient base.} we get
\begin{equation}
    \ln N(\epsilon) \approx d\ln\left(\frac{1}{\epsilon}\right) + \ln(c\mu(S))
    \label{eq:box-counting-line-regression}
\end{equation}
Solving for $d$,
\begin{equation}
    d \approx \frac{\ln N(\epsilon) - \ln(c\mu(S))}{\ln(1/\epsilon)}
    \label{eq:box-counting-approximation}
\end{equation}
Now, if we let $\epsilon$ approach 0, this box-counting method will capture all of the fine structure and detail of $S$, so in the limit, $N(\epsilon)$ is exactly proportional to the measure of the scaled set. Because $\ln N(\epsilon)$ and $\ln(1/\epsilon)$ grow without bound, the $\ln(c\mu(S))$ term becomes insignificant in the limit, so we can rewrite Equation \ref{eq:box-counting-approximation} as
\begin{equation}
    d = \lim_{\epsilon\to 0}\frac{\ln N(\epsilon)}{\ln(1/\epsilon)}
    \label{eq:box-counting-dimension-exact}
\end{equation}
This fractal dimension $d$ is known as the box-counting or Minkowski–Bouligand dimension \cite[p. 70]{ott}.\footnote{There are many different other kinds of fractal dimensions that capture the geometry of fractals in different ways, like the information dimension, correlation dimension, Rényi dimension, and Hausdorff dimension, to name a few, but the box-counting dimension is a simple one that works well intuitively and computationally.}

To get a better understanding of box-counting, we will once again examine the Hénon attractor. Obviously, it is impractical to divide all of state space into an infinite number of infinitesimal boxes to determine the dimension of a geometric object, so instead, our strategy is to utilize Equation \ref{eq:box-counting-line-regression}. Namely, if we numerically calculate a bunch of $N(\epsilon)$ values for different box sizes and plot the values on a graph of $N(\epsilon)$ vs. $1/\epsilon$, we should expect these points to approximately fall along a power function described by Equation \ref{eq:box-counting-power-function}. Then, by Equation \ref{eq:box-counting-line-regression}, if we plot these points on a graph of $\ln N(\epsilon)$ vs. $\ln(1/\epsilon)$, we should see the points approximately fall along a line with a slope of $d$. Therefore, if we were to calculate the linear regression of these points $(\ln(1/\epsilon),\,\ln N(\epsilon))$, its slope would be approximately equal to the set's fractal dimension $d$.

\begin{table}[t]
    \centering
    \begin{tabular}{c|c}
        $\epsilon$ & $N(\epsilon)$ \\
        \hline
        1/4 & 28 \\
        1/8 & 63 \\
        1/16 & 148 \\
        1/32 & 349 \\
        1/64 & 832 \\
        1/128 & 1921 \\
        1/256 & 4519 \\
    \end{tabular}
    \caption{Some $N(\epsilon)$ values of the Hénon attractor, calculated using the code in Appendix \ref{henon-box-counting-code}}
    \vspace{0.5cm}
    \label{tab:box_henon_values}
\end{table}

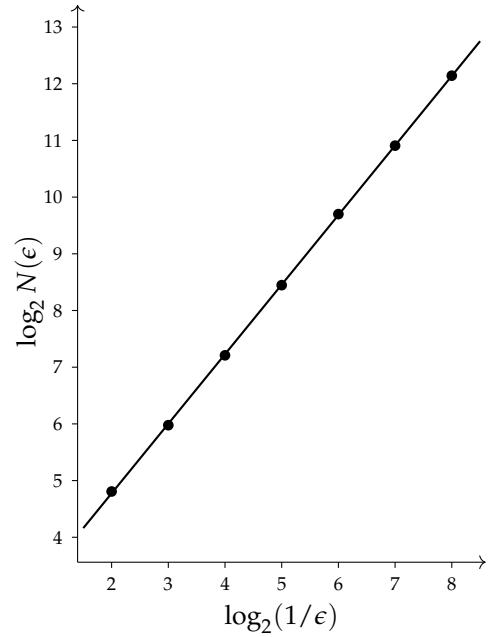
\begin{figure}
    \centering
    \begin{tikzpicture}[scale=0.75]
        \draw[->] (1.4,3.56)--(8.6,3.56);
        \path (0,3.56)--(10,3.56);
        \draw[->] (1.4,3.56)--(1.4,13.39);
        \draw[thick, domain=1.5:8.5]plot(\x,{1.227*\x+2.321});
        \filldraw[thick] (2,4.8074) circle [radius=0.075];
        \filldraw[thick] (3,5.9773) circle [radius=0.075];
        \filldraw[thick] (4,7.2095) circle [radius=0.075];
        \filldraw[thick] (5,8.4471) circle [radius=0.075];
        \filldraw[thick] (6,9.7004) circle [radius=0.075];
        \filldraw[thick] (7,10.908) circle [radius=0.075];
        \filldraw[thick] (8,12.142) circle [radius=0.075];
        \node[below] at (5,3) {$\log_2(1/\epsilon)$};
        \node[left, rotate=90] at (0.55,9.5) {$\log_2 N(\epsilon)$};
        \draw (2, 3.56)--(2, 3.46); 
        \draw (3, 3.56)--(3, 3.46);
        \draw (4, 3.56)--(4, 3.46);
        \draw (5, 3.56)--(5, 3.46); 
        \draw (6, 3.56)--(6, 3.46);
        \draw (7, 3.56)--(7, 3.46);
        \draw (8, 3.56)--(8, 3.46);
        \node[below] at (2, 3.46) {\scriptsize $2$};
        \node[below] at (3, 3.46) {\scriptsize $3$};
        \node[below] at (4, 3.46) {\scriptsize $4$};
        \node[below] at (5, 3.46) {\scriptsize $5$};
        \node[below] at (6, 3.46) {\scriptsize $6$};
        \node[below] at (7, 3.46) {\scriptsize $7$};
        \node[below] at (8, 3.46) {\scriptsize $8$};
        \draw (1.4, 4)--(1.3, 4);
        \draw (1.4, 5)--(1.3, 5);
        \draw (1.4, 6)--(1.3, 6);
        \draw (1.4, 7)--(1.3, 7);
        \draw (1.4, 8)--(1.3, 8);
        \draw (1.4, 9)--(1.3, 9);
        \draw (1.4, 10)--(1.3, 10);
        \draw (1.4, 11)--(1.3, 11);
        \draw (1.4, 12)--(1.3, 12);
        \draw (1.4, 13)--(1.3, 13);
        \node[left] at (1.3, 4) {\scriptsize $4$};
        \node[left] at (1.3, 5) {\scriptsize $5$};
        \node[left] at (1.3, 6) {\scriptsize $6$};
        \node[left] at (1.3, 7) {\scriptsize $7$};
        \node[left] at (1.3, 8) {\scriptsize $8$};
        \node[left] at (1.3, 9) {\scriptsize $9$};
        \node[left] at (1.3, 10) {\scriptsize $10$};
        \node[left] at (1.3, 11) {\scriptsize $11$};
        \node[left] at (1.3, 12) {\scriptsize $12$};
        \node[left] at (1.3, 13) {\scriptsize $13$};
    \end{tikzpicture}
    \caption{A plot of the points in Table \ref{tab:box_henon_values} on a graph of $\log_2 N(\epsilon)$ vs. $\log_2(1/\epsilon)$ and their line of best-fit}
    \label{fig:henon-fractal-dimension-linear-reg}
\end{figure}

We follow this method to calculate the fractal dimension of the Hénon map by first using the code in Appendix \ref{henon-box-counting-code} to generate values of $N(\epsilon)$ for various values of $\epsilon=2^{-k}$ for numerical simplicity. These values are shown in Table \ref{tab:box_henon_values}. Since we have $\epsilon$ values that are integer powers of 2, we take base 2 logs instead of natural logs and plot the points $(\log_2(1/\epsilon),\,\log_2 N(\epsilon))$ in Figure \ref{fig:henon-fractal-dimension-linear-reg}. Then, numerically determining the best-fit line of these points gives us
\begin{equation}
    \log_2 N(\epsilon) = 1.23\log_2\left(\frac{1}{\epsilon}\right) + 2.32
\end{equation}
with an $R^2$ value of $0.9999$. We can therefore conclude that the Hénon attractor is approximately 1.23-dimensional.

Intuitively, this makes sense. Although the Hénon attractor lives in two-dimensional space, it is composed of curves that appear to be one-dimensional (see Figures \ref{fig:henon-attractor} and \ref{fig:henon-zooms}). Therefore, it should scale similarly to a one-dimensional object. However, it has a fine structure and self-similarity that touches more small-$\epsilon$ boxes than a simple curve would, so its dimension $d$ is slightly higher than 1. With this intuition, we are now prepared to give a formal definition of a fractal: a fractal is a geometric object with a non-integer dimension \cite{mandelbrot2}. 

This idea of a non-integer dimension captures our intuitive understanding of the detail and roughness of fractals, objects that don't have length, area, or volume, but rather, a positive measure of something in between. The ``rougher'' a geometric object is, the more boxes it needs to capture its fine detail at a small scale, so the higher its fractal dimension is. In fact, physical objects don't have integer dimensions because they aren't perfectly smooth geometric objects; they all have some roughness, and thus, they have non-integer dimensions and a fractal-like structure.\footnote{For more information about how fractals permeate the natural world, see the book by Mandelbrot \cite{mandelbrot2} for more details.}

\subsubsection{The Kaplan-Yorke Conjecture}

In a paper by Kaplan and Yorke \cite{kaplan-yorke}, a quantity now known as the Lyapunov dimension $d_l$ is introduced. To calculate this quantity, which we will soon discover has a strong connection to the fractal structure of attractors, recall from Section \ref{quantification} that a Lyapunov spectrum's exponents are ordered such that $\lambda_1\geq\lambda_2\geq\hdots\geq\lambda_n$. Now, let $\kappa$ be the largest index such that
\begin{equation}
    \sum_{i=1}^{\kappa}\lambda_i\geq 0
    \label{eq:kappa}
\end{equation}
Using this, we define the Lyapunov dimension $d_l$ as
\begin{equation}
    d_l = \kappa + \frac{1}{|\lambda_{\kappa+1}|}\sum_{i=1}^{\kappa}\lambda_i
    \label{eq:dl}
\end{equation}
Referring back to the Lyapunov spectrum of the Hénon map that we calculated earlier in this section, $\{\lambda_1,\,\lambda_2\}\approx\{0.419,\,-1.623\}$, we can see that the index $\kappa$ is equal to 1 since $\lambda_1+\lambda_2<0$, so the Lyapunov dimension $d_l$ of the Hénon attractor is
\begin{equation}
    d_l = 1+\frac{\lambda_1}{|\lambda_2|} \approx 1+\frac{0.419}{|-1.623|} \approx 1.26
\end{equation}

Notice that this Lyapunov dimension $d_l$ is surprisingly close to the attractor's fractal dimension $d\approx 1.23$, which is off by less than 3\%, well within the range of numerical error in calculating the fractal dimension and Lyapunov exponents. This is a result of the Kaplan-Yorke conjecture, which states that the Lyapunov dimension of an attractor is equal to its fractal dimension for some systems \cite{ott-attractor-dim}.\footnote{More specifically, the Kaplan-Yorke conjecture states that the Lyapunov dimension is equal to an object's information dimension, which is calculated in a slightly different way to the box-counting dimension, but is, in almost all cases, close to or equal to it.} In other words,
\begin{equation}
    d_l = d
\end{equation}
The Kaplan-Yorke conjecture is rather extraordinary because it relates the dynamics on an attractor to the attractor's geometry and structure, and it is a strong support of the fact that chaos and certain geometrical properties like fractal geometry are linked. Although some of the systems we examine in this paper do not actually obey the Kaplan-Yorke conjecture for various reasons, we will still use it as an indication of whether the attractor of a system is fractal.

\subsection{Basins of Attraction}
\label{basins-of-attraction}

Recall from Section \ref{nonchaoticattractors} that Property 2 of attractors says that $A$ is an attractor if it attracts an open set of initial conditions $U$, where for all initial conditions $\mathbf{x}_0\in U$, this distance from $\mathbf{f}^t(\mathbf{x}_0)$ to $A$ goes to 0. The largest $U$ with this property is called the basin of attraction of $A$ \cite[p. 332]{strogatz}, which we will denote as $\hat{A}$. Essentially, $\hat{A}$ is a region in state space composed of all the initial conditions that approach $A$ in the long term. The geometrical structure of different basins of attraction varies greatly among different dynamical systems, and depending on what type of basin an initial state belongs to, its long-term behavior can be fundamentally different \cite{ottboa}. For this reason, our first priority concerning basins of attraction is establishing a method to classify different basins and quantify their size.

To do this, we will follow a similar approach to the one outlined by Sprott and Xiong \cite{sprott}. In this method, rather than dealing with the complex geometry of different attractors, we instead capture the general location and size of the attractor with two numbers. The first number is the attractor's mean or center of mass $\langle A\rangle$, which is the average of all the attractor's points. For a fractal attractor composed of an infinite number of points,\footnote{We concern our analysis with attractors like these because our equations can be trivially altered for attractors of a finite number of points.} let us say $A=\{\mathbf{a}_1,\,\mathbf{a}_2,\,\hdots\}$, where the states $\mathbf{a}_i$ are in no particular order. Then, the mean of $A$ is
\begin{equation}
    \langle A\rangle = \lim_{j\to\infty}\frac{1}{j}\sum_{i=1}^j\mathbf{a}_i
    \label{eq:mean}
\end{equation}
Our second number of interest is the attractor's standard deviation $\sigma_A$, or the variation of the attractors' points around its mean. From statistics \cite[p. 56]{stats}, we know this is calculated by
\begin{equation}
    \sigma_A = \sqrt{\lim_{j\to\infty}\frac{1}{j}\sum_{i=1}^j|\mathbf{a}_i-\langle A\rangle|^2}
    \label{eq:standard-dev}
\end{equation}
Using this, we can define a quantity $\xi$ that we will call ``normalized distance'' from an attractor, which is the Euclidean distance between a state $\mathbf{x}$ and the attractor's mean, normalized by the attractor's standard deviation:
\begin{equation}
    \xi = \frac{|\mathbf{x} - \langle A\rangle|}{\sigma_A}
    \label{eq:normalized-distance}
\end{equation}
This notion of distance accounts for the fact that some attractors are bigger than others, and it will allow us to quantify basins relative to their attractor's size.

To motivate our classification method, we will introduce two more useful sets. First, let us say $S(\xi)$ is the set of all states that lie in an open $n$-dimensional ball\footnote{We define a one-dimensional ball to be a line segment, a two-dimensional ball to be a disk, a three-dimensional ball to be a filled-in sphere, and so on.} of radius $\xi$ centered at $\langle A\rangle$, where $n$ is the dimension of the state space $A$ lives in. In other words, $S(\xi)$ contains all the states less than $\xi$ away from $\langle A\rangle$. Mathematically, this definition is
\begin{equation}
    S(\xi) = \{\mathbf{x}:\:|\mathbf{x}-\langle A\rangle|<\xi\}
\end{equation}
where $:$ means ``such that.'' Our second useful set is $\hat{A}(\xi)$, which is the subset of the basin of attraction $\hat{A}$ containing all the states in the $n$-dimensional ball of radius $\xi$ centered at $\langle A\rangle$ that get attracted to $A$. In other words, it is the intersection of $\hat{A}$ and $S(\xi)$:\footnote{The intersection of $S_1$ and $S_2$, denoted $S_1\cap S_2$, is the set of all the elements contained in both $S_1$ and $S_2$.}
\begin{equation}
    \hat{A}(\xi) = \hat{A}\cap S(\xi)
    \label{eq:a-hat-xi-def}
\end{equation}

Now, we are ready to introduce the function that our basin classification method is based around: $P(\xi)$, which is the probability that an initial state $\mathbf{x}_0\in S(\xi)$ is in the basin of attraction of $A$. In other words, it is the fraction of states in the $n$-dimensional ball with radius $\xi$ centered at $\langle A\rangle$ that get attracted to $A$. Using the measure function, we can write this as
\begin{equation}
    P(\xi) = \frac{\mu(\hat{A}(\xi))}{\mu(S(\xi))}
    \label{eq:p-xi-def}
\end{equation}
In the limit $\xi\to\infty$, $P(\xi)$ follows a power law \cite{sprott}:
\begin{equation}
    P(\xi) = \frac{P_0}{\xi^{\gamma}}
    \label{eq:probability-function-power-law}
\end{equation}

Based on these parameters $P_0$ and $\gamma$, we can now divide basins of attraction into four distinct classes, ordered from largest to smallest:
\begin{enumerate}
    \item Class 1 basins have $P_0=1$ and $\gamma=0$. Then, the power law is $P(\xi)=1$, which means the probability of a state in any $n$-dimensional ball being in the basin of attraction is 100\%. Therefore, these basins include all of state space (except perhaps a set of finite measure).
    \item Class 2 basins have $P_0<1$ and $\gamma=0$. Then, the power law is $P(\xi)=P_0$, which is independent of $\gamma$. Therefore, all $n$-dimensional balls centered at $\langle A\rangle$ contain the same fraction of states that go to $A$, meaning these basins must occupy a fixed fraction of state space.
    \item Class 3 basins have $0<\gamma<n$, where $n$ is the dimension of the state space the basin belongs to. Notice that, as we scale $\xi$, $P(\xi)$ drops off slower than $\mu(S(\xi))$ grows since as we recall from Section \ref{strangeattractors} and our discussion of dimension, the measure of an $n$-dimensional ball scales according to $\xi^n$ while $P(\xi)$ scales according to $\xi^{-\gamma}$. Therefore, these basins extend to infinity in some directions but take up an increasingly small fraction of state space as we move out.
    \item Class 4 basins have $\gamma=n$. In this case, since $P(\xi)$ drops off at the same rate $\mu(S(\xi))$ grows, we can conclude that these basins occupy a finite region of state space and therefore have a well-defined size relative to their attractors. The linear, normalized size of a Class 4 basin is given by 
    \begin{equation}
        \xi_0 = P_0^{1/n}
    \end{equation}
\end{enumerate}
It is worth noting that although using the normalized distance $\xi$ doesn't affect the basin class or the value of $\gamma$, it does affect $P_0$, and thus, the relative size of a Class 4 basin. See the paper by Sprott and Xiong \cite{sprott} for examples of all these kinds of basins in both two and three-dimensional state space.

Here, the system we will explore is once again the Hénon map, whose basin we will find and classify to demonstrate our numerical method of finding $P(\xi)$. Since points that do not end up in the Hénon attractor diverge to infinity \cite{henon}, we can calculate $P(\xi)$ for the Hénon attractor easily by saying that if an initial state $\xi$ away from $\langle A\rangle$ eventually maps to a point greater than $10000\xi$ away from $\langle A\rangle$, it isn't in $\hat{A}$. Because $P(\xi)$ is a power function, we can find $\gamma$ using a similar method to the one we used to find the fractal dimension $d$ in Section \ref{strangeattractors}. Namely, taking logs on both sides of Equation \ref{eq:probability-function-power-law}, we get that
\begin{equation}
    \ln P(\xi) = -\gamma\ln\xi + \ln P_0
    \label{eq:log-log-xi-relationship}
\end{equation}
where we purposefully take a $-\gamma$ out rather than leaving a $1/\xi$ inside the log because we are interested in the limit $\xi\to\infty$. Then, if we numerically find $P(\xi)$ for various $\xi = 2^k$, plotting $\log_2 P(\xi)$ vs. $\log_2 \xi$ will give us points along a line with slope $-\gamma$.

We can approximate $P(\xi)$ for a given $\xi$ by choosing a bunch of random states in $S(\xi)$ and iterating them to see if they end up in the basin. However, a problem arises in the case where there is a lot of space in $S(2^k)$ that doesn't contain the basin. Then, $P(2^k)$ may be very close to 0, so a numerical calculation testing only a finite number of random states may not be accurate. To counteract this issue, we detail a shell method for calculating values of $P(\xi)$ in Appendix \ref{iterative-shell-method-derivation}. The concept behind this method is an iterative calculation of $P(2^{k+1})$ using $P(2^k)$ and $\Delta P(2^{k})$, which is the probability that a state lying in the shell centered at $\langle A\rangle$ with inner radius $\xi=2^k$ and outer radius $\xi=2^{k+1}$ is also in the basin of attraction $\hat{A}$. We show in Appendix \ref{iterative-shell-method-derivation} that this can be accomplished using 
\begin{equation}
    P(2^{k+1}) = \frac{P(2^k)}{2^n} + \left(1-\frac{1}{2^n}\right)\Delta P(2^k)
    \label{eq:iteration-of-shell-method}
\end{equation}
We can use this equation to generate a data set of $P(2^k)$ values using a Monte Carlo algorithm. First, we estimate $P(1)$ by first randomly picking initial states in the sphere of radius $\xi=1$ centered at $\langle A\rangle$ and checking whether or not they are in the basin by generating their orbits. Then, we get $P(1)$ by dividing the number of states that were in the basin of attraction by the total number of states we tested. Using a similar approach, we can estimate $\Delta P(1)$ by randomly picking states in the shell with inner radius $\xi=1$ and outer radius $\xi=2$ centered at $\langle A\rangle$. We can then calculate $P(2)$ using Equation \ref{eq:iteration-of-shell-method}:
\begin{equation}
    P(2) = \frac{P(1)}{2^n} + \left(1-\frac{1}{2^n}\right)\Delta P(1)
\end{equation}
We can then numerically estimate $\Delta P(2)$ to calculate $P(4)$, and so on. 

This seemingly roundabout way of finding $P(\xi)$ is useful because it allows us to distinguish $P(\xi)$ from 0 for large values of $\xi$. For example, even if $\Delta P(2^k)$ is 0 because the attractor lies entirely inside of the shell's inner radius $\xi=2^k$, calculating $P(2^{k+1})$ using Equation \ref{eq:iteration-of-shell-method} will give a non-zero result because we iterate up to it. However, if we were to calculate $P(2^{k+1})$ directly, our Monte Carlo algorithm may fail to give an accurate result due to the vanishingly small basin within $S(2^{k+1})$ that a computationally reasonable random selection of points wouldn't pick up on. Thus, our shell method allows us to find an accurate power function $P(\xi)$ in a computationally efficient way. 

\begin{table}[t]
    \centering
    \begin{tabular}{c|c}
        $\xi$ & $P(\xi)$ \\
        \hline \\ [-10px]
        $2^0=1$ & 0.8751 \\
        $2^1=2$ & 0.6976 \\
        $2^2=4$ & 0.2708 \\
        $2^3=8$ & 0.09465 \\
        $2^4=16$ & 0.03085 \\
        $2^5=32$ & 0.01000 \\
        $2^6=64$ & 0.003124 \\
    \end{tabular}
    \hspace{0.25cm}
    \begin{tabular}{c|c}
        $\xi$ & $P(\xi)$ \\
        \hline \\ [-10px]
        $2^7=128$ & \num{9.220e-4} \\
        $2^8=256$ & \num{2.598e-4} \\
        $2^9=512$ & \num{7.469e-5} \\
        $2^{10}=1024$ & \num{2.167e-5} \\
        $2^{11}=2048$ & \num{6.918e-6} \\
        $2^{12}=4096$ & \num{1.730e-6} \\
        $2^{13}=8192$ & \num{4.324e-7} \\
    \end{tabular}
    \caption{Some approximate $P(\xi)$ values of the Hénon basin, calculated using the code in Appendix \ref{classifying-henon-basin-code}}
    \label{tab:p_function_henon_values}
\end{table}

\begin{figure}
    \centering
    \begin{tikzpicture}[scale=0.425]
        \draw[->] (-0.5,3/2)--(13.5,3/2);
        \path (-1,0)--(15, 0);
        \draw[->] (-0.5,3/2)--(-0.5,-24/2);
        \draw[thick, domain=-0.25:13.25]plot(\x,{-1.7503/2*\x+1.9694/2});
        \filldraw[thick] (2,-1.885/2) circle [radius=0.15];
        \filldraw[thick] (3,-3.401/2) circle [radius=0.15];
        \filldraw[thick] (4,-5.019/2) circle [radius=0.15];
        \filldraw[thick] (5,-6.644/2) circle [radius=0.15];
        \filldraw[thick] (6,-8.322/2) circle [radius=0.15];
        \filldraw[thick] (7,-10.083/2) circle [radius=0.15];
        \filldraw[thick] (8,-11.91/2) circle [radius=0.15];
        \filldraw[thick] (9,-13.71/2) circle [radius=0.15];
        \filldraw[thick] (10,-15.49/2) circle [radius=0.15];
        \filldraw[thick] (11,-17.14/2) circle [radius=0.15];
        \filldraw[thick] (12,-19.14/2) circle [radius=0.15];
        \filldraw[thick] (13,-21.1411/2) circle [radius=0.15];
        \filldraw[thick] (0,-0.192/2) circle [radius=0.15];
        \filldraw[thick] (1,-0.52/2) circle [radius=0.15];
        \node[above] at (6.25,2.5) {$\log_2\xi$};
        \node[left, rotate=90] at (-2.9,-3) {$\log_2 P(\xi)$};
        \draw (0, 1.5)--(0, 1.7); 
        \draw (1, 1.5)--(1, 1.7);
        \draw (2, 1.5)--(2, 1.7); 
        \draw (3, 1.5)--(3, 1.7);
        \draw (4, 1.5)--(4, 1.7); 
        \draw (5, 1.5)--(5, 1.7);
        \draw (6, 1.5)--(6, 1.7); 
        \draw (7, 1.5)--(7, 1.7);
        \draw (8, 1.5)--(8, 1.7); 
        \draw (9, 1.5)--(9, 1.7);
        \draw (10, 1.5)--(10, 1.7); 
        \draw (11, 1.5)--(11, 1.7);
        \draw (12, 1.5)--(12, 1.7); 
        \draw (13, 1.5)--(13, 1.7);
        \node[above] at (0, 1.7) {\scriptsize $0$};
        \node[above] at (2, 1.7) {\scriptsize $2$};
        \node[above] at (4, 1.7) {\scriptsize $4$};
        \node[above] at (6, 1.7) {\scriptsize $6$};
        \node[above] at (8, 1.7) {\scriptsize $8$};
        \node[above] at (10, 1.7) {\scriptsize $10$};
        \node[above] at (12, 1.7) {\scriptsize $12$};
        \draw (-0.5, 0)--(-0.7, 0);
        \draw (-0.5, 1/2)--(-0.7, 1/2);
        \draw (-0.5, 2/2)--(-0.7, 2/2);
        \draw (-0.5, -1/2)--(-0.7, -1/2);
        \draw (-0.5, -2/2)--(-0.7, -2/2);
        \draw (-0.5, -3/2)--(-0.7, -3/2);
        \draw (-0.5, -4/2)--(-0.7, -4/2);
        \draw (-0.5, -5/2)--(-0.7, -5/2);
        \draw (-0.5, -6/2)--(-0.7, -6/2);
        \draw (-0.5, -7/2)--(-0.7, -7/2);
        \draw (-0.5, -8/2)--(-0.7, -8/2);
        \draw (-0.5, -9/2)--(-0.7, -9/2);
        \draw (-0.5, -10/2)--(-0.7, -10/2);
        \draw (-0.5, -11/2)--(-0.7, -11/2);
        \draw (-0.5, -12/2)--(-0.7, -12/2);
        \draw (-0.5, -13/2)--(-0.7, -13/2);
        \draw (-0.5, -14/2)--(-0.7, -14/2);
        \draw (-0.5, -15/2)--(-0.7, -15/2);
        \draw (-0.5, -16/2)--(-0.7, -16/2);
        \draw (-0.5, -17/2)--(-0.7, -17/2);
        \draw (-0.5, -18/2)--(-0.7, -18/2);
        \draw (-0.5, -19/2)--(-0.7, -19/2);
        \draw (-0.5, -20/2)--(-0.7, -20/2);
        \draw (-0.5, -21/2)--(-0.7, -21/2);
        \draw (-0.5, -22/2)--(-0.7, -22/2);
        \draw (-0.5, -23/2)--(-0.7, -23/2);
        \node[left] at (-0.7, 0) {\scriptsize $0$};
        \node[left] at (-0.7, -4/2) {\scriptsize $-4$};
        \node[left] at (-0.7, -8/2) {\scriptsize $-8$};
        \node[left] at (-0.7, -12/2) {\scriptsize $-12$};
        \node[left] at (-0.7, -16/2) {\scriptsize $-16$};
        \node[left] at (-0.7, -20/2) {\scriptsize $-20$};
    \end{tikzpicture}
    \caption{A plot of the points in Table \ref{tab:p_function_henon_values} on a graph of $\log_2 P(\xi)$ vs. $\log_2\xi$ and their line of best-fit}
    \label{fig:henon-p-xi-function-linear-reg}
\end{figure}
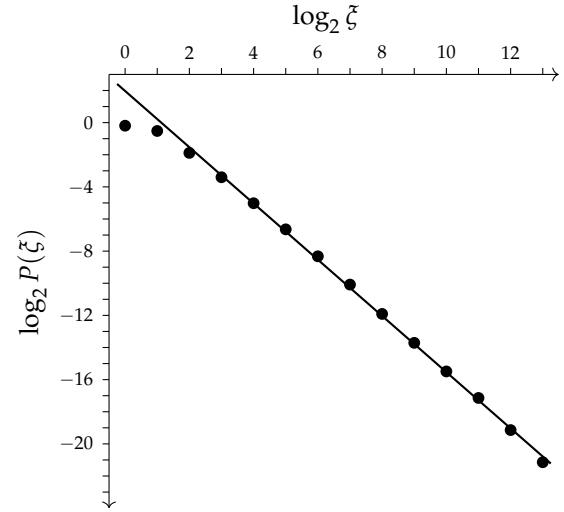

The code in Appendix \ref{classifying-henon-basin-code} implements this method to find $P(2^k)$ values for the basin of the Hénon attractor, which are displayed in Table \ref{tab:p_function_henon_values}. Using this code, we find the Hénon attractor's mean is approximately $\langle A\rangle\approx \langle 0.2571,\,0.0771 \rangle$ and its standard deviation is approximately $\sigma_A\approx 0.7526$, both of which are visually consistent with Figure \ref{fig:henon-attractor}. In Figure \ref{fig:henon-p-xi-function-linear-reg}, we plot the values in Table \ref{tab:p_function_henon_values} on a $\log_2 P(\xi)$ vs. $\log_2\xi$ plot and take a linear regression,\footnote{We neglect the $P(1)$ and $P(2)$ values when taking the linear regression because we are interested in the limit $\xi\to\infty$.} which gives us,
\begin{equation}
    \log_2P(\xi) = -1.750\log_2\xi + 1.969
\end{equation}
with an $R^2$ value of 0.999. By Equation \ref{eq:log-log-xi-relationship}, $\gamma\approx-1.750$ and $P_0\approx3.196$, which indicates that the Hénon attractor has a Class 3 basin.

\begin{figure}
    \centering
    \includegraphics[scale=0.028]{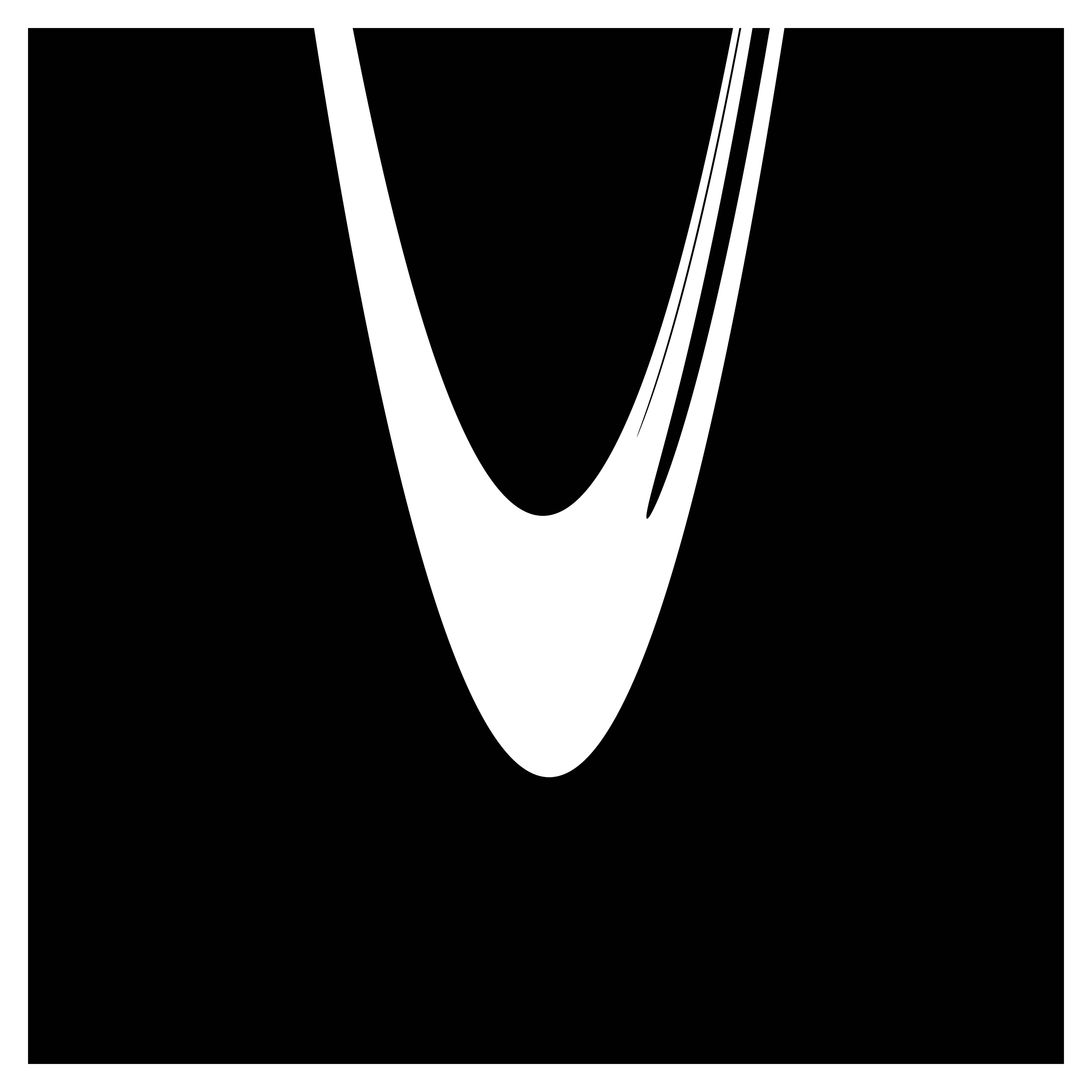}
    \caption{The basin of the Hénon attractor for $-5\leq x\leq 5$ and $-5\leq y\leq 5$ using the parameters $a=1.4$ and $b=0.3$, graphed with the code in Appendix \ref{visualizing-henon-basin-code}}
    \label{fig:henon-basin-nonfractal}
\end{figure}

In Figure \ref{fig:henon-basin-nonfractal}, we visualize the Hénon basin using the code in Appendix \ref{visualizing-henon-basin-code}. Here, points in the basin of the Hénon attractor are shown in white, and points that diverge to infinity are shown in black. In this regard, we can consider infinity the other attractor of the Hénon map, meaning the white region is the basin of the strange attractor in Figure \ref{fig:henon-attractor} and the black region is the basin of infinity. Visually, we can see that Figure \ref{fig:henon-basin-nonfractal} aligns with the fact that the Hénon basin is Class 3: it has a quasi-parabolic shape that takes up less and less space as we extend outward.

The fact that the Hénon map has two attractors, the strange attractor and infinity, makes the Hénon map a bistable system, or a system with two distinct, coexisting attractors. Bistability is the simplest case of multistability, which is defined as the coexistence of several attractors for a given set of system parameters \cite[pg. 6]{pisarchikbook}. Each attractor of a multistable system has an associated basin of attraction, and the set of points that separate two or more basins is called the boundary of the basins. As we can see, the basin boundary of the Hénon map with our standard parameters $a=1.4$ and $b=0.3$ is smooth, but by varying the parameters, we can get more interesting geometries. Many dynamical systems exhibit complex and counterintuitive geometrical properties resulting in sensitive dependence on initial conditions in their basins of attractions and basin boundaries. We will explore some of these geometrical properties in this section.

\begin{figure}
    \centering
    \begin{tikzpicture}[scale=0.75]
        \draw (0, 0)--(10, 0);
        \draw (10, 0)--(10, 4);
        \draw (10, 4)--(0, 4);
        \draw (0, 4)--(0, 0);
        \draw[very thick] (6, 4)--(4, 0);
        \node[below right] at (0, 4) {\large $\hat{A}$};
        \node[below left] at (10, 4) {\large $\hat{C}$};
        \filldraw (4.75, 1) circle [radius=1.5pt];
        \draw[thick] (4.75, 1) circle [radius=0.75];
        \node[below] at (4.75, 1) {\scriptsize $\mathbf{b}$};
        \draw[->, thick] (4.75, 1)--(4.75+0.866*0.75, 1+0.75/2);
        \node at (4.75+0.866*0.375-0.15, 1+0.375/2+0.15) {\scriptsize $\epsilon$};
        \filldraw (2, 2.5) circle [radius=1.5pt];s
        \draw[thick] (2, 2.5) circle [radius=0.75];
        \node[below] at (2, 2.5) {\scriptsize $\mathbf{a}$};
        \draw[->, thick] (2, 2.5)--(2+0.866*0.75, 2.5+0.75/2);
        \node at (2+0.866*0.375-0.15, 2.5+0.375/2+0.15) {\scriptsize $\epsilon$};
        \filldraw (8.5, 1.5) circle [radius=1.5pt];
        \draw[thick] (8.5, 1.5) circle [radius=0.75];
        \node[below] at (8.5, 1.5) {\scriptsize $\mathbf{c}$};
        \draw[->, thick] (8.5, 1.5)--(8.5+0.866*0.75, 1.5+0.75/2);
        \node at (8.5+0.866*0.375-0.15, 1.5+0.375/2+0.15) {\scriptsize $\epsilon$};
        \draw[-Stealth] (6.5, 3)--(5.5, 3);
        \node[right] at (6.5, 3) {\large $\Sigma$};
    \end{tikzpicture}
    \caption{Two basins of attractions $\hat{A}$ and $\hat{C}$ divided by smooth basin boundary $\Sigma$ and example initial states $\mathbf{a}$, $\mathbf{b}$, and $\mathbf{c}$ with uncertainty $\epsilon$}
    \label{fig:basin-boundary-uncertainty-schematic}
\end{figure}
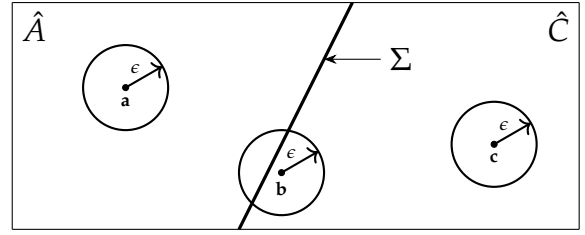

\subsubsection{Fractal Basin Boundaries}

A given point $\mathbf{x}$ is on the boundary of a basin $\hat{A}$ if and only if every
neighborhood of $\mathbf{x}$ contains both points in $\hat{A}$ and points not in $\hat{A}$. Let us denote a basin boundary with $\Sigma$. Like before, it will be useful to quantify basin boundaries in a way that allows us to get a grasp on their complex geometry. To illustrate our method of quantifying basin boundaries, let us consider a region of the state space of a two-dimensional system with two attractors $A$ and $C$ (like the Hénon map). This region of state space contains a subset of the basins of both $A$ and $C$, which are divided by a basin boundary $\Sigma$. Therefore, if we were to choose a random initial state $\mathbf{x}_0$ in this region, it would lie in either basin $\hat{A}$ or $\hat{C}$ with a 100\% probability \cite{ottboa}. This is because the basin boundary $\Sigma$ has zero measure in two-dimensional state space as it is a one-dimensional object; from Property 4 of measures in Section \ref{strangeattractors}, $\mu^2(\Sigma)=0$.

Now, let us say that this state $\mathbf{x}_0$ has some uncertainty $\epsilon$. Specifically, let us say that the actual state may be anywhere in a disk with radius $\epsilon$ centered at $\mathbf{x}_0$, or $|\mathbf{x}-\mathbf{x}_0|<\epsilon$.\footnote{For a general system, this would mean that the state may be anywhere in an $n$-dimensional ball with radius $\epsilon$ centered at $\mathbf{x}_0$.} In Figure \ref{fig:basin-boundary-uncertainty-schematic}, we display a schematic diagram of this region of state space with three initial states $\mathbf{a}$, $\mathbf{b}$, and $\mathbf{c}$. We can see that initial state the $\mathbf{a}$, despite having some uncertainty, will always be attracted to attractor $A$. Similarly, the initial state $\mathbf{c}$ will always be attracted to attractor $C$. However, we can see that the uncertainty $\epsilon$ causes the initial state $\mathbf{b}$ to have the possibility of being attracted to either $A$ or $C$. Picking a random state in our region of state space, we will now consider what the probability $\varrho(\epsilon)$ is that the uncertainty $\epsilon$ will cause us to make a mistake in predicting which attractor our random state will go to. As we can see in Figure \ref{fig:basin-boundary-uncertainty-schematic}, this probability $\varrho(\epsilon)$ is identical to the fraction of our region of state space that lies a distance $\epsilon$ from the basin boundary $\Sigma$. It is clear from Figure \ref{fig:basin-boundary-uncertainty-schematic} that this subset of states that lie a distance $\epsilon$ from $\Sigma$ is a strip of width $2\epsilon$ centered at $\Sigma$. Therefore, the fraction $\varrho(\epsilon)$ scales proportionally to $\epsilon$:
\begin{equation}
    \varrho(\epsilon)\sim\epsilon
    \label{eq:smooth-boundary-exponent}
\end{equation}
where $\sim$ means ``is proportional to.'' 

However, Equation \ref{eq:smooth-boundary-exponent} only applies to smooth boundaries, like the ones in Figures \ref{fig:henon-basin-nonfractal} and \ref{fig:basin-boundary-uncertainty-schematic}. For fractal basin boundaries, namely, those with a non-integer dimension $d$, $\varrho(\epsilon)$ follows the power law
\begin{equation}
    \varrho(\epsilon)\sim\epsilon^{\mathfrak{u}}
    \label{eq:fractal-boundary-exponent}
\end{equation}
where $\mathfrak{u}$ is a number less than 1 known as the uncertainty exponent \cite{ottboa}. McDonald \textit{et al.} \cite{mcdonald} prove that the uncertainty exponent is related to a basin boundary's dimension $d$ by the following relation:
\begin{equation}
    \mathfrak{u} = n-d
    \label{eq:uncertainty-exp-frac-dim}
\end{equation}
where $n$ is the dimension of the system's state space. When $\mathfrak{u}=1$, Equation \ref{eq:uncertainty-exp-frac-dim} gives us $d=n-1$, but when $\mathfrak{u}<1$, the boundary is fractal because $d$ is then not an integer. When $\mathfrak{u}$ is substantially less than 1, a significant improvement in initial uncertainty $\epsilon$ causes only a marginal reduction in final state uncertainty, or which attractor a state will actually end up in \cite{grebogi-final-state}. For example, if $\mathfrak{u}=0.1$, to reduce the final uncertainty $\varrho(\epsilon)$ by a factor of 10, we will have to reduce our initial uncertainty $\epsilon$ by a factor of $10^{10}$. Thus, an uncertainty exponent $\mathfrak{u}$ less than 1, indicating the presence of a fractal basin boundary, directly results in sensitive dependence on initial conditions \cite[p. 151]{ott}, as it becomes increasingly difficult to predict where initial states will end up as we near the fractal boundary.

We are specifically interested in the behavior of $\varrho(\epsilon)$ when $\epsilon$ is small \cite{mcdonald}, so to numerically calculate uncertainty exponents, we can find a bunch of $\varrho(\epsilon)$ values for $\epsilon = 2^{-k}$ and graph them on a plot of $\log_2\varrho(\epsilon)$ vs. $\log_2\epsilon$. Specifically, taking logs on both sides of Equation \ref{eq:fractal-boundary-exponent}, we get
\begin{equation}
    \ln\varrho(\epsilon) = \mathfrak{u}\ln\epsilon + c
\end{equation}
where $c$ is the natural log of the proportionality constant. We approximate a given $\varrho(2^{-k})$ value for the Hénon boundary using the code in Appendix \ref{uncertainty_exponent_henon_basin_boundary_code}, which implements a Monte Carlo algorithm. Namely, we choose a random initial state in $-5\leq x\leq 5$ and $-5\leq y\leq 5$ (shown in Figure \ref{fig:henon-basin-nonfractal}), which we will denote as $\langle x_0,\, y_0 \rangle$, then test whether it belongs to the basin of the strange attractor or the basin of infinity by iterating it. Then, we test four perturbed states $\langle x_0+\epsilon,\, y_0 \rangle$, $\langle x_0-\epsilon,\, y_0 \rangle$, $\langle x_0,\, y_0+\epsilon \rangle$, and $\langle x_0,\, y_0-\epsilon \rangle$. If any of these states end up in a different attractor, then our initial state was uncertain. Using our standard parameters of the Hénon map ($a=1.4$ and $b=0.3$), the code in Appendix \ref{uncertainty_exponent_henon_basin_boundary_code} gives us $\mathfrak{u}=1$, which makes sense from Figure \ref{fig:henon-basin-nonfractal} as the boundary appears smooth.

\begin{figure}
    \centering
    \includegraphics[scale=0.028]{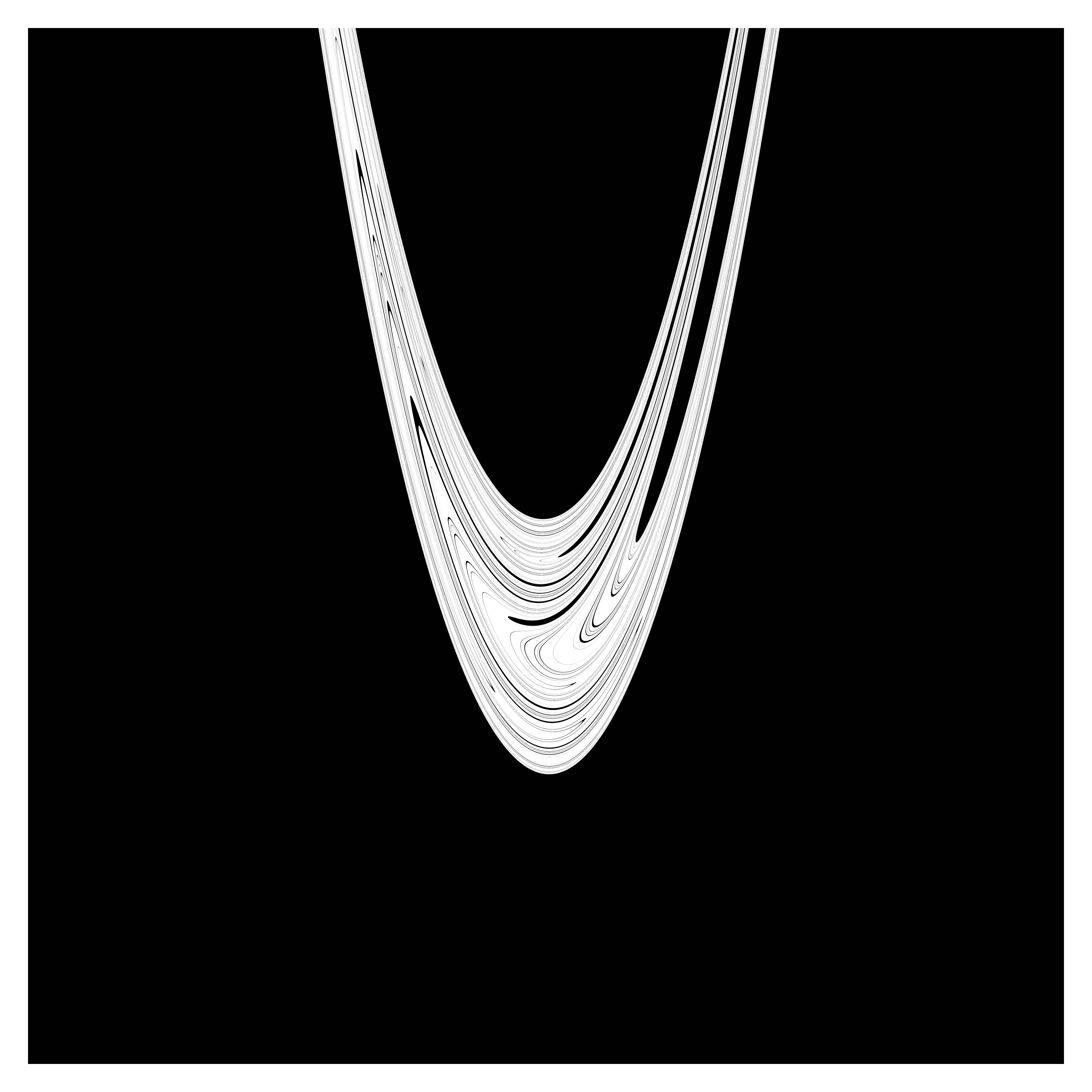}
    \caption{The basin of the Hénon attractor for $-5\leq x\leq 5$ and $-5\leq y\leq 5$ using the parameters $a=1.45$ and $b=0.3$, graphed with the code in Appendix \ref{visualizing-henon-basin-code}}
    \label{fig:henon-basin-fractal-boundary}
\end{figure}
\begin{figure*}
    \centering
    \hfill
    \begin{subfigure}[b]{0.45\textwidth}
        \centering
        \includegraphics[scale=0.275]{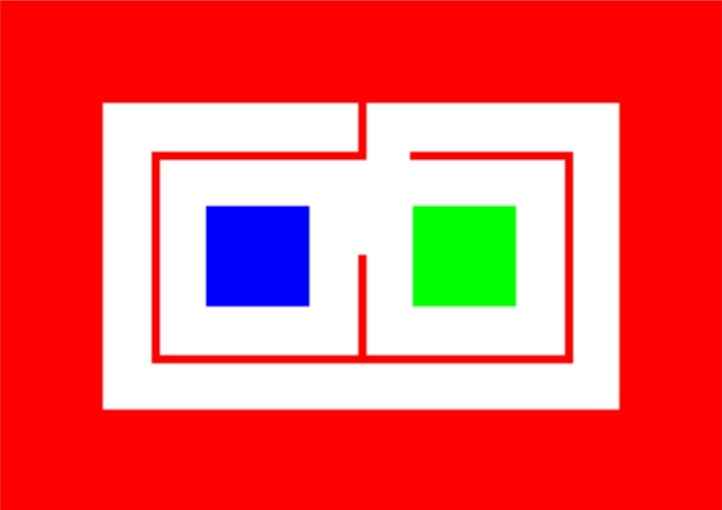}
        \vspace{4px}
        \caption{Step 1}
        \label{fig:wada-step-1}
    \end{subfigure}
    \hfill
    \begin{subfigure}[b]{0.45\textwidth}
        \centering
        \includegraphics[scale=0.275]{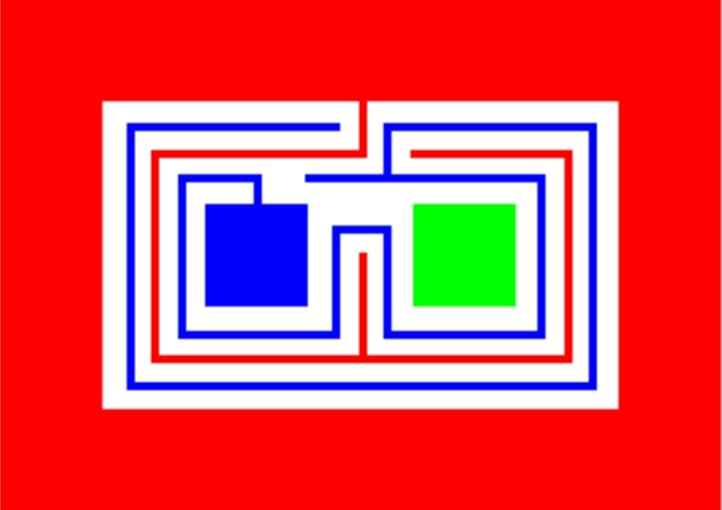}
        \vspace{4px}
        \caption{Step 2}
        \label{fig:wada-step-2}
    \end{subfigure}
    \hfill
    \vspace{2px}    
    \caption{Two steps in the construction of the lakes of Wada, from \cite{pisarchikbook}}
    \label{fig:wada-lakes-construction}
\end{figure*}

However, if we consider the Hénon map with the parameters $a=1.45$ and $b=0.3$, we can see in Figure \ref{fig:henon-basin-fractal-boundary} that its basin boundary is far more complex. Using the code in Appendix \ref{uncertainty_exponent_henon_basin_boundary_code} to calculate $\varrho(2^{-k})$ values and taking a linear regression (neglecting the first four points as we are interested in small values of $\epsilon$), we get that
\begin{equation}
    \log_2\varrho(\epsilon) = 0.06954\log_2\epsilon - 3.050
\end{equation}
with an $R^2$ value of 0.984. Thus, $\mathfrak{u}\approx0.0695$ for our choice of parameters, indicating a large amount of final state uncertainty and ``geometrical'' chaos on the Hénon boundary. Visually, we can see in Figure \ref{fig:henon-basin-fractal-boundary} why this is the case: any uncertainty near the boundary will significantly affect where an initial state will be attracted. From Equation \ref{eq:uncertainty-exp-frac-dim}, we can find that the dimension $d$ of the Hénon boundary for $a=1.45$ and $b=0.3$ is
\begin{equation}
    d = n-\mathfrak{u} \approx 2-0.0695 = 1.9305
\end{equation}
which is significantly larger than 1. This indicates that the Hénon boundary for these parameters has a large amount of ``roughness'' and detail that significantly increases its fractal dimension from the $d=1$ dimension for the Hénon map's standard parameters, which is immediately obvious from Figures \ref{fig:henon-basin-nonfractal} and \ref{fig:henon-basin-fractal-boundary}.

\subsubsection{Wada Basins}

Let us consider an $n$-dimensional system with $w\geq3$ attractors $A_1,\,A_2,\,\hdots\,,\,A_w$ and associated basins $\hat{A}_1,\,\hat{A}_2,\,\hdots\,,\,\hat{A}_w$. In a system like this, it is possible for the basins to satisfy the counterintuitive Wada property, for which 3 or more basins $\hat{A}_1,\,\hat{A}_2,\,\hdots\,,\,\hat{A}_w$ share the same boundary $\Sigma$. Specifically, the basins are Wada if each boundary point of every basin is also a boundary point of every other basin \cite{kennedy}. It is immediately evident that the presence of Wada basins is an even larger barrier to final state predictability than a standard fractal boundary because an initial uncertainty $d\mathbf{x}_0$ could potentially cause an initial state to end up in one of many distinct attractors. Thus, we want to be able to detect when basins exhibit the Wada property because it directly results in extreme sensitivity to initial conditions \cite{zhang}.

To understand how it is possible for three basins to be Wada, we will use the lakes of Wada example described by Kennedy and Yorke \cite{kennedy}. Imagine a white island surrounded by a red ocean. On the island, there are two lakes, one with blue water and one with green water. To construct the lakes of Wada, our first step, which is shown in Figure \ref{fig:wada-step-1}, is to dig a canal from the ocean into the island such that once the canal is dug, there are no points of land further than 1 unit away from red water. The second step, shown in Figure \ref{fig:wada-step-2}, is to dig a canal from the blue lake into the island such that each point in the remaining land is no further than $1/2$ units away from blue water. Similarly, the third step is to dig a canal from the green lake so that the remaining land is within $1/4$ units away from green water. After that, we return to the red ocean, digging a canal that makes the remaining land within $1/8$ units from red water, and so on, cycling through the three bodies of water. After an infinite number of steps, each point on the remaining land is arbitrarily close to all three bodies of water, so the remaining land is the boundary of all three. In other words, since every neighborhood of each land point contains points in all three bodies of water, the land is a Wada boundary of the three bodies.

This example clearly demonstrates how Wada boundaries are indicative of ``geometrical'' chaos due to extreme sensitivity to initial conditions. For this reason, we would like to have a method to detect the presence of the Wada property in any $n$-dimensional dynamical system with $w\geq3$ attractors. We will base our method of detecting Wada basins on the algorithm presented by Daza \textit{et al.} \cite{daza}, which uses a grid approach. This method goes as follows:
\begin{enumerate}
    \item Given an $n$-dimensional dynamical system with $w\geq3$ attractors, consider a ``cubical'' region of state space\footnote{We define a one-dimensional ``cubical'' region to be a line segment, a two-dimensional ``cubical'' region to be a square, a three-dimensional ``cubical'' region to be a standard cube, and so on.} that includes subsets of all the system's basins $\hat{A}_1,\,\hat{A}_2,\,\hdots\,,\,\hat{A}_w$.
    \item Cover the region with an $s\times s\times\hdots\times s$ (with $s$ being repeated $n$ times) grid $G$ composed of $s^n$ closed boxes $\Box_i$, where $i=1,\,2,\,\hdots\,,\,s^n$. It is beneficial to think of each box $\Box_i$ as an element of the set $G$. Mathematically, this means
    \begin{equation}
        G=\bigcup_{i=1}^{s^n}\Box_i
    \end{equation}
    In a two-dimensional system, $G$ is an $s\times s$ grid, where the boxes $\Box_i$ are indexed from left to right, top to bottom. 
    \item Define a function $C(\mathbf{x})$ that tells us which basin a given state belongs to. Specifically, for a state $\mathbf{x}\in\hat{A}_j$, define $C(\mathbf{x})=j$.
    \item For a given box $\Box_i$, define $C(\Box_i)=C(\mathbf{x})$, where $\mathbf{x}$ is the point at the center of $\Box_i$. Practically, it is useful to think of $C(\Box_i)$ as the color of $\Box_i$.
    \item Define $b(\Box_i)$ to be the set of boxes consisting of $\Box_i$ and all the boxes sharing at least 1 boundary point with $\Box_i$. Therefore, for a two-dimensional system, $b(\Box_i)$ is a $3\times3$ collection of boxes with $\Box_i$ as the central box:
    \begin{equation}
        \begin{split}
            b(\Box_i) &= \bigcup_{j=-1}^1\bigcup_{k=-1}^1\Box_{i+js+k} \\
            &= \Box_{i-s-1}\cup\Box_{i-s}\cup\Box_{i-s+1} \\
            &\mathrel{\phantom{=}}\cup\,\Box_{i-1}\cup\Box_{i}\cup\Box_{i+1} \\
            &\mathrel{\phantom{=}}\cup\,\Box_{i+s-1}\cup\Box_{i+s}\cup\Box_{i+s+1}
        \end{split}
    \end{equation}
    \item Define the function $K(\Box_i)$ to be the number of distinct colors $C(\Box)$ in $b(\Box_i)$ and calculate $K(\Box_i)$ for each box $\Box_i\in G$
    \item For every box with a $K(\Box_i)$ that is not 1 or $w$, draw a line segment from the center point of $\Box_i$ to the center point of a box in $b(\Box_i)$ that has a different color than $\Box_i$. Now, step $p=1$ is to calculate the color of the midpoint of this line segment. If this is a different color from all the other colors of the boxes in $b(\Box_i)$, there is now one more color in $b(\Box_i)$, so $K(\Box_i)$ increases by one. If $K(\Box_i)=w$ now, stop and move on to the next box with a $K(\Box_i)$ that is not 1 or $w$. Otherwise, move on to step $p=2$, which is to calculate the colors of two more points, namely, the points $1/4$ and $3/4$ up the line segment. Adjust $K(\Box_i)$ accordingly if either of these are new colors. Continue this process, step $p=3$ being dividing the line segment into eighths, step $p=3$ being dividing the line segment into sixteenths, and so on, until either $K(\Box_i)=w$ or the number of calculated points gets sufficiently large enough to assume that $K(\Box_i)<w$.
    \item For each step $p$ in 7. and for $k=1,\,2,\,\hdots\,,\,w$, define $G_k^p\subset G$ to be the set of all $\Box_i$ for which $K(\Box_i)=k$ immediately after step $p$. In other words, if $K(\Box_i)$ is calculated immediately after step $p$, then
    \begin{equation}
        G_k^p = \{\Box_i:\:K(\Box_i)=k\}
    \end{equation}
\end{enumerate}

After a given step $p$, this method sorts all the boxes in the grid $G$ into distinct sets:
\begin{equation}
    G = \bigcup_{k=1}^w G_k^p
\end{equation}
By definition, the set $G_k^p$ that a given box $\Box_i$ belongs to tells us how many different colors $k$ are within $b(\Box_i)$ after $p$ steps. This means that, after an infinite number of steps, it indicates whether the box is inside a basin or on (or sufficiently near) a certain basin boundary. Specifically, $G_1^{\infty}$ is composed of boxes $\Box_i$ that are in the interior of a basin, as $k=1$ indicates that all the center points of all the boxes in a given $b(\Box_i)$ end up in the same attractor. $G_2^{\infty}$ is composed of boxes $\Box_i$ that are on or sufficiently near a boundary of two attractors since an infinite number of points in a neighborhood of the center point of $\Box_i$ led to two distinct attractors. Similarly, $G_3^{\infty}$ is composed of boxes that are on or sufficiently near a boundary of three attractors, and so on. 

By definition, a system with $w$ attractors exhibits the Wada property if every basin boundary point is a boundary point of all $w$ basins of attraction. This means that for a system with Wada basins, every point in state space is either inside a basin or on a Wada boundary, meaning every box in the grid $G$ is either an element of $G_1^{\infty}$ or $G_w^{\infty}$. Mathematically, we can write this condition as
\begin{equation}
    \lim_{p\to\infty}\sum_{k=2}^{w-1}\left|G_k^p\right|=0
    \label{eq:wada-condition}
\end{equation}
where $\left|G_k^p\right|$ represents the number of boxes in, or the cardinality of, the set $G_k^p$.

We can label a system as Wada or not Wada by whether or not Equation \ref{eq:wada-condition} is satisfied, but it will be useful to have a parameter to determine whether a system is partial Wada. A system with partially Wada basins contains some basin boundary points that are shared by all $w$ basins, but at least some boundary points that are not \cite{daza}. To quantify this, we can define a parameter $W_k$ that tells us the fraction of boxes on or sufficiently near a boundary that belong to a boundary shared by $k$ basins:
\begin{equation}
    W_k=\lim_{p\to\infty}\frac{\left|G_k^p\right|}{\sum_{j=2}^w\left|G_j^p\right|}
\end{equation}
since all $G_j^{\infty}$ for $j\geq2$ contain only boxes on or sufficiently near a boundary. Clearly, if $W_w=1$, then the system is Wada, but if $0<W_w<1$, then the system is partial Wada by our established definition.

\begin{figure*}
    \centering
    \includegraphics[scale=0.06]{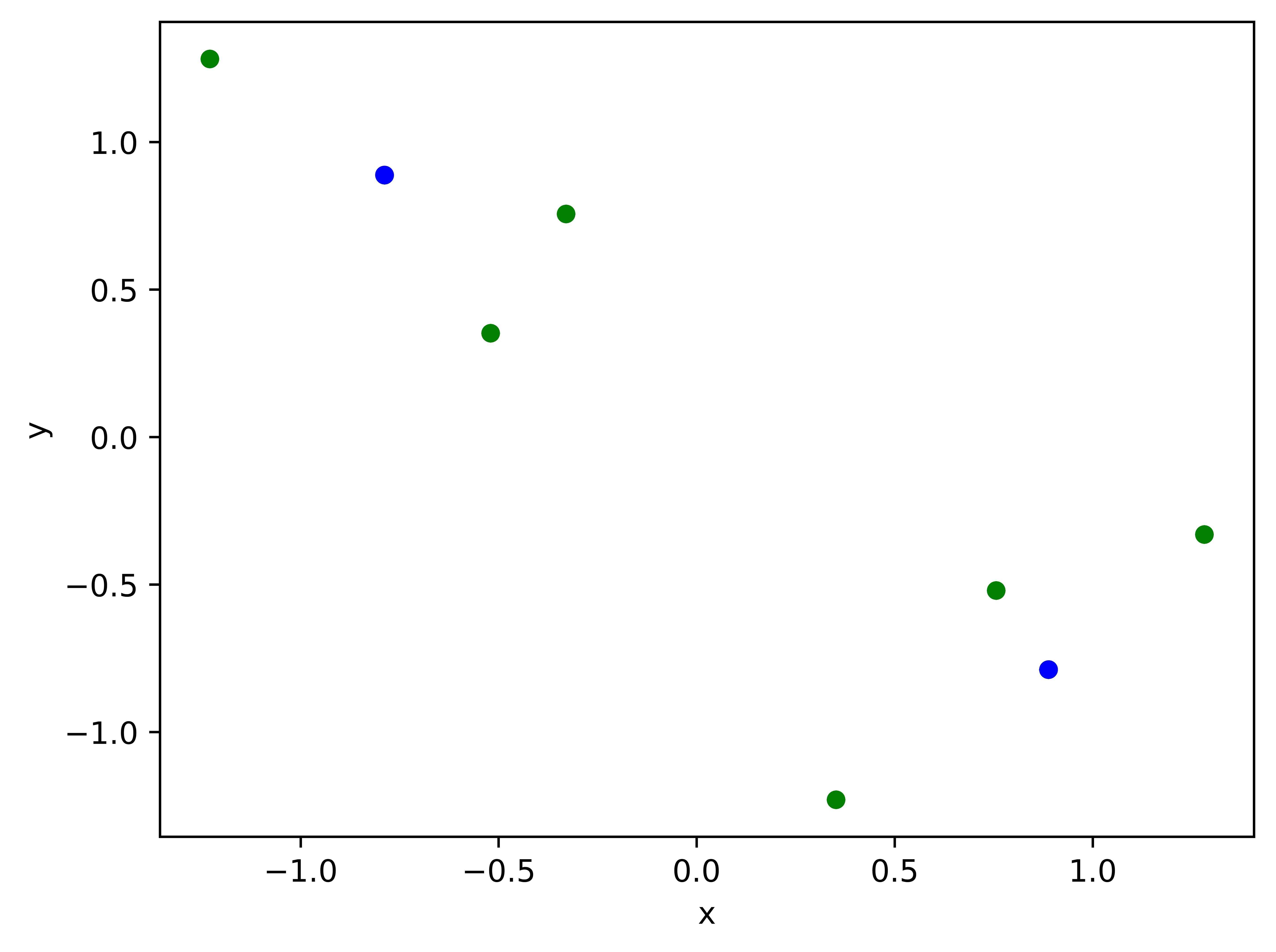}
    \caption{Attractors of the map in Equation \ref{eq:alternative-henon-map} with parameters $a=0.71$ and $b=0.9$, where the blue points are a 2-cycle attractor and the green points are a 6-cycle attractor, graphed with the code in Appendix \ref{alternate-henon-map-attractors-code}}
    \label{fig:alternate-henon-attractors}
\end{figure*}
\begin{figure*}
    \centering
    \begin{subfigure}{0.495\textwidth}
    \centering
        \includegraphics[scale=0.06]{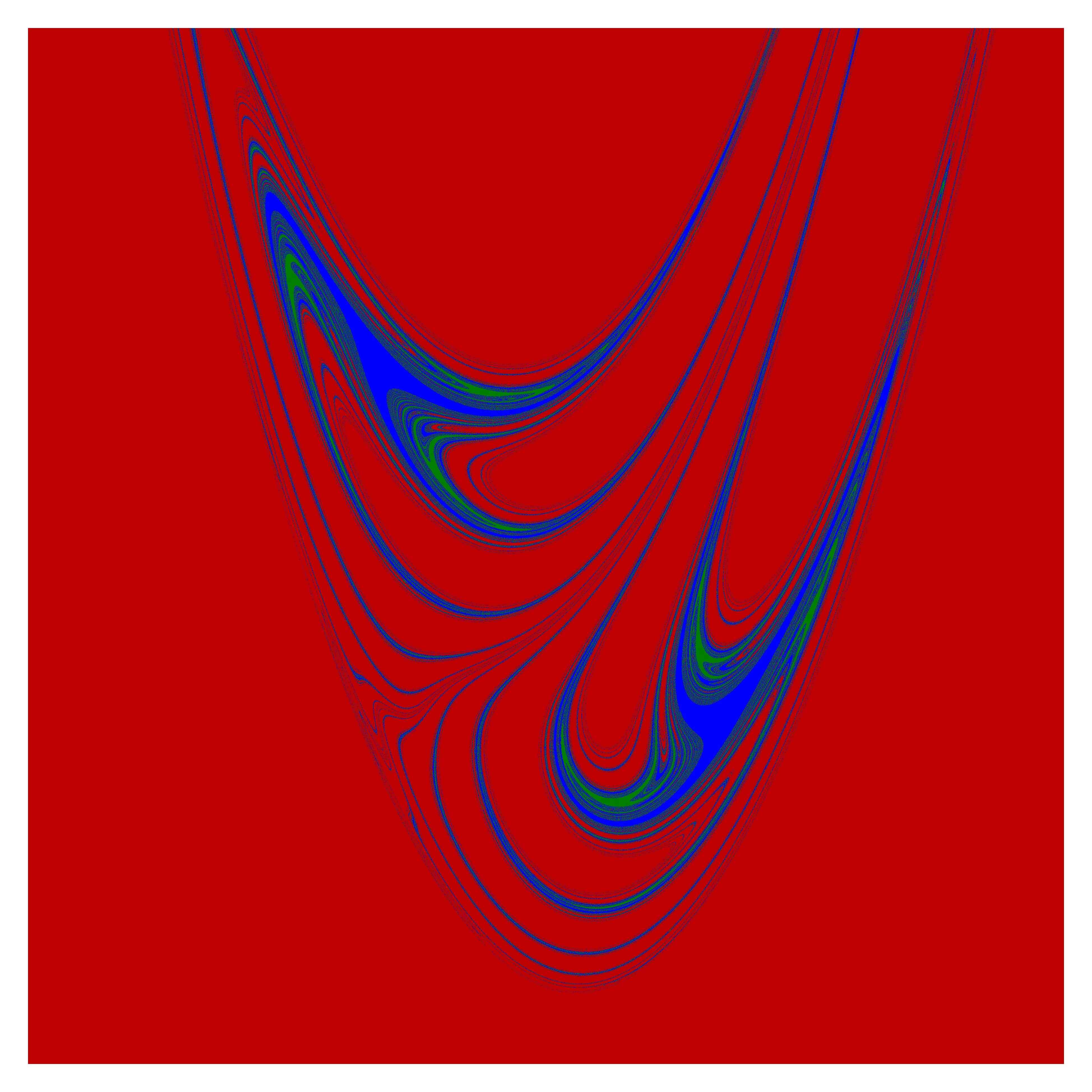}
        \caption{$-2.5\leq x\leq 2.5$, $-2.5\leq y\leq 2.5$}
        \label{fig:henon-wada-basins-1}
    \end{subfigure}
    \hfill
    \begin{subfigure}{0.495\textwidth}
        \centering
        \includegraphics[scale=0.075]{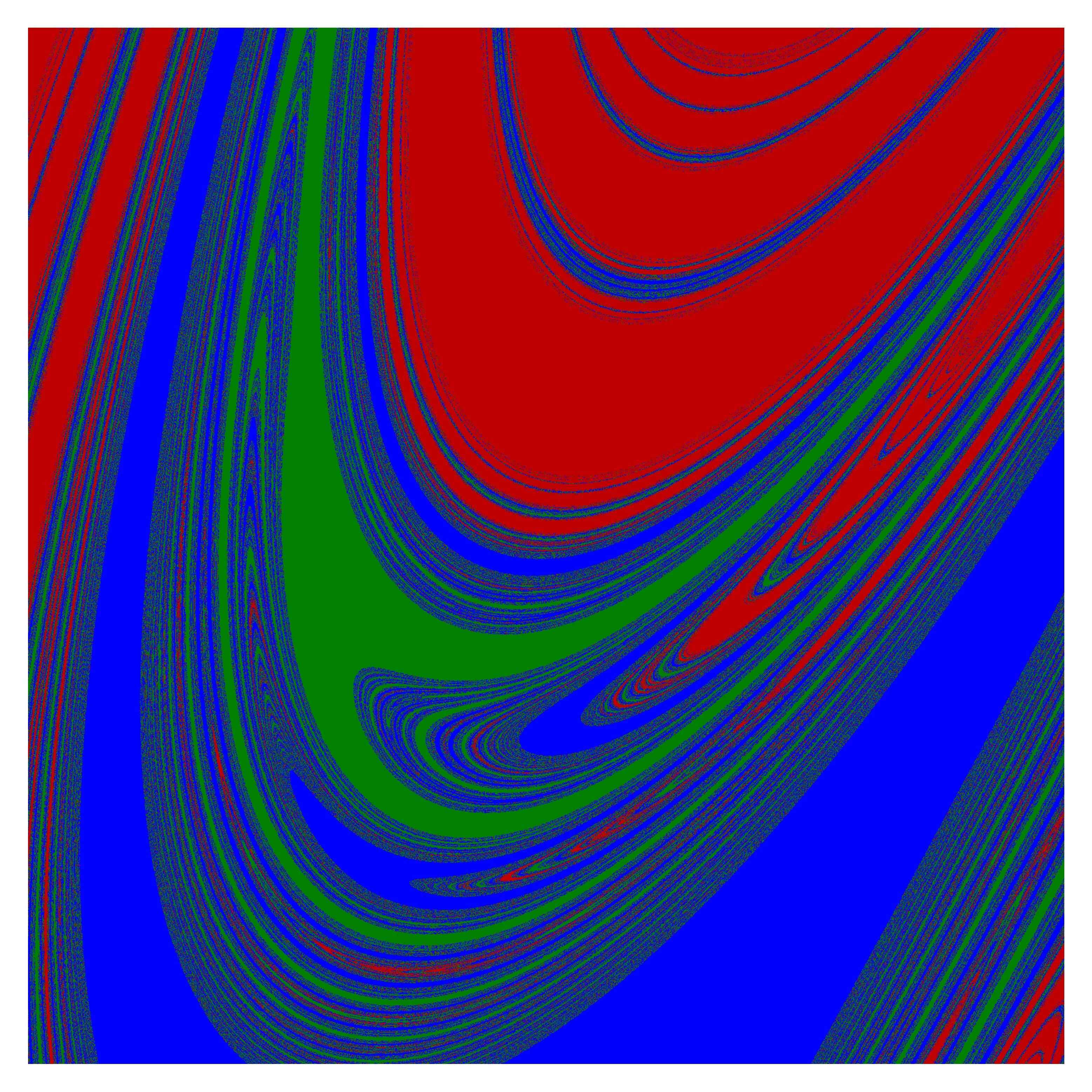}
        \caption{$0.6\leq x\leq 1.1$, $-0.75\leq y\leq -0.25$}
        \label{fig:henon-wada-basins-2}
    \end{subfigure}
    \caption{Basins of attraction of the map in Equation \ref{eq:alternative-henon-map} with parameters $a=0.71$ and $b=0.9$, where the red area represents the basin of infinity, the blue area represents the basin of a 2-cycle attractor, and the green area represents the basin of a 6-cycle attractor, graphed with the code in Appendix \ref{visualizing-wada-henon-code}}
    \label{fig:henon-wada-basins}
\end{figure*}

\begin{table}[b]
    \centering
    \hfill
    \begin{tabular}{c|c|c|c}
        $p$ & $G_1^p$ & $G_2^p$ & $G_3^p$ \\ [2px]
        \hline
        1 & 7186 & 1286 & 1528 \\
        2 & 7186 & 1117 & 1697 \\
        3 & 7186 & 892 & 1922 \\
        4 & 7186 & 732 & 2082 \\
        5 & 7186 & 581 & 2233 \\
        6 & 7186 & 512 & 2302 \\
        7 & 7186 & 475 & 2339 \\
    \end{tabular}
    \hfill
    \begin{tabular}{c|c|c|c}
        $p$ & $G_1^p$ & $G_2^p$ & $G_3^p$ \\ [2px]
        \hline
        8 & 7186 & 446 & 2368 \\
        9 & 7186 & 410 & 2404 \\
        10 & 7186 & 385 & 2429 \\
        11 & 7186 & 347 & 2467 \\
        12 & 7186 & 324 & 2490 \\
        13 & 7186 & 280 & 2534 \\
        14 & 7186 & 250 & 2564 \\
    \end{tabular}
    \hfill
    \caption{Some $G_k^p$ values for a $100\times 100$ grid over the region in Figure \ref{fig:henon-wada-basins-1}, calculated using the code in Appendix \ref{detecting-wada-alternate-henon-code}}
    \label{tab:G_k^p-values-alternate-henon}
\end{table}
\begin{figure}[hb!]
    \centering
    \begin{tikzpicture}[scale=0.4]
        \draw[->] (0,0)--(0,15.5);
        \draw[->] (0,0)--(15,0);
        \path (-2, 0)--(18, 0);
        \filldraw[thick] (0,15.04) circle [radius=0.15];
        \filldraw[thick] (1,12.86) circle [radius=0.15];
        \filldraw[thick] (2,11.17) circle [radius=0.15];
        \filldraw[thick] (3,8.92) circle [radius=0.15];
        \filldraw[thick] (4,7.32) circle [radius=0.15];
        \filldraw[thick] (5,5.81) circle [radius=0.15];
        \filldraw[thick] (6,5.12) circle [radius=0.15];
        \filldraw[thick] (7,4.75) circle [radius=0.15];
        \filldraw[thick] (8,4.46) circle [radius=0.15];
        \filldraw[thick] (9,4.10) circle [radius=0.15];
        \filldraw[thick] (10,3.85) circle [radius=0.15];
        \filldraw[thick] (11,3.47) circle [radius=0.15];
        \filldraw[thick] (12,3.24) circle [radius=0.15];
        \filldraw[thick] (13,2.80) circle [radius=0.15];
        \filldraw[thick] (14,2.50) circle [radius=0.15];
        \node[below] at (8,-1.5) {$p$};
        \node[left, rotate=90] at (-2.25,7.85) {$G_2^p$};
        \draw (0, 0)--(0, -0.2); 
        \draw (1, 0)--(1, -0.2);
        \draw (2, 0)--(2, -0.2);
        \draw (3, 0)--(3, -0.2); 
        \draw (4, 0)--(4, -0.2);
        \draw (5, 0)--(5, -0.2);
        \draw (6, 0)--(6, -0.2);
        \draw (7, 0)--(7, -0.2); 
        \draw (8, 0)--(8, -0.2);
        \draw (9, 0)--(9, -0.2);
        \draw (10, 0)--(10, -0.2); 
        \draw (11, 0)--(11, -0.2);
        \draw (12, 0)--(12, -0.2);
        \draw (13, 0)--(13, -0.2);
        \draw (14, 0)--(14, -0.2);
        \node[below] at (0, -0.2) {\scriptsize $0$};
        \node[below] at (4, -0.2) {\scriptsize $4$};
        \node[below] at (8, -0.2) {\scriptsize $8$};
        \node[below] at (12, -0.2) {\scriptsize $12$};
        \draw (0, 0)--(-0.2, 0);
        \draw (0, 1)--(-0.2, 1);
        \draw (0, 2)--(-0.2, 2);
        \draw (0, 3)--(-0.2, 3);
        \draw (0, 4)--(-0.2, 4);
        \draw (0, 5)--(-0.2, 5);
        \draw (0, 6)--(-0.2, 6);
        \draw (0, 7)--(-0.2, 7);
        \draw (0, 8)--(-0.2, 8);
        \draw (0, 9)--(-0.2, 9);
        \draw (0, 10)--(-0.2, 10);
        \draw (0, 11)--(-0.2, 11);
        \draw (0, 12)--(-0.2, 12);
        \draw (0, 13)--(-0.2, 13);
        \draw (0, 14)--(-0.2, 14);
        \draw (0, 15)--(-0.2, 15);
        \node[left] at (-0.2, 0) {\scriptsize $0$};
        \node[left] at (-0.2, 2) {\scriptsize $200$};
        \node[left] at (-0.2, 4) {\scriptsize $400$};
        \node[left] at (-0.2, 6) {\scriptsize $600$};
        \node[left] at (-0.2, 8) {\scriptsize $800$};
        \node[left] at (-0.2, 10) {\scriptsize $1000$};
        \node[left] at (-0.2, 12) {\scriptsize $1200$};
        \node[left] at (-0.2, 14) {\scriptsize $1400$};
    \end{tikzpicture}
    \caption{A plot of some of the points in Table \ref{tab:G_k^p-values-alternate-henon} on a graph of $G_2^p$ vs. $p$, showing that $G_2^p\to0$ as $p\to\infty$}
    \label{fig:G_k^p-graph-alternate-henon}
\end{figure}
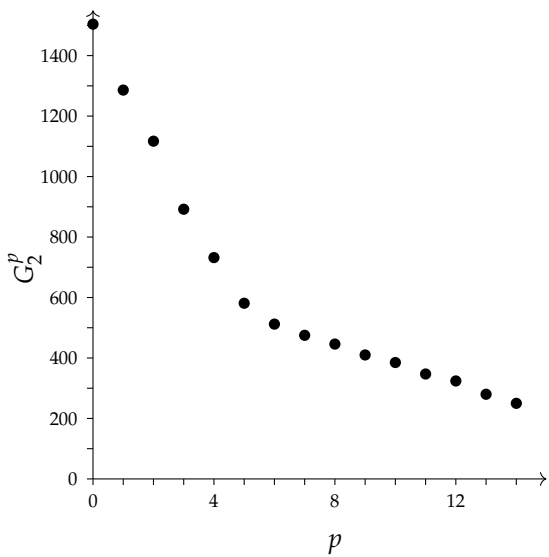

As an example of a dynamical system with three attractors, let us consider an alternative Hénon map defined by the function
\begin{equation}
    \mathbf{f}(\mathbf{x};\,a,\,b) = \begin{pmatrix}
        a-x^2+by \\
        x
    \end{pmatrix}
    \label{eq:alternative-henon-map}
\end{equation}
with parameters $a=0.71$ and $b=0.9$. In this system, there are three attractors: a 2-cycle, a 6-cycle, and infinity \cite{nusse}. The two periodic orbit attractors are shown in Figure \ref{fig:alternate-henon-attractors} using the code in Appendix \ref{alternate-henon-map-attractors-code}, with the 2-cycle being represented by blue points and the 6-cycle being represented by green points. In Figure \ref{fig:henon-wada-basins-1}, we visualize the basins of this map using the code in Appendix \ref{visualizing-wada-henon-code}, with the red area being the basin of infinity, the blue area being the basin of the 2-cycle, and the green area being the basin of the 6-cycle. Zooming in (Figure \ref{fig:henon-wada-basins-2}), we can see complex geometry on the boundary of the basins, not only being fractal, but also red, blue, and green points being intermingled with each other. For this reason, we suspect the basins might be Wada, so we implement our grid algorithm for detecting the Wada property in Appendix \ref{detecting-wada-alternate-henon-code} using a $100\times 100$ grid layered over the region in Figure \ref{fig:henon-wada-basins-1}. The results are shown in Table \ref{tab:G_k^p-values-alternate-henon}, and a plot $G_2^p$ against $p$ is shown in Figure \ref{fig:G_k^p-graph-alternate-henon}. We can see that $G_2^p$ is tending to 0, so we can assume that Equation \ref{eq:wada-condition} is satisfied. Therefore, all boundary points are Wada points, meaning this system exhibits the Wada property.

\subsubsection{Milnor Attractors and Riddled Basins}

In order to discuss the concept of riddled basins of attraction, we will first need to establish Milnor's definition of an attractor \cite{milnor}, which is different from the definition we established in Section \ref{nonchaoticattractors}. We will consider a subset of state space $\mathcal{A}\subset\mathbb{R}^n$ to be a Milnor attractor if it satisfies the following two properties:
\begin{enumerate}
    \item The set of all points that $\mathcal{A}$ attracts, or its basin of attraction $\hat{\mathcal{A}}$, has a strictly positive measure. That is, $\mu(\hat{\mathcal{A}})>0$.
    \item There is no $\mathcal{A}'\subset\mathcal{A}$ such that $\hat{\mathcal{A}}'=\hat{\mathcal{A}}$.
\end{enumerate}
Recall from Section \ref{nonchaoticattractors} that we required a standard attractor $A$ to attract an open set of initial conditions $U$. In Property 1 of Milnor attractors, we loosen this requirement to say that a Milnor attractor $\mathcal{A}$ attracts a region with positive measure, meaning the basin of $\mathcal{A}$ takes up a non-zero ``volume'' of state space but is not necessarily an open set. Another way of viewing this is that if $\mathcal{A}$ is a Milnor attractor, there exists a bounded region of state space such that if a state is randomly chosen from the region, there is a non-zero probability that the state will be attracted to $\mathcal{A}$ \cite{ott-riddled}. Property 2 just ensures that every part of $\mathcal{A}$ plays a role in the attractor, meaning we can't take out a part of $\mathcal{A}$ and still attract the same set of initial conditions \cite{milnor-wiki}.

When a certain kind of symmetry exists in a chaotic system, Milnor attractors of the system may be associated with basins that are riddled. Riddled basins of attraction occur in systems with, because of some kind of symmetry, a smooth invariant manifold, which is essentially a smooth surface in state space with the property that any initial condition in the surface will have a forward orbit that stays in the surface \cite[p. 15]{pisarchikbook}. A riddled basin of attraction $\hat{\mathcal{A}}$ has the property that every point $\mathbf{a}\in\hat{\mathcal{A}}$ has points arbitrarily close to it belonging to another basin $\hat{\mathcal{C}}$ \cite{ott-riddled}. Specifically, for any $\epsilon>0$ and any $\mathbf{a}\in\hat{\mathcal{A}}$, an $n$-dimensional ball centered at $\mathbf{a}$ with radius $\epsilon$, namely, $|\mathbf{x}-\mathbf{a}|<\epsilon$, will always contain a positive measure of points that get attracted to $\mathcal{C}$ instead of $\mathcal{A}$. In other words, even though $\hat{\mathcal{A}}$ has a positive measure, the set $\hat{\mathcal{A}}$ and its boundary set are the same because every neighborhood of every point in $\hat{\mathcal{A}}$ contains points both inside and outside of the set $\hat{\mathcal{A}}$ \cite{ottboa}. It is now clear why riddled basins can only be associated with a Milnor attractor and an attractor by our previous definition. Since $\hat{\mathcal{A}}$ is not an open set, it doesn't qualify as the basin of a standard attractor $A$. 

The practical implication of the riddled basin property is that we can never be sure whether a point $\mathbf{a}\in\hat{\mathcal{A}}$ will be attracted to $\mathcal{A}$ or some other attractor $\mathcal{C}$ because no matter how small we make the error $\delta\mathbf{a}$, there will always be a chance $\mathbf{a} + \delta\mathbf{a}$ lies in $\hat{\mathcal{C}}$ rather than $\hat{\mathcal{A}}$. For this reason, the uncertainty exponent of a riddled basin is $\mathfrak{u}=0$, so the presence of a riddled basin is the most extreme version of sensitive dependence on initial conditions resulting from a geometrical property of a chaotic system that we have discussed so far.

However, an even stronger geometrical property leading to sensitive dependence on initial conditions in a system is the presence of intermingled basins. Intermingled basins occur in $n$-dimensional systems with $i\geq 2$ Milnor attractors $\mathcal{A}_1,\, \mathcal{A}_2,\,\hdots\,,\, \mathcal{A}_i$. If all the system's basins of attractions $\hat{\mathcal{A}}_1,\, \hat{\mathcal{A}}_2,\,\hdots\,,\, \hat{\mathcal{A}}_i$ are riddled with each other, then we say that the basins are intermingled. More rigorously, this means that for any point $\mathbf{a}\in\mathbb{R}^n$, any $n$-dimensional ball centered around $\mathbf{a}$ will contain a positive measure of points in each of the basins $\hat{\mathcal{A}}_1,\, \hat{\mathcal{A}}_2,\,\hdots\,,\, \hat{\mathcal{A}}_i$, meaning, for each attractor $\mathcal{A}_1,\, \mathcal{A}_2,\,\hdots\,,\, \mathcal{A}_i$, there is a non-zero probability that a random initial state in the ball $|\mathbf{x}-\mathbf{a}|<\epsilon$ will go to that attractor \cite{alexander}.

Proving that a given basin $\hat{\mathcal{A}}$ is riddled requires two steps \cite{alexander}:
\begin{enumerate}
    \item Show that enough points are attracted to $\mathcal{A}$ to say that the probability of that happening is positive. In the paper by Alexander \textit{et al.} \cite{alexander}, a method is described to do this using Lyapunov exponents.
    \item Show that sufficiently many points are repelled from $\mathcal{A}$. According to Alexander \textit{et al.} \cite{alexander}, in systems with riddled basins, this is usually implied by symmetry.
\end{enumerate}
Riddled basins can be found in models for a variety of different systems in many different fields. Some of these include electronic circuits, coupled chaotic oscillators, learning and mechanical systems, and interacting populations \cite[p. 17]{pisarchikbook}. However, most ordinary systems do not contain smooth invariant manifolds, which are the requirement for riddled basins to exist, so riddled basins do not usually appear in practice \cite{ottboa}. For this reason, we will not discuss in detail how to implement the two steps above, nor the mathematical theory behind smooth invariant manifolds. Our main purpose in discussing the basics behind riddled basins was simply to introduce the most extreme version of sensitivity to initial conditions emerging from geometrical aspects of dynamical systems. However, they will not be relevant in our analysis and discussion of the Rulkov maps.

\section{Slow-Fast Systems and Dynamics}
\label{slow-fast-systems-and-dynamics}

Slow-fast systems are dynamical systems that involve several variables evolving on different time scales \cite{zheng}. Dynamical systems with separated slow and fast variables are invaluable for modeling many real-world applications, including lasers, chemical reactions, optoelectronic systems, and ecological systems \cite{omelchenko, zheng}. For our purposes, however, we are interested in using slow-fast systems to model the behavior of biological neurons. 

We will first discuss the continuous-time case of a two-variable slow-fast system. From Equation \ref{eq:diffeq-general}, a two-dimensional continuous-time dynamical system can be represented as
\begin{equation}
    \begin{pmatrix}
        \dot x \\
        \dot y
    \end{pmatrix}
    =
    \begin{pmatrix}
        \frac{dx}{dt} \\[3pt]
        \frac{dy}{dt}
    \end{pmatrix}
    =
    \begin{pmatrix}
        g^{[1]}(x,\,y) \\
        g^{[2]}(x,\,y) \\
    \end{pmatrix}
\end{equation}
This system can be transformed into the standard form of a slow-fast system as shown by Ginoux \cite[p. 70]{ginoux} by defining $g^{[1]}(x,\,y) = \chi(x,\,y)$ and $g^{[2]}(x,\,y) = \eta \omega(x,\,y)$, where $0<\eta\ll 1$. Then,
\begin{equation}
    \begin{pmatrix}
        \dot x \\
        \dot y
    \end{pmatrix}
    =
    \begin{pmatrix}
        \chi(x,\,y) \\
        \eta \omega(x,\,y) \\
    \end{pmatrix}
    \label{eq:slow-fast-def-fast-time}
\end{equation}
Assuming $\chi$ and $\omega$ are approximately on the same time scale, since $\eta$ is a small number, $\dot y$ is small at any given time, meaning $y$ changes slowly. For this reason, we call $x$ the fast variable and $y$ the slow variable. Often, we can approximate the dynamics of a slow-fast system by analyzing the fast and slow dynamics separately, treating $y$ as a slowly drifting parameter of $\frac{dx}{dt}$. However, to accurately describe the system's dynamics, we must analyze the slow and fast dynamics together as a two-dimensional map.

While slow-fast dynamics in continuous-time dynamical systems have been studied extensively, discrete-time slow-fast systems have received far less attention. In fact, an agreed-upon form of this type of system has not yet even been established. However, we will define a slow-fast discrete-time dynamical system as one governed by the iteration function
\begin{equation}
    \begin{pmatrix}
        x_{k+1} \\
        y_{k+1}
    \end{pmatrix}
    =
    \begin{pmatrix}
        f(x_k,\,y_k) \\
        y_k + \eta g(x_k,\,y_k)
    \end{pmatrix}
\end{equation}
where $0<\eta\ll 1$. In this case, $x$ is the fast variable and $y$ is the slow variable, as each iteration changes $y$ by only a small amount. In this section, we will explore the dynamics and relevance of both continuous-time and discrete-time slow-fast systems, with a specific application to the dynamics and properties of neuronal models.

\subsection{Neimark-Sacker Bifurcations}
\label{neimark-sacker-bifurcations}

A type of bifurcation known as the Andronov-Hopf bifurcation occurs commonly in continuous-time slow-fast systems, as well as dynamical systems governed by ordinary differential equations in general. The sister to the Andronov-Hopf bifurcation for discrete-time dynamical systems is known as the Neimark-Sacker bifurcation. In this section, we will present a brief overview of the dynamics and qualitative geometry of the Neimark-Sacker bifurcation, which will be useful in analyzing our discrete-time slow-fast neuron systems.

To illustrate how the Neimark-Sacker bifurcation arises, consider a two-dimensional discrete-time dynamical system with a dependence on some parameter $\alpha$:
\begin{equation}
    \mathbf{x}_{k+1} = \mathbf{f}(\mathbf{x}_k;\,\alpha) = \begin{pmatrix}
        f^{[1]}(x_k,\,y_k;\,\alpha) \\
        f^{[2]}(x_k,\,y_k;\,\alpha)
    \end{pmatrix}
    \label{eq:alpha-parameter-2d-map}
\end{equation}
Recall from Section \ref{quantification} that the Jacobian matrix of this system is
\begin{equation}
    J(\mathbf{x}) = \begin{pmatrix}
        \frac{\partial f^{[1]}}{\partial x} & \frac{\partial f^{[1]}}{\partial y} \\[4pt]
        \frac{\partial f^{[2]}}{\partial x} & \frac{\partial f^{[2]}}{\partial y}
    \end{pmatrix}
\end{equation}
Now, assume the system has a fixed point $\mathbf{x}_s(\alpha)$. Recall from Section \ref{nonchaoticattractors} that $\mathbf{x}_s(\alpha)$ is an attractor if $|\nu_{1,\,2}(\alpha)|<1$, but it is a repeller if $|\nu_{1,\,2}(\alpha)|>1$, where $\nu_{1,\,2}(\alpha)$ are the eigenvalues of $J(\mathbf{x}_s(\alpha))$. This set of criteria also works if $\nu_{1,\,2}(\alpha)$ are complex numbers, where $|\nu_{1,\,2}(\alpha)|$ represents the modulus of the eigenvalues.\footnote{For a brief review of the complex algebra used in this section, see Appendix \ref{complex-algebra}.}
A Neimark-Sacker bifurcation can occur in the case where $J(\mathbf{x}_s(\alpha))$ has a complex conjugate pair of eigenvalues \cite{kuznetsov-ns}:
\begin{equation}
    \nu_{1,\,2}(\alpha) = r(\alpha)e^{\pm i\varphi(\alpha)}
    \label{eq:complex-conj-eigenvals}
\end{equation}
where $r(\alpha) = |\nu_{1,\,2}(\alpha)|$ and $\varphi(\alpha) = \Arg(\nu_{1}(\alpha))$. Just like the period-doubling bifurcation, the Neimark-Sacker bifurcation involves the fixed point $\mathbf{x}_s(\alpha)$ changing its stability, either from an attractor to a repeller or from a repeller to an attractor. Specifically, a Neimark-Sacker bifurcation occurs at a given $\alpha = \alpha_0$ if the following two conditions are satisfied \cite{kuznetsov-ns}:\footnote{The requirements are actually more stringent than this, but we only cover the basics of the Neimark-Sacker bifurcation in this paper. For more details, an interested reader is recommended to see the original chapter detailing this kind of bifurcation in the dissertation by Sacker \cite{sacker}.}
\begin{enumerate}
    \item Following directly from the criteria for fixed point attractiveness,
    \begin{equation}
        r(\alpha_0) = 1
    \end{equation}
    In other words, the eigenvalues lie on the unit circle $|z|=1$ at $\alpha_0$.
    \item Also from the criteria for fixed point attractiveness,
    \begin{equation}
        \frac{dr}{d\alpha}\bigg|_{\alpha=\alpha_0} \neq 0
    \end{equation}
    This ensures that the eigenvalues $\nu_{1,\,2}(\alpha)$ are passing through the unit circle $z=1$ at a non-zero speed, which is necessary for the fixed point to change its stability.
\end{enumerate}
Therefore, the eigenvalues of the Jacobian at a fixed point undergoing a Neimark-Sacker bifurcation are, applying Condition 1 to Equation \ref{eq:complex-conj-eigenvals},
\begin{equation}
    \nu_{1,\,2}(\alpha_0) = e^{\pm i\varphi}
    \label{eq:eigenvalues-at-ns-bifurcation}
\end{equation}

In addition to a fixed point changing stability, Neimark-Sacker bifurcations are also characterized by the birth or death of an attracting or repelling periodic orbit at the fixed point $\mathbf{x}_s$. Specifically, there are two main types of Neimark-Sacker bifurcations: supercritical and subcritical. A supercritical bifurcation involves the birth or death of an attracting periodic orbit, while a subcritical bifurcation involves the birth or death of a repelling periodic orbit. We will demonstrate these Neimark-Sacker bifurcation types for the case where, from Condition 2 of Neimark-Sacker bifurcations, $r'(\alpha_0) > 0$. This means that $|\nu_{1,\,2}(\alpha_0)|$ is less than 1 before the bifurcation and greater than 1 after, indicating that the fixed point turns from an attractor to a repeller.

\begin{figure*}
    \centering
    \begin{subfigure}[b]{0.25\textwidth}
        \centering
        \includegraphics[scale=0.3]{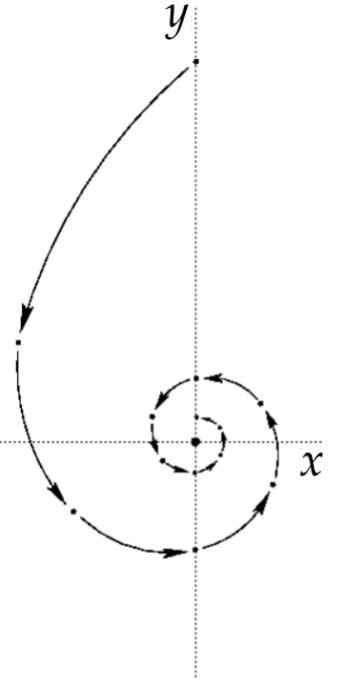}
        \vspace{4px}
        \caption{$\alpha < \alpha_0$}
        \label{fig:supercrit_a<0}
    \end{subfigure}
    \hfill
    \begin{subfigure}[b]{0.3\textwidth}
        \centering
        \includegraphics[scale=0.3]{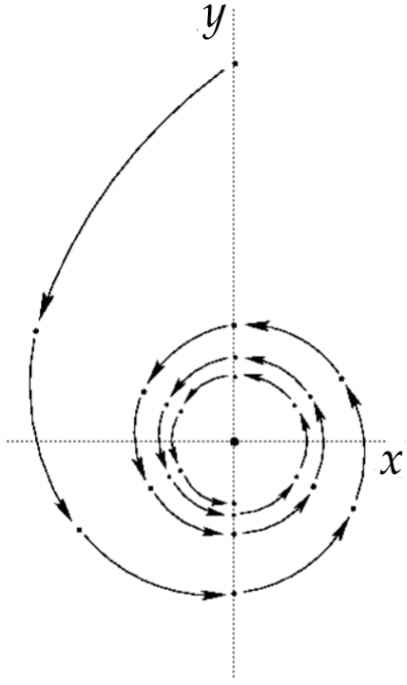}
        \vspace{4px}
        \caption{$\alpha = \alpha_0$}
        \label{fig:supercrit_a=0}
    \end{subfigure}
    \hfill
    \begin{subfigure}[b]{0.35\textwidth}
        \centering
        \includegraphics[scale=0.3]{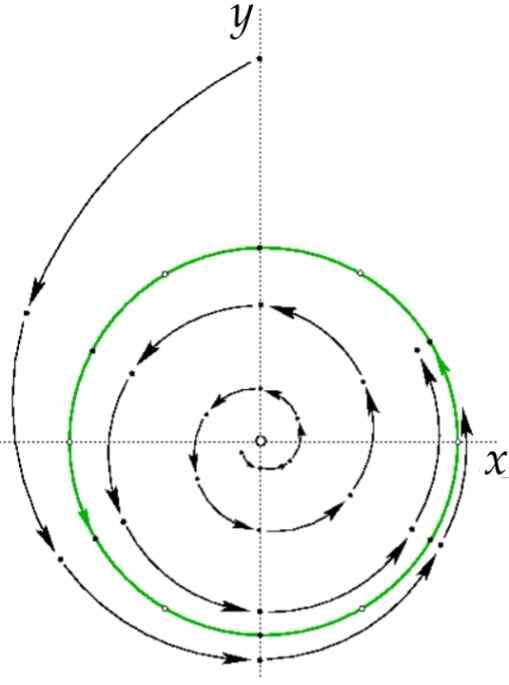}
        \vspace{4px}
        \caption{$\alpha > \alpha_0$}
        \label{fig:supercrit_a>0}
    \end{subfigure}
    \vspace{2px}
    \caption{Diagrams before, at, and after a supercritical Neimark-Sacker bifurcation of the fixed point $\mathbf{x}_s = \langle 0,\, 0 \rangle$ in a two-dimensional map, showing the birth of an attracting periodic orbit, from \cite{kuznetsov-ns}}
    \label{fig:supercritical-ns-bifurcation}
\end{figure*}

Figure \ref{fig:supercritical-ns-bifurcation} shows a graphical illustration of a supercritical Neimark-Sacker bifurcation of the fixed point $\mathbf{x}_s = \langle 0,\, 0 \rangle$. Before the bifurcation, we can see in Figure \ref{fig:supercrit_a<0} that $\mathbf{x}_s$ is attracting. At the bifurcation $\alpha = \alpha_0$, we can see in Figure \ref{fig:supercrit_a=0} that  $\mathbf{x}_s$ is still attracting, but much more weakly so. Finally, after the bifurcation, $\mathbf{x}_s$ is repelling, and an attracting periodic orbit is born from $\mathbf{x}_s$, expanding out from the fixed point as $\alpha$ increases. In Figure \ref{fig:supercrit_a>0}, we can see how an initial state starting near $\mathbf{x}_s$ is repelled from the fixed point and attracted to the green periodic orbit.

\begin{figure*}
    \centering
    \begin{subfigure}[b]{0.35\textwidth}
        \centering
        \includegraphics[scale=0.3]{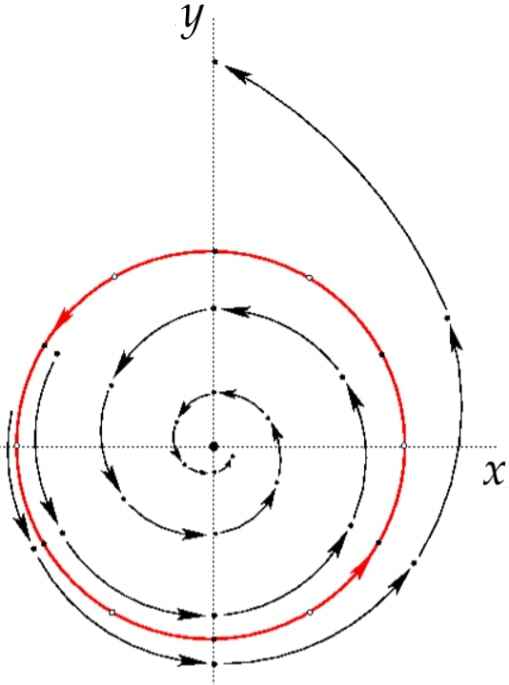}
        \vspace{4px}
        \caption{$\alpha < \alpha_0$}
        \label{fig:subcrit_a<0}
    \end{subfigure}
    \hfill
    \begin{subfigure}[b]{0.3\textwidth}
        \centering
        \includegraphics[scale=0.3]{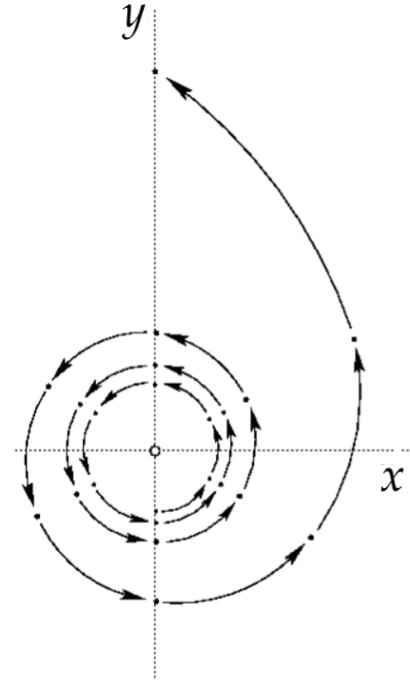}
        \vspace{4px}
        \caption{$\alpha = \alpha_0$}
        \label{fig:subcrit_a=0}
    \end{subfigure}
    \hfill
    \begin{subfigure}[b]{0.25\textwidth}
        \centering
        \includegraphics[scale=0.3]{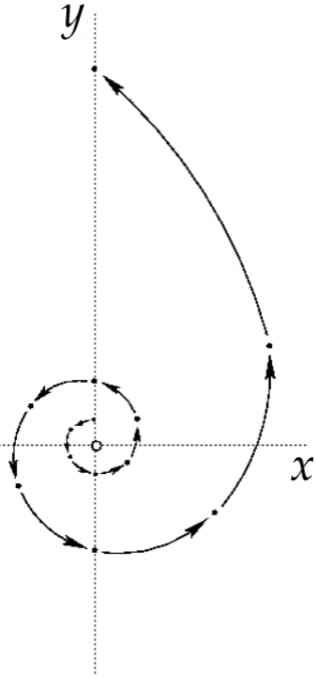}
        \vspace{4px}
        \caption{$\alpha > \alpha_0$}
        \label{fig:subcrit_a>0}
    \end{subfigure}
    \vspace{2px}
    \caption{Diagrams before, at, and after a subcritical Neimark-Sacker bifurcation of the fixed point $\mathbf{x}_s = \langle 0,\, 0 \rangle$ in a two-dimensional map, showing the death of a repelling periodic orbit, from \cite{kuznetsov-ns}}
    \label{fig:subcritical-ns-bifurcation}
\end{figure*}

Examining a similar system, Figure \ref{fig:subcritical-ns-bifurcation} shows a graphical illustration of a subcritical Neimark-Sacker bifurcation of the same fixed point. Before the bifurcation, we can see in Figure \ref{fig:subcrit_a<0} that there is a repelling periodic orbit, shown in red. An initial condition that starts outside the red periodic orbit is repelled to infinity, and an initial condition that starts inside the red periodic orbit is repelled from it and attracted to the fixed point $\mathbf{x}_s$. As $\alpha$ increases to $\alpha_0$, the repelling period orbit tightens around $\mathbf{x}_s$ like a noose, until at $\alpha_0$, the repelling orbit closes in on $\mathbf{x}_s$, which effectively makes $\mathbf{x}_s$ a repelling fixed point. We can see this in Figure \ref{fig:subcrit_a=0}, where the fixed point is weakly repelling initial conditions. Finally, in Figure \ref{fig:subcrit_a>0}, we can see $\mathbf{x}_s$ acting as a true repeller, sending initial conditions off to infinity (or some other far away attractor).

Although there are some ways to analytically determine whether a Neimark-Sacker bifurcation is supercritical or subcritical,\footnote{See the article by Kuznetsov and Sacker \cite{kuznetsov-ns} for a brief overview on how to do this.} it is beyond the scope of this paper. However, a quick and easy way to infer whether a given Neimark-Sacker bifurcation is supercritical or subcritical is to model the system numerically \cite[p. 256]{strogatz}.

It is worthwhile to note that, when there is a Neimark-Sacker bifurcation at $\alpha=\alpha_0$, the determinant\footnote{The determinant of a $2\times 2$ matrix is the product of its off-diagonal entries subtracted from the product of its diagonal entries.} of the Jacobian of the fixed point at $\alpha=\alpha_0$ is
\begin{equation}
    \det J(\mathbf{x}_s(\alpha_0)) = 1
    \label{eq:det-jacobian-ns-bif}
\end{equation}
and its trace\footnote{The trace of a matrix is the sum of its diagonal entries.} is
\begin{equation}
    \tr J(\mathbf{x}_s(\alpha_0)) = 2\cos\varphi
    \label{eq:trace-jacobian-ns-bif}
\end{equation}
We derive these Jacobian properties in Appendix \ref{det-and-trace-ns}.

\subsection{Behavior and Modeling of Biological Neurons}
\label{behavior_and_modeling_of_biological_neurons}

One of the many applications of dynamical systems theory is in modeling the behavior of biological neurons, which is often done using a slow-fast system. Neurons are cells within the nervous system that transmit and propagate messages to and from the brain via electrical and chemical signals. They can be broadly classified into three categories: sensory neurons, motor neurons, and interneurons, which allow living organisms to receive chemical and physical sensory inputs, voluntarily and involuntarily move, and respond to sensory inputs by transmitting signals and producing a motor response, respectively.

\subsubsection{Spiking Behavior}

Neurons are said to ``fire'' when an action potential takes place. These action potentials, or spikes, are sharp electric potentials across a cell's membrane that propagate signals to other neurons. Mimicking the behavior of biological neurons, mathematical neuron models represent the electric potential across a given neuron's cell membrane and depict its action potentials as spikes in the voltage. The action potentials occur as the result of an accumulation of electrical inputs which are referred to as synaptic potentials, with the name being derived from the synapse, or the gap between two neuronal cells across which neurotransmitters or electrical impulses are transmitted. 

Synaptic potentials are transmitted across the synapse to receiving cells, where they are then known as postsynaptic potentials. These inputs may either be excitatory, encouraging the firing of a cell, or inhibitory, which prevents neuronal firing. If the sum of all excitatory and inhibitory signals raises the membrane potential by enough to meet a given threshold potential for that cell, then an action potential will occur. The resting membrane potential of a neuronal cell is determined by the concentration gradient of ions across its cell membrane. The distribution of these ions results in a potential difference across the membrane that makes the cell polarized in this resting state. A neuron remains at this resting membrane potential until electrical signals depolarize it to its threshold potential. Synaptic potentials can generate an action potential in one of two ways: spatial summation or temporal summation. Spatial summation occurs when excitatory impulses from numerous synapses instantaneously converge on the same postsynaptic cell. Temporal summation, on the other hand, comes from consecutive impulses from one synapse to the postsynaptic neuron. When these impulses depolarize the cell, the action potential either will or won't take place depending on whether or not the voltage reaches the threshold potential. There is no phenomenon where a neuron only partially fires; neurons always fire at full strength.

\subsubsection{Neuron Electrophysiology} Ion currents through a given neuron's cell membrane govern all electrical behavior within it. The four main ions whose movements drive action potentials are sodium (Na$^+$), potassium (K$^+$), calcium (Ca$^{2+}$), and chloride (Cl$^-$). The electrochemical gradient across a cell's membrane is typically caused by a high concentration of Na$^+$, Cl$^-$, and Ca$^{2+}$ outside the cell's membrane, as well as a high concentration of K$^+$ and A$^-$ (other anions) within the cell's membrane \cite[p. 26]{izhikevich-book}. Together, the concentration gradient caused by the varying amounts of these ions on either side of the membrane and the electric field gradient caused by the differing charges of the ions create an electrochemical gradient. In a quiescent state, both active and passive transport allow this gradient to be maintained. 

Channel proteins within the cell membrane facilitate the movement of ions; however, Na$^+$ and Ca$^{2+}$ movement is relatively insignificant in the resting state. Additionally, the A$^-$ anions cannot diffuse through the membrane via channel proteins, but they enable passive transport by attracting K$^+$ ions into the cell and repelling Cl$^-$ ions out of the cell by Coulomb's law. In addition to this passive transport, concentration gradients are maintained by active transport from ion pumps such as the Na$^+$-K$^+$ pump which moves 3 Na$^+$ ions out of the cell and brings 2 K$^+$ ions into the cell.

Once a sensory or electrical input stimulates the neuron by depolarizing it to its threshold potential, the neuron will fire. The firing is caused by the opening of voltage-gated ion channels, which allow for specific ions to flow freely through the cell membrane. Once the cell is depolarized to the threshold potential, Na$^+$ and Ca$^{2+}$ channels will allow for the rapid flow of Na$^+$ ions and a mildly increased flow of Ca$^{2+}$ ions into the cell. This results in further depolarization of the cell at a very fast rate, causing the spike. Once the spike reaches its apex, the cell rapidly decreases in voltage since the higher voltage of the cell results in the inactivation of the Na$^+$ and Ca$^{2+}$ channels and triggers the voltage-gated outward K$^+$ and Cl$^-$ currents \cite[p. 25]{izhikevich-book}. These currents make the neuron's voltage rapidly decrease, completing the spike. Once the cell returns to its resting membrane potential, these gates take time to close, so the cell becomes hyperpolarized past the resting voltage. This increased negativity of the membrane potential prevents another action potential from occurring in the same place because a greater amount of stimulus is required to reach the threshold potential. As a result, an action potential will take place further along the cell membrane. The membrane potential returns to its resting voltage from the hyperpolarized state due to the membrane's natural permeability to K$^+$ and Na$^+$, which work to balance out the concentrations of the ions after the K$^+$ channels close \cite[p. 25]{izhikevich-book}.

\subsubsection{Nernst Potential}

The movement of each ion across the cell membrane is driven by the concentration gradient and electric potential gradient of each ion. For example, the concentration of K$^+$ ions is greater inside the cell than outside the cell, resulting in the diffusion of K$^+$ ions outwards in the resting state. As the positively charged ions exit the cell, a net negative charge is left inside the cell, making it more difficult for the K$^+$ ions to continue exiting the cell given the newly created charge gradient. Specifically, the positively charged ions will be more attracted to the negatively charged interior rather than the positively charged exterior. As the concentration and electrical gradients continue to act on the K$^+$ ions, an equilibrium state is achieved once the force of the concentration gradient is equal and opposite to the force of the electrical gradient. This equilibrium potential varies between different ions and is calculated using the Nernst equation \cite[p. 26]{izhikevich-book}:
\begin{equation}
    E_{\text{ion}}=\frac{RT}{zF}\ln{\frac{\text{[Ion]}_{\text{out}}}{\text{[Ion]}_{\text{in}}}}
    \label{eq:nernst}
\end{equation}
In this equation, $R$ is the universal gas constant (8,315 mJ/(K $\cdot$ mol)), $T$ is the temperature in Kelvin, $F$ is Faraday's constant (96,480 coulombs/mol), and $z$ is the charge of each ion. Using body temperature $T$ = 310 K and standard ion concentrations, we can calculate equilibrium potentials for all four ions with the value of $z$ being 1 for Na$^+$ and K$^+$, $-1$ for Cl$^-$, and 2 for Ca$^{2+}$. This yields a range of 61 to 90 mV for Na$^+$, whose extracellular concentration is 145 mM, while the intracellular concentration ranges from 5 to 15 mM. For K$^+$, whose average extracellular concentration is 5 mM and average intracellular concentration is 140 mM, the equilibrium potential is approximately $-90$ mV. Cl$^-$ has an extracellular concentration of 110 mM and an intracellular concentration of 4 mM, resulting in an equilibrium potential of $-89$ mV. Finally, the concentration of Ca$^{2+}$ outside the cell is between 2.5 and 5 mM and the concentration within the cell is 0.1 $\mu$M, yielding an equilibrium potential between 136 and 146 mV \cite[p. 26]{izhikevich-book}. We will refer to these equilibrium potentials as $E_{\mathrm{Na}}$, $E_{\mathrm{K}}$, $E_{\mathrm{Cl}}$, and $E_{\mathrm{Ca}}$, respectively.

When the membrane potential of any given ion is equal to the equilibrium potential for that ion, the current is zero by the definition of the Nernst equilibrium potential. The current through the cell membrane of any ion is proportional to the difference between these two potentials, where the membrane potential can be thought of as trying to ``chase'' the equilibrium potential at all times. The following equation represents this relationship:
\begin{equation}
    I_{\text{ion}}=g_{\text{ion}}(v-E_{\text{ion}})
\end{equation}
where $I_{\text{ion}}$ represents the current of a given ion, $v$ is the membrane potential, and $E_{\text{ion}}$ is, as mentioned previously, the equilibrium potential. The constant $g_{\text{ion}}$ represents the conductance (the reciprocal of resistance) of the ion and is measured in millisiemens per squared centimeter (mS/cm$^2$) \cite[p. 27]{izhikevich-book}.

\subsubsection{Hodgkin-Huxley Model}

The Hodgkin-Huxley model describes the dynamics of membrane potential changes in response to electrical stimuli. Through experimentation, Hodgkin and Huxley found that a neuron carries three major types of currents: the voltage-gated persistent K$^+$ current with four activation gates (resulting in the $n^4$ term in the equation below), the voltage-gated transient Na$^+$ current with three activation gates and one inactivation gate (resulting in the $m^3h$ term below), and the Ohmic leak current, which is carried mostly by Cl$^-$ ions \cite[p. 37]{izhikevich-book}. Their model is a system of differential equations:
\begin{align}
    \begin{split}
        C\dot V {}&= I - \overline{g}_{\mathrm{K}}n^4(V-E_{\mathrm{K}}) \\
        &\quad -\overline{g}_{\mathrm{Na}}m^3h(V-E_{\mathrm{Na}})-\overline{g}_L(V-E_{\mathrm{L}})
    \end{split}\\
    \dot n {}&= \alpha_n(V)(1-n)-\beta_n(V)n\\
    \dot m {}&= \alpha_m(V)(1-m)-\beta_m(V)m\\
    \dot h {}&= \alpha_h(V)(1-h)-\beta_h(V)h
\end{align}
where
\begin{align}
    \alpha_n(V) &= \frac{0.01(10-V)}{\exp[(10-V)/10]-1}\\
    \beta_n(V) &= 0.125\exp\left(-\frac{V}{80}\right)\\
    \alpha_m(V) &= \frac{0.1(25-V)}{\exp[(25-V)/10]-1}\\
    \beta_m(V) &= 4\exp\left(-\frac{V}{18}\right)\\
    \alpha_h(V) &= 0.07\exp\left(-\frac{V}{20}\right)\\
    \beta_h(V) &= \frac{1}{\exp[(30-V)/10]+1}
\end{align}
The Hodgkin-Huxley model is designed to represent the neuron as a basic circuit, describing the conductance changes of Na$^+$ and K$^+$ channels over time by modeling the voltage-dependent changes in ion channel conductance and the resulting membrane currents during an action potential. Here, $\overline{g}$ represents the maximum conductance of an ion and $C$ represents the capacitance of the neuron. In these equations, $n$, $m$, and $h$ are dimensionless variables that represent the activation and inactivation of the different voltage-gated ion channels. The variable $n$ is the activation gating variable for the K$^+$ channels, reflecting the proportion of K$^+$ channels that are open and available for conducting ions. As the membrane potential depolarizes, $n$ increases, leading to the opening of more K$^+$ channels and an increase in K$^+$ conductance. The variable $m$ is the activation gating variable for the Na$^+$ channels, reflecting the proportion of Na$^+$ channels that are open. Specifically, it is associated with the rapid activation of Na$^+$ channels in response to depolarization. Finally, $h$ is the inactivation gating variable for the Na$^+$ channels. Unlike $n$ and $m$, $h$ describes the proportion of Na$^+$ channels that are inactivated or unable to conduct ions, which typically follows their initial activation. As the cell depolarizes, $h$ decreases, allowing for the recovery of Na$^+$ channels from inactivation. In the equations, the parameters reflect a voltage that has been shifted by 65 mV in order to maintain a resting membrane potential of approximately 0 \cite[p. 37]{izhikevich-book}.

\subsubsection{Izhikevich Model}

The Izhikevich model of spiking behavior is a continuous-time neuron model governed by a two-dimensional system of ordinary differential equations. It is designed to satisfy and balance two seemingly mutually exclusive criteria defined by Izhikevich \cite{izhikevich-article}:
\begin{enumerate}
    \item The model must be computationally simple and efficient.
    \item The model must be capable of accurately capturing the dynamics and firing patterns of a biological neuron.
\end{enumerate}
The model, defined by the following differential equations, is designed to incorporate elements of the biophysically accurate Hodgkin-Huxley model while satisfying the first criterion of simplicity:
\begin{equation}
    \begin{pmatrix}
        \dot v \\
        \dot u
    \end{pmatrix}
    =
    \begin{pmatrix}
        0.04v^2+5v+140-u+I \\
        a(bv-u)
    \end{pmatrix}
    \label{eq:izh-differentials}
\end{equation}
with the resetting behavior
\begin{equation}
    \text{if } v\geq30 \text{ mV, then }
    \begin{cases}
        v \leftarrow c\\
        u \leftarrow u + d
    \end{cases}
    \label{eq:izh-resetting}
\end{equation}
In this system, the fast voltage variable $v(t)$ represents the membrane potential in millivolts mV, and the slow recovery variable $u(t)$ describes the recovery after an action potential by modeling the activation of the K$^+$ current, inactivation of the Na$^+$ current, or a combination of both. In this model, both $u$ and $v$ are differentiated with respect to time $t$, which is being measured in milliseconds. The parameter $I$ represents the direct current injected into the neuron, and $a$, $b$, $c$, and $d$ are parameters. Once the spike reaches its apex at 30 mV, the model is reset by Equation \ref{eq:izh-resetting}. Defined by Izhikevich \cite{izhikevich-model}, the resting membrane potential of the model is between $-70$ and $-65$ mV depending on the value of parameter $b$. Additionally, like in real biological neurons, there is no fixed threshold potential; it can be as low as $-55$ mV and as high as $-40$ mV depending on the membrane potential prior to the spike. 

The parameters of the Izhikevich model $a$, $b$, $c$, and $d$ are interpreted as follows:
\begin{enumerate}
    \item The parameter $a$ is synonymous with $\eta$ (Equation \ref{eq:slow-fast-def-fast-time}), so it is bounded by $0<\eta\ll 1$. Smaller values of $a$ result in a slower recovery. Izhikevich \cite{izhikevich-model} notes a typical value of $a$ to be 0.02.
    \item The parameter $b$ represents the sensitivity of the recovery variable $u$ to fluctuations in voltage $v$ below the threshold potential. Therefore, it more strongly couples $u$ and $v$ as we increase its value. Izhikevich \cite{izhikevich-model} notes a typical value of $b$ to be 0.2. 
    \item The parameter $c$ represents the reset value of the voltage $v$ after a spike, meaning it controls the resting membrane potential. Izhikevich \cite{izhikevich-model} notes a typical value of $c$ to be $-65$ mV.
    \item The parameter $d$ represents the rest value of the recovery variable $u$. In general, the variable $u$ resets according to $d$ then decays again with rate $a$. Izhikevich \cite{izhikevich-model} notes a typical value of $d$ to be 2.
\end{enumerate}
By adjusting these individual parameters to model properties of neurons such as a lower or higher resting voltage (controlled by $c$) or a larger or smaller after-spike jump in $u$ (controlled by $d$), the Izhikevich model can simulate a variety of different neuronal behaviors.

Neuron activity may be described as bursting when ``neuron activity alternates between a quiescent state and repetitive spiking'' \cite{izhikevich-article}. The quiescent state is a period when a neuron is not generating action potentials, instead remaining under the threshold potential in a resting state. In the Izhikevich model, bursting can be achieved with a high voltage rest and large after-spike jump of $u$, meaning a higher relative value for $c$ and $d$, with the standard example being $c=-55$ mV and $d=4$ \cite{izhikevich-model}.

\section{The Rulkov Maps}
\label{rulkov-maps}

Although the two-dimensional, continuous-time Izhikevich model is able to capture the complex dynamics of a biological neuron in a simple and condensed way, we would like to explore even further simplifications using a low-dimension discrete-time system. However, despite our interest in a simpler neuron model, we still want the system capable of modeling all of a neuron's oscillatory behavior. Specifically, we are interested in a map that can model both regular and irregular spiking and bursting.

From our discussion of the Izhikevich model, we know that we need a slow-fast system in order to model fast bursts of spikes on top of slow oscillations. To accomplish this, Rulkov \cite{rulkov} constructed a simple slow-fast map based on the Izhikevich model that accurately models the dynamics of biological neuron:\footnote{In the original paper by Rulkov \cite{rulkov} that introduces Rulkov map 1, the parameter $\sigma'=\sigma+1$ is used, but in this paper, we use the slightly modified form from the review by Ibarz, Casado, and Sanjuán \cite{ibarz}. Similarly for Rulkov map 2 \cite{rulkov2}, $\sigma'=\eta$ and $\beta'=-\eta\sigma$ are used originally.}
\begin{equation}
    \begin{pmatrix}
        x_{k+1} \\
        y_{k+1}
    \end{pmatrix}
    =
    \begin{pmatrix}
        f(x_k,\,y_k;\,\alpha) \\
        y_k - \eta(x_k - \sigma)
    \end{pmatrix}
    \label{eq:rulkov-map}
\end{equation}
where $\alpha$, $\sigma$, and $\eta$ are parameters. To make $y$ a slow variable, we need $0<\eta\ll1$, so we choose $\eta=0.001$ for the rest of this paper. This form of the map follows our definition of a slow-fast discrete-time dynamical system in Section \ref{slow-fast-systems-and-dynamics}. The physical interpretation of the Rulkov map parameters will be discussed in Section \ref{injection-of-current}.\footnote{As we know from Section \ref{behavior_and_modeling_of_biological_neurons}, the Izhikevich model references variables like $v$, $u$, and $t$ that have direct physical representations ($v$ represent membrane potential with units mV and $t$ represents time with units ms). However, the Rulkov maps take a phenomenological approach to modeling neuronal behaviors for the sake of simplicity, stepping away from biology and towards the realm of mathematical physics, using variables $x$, $y$, and $k$, as well as parameters $\alpha$, $\sigma$, and $\eta$, which are not direct representations of physical values with units. Of course, $x$ is related to membrane potential, values of $k$ are even time steps, and as we will find out in Section \ref{injection-of-current}, $\sigma$ relates to an injected DC, but we do not make any claim to the physical values and units associated with these variables and parameters. In order to alter this simplified model's results for experimental use, one must only make the simple translation of scaling and adjusting values to correspond with neuronal behavior.} 

Our research is focused on two different forms of the function $f(x,\,y;\,\alpha)$ from Equation \ref{eq:rulkov-map} associated with two different maps, which we will call Rulkov map 1 and Rulkov map 2.\footnote{In the literature, these maps are commonly called the non-chaotic and chaotic Rulkov maps, respectively. However, we avoid this terminology because of the chaotic dynamics that both maps exhibit.} Introduced in a paper by Rulkov \cite{rulkov}, the function associated with Rulkov map 1 is
\begin{equation}
    f_1(x,\,y;\,\alpha) = 
    \begin{cases}
        \alpha/(1-x) + y, & x\leq 0 \\
        \alpha + y, & 0 < x < \alpha + y \\
        -1, & x\geq \alpha + y
    \end{cases}
    \label{eq:rulkov_1_fast_equation}
\end{equation}
Rulkov map 1 is capable of producing periodic spikes and non-chaotic bursts, but it also exhibits chaotic behavior for some parameter values. It is immediately evident that the piecewise structure of the function $f_1(x,\,y;\,\alpha)$ incorporates a built-in resetting mechanism similar to the Izhikevich model. In a different paper by Rulkov \cite{rulkov2}, the function associated with what we call Rulkov map 2 is introduced:
\begin{equation}
    f_2(x,\,y;\,\alpha) = \frac{\alpha}{1+x^2} + y
    \label{eq:rulkov_2_fast_equation}
\end{equation}
Rulkov map 2 is a chaotic model, producing irregular spikes and bursts. In this section, we will first examine the spiking and bursting behavior, bifurcation theory, and chaotic dynamics of Rulkov map 1. Then, we will discuss the complex dynamics of Rulkov map 2.

\subsection{Individual Dynamics of Rulkov Map 1}
\label{individual-dynamics-of-rulkov-map-1}

Recall from Section \ref{slow-fast-systems-and-dynamics} that for a small value of $\eta$, we can approximate a slow-fast system by splitting it into fast and slow motions in the limit $\eta\to 0$. Specifically, for Rulkov map 1, we will approximate the time evolution of the fast variable $x$ by treating $y$ as a slowly drifting parameter of the fast map. In other words, from Equation \ref{eq:rulkov-map},
\begin{equation}
    x_{k+1} = f_1(x_k;\,y,\,\alpha)
\end{equation}
where we put $y$ after the semicolon to indicate that we are treating it as a parameter. By Equation \ref{eq:rulkov-map}, the $y$ drifts slowly according to
\begin{equation}
    y_{k+1} = y_k-\eta(x_k-\sigma)
    \label{eq:rulkov_1_slow_map}
\end{equation}
We will first consider the slow evolution of $y$.

\subsubsection{Slow Map}

From Equation \ref{eq:rulkov_1_slow_map}, it is easy to see that the value of $y$ remains fixed only when
\begin{equation}
    \eta(x-\sigma)=0
\end{equation}
Therefore, the value of $x$ that leaves $y$ unchanged, which we will denote as $x_{s,\,\mathrm{slow}}$, is
\begin{equation}
    x_{s,\,\mathrm{slow}} = \sigma
    \label{eq:rulkov_1_slow_map_fixed_y}
\end{equation}
From Equation \ref{eq:rulkov_1_slow_map}, we can see that if $x<x_{s,\,\mathrm{slow}}$, $y$ will slowly increase, but if $x>x_{s,\,\mathrm{slow}}$, $y$ will slowly decrease. The way $y$ changes in accordance with the value of $x$ is important in understanding the dynamics of the fast map when treating $y$ as a parameter.

\begin{figure}
    \centering
    \begin{tikzpicture}[scale=0.9]
            \begin{axis}[
                    axis lines = left,
                    x=1cm,
                    y=1cm,
                    xlabel =\large \(x_k\),
                    xtick align=outside,
                    xtick pos=left,
                    xmin=-3, xmax=3.5,
                    xtick={-3,-2,-1,0,1,2,3},
                    minor xtick={-2.5,-1.5,-0.5,0.5,1.5,2.5},
                    ylabel = \large \(x_{k+1}\),
                    ytick align=outside,
                    ytick pos=left,
                    ymin=-3, ymax=3.5,
                    ytick={-3,-2,-1,0,1,2,3},
                    minor ytick={-2.5,-1.5,-0.5,0.5,1.5,2.5},
                ]
            \addplot [
                domain=-3:0, 
                samples=100,
                color=blue!80!black,
                very thick
            ]
            {6/(1-x)-3.93};
            \addplot [
                domain=0:6-3.93, 
                color=blue!80!black,
                very thick
            ]
            {6-3.93};
            \addplot [
                domain=6-3.93:3.5, 
                color=blue!80!black,
                very thick
            ]
            {-1};
            \addplot [
                domain=-3:3.5, 
                color=black,
                thick,
                dashed
            ]
            {x};
            \addplot[color=green!80!black, mark=*] coordinates {(-1.741,-1.741)};
            \node[anchor=north west] at (axis cs:-1.741,-1.741) {\small $x_{s,\,\mathrm{fast,\,stable}}$};
            \addplot[color=red, mark=*] coordinates {(-1.189,-1.189)};
            \node[anchor=north west] at (axis cs:-1.189,-1.189) {\small $x_{s,\,\mathrm{fast,\,unstable}}$};
            \addplot[color=blue, mark=o] coordinates {(2.07,2.07)};
            \addplot[color=blue, mark=*] coordinates {(2.07,-1)};
            \addplot+[
                color=green!80!black,
                mark=none,
                densely dotted,
                very thick
            ]
            coordinates
            {(2.07,-1) (-1,-1) (-1,-0.93) (-0.93,-0.93) (-0.93,-0.821) (-0.821,-0.821) (-0.821,-0.635) (-0.635,-0.635) (-0.635,-0.26) (-0.26,-0.26) (-0.26,0.832) (0.832,0.832) (0.832,2.07) (2.07,2.07) (2.07,-1)};
            \node[anchor=south] at (axis cs:0.535,-1) {\small $O^q(-1)$};
            \end{axis}
        \end{tikzpicture}
    \caption{The function $x_{k+1}=f_1(x_k;\,y,\,\alpha)$ graphed in blue for $y=-3.93$ and $\alpha=6$, with the fixed points $x_{s,\,\mathrm{fast,\,stable}}$ and $x_{s,\,\mathrm{fast,\,unstable}}$ at the intersection between the function and the dashed black line $x_{k+1} = x_k$ and the periodic orbit $O^q(-1)$ shown with a dotted green line}
    \label{fig:rulkov_fast_alpha6_y-3.93-example}
\end{figure}
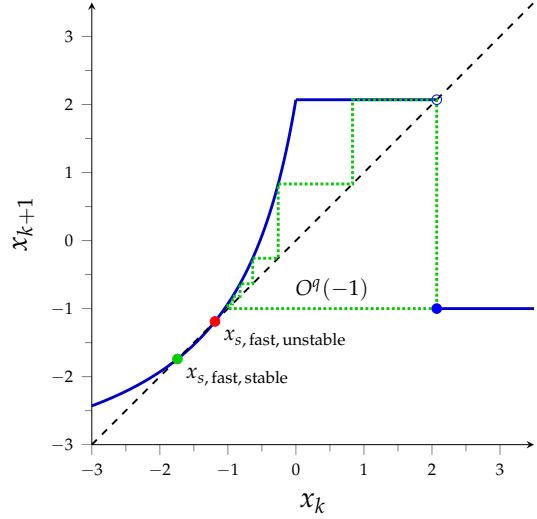

\subsubsection{Fast Map}

In Figure \ref{fig:rulkov_fast_alpha6_y-3.93-example}, we graph the shape of an iteration of the fast map $x_{k+1}=f_1(x_k;\,y,\,\alpha)$ for $y=-3.93$ and $\alpha=6$. We can see the piecewise structure of this function from Equation \ref{eq:rulkov_1_fast_equation}: it follows a curve for $x\leq0$ up to a horizontal line for $0<x<\alpha+y$ then reaches a discontinuity and jumps down to $-1$ for $x\geq\alpha+y$. It follows from Equation \ref{eq:rulkov_1_fast_equation} that changing the value of $y$ will move the graph up or down, except for the piece where $x\geq\alpha+y$ which always stays at $-1$. In the figure, we can also see two fixed points, one stable and one unstable, at the intersection of this function with the line $x_{k+1} = x_k$ and a periodic orbit $O^q(x_p)$ with $x_p=-1$. We visualize this orbit $O^q(-1)$ by taking a point on the curve $(x_k,\,x_{k+1})$ and tracing horizontally from that point to the line $x_{k+1} = x_k$, yielding the point $(x_{k+1},\,x_{k+1})$. We then trace vertically from that point to the function, which yields the next point in our periodic orbit $(x_{k+1},\,x_{k+2})$, from which we continue this process until we get back to the original point. This type of visualization is known as a cobweb orbit \cite[p. 358]{strogatz}, and we show it in Figure \ref{fig:rulkov_fast_alpha6_y-3.93-example} using the code in Appendix \ref{rulkov_1_and_cobweb_code}.

From Figure \ref{fig:rulkov_fast_alpha6_y-3.93-example}, it is clear that fixed points can only appear on the curve where $x\leq 0$. Therefore, when considering fixed points, we can treat the function $f_1(x;\,y,\,\alpha)$ as the first piece of Equation \ref{eq:rulkov_1_fast_equation}:
\begin{equation}
    f_1(x;\,y,\,\alpha) = \frac{\alpha}{1-x} + y
    \label{eq:rulkov_1_fast_map_less_0}
\end{equation}
We know from Section \ref{background} (Equation \ref{eq:stationary}) that a fixed point of the fast map $x_{s,\,\mathrm{fast}}$ follows $x_{s,\,\mathrm{fast}} = f_1(x_{s,\,\mathrm{fast}};\, y,\, \alpha)$. Substituting into Equation \ref{eq:rulkov_1_fast_map_less_0} and rearranging, we get the equation for the fixed points of the fast map:
\begin{equation}
    y = x_{s,\,\mathrm{fast}} - \frac{\alpha}{1-x_{s,\,\mathrm{fast}}}
    \label{eq:rulkov_fast_map_fixed_points}
\end{equation}

Using the criteria for the attractiveness of fixed points from Section \ref{nonchaoticattractors}, we can now find the values of $x$ for which a given $x_{s,\,\mathrm{fast}}$ is attracting or repelling. We know from Section \ref{nonchaoticattractors} that to determine the attractiveness of $x_{s,\,\mathrm{fast}}$, we are interested in the function $|f_1'(x;\,y,\,\alpha)|$. Using elementary calculus (and keeping in mind we are treating $y$ as a parameter), we get
\begin{equation}
    f_1'(x;\,y,\,\alpha) = \frac{\alpha}{(1-x)^2}
    \label{eq:rulkov_fast_map_derivative}
\end{equation}
For our purposes, we are interested in values of $\alpha$ greater than 0; for $\alpha<0$, the map doesn't produce our desired behavior. In this case,
\begin{equation}
    |f_1'(x;\,y,\,\alpha)| = \frac{\alpha}{(1-x)^2}
\end{equation}
because $\alpha$ and $(1-x)^2$ are always positive. From the criteria for fixed point attractiveness, the stable fixed points of the fast map $x_{s,\,\mathrm{fast,\,stable}}$ are given by
\begin{equation}
    |f_1'(x;\,y,\,\alpha)| = \frac{\alpha}{(1-x)^2} < 1
\end{equation}
Multiplying by $(1-x)^2$ on both sides and taking a square root, we get either
\begin{equation}
    \sqrt{\alpha} < 1-x \quad \mathrm{or} \quad -\sqrt{\alpha} > 1-x
\end{equation}
However, since we are working with values $x\leq0$, this leaves us with only the first inequality. Solving this gives us that the values of $x_{s,\,\mathrm{fast,\,stable}}$ are on the interval
\begin{equation}
    x_{s,\,\mathrm{fast,\,stable}} < 1-\sqrt{\alpha}
    \label{eq:rulkov_1_stable_fast_int}
\end{equation}
For unstable fixed points of the fast map $x_{s,\,\mathrm{fast,\,unstable}}$, we need
\begin{equation}
    |f_1'(x;\,y,\,\alpha)| = \frac{\alpha}{(1-x)^2} > 1
\end{equation}
This gives us
\begin{equation}
    1-\sqrt{\alpha} < x < 1+\sqrt{\alpha}
\end{equation}
Taking into account the region on which fixed points are defined ($x\leq 0$), this leaves us with
\begin{equation}
    1-\sqrt{\alpha} < x_{s,\,\mathrm{fast,\,unstable}} \leq 0
    \label{eq:rulkov_1_unstable_fast_int}
\end{equation}
These two simple inequalities in Equations \ref{eq:rulkov_1_stable_fast_int} and \ref{eq:rulkov_1_unstable_fast_int} allow us to label any fixed point satisfying Equation \ref{eq:rulkov_fast_map_fixed_points} as either stable or unstable.

\begin{figure*}[ht!]
    \centering
    \hfill
    \begin{subfigure}[t]{0.475\textwidth}
        \centering
        \begin{tikzpicture}[scale=0.95]
            \begin{axis}[
                xlabel = \normalsize \(y\),
                xtick align=outside,
                xtick pos=left,
                xmin=-3.5, xmax=-2.5,
                xtick={-3.5,-3,-2.5},
                minor xtick={-3.4,-3.3,-3.2,-3.1,-2.9,-2.8,-2.7,-2.6},
                ylabel = \normalsize \(x\),
                ytick align=outside,
                ytick pos=left,
                ymin=-2.5, ymax=0.5,
                ytick={-2,-1,0},
                minor ytick={0.4,0.2,-0.2,-0.4,-0.6,-0.8,-1.2,-1.4,-1.6,-1.8,-2.2,-2.4},
                scatter/use mapped color={
                    draw=green!80!black,
                    fill=green!80!black,
                },
            ]
            \addplot [
                domain=-3.5:-3, 
                samples=200,
                color=green!80!black,
                very thick
            ]
            {(1+x-sqrt((x-1)^2-16))/2};
            \addplot [
                domain=-3.5:-3, 
                samples=200,
                color=red,
                very thick,
                dashed
            ]
            {(1+x+sqrt((x-1)^2-16))/2};
            \addplot+[
                only marks,
                scatter,
                mark size=0.5pt,
            ]
            table 
            {data/rulkov_alpha4.dat};
            \addplot+[
                only marks,
                scatter,
                mark size=0.5pt,
            ]
            table 
            {data/rulkov_alpha4_2.dat};
            \addplot[color=green!80!black, mark=*, mark size=2.5pt] coordinates {(-3.1,-1.5)};
            \node[anchor=north west] at (axis cs:-3.1,-1.5) {\small $\mathbf{x}_s$};
            \addplot [
                domain=-3.5:-2.5, 
                thick,
                densely dotted
            ]
            {-1.5};
            \node[anchor=north west] at (axis cs:-3.4,-2.117) {\small $B_{\mathrm{stable}}$};
            \node[anchor=south west] at (axis cs:-3.4,-0.283) {\small $B_{\mathrm{unstable}}$};
            \node at (axis cs:-2.8,0.2) {\small $B_{\mathrm{spikes}}$};
            \node[anchor=south east] at (axis cs:-2.5,-1.5) {\small $x_{s,\,\mathrm{slow}}$};
            \draw[->, thick, color=blue] (axis cs:-2.7,-0.3) -- (axis cs:-2.9,-0.6);
            \draw[->, thick, color=blue] (axis cs:-2.975,-1.15) -- (axis cs:-3.025,-1.4);
            \draw[->, thick, color=blue] (axis cs:-3.25,-2) -- (axis cs:-3.15,-1.775);
            \end{axis}
        \end{tikzpicture}
        \caption{Stable ($B_{\mathrm{stable}}$, shown in green), unstable ($B_{\mathrm{unstable}}$, shown in red), and spiking ($B_{\mathrm{spikes}}$, shown in green, approximated with the code in Appendix \ref{spiking_branch_of_rulkov_1_code}) branches of Rulkov map 1 in two-dimensional state space $\langle y,\, x \rangle$, with the values of $x$ that keep the slow map fixed ($x_{s,\,\mathrm{slow}}$, shown in black), the fixed point of the map ($\mathbf{x}_s$), and arrows representing typical motion (shown in blue)}
        \label{fig:rulkov_1_state_space_diagram_alpha4}
        \vspace{4px}
    \end{subfigure}
    \hfill
    \begin{subfigure}[t]{0.475\textwidth}
        \centering
        \vspace{-6.5cm}
        \includegraphics[scale=0.1125]{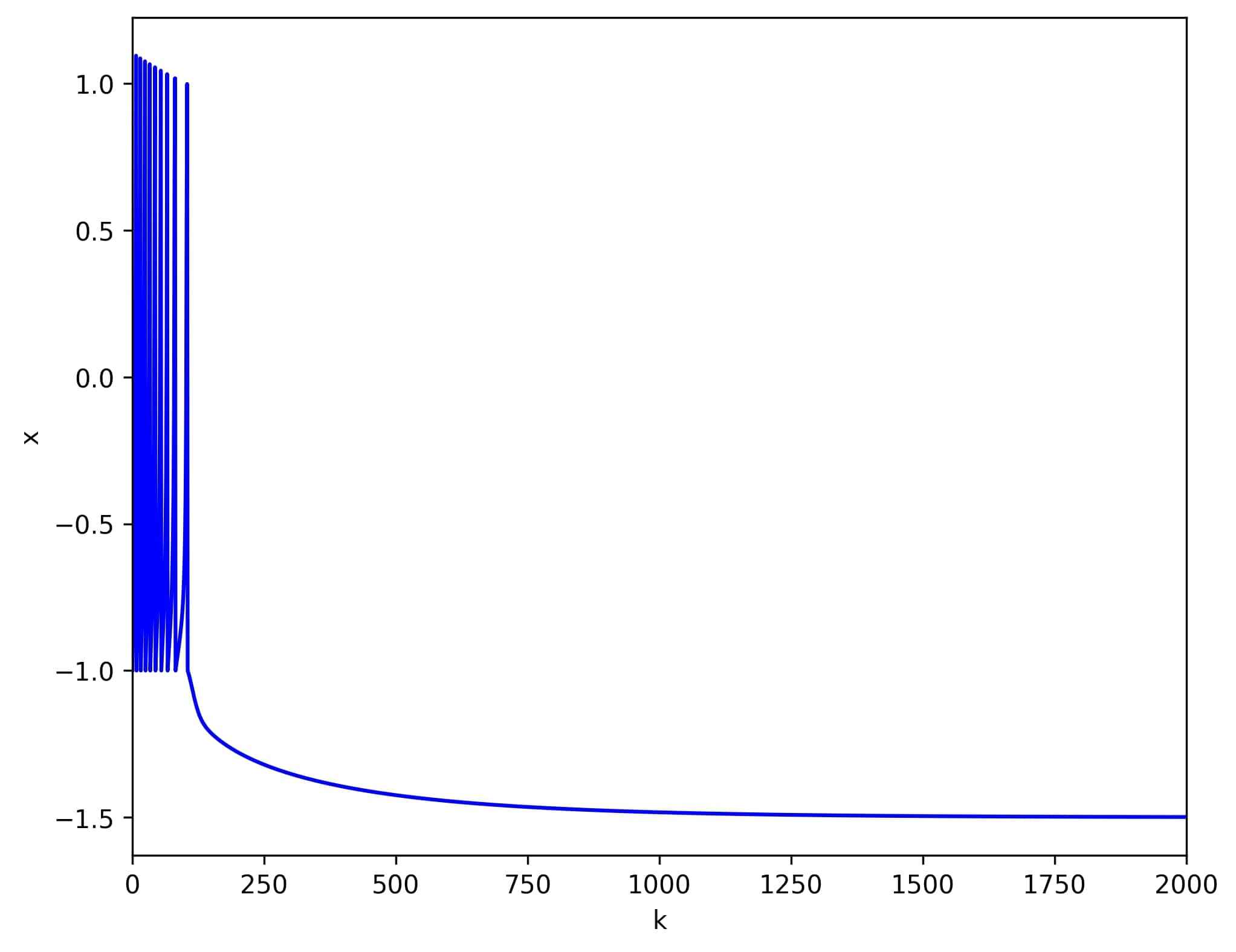}
        \vspace{0.5cm}
        \caption{Graph of the fast variable $x_k$ displaying spiking to silence behavior generated by Rulkov map 1, graphed with the code in Appendix \ref{fast_var_orbs_of_rulkov_1_code}}
        \label{fig:rulkov_x_vs_k_graph_alpha4}
    \end{subfigure}
    \hfill
    \caption{A state space diagram and a graph of $x_k$ for parameters $\alpha=4$ and $\sigma=-1.5$ of Rulkov map 1}
    \label{fig:rulkov_1_alpha4_sigma-1.5_graphs}
\end{figure*}
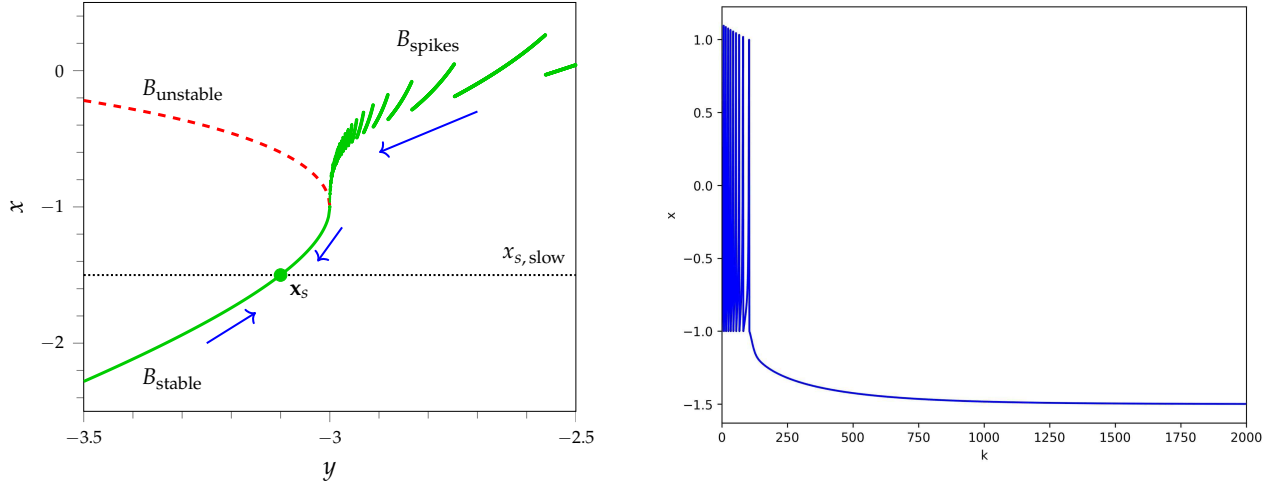

Because we are treating $y$ as a parameter, it will be useful to rearrange Equation \ref{eq:rulkov_fast_map_fixed_points} to get the fixed points of the fast map $x_{s,\,\mathrm{fast}}$ as a function of $y$. Multiplying both sides of Equation \ref{eq:rulkov_fast_map_fixed_points} by $1-x$ and rearranging gives us
\begin{equation}
    x^2 - (1+y)x + (\alpha+y) = 0
\end{equation}
Then, by the quadratic formula, we get
\begin{equation}
    x_{s,\,\mathrm{fast}} = \frac{1+y\pm\sqrt{(y-1)^2-4\alpha}}{2}
    \label{eq:rulkov_1_fixed_points_func_y}
\end{equation}
To find the $x_{s,\,\mathrm{fast,\,stable}}$ points as a function of $y$, we simply consider the part of Equation \ref{eq:rulkov_1_fixed_points_func_y} that is defined on $x<1-\sqrt{\alpha}$ (Equation \ref{eq:rulkov_1_stable_fast_int}). We get this if we take the negative branch of the square root, so we can define the stable branch of Rulkov map 1 in two-dimensional state space, with $x$ a function of $y$, as
\begin{equation}
    B_{\mathrm{stable}}(y;\,\alpha) = \frac{1+y-\sqrt{(y-1)^2-4\alpha}}{2}
\end{equation}
The stable branch is a branch of slow motion in two-dimensional state space since $x$ stays on the branch as $y$ slowly drifts. Similarly, we can define the unstable branch of Rulkov map 1 as
\begin{equation}
    B_{\mathrm{unstable}}(y;\,\alpha) = \frac{1+y+\sqrt{(y-1)^2-4\alpha}}{2}
\end{equation}
which is defined on $1-\alpha<x\leq0$. The domain of both the stable and unstable branches can also be found by analysis of the discriminant. Namely, we need $(y-1)^2-4\alpha>0$, so the domains of $B_{\mathrm{stable}}(y;\,\alpha)$ and $B_{\mathrm{unstable}}(y;\,\alpha)$ are $y<1-2\sqrt{\alpha}$. These stable and unstable branches are graphed for different values of $\alpha$ in Figures \ref{fig:rulkov_1_state_space_diagram_alpha4} and \ref{fig:rulkov_1_state_space_diagram_alpha6}.

We will now discuss the periodic orbits of the fast map, which represent oscillations or spikes. From Figure \ref{fig:rulkov_fast_alpha6_y-3.93-example}, we can see that for a given $y$, there will be only one periodic orbit of $x_{k+1} = f_1(x_k;\,y,\,\alpha)$. This is because initial conditions that don't end up at the fixed point $x_{s,\,\mathrm{fast,\,stable}}$ will eventually end up at $-1$, where a periodic orbit is born. To see how this periodic orbit arises, consider starting from $x_p=-1$ and iterating. If this $f(x_p) > x_{s,\,\mathrm{fast,\,unstable}}$, it won't get attracted to $x_{s,\,\mathrm{fast,\,stable}}$ (see Figure \ref{fig:rulkov_fast_alpha6_y-3.93-example}). Instead, it will get repelled up the curve $f_1(x;\,y,\,\alpha)$, $x$ increasing until it is greater than $\alpha+y$, which brings us back to $-1$. Then, for a given $y$ and $\alpha$, we can denote the periodic orbit with periodic point $x_p=-1$ as $O^q(x_p;\,y,\,\alpha)$, or $O^q(-1)$ for short. This periodic orbit is stable because orbits that don't get attracted to the stable fixed point $x_{s,\,\mathrm{fast,\,stable}}$ eventually end up at $-1$.

To visualize the location of the spiking branch $B_{\mathrm{spikes}}(y;\,\alpha)$ in two-dimensional state space, we can approximate the $x$ value of a given $O^q(x_p;\,y,\,\alpha)$ by calculating the mean value of $x$ in the periodic orbit:
\begin{equation}
    \begin{split}
        B_{\mathrm{spikes}}(y;\,\alpha) &= \langle O^q(x_p;\,y,\,\alpha)\rangle \\
        &= \frac{1}{q}\sum_{i=1}^q f^q(x_p;\,y,\,\alpha)
    \end{split}
\end{equation}
In Figures \ref{fig:rulkov_1_state_space_diagram_alpha4} and \ref{fig:rulkov_1_state_space_diagram_alpha6}, we graph the spiking branch $B_{\mathrm{spikes}}$ for different values of $\alpha$ using the code in Appendix \ref{spiking_branch_of_rulkov_1_code}. As we can see, there are many discontinuities in the branch. These appear because of the bifurcations that occur when $O^q(x_p;\,y,\,\alpha)$ contains the point $x=0$. Here, the value of the period $q$ changes, resulting in a discontinuity in $\langle O^q(x_p;\,y,\,\alpha)\rangle$.

\subsubsection{Combining the Individual Maps}

We can now see that if the value of $x$ that leaves $y$ unchanged $x_{s,\,\mathrm{slow}}$ is in the stable branch $B_{\mathrm{stable}}(y;\,\alpha)$, then a stable fixed point of the map $\mathbf{x}_s$ exists at the intersection of $x=x_{s,\,\mathrm{slow}}$ and $x=B_{\mathrm{stable}}(y;\,\alpha)$. This is because at $x_{s,\,\mathrm{slow}}$, $y$ doesn't change, and on $B_{\mathrm{stable}}(y;\,\alpha)$, $x$ doesn't change, so the point where these intersect is stationary. However, if $x_{s,\,\mathrm{slow}}$ is in the unstable branch $B_{\mathrm{unstable}}(y;\,\alpha)$, then an unstable fixed point of the map $\mathbf{x}_s$ exists and oscillations will occur. We will discuss both of these possibilities in more detail soon.

Because fixed point stability occurs when $x_{s,\,\mathrm{slow}}$ is in the stable branch but stops when it is in the unstable branch, we call the point where $x_{s,\,\mathrm{slow}}$ is on the border of the branches, which we know from Equations \ref{eq:rulkov_1_stable_fast_int} and \ref{eq:rulkov_1_unstable_fast_int} is $x_{s,\,\mathrm{slow}}=1-\sqrt{\alpha}$, the threshold of excitation of Rulkov map 1 \cite{rulkov}. Recalling Equation \ref{eq:rulkov_1_slow_map_fixed_y}, the threshold of excitation corresponds to the values of $\sigma$
\begin{equation}
    \sigma_{\mathrm{th}} = 1-\sqrt{\alpha}
    \label{eq:sigma-threshold-of-excitation}
\end{equation}

\begin{figure*}[hp!]
    \centering
    \begin{subfigure}[b]{0.475\textwidth}
        \centering
        \begin{tikzpicture}[scale=0.75]
            \begin{axis}[
                    axis lines = left,
                    x=1cm,
                    y=1cm,
                    xlabel =\large \(x_k\),
                    xtick align=outside,
                    xtick pos=left,
                    xmin=-3, xmax=3.5,
                    xtick={-3,-2,-1,0,1,2,3},
                    minor xtick={-2.5,-1.5,-0.5,0.5,1.5,2.5},
                    ylabel = \large \(x_{k+1}\),
                    ytick align=outside,
                    ytick pos=left,
                    ymin=-3, ymax=3.5,
                    ytick={-3,-2,-1,0,1,2,3},
                    minor ytick={-2.5,-1.5,-0.5,0.5,1.5,2.5},
                ]
            \addplot [
                domain=-3:0, 
                samples=100,
                color=blue!80!black,
                very thick
            ]
            {6/(1-x)-3.93};
            \addplot [
                domain=0:6-3.93, 
                color=blue!80!black,
                very thick
            ]
            {6-3.93};
            \addplot [
                domain=6-3.93:3.5, 
                color=blue!80!black,
                very thick
            ]
            {-1};
            \addplot [
                domain=-3:3.5, 
                color=black,
                thick,
                dashed
            ]
            {x};
            \addplot[color=green!80!black, mark=*] coordinates {(-1.741,-1.741)};
            \node[anchor=north west] at (axis cs:-1.741,-1.741) {\small $x_{s,\,\mathrm{fast,\,stable}}$};
            \addplot[color=red, mark=*] coordinates {(-1.189,-1.189)};
            \node[anchor=north west] at (axis cs:-1.189,-1.189) {\small $x_{s,\,\mathrm{fast,\,unstable}}$};
            \addplot[color=blue, mark=o] coordinates {(2.07,2.07)};
            \addplot[color=blue, mark=*] coordinates {(2.07,-1)};
            \addplot+[
                color=green!80!black,
                mark=none,
                densely dotted,
                very thick
            ]
            coordinates
            {(2.07,-1) (-1,-1) (-1,-0.93) (-0.93,-0.93) (-0.93,-0.821) (-0.821,-0.821) (-0.821,-0.635) (-0.635,-0.635) (-0.635,-0.26) (-0.26,-0.26) (-0.26,0.832) (0.832,0.832) (0.832,2.07) (2.07,2.07) (2.07,-1)};
            \node[anchor=south] at (axis cs:0.535,-1) {\small $O^q(-1)$};
            \end{axis}
        \end{tikzpicture}
        \caption{$y=-3.93$}
        \label{fig:rulkov_fast_alpha6_y-3.93}
        \vspace{10px}
    \end{subfigure}
    \begin{subfigure}[b]{0.475\textwidth}
        \centering
        \begin{tikzpicture}[scale=0.75]
            \begin{axis}[
                    axis lines = left,
                    x=1cm,
                    y=1cm,
                    xlabel =\large \(x_k\),
                    xtick align=outside,
                    xtick pos=left,
                    xmin=-3, xmax=3.5,
                    xtick={-3,-2,-1,0,1,2,3},
                    minor xtick={-2.5,-1.5,-0.5,0.5,1.5,2.5},
                    ylabel = \large \(x_{k+1}\),
                    ytick align=outside,
                    ytick pos=left,
                    ymin=-3, ymax=3.5,
                    ytick={-3,-2,-1,0,1,2,3},
                    minor ytick={-2.5,-1.5,-0.5,0.5,1.5,2.5},
                ]
            \addplot [
                domain=-3:0, 
                samples=100,
                color=blue!80!black,
                very thick
            ]
            {6/(1-x)-3.99};
            \addplot [
                domain=0:6-3.99, 
                color=blue!80!black,
                very thick
            ]
            {6-3.99};
            \addplot [
                domain=6-3.99:3.5, 
                color=blue!80!black,
                very thick
            ]
            {-1};
            \addplot [
                domain=-3:3.5, 
                color=black,
                thick,
                dashed
            ]
            {x};
            \addplot[color=green!80!black, mark=*] coordinates {(-1.969,-1.969)};
            \node[anchor=north west] at (axis cs:-1.969,-1.969) {\small $x_{s,\,\mathrm{fast,\,stable}}$};
            \addplot[color=red, mark=*] coordinates {(-1.021,-1.021)};
            \node[anchor=north west] at (axis cs:-1.021,-1.021) {\small $x_{s,\,\mathrm{fast,\,unstable}}$};
            \addplot[color=blue, mark=o] coordinates {(2.01,2.01)};
            \addplot[color=blue, mark=*] coordinates {(2.01,-1)};
            \addplot+[
                color=green!80!black,
                mark=none,
                densely dotted,
                very thick
            ]
            table 
            {data/rulkov_periodic_orbit_cobweb_alpha6_y-3.99.dat};
            \node[anchor=south] at (axis cs:0.5,-1) {\small $O^q(-1)$};
            \end{axis}
        \end{tikzpicture}
        \caption{$y=-3.99$}
        \label{fig:rulkov_fast_alpha6_y-3.99}
        \vspace{10px}
    \end{subfigure}
    \begin{subfigure}[b]{0.475\textwidth}
        \centering
        \begin{tikzpicture}[scale=0.75]
            \begin{axis}[
                    axis lines = left,
                    x=1cm,
                    y=1cm,
                    xlabel =\large \(x_k\),
                    xtick align=outside,
                    xtick pos=left,
                    xmin=-3, xmax=3.5,
                    xtick={-3,-2,-1,0,1,2,3},
                    minor xtick={-2.5,-1.5,-0.5,0.5,1.5,2.5},
                    ylabel = \large \(x_{k+1}\),
                    ytick align=outside,
                    ytick pos=left,
                    ymin=-3, ymax=3.5,
                    ytick={-3,-2,-1,0,1,2,3},
                    minor ytick={-2.5,-1.5,-0.5,0.5,1.5,2.5},
                ]
            \addplot [
                domain=-3:0, 
                samples=100,
                color=blue!80!black,
                very thick
            ]
            {6/(1-x)-4.01};
            \addplot [
                domain=0:6-4.01, 
                color=blue!80!black,
                very thick
            ]
            {6-4.01};
            \addplot [
                domain=6-4.01:3.5, 
                color=blue!80!black,
                very thick
            ]
            {-1};
            \addplot [
                domain=-3:3.5, 
                color=black,
                thick,
                dashed
            ]
            {x};
            \addplot[color=green!80!black, mark=*] coordinates {(-2.029,-2.029)};
            \node[anchor=north west] at (axis cs:-2.029,-2.029) {\small $x_{s,\,\mathrm{fast,\,stable}}$};
            \addplot[color=red, mark=*] coordinates {(-0.981,-0.981)};
            \node[anchor=north west] at (axis cs:-0.981,-0.981) {\small $x_{s,\,\mathrm{fast,\,unstable}}$};
            \addplot[color=blue, mark=o] coordinates {(1.99,1.99)};
            \addplot[color=blue, mark=*] coordinates {(1.99,-1)};
            \addplot+[
                color=green!80!black,
                mark=none,
                densely dotted,
                very thick
            ]
            table 
            {data/rulkov_attraction_cobweb_alpha6_y-4.01.dat};
            \end{axis}
        \end{tikzpicture}
        \caption{$y=-4.01$}
        \label{fig:rulkov_fast_alpha6_y-4.01}
        \vspace{4px}
    \end{subfigure}
    \begin{subfigure}[b]{0.475\textwidth}
        \centering
        \begin{tikzpicture}[scale=0.75]
            \begin{axis}[
                    axis lines = left,
                    x=1cm,
                    y=1cm,
                    xlabel =\large \(x_k\),
                    xtick align=outside,
                    xtick pos=left,
                    xmin=-3, xmax=3.5,
                    xtick={-3,-2,-1,0,1,2,3},
                    minor xtick={-2.5,-1.5,-0.5,0.5,1.5,2.5},
                    ylabel = \large \(x_{k+1}\),
                    ytick align=outside,
                    ytick pos=left,
                    ymin=-3, ymax=3.5,
                    ytick={-3,-2,-1,0,1,2,3},
                    minor ytick={-2.5,-1.5,-0.5,0.5,1.5,2.5},
                ]
            \addplot [
                domain=-3:0, 
                samples=100,
                color=blue!80!black,
                very thick
            ]
            {6/(1-x)-3.899};
            \addplot [
                domain=0:6-3.899, 
                color=blue!80!black,
                very thick
            ]
            {6-3.899};
            \addplot [
                domain=6-3.899:3.5, 
                color=blue!80!black,
                very thick
            ]
            {-1};
            \addplot [
                domain=-3:3.5, 
                color=black,
                thick,
                dashed
            ]
            {x};
            \addplot[color=red, mark=*] coordinates {(-1.449,-1.449)};
            \node[anchor=north west] at (axis cs:-1.449,-1.449) {\small $x_{s,\,\mathrm{fast,\,unstable}}$};
            \addplot[color=blue, mark=o] coordinates {(2.101,2.101)};
            \addplot[color=blue, mark=*] coordinates {(2.101,-1)};
            \addplot+[
                color=green!80!black,
                mark=none,
                densely dotted,
                very thick
            ]
            table 
            {data/rulkov_periodic_orbit_cobweb_alpha6_y-3.899.dat};
            \node[anchor=south] at (axis cs:0.55,-1) {\small $O^q(-1)$};
            \end{axis}
        \end{tikzpicture}
        \caption{$y=-3.899$}
        \label{fig:rulkov_fast_alpha6_y-3.899}
        \vspace{4px}
    \end{subfigure}
    \caption{The function $x_{k+1}=f_1(x_k;\,y,\,\alpha)$ graphed for $\alpha=6$, showing a bursting orbit}
    \label{fig:rulkov_fast_maps_alpha6}
\end{figure*}
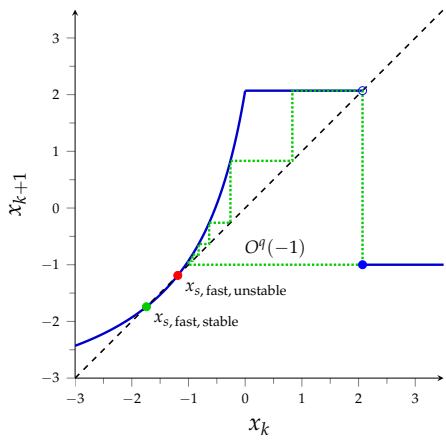
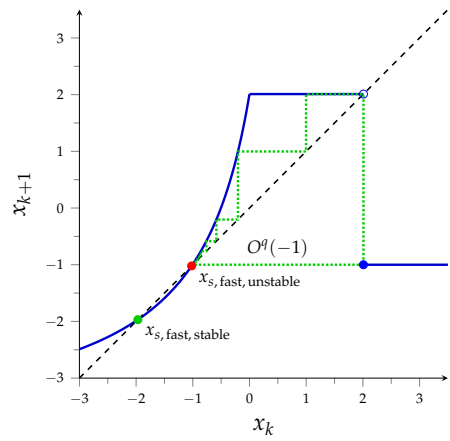
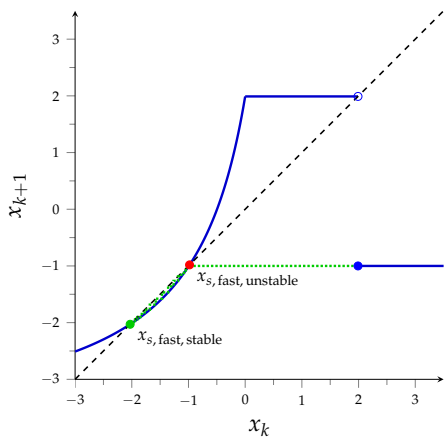
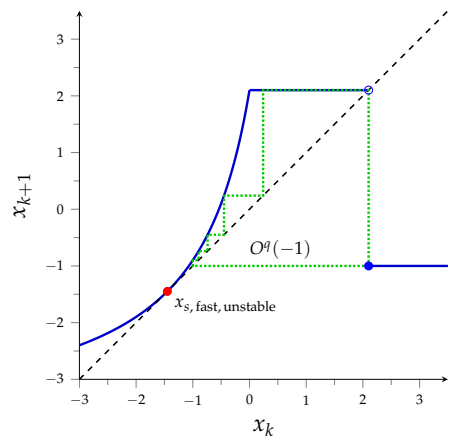

When combining the slow and fast maps, a bifurcation occurs when $y$ decreases to the point where $x_{s,\,\mathrm{fast,\,unstable}}$ passes through $x=-1$ (see Figures \ref{fig:rulkov_fast_alpha6_y-3.99} and \ref{fig:rulkov_fast_alpha6_y-4.01}), during which dynamics change abruptly from a periodic orbit to a fixed point after being attracted to $x_{s,\,\mathrm{fast,\,stable}}$.\footnote{This type of bifurcation results from the formation of a homoclinic orbit at $x=-1$, which we will not discuss in this paper. For more information, see Chapters 6 and 8 in the book by Strogatz \cite{strogatz}.} We know from Equation \ref{eq:rulkov_1_unstable_fast_int} that $x_{s,\,\mathrm{fast,\,unstable}}$ only exists on $1-\sqrt{\alpha} < x_{s,\,\mathrm{fast,\,unstable}} \leq 0$. Therefore, this type of bifurcation can only occur when $1-\sqrt{\alpha} < -1$, or $\alpha > 4$. When $\alpha<4$, combining the individual maps leads to two types of behavior: silence and spiking. However, this bifurcation leads to the possibility of bursting when $\alpha>4$. We will now discuss each of these types of dynamics in turn.

\begin{figure*}[ht!]
    \centering
    \begin{subfigure}[b]{0.475\textwidth}
        \centering
        \begin{tikzpicture}[scale=0.9]
            \begin{axis}[
                    axis lines = left,
                    x=1cm,
                    y=1cm,
                    xlabel =\large \(x_k\),
                    xtick align=outside,
                    xtick pos=left,
                    xmin=-3, xmax=3.5,
                    xtick={-3,-2,-1,0,1,2,3},
                    minor xtick={-2.5,-1.5,-0.5,0.5,1.5,2.5},
                    ylabel = \large \(x_{k+1}\),
                    ytick align=outside,
                    ytick pos=left,
                    ymin=-3, ymax=3.5,
                    ytick={-3,-2,-1,0,1,2,3},
                    minor ytick={-2.5,-1.5,-0.5,0.5,1.5,2.5},
                ]
            \addplot [
                domain=-3:0, 
                samples=100,
                color=blue!80!black,
                very thick
            ]
            {4/(1-x)-2.7};
            \addplot [
                domain=0:4-2.7, 
                color=blue!80!black,
                very thick
            ]
            {4-2.7};
            \addplot [
                domain=4-2.7:3.5, 
                color=blue!80!black,
                very thick
            ]
            {-1};
            \addplot [
                domain=-3:3.5, 
                color=black,
                thick,
                dashed
            ]
            {x};
            \addplot[color=blue, mark=o] coordinates {(4-2.7,4-2.7)};
            \addplot[color=blue, mark=*] coordinates {(4-2.7,-1)};
            \addplot+[
                color=green!80!black,
                mark=none,
                densely dotted,
                very thick
            ]
            table 
            {data/rulkov_periodic_orbit_cobweb_alpha4_y-2.7.dat};
            \node[anchor=south] at (axis cs:0.15,-1) {\small $O^q(-1)$};
            \end{axis}
        \end{tikzpicture}
        \caption{$y=-2.7$}
        \label{fig:rulkov_fast_alpha4_y-2.7}
    \end{subfigure}
    \begin{subfigure}[b]{0.475\textwidth}
        \centering
        \begin{tikzpicture}[scale=0.9]
            \begin{axis}[
                    axis lines = left,
                    x=1cm,
                    y=1cm,
                    xlabel =\large \(x_k\),
                    xtick align=outside,
                    xtick pos=left,
                    xmin=-3, xmax=3.5,
                    xtick={-3,-2,-1,0,1,2,3},
                    minor xtick={-2.5,-1.5,-0.5,0.5,1.5,2.5},
                    ylabel = \large \(x_{k+1}\),
                    ytick align=outside,
                    ytick pos=left,
                    ymin=-3, ymax=3.5,
                    ytick={-3,-2,-1,0,1,2,3},
                    minor ytick={-2.5,-1.5,-0.5,0.5,1.5,2.5},
                ]
            \addplot [
                domain=-3:0, 
                samples=100,
                color=blue!80!black,
                very thick
            ]
            {4/(1-x)-3.05};
            \addplot [
                domain=0:4-3.05, 
                color=blue!80!black,
                very thick
            ]
            {4-3.05};
            \addplot [
                domain=4-3.05:3.5, 
                color=blue!80!black,
                very thick
            ]
            {-1};
            \addplot [
                domain=-3:3.5, 
                color=black,
                thick,
                dashed
            ]
            {x};
            \addplot[color=green!80!black, mark=*] coordinates {(-1.342,-1.342)};
            \node[anchor=north west] at (axis cs:-1.342,-1.342) {\small $x_{s,\,\mathrm{fast,\,stable}}$};
            \addplot[color=red, mark=*] coordinates {(-0.708,-0.708)};
            \node[anchor=east] at (axis cs:-0.708,-0.708) {\small $x_{s,\,\mathrm{fast,\,unstable}}$};
            \addplot[color=blue, mark=o] coordinates {(4-3.05,4-3.05)};
            \addplot[color=blue, mark=*] coordinates {(4-3.05,-1)};
            \end{axis}
        \end{tikzpicture}
        \caption{$y=-3.05$}
        \label{fig:rulkov_fast_alpha4_y-3.05}
    \end{subfigure}
    \caption{The function $x_{k+1}=f_1(x_k;\,y,\,\alpha)$ graphed for $\alpha=4$, showing the attraction of a periodic orbit in $B_{\mathrm{spikes}}$ to the stable fixed point $\mathbf{x}_s$}
    \label{fig:rulkov_fast_maps_alpha4}
\end{figure*}
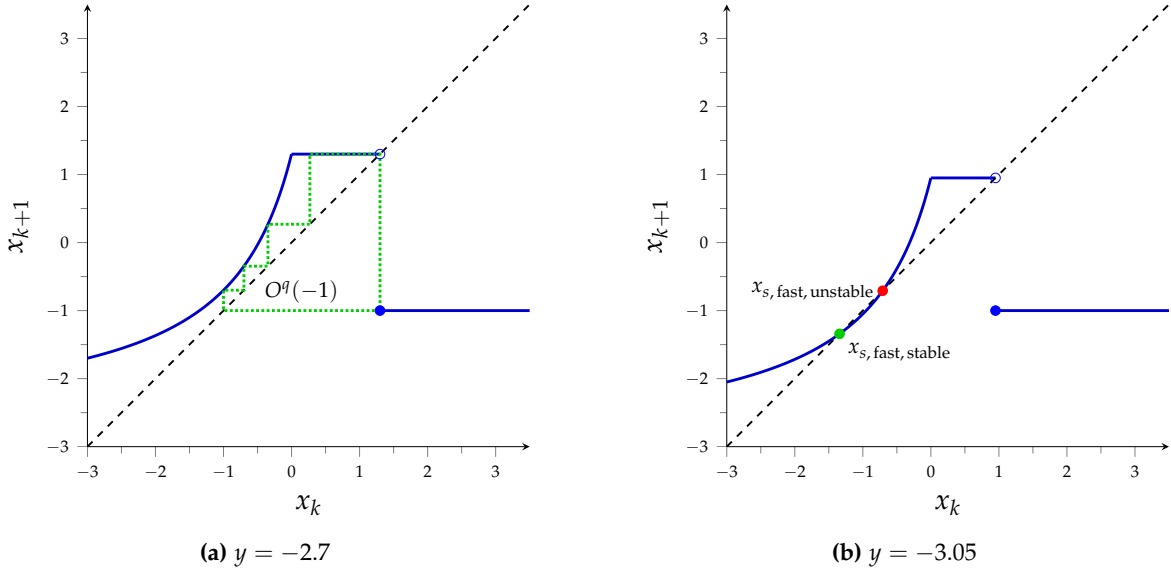

\subsubsection{Silence}

In Rulkov map 1, silence occurs when the state $\mathbf{x}$ settles at the fixed point of the map $\mathbf{x}_s$. In order for this to occur, $\mathbf{x}_s$ must be stable, meaning $x=x_{s,\,\mathrm{slow}}$ must intersect the stable branch of the fast map $B_{\mathrm{stable}}(y;\,\alpha)$. In Figure \ref{fig:rulkov_1_state_space_diagram_alpha4}, we show a diagram of state space for $\alpha=4$ and $\sigma=-1.5$, which are parameters for which silence will always eventually occur due to the global stability of $\mathbf{x}_s$. For this example, the value of $x$ that leaves $y$ fixed is, by Equation \ref{eq:rulkov_1_slow_map_fixed_y},
\begin{equation}
    x_{s,\,\mathrm{slow}} = \sigma = -1.5
\end{equation}
The fixed value of $y$, or $y_s$, lies at the intersection between $B_{\mathrm{stable}}$ and $x_{s,\,\mathrm{slow}}$ (see Figure \ref{fig:rulkov_1_state_space_diagram_alpha4}), so by Equation \ref{eq:rulkov_fast_map_fixed_points},
\begin{equation}
    y_s = x_{s,\,\mathrm{slow}} - \frac{\alpha}{1-x_{s,\,\mathrm{slow}}} = -3.1
\end{equation}
Therefore, the fixed point of the map is
\begin{equation}
    \mathbf{x}_s = \begin{pmatrix}
        x_{s,\,\text{slow}} \\
        y_s
    \end{pmatrix}
    \begin{pmatrix}
        -1.5 \\
        -3.1
    \end{pmatrix}
\end{equation}

To demonstrate how silence arises, we will consider two cases: $y$ starting above $y_s$ and $y$ starting below $y_s$. Let us first consider the slightly more complex case of $y$ starting above $y_s$. Observing Figure \ref{fig:rulkov_1_state_space_diagram_alpha4}, this means $x$ will either be in the stable branch or the spiking branch. In Figure \ref{fig:rulkov_fast_alpha4_y-2.7}, we pick an initial value of $y_0=-2.7$, where matching this to Figure \ref{fig:rulkov_1_state_space_diagram_alpha4}, we can see that $x$ starts is in the spiking branch with an average value much larger than $x_{s,\,\mathrm{slow}}=-1.5$. Therefore, we know from our discussion of the slow map that this indicates $y$ will decrease, moving us leftward along the blue arrow in Figure \ref{fig:rulkov_1_state_space_diagram_alpha4} along the spiking branch. In Figure \ref{fig:rulkov_fast_alpha4_y-2.7}, we know from our discussion of the fast map that the decrease in $y$ will cause the curve to move down until the two fixed points $x_{s,\,\mathrm{fast,\,stable}}$ and $x_{s,\,\mathrm{fast,\,unstable}}$ are born. We can see in Figure \ref{fig:rulkov_fast_alpha4_y-3.05} that $x$ will be attracted to $x_{s,\,\mathrm{fast,\,stable}}$ as soon as the point is born, so $y$ will just slowly decrease until $x_{s,\,\mathrm{fast,\,stable}}=x_{s,\,\mathrm{slow}}$ and $y=y_s$, which corresponds to slow motion along the stable branch (see Figure \ref{fig:rulkov_1_state_space_diagram_alpha4}). We can see these dynamics occur in the graph of $x_k$ in Figure \ref{fig:rulkov_x_vs_k_graph_alpha4}, which shows repeated spiking followed by a slow, continuous decrease to $x_{s,\,\mathrm{slow}}$.

In the case where $y$ begins less than $y_s$, $x$ will be attracted to $x_{s,\,\mathrm{fast,\,stable}}<x_{s,\,\mathrm{slow}}$ (see Figure \ref{fig:rulkov_1_state_space_diagram_alpha4}). Then, $y$ will increase until it reaches $y_s$, $x$ slowly moving along the stable branch until it reaches $\mathbf{x}_s$. 

\begin{figure*}[ht!]
    \centering
    \begin{subfigure}[t]{0.475\textwidth}
        \centering
        \includegraphics[scale=0.1125]{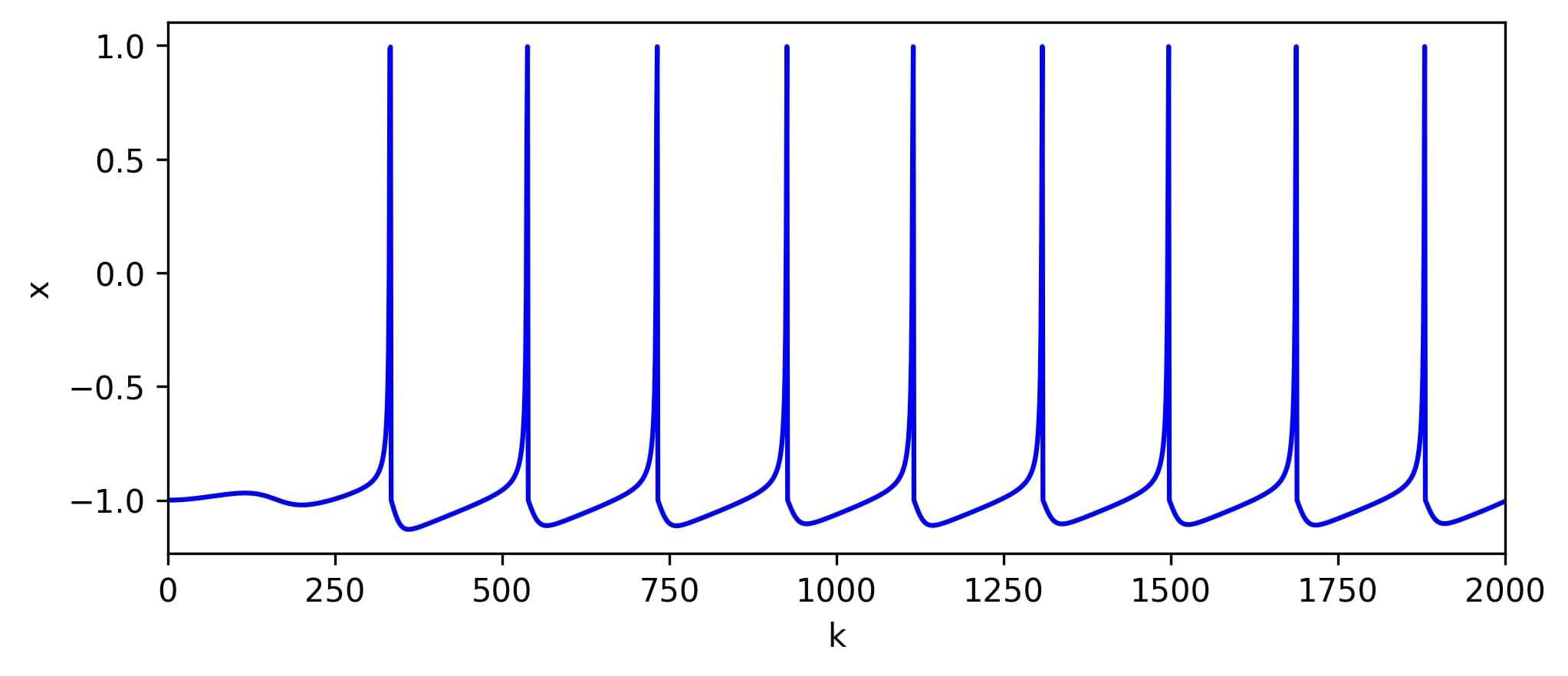}
        \caption{$\sigma=-0.99$}
        \label{fig:rulkov_1_spiking_sigma-0.99}
    \end{subfigure}
    \begin{subfigure}[t]{0.475\textwidth}
        \centering
        \includegraphics[scale=0.1125]{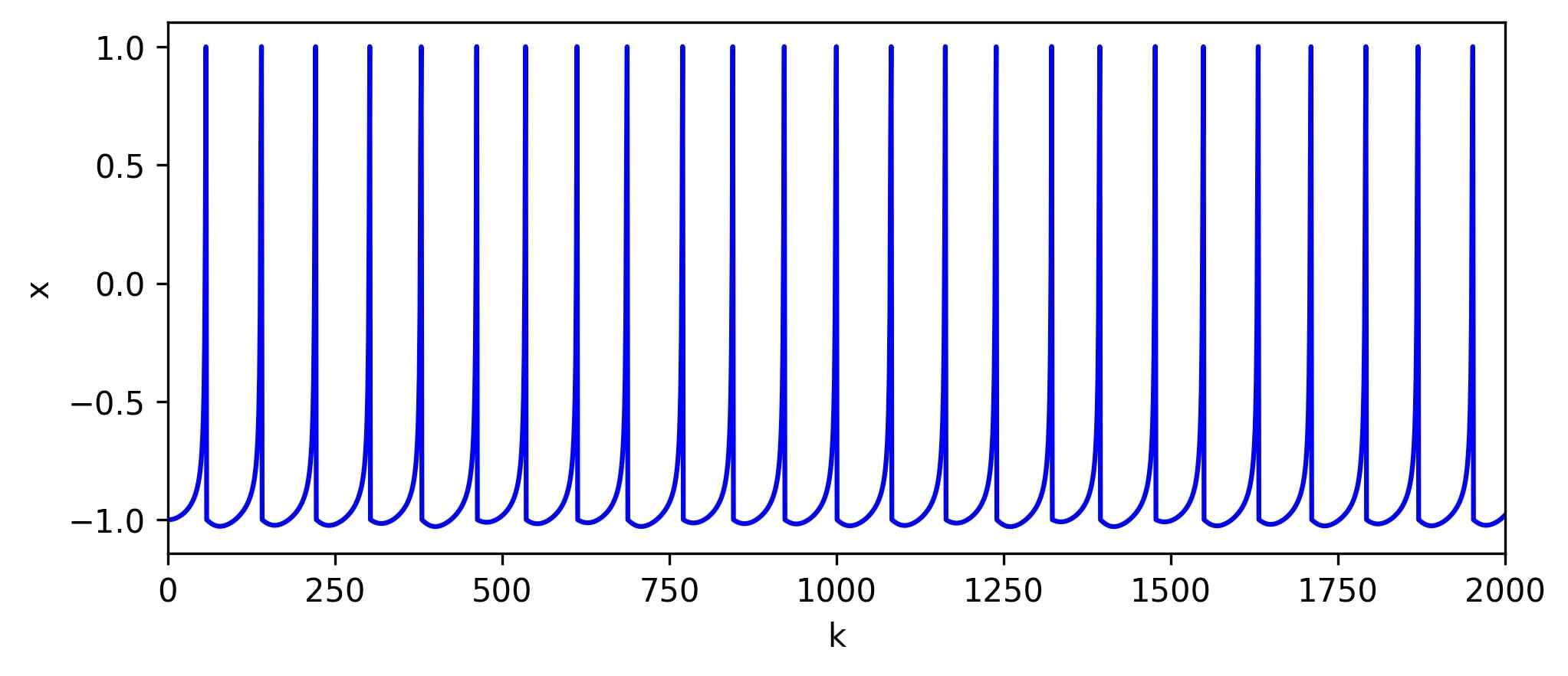}
        \caption{$\sigma=-0.9$}
        \label{fig:rulkov_1_spiking_sigma-0.9}
    \end{subfigure}
    \vspace{10px} \\
    \begin{subfigure}[t]{0.475\textwidth}
        \centering
        \includegraphics[scale=0.1125]{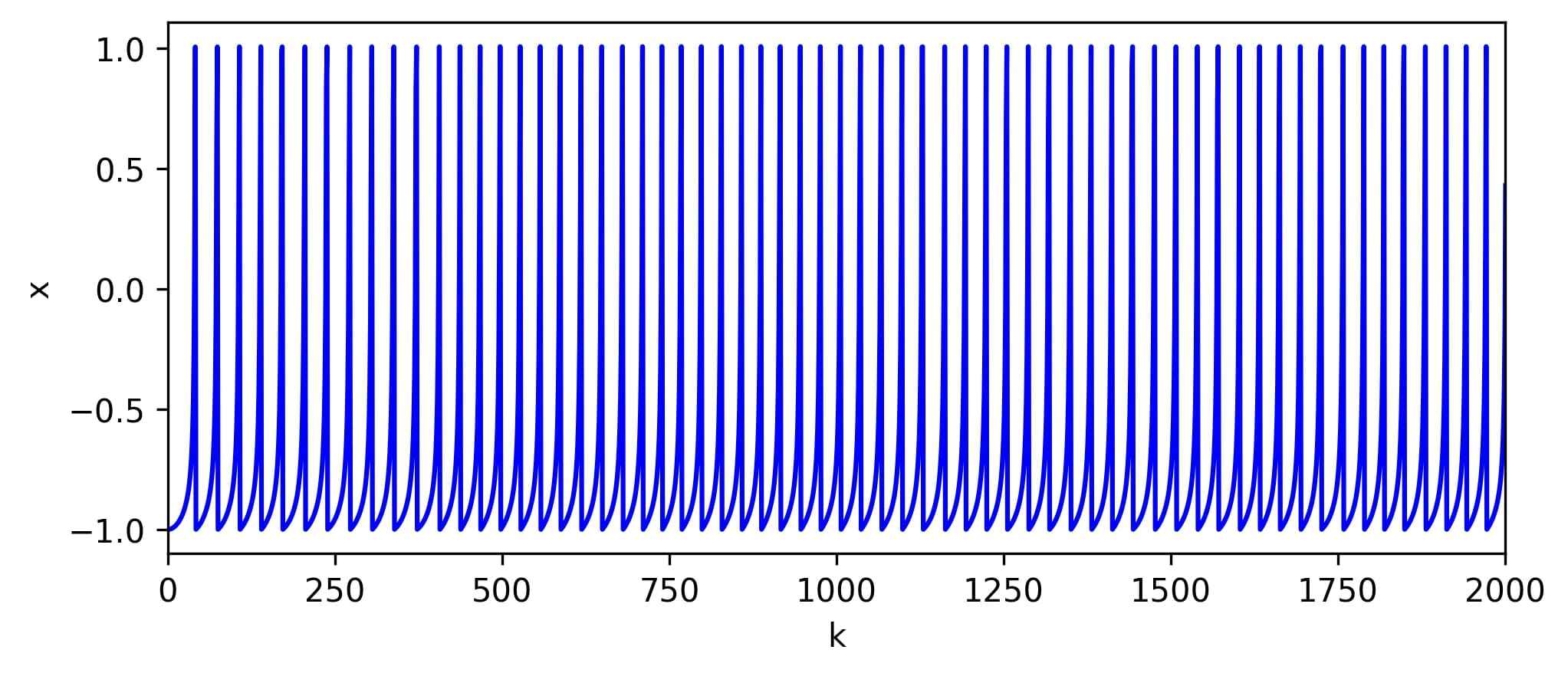}
        \caption{$\sigma=-0.75$}
        \label{fig:rulkov_1_spiking_sigma-0.75}
    \end{subfigure}
    \begin{subfigure}[t]{0.475\textwidth}
        \centering
        \includegraphics[scale=0.1125]{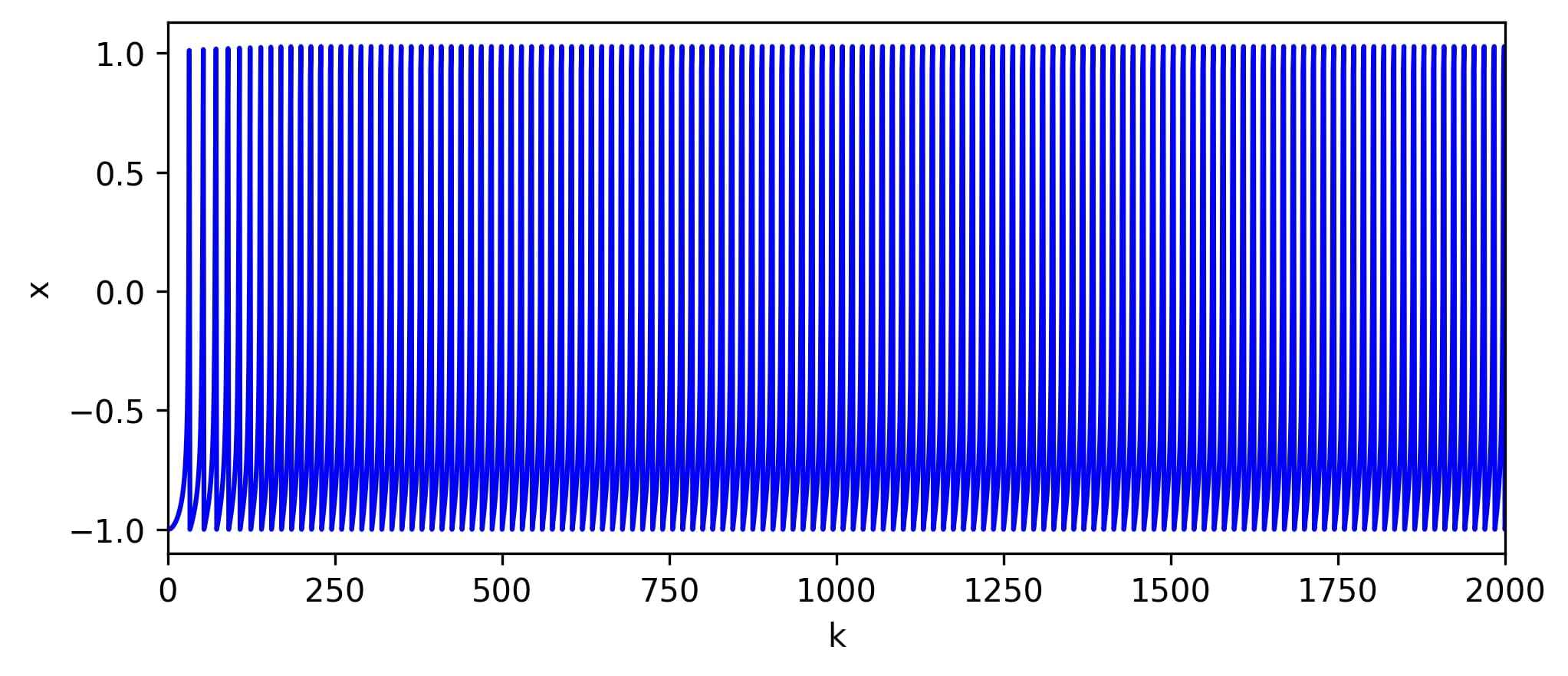}
        \caption{$\sigma=-0.5$}
        \label{fig:rulkov_1_spiking_sigma-0.5}
    \end{subfigure}
    \caption{Graphs of $x_k$ displaying spiking behavior with increasing frequency generated by Rulkov map 1 systems with $\alpha=4$, graphed with the code in Appendix \ref{fast_var_orbs_of_rulkov_1_code}}
    \label{fig:rulkov_1_spiking_graphs}
\end{figure*}

\begin{figure*}[hp!]
    \centering
    \hfill
    \begin{subfigure}[t]{0.475\textwidth}
        \centering
        \begin{tikzpicture}[scale=0.95]
            \begin{axis}[
                xlabel =\normalsize \(y\),
                xtick align=outside,
                xtick pos=left,
                xmin=-4.02, xmax=-3.83,
                xtick={-4,-3.95,-3.9,-3.85},
                minor xtick={-4.02, -4.01, -3.99,-3.98,-3.97,-3.96,-3.94,-3.93,-3.92,-3.91,-3.89,-3.88,-3.87,-3.86, -3.84, -3.83},
                ylabel = \normalsize \(x\),
                ytick align=outside,
                ytick pos=left,
                ymin=-2.5, ymax=0.5,
                ytick={-2,-1,0},
                minor ytick={0.4,0.2,-0.2,-0.4,-0.6,-0.8,-1.2,-1.4,-1.6,-1.8,-2.2,-2.4},
                scatter/use mapped color={
                    draw=green!80!black,
                    fill=green!80!black,
                },
            ]
            \addplot [
                domain=-4.02:-3.83, 
                samples=300,
                color=green!80!black,
                very thick
            ]
            {(1+x-sqrt((x-1)^2-24))/2};
            \addplot [
                domain=-4.02:-3.83, 
                samples=300,
                color=red,
                very thick,
                dashed
            ]
            {(1+x+sqrt((x-1)^2-24))/2};
            \addplot+[
                only marks,
                scatter,
                mark size=0.5pt,
            ]
            table 
            {data/rulkov_alpha6.dat};
            \addplot+[
                only marks,
                scatter,
                mark size=0.5pt,
            ]
            table 
            {data/rulkov_alpha6_2.dat};
            \addplot+[
                only marks,
                scatter,
                mark size=0.5pt,
            ]
            table 
            {data/rulkov_alpha6_3.dat};
            \addplot[color=red, mark=*, mark size=2.5pt] coordinates {(-3.917,-1.25)};
            \node[anchor=south west] at (axis cs:-3.917,-1.25) {\small $\mathbf{x}_s$};
            \addplot [
                domain=-4.02:-3.83, 
                thick,
                densely dotted
            ]
            {-1.25};
            \node[anchor=north west] at (axis cs:-3.98,-1.937) {\small $B_{\mathrm{stable}}$};
            \node[anchor=south west] at (axis cs:-3.98,-1.043 ) {\small $B_{\mathrm{unstable}}$};
            \node at (axis cs:-3.94,0.2) {\small $B_{\mathrm{spikes}}$};
            \node[anchor=south east] at (axis cs:-3.83,-1.25) {\small $x_{s,\,\mathrm{slow}}$};
            \draw[->, thick, color=blue] (axis cs:-3.92,-0.4) -- (axis cs:-3.97,-0.6);
            \draw[->, thick, color=blue] (axis cs:-4,-1.1) -- (axis cs:-4,-1.9);
            \draw[->, thick, color=blue] (axis cs:-3.98,-1.8) -- (axis cs:-3.93,-1.6);
            \draw[->, thick, color=blue] (axis cs:-3.895,-1.449) -- (axis cs:-3.895,-0.3);
            \end{axis}
        \end{tikzpicture}
        \caption{Branches of Rulkov map 1 in two-dimensional state space $\langle y,\, x \rangle$, with $x_{s,\,\mathrm{slow}}$, $\mathbf{x}_s$, and arrows representing typical motion}
        \label{fig:rulkov_1_state_space_diagram_alpha6}
    \end{subfigure}
    \hfill
    \begin{subfigure}[t]{0.475\textwidth}
        \centering
        \vspace{-6.6cm}
        \includegraphics[scale=0.1125]{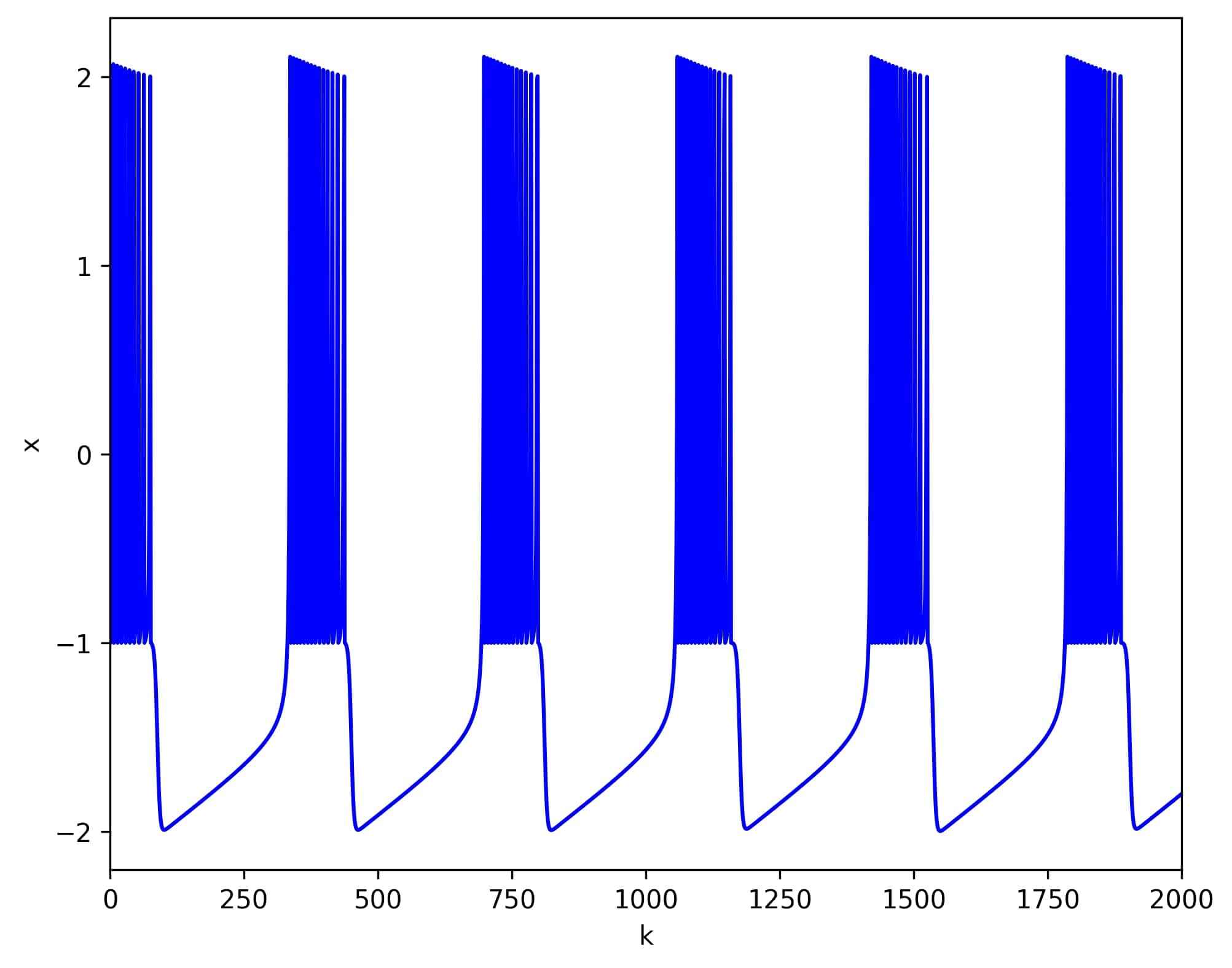}
        \vspace{0.6cm}
        \caption{Graph of the fast variable $x_k$ displaying bursting behavior, graphed with the code in Appendix \ref{fast_var_orbs_of_rulkov_1_code}}
        \label{fig:rulkov_x_vs_k_graph_alpha6}
    \end{subfigure}
    \hfill
    \caption{A state space diagram and a graph of $x_k$ for parameters $\alpha=6$ and $\sigma=-1.25$ of Rulkov map 1}
    \label{fig:rulkov_1_alpha6_sigma-1.25_graphs}
\end{figure*}

\subsubsection{Spiking}

Spiking behavior follows logically from silence, the only difference being that the fixed point of the map $\mathbf{x}_s$ is in the unstable branch due to a higher $x_{s,\,\mathrm{slow}}=\sigma$ value. For this reason, the state space diagram for spiking behavior is identical to the one for silence in Figure \ref{fig:rulkov_1_state_space_diagram_alpha4} except that the black $x_{s,\,\mathrm{slow}}$ line is moved up to intersect $B_{\mathrm{unstable}}$ rather than $B_{\mathrm{stable}}$. Then, if $y$ starts low enough to be in the domain of the stable and unstable branches, $x$ will slowly move up the stable branch since $x$ is less than $x_{s,\,\mathrm{slow}}$. Once $x$ increases to the point of leaving the stable branch, $x$ will move along the spiking branch until it reaches some periodic orbit $O^q(y;\,\alpha)$ that has $\langle O^q(y;\,\alpha)\rangle = x_{s,\,\mathrm{slow}}$. It is worth noting that due to the discontinuous nature of $B_{\mathrm{spikes}}$, there are multiple different values of $y$ that correspond to a stable spiking orbit because $x_{s,\,\mathrm{slow}}$ intersects $B_{\mathrm{spikes}}$ at multiple points, so the value of $y$ that a periodic orbit will end up oscillating around depends on the initial conditions of the neuron. Specifically, starting with values of $y$ on different discontinuous branches of the spiking branch $B_{\mathrm{spikes}}$ may cause the different initial conditions to be attracted to different periodic orbits.

A general trend that occurs with spiking behavior in Rulkov map 1 is that a higher value of $\sigma$ corresponds to a higher frequency of spikes. In the graph of state space (Figure \ref{fig:rulkov_1_state_space_diagram_alpha4}), we can see that this is the case because a higher $x_{s,\,\mathrm{slow}}$ corresponds to a higher value of $y$, and each discontinuity in $B_{\mathrm{spikes}}$ as we increase $y$ is a decrease in the period $q$ and an increase in the frequency of $O^q(-1)$. We can also see why this is the case in the fast map: a higher value of $y$ corresponds to a higher curve, resulting in fewer iterations of the map before returning to the original value of $-1$. Of course, a higher value of $\sigma$ is directly associated with a higher value of $x_{s,\,\mathrm{slow}}$, leading to a higher value of $y$ and a higher spike frequency. In Figure \ref{fig:rulkov_1_spiking_graphs}, we can see this increase in the frequency of spikes by increasing the parameter $\sigma$.

\subsubsection{Bursting}

Like spiking, bursting occurs when $\mathbf{x}_s$ is in the unstable branch. However, unlike spiking, we know from our discussion of combining the individual maps that bursting can only occur when $\alpha>4$, where $x_{s,\,\mathrm{fast,\,unstable}}$ can pass upwards through $x=-1$. Bursting is more complex than both silence and spiking, so we will carefully go over its dynamics.

In Figure \ref{fig:rulkov_1_state_space_diagram_alpha6}, we show a diagram of state space for $\alpha=6$ and $\sigma=-1.25$, which we can immediately see is fundamentally different from our earlier diagram in Figure \ref{fig:rulkov_1_state_space_diagram_alpha4}. To see how bursting arises, let us consider an initial condition where $x>x_{s,\,\mathrm{slow}}=-1.25$, meaning it is in the spiking branch. We give one example of this in Figure \ref{fig:rulkov_fast_alpha6_y-3.93}, where we can see a periodic orbit exists simultaneously with the two fixed points, which is something that doesn't occur when $\alpha\leq 4$. In Figure \ref{fig:rulkov_1_state_space_diagram_alpha6}, we can see this as $B_{\mathrm{stable}}$, $B_{\mathrm{unstable}}$, and $B_{\mathrm{spikes}}$ all being defined for some values of $y$. For the periodic orbit in Figure \ref{fig:rulkov_fast_alpha6_y-3.93} $O^q(-1;\,-3.93,\,6)$, the average value of $x$ is greater than $x_{s,\,\mathrm{slow}}$ (see Figure \ref{fig:rulkov_1_state_space_diagram_alpha6}), so $y$ will decrease. The next step is displayed in Figures \ref{fig:rulkov_fast_alpha6_y-3.99} and \ref{fig:rulkov_fast_alpha6_y-4.01}, showing that soon as $y$ gets small enough that $x_{s,\,\mathrm{fast,\,unstable}}$ passes through $-1$, the state gets repelled down from $x_{s,\,\mathrm{fast,\,unstable}}$ and attracted to $x_{s,\,\mathrm{fast,\,stable}}$.\footnote{In Figure \ref{fig:rulkov_fast_alpha6_y-4.01}, for clarity, we show a $y$ value past the point where $x$ will be attracted to $x_{s,\,\mathrm{fast,\,stable}}$, which is when $y$ is infinitesimally less than $-4$.} In Figure \ref{fig:rulkov_1_state_space_diagram_alpha6}, we can see this as a jump from $B_{\mathrm{spikes}}$ to $B_{\mathrm{stable}}$. Because $x_{s,\,\mathrm{slow}}$ intersects the unstable branch, $x$ is now less than $x_{s,\,\mathrm{slow}}$ (see Figure \ref{fig:rulkov_1_state_space_diagram_alpha6}), so $y$ slowly increases, moving dynamics slowly along the stable branch. Finally, once $y$ reaches the edge of the domain of the stable and unstable branches (see Figure \ref{fig:rulkov_fast_alpha6_y-3.899}), $x$ will be repelled away into the spiking branch, starting the cycle over again. This slow repeated oscillation between periodic spikes and silence is what we call bursting, which we can see graphed in Figure \ref{fig:rulkov_x_vs_k_graph_alpha6} using the code in Appendix \ref{fast_var_orbs_of_rulkov_1_code}.

\begin{figure}
    \centering
    \begin{tikzpicture}[scale=1]
        \draw (-6,2)--(0,2);
        \draw (-6,2)--(-6,8);
        \draw[thick, domain=2:8, variable=\y, samples=100]plot({3-3*sqrt(\y)},\y);
        \draw[thick] (-1*3, 4)--(-0.72*3, 5);
        \draw[thick] (-0.72*3, 5)--(-0.62*3, 6);
        \draw[thick] (-0.62*3, 6)--(-0.53*3, 7);
        \draw[thick] (-0.53*3, 7)--(-0.37*3, 8);
        \node[below] at (-3,1.5) {\Large $\sigma$};
        \node[left, rotate=90] at (-6.7,5.25) {\Large $\alpha$};
        \draw (-2*3, 2)--(-2*3, 1.85); 
        \draw (-1.5*3, 2)--(-1.5*3, 1.85); 
        \draw (-1*3, 2)--(-1*3, 1.85);  
        \draw (-0.5*3, 2)--(-0.5*3, 1.85); 
        \draw (-0*3, 2)--(-0*3, 1.85);  
        \node[below] at (-2*3, 1.85) {\scriptsize $-2$};
        \node[below] at (-1.5*3, 1.85) {\scriptsize $-1.5$};
        \node[below] at (-1*3, 1.85) {\scriptsize $-1$};
        \node[below] at (-0.5*3, 1.85) {\scriptsize $-0.5$};
        \node[below] at (0*3, 1.85) {\scriptsize $0$};
        \draw (-6, 2)--(-6.15, 2);
        \draw (-6, 3)--(-6.15, 3);
        \draw (-6, 4)--(-6.15, 4);
        \draw (-6, 5)--(-6.15, 5);
        \draw (-6, 6)--(-6.15, 6);
        \draw (-6, 7)--(-6.15, 7);
        \draw (-6, 8)--(-6.15, 8);
        \node[left] at (-6.15, 2) {\scriptsize $2$};
        \node[left] at (-6.15, 4) {\scriptsize $4$};
        \node[left] at (-6.15, 6) {\scriptsize $6$};
        \node[left] at (-6.15, 8) {\scriptsize $8$};
        \node at (-1.5*3, 3.5) {\large silence};
        \node[align = center] at (-1.05*3, 7) {\large bursts \\ \large of spikes};
        \node at (-0.25*3, 4) {\large spikes};
        \node[below left] at (-1.449*3, 6) {\large $\sigma_{\mathrm{th}}$};
        \node[below right] at (-0.62*3, 6) {\large $C_{\mathrm{bs}}$};
    \end{tikzpicture}
    \vspace{2px}
    \caption{Bifurcation diagram of Rulkov map 1 in parameter space $\langle \sigma,\, \alpha \rangle$}
    \label{fig:rulkov-1-bifurc-diag-param-space}
\end{figure}
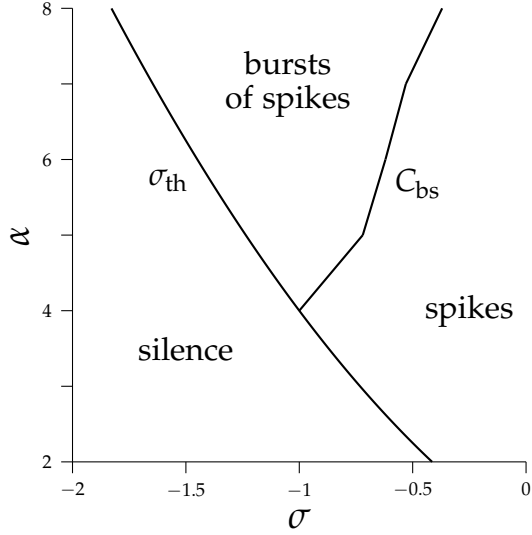

\subsection{Bifurcation Analysis of Rulkov Map 1}
\label{bifurcation-analysis-rulkov-map-1}

We have already discussed some of the bifurcations that occur in Rulkov map 1, such as the bifurcations when $0\in O^q(-1)$ that result in the discontinuities in $B_{\mathrm{spikes}}$ and the bifurcations when $x_{s,\,\mathrm{fast,\,unstable}}=-1$ that result in bursting behavior. However, in this section, we will conduct further analysis of the bifurcations from Sections \ref{bifurcations} and \ref{neimark-sacker-bifurcations} that appear in Rulkov map 1. We will also consider Rulkov map 1 without our approximation from before of splitting up the map into slow and fast motions. Instead, we will treat the map as it is shown in Equation \ref{eq:rulkov-map}, with the mapping of a two-dimensional state vector where $\eta$ is a small and finite parameter. From this perspective, we get a glimpse into the origin of chaotic dynamics in Rulkov map 1, which we further develop in Section \ref{chaotic-dynamics-rulkov-map-1}.

The simplest bifurcation in Rulkov map 1 occurs in its fast map. Specifically, we can see from Equation \ref{eq:rulkov_1_fixed_points_func_y} that no fixed points of the fast map exist for $y>1-2\sqrt{\alpha}$, one fixed point of the fast map exists for $y=1-2\sqrt{\alpha}$, and two fixed points of the fast map exist for $y<1-2\sqrt{\alpha}$. Recalling Section \ref{bifurcations}, this indicates that a saddle-node bifurcation occurs in the fast map when the parameter $y=1-2\sqrt{\alpha}$. We can see this visually in the intersection of the iteration function of the fast map $x_{k+1}=f_1(x_k;\,y,\,\alpha)$ with the line $x_{k+1}=x_k$ in Figures \ref{fig:rulkov_fast_maps_alpha4} and \ref{fig:rulkov_fast_maps_alpha6}, as well as the diagrams of state space showing the stable and unstable branches in Figures \ref{fig:rulkov_1_state_space_diagram_alpha4} and \ref{fig:rulkov_1_state_space_diagram_alpha6}, which can be thought of as a bifurcation diagram with $y$ as a parameter.

We will now consider bifurcations in the combined individual maps. The parameter space of a dynamical system is defined as the space of all its possible parameter values. For example, the parameter space of the logistic map introduced in Section \ref{bifurcations} is the space of all possible values of the parameter $r$. Because $r$ can take on any real value, the parameter space of the logistic map is therefore $\mathbb{R}$. Similarly, for Rulkov map 1, the two parameters $\sigma$ and $\alpha$ can both take on any real value, so the parameter space of Rulkov map 1 is $\mathbb{R}^2$. For our purposes, we will consider the subset of parameter space where $-2\leq\sigma\leq 0$ and $2\leq\alpha\leq 8$.

From the previous section (Section \ref{individual-dynamics-of-rulkov-map-1}), we know that the fixed point of Rulkov map 1 changes from stable to unstable at the threshold of excitation, which is the transition between silence and spiking-bursting behavior. Reiterating Equation \ref{eq:sigma-threshold-of-excitation}, the threshold of excitation curve $\sigma$ as a function of $\alpha$ is
\begin{equation}
    \sigma_{\mathrm{th}} = 1-\sqrt{\alpha}
\end{equation}
The bifurcation curve between bursting and spiking, which we will denote as $C_{\mathrm{bs}}$ is not as easily defined. In this paper, we rely on a numerical method of approximating it using the code in Appendix \ref{fast_var_orbs_of_rulkov_1_code}. Now, with these two bifurcation curves $\sigma_{\mathrm{th}}$ and $C_{\mathrm{bs}}$, one calculated exactly and one calculated numerically, we can graph this bifurcation diagram in our chosen region of parameter space $-2\leq\sigma\leq 0$ and $2\leq\alpha\leq 8$, which we do in Figure \ref{fig:rulkov-1-bifurc-diag-param-space}. This bifurcation diagram reflects the fact we know from Section \ref{individual-dynamics-of-rulkov-map-1} that bursts of spikes can only occur when $\alpha>4$, represented in Figure \ref{fig:rulkov-1-bifurc-diag-param-space} with bursting behavior occurring in the ``triangle'' formed by $\sigma_{\mathrm{th}}$ and $C_{\mathrm{bs}}$ above the $\alpha=4$ mark. 

It is also critical to note that this bifurcation diagram does not contain information about chaotic spiking and bursting that occurs for the values of $y$ where the spiking branch is densely folded. Specifically, in Figure \ref{fig:rulkov_1_state_space_diagram_alpha4}, approaching the point where the domain of $B_{\mathrm{spikes}}$ ends ($y=-3$), we can see increasingly close discontinuities in $B_{\mathrm{spikes}}$. Noting that the transition between silence and spiking occurs when $x_{s,\,\mathrm{slow}}$ goes from intersecting the stable branch to the unstable branch (see Figure \ref{fig:rulkov_1_state_space_diagram_alpha4}), we can see that in the vicinity of this transition, $x_{s,\,\mathrm{slow}}$ intersects $B_{\mathrm{spikes}}$ in this densely folded region. This results in dynamics sensitive to initial conditions and instability around some parts of $\sigma_{\mathrm{th}}$ because, recalling from Section \ref{individual-dynamics-of-rulkov-map-1}, spiking orbits depend on initial conditions due to the spiking branch's discontinuities. Because $x_{s,\,\mathrm{slow}}$ only intersects $B_{\mathrm{spikes}}$ in the spiking region of Figure \ref{fig:rulkov-1-bifurc-diag-param-space}, we predict that this chaotic spiking occurs only on the right side of $\sigma_{\mathrm{th}}$. Similarly, we can see this densely folded region of the spiking branch mirrored in Figure \ref{fig:rulkov_1_state_space_diagram_alpha6} when $y$ reaches the point where $x_{s,\,\text{fast, unstable}}=-1$ and the dynamics jump to the stable branch. In this case, the transition between bursting and spiking occurs when $x_{s,\,\mathrm{slow}}$ intersects the densely folded spiking branch. For the same reason then, we conjecture that chaotic spiking-bursting occurs around and mainly to the right of $C_{\mathrm{bs}}$ as well. We will elaborate on these chaotic dynamics further in Section \ref{chaotic-dynamics-rulkov-map-1}, but for now, this discussion emphasizes that the bifurcation diagram in Figure \ref{fig:rulkov-1-bifurc-diag-param-space} certainly doesn't tell the full story, and it indicates why the curve $C_{\mathrm{bs}}$ cannot be calculated exactly as the boundary between bursting and spiking isn't clear cut.

We will now consider Rulkov map 1 as a two-dimensional system without separating it into slow and fast variables. Reiterating Equation \ref{eq:rulkov-map}, Rulkov map 1 is defined by the iteration function
\begin{equation}
    \mathbf{x}_{k+1} = \mathbf{f}_1(\mathbf{x}_k;\,\sigma,\,\alpha,\,\eta) =
    \begin{pmatrix}
        f_1(x_k,\,y_k;\,\alpha) \\
        y_k - \eta(x_k - \sigma)
    \end{pmatrix}
\end{equation}
where, from Equation \ref{eq:rulkov_1_fast_equation}, 
\begin{equation}
    f_1(x,\,y;\,\alpha) = 
    \begin{cases}
        \alpha/(1-x) + y, & x\leq 0 \\
        \alpha + y, & 0 < x < \alpha + y \\
        -1, & x\geq \alpha + y
    \end{cases}
\end{equation}
The fixed point of Rulkov map 1 exists at, by Equations \ref{eq:rulkov_1_slow_map_fixed_y} and \ref{eq:rulkov_fast_map_fixed_points},
\begin{equation}
    \mathbf{x}_s = \begin{pmatrix}
        x_{s,\,\mathrm{slow}} \\
        y_s
    \end{pmatrix}
    =
    \begin{pmatrix}
        \sigma \\
        \sigma - \alpha/(1-\sigma)
    \end{pmatrix}
    \label{eq:rulkov_fixed_point_location}
\end{equation}
We know from Section \ref{individual-dynamics-of-rulkov-map-1} that in our approximation of splitting up the map in the limit $\eta\to 0$, $\mathbf{x}_s$ changes its stability at the threshold of excitation $\sigma_{\mathrm{th}}$. Therefore, by continuity, $\mathbf{x}_s$ must also change its stability for small and finite $\eta$. We know of three bifurcations that result in a change in the stability of a fixed point: a saddle-node bifurcation, a period-doubling bifurcation, and a Neimark-Sacker bifurcation. It is immediately obvious that the change in stability of the fixed point of the map $\mathbf{x}_s$ cannot be attributed to a saddle-node bifurcation because a saddle-node bifurcation requires two fixed points: one stable and one unstable. Now, recall from Section \ref{bifurcations} that a period-doubling bifurcation for a one-dimensional map occurs when $f'(x_s)=-1$ (Equation \ref{eq:period-doubling-bifurcation-requirement}). From our approximation in Section \ref{individual-dynamics-of-rulkov-map-1}, Equation \ref{eq:rulkov_fast_map_derivative} tells us that the derivative of the fast map's iteration function for $x\leq 0$ is
\begin{equation}
    f_1'(x;\,y,\,\alpha) = \frac{\alpha}{(1-x)^2}
\end{equation}
We know from Equations \ref{eq:rulkov_1_stable_fast_int} and \ref{eq:rulkov_1_unstable_fast_int} that $\mathbf{x}_s$ changes stability at $x_{s,\,\mathrm{fast}} = 1-\sqrt{\alpha}$. Substituting yields
\begin{equation}
    f_1'(1-\sqrt{\alpha};\,y,\,\alpha) = \frac{\alpha}{[1-(1-\sqrt{\alpha})]^2} = 1
\end{equation}
which is not $-1$. Therefore, by continuity, since our approximation in the limit $\eta\to 0$ doesn't give us a period-doubling bifurcation, it also cannot when $\eta$ is small and finite. This leaves us with the change in stability of $\mathbf{x}_s$ resulting from a Neimark-Sacker bifurcation, which we know from Section \ref{neimark-sacker-bifurcations} occurs when the eigenvalues of the Jacobian matrix of the map at the fixed point $J(\mathbf{x}_s)$ has a pair of complex conjugate eigenvalues with modulus 1: $\nu_{1,\,2}=e^{\pm i\varphi}$.

The Jacobian matrix of Rulkov map 1 $J(\mathbf{x})$ for $x\leq 0$ is
\begin{equation}
    J(\mathbf{x}) = 
    \begin{pmatrix}
        \frac{\partial f^{[1]}_1}{\partial x} & \frac{\partial f^{[1]}_1}{\partial y} \\[4px]
        \frac{\partial f^{[2]}_1}{\partial x} & \frac{\partial f^{[2]}_1}{\partial y}
    \end{pmatrix}
    =
    \begin{pmatrix}
        \frac{\alpha}{(1-x)^2} & 1 \\
        -\eta & 1
    \end{pmatrix}
\end{equation}
where $f^{[1]}_1=f_1$ and $f^{[2]}_1$ is the slow map iteration function. Substituting Equation \ref{eq:rulkov_fixed_point_location}, the Jacobian at the fixed point is
\begin{equation}
    J(\mathbf{x}_s) = \begin{pmatrix}
        \frac{\alpha}{(1-\sigma)^2} & 1 \\
        -\eta & 1
    \end{pmatrix}
\end{equation}
From Section \ref{neimark-sacker-bifurcations}, we know that when the Neimark-Sacker bifurcation occurs, the determinant and trace of $J(\mathbf{x}_s)$ are 1 and $2\cos\varphi$, respectively, which we prove in Appendix \ref{det-and-trace-ns}. Therefore, first considering the determinant of $J(\mathbf{x}_s)$, the Neimark-Sacker bifurcation of $\mathbf{x}_s$ occurs when 
\begin{equation}
    \det J(\mathbf{x}_s) = \frac{\alpha}{(1-\sigma)^2}+\eta = 1
\end{equation}
Solving for $\sigma$, we get
\begin{equation}
    \sigma_{\text{n-s bif}} = 1-\sqrt{\frac{\alpha}{1-\eta}}
    \label{eq:rulkov-sigma-n-s-bif}
\end{equation}
It is clear that, for small values of $\eta$, the Neimark-Sacker bifurcation curve $\sigma_{\text{n-s bif}}$ is almost exactly the same as the threshold of excitation curve $\sigma_{\mathrm{th}} = 1-\sqrt{\alpha}$ (Equation \ref{eq:sigma-threshold-of-excitation}) we calculated in our approximation:
\begin{equation}
    \sigma_{\text{n-s bif}} \approx \sigma_{\mathrm{th}}
\end{equation}
This confirms that the fixed point $\mathbf{x}_s$ indeed changes from stable to unstable through a Neimark-Sacker bifurcation, and it verifies the validity of our approximation.

We are now interested in finding the eigenvalues of the Jacobian at the fixed point on the Neimark-Sacker bifurcation curve $J(\mathbf{x}_s(\sigma_{\text{n-s bif}}))$, which should fit the form $\nu_{1,\,2}=e^{\pm i\varphi}=\cos\varphi\pm i\sin\varphi$. From linear algebra, we know that we can find these eigenvalues using
\begin{equation}
    \det(J(\mathbf{x}_s(\sigma_{\text{n-s bif}})) - \nu I) = 0
\end{equation}
Evaluating this determinant by substituting Equation \ref{eq:rulkov-sigma-n-s-bif} gives us the characteristic equation:
\begin{equation}
    \begin{split}
        \det(J(&\mathbf{x}_s(\sigma_{\text{n-s bif}}))-\nu I) \\
        &= \det \begin{pmatrix}
            \frac{\alpha}{(1-\sigma_{\text{n-s bif}})^2}-\nu & 1 \\[4px]
            -\eta & 1-\nu
        \end{pmatrix} \\
        &= \det \begin{pmatrix}
            1-\eta-\nu & 1 \\
            -\eta & 1-\nu
        \end{pmatrix} \\
        &= (1-\eta-\nu)(1-\nu) + \eta \\
        &= \nu^2 + (\eta-2)\nu + 1 = 0
    \end{split}
\end{equation}
Solving this quadratic by keeping in mind that $0<\eta\ll 1$ gives us  
\begin{equation}
    \nu_{1,\,2} = \frac{2-\eta}{2}\pm\frac{i}{2}\sqrt{\eta(4-\eta)}
\end{equation}
and since we know that $\nu_{1,\,2}=\cos\varphi\pm i\sin\varphi$, we can easily verify that $\tr J(\mathbf{x}_s(\sigma_{\text{n-s bif}})) = 2\cos\varphi$. It is also interesting to note that the eigenvalues depend only on $\eta$. Extending on this analysis, Shilnikov and Rulkov \cite{shilnikov} rigorously prove that this Neimark-Sacker bifurcation is subcritical, which is something that is outside the scope of this paper but that we can easily verify computationally. This indicates that an unstable periodic orbit closes in on the fixed point $\mathbf{x}_s$, which arises due to the collision of the map's stable and unstable invariant manifolds forming a periodic orbit known as a canard. This phenomenon is what leads to chaotic spiking near the Neimark-Sacker bifurcation curve: close initial conditions near the stable manifold are blown up and separated by the unstable manifold, but they get mixed again by the resetting mechanism of Rulkov map 1. For more details into the origin of chaos in Rulkov map 1, we recommend an interested reader to see the paper by Shilnikov and Rulkov \cite{shilnikov}. 

\begin{figure*}[ht!]
    \centering
    \hfill
    \begin{subfigure}[t]{0.475\textwidth}
        \centering
        \includegraphics[scale=0.1325]{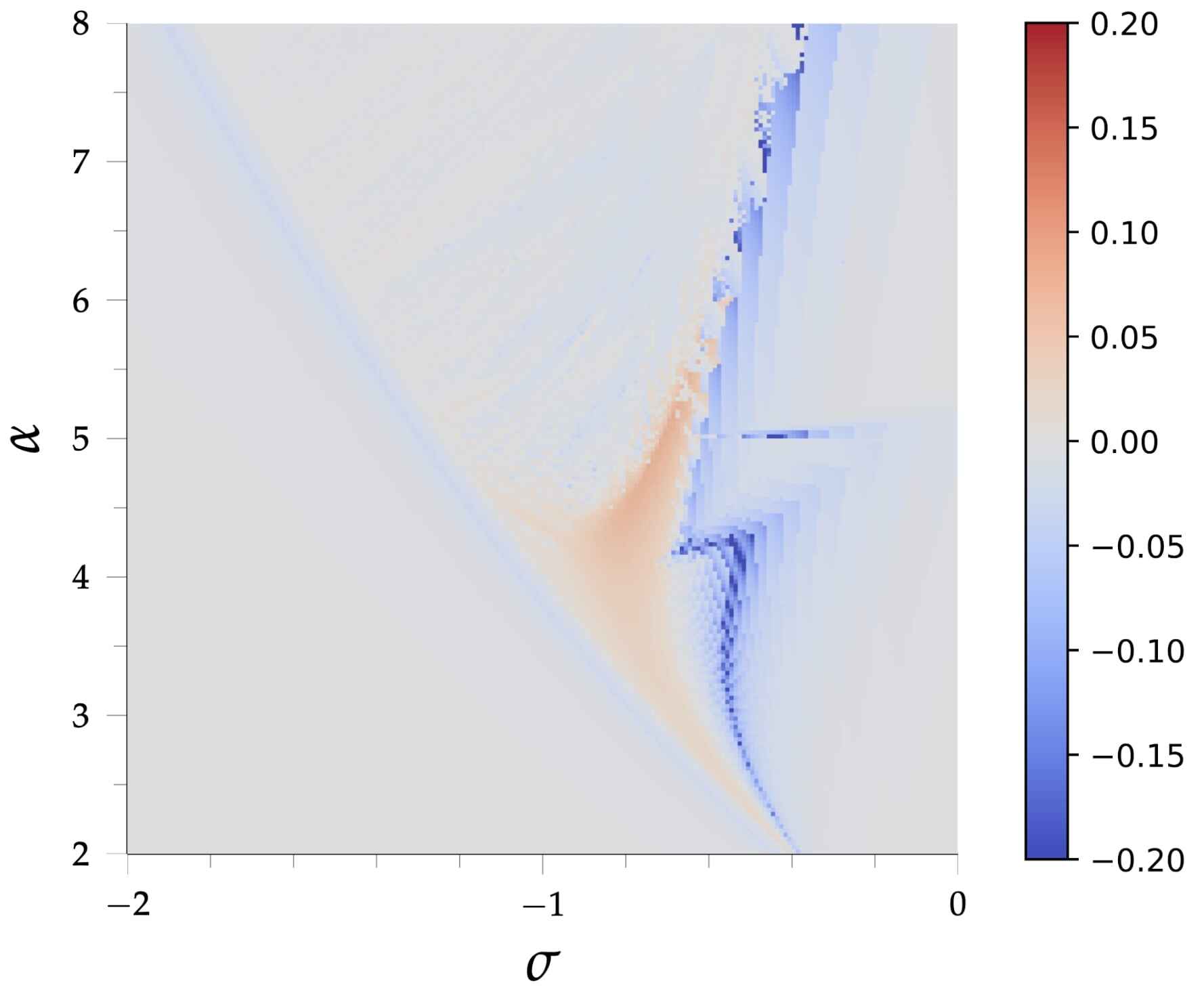}
        \caption{Color map in parameter space, with $\sigma$ on the $x$-axis and $\alpha$ on the $y$-axis}
        \label{fig:rulkov_1_lyapunov_exponents_parameter_space}
    \end{subfigure}
    \hfill
    \begin{subfigure}[t]{0.475\textwidth}
        \centering
        \includegraphics[scale=0.03]{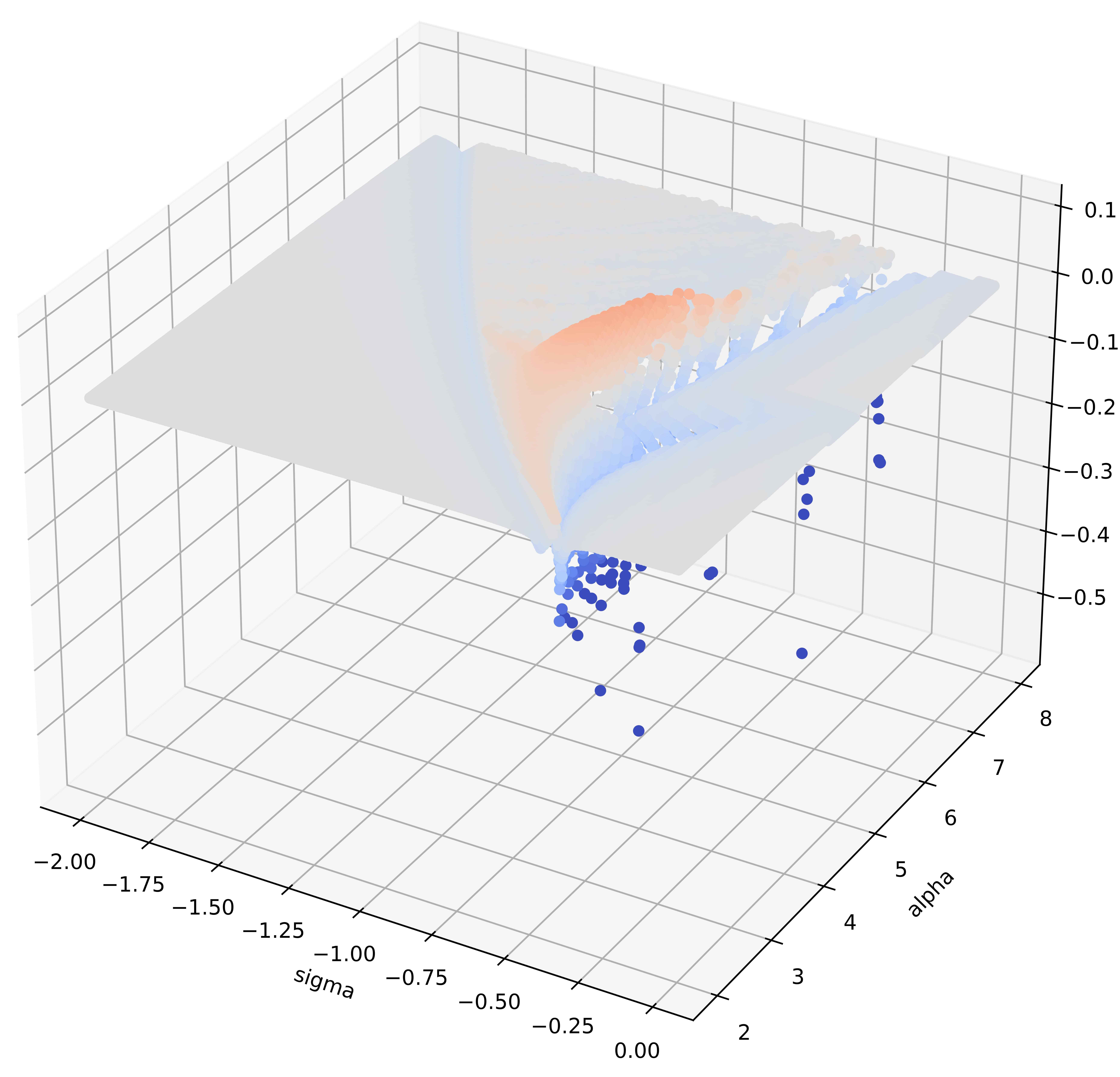}
        \vspace{2px}
        \caption{Three-dimensional graph, with $\sigma$ on the $x$-axis, $\alpha$ on the $y$-axis, and with maximal Lyapunov exponents on the $z$-axis}
        \label{fig:rulkov_1_lyapunov_exponents_3d_graph}
    \end{subfigure}
    \hfill
    \caption{Visualizations of the maximal Lyapunov exponent $\lambda_1$ of Rulkov map 1 in the subset of parameter space $-2\leq\sigma\leq 0$ and $2\leq\alpha\leq 8$, visualized using the code in Appendix \ref{lyapunov_spectrum_rulkov_map_1_code}}
    \label{fig:rulkov_1_lyapunov_exponents_visualizations}
    \vspace{4px}
\end{figure*}

\begin{figure*}[ht!]
    \centering
    \hfill
    \begin{subfigure}[t]{0.475\textwidth}
        \centering
        \includegraphics[scale=0.115]{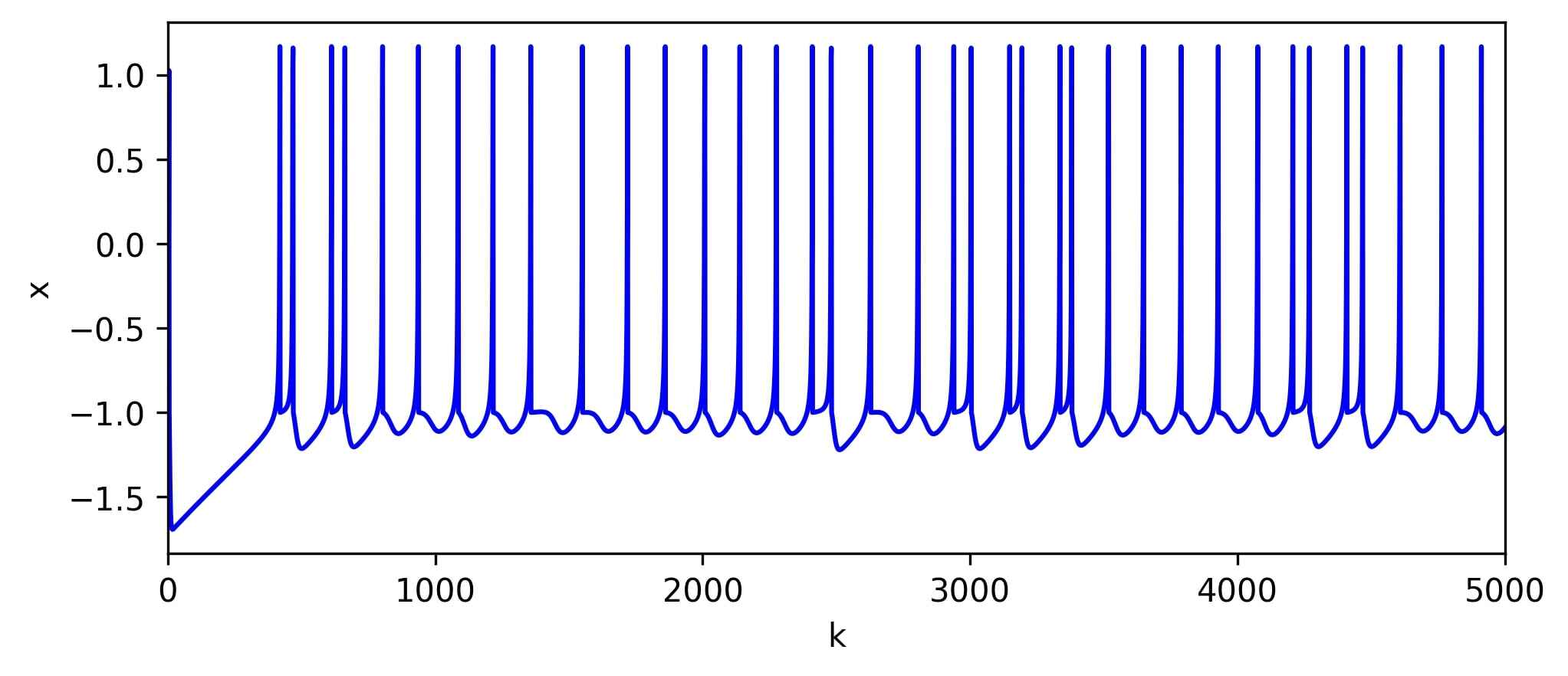}
        \caption{$\alpha=4.327$, $\sigma=-1.01$, $\lambda_1\approx 0.0260$}
        \label{fig:rulkov_1_chaotic_alpha4.327_sigma-1.01}
        \vspace{6px}
    \end{subfigure}
    \hfill
    \begin{subfigure}[t]{0.475\textwidth}
        \centering
        \includegraphics[scale=0.115]{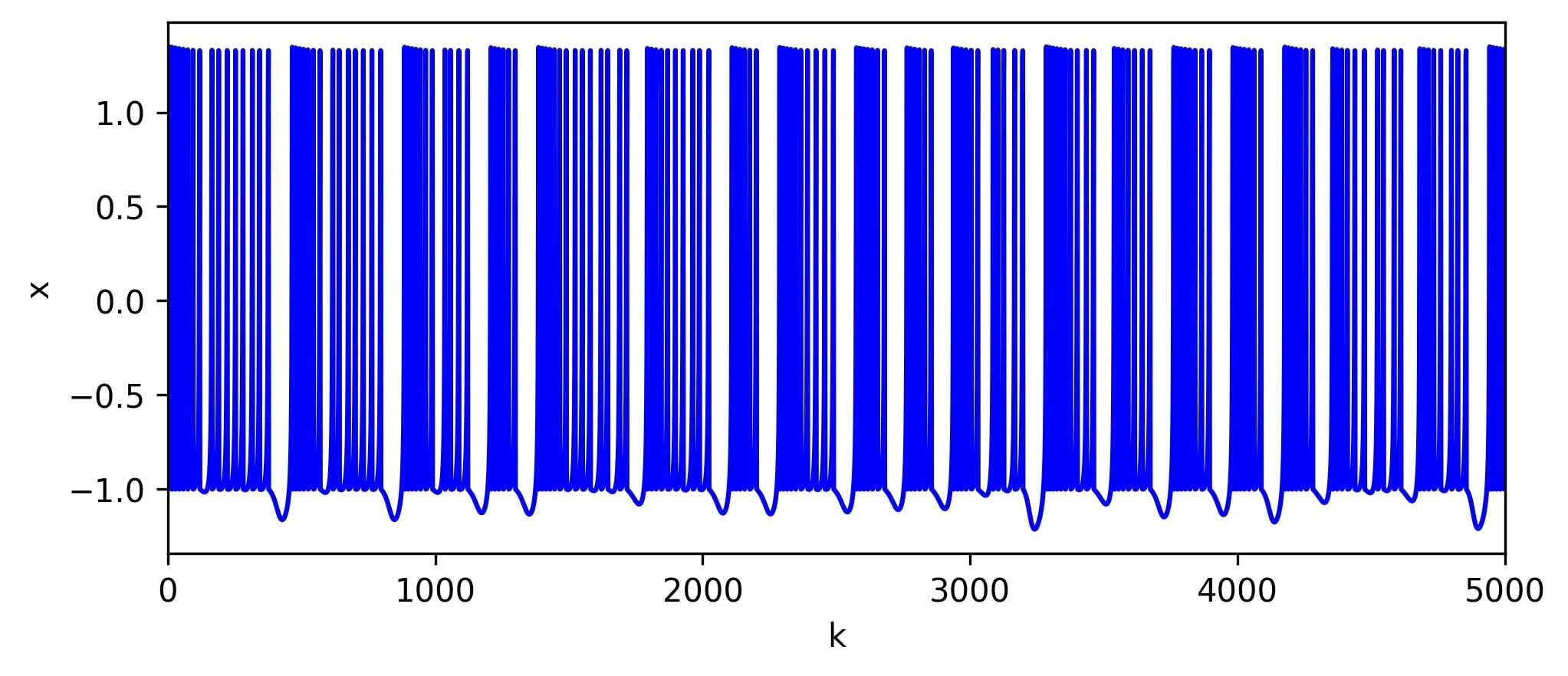}
        \caption{$\alpha=4.65$, $\sigma=-0.8$, $\lambda_1\approx 0.0443$}
        \label{fig:rulkov_1_chaotic_alpha4.65_sigma-0.8}
        \vspace{6px}
    \end{subfigure}
    \hfill
    \caption{Chaotic spiking and bursting behavior in Rulkov map 1 graphed with the code in Appendix \ref{fast_var_orbs_of_rulkov_1_code}, with the maximal Lyapunov exponent $\lambda_1$ calculated using the code in Appendix \ref{lyapunov_spectrum_rulkov_map_1_code}}
    \label{fig:rulkov_1_chaotic_spiking_bursting_graphs}
\end{figure*}

\subsection{Chaotic Dynamics of Rulkov Map 1}
\label{chaotic-dynamics-rulkov-map-1}

Now that we have qualitatively and analytically discussed the origin of chaos in Rulkov map 1, we are now ready to analyze and quantify the chaotic dynamics along the bifurcation curves $\sigma_{\mathrm{th}}$ and $C_{\mathrm{bs}}$. To do this, we will return to our method of quantifying chaos using Lyapunov exponents from Section \ref{quantification}, specifically, the QR factorization method we used to calculate the Lyapunov spectrum of the Hénon map in Section \ref{strangeattractors}. We know from Section \ref{quantification} that calculating the Lyapunov spectrum for a multi-dimensional system relies on the system's Jacobian matrix. From Equations \ref{eq:rulkov-map} and \ref{eq:rulkov_1_fast_equation}, we find that the Jacobian matrix of Rulkov map 1 is
\begin{equation}
    J(\mathbf{x}) = \begin{cases}
        \begin{pmatrix}
            \frac{\alpha}{(1-x)^2} & 1 \\
            -\eta & 1
        \end{pmatrix}, & x\leq 0 \\[0.5cm]
        \begin{pmatrix}
            0 & 1 \\
            -\eta & 1
        \end{pmatrix}, & 0 < x < \alpha + y \\[0.5cm]
        \begin{pmatrix}
            0 & 0 \\
            -\eta & 1
        \end{pmatrix}, & x\geq \alpha + y
    \end{cases}
\end{equation}
which is a piecewise function due to the piecewise structure of $f_1(x,\,y;\,\alpha)$. Using this, we implement the QR factorization method outlined in Appendix \ref{qr-meth-lyap-spec-calc} into the code in Appendix \ref{lyapunov_spectrum_rulkov_map_1_code}.

In Figure \ref{fig:rulkov_1_lyapunov_exponents_visualizations}, we visualize the maximal Lyapunov exponents of Rulkov map 1 in our standard subset of parameter space ($-2\leq\sigma\leq 0$ and $2\leq\alpha\leq 8$) using the code in Appendix \ref{lyapunov_spectrum_rulkov_map_1_code}. Specifically, in Figure \ref{fig:rulkov_1_lyapunov_exponents_parameter_space}, we show the maximal Lyapunov exponent for a given point in parameter space using color, red points being associated with a positive maximal Lyapunov exponent, which we know from Section \ref{quantification} indicates chaotic dynamics, and blue points being associated with a negative maximal Lyapunov exponent, which indicates non-chaotic dynamics. In Figure \ref{fig:rulkov_1_lyapunov_exponents_3d_graph}, we show the maximal Lyapunov exponents in parameter space on the $z$-axis, in addition to the same color scheme as Figure \ref{fig:rulkov_1_lyapunov_exponents_parameter_space}. Comparing Figure \ref{fig:rulkov_1_lyapunov_exponents_visualizations} with our crude bifurcation diagram in Figure \ref{fig:rulkov-1-bifurc-diag-param-space}, we can immediately see the threshold of excitation $\sigma_{\mathrm{th}}$ marked with a curve of blue stability in both of the figures. Additionally, as we predicted in Section \ref{bifurcation-analysis-rulkov-map-1}, we can see a prominent region of chaotic dynamics present to the right and near the bottom of the curves $\sigma_{\mathrm{th}}$ and $C_{\mathrm{bs}}$. Specifically, this region of chaotic dynamics is present around the intersection of $\sigma_{\mathrm{th}}$ and $C_{\mathrm{bs}}$ where the spiking branch is densely folded. In the ``triangle'' of bursting behavior from the bifurcation diagram in Figure \ref{fig:rulkov-1-bifurc-diag-param-space}, we can see light strips of chaotic and non-chaotic dynamics in Figure \ref{fig:rulkov_1_lyapunov_exponents_parameter_space}. In Figures \ref{fig:rulkov_1_lyapunov_exponents_parameter_space} and \ref{fig:rulkov_1_lyapunov_exponents_3d_graph}, we can also see that past the regions to the right of $C_{\mathrm{bs}}$, which we know from Section \ref{individual-dynamics-of-rulkov-map-1} is associated with low frequency (small $\sigma$) spiking, contains the most stable dynamics in this subset of state space. We can also see strips of varying blue stability in this low-frequency spiking region, which we conjecture arises from the discontinuities in the spiking branch, or in other words, the varying periods of spiking orbits.

We have already seen examples of non-chaotic silence, spiking, and bursting behaviors in Section \ref{individual-dynamics-of-rulkov-map-1} (see Figures \ref{fig:rulkov_x_vs_k_graph_alpha4}, \ref{fig:rulkov_1_spiking_graphs}, and \ref{fig:rulkov_x_vs_k_graph_alpha6}), but our analysis of the maximal Lyapunov exponents in parameter space allowed us to pinpoint a red region of chaotic dynamics in Rulkov map 1. Graphs of the evolution of $x$ for two sets of parameters in this red region are shown in Figure \ref{fig:rulkov_1_chaotic_spiking_bursting_graphs}. Both of these graphs have a maximal Lyapunov exponent $\lambda_1$ that is larger than 0, exhibiting chaotic, irregular dynamics. In Figure \ref{fig:rulkov_1_chaotic_alpha4.327_sigma-1.01}, we display a system exhibiting chaotic spiking, where the spikes are spaced unevenly and occasionally burst twice. In Figure \ref{fig:rulkov_1_chaotic_alpha4.65_sigma-0.8}, we display a system exhibiting chaotic bursting, with unevenly spaced bursts of different durations.

It is critical to note that, although a Rulkov map 1 system will be attracted to a chaotic attractor for some parameter values, this attractor is not fractal due to the resetting mechanism of Rulkov map 1. Specifically, this is because, although infinitesimally close orbits will diverge from each other in the short term, their fast variables will always be sent back to $-1$ together once they are larger than $\alpha+y$ (Equation \ref{eq:rulkov_1_fast_equation}). This leads to the Rulkov map 1 chaotic attractor not being a true strange attractor, which we can easily verify using the Kaplan-Yorke conjecture.\footnote{See the end of Section \ref{strangeattractors}.} As we can confirm using the code in Appendix \ref{lyapunov_spectrum_rulkov_map_1_code}, the Lyapunov spectrum of any Rulkov 1 neuron will always have the form $\{\lambda_1,\, -\infty\}$, where $\lambda_1$ is some finite real number. $\lambda_2$ being negative infinity captures the fact that perturbations get instantly collapsed to 0 as soon as their $x$ values get bigger than $\alpha+y$. For a chaotic spiking-bursting neuron, $\lambda_1>0$. In this case, because $\kappa=1$ is the largest index such that 
\begin{equation}
    \sum_{i=1}^{\kappa}\lambda_i\geq 0
\end{equation}
the Kaplan-Yorke conjecture tells us that the Lyapunov dimension of the Rulkov map 1 chaotic attractor is
\begin{equation}
        d_l = \kappa + \frac{1}{|\lambda_{\kappa+1}|}\sum_{i=1}^{\kappa}\lambda_i = 1 + \frac{\lambda_1}{|-\infty|} = 1
\end{equation}
which is an integer, indicating that the chaotic attractor is non-fractal. However, although a single Rulkov 1 neuron alone isn't fractal, we will discover fractal pseudo-attractors generated by Rulkov map 1 when we analyze an asymmetrically electrically coupled Rulkov 1 neuron system in Section \ref{asym-elec-coup-two-rulkov-1-neurons-geometry}, as well as true high-dimensional fractal attractors in our ring lattice system analysis in Section \ref{ring-lattice-geometry}. For now, however, this discussion of chaotic spiking-bursting in Rulkov map 1 provides a good transition to the extreme chaotic dynamics characteristic of Rulkov map 2.

\subsection{Dynamics of Rulkov Map 2}
\label{dynamics-rulkov-map-2}

Rulkov map 2 is modeled by Equation \ref{eq:rulkov-map} with Equation \ref{eq:rulkov_2_fast_equation} modeling the fast variable:
\begin{equation}
    \mathbf{x}_{k+1} = \mathbf{f}_2(\mathbf{x}_k)
\end{equation}
or, explicitly, 
\begin{equation}
    \begin{split}
        \begin{pmatrix}
            x_{k+1} \\
            y_{k+1}
        \end{pmatrix} &= 
        \begin{pmatrix}
            f_2\e{1}(x_k,\,y_k;\,\sigma,\,\alpha,\,\eta) \\[4px]
            f_2\e{2}(x_k,\,y_k;\,\sigma,\,\alpha,\,\eta)
        \end{pmatrix} \\
        &= \begin{pmatrix}
            f_2(x_k,\,y_k;\,\alpha) \\
            y_k - \eta(x_k - \sigma)
        \end{pmatrix} \\
        &= \begin{pmatrix}
            \frac{\alpha}{(1+x_k)^2} + y_k \\
            y_k - \eta(x_k-\sigma)
        \end{pmatrix}
    \end{split}
\end{equation}
This map is a chaotic model, but we will not rigorously analyze its dynamics due to its algebraic complexity and similarities to the Rulkov map 1.\footnote{For a rigorous treatment of this system's dynamics, see the paper by Rulkov \cite{rulkov2} or the review by Ibarz, Casado, and Sanjuán \cite{ibarz}.} Instead, we will briefly and qualitatively explore certain characteristics of the map and its behavior in order to gain insight into emergent geometrical properties of interest.

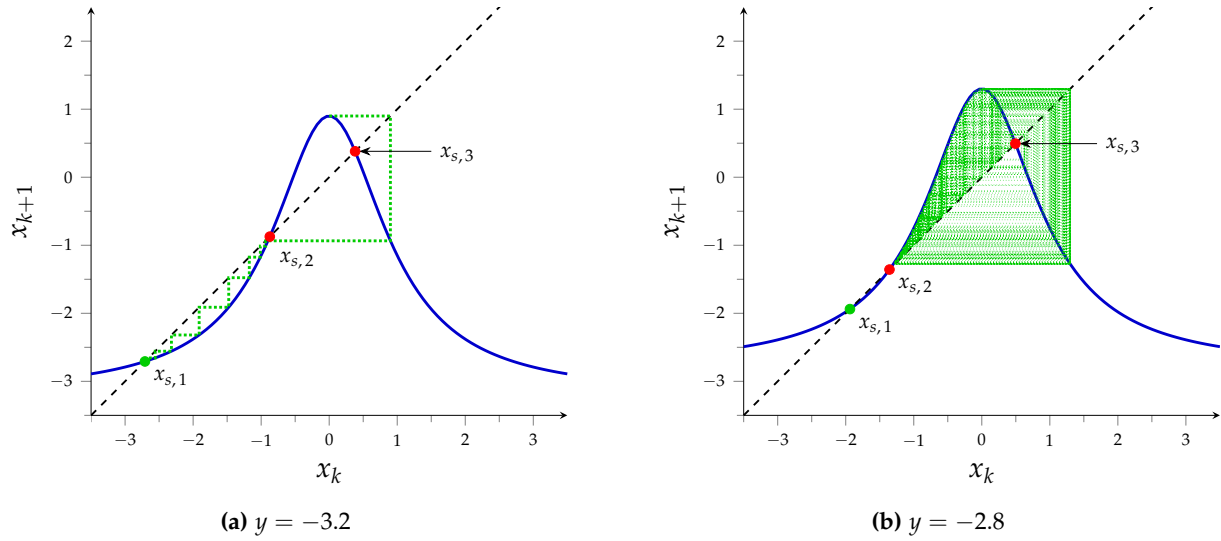
\begin{figure*}[htp!]
    \begin{subfigure}{0.475\textwidth}
        \centering
        \begin{tikzpicture}[scale=0.9]
            \begin{axis}[
                    axis lines = left,
                    x=1cm,
                    y=1cm,
                    xlabel =\large \(x_k\),
                    xtick align=outside,
                    xtick pos=left,
                    xmin=-3.5, xmax=3.5,
                    xtick={-3,-2,-1,0,1,2,3},
                    minor xtick={-3.5,-2.5,-1.5,-0.5,0.5,1.5,2.5},
                    ylabel = \large \(x_{k+1}\),
                    ytick align=outside,
                    ytick pos=left,
                    ymin=-3.5, ymax=2.5,
                    ytick={-3,-2,-1,0,1,2},
                    minor ytick={-3.5,-2.5,-1.5,-0.5,0.5,1.5},
                ]
            \addplot [
                domain=-3.5:3.5, 
                samples=100,
                color=blue!80!black,
                very thick
            ]
            {4.1/(1 + x^2) - 3.2};
            \addplot [
                domain=-3.5:3.5, 
                color=black,
                thick,
                dashed
            ]
            {x};
            \addplot+[
                color=green!80!black,
                mark=none,
                densely dotted,
                very thick
            ]
            table 
            {data/rulkov2_orbit_cobweb_y-3.2.dat};
    
            \addplot[color=green!80!black, mark=*] coordinates {(-2.708,-2.708)};
            \node[anchor=north west] at (axis cs:-2.708,-2.708) {\small $x_{s,\,1}$};
    
            \addplot[color=red, mark=*] coordinates {(-0.873,-0.873)};
            \node[anchor=north west] at (axis cs:-0.853,-0.963) {\small $x_{s,\,2}$};
            
            \addplot[color=red, mark=*] coordinates {(0.381,0.381)};
            \node[anchor=west] at (axis cs:1.501,0.331) {\small $x_{s,\,3}$};
    
            \draw[-Stealth] (axis cs:1.501,0.381)--(axis cs:0.441,0.381);
    
            \end{axis}
        \end{tikzpicture}
        \caption{$y=-3.2$}
        \label{fig:rulkov2_fast_alpha4.1_y-3.2}
    \end{subfigure}
    \begin{subfigure}{0.495\textwidth}
        \centering
        \begin{tikzpicture}[scale=0.9]
            \begin{axis}[
                    axis lines = left,
                    x=1cm,
                    y=1cm,
                    xlabel =\large \(x_k\),
                    xtick align=outside,
                    xtick pos=left,
                    xmin=-3.5, xmax=3.5,
                    xtick={-3,-2,-1,0,1,2,3},
                    minor xtick={-3.5,-2.5,-1.5,-0.5,0.5,1.5,2.5},
                    ylabel = \large \(x_{k+1}\),
                    ytick align=outside,
                    ytick pos=left,
                    ymin=-3.5, ymax=2.5,
                    ytick={-3,-2,-1,0,1,2},
                    minor ytick={-3.5,-2.5,-1.5,-0.5,0.5,1.5},
                ]
            \addplot [
                domain=-3.5:3.5, 
                samples=100,
                color=blue!80!black,
                very thick
            ]
            {4.1/(1 + x^2) - 2.8};
            \addplot [
                domain=-3.5:3.5, 
                color=black,
                thick,
                dashed
            ]
            {x};
            \addplot+[
                color=green!80!black,
                mark=none,
                densely dotted,
                very thin
            ]
            table 
            {data/rulkov2_orbit_cobweb_y-2.8.dat};
            
            \addplot[color=green!80!black, mark=*] coordinates {(-1.938,-1.938)};
            \node[anchor=north west] at (axis cs:-1.938,-1.938) {\small $x_{s,\,1}$};
    
            \addplot[color=red, mark=*] coordinates {(-1.357,-1.357)};
            \node[anchor=north west] at (axis cs:-1.357,-1.357) {\small $x_{s,\,2}$};
    
            \addplot[color=red, mark=*] coordinates {(0.494,0.494)};
            \node[anchor=west] at (axis cs:1.694,0.464) {\small $x_{s,\,3}$};
    
            \draw[-Stealth] (axis cs:1.694,0.494)--(axis cs:0.554,0.494);
            
            \end{axis}
        \end{tikzpicture}
        \caption{$y=-2.8$}
        \label{fig:rulkov2_fast_alpha4.1_y-2.8}
    \end{subfigure}
    \caption{The function $x_{k+1}=f_2(x_k;\,y,\,\alpha)$ graphed in blue for $\alpha=4.1$, with the stable fixed point $x_{s,\,1}$ shown in green and unstable fixed points $x_{s,\,2}$ and $x_{s,\,3}$ shown in red at the intersection between the function and the dashed black line $x_{k+1}=x_k$ and cobweb orbits (produced with the code in Appendix \ref{rulkov_2_graphs_and_cobweb_code}) shown with a dotted green line}
    \label{fig:rulkov2_cobwebs}
\end{figure*}

\begin{figure*}[htp!]
    \centering    \includegraphics[scale = 0.25]{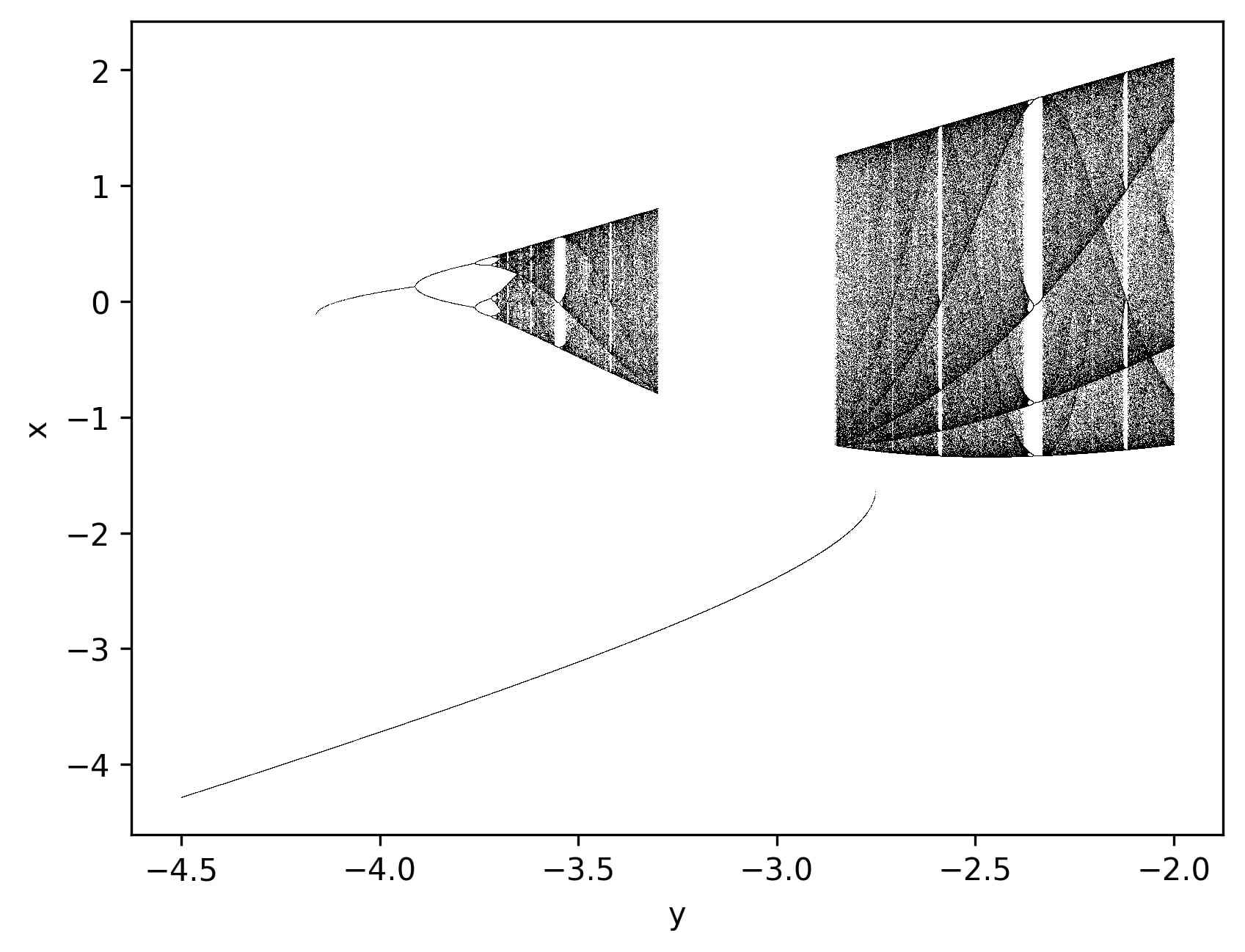}
    \caption{Bifurcation diagram of the fast map of Rulkov map 2 for $\alpha=4.1$ and $-4.5\leq y\leq -2$, graphed with the code in Appendix \ref{bifurcation_diagram_rulkov_2_code}}
    \label{fig:bifurc-rulkov-2}
\end{figure*}

We will begin by making the approximation of splitting up the map into slow and fast motions as we did for Rulkov map 1, treating $y$ as a slowly drifting parameter of $x$. In Figure \ref{fig:rulkov2_cobwebs}, we graph the behavior of the fast map $x_{k+1}=f_2(x_k;\,y,\,\alpha)$ for parameter values $\alpha=4.1$ and both $y=-3.2$ and $y=-2.8$ using the code in Appendix \ref{rulkov_2_graphs_and_cobweb_code}. For any value of $y$ where the fixed points $x_{s,\,1}$ and $x_{s,\,2}$ exist, $x_{s,\,1}$ will always be stable and $x_{s,\,2}$ will always be unstable. The fixed point $x_{s,\,3}$ can either be stable or unstable, but in both of these cases, it is unstable \cite{rulkov2}. Figure \ref{fig:rulkov2_fast_alpha4.1_y-3.2} shows the fast map with $y=-3.2$, where the initial state is attracted to the point $x_{s,\,1}$. This can also be seen in the bifurcation diagram displayed in Figure \ref{fig:bifurc-rulkov-2} because there is an island of stability when $y=-3.2$, where any initial state will be attracted to $x_{s,\,1}$. In Figure \ref{fig:rulkov2_fast_alpha4.1_y-2.8}, however, we can see that our initial state produces a chaotic orbit. For this $y$ value though, a different initial state could have also been attracted to $x_{s,\,1}$ instead of a chaotic orbit. This can be seen in the bifurcation diagram at the value of $y=-2.8$, where an initial state can either be attracted to the fixed point towards the bottom of the diagram or the chaotic orbit above it. In this bifurcation diagram, making a connection with Rulkov map 1, the curve at the bottom corresponds with the stable branch $B_{\text{stable}}$ and the chaotic orbits at the top correspond with the spiking branch $B_{\text{spikes}}$.\footnote{See Section \ref{individual-dynamics-of-rulkov-map-1} and Figures \ref{fig:rulkov_1_state_space_diagram_alpha4} and \ref{fig:rulkov_1_state_space_diagram_alpha6}.}

\begin{figure*}[htp!]
    \centering
    \begin{subfigure}{0.9\textwidth}
        \centering
        \includegraphics[scale=0.1875]{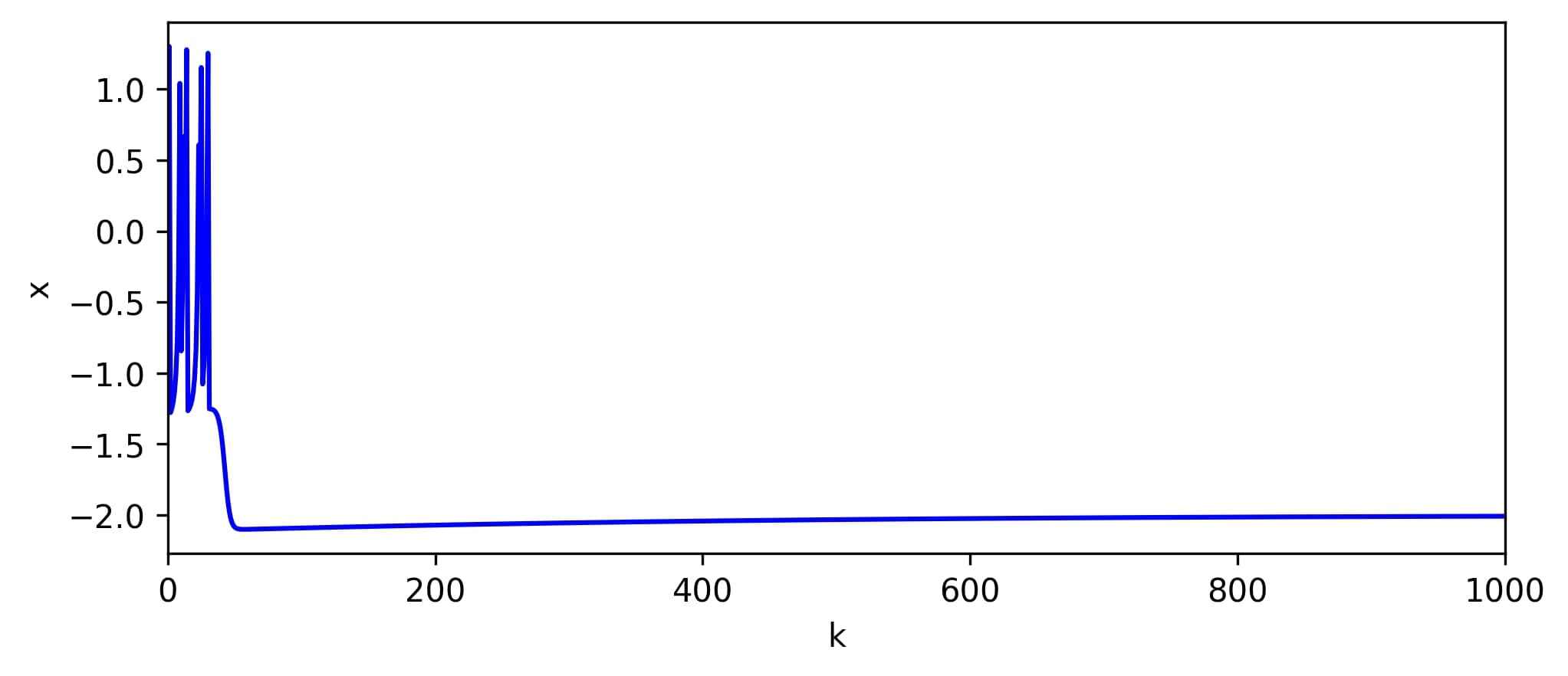}
        \caption{$\alpha=4.1$, $\sigma=-2$, $\lambda\approx\{-2.955\times 10^{-3},\,-0.4171\}$}
        \label{fig:rulkov_2_silence}
        \vspace{10px}
    \end{subfigure}
    \begin{subfigure}{0.9\textwidth}
        \centering
        \includegraphics[scale=0.1875]{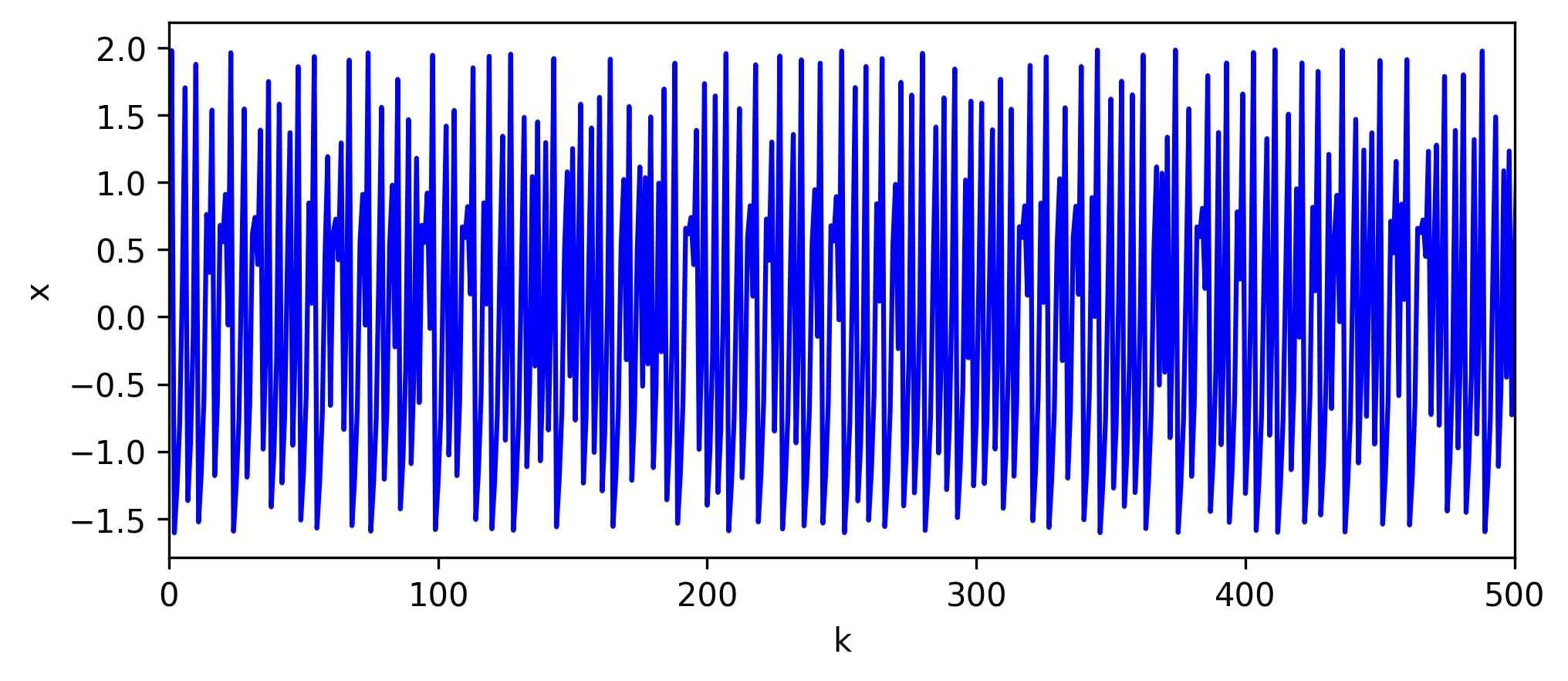}
        \caption{$\alpha=4.5$, $\sigma=0$, $\lambda\approx\{0.5449,\, -2.070\times 10^{-4}\}$}
        \label{fig:rulkov_2_spiking}
        \vspace{10px}
    \end{subfigure}
     \begin{subfigure}{0.9\textwidth}
        \centering
        \includegraphics[scale=0.1875]{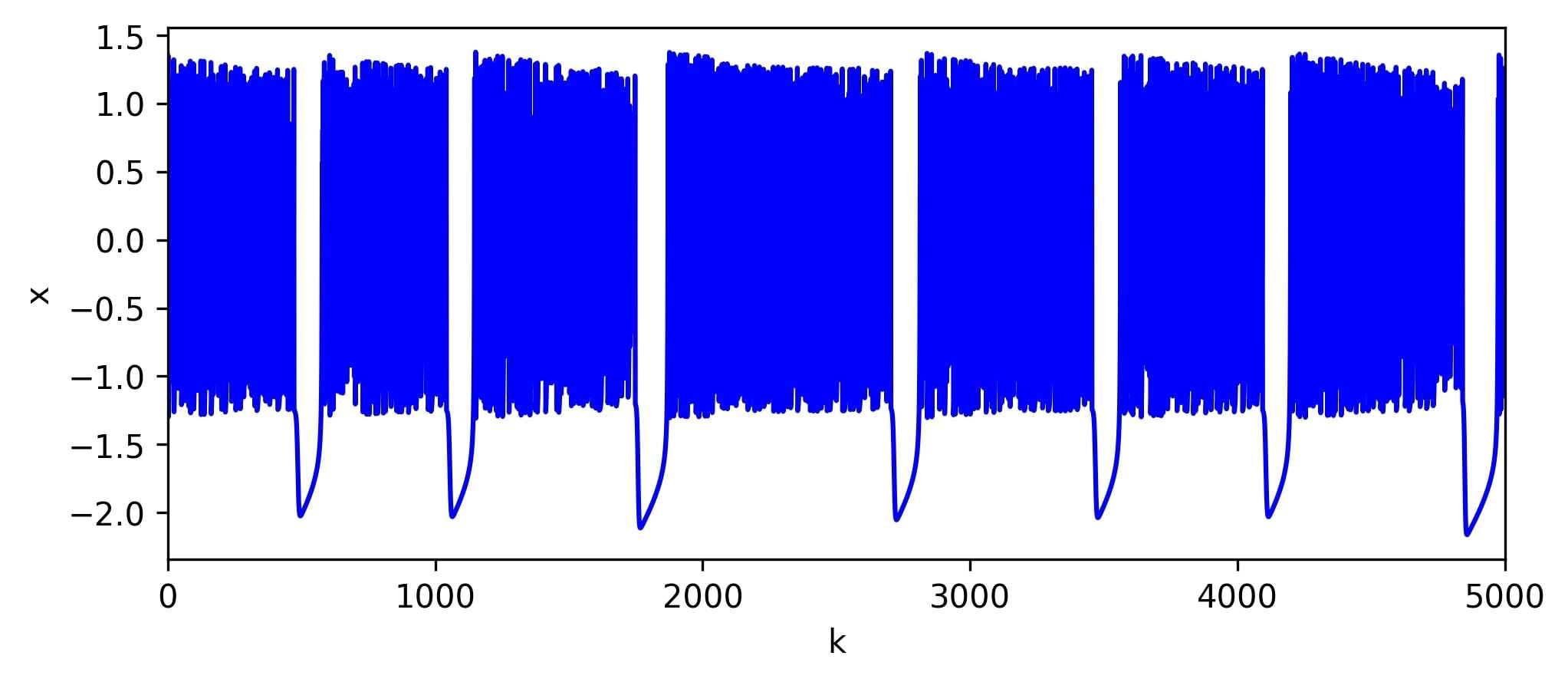}
        \caption{$\alpha=4.1$, $\sigma=-0.5$, $\lambda \approx \{0.5025,\, -0.03376\}$}
        \label{fig:rulkov_2_bursting}
        \vspace{10px}
    \end{subfigure}
    \caption{Graphs of $x_k$ for Rulkov map 2 displaying silence, spiking, and bursting behavior, visualized using the code in Appendix \ref{rulkov_2_graphs_and_cobweb_code}}
    \label{fig:rulkov_2_graphs}
\end{figure*}

In Figure \ref{fig:rulkov_2_graphs}, we graph silence, spiking, and bursting orbits of Rulkov map 2. In these graphs, we show the fast variable orbits of the combined map. As seen in Figure \ref{fig:rulkov_2_silence}, silence is achieved with parameter values of $\alpha=4.1$ and $\sigma=-2$, where the state is attracted to the fixed point of the map $\mathbf{x}_s$ after some chaotic spiking. 
Chaotic spiking behavior can be seen in the next graph, Figure \ref{fig:rulkov_2_spiking}, where we show the Rulkov 2 system with parameters $\alpha=4.5$ and $\sigma=0$. Here, its irregular spiking contrasts strongly with the calm and consistent spiking behavior seen in Rulkov map 1 (see Figure \ref{fig:rulkov_1_spiking_graphs}). Lastly, Figure \ref{fig:rulkov_2_bursting} demonstrates the bursting behavior of a Rulkov 2 neuron achieved with parameter values $\alpha=4.1$ and $\sigma=-0.5$. Here, the graph alternates between chaotic bursts of spikes and silence, contrasting with the slow oscillation between regular bursts of spikes and silence in Rulkov map 1 (see Figure \ref{fig:rulkov_x_vs_k_graph_alpha6}).

We are now interested in calculating the Lyapunov exponents of Rulkov map 2. To obtain the Lyapunov spectrum for some Rulkov 2 neuron, we first find that the Jacobian of the combined map is
\begin{equation}
    J(\mathbf{x}) = \begin{pmatrix}
        \frac{\partial f^{[1]}_2}{\partial x} & \frac{\partial f^{[1]}_2}{\partial y} \\[4px]
        \frac{\partial f^{[2]}_2}{\partial x} & \frac{\partial f^{[2]}_2}{\partial y}
    \end{pmatrix}
    =
    \begin{pmatrix}
        \frac{-2x\alpha}{(1-x^2)^2} & 1 \\
        -\eta & 1
    \end{pmatrix}
\end{equation}
which is, thankfully, not piecewise. We then implement this Jacobian into the code in Appendix \ref{rulkov_2_graphs_and_cobweb_code}, which allows us to find the Lyapunov exponents for the systems in Figure \ref{fig:rulkov_2_graphs}. The negative Lyapunov exponents $\lambda\approx\{-2.955\times 10^{-3},\,-0.4171\}$ for the system in Figure \ref{fig:rulkov_2_silence} indicate that there is no chaotic behavior, supporting the fact that the neuron is in a resting state. In Figures \ref{fig:rulkov_2_spiking} and \ref{fig:rulkov_2_bursting}, the Lyapunov spectrums are $\lambda\approx\{0.5449,\, -2.070\times 10^{-4}\}$ and $\lambda \approx \{0.5025,\, -0.03376\}$, respectively. The presence of a positive maximal Lyapunov exponent in both shows that the spiking and bursting of Rulkov map 2 are indeed chaotic.

\section{Modeling the Injection of Direct Current into Rulkov Neurons}
\label{injection-of-current}

In experiments, biologists can alter the behavior of biological neurons by injecting the cell with a direct electrical current through an electrode \cite{rulkov}. To model an injection of current from a DC voltage source, we can make a slight modification of the Rulkov mapping equation shown in Equation \ref{eq:rulkov-map}:
\begin{equation}
    \begin{pmatrix}
        x_{k+1} \\
        y_{k+1}
    \end{pmatrix}
    =
    \begin{pmatrix}
        f(x_k,\,y_k + \beta_k;\,\alpha) \\
        y_k - \eta(x_k - \sigma_k)
    \end{pmatrix}
    \label{eq:rulkov-map-injected-current}
\end{equation}
where the parameters $\beta_k$ and $\sigma_k$ model a time-varying injected current $I_k$ analogous to the constant parameter $I$ in the Hodgkin-Huxley and Izhikevich models.\footnote{As we discussed in Section \ref{rulkov-maps}, because the Rulkov maps are phenomenological models, the units and magnitude of $I_k$ don't have a direct physical interpretation.} Specifically,
\begin{align}
    \beta_k &= \beta^c I_k \label{eq:rulkov_1_beta_k} \\
    \sigma_k &= \sigma + \sigma^c I_k \label{eq:rulkov_1_sigma_k}
\end{align}
where $\beta^c$ and $\sigma^c$ are coefficients selected to achieve the desired response behavior. We can immediately see that $\sigma_k$ effectively changes the value of $\sigma$ in the slow map and $\beta_k$ effectively changes the value of $y$ in the fast map. In this section, we will discuss in more detail how pulses of current $I_k$ and different coefficients $\beta^c$ and $\sigma^c$ influence the behavior of neurons governed by Rulkov map 1. This investigation into the various responses from a simple injection of current with simple coefficient values $\sigma^c$ and $\beta^c$ will provide us with a general understanding of how the dynamics of Rulkov map 1 are influenced by an injection of current. This will be useful before we investigate the significantly more complex flow of current that occurs between coupled neurons in Section \ref{coupling-of-rulkov-neurons}.

Plugging the fast variable iteration function of Rulkov map 1 into Equation \ref{eq:rulkov-map-injected-current} and recalling the bifurcation diagram in Figure \ref{fig:rulkov-1-bifurc-diag-param-space}, we can see that if we want a given Rulkov neuron to be able to produce silence and spiking behaviors only, we should set $\alpha$ to be less than 4. Then, no matter the amount of injected current $I$, the model will not show bursts of spikes. 

To further our qualitative understanding of the behavior of Rulkov map 1 with an injection of current, we will consider a simple form of the time-varying injected current function $I_k$, namely, a current pulse. We will say that a current pulse $I_k$ with a start time of $t_{\mathrm{start}}$, a duration of $t_{\mathrm{duration}}$, and a magnitude of $a$ has the form
\begin{equation}
    I_k = \begin{cases}
        0, & 0\leq k<t_{\mathrm{start}} \\
        \pm a, & t_{\mathrm{start}}\leq k< t_{\mathrm{start}} + t_{\mathrm{duration}} \\
        0 & k \geq t_{\mathrm{start}} + t_{\mathrm{duration}}
    \end{cases}
    \label{eq:pulse-of-current}
\end{equation}
where the plus or minus indicates the direction of current in or out of the neuron. In this section, we will first explore how a low-frequency spiking Rulkov map 1 system with sigma coefficient $\sigma^c = 1$ and beta coefficient $\beta^c=0$ responds to a positive and negative pulse of current $I_k$ with magnitude 1. Then, we will examine $\beta$-dependence in Rulkov map 1 by adding on a beta coefficient $\beta^c=1$. Bringing us out of the realm of theory, the qualitative dynamics we observe in this section also observed in experiments with real biological neurons \cite{rulkov}.

\begin{figure*}[ht!]
    \centering
    \hfill
    \begin{subfigure}[t]{0.475\textwidth}
        \centering
        \includegraphics[scale=0.1125]{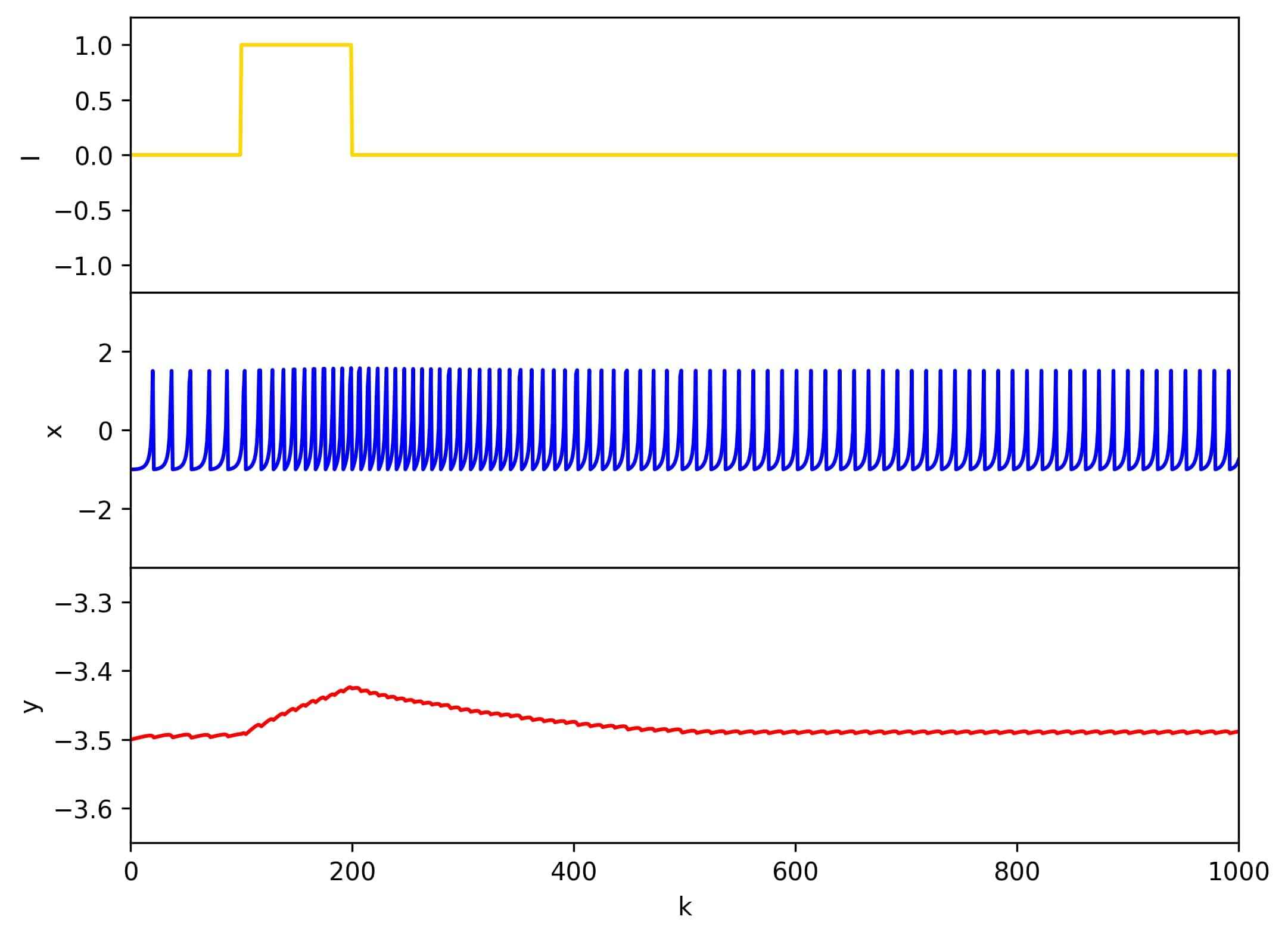}
        \caption{Positive pulse of $I_k$}
        \label{fig:inject_current_rulkov_1_I+1_sc1_bc0}
    \end{subfigure}
    \hfill
    \begin{subfigure}[t]{0.475\textwidth}
        \centering
        \includegraphics[scale=0.1125]{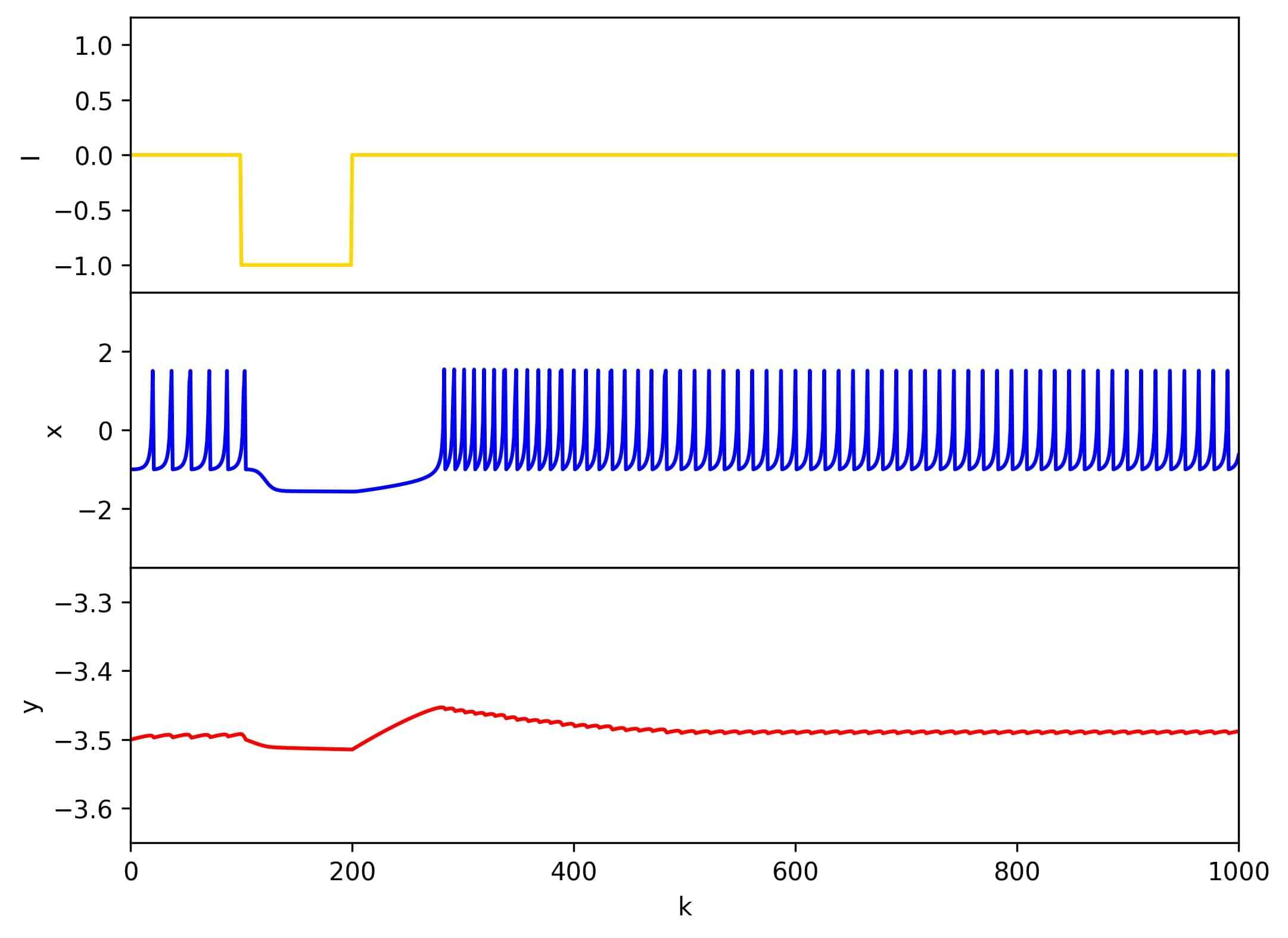}
        \caption{Negative pulse of $I_k$}
        \label{fig:inject_current_rulkov_1_I-1_sc1_bc0}
    \end{subfigure}
    \hfill
    \caption{Graphs of the response of Rulkov map 1 to pulses of current $I_k$ with magnitude 1, displaying visualizations of $I_k$, $x_k$, and $y_k$ with parameters $\alpha=5$, $\sigma=-0.6$ and coefficients $\sigma^c=1$, $\beta^c=0$, graphed with the code in Appendix \ref{orbits_rulkov_1_current_pulse_injection_code}}
    \label{fig:inject_current_rulkov_sc1_bc0_graphs}
    \vspace{4px}
\end{figure*}

\subsection{Current Pulse Response with Positive \texorpdfstring{$\sigma^c$}{TEXT}}
\label{current-pulse-response-positive-sigmac}

To explore the effects of a current pulse into a positive $\sigma^c$ Rulkov map 1 system, we will consider the system governed by the parameters $\alpha=5$ and $\sigma=-0.6$, which we can see from our maximal Lyapunov exponents diagram in parameter space (Figure \ref{fig:rulkov_1_lyapunov_exponents_parameter_space}) puts us comfortably in the blue non-chaotic spiking strip. Let us set the sigma coefficient $\sigma^c$ equal to 1 and the beta coefficient $\beta^c$ equal to 0. Then, we will subject this spiking system to a current pulse $I_k$ with magnitude 1 starting at $t=100$ and lasting for 100 time steps. We will now qualitatively analyze the responses of this system to both a positive and negative current pulse, the dynamics of which are prominently displayed in Figure \ref{fig:inject_current_rulkov_sc1_bc0_graphs}.

First, let us consider a positive pulse of current. Explicitly, from Equation \ref{eq:pulse-of-current}, our injected current is
\begin{equation}
    I_k = \begin{cases}
        0, & 0\leq k<100 \\
        1, & 100\leq k< 200 \\
        0 & k \geq 200
    \end{cases}
    \label{eq:positive_pulse_of_current}
\end{equation}
which is represented by the gold curve at the top of Figure \ref{fig:inject_current_rulkov_1_I+1_sc1_bc0}. During the current pulse, because $\sigma^c=1$, Equation \ref{eq:rulkov_1_sigma_k} tells us that $\sigma_k$ shoots up by 1, which raises the effective value of $x_{s,\,\mathrm{slow}} = \sigma_k$ (Equation \ref{eq:rulkov_1_slow_map_fixed_y}) by 1. Now that the average value of $x$ is less than $x_{s,\,\mathrm{slow}}$, we know from Section \ref{individual-dynamics-of-rulkov-map-1} that $y$ will begin to slowly increase, which we see reflected in the graph of $y_k$ at the bottom of Figure \ref{fig:inject_current_rulkov_1_I+1_sc1_bc0} for $100\leq k<200$. This increasing $y$ raises the fast map (see Figure \ref{fig:rulkov_fast_alpha6_y-3.93-example}), leading to an increasing spiking frequency. This is reflected in the graph of $x_k$ in the middle of Figure \ref{fig:inject_current_rulkov_1_I+1_sc1_bc0}, where the spikes begin to get faster and faster starting at $k=100$. Once the pulse is over, $\sigma_k$ returns to the $\sigma=-0.6$ value, so the $y$ value gradually decreases to its original value, and the spikes slow down to their original frequency.

Now, let us consider a negative pulse of current. The function $I_k$ is the exact same as Equation \ref{eq:positive_pulse_of_current} except that its value is $-1$ for $100\leq k<200$. Mirroring the positive pulse of current, the effective value of $x_{s,\,\mathrm{slow}}$ and therefore the fixed point of the map $\mathbf{x}_s$ will be pushed down in two-dimensional state space $\langle y,\, x \rangle$. In our example, the magnitude of $I_k$ is big enough that $\mathbf{x}_s$ is pushed down into the stable branch $B_{\mathrm{stable}}$, so $y$ decreases until dynamics jump from $B_{\mathrm{spikes}}$ to $B_{\mathrm{stable}}$ when $x_{s,\,\mathrm{fast,\,unstable}}$ passes up through $-1$ analogous to the spiking to silence step in a bursting orbit. Unlike a bursting orbit, however, the new $\mathbf{x}_s$ lies on the stable branch, so dynamics move along it, $x$ and $y$ both decreasing until comfortable silence is reached. Once the pulse is over, $x_{s,\,\mathrm{slow}}$ intersects the spiking branch once more, so dynamics now move up the stable branch, but in doing so, $y$ overshoots its original value, analogous to the jump up to $B_{\text{spikes}}$ towards the end of a periodic bursting orbit. Therefore, the spiking frequency is too high when dynamics jump up to $B_{\text{spikes}}$, so the spikes calm down as $y$ decreases back to its original value. All of this qualitative analysis is reflected visually in Figure \ref{fig:inject_current_rulkov_1_I-1_sc1_bc0}.

\begin{figure*}[ht!]
    \centering
    \hfill
    \begin{subfigure}[t]{0.475\textwidth}
        \centering
        \includegraphics[scale=0.1125]{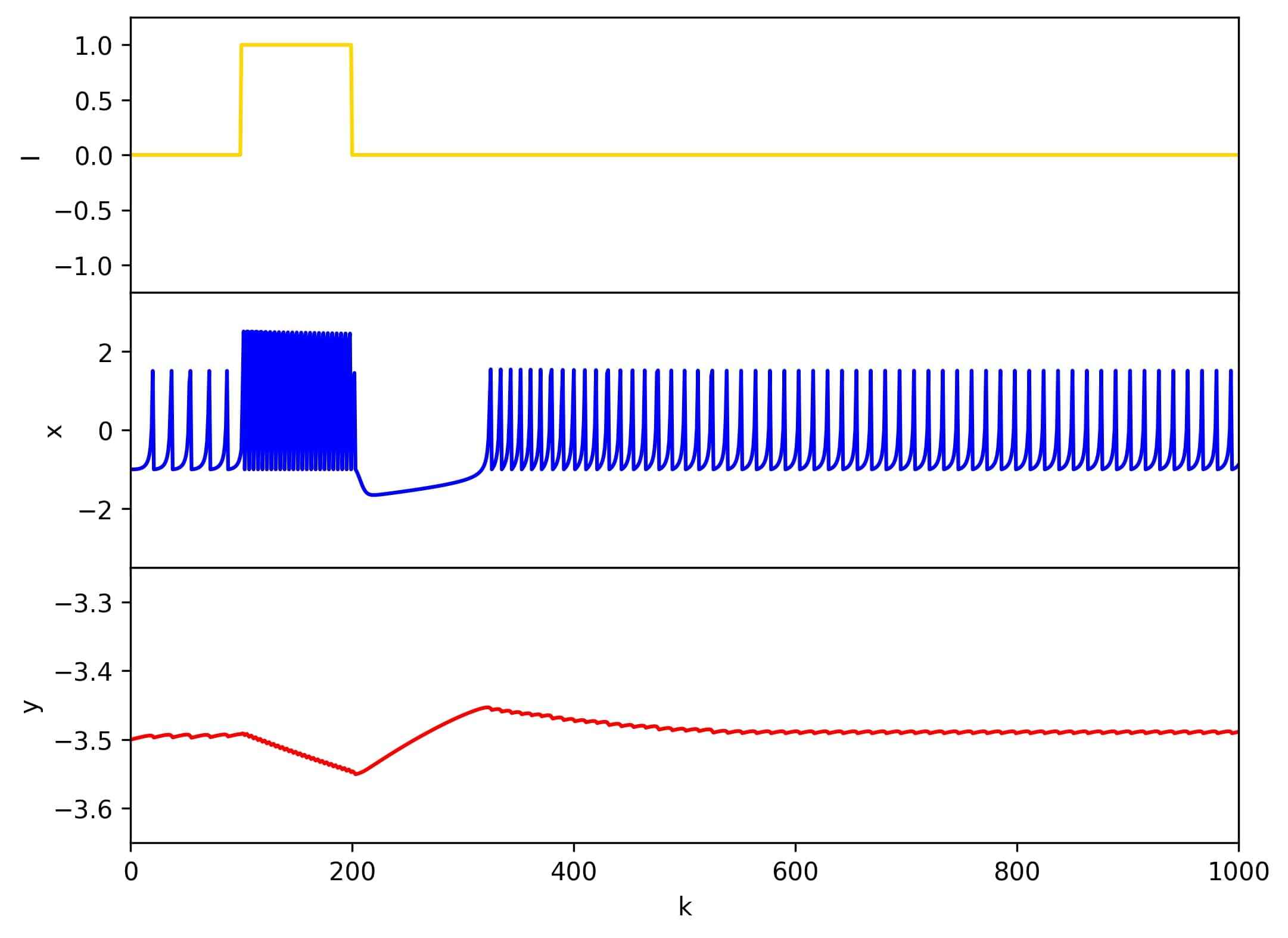}
        \caption{Positive pulse of $I_k$}
        \label{fig:inject_current_rulkov_1_I+1_sc1_bc1}
    \end{subfigure}
    \hfill
    \begin{subfigure}[t]{0.475\textwidth}
        \centering
        \includegraphics[scale=0.1125]{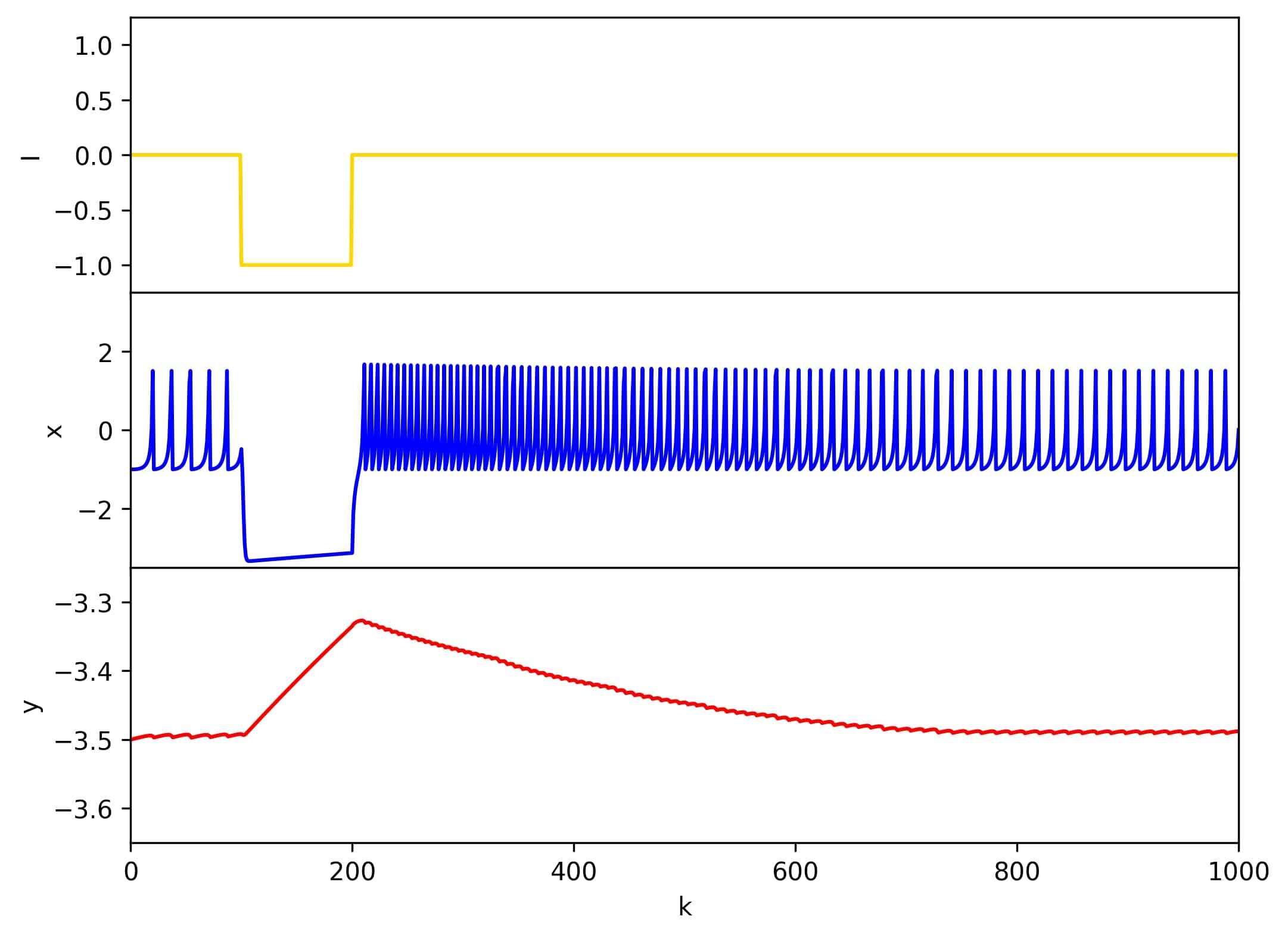}
        \caption{Negative pulse of $I_k$}
        \label{fig:inject_current_rulkov_1_I-1_sc1_bc1}
    \end{subfigure}
    \hfill
    \caption{Graphs of the response of Rulkov map 1 to pulses of current $I_k$ with magnitude 1, displaying visualizations of $I_k$, $x_k$, and $y_k$ with parameters $\alpha=5$, $\sigma=-0.6$ and coefficients $\sigma^c=1$, $\beta^c=1$, graphed with the code in Appendix \ref{orbits_rulkov_1_current_pulse_injection_code}}
    \label{fig:inject_current_rulkov_sc1_bc1_graphs}
    \vspace{4px}
\end{figure*}

\subsection{Current Pulse Response with Positive \texorpdfstring{$\sigma^c$}{TEXT} and \texorpdfstring{$\beta^c$}{TEXT}}

We will now consider what happens when the beta coefficient $\beta^c$ is not equal to zero. For this, we will use the same system as Section \ref{current-pulse-response-positive-sigmac} except with $\beta^c=1$. Immediately comparing Figures \ref{fig:inject_current_rulkov_sc1_bc0_graphs} and \ref{fig:inject_current_rulkov_sc1_bc1_graphs}, we can see that this positive beta coefficient results in significant changes in dynamics. 

Just like before, let us first consider a positive pulse of current. In this case, the effective value of $x_{s,\,\mathrm{slow}}=\sigma_k$ is similarly bumped up like before, but the effective value of $y$, which from Equation \ref{eq:rulkov-map-injected-current} is $y_k+\beta_k$, also increases by one, shooting the fast map (Figure \ref{fig:rulkov_fast_alpha6_y-3.93-example}) up and propelling the dynamics far into the spiking branch. This significantly and instantaneously increases both the frequency of the spikes and the average value of $x$, which is clearly reflected in Figure \ref{fig:inject_current_rulkov_1_I+1_sc1_bc1}. Now, the average value of $x$ is larger than the increased value of $x_{s,\,\mathrm{slow}}$, so $y$ decreases to compensate, slightly decreasing the spiking frequency and average value of $x$. When the pulse finishes, $x_{s,\,\mathrm{slow}}$ goes back to its original value, but suddenly $y$ is now too low from its adjustment during the pulse, so the orbit is immediately attracted to $x_{s,\,\mathrm{fast,\,stable}}$. Now, dynamics move along the stable branch before being propelled up into the spiking branch, where $y$ overshot its original value and gradually makes its way back down. This is qualitatively identical to the aftermath of the pulse from the second example in Section \ref{current-pulse-response-positive-sigmac} (Figure \ref{fig:inject_current_rulkov_1_I-1_sc1_bc0}), which we can see visually by comparing it with Figure \ref{fig:inject_current_rulkov_1_I+1_sc1_bc1}.

A negative current pulse mirrors the positive pulse as it did for the system with $\beta^c=0$. When the pulse hits, the fixed point of the map $\mathbf{x}_s$ shifts down to the stable branch, and the effective value of $y$ shifts down by 1, which pushes the fast map down significantly. This immediately propels the state far down along the stable branch, making $x$ is less than the new $x_{s,\,\mathrm{slow}}$. As a result, $y$ increases to compensate, and dynamics move slowly up the stable branch as a result. When the pulse is over, the value of $y$ is now much higher than its original value, meaning the average $x$ value is greater than the original value of $x_{s,\,\mathrm{slow}}$. As a result, dynamics slowly drift back along the spiking branch before reaching the stable spiking orbit. 

As we can see from our simple examples, the injection of current into Rulkov map 1 can produce varying response behaviors and interesting dynamics. For any current function $I_k$ and coefficients $\sigma^c$ and $\beta^c$,\footnote{Although we did not discuss it explicitly, it is easy to see from Equations \ref{eq:rulkov_1_beta_k} and \ref{eq:rulkov_1_sigma_k} that a negative $\sigma^c$ or a negative $\beta^c$ will effectively flip the direction of current $I_k$.} although we can computationally model the map's resulting behavior using the code in Appendix \ref{orbits_rulkov_1_current_pulse_injection_code}, we can also do a simple qualitative analysis as we did in this section.

\section{Electrical Coupling of Rulkov Neurons}
\label{coupling-of-rulkov-neurons}

Now that we have an understanding of how the dynamics of Rulkov neurons are affected by an injection of current, we are now interested in electrically coupling neurons with a flow of current between them. Specifically, let us say we have some number of coupled Rulkov neurons with states $\mathbf{x}_i$, where we use $i$ to index the neurons. The dynamics of the $i$th Rulkov neuron then have some dependence on coupling parameters $\mathfrak{C}_{i,\,x}(t)$ and $\mathfrak{C}_{i,\,y}(t)$. Specifically, mirroring Equation \ref{eq:rulkov-map-injected-current}, we will define the mapping function of the $i$th coupled neuron as
\begin{equation}
    \begin{pmatrix}
        x_{i,\,k+1} \\
        y_{i,\,k+1}
    \end{pmatrix} = 
    \begin{pmatrix}
        f(x_{i,\,k},\, y_{i,\,k} + \mathfrak{C}_{i,\,x}(k);\,\alpha_i) \\[2px]
        y_{i,\,k} - \eta x_{i,\,k} + \eta[\sigma_i + \mathfrak{C}_{i,\,y}(k)]
    \end{pmatrix}
    \label{eq:rulkov_coupled_mapping}
\end{equation}
where $\mathbf{x}_{i,\,k}$ is the state of the neuron $\mathbf{x}_i$ at the time step $k$. The coupling parameters $\mathfrak{C}_{i,\,x}(t)$ and $\mathfrak{C}_{i,\,y}(t)$ depend on factors like the structural arrangement of the system's neurons in physical space and the electrical coupling strength, which we denote as $g^e$.\footnote{We denote the coupling strength constant in this way because it is representative of the coupling conductance of a given neuron, analogous to the ion channel conductance that we discussed in Section \ref{behavior_and_modeling_of_biological_neurons}. For this reason, a higher $g^e$ indicates a higher current flow for a given voltage difference.} In this section, we will start by overviewing the simple case of electrically coupling two neurons, the structure of which is given by Rulkov \cite{rulkov}. Then, we will examine a significantly more complex system: an electrically coupled neuron ring lattice.

\subsection{Two Electrically Coupled Rulkov 1 Neurons}
\label{two-electrically-coupled-rulkov-1-neurons}

In electrically coupled neuron systems, the difference in the voltages, or fast variables, of two adjacent neurons $\mathbf{x}_i$ and $\mathbf{x}_j$ is what results in a flow of current between them. For this reason, we model the electrical coupling parameters $\mathfrak{C}_{i,\,x}(t)$ and $\mathfrak{C}_{i,\,y}(t)$ to be proportional to the difference between the voltage of a given neuron $\mathbf{x}_i$ and the voltages of its adjacent neurons $\mathbf{x}_j$. Specifically, we define the electrical coupling parameters of the neuron $\mathbf{x}_i$ to be
\begin{align}
    \mathfrak{C}_{i,\,x}(t) = \frac{\beta^c_i}{|\mathcal{N}_i|} \sum_{j\in\mathcal{N}_i} g^e_{ji}(x_{j,\,t}-x_{i,\,t}) \label{eq:electrical_coupling_parameter_general_x} \\
    \mathfrak{C}_{i,\,y}(t) = \frac{\sigma^c_i}{|\mathcal{N}_i|} \sum_{j\in\mathcal{N}_i} g^e_{ji}(x_{j,\,t}-x_{i,\,t}) \label{eq:electrical_coupling_parameter_general_y}
\end{align}
where $\mathcal{N}_i$ is the set of neurons that are adjacent to $\mathbf{x}_i$, $|\mathcal{N}_i|$ is the cardinality or number of elements of $\mathcal{N}_i$, $g^e_{ji}$ is the electrical coupling strength or coupling conductance from $\mathbf{x}_j$ to $\mathbf{x}_i$, and the notation $\sum_{j\in\mathcal{N}_i}$ indicates to sum over all the elements of $\mathcal{N}_i$. To calculate $\mathfrak{C}_{i,\,x}(t)$ and $\mathfrak{C}_{i,\,y}(t)$, Equations \ref{eq:electrical_coupling_parameter_general_x} and \ref{eq:electrical_coupling_parameter_general_y} simply average over all of the coupling-strength-scaled voltage differences between $\mathbf{x}_i$ and its adjacent neurons, then multiply that average by the neuron's beta or sigma coefficients for substitution into Equation \ref{eq:rulkov_coupled_mapping} as an injection of current.

\subsubsection{Symmetrical Coupling}

We will first consider the simplest coupling case: a symmetrical electrical coupling of two Rulkov 1 neurons $\mathbf{x}_{1}$ and $\mathbf{x}_{2}$. In this case, because we are dealing with a symmetrical coupling, we will write $g^e = g^e_{12} = g^e_{21}$. Additionally, because $\mathbf{x}_{1}$ and $\mathbf{x}_{2}$ are the only two neurons, $\mathcal{N}_1 = \{\mathbf{x}_{2}\}$ and $\mathcal{N}_2 = \{\mathbf{x}_{1}\}$. Finally, let us set $\beta^c_1=\sigma^c_1=\beta^c_2=\sigma^c_2=1$. In this extremely simple case, substituting into Equations \ref{eq:electrical_coupling_parameter_general_x} and \ref{eq:electrical_coupling_parameter_general_y} yields
\begin{align}
    \mathfrak{C}_1(t) &= \mathfrak{C}_{1,\,x}(t) = \mathfrak{C}_{1,\,y}(t) = g^e(x_{2,\,t} - x_{1,\,t}) \label{eq:rulkov_1_sym_coup_param_1} \\
    \mathfrak{C}_2(t) &=\mathfrak{C}_{2,\,x}(t) = \mathfrak{C}_{2,\,y}(t) = g^e(x_{1,\,t} - x_{2,\,t}) \label{eq:rulkov_1_sym_coup_param_2}
\end{align}
It is immediately clear that the coupling parameters of the two neurons are negatives of each other: $\mathfrak{C}_1(t) = -\mathfrak{C}_2(t)$. 

In order to model two coupled neurons, each with two variables, we need a four-dimensional system with the state vector
\begin{equation}
    \mathbf{X} = 
    \begin{pmatrix}
        X^{[1]} \\[2px]
        X^{[2]} \\[2px]
        X^{[3]} \\[2px]
        X^{[4]}
    \end{pmatrix}
    = \begin{pmatrix}
        x_{1} \\
        y_{1} \\
        x_{2} \\
        y_{2}
    \end{pmatrix}
\end{equation}
and the iteration function
\begin{equation}
    \mathbf{X}_{k+1} = \mathbf{F}(\mathbf{X}_k)
\end{equation}
or, explicitly,
\begin{equation}
    \begin{pmatrix}
        x_{1,\,k+1} \\
        y_{1,\,k+1} \\
        x_{2,\,k+1} \\
        y_{2,\,k+1}
    \end{pmatrix}
    = \begin{pmatrix}
        F^{[1]}(x_{1,\,k},\,y_{1,\,k},\,x_{2,\,k},\,y_{2,\,k}) \\[4px]
        F^{[2]}(x_{1,\,k},\,y_{1,\,k},\,x_{2,\,k},\,y_{2,\,k}) \\[4px]
        F^{[3]}(x_{1,\,k},\,y_{1,\,k},\,x_{2,\,k},\,y_{2,\,k}) \\[4px]
        F^{[4]}(x_{1,\,k},\,y_{1,\,k},\,x_{2,\,k},\,y_{2,\,k})
    \end{pmatrix}
\end{equation}
Plugging our coupling parameters from Equations \ref{eq:rulkov_1_sym_coup_param_1} and \ref{eq:rulkov_1_sym_coup_param_2} into our general iteration function for coupled Rulkov maps (Equation \ref{eq:rulkov_coupled_mapping}), we get the iteration function for our system of two symmetrically electrically coupled Rulkov 1 neurons:
\begin{equation}
    \begin{split}
        \mathbf{F}(\mathbf{X}) &= \begin{pmatrix}
            F^{[1]}(x_{1},\,y_{1},\,x_{2},\,y_{2}) \\[4px]
            F^{[2]}(x_{1},\,y_{1},\,x_{2},\,y_{2}) \\[4px]
            F^{[3]}(x_{1},\,y_{1},\,x_{2},\,y_{2}) \\[4px]
            F^{[4]}(x_{1},\,y_{1},\,x_{2},\,y_{2})
        \end{pmatrix} \\
        &= \begin{pmatrix}
            f_1(x_{1},\, y_{1}+g^e(x_{2} - x_{1});\, \alpha_1) \\[2px]
            y_{1} - \eta x_{1} + \eta[\sigma_1 + g^e(x_{2} - x_{1})] \\[2px]
            f_1(x_{2},\, y_{2}+g^e(x_{1} - x_{2});\, \alpha_2) \\[2px]
            y_{2} - \eta x_{2} + \eta[\sigma_2 + g^e(x_{1} - x_{2})]
        \end{pmatrix}
    \end{split}
    \label{eq:rulkov_1_sym_coup_iter_func}
\end{equation}
Unfortunately, because the state space of this system is four-dimensional, we cannot visualize it anymore.\footnote{That is, unless the reader is a higher-dimensional creature.} However, we established in Sections \ref{background} and \ref{geometry_dynamics_chaos} that the techniques that we use to analyze lower-dimensional systems work for systems of any number of dimensions, so we can still apply our methods to this system. For example, we can use Lyapunov exponents to quantify the system's chaotic behavior. The Jacobian of this system $J(\mathbf{X})$ is a $4\times 4$ matrix:
\begin{equation}
    J(\mathbf{X}) = \begin{pmatrix}
        \frac{\partial F^{[1]}}{\partial x_{1}} & \frac{\partial F^{[1]}}{\partial y_{1}} & \frac{\partial F^{[1]}}{\partial x_{2}} & \frac{\partial F^{[1]}}{\partial y_{2}} \\[6px]
        \frac{\partial F^{[2]}}{\partial x_{1}} & \frac{\partial F^{[2]}}{\partial y_{1}} & \frac{\partial F^{[2]}}{\partial x_{2}} & \frac{\partial F^{[2]}}{\partial y_{2}} \\[6px]
        \frac{\partial F^{[3]}}{\partial x_{1}} & \frac{\partial F^{[3]}}{\partial y_{1}} & \frac{\partial F^{[3]}}{\partial x_{2}} & \frac{\partial F^{[3]}}{\partial y_{2}} \\[6px]
        \frac{\partial F^{[4]}}{\partial x_{1}} & \frac{\partial F^{[4]}}{\partial y_{1}} & \frac{\partial F^{[4]}}{\partial x_{2}} & \frac{\partial F^{[4]}}{\partial y_{2}} 
    \end{pmatrix} \\
\end{equation}
Evaluating the partial derivatives of the functions in Equation \ref{eq:rulkov_1_sym_coup_iter_func} yields an impressively complex-looking result due to the piecewise nature of the two functions $f_1$ involved in $\mathbf{F}(\mathbf{X})$:
\onecolumn
\begin{equation}   
    \def\stackalignment{l}
    J(\mathbf{X}) = \begin{cases}
        \begin{cases}
            \begin{pmatrix}
                \frac{\alpha}{(1-x_{1})^2} - g^e & 1 & g^e & 0 \\
                -\eta(1+g^e) & 1 & \eta g^e & 0 \\
                g^e & 0 & \frac{\alpha_2}{(1-x_{2})^2} - g^e & 1 \\
                \eta g^e & 0 & -\eta(1+g^e) & 1
            \end{pmatrix}, & \text{if $x_{1}\leq 0$}, \\[1cm]
            \begin{pmatrix}
                -g^e & 1 & g^e & 0 \\
                -\eta(1+g^e) & 1 & \eta g^e & 0 \\
                g^e & 0 & \frac{\alpha_2}{(1-x_{2})^2} - g^e & 1 \\
                \eta g^e & 0 & -\eta(1+g^e) & 1
            \end{pmatrix}, & 
            \stackunder{\text{if $0<x_{1}<\alpha_1$}}{\text{\phantom{if} $+y_{1}+\mathfrak{C}_1,$}} \\[1cm]
            \begin{pmatrix}
                0 & 0 & 0 & 0 \\
                -\eta(1+g^e) & 1 & \eta g^e & 0 \\
                g^e & 0 & \frac{\alpha_2}{(1-x_{2})^2} - g^e & 1 \\
                \eta g^e & 0 & -\eta(1+g^e) & 1
            \end{pmatrix}, & 
            \stackunder{\text{if $x_{1}\geq\alpha_1$}}{\text{\phantom{if} $+y_{1}+\mathfrak{C}_1,$}}
        \end{cases}
        & \hspace{0.5cm}\text{and $x_{2}\leq 0$} \vspace{0.5cm} \\
        \begin{cases}
            \begin{pmatrix}
                \frac{\alpha}{(1-x_{1})^2} - g^e & 1 & g^e & 0 \\
                -\eta(1+g^e) & 1 & \eta g^e & 0 \\
                g^e & 0 & -g^e & 1 \\
                \eta g^e & 0 & -\eta(1+g^e) & 1
            \end{pmatrix}, & 
            \hspace{0.25cm}\text{if $x_{1}\leq 0$}, \\[1cm]
            \begin{pmatrix}
                -g^e & 1 & g^e & 0 \\
                -\eta(1+g^e) & 1 & \eta g^e & 0 \\
                g^e & 0 & -g^e & 1 \\
                \eta g^e & 0 & -\eta(1+g^e) & 1
            \end{pmatrix}, & 
            \hspace{0.25cm}\stackunder{\text{if $0<x_{1}<\alpha_1$}}{\text{\phantom{if} $+y_{1}+\mathfrak{C}_1,$}} \\[1cm]
            \begin{pmatrix}
                0 & 0 & 0 & 0 \\
                -\eta(1+g^e) & 1 & \eta g^e & 0 \\
                g^e & 0 & -g^e & 1 \\
                \eta g^e & 0 & -\eta(1+g^e) & 1
            \end{pmatrix}, & 
            \hspace{0.25cm}\stackunder{\text{if $x_{1}\geq\alpha_1$}}{\text{\phantom{if} $+y_{1}+\mathfrak{C}_1,$}} \\[1cm]
        \end{cases} & \hspace{0.5cm}\stackunder{\text{and $0<x_{2}<\alpha_2$}}{\text{\phantom{and} $+y_{1}+\mathfrak{C}_2$}} \vspace{0.5cm} \\
        \begin{cases}
            \begin{pmatrix}
                \frac{\alpha}{(1-x_{1})^2} - g^e & 1 & g^e & 0 \\
                -\eta(1+g^e) & 1 & \eta g^e & 0 \\
                0 & 0 & 0 & 0 \\
                \eta g^e & 0 & -\eta(1+g^e) & 1
            \end{pmatrix}, & 
            \hspace{0.25cm}\text{if $x_{1}\leq 0$}, \\[1cm]
            \begin{pmatrix}
                -g^e & 1 & g^e & 0 \\
                -\eta(1+g^e) & 1 & \eta g^e & 0 \\
                0 & 0 & 0 & 0 \\
                \eta g^e & 0 & -\eta(1+g^e) & 1
            \end{pmatrix}, & 
            \hspace{0.25cm}\stackunder{\text{if $0<x_{1}<\alpha_1$}}{\text{\phantom{if} $+y_{1}+\mathfrak{C}_1,$}} \\[1cm]
            \begin{pmatrix}
                0 & 0 & 0 & 0 \\
                -\eta(1+g^e) & 1 & \eta g^e & 0 \\
                0 & 0 & 0 & 0 \\
                \eta g^e & 0 & -\eta(1+g^e) & 1
            \end{pmatrix}, & 
            \hspace{0.25cm}\stackunder{\text{if $x_{1}\geq\alpha_1$}}{\text{\phantom{if} $+y_{1}+\mathfrak{C}_1,$}}
        \end{cases} & \hspace{0.5cm}\stackunder{\text{and $x_{2}\geq\alpha_2$}}{\text{\phantom{and} $+y_{1}+\mathfrak{C}_2,$}}
    \end{cases}
    \label{eq:rulkov_1_extremely_big_jacobian_matrix}
\end{equation}
\twocolumn
\noindent This is obviously extremely unwieldy to work with, so in Appendix \ref{partioning-jacobian-matrix-for-two-coup-rulkov-1-neurons}, we make some substitutions to simplify the entirety of this Jacobian matrix into
\begin{equation}
    J(\mathbf{X}) = \begin{pmatrix}
        J_{\text{dg},\,a}(x_{1},\,\alpha_1,\,g^e) & J_{\text{odg},\,b}(g^e) \\
        J_{\text{odg},\,c}(g^e) & J_{\text{dg},\,d}(x_{2},\,\alpha_2,\,g^e)
    \end{pmatrix}
\end{equation}
which we recognize is extremely economical compared to its original expanded form. An additional benefit to this simplification is that we can use this same Jacobian structure for different systems by adjusting the inputs into the Jacobian submatrices. We implement this Jacobian partitioning method of calculating the Lyapunov spectrum into the code in Appendix \ref{sym-elec-coup-rulkov-1-neurons-code}.

\begin{figure}[ht!]
    \centering
    \begin{subfigure}[t]{0.475\textwidth}
        \centering
        \includegraphics[scale=0.09]{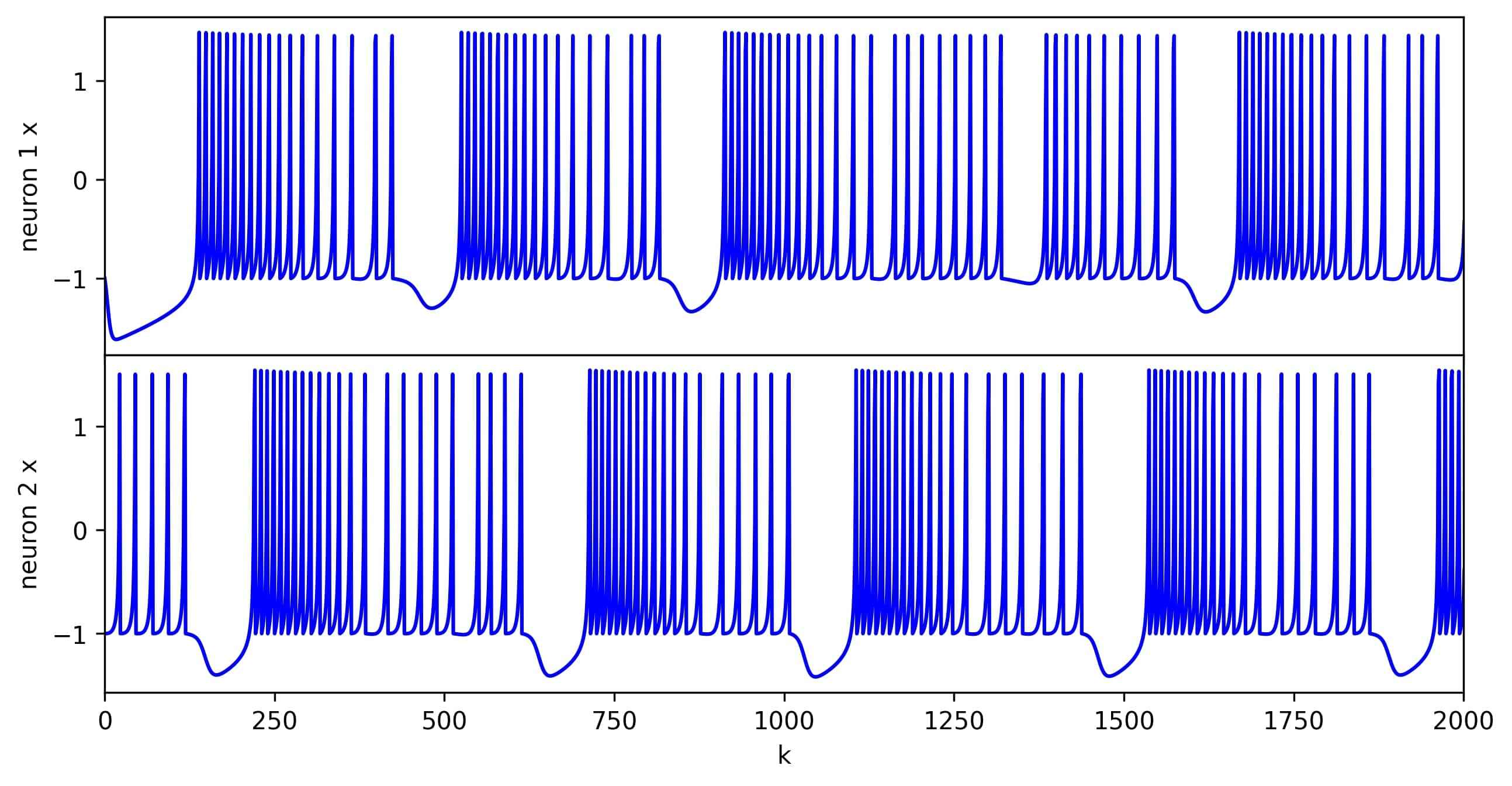}
        \caption{$g^e=0$, $\lambda_1\approx 0.0374$}
        \label{fig:sym_coup_rulkov_1_ge0}
        \vspace{8px}
    \end{subfigure}
    \begin{subfigure}[t]{0.475\textwidth}
        \centering
        \includegraphics[scale=0.09]{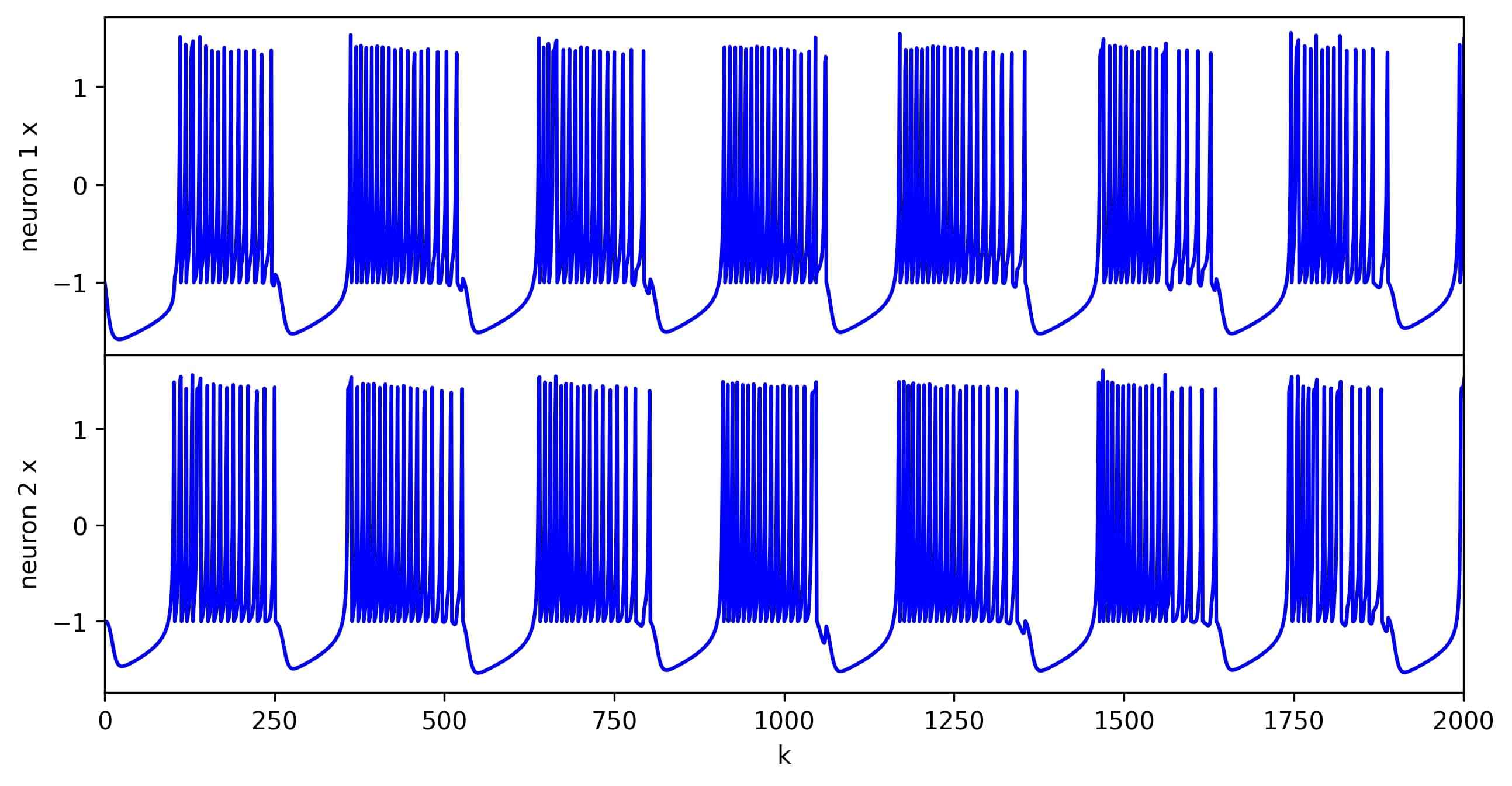}
        \caption{$g^e=0.05$, $\lambda_1\approx 0.0339$}
        \label{fig:sym_coup_rulkov_1_ge0.05}
        \vspace{8px}
    \end{subfigure}
    \begin{subfigure}[t]{0.475\textwidth}
        \centering
        \includegraphics[scale=0.09]{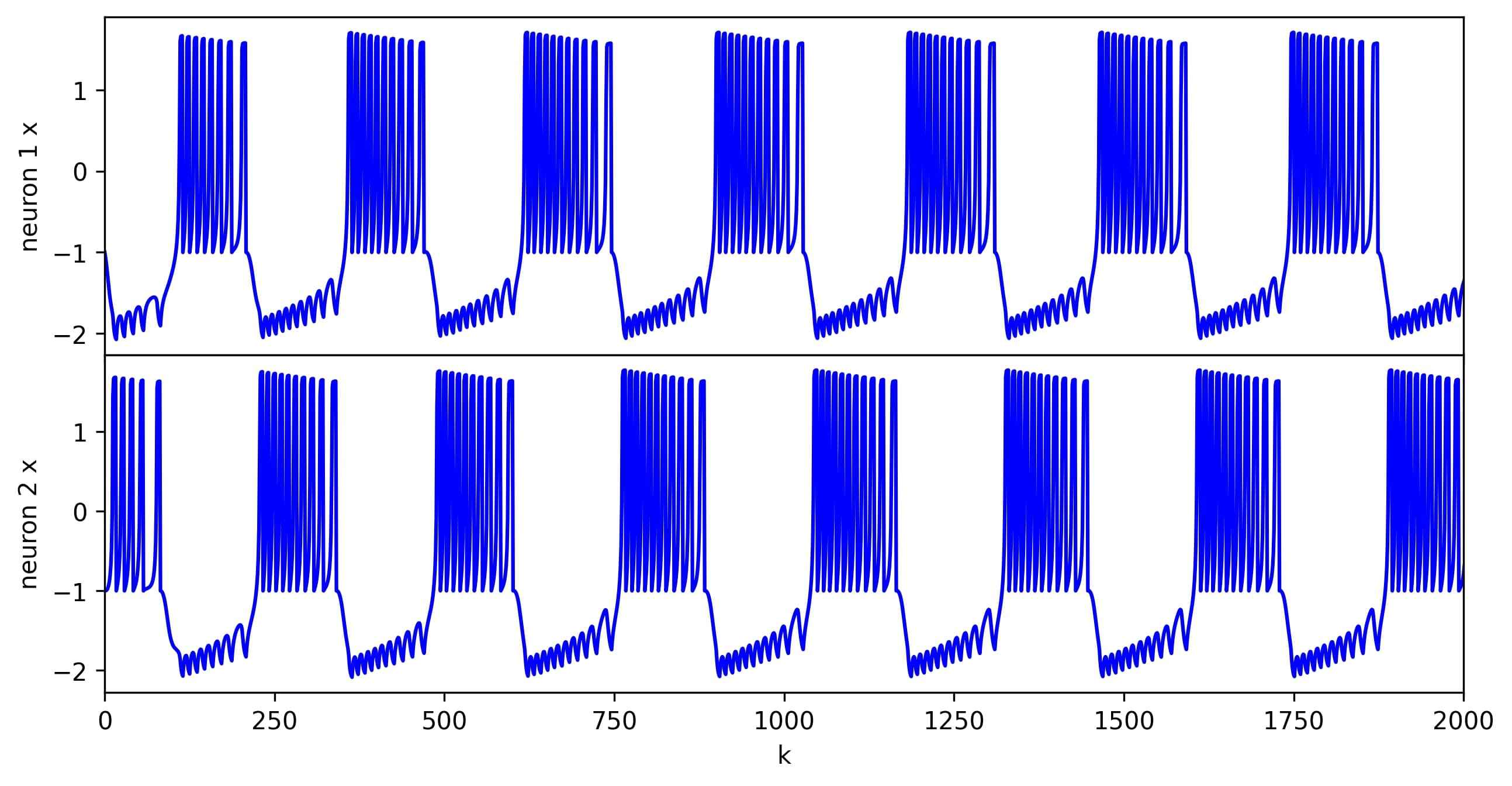}
        \caption{$g^e=-0.05$, $\lambda_1\approx-0.0080$}
        \label{fig:sym_coup_rulkov_1_ge-0.05}
        \vspace{8px}
    \end{subfigure}
    \caption{Graphs of $x_{1,\,k}$ and $x_{2,\,k}$ for two symmetrically electrically coupled Rulkov 1 neurons with initial states $\mathbf{x}_{1,\,0}=\mathbf{x}_{2,\,0}=\langle -1,\, -3.5 \rangle$ and parameters $\sigma_1=-0.75$, $\sigma_2=-0.76$, $\alpha_1=4.9$, $\alpha_2=5.0$, graphed and maximal Lyapunov exponents calculated using the code in Appendix \ref{sym-elec-coup-rulkov-1-neurons-code}}
    \label{fig:sym_coup_rulkov_1_graphs}
\end{figure}

We will now consider an example of this kind of system: two symmetrically electrically coupled neurons operating under the parameters $\sigma_1=-0.75$, $\sigma_2=-0.76$, $\alpha_1=4.9$, $\alpha_2=5.0$ and identical initial conditions $\mathbf{x}_{1,\,0}=\mathbf{x}_{2,\,0}=\langle -1,\, -3.5 \rangle$. First, let us consider the case with no coupling $g^e=0$. Matching the parameters of the uncoupled maps to the maximal Lyapunov exponent visualization in Figure \ref{fig:rulkov_1_lyapunov_exponents_visualizations}, we see that these parameters place us in the chaotic bursting range. In Figure \ref{fig:sym_coup_rulkov_1_ge0}, we graph the fast variable orbits $x_{1,\,k}$ and $x_{2,\,k}$ of both uncoupled neurons, where we can see the unsynchronized chaotic bursting occur. The maximal Lyapunov exponent of the four-dimensional uncoupled system is $\lambda_1\approx 0.0374$, confirming it is still chaotic. 

Now, if we increase the strength of the electric coupling, these chaotic bursts will synchronize with each other. This is because a difference in voltage between the two neurons will cause current to flow between them and equalize their voltages over time. We can see this with the electrical coupling strength $g^e=0.05$ in Figure \ref{fig:sym_coup_rulkov_1_ge0.05}, where the behavior of the neurons remains chaotic with a positive maximal Lyapunov exponent $\lambda_1\approx 0.0339$. Although the spiking and silence periods of the two neurons happen at the same time, the spikes within the bursts are still chaotic and unsynchronized.

Finally, we will consider a negative electrical coupling strength. In Figure \ref{fig:sym_coup_rulkov_1_ge-0.05}, we graph our system with $g^e=-0.05$, which results in the neurons bursting in antiphase with each other. This result makes sense from our understanding of how injection of current works, but an interesting observation is that these antiphase bursts are not chaotic, having a negative maximal Lyapunov exponent $\lambda_1\approx -0.0080$. This can be justified by considering the effects of the coupling parameters $\mathfrak{C}_i(t)$ on the graph of the fast map (Figure \ref{fig:rulkov_fast_alpha6_y-3.93-example}). Specifically, we know from Section \ref{injection-of-current} that the $i$th neuron's coupling parameter $\mathfrak{C}_i(t)$ is positive when the neuron is in a burst and negative when the neuron is between the bursts. Therefore, because $\mathfrak{C}_i(t)$ is analogous to the DC parameters $\beta_k$ and $\sigma_k$, the negative coupling forces each neuron's dynamics to stay high on the spiking branch $B_{\mathrm{spikes}}$ during a burst and low on the stable branch $B_{\mathrm{stable}}$ between bursts. Since these regions are outside of the densely folded region of $B_{\text{spikes}}$ (Figure \ref{fig:rulkov_1_state_space_diagram_alpha6}), dynamics remain non-chaotic.

Similar behavior occurs for neurons in the regime of non-chaotic spiking: a positive $g^e$ will cause the spikes to synchronize with each other. According to Rulkov \cite{rulkov}, the discrete-time nature of Rulkov map 1 can cause the periodic spikes to lock into different ratios, which results in multistability. Rather than examining this, however, we will explore a novel case of multistability in a system of two identical Rulkov 1 neurons with an asymmetrical electrical coupling.

\subsubsection{Asymmetrical Coupling}

To calculate the coupling parameters for an asymmetrical electrical coupling of two Rulkov 1 neurons, we plug $\beta^c_1=\sigma^c_1=\beta^c_2=\sigma^c_2=1$, $\mathcal{N}_1=\{\mathbf{x}_{2}\}$, $\mathcal{N}_2=\{\mathbf{x}_{1}\}$, $g^e_1=g^e_{21}$, and $g^e_2=g^e_{12}$ into Equations \ref{eq:electrical_coupling_parameter_general_x} and \ref{eq:electrical_coupling_parameter_general_y}:
\begin{align}
    \mathfrak{C}_1(t) &= g^e_1(x_{2,\,t}-x_{1,\,t}) \\
    \mathfrak{C}_2(t) &= g^e_2(x_{1,\,t}-x_{2,\,t})
\end{align}
In this asymmetrical case, the two electrical coupling strength constants are not equal: $g^e_1\neq g^e_2$. Therefore, the coupling parameters are not negatives of each other in this case. Altering Equation \ref{eq:rulkov_1_sym_coup_iter_func} for these coupling parameters, we get
\begin{equation}
    \mathbf{F}(\mathbf{X}) = \begin{pmatrix}
        f_1(x_{1},\, y_{1}+g^e_1(x_{2} - x_{1});\, \alpha_1) \\[2px]
        y_{1} - \eta x_{1} + \eta[\sigma_1 + g^e_1(x_{2} - x_{1})] \\[2px]
        f_1(x_{2},\, y_{2}+g^e_2(x_{1} - x_{2});\, \alpha_2) \\[2px]
        y_{2} - \eta x_{2} + \eta[\sigma_2 + g^e_2(x_{1} - x_{2})]
    \end{pmatrix}
\end{equation}
Using the same conditions for assigning values to the variables $a$, $b$, $c$, and $d$ from Equations \ref{eq:jacobian-abcd-1} and \ref{eq:jacobian-abcd-2}, we can write the Jacobian of our asymmetrically electrically coupled Rulkov 1 neuron system as 
\begin{equation}
    J(\mathbf{X}) = \begin{pmatrix}
        J_{\text{dg},\,a}(x_{1},\,\alpha_1,\,g^e_1) & J_{\text{odg},\,b}(g^e_1) \\
        J_{\text{odg},\,c}(g^e_2) & J_{\text{dg},\,d}(x_{2},\,\alpha_2,\,g^e_2)
    \end{pmatrix}
\end{equation}

\begin{figure}[ht!]
    \centering
    \begin{subfigure}[t]{0.475\textwidth}
        \centering
        \includegraphics[scale=0.09]{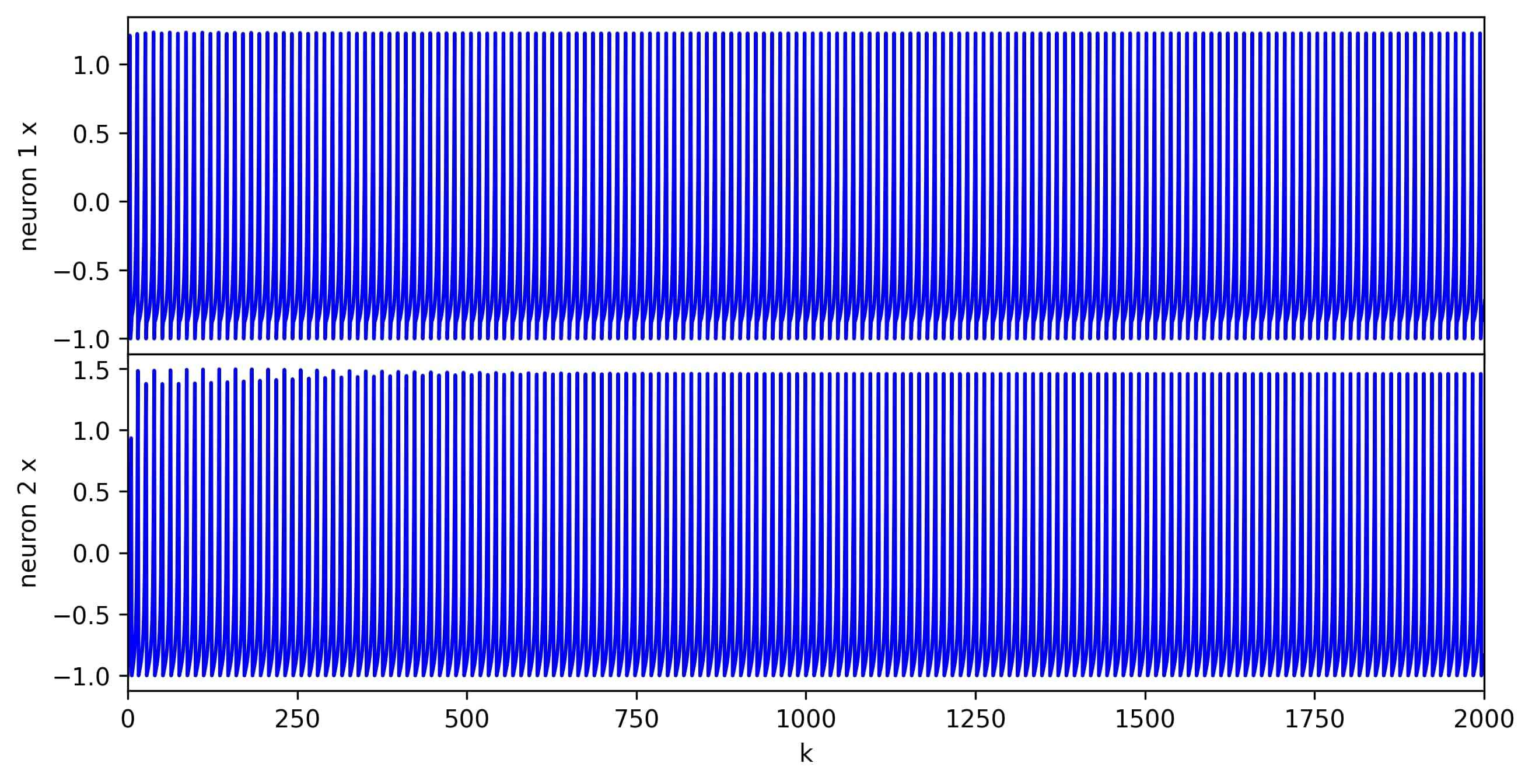}
        \caption{$\mathbf{x}_{1,\,0} = \langle -0.54,\, -3.25 \rangle$, $\mathbf{x}_{2,\,0} = \langle -1,\, -3.25 \rangle$, $\lambda_1\approx -0.0057$}
        \label{fig:asym_coup_rulkov_1_spiking}
        \vspace{8px}
    \end{subfigure}
    \begin{subfigure}[t]{0.475\textwidth}
        \centering
        \includegraphics[scale=0.09]{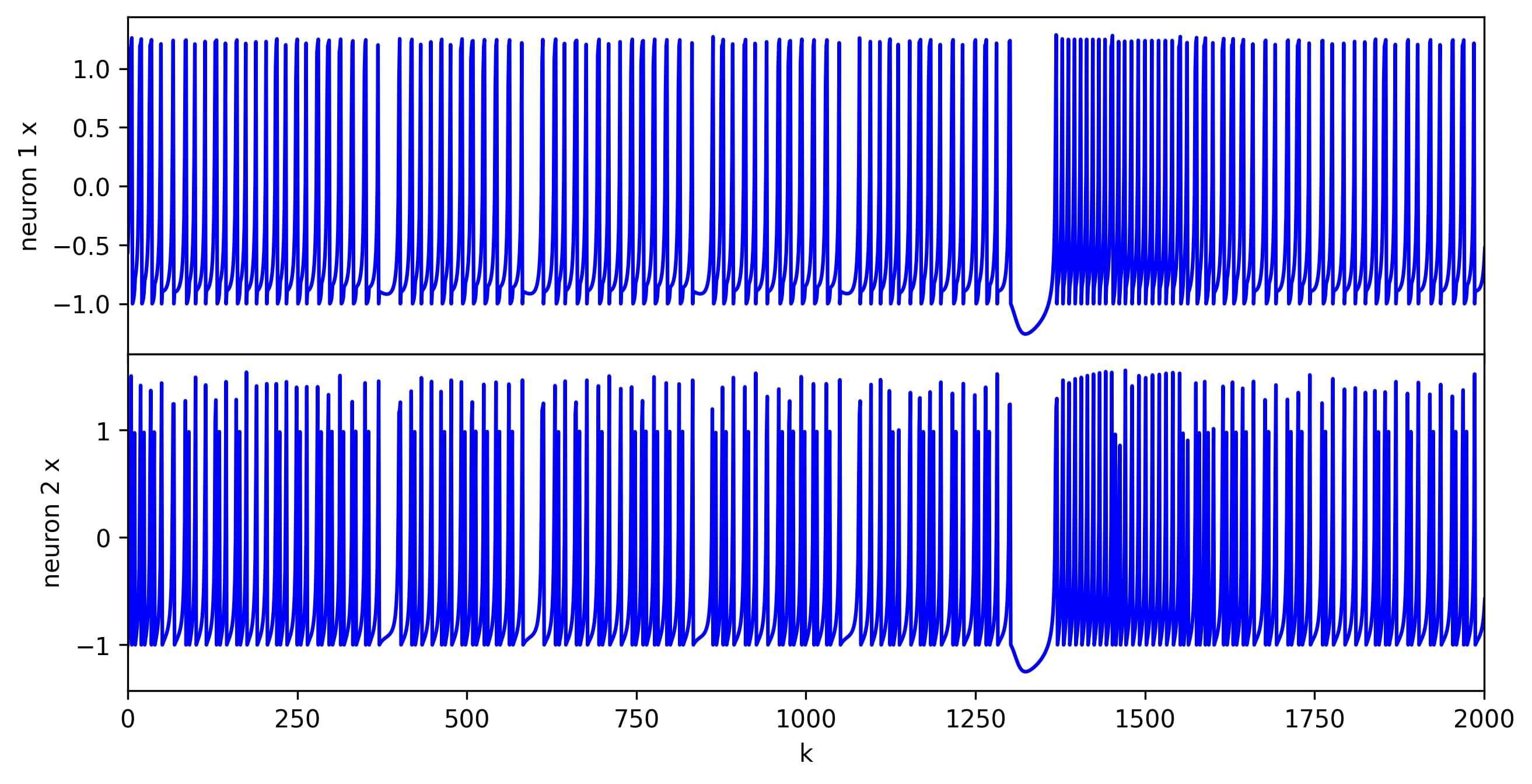}
        \caption{$\mathbf{x}_{1,\,0} = \langle -0.56,\, -3.25 \rangle$, $\mathbf{x}_{2,\,0} = \langle -1,\, -3.25 \rangle$, $\lambda_1\approx 0.0144$}
        \label{fig:asym_coup_rulkov_1_chaos}
        \vspace{8px}
    \end{subfigure}
    \caption{Multistability displayed in graphs of $x_{1,\,k}$ and $x_{2,\,k}$ for two asymmetrically electrically coupled Rulkov 1 neurons with parameters $\sigma_1=\sigma_2=-0.5$, $\alpha_1=\alpha_2=4.5$ and coupling strengths $g^e_1=0.05$, $g^e_2=0.25$, graphed and maximal Lyapunov exponents calculated using the code in Appendix \ref{asym-elec-coup-rulkov-1-neurons-code}}
    \label{fig:asym_coup_rulkov_1_graphs}
\end{figure}

The parallels between this system and our symmetrically coupled system are evident, so in Appendix \ref{asym-elec-coup-rulkov-1-neurons-code}, we use the same functions from Appendix \ref{sym-elec-coup-rulkov-1-neurons-code} to plot fast variable orbits and calculate the Lyapunov spectrum of two asymmetrically coupled Rulkov 1 neurons with identical parameters $\sigma_1=\sigma_2=-0.5$ and $\alpha_1=\alpha_2=4.5$. The coupling strengths we use are $g^e_1=0.05$ and $g^e_2=0.25$, which results in neuron $\mathbf{x}_2$ ``feeling'' the difference in the voltages of the two neurons more. In Figure \ref{fig:asym_coup_rulkov_1_graphs}, we plot the fast variable orbits of this system for two slightly different initial conditions. In the first graph (Figure \ref{fig:asym_coup_rulkov_1_spiking}), we can see non-chaotic, synchronized spiking with a negative maximal Lyapunov exponent $\lambda_1\approx-0.0057$. However, in the second graph (Figure \ref{fig:asym_coup_rulkov_1_chaos}), after changing the initial $\mathbf{x}_1$ voltage value slightly, chaotic spiking-bursting occurs in both neurons ($\lambda_1\approx0.0025$). This is a clear example of multistability and dependence on initial conditions, so we will explore the geometries of this system's state space in Section \ref{asym-elec-coup-two-rulkov-1-neurons-geometry}.

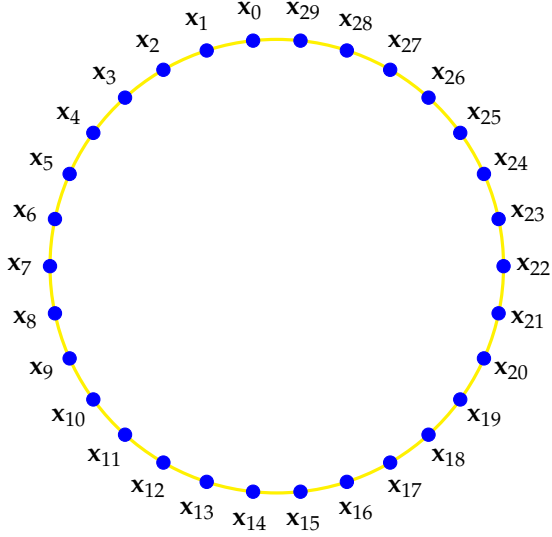
\begin{figure}
    \centering
    \begin{tikzpicture}[scale=1]
        \draw[yellow, very thick] (0, 0) circle [radius=3cm];
        \foreach \a in {0,1,...,29}{
            \filldraw[blue] (\a*360/30: 3cm) circle [radius=2.5pt];
            \draw (\a*360/30+96: 3.4cm) node{$\mathbf{x}_{\a}$};
        }
    \end{tikzpicture}
    \caption{Visualization of a ring of $\zeta=30$ Rulkov 1 neurons}
    \label{fig:ring-30-rulkov-1-neurons}
\end{figure}

\subsection{Neuron Ring Lattice}
\label{neuron-ring-lattice}

A lattice of neurons is a specific arrangement of neurons in physical space with defined physical connections between them. The simplicity of the Rulkov maps allows us to examine the dynamics of complex neuron lattice systems in a way that a more complex model, such as the Hodgkin-Huxley model, couldn't. In this section, we will examine a neuron lattice structure inspired by statistical mechanics, namely, a ring of $\zeta$ electrically coupled Rulkov 1 neurons $\mathbf{x}_0,\,\mathbf{x}_1,\,\hdots\,,\,\mathbf{x}_{\zeta-1}$, each having a flow of current with its neighbor. This lattice structure is visualized in Figure \ref{fig:ring-30-rulkov-1-neurons} for $\zeta=30$, where neurons are represented by blue points and the electric coupling connections are shown in gold. 

We would now like to calculate the coupling parameters for each of these neurons. Again, we will let $\beta^c_i=\sigma^c_i=1$ for simplicity. We will also assume that all couplings are symmetric: $g^e=g^e_{ji}$ for all $i\neq j$. Because of the circular nature of this lattice system, we can write $\mathcal{N}_i$, the set of neurons that are adjacent to the neuron $\mathbf{x}_i$, in a compact form using modular arithmetic:
\begin{equation}
    \mathcal{N}_i = \{\mathbf{x}_{(i-1)\bmod\zeta},\,\mathbf{x}_{(i+1)\bmod\zeta}\}
\end{equation}
This accounts for the fact that $\mathcal{N}_0 = \{\mathbf{x}_{\zeta-1},\,\mathbf{x}_1\}$ and $\mathcal{N}_{\zeta-1} = \{\mathbf{x}_{\zeta-2},\,\mathbf{x}_0\}$. Plugging all this into Equations \ref{eq:electrical_coupling_parameter_general_x} and \ref{eq:electrical_coupling_parameter_general_y}, we can write the coupling parameters of this ring system as
\begin{equation}
    \begin{split}
        \mathfrak{C}_{i} &= \mathfrak{C}_{i,\,x} = \mathfrak{C}_{i,\,y} \\
        &= \frac{g^e}{2}[(x_{(i-1)\bmod\zeta}-x_i)+(x_{(i+1)\bmod\zeta}-x_i)] \\
        &= \frac{g^e}{2}[x_{(i-1)\bmod\zeta}+x_{(i+1)\bmod\zeta}-2x_i]
    \end{split}
    \label{eq:ring-coup-params}
\end{equation}
Considering all $\zeta$ neurons together, the state vector of this entire ring system can be written as
\begin{equation}
    \mathbf{X} = \begin{pmatrix}
        X\e{1} \\
        X\e{2} \\
        X\e{3} \\
        X\e{4} \\
        \vdots \\[3px]
        X\e{2\zeta-1} \\
        X\e{2\zeta}
    \end{pmatrix} = 
    \begin{pmatrix}
        x_0 \\
        y_0 \\
        x_1 \\
        y_1 \\
        \vdots \\
        x_{\zeta-1} \\
        y_{\zeta-1}
    \end{pmatrix}
    \label{eq:ring-system-state-vector}
\end{equation}
The state space of this ring lattice system is $2\zeta$-dimensional since we have one slow variable and one fast variable for each of the $\zeta$ neurons in the ring. Plugging the coupling parameters from Equation \ref{eq:ring-coup-params} into the general iteration function for coupled Rulkov maps (Equation \ref{eq:rulkov_coupled_mapping}) for each neuron in the ring yields the impressive $2\zeta$-dimensional iteration function for our ring system:
\begin{equation}
    \begin{gathered}
        \mathbf{F}(\mathbf{X})
        = \begin{pmatrix}
            F\e{1}(x_0,\,y_0,\,x_1,\,y_1,\,\hdots\,,\,x_{\zeta-1},\,y_{\zeta-1}) \\[4px]
            F\e{2}(x_0,\,y_0,\,x_1,\,y_1,\,\hdots\,,\,x_{\zeta-1},\,y_{\zeta-1}) \\[4px]
            F\e{3}(x_0,\,y_0,\,x_1,\,y_1,\,\hdots\,,\,x_{\zeta-1},\,y_{\zeta-1}) \\[4px]
            F\e{4}(x_0,\,y_0,\,x_1,\,y_1,\,\hdots\,,\,x_{\zeta-1},\,y_{\zeta-1}) \\[4px]
            \vdots \\[4px]
            F\e{2\zeta-1}(x_0,\,y_0,\,x_1,\,y_1,\,\hdots\,,\,x_{\zeta-1},\,y_{\zeta-1}) \\[4px]
            F\e{2\zeta}(x_0,\,y_0,\,x_1,\,y_1,\,\hdots\,,\,x_{\zeta-1},\,y_{\zeta-1})
        \end{pmatrix} \\[0.25cm]
        = \begin{pmatrix}
            f_1\Big(x_0,\,y_0+\frac{g^e}{2}(x_{\zeta-1}+x_1-2x_0);\,\alpha_0\Big) \\[4px]
            y_0 - \eta x_0 + \eta\Big[\sigma_0 + \frac{g^e}{2}(x_{\zeta-1}+x_1-2x_0)\Big] \\[4px]
            f_1\Big(x_1,\,y_1+\frac{g^e}{2}(x_0+x_2-2x_1);\,\alpha_1\Big) \\[4px]
            y_1 - \eta x_1 + \eta\Big[\sigma_1 + \frac{g^e}{2}(x_0+x_2-2x_1)\Big] \\[4px]
            \vdots \\[4px]
            f_1\Big(x_{\zeta-1},\,y_{\zeta-1}+\frac{g^e}{2}(x_{\zeta-2}+x_0-2x_{\zeta-1});\,\alpha_{\zeta-1}\Big) \\[4px]
            y_{\zeta-1} - \eta x_{\zeta-1} + \eta\Big[\sigma_{\zeta-1} + (x_{\zeta-2}+x_0-2x_{\zeta-1})\Big] 
        \end{pmatrix}
        \vspace{0.25cm}
    \end{gathered}
    \label{eq:ring-iteration-func}
\end{equation}
\vspace{0.1cm}

We now have all the tools we need to start looking into the dynamics of this ring system, but before we do that, we want to have a way to apply our classic method of quantifying chaos, Lyapunov exponents, to this significantly more complex system. Although in the previous section we hid some of the hairy details of calculating the Lyapunov exponents of the asymmetrically electrically coupled Rulkov 1 system in an appendix,\footnote{See Appendix \ref{partioning-jacobian-matrix-for-two-coup-rulkov-1-neurons}.} out of respect for the complexity of this system, as well as this system's applicability to more complicated arrangements of neurons in physical space, we will work through the derivation of the Lyapunov spectrum of this ring lattice here.\footnote{As the reader will certainly notice, and perhaps amusingly, the complexity of this calculation forces us to enter a one-column format. If the reader does not wish to go through this derivation with us, they need only skip to the point where the paper enters two columns again.} As always, the first and (for these more complex systems) most difficult step to calculating the Lyapunov spectrum of any system is to find its Jacobian matrix, so this is where we will begin our journey. Written out in partial derivative form, the Jacobian matrix of this ring lattice is
\onecolumn
\begin{equation}
    J(\mathbf{X}) = \begin{pmatrix}
        \frac{\partial F\e{1}}{\partial X\e{1}} & \frac{\partial F\e{1}}{\partial X\e{2}} & \hdots & \frac{\partial F\e{1}}{\partial X\e{2\zeta-1}} & \frac{\partial F\e{1}}{\partial X\e{2\zeta}} \\[6pt]
        \frac{\partial F\e{2}}{\partial X\e{1}} & \frac{\partial F\e{2}}{\partial X\e{2}} & \hdots & \frac{\partial F\e{2}}{\partial X\e{2\zeta-1}} & \frac{\partial F\e{2}}{\partial X\e{2\zeta}} \\[6pt]
        \vdots & \vdots & \ddots & \vdots & \vdots \\[4pt]
        \frac{\partial F\e{2\zeta-1}}{\partial X\e{1}} & \frac{\partial F\e{2\zeta-1}}{\partial X\e{2}} & \hdots & \frac{\partial F\e{2\zeta-1}}{\partial X\e{2\zeta-1}} & \frac{\partial F\e{2\zeta-1}}{\partial X\e{2\zeta}} \\[6pt]
        \frac{\partial F\e{2\zeta}}{\partial X\e{1}} & \frac{\partial F\e{2\zeta}}{\partial X\e{2}} & \hdots & \frac{\partial F\e{2\zeta}}{\partial X\e{2\zeta-1}} & \frac{\partial F\e{2\zeta}}{\partial X\e{2\zeta}}
    \end{pmatrix}
    = \begin{pmatrix}
        \frac{\partial F\e{1}}{\partial x_0} & \frac{\partial F\e{1}}{\partial y_0} & \hdots & \frac{\partial F\e{1}}{\partial x_{\zeta-1}} & \frac{\partial F\e{1}}{\partial y_{\zeta-1}} \\[6pt]
        \frac{\partial F\e{2}}{\partial x_0} & \frac{\partial F\e{2}}{\partial y_0} & \hdots & \frac{\partial F\e{2}}{\partial x_{\zeta-1}} & \frac{\partial F\e{2}}{\partial y_{\zeta-1}} \\[6pt]
        \vdots & \vdots & \ddots & \vdots & \vdots \\[4pt]
        \frac{\partial F\e{2\zeta-1}}{\partial x_0} & \frac{\partial F\e{2\zeta-1}}{\partial y_0} & \hdots & \frac{\partial F\e{2\zeta-1}}{\partial x_{\zeta-1}} & \frac{\partial F\e{2\zeta-1}}{\partial y_{\zeta-1}} \\[6pt]
        \frac{\partial F\e{2\zeta}}{\partial x_0} & \frac{\partial F\e{2\zeta}}{\partial y_0} & \hdots & \frac{\partial F\e{2\zeta}}{\partial x_{\zeta-1}} & \frac{\partial F\e{2\zeta}}{\partial y_{\zeta-1}}
    \end{pmatrix}
    \label{eq:ring-jacobian-partial-derivs}
\end{equation}
If we wanted to write out this Jacobian matrix explicitly like we did in Equation \ref{eq:rulkov_1_extremely_big_jacobian_matrix}, it would have to be written as $3^{\zeta}$ $2\zeta\times 2\zeta$ matrices due to the piecewise nature of Rulkov map 1. To put this into context, if we were to write $J(\mathbf{X})$ for our $\zeta=30$ neuron ring system in Figure \ref{fig:ring-30-rulkov-1-neurons} explicitly, we would need on the order of $10^{14}$ $60\times 60$ matrices. This is obviously absurd and not feasible, so instead, we can use indices to compactify the entirety of the Jacobian matrix in Equation \ref{eq:ring-jacobian-partial-derivs} into
\begin{equation}
    J_{mj}(\mathbf{X}) = \frac{\partial F\e{m}}{\partial X\e{j}}
    \label{eq:compactified-ring-jacobian}
\end{equation}
where $J_{mj}(\mathbf{X})$ is the entry in the $m$th row and $j$th column of $J(\mathbf{X})$. This makes the Jacobian matrix immediately look a lot less scary, and it makes it much easier to work with. We will now turn our attention to systematically calculating this derivative for all the different values of $m$ and $j$.

The first useful realization to make about the Jacobian in this form is the fundamental difference between an even value of $j$ and an odd value of $j$ (the column number of $J(\mathbf{X})$). Looking carefully at Equation \ref{eq:ring-jacobian-partial-derivs}, we can see that when $j$ is odd, we are differentiating functions with respect to fast variables $x$, but when $j$ is even, we are differentiating functions with respect to slow variables $y$. Playing around with this a little bit, we can find that if $j$ is odd, the neuron index of the fast variable we are differentiating with respect to is $i=(j-1)/2$. Similarly, if $j$ is even, the neuron index of the slow variable we are differentiating with respect to is $i=j/2-1$. If the reader desires, they may easily check this with a few test values in the written-out Jacobian matrix. Of course, an odd $j$ will satisfy $j\bmod 2=1$ and an even $j$ will satisfy $j\bmod 2=0$, so we can write Equation \ref{eq:compactified-ring-jacobian} in the more revealing form
\begin{equation}
    J_{mj}(\mathbf{X}) = \frac{\partial F\e{m}}{\partial X\e{j}} = \begin{cases}
        \frac{\partial F\e{m}}{\partial x_{(j-1)/2}}, & \text{if }j\bmod 2=1 \\
        \frac{\partial F\e{m}}{\partial y_{j/2-1}}, & \text{if }j\bmod 2=0 \\
    \end{cases}
    \label{eq:ring-jacob-j-split}
\end{equation}

Now, let us consider the difference between even and odd values of $m$, the dimension index of the function $\mathbf{F}(\mathbf{X})$. Referring back to Equation \ref{eq:ring-iteration-func}, we can see that when $m$ is odd, we are differentiating the piecewise fast map function $f_1$, and when $m$ is even, we are differentiating the Rulkov slow map function. Because it is easier, let us first consider the even values of $m$, where $m\bmod 2=0$. We know from Equation \ref{eq:rulkov_coupled_mapping} that the slow variable of the $i$th coupled Rulkov neuron iterates according to
\begin{equation}
    \begin{split}
        y_{i,\,k+1} &= y_{i,\,k} - \eta x_{i,\,k} + \eta[\sigma_i + \mathfrak{C}_i(k)] \\
         &= y_{i,\,k} - \eta x_{i,\,k} + \eta\Bigg[\sigma_i + \frac{g^e}{2}(x_{(i-1)\bmod\zeta,\,k}+x_{(i+1)\bmod\zeta,\,k}-2x_{i,\,k}) \Bigg]
    \end{split}
\end{equation}
In the case where $m\bmod 2=0$, mirroring when $k\bmod 2=0$, the neuron index of the iteration function we are taking the derivative of is $i=m/2-1$. Therefore, we can write the iteration function $F\e{m}$ for the case where $m$ is even as
\begin{equation}
    F\e{m} = y_{m/2-1}-\eta x_{m/2-1} + \eta\Bigg[\sigma_{m/2-1} + \frac{g^e}{2}(x_{(m/2-2)\bmod\zeta} + x_{(m/2)\bmod\zeta} - 2x_{m/2-1})\Bigg]
    \label{eq:even-m-ring-function}
\end{equation}
Looking carefully at this function, we can see that the only variables present are $y_{m/2-1}$, $x_{m/2-1}$, $x_{(m/2-2)\bmod\zeta}$, and $x_{(m/2)\bmod\zeta}$. Differentiating Equation \ref{eq:even-m-ring-function} with respect to each of these variables in turn gives us
\begin{align}
    \frac{\partial F\e{m}}{\partial y_{m/2-1}} &= 1 \label{eq:first-non-zero-deriv} \\
    \frac{\partial F\e{m}}{\partial x_{m/2-1}} &= -\eta-\eta g^e = -\eta(1+g^e) \label{eq:sec-non-zero-deriv} \\
    \frac{\partial F\e{m}}{\partial x_{(m/2-2)\bmod\zeta}} &= \frac{\eta g^e}{2} \label{eq:third-non-zero-deriv}\\
    \frac{\partial F\e{m}}{\partial x_{(m/2)\bmod\zeta}} &= \frac{\eta g^e}{2} \label{eq:fourth-non-zero-deriv}
\end{align}
Differentiating with respect to any other variable will give us 0. The tricky part now is to determine the values of $j$ that give us these non-zero derivatives. The important thing to remember in order to make this easier is that this function $F\e{m}$ is the iteration function for the state $X\e{m}$:
\begin{equation}
    X\e{m}_{k+1} = F\e{m}(X\e{1}_k,\,X\e{2}_k,\,X\e{3}_k,\,X\e{4}_k,\,\hdots\,,\, X\e{2\zeta - 1}_k,\,X\e{2\zeta}_k)
    \label{eq:iterate-one-ring-dimension}
\end{equation}
For even $m$, we know from Equation \ref{eq:ring-jacob-j-split} that $X\e{m} = y_{m/2-1}$. Then, substituting the state vector entries from Equation \ref{eq:ring-system-state-vector} into the arguments of $F\e{m}$ allows us to write Equation \ref{eq:iterate-one-ring-dimension} as
\begin{equation}
    y_{m/2-1,\,k+1} = F\e{m}(x_{0,\,k},\,y_{0,\,k},\,x_{1,\,k},\,y_{1,\,k},\,\hdots\,,\,x_{\zeta-1,\,k},\,y_{\zeta-1,\,k})
\end{equation}
We can then more easily determine the values of $j$ that will give us the non-zero values of $\frac{\partial F\e{m}}{\partial X\e{j}}$ by thinking up and down the dimensions of the state vector $\mathbf{X}$ from the value where $X\e{j} = X\e{m} = y_{m/2-1}$. This might sound a little confusing, so we will work through these next examples slowly. Starting with the first non-zero derivative in Equation \ref{eq:first-non-zero-deriv}, we want to find $j$ such that $X\e{j} = y_{m/2-1}$. We know that $X\e{m} = y_{m/2-1}$, so equating the dimension indices gives us that $\frac{\partial F\e{m}}{\partial X\e{j}}=\frac{F\e{m}}{\partial y_{m/2-1}}=1$ if $j=m$ (given $m$ is even). Next, for the second non-zero derivative in Equation \ref{eq:sec-non-zero-deriv}, we want the value of $j$ such that $X\e{j} = x_{m/2-1}$. We know that $x_{m/2-1}$ is the fast variable of the same neuron associated with the slow variable $y_{m/2-1}$. Taking a look at the state vector in Equation \ref{eq:ring-system-state-vector}, we see that $X\e{j} = x_{m/2-1}$ is one dimension up from $X\e{m} = y_{m/2-1}$, or in other words, $X\e{m-1} = x_{m/2-1}$. Matching dimension indices, this means that $\frac{\partial F\e{m}}{\partial X\e{j}}=-\eta(1+g^e)$ if $j=m-1$. Moving swiftly along to the third non-zero derivative in Equation \ref{eq:third-non-zero-deriv}, we want the value of $j$ such that $X\e{j} = x_{(m/2-2)\bmod\zeta}$, which poses a slightly different challenge due to the presence of the $\bmod$. However, we can easily disarm this challenge by remembering that the $\bmod$ only becomes relevant when $m=2$, which is when we are working with neuron $i = m/2-1 = 0$. In this case, the ``clockwise'' neighbor (see Figure \ref{fig:ring-30-rulkov-1-neurons}) to the neuron $\mathbf{x}_0$ is $\mathbf{x}_{\zeta-1}$. Referring back to the state vector in Equation \ref{eq:ring-system-state-vector}, the fast variable associated with the neuron $\mathbf{x}_{\zeta-1}$, which is the variable $X\e{j} = x_{(m/2-2)\bmod\zeta}$ for $m=2$, is $X\e{2\zeta-1} = x_{\zeta-1}$. For any other even $m$ besides $m=2$, the $\bmod$ doesn't apply, so we can proceed normally. It is easy to see that the fast variable $X\e{j} = x_{m/2-2}$ is one neuron up from the slow variable of $X\e{m} = y_{m/2-1}$. Looking at the state vector in Equation \ref{eq:ring-system-state-vector}, we can see that going from the slow variable of one neuron to the fast variable of its ``clockwise neighbor'' (one neuron up from it) is equivalent to going three dimensions up the state vector: $X\e{m-3} = x_{m/2-2}$. We can then conclude that $\frac{\partial F\e{m}}{\partial X\e{j}} = ng^e/2$ if $j=m-3$ and $m\neq 2$, or $j=2\zeta - 1$ and $m=2$. Finding the $j$ values for the non-zero derivative in Equation \ref{eq:fourth-non-zero-deriv}, which concerns the ``counterclockwise'' neighbor of $\mathbf{x}_{m/2-1}$, follows directly from the previous examples: $\frac{\partial F\e{m}}{\partial X\e{j}} = ng^e/2$ if $j=m+1$ and $m\neq 2\zeta$, or $j=1$ and $m=2\zeta$. Now, we can summarize all the results from this paragraph into one piecewise function for the derivative $\frac{\partial F\e{m}}{\partial X\e{j}}$ when $m\bmod 2=0$: 
\begin{equation}
    \frac{\partial F\e{m}}{\partial X\e{j}} = 
    \begin{cases}
        1, & \hspace{0.325cm}\text{if }j=m \\
        -\eta(1+g^e), & \hspace{0.325cm}\text{if }j=m-1 \\
        \eta g^e/2, & \begin{cases}
            \text{if }j=m-3\text{ and }m\neq 2 \\
            \text{or }j=2\zeta-1\text{ and }m=2 \\
            \text{or }j=m+1\text{ and }m\neq 2\zeta \\
            \text{or }j=1\text{ and }m=2\zeta
        \end{cases} \\
        0, & \hspace{0.325cm}\text{otherwise}
    \end{cases}
    \label{eq:ring-jacobian-part-1}
\end{equation}

Now, let us move on to the harder case $m\bmod 2=1$, where we are differentiating the Rulkov 1 fast map function. We know from Equation \ref{eq:rulkov_coupled_mapping} that the fast variable of the $i$th coupled Rulkov neuron iterates according to
\begin{equation}
    \begin{split}
        x_{i,\,k+1} &= f_1(x_{i,\,k},\, y_{i,\,k} + \mathfrak{C}_{i}(k);\,\alpha_i) \\
         &= f_1\Bigg(x_{i,\,k},\, y_{i,\,k} + \frac{g^e}{2}(x_{(i-1)\bmod\zeta,\,k}+x_{(i+1)\bmod\zeta,\,k}-2x_{i,\,k});\,\alpha_i\Bigg)
    \end{split}
\end{equation}
or, written out explicitly using the piecewise form in Equation \ref{eq:rulkov_1_fast_equation},
\begin{equation}
    x_{i,\,k+1} = \begin{cases}
        \frac{\alpha}{1-x_{i,\,k}} + y_{i,\,k} + \frac{g^e}{2}(x_{(i-1)\bmod\zeta,\,k}+x_{(i+1)\bmod\zeta,\,k}-2x_{i,\,k}), & x_{i,\,k}\leq 0 \\
        \alpha + y_{i,\,k} + \frac{g^e}{2}(x_{(i-1)\bmod\zeta,\,k}+x_{(i+1)\bmod\zeta,\,k}-2x_{i,\,k}), & 0 < x_{i,\,k} < \alpha + y_{i,\,k} + \mathfrak{C}_{i}(k) \\
        -1, & x_{i,\,k}\geq \alpha + y_{i,\,k} + \mathfrak{C}_{i}(k)
    \end{cases}
\end{equation}
For odd $m$, mirroring odd values of $j$ from Equation \ref{eq:ring-jacob-j-split}, we know that we are dealing with the fast variable iteration function of the neuron with $i = (m-1)/2$. Therefore, we can write the iteration function $F\e{m}$ for the case where $m\bmod 2=1$ as
\begin{equation}
    \begin{split}
        F\e{m} &= f_1(x_{(m-1)/2},\,y_{(m-1)/2} + \mathfrak{C}_{(m-1)/2};\,\alpha_{(m-1)/2}) \\[0.25cm]
        &= \begin{cases}
            \frac{\alpha_{(m-1)/2}}{1-x_{(m-1)/2}} + y_{(m-1)/2} + \frac{g^e}{2}(x_{[(m-3)/2] \bmod\zeta} + x_{[(m+1)/2] \bmod\zeta} - 2x_{(m-1)/2}), & x_{(m-1)/2} \leq 0 \vspace{0.5cm} \\
            \alpha_{(m-1)/2} + y_{(m-1)/2} + \frac{g^e}{2}(x_{[(m-3)/2] \bmod\zeta} + x_{[(m+1)/2] \bmod\zeta} - 2x_{(m-1)/2}), & 0 < x_{(m-1)/2} < \alpha_{(m-1)/2} \\
             & + y_{(m-1)/2} + \mathfrak{C}_{(m-1)/2} \vspace{0.5cm} \\
            -1, & x_{(m-1)/2} \geq \alpha_{(m-1)/2} \\
            & + y_{(m-1)/2} + \mathfrak{C}_{(m-1)/2} \\
        \end{cases}
    \end{split}
    \label{eq:piecewise-ring-function}
\end{equation}
Because there are three distinct functions within this piecewise form of $F\e{m}$, we will consider each of the pieces in turn.

First, let us investigate the case where $x_{(m-1)/2} \leq 0$. Taking a look at this piece of the function, we see that the only variables present are $y_{(m-1)/2}$, $x_{(m-1)/2}$, $x_{[(m-3)/2]\bmod\zeta}$, and $x_{[(m+1)/2]\bmod\zeta}$. Then, the only non-zero derivatives of this piece of $F\e{m}$ are
\begin{align}
    \frac{\partial F\e{m}}{\partial y_{(m-1)/2}} &= 1 \label{eq:piece-1-first-deriv} \\
    \frac{\partial F\e{m}}{\partial x_{(m-1)/2}} &= \frac{\alpha_{(m-1)/2}}{(1-x_{(m-1)/2})^2} - g^e \label{eq:piece-1-sec-deriv} \\
    \frac{\partial F\e{m}}{\partial x_{[(m-3)/2]\bmod\zeta}} &= \frac{g^e}{2} \label{eq:piece-1-third-deriv} \\
    \frac{\partial F\e{m}}{\partial x_{[(m+1)/2]\bmod\zeta}} &= \frac{g^e}{2} \label{eq:piece-1-fourth-deriv}
\end{align}
We will now once again systematically determine the values of $j$ that give us these non-zero derivatives. For this case where $m$ is odd, checking the state vector in Equation \ref{eq:ring-system-state-vector} indicates that $F\e{m}$ is the iteration function for $x_{(m-1)/2}$:
\begin{equation}
    x_{(m-1)/2,\,k+1} = F\e{m}(x_{0,\,k},\,y_{0,\,k},\,x_{1,\,k},\,y_{1,\,k},\,\hdots\,,\,x_{\zeta-1,\,k},\,y_{\zeta-1,\,k})
\end{equation}
meaning $X\e{m} = x_{(m-1)/2}$. For the first non-zero derivative in Equation \ref{eq:piece-1-first-deriv}, we want to find $j$ such that $X\e{j} = y_{(m-1)/2}$, which is one dimension down from its associated fast variable $x_{(m-1)/2}$. In other words, $X\e{m+1} = y_{(m-1)/2}$, so $\frac{\partial F\e{m}}{\partial X\e{j}}=1$ if $j=m+1$ (given $m$ is odd). For the second non-zero derivative in Equation \ref{eq:piece-1-sec-deriv}, we are differentiating with respect to $X\e{j} = x_{(m-1)/2}$, meaning $\frac{\partial F\e{m}}{\partial X\e{j}}=\alpha_{(m-1)/2}(1-x_{(m-1)/2})^{-2} - g^e$ if $j=m$. The slightly more challenging $j$ to find comes with the third non-zero derivative in Equation \ref{eq:piece-1-third-deriv}, where $X\e{j} = x_{[(m-3)/2]\bmod\zeta}$, or the ``clockwise'' neighbor. Similar to before, let us first consider the special case where $X\e{m} = x_{(m-1)/2}$ is the fast variable of the neuron with index $i=0$, which happens when $m=1$. In this case, the ``clockwise'' neighbor to $\mathbf{x}_0$ is $\mathbf{x}_{\zeta-1}$, which is associated with the fast variable $X\e{2\zeta-1} = x_{[(m-3)/2]\bmod\zeta} = x_{\zeta-1}$ for $m=1$. For the ``normal'' cases, going from $X\e{m} = x_{(m-1)/2}$ to its clockwise neighbor $X\e{j} = x_{(m-3)/2}$ requires moving up two dimensions: $X\e{m-2} = x_{(m-3)/2}$. Therefore, $\frac{\partial F\e{m}}{\partial X\e{j}}=g^e/2$ if $j=2\zeta-1$ and $m=1$, or $j=m-2$ and $m\neq 1$. The fourth non-zero derivative, concerning the ``counterclockwise'' neighbor $X\e{j} = x_{[(m+1)/2]\bmod\zeta}$, follows naturally from this, yielding that $\frac{\partial F\e{m}}{\partial X\e{j}}=g^e/2$ if $j=1$ and $m=2\zeta-1$, or $j=m+2$ and $m\neq 2\zeta-1$. We can now summarize the results from this paragraph into one piecewise function for the derivative $\frac{\partial F\e{m}}{\partial X\e{j}}$ that holds for $x_{(m-1)/2}\leq 0$ when $m\bmod 2=1$:
\begin{equation}
    \frac{\partial F\e{m}}{\partial X\e{j}} = 
    \begin{cases}
        1, & \hspace{0.325cm}\text{if }j=m+1 \\
        \frac{\alpha_{(m-1)/2}}{(1-x_{(m-1)/2})^2} - g^e, & \hspace{0.325cm}\text{if }j=m\\
        g^e/2, & \begin{cases}
            \text{if }j=m-2\text{ and }m\neq 1 \\
            \text{or }j=2\zeta-1\text{ and }m=1 \\
            \text{or }j=m+2\text{ and }m\neq 2\zeta - 1 \\
            \text{or }j=1\text{ and }m=2\zeta - 1
        \end{cases} \\
        0, & \hspace{0.325cm}\text{otherwise}
    \end{cases}
    \label{eq:ring-jacobian-part-2}
\end{equation}
Luckily, this is the peak of the difficulty mountain for this derivation of the Jacobian of our ring system, and it only goes downhill from here in complexity.

Continuing to dismantle the complex piecewise function in Equation \ref{eq:piecewise-ring-function}, let us next consider the case where $0 < x_{(m-1)/2} < \alpha_{(m-1)/2} + y_{(m-1)/2} + \mathfrak{C}_{(m-1)/2}$. Taking a look at this piece of Equation \ref{eq:piecewise-ring-function}, we can see that it contains the same variables as the piece where $x_{(m-1)/2}\leq 0$, namely, $y_{(m-1)/2}$, $x_{(m-1)/2}$, $x_{[(m-3)/2]\bmod\zeta}$, and $x_{[(m+1)/2]\bmod\zeta}$. The function itself, however, is evidently different, so the non-zero derivatives are also different:
\begin{align}
    \frac{\partial F\e{m}}{\partial y_{(m-1)/2}} &= 1 \label{eq:piece-2-first-deriv} \\
    \frac{\partial F\e{m}}{\partial x_{(m-1)/2}} &= - g^e \label{eq:piece-2-sec-deriv}\\
    \frac{\partial F\e{m}}{\partial x_{[(m-3)/2]\bmod\zeta}} &= \frac{g^e}{2} \label{eq:piece-2-third-deriv}\\
    \frac{\partial F\e{m}}{\partial x_{[(m+1)/2]\bmod\zeta}} &= \frac{g^e}{2} \label{eq:piece-2-fourth-deriv}
\end{align}
We might think that we now have to find the values of $j$ that give us these non-zero partial derivatives once again, but thankfully, we've lucked out in this case because we are dealing with the same state $X\e{m}=x_{(m-1)/2}$ and the same variables as the first piece of $F\e{m}$: $y_{(m-1)/2}$, $x_{(m-1)/2}$, $x_{[(m-3)/2]\bmod\zeta}$, and $x_{[(m+1)/2]\bmod\zeta}$. All of the relationships between these variables in state space are the same as before because the piece of $F\e{m}$ we are working with has no influence on the variables we are connecting. Therefore, all we have to do is plug the non-zero derivatives from Equations \ref{eq:piece-2-first-deriv}, \ref{eq:piece-2-sec-deriv}, \ref{eq:piece-2-third-deriv}, and \ref{eq:piece-2-fourth-deriv} into the appropriate locations of the general form of the derivative in displayed in Equation \ref{eq:ring-jacobian-part-2}. Making these substitutions yields the piecewise function for the derivative $\frac{\partial F\e{m}}{\partial X\e{j}}$ that holds for $0 < x_{(m-1)/2} < \alpha_{(m-1)/2} + y_{(m-1)/2} + \mathfrak{C}_{(m-1)/2}$ when $m\bmod 2=1$:
\begin{equation}
    \frac{\partial F\e{m}}{\partial X\e{j}} = 
    \begin{cases}
        1, & \hspace{0.325cm}\text{if }j=m+1 \\
        - g^e, & \hspace{0.325cm}\text{if }j=m\\
        g^e/2, & \begin{cases}
            \text{if }j=m-2\text{ and }m\neq 1 \\
            \text{or }j=2\zeta-1\text{ and }m=1 \\
            \text{or }j=m+2\text{ and }m\neq 2\zeta - 1 \\
            \text{or }j=1\text{ and }m=2\zeta - 1
        \end{cases} \\
        0, & \hspace{0.325cm}\text{otherwise}
    \end{cases}
    \label{eq:ring-jacobian-part-3}
\end{equation}

Finally, for the last part of the piecewise function $F\e{m}$ in Equation \ref{eq:piecewise-ring-function}, where $x_{(m-1)/2} \geq \alpha_{(m-1)/2} + y_{(m-1)/2} + \mathfrak{C}_{(m-1)/2}$, we have been blessed with an extremely easy function to differentiate: $F\e{m}=-1$. Obviously, the derivative of $-1$ with respect to any variable is 0. Therefore, for the case where $x_{(m-1)/2} \geq \alpha_{(m-1)/2} + y_{(m-1)/2} + \mathfrak{C}_{(m-1)/2}$ for $m\bmod 2=1$, we can write the derivative $\frac{\partial F\e{m}}{\partial X\e{j}}$ very simply as
\begin{equation}
    \frac{\partial F\e{m}}{\partial X\e{j}} = 0
    \label{eq:ring-jacobian-part-4}
\end{equation}
which concludes all the possible cases of the derivative $\frac{\partial F\e{m}}{\partial X\e{j}}$.

So now, for the grand finale, we combine all of these calculations into one partial derivative to rule them all: $\frac{\partial F\e{m}}{\partial X\e{j}} = J_{mj}(\mathbf{X})$. This derivative represents the $m$th row and the $j$th column of the Jacobian of any state $\mathbf{X}$ of this symmetrically coupled ring lattice of Rulkov 1 neurons for any $m$ or $j$ between $1$ and $2\zeta$. To do this, we will combine all of the cases for the different values of $m$ and $j$ we have discussed: for even values of $m$ and all values of $j$, Equation \ref{eq:ring-jacobian-part-1}; for odd values of $m$, values $x_{(m-1)/2}\leq 0$, and all values of $j$, Equation \ref{eq:ring-jacobian-part-2}; for odd values of $m$, values $0 < x_{(m-1)/2} < \alpha_{(m-1)/2} + y_{(m-1)/2} + \mathfrak{C}_{(m-1)/2}$, and all values of $j$, Equation \ref{eq:ring-jacobian-part-3}; and for odd values of $m$, values $x_{(m-1)/2} \geq \alpha_{(m-1)/2} + y_{(m-1)/2} + \mathfrak{C}_{(m-1)/2}$, and all values of $j$, Equation \ref{eq:ring-jacobian-part-4}. Putting all of this together yields the generalized entry of the Jacobian $J_{mj}(\mathbf{X})$ shown on the next page. It is worth noting that when using this equation in practice, it should be read from right to left. And now, without further ado, here it is:
\newpage
\begin{equation}
    J_{mj}(\mathbf{X}) = 
    \begin{cases}
        \begin{cases}
        
            \begin{cases}
                1, & \hspace{0.325cm}\text{if }j=m+1, \\
                \frac{\alpha_{(m-1)/2}}{(1-x_{(m-1)/2})^2} - g^e, & \hspace{0.325cm}\text{if }j=m,\\
                 g^e/2, & \begin{cases}
                    \text{if }j=m-2,\\
                    \phantom{if }\text{ and }m\neq 1, \\
                    \text{or }j=2\zeta-1,\\
                    \phantom{if }\text{ and }m=1, \\
                    \text{or }j=m+2,\\
                    \phantom{if }\text{ and }m\neq 2\zeta - 1, \\
                    \text{or }j=1,\\
                    \phantom{if }\text{ and }m=2\zeta - 1,
                \end{cases} \\
                0, & \hspace{0.325cm}\text{otherwise,}
            \end{cases} \text{for } x_{(m-1)/2} \leq 0, \\
            \\
            \\
            \begin{cases}
                1, & \hspace{2.225cm}\text{if }j=m+1, \\
                - g^e, & \hspace{2.225cm}\text{if }j=m,\\
                g^e/2, & \hspace{1.9cm}\begin{cases}
                    \text{if }j=m-2,\\
                    \phantom{if }\text{ and }m\neq 1, \\
                    \text{or }j=2\zeta-1,\\
                    \phantom{if }\text{ and }m=1, \\
                    \text{or }j=m+2,\\
                    \phantom{if }\text{ and }m\neq 2\zeta - 1, \\
                    \text{or }j=1,\\
                    \phantom{if }\text{ and }m=2\zeta - 1,
                \end{cases} \stackunder{\text{for $0 < x_{(m-1)/2} < \alpha_{(m-1)/2}$}}{\text{$+ y_{(m-1)/2} + \mathfrak{C}_{(m-1)/2},$}} \\
                0, & \hspace{2.225cm}\text{otherwise,}
            \end{cases}\\
            \\
            \\
            0, \hspace{6.685cm} \text{for} \stackunder{x_{(m-1)/2} \geq \alpha_{(m-1)/2}}{+ y_{(m-1)/2} + \mathfrak{C}_{(m-1)/2},}\\
            
        \end{cases} & \text{when } m\bmod 2=1 \\
        \\
        \\
        \begin{cases}
            1, & \hspace{0.67cm}\text{if }j=m \\
            -\eta(1+g^e), & \hspace{0.67cm}\text{if }j=m-1 \\
            \eta g^e/2, & \hspace{0.345cm}
            \begin{cases}
                \text{if }j=m-3\\
                    \phantom{if }\text{ and }m\neq 2 \\
                \text{or }j=2\zeta-1\\
                    \phantom{if }\text{ and }m=2 \\
                \text{or }j=m+1\\
                    \phantom{if }\text{ and }m\neq 2\zeta \\
                \text{or }j=1\\
                    \phantom{if }\text{ and }m=2\zeta
            \end{cases} \\
        0, & \hspace{0.67cm}\text{otherwise,}
        \end{cases} & \text{when } m\bmod 2=0
    \end{cases}
    \label{eq:THE-jacobian-entry}
\end{equation}

\twocolumn

Now that we have completed this Herculean task, there comes the relatively straightforward process of coding Equation \ref{eq:THE-jacobian-entry} into Python, which we do in Appendix \ref{ring-lattice-code}. Along with a modification of the Lyapunov QR factorization code to account for the potentially much larger Jacobian matrices, this code calculates the Lyapunov spectrum of a ring lattice system orbit $O(\mathbf{X}_0)$ by using Equation \ref{eq:THE-jacobian-entry} to determine the Jacobian matrices $J(\mathbf{X})$ for all $\mathbf{X}\in O(\mathbf{X}_0)$. Now that we have established all the tools we need, we will apply them to exploring the system displayed in Figure \ref{fig:ring-30-rulkov-1-neurons}: a ring of $\zeta=30$ Rulkov 1 neurons. Because each of these neurons has one slow and one fast variable (see Equation \ref{eq:ring-system-state-vector}), we are dealing with a 60-dimensional state space, which is far bigger than any space we have dealt with so far in this paper. Within this ring system, we will explore the dynamics that emerge from three specific initial states: one where different neurons have different initial fast variable values $x_{i,\,0}$ but all have the same $y_{i,\,0}$, $\sigma_i$, and $\alpha_i$ values, one where different neurons have different $x_{i,\,0}$ and $\sigma_i$ values but the same $y_{i,\,0}$ and $\alpha_i$ values, and one where different neurons have different $x_{i,\,0}$, $\sigma_i$, and $\alpha_i$ values but the same $y_{i,\,0}$ values. We do not consider the case where different neurons start with different $y_{i,\,0}$ values because different evolutions of the slow variable are accounted for by different values of $\sigma_i$ (since we know from Section \ref{individual-dynamics-of-rulkov-map-1} $\sigma_i$ determines the value of $x_i$ that will keep $y_i$ steady).

Before we start, let us quickly take care of some brief notational business. Since, in this section, we deal with many different distinct values of the parameters $\sigma_i$ and $\alpha_i$, let us collect all of the values of $\sigma_i$ and $\alpha_i$ into $\zeta$-dimensional vectors. We will denote the vector of all $\sigma_i$ values as
\begin{equation}
    \boldsymbol{\sigma} = \begin{pmatrix}
        \sigma_0 \\
        \sigma_1 \\
        \vdots \\
        \sigma_{\zeta-1}
    \end{pmatrix}
\end{equation}
and the vector of all $\alpha_i$ values as
\begin{equation}
    \boldsymbol{\alpha} = \begin{pmatrix}
        \alpha_0 \\
        \alpha_1 \\
        \vdots \\
        \alpha_{\zeta-1}
    \end{pmatrix}
\end{equation}
An interesting realization to make is that the parameter space of this ring lattice system, which is the space of all $\sigma_i$ and $\alpha_i$ values (not including $\eta$), is isomorphic to the state space of the system because they both can be represented in $\mathbb{R}^{2\zeta}$. For convenience, we will also introduce the one vector, $\mathbf{1}$, as a vector of all ones 
\begin{equation}
    \mathbf{1} = \begin{pmatrix}
        1 \\
        1 \\
        \vdots \\
        1
    \end{pmatrix} \in \mathbb{R}^n
\end{equation}
for some arbitrary number of dimensions $n$ that will be clear from context.

\begin{figure*}[hp!]
    \centering
    \begin{subfigure}{0.475\textwidth}
        \centering
        \includegraphics[scale=0.08]{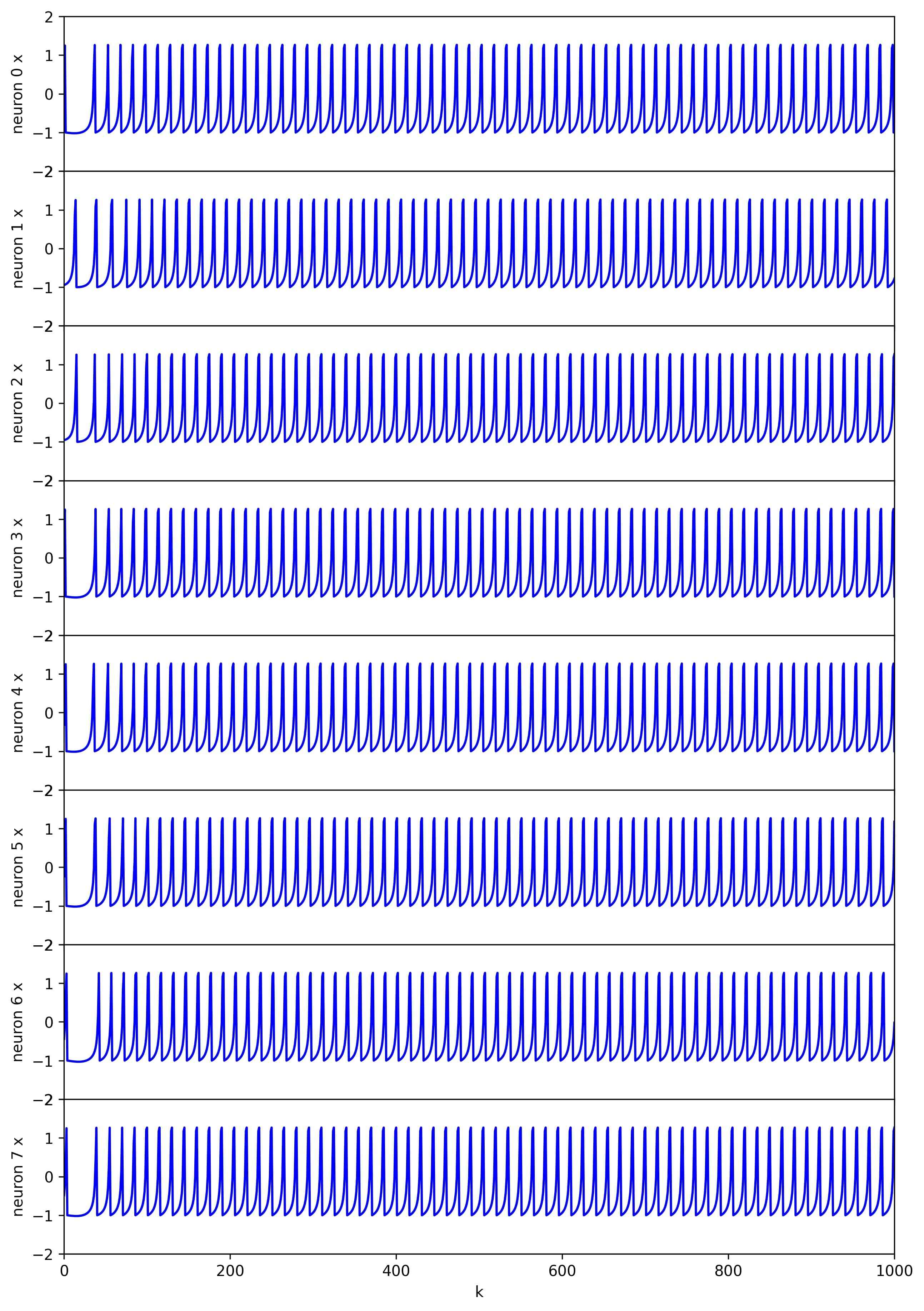}
        \vspace{-0.2cm}
        \caption{$g^e=0$, $\lambda_1\approx -0.0938$}
        \label{fig:random_x_ge0}
        \vspace{0.2cm}
    \end{subfigure}
    \begin{subfigure}{0.475\textwidth}
        \centering
        \includegraphics[scale=0.08]{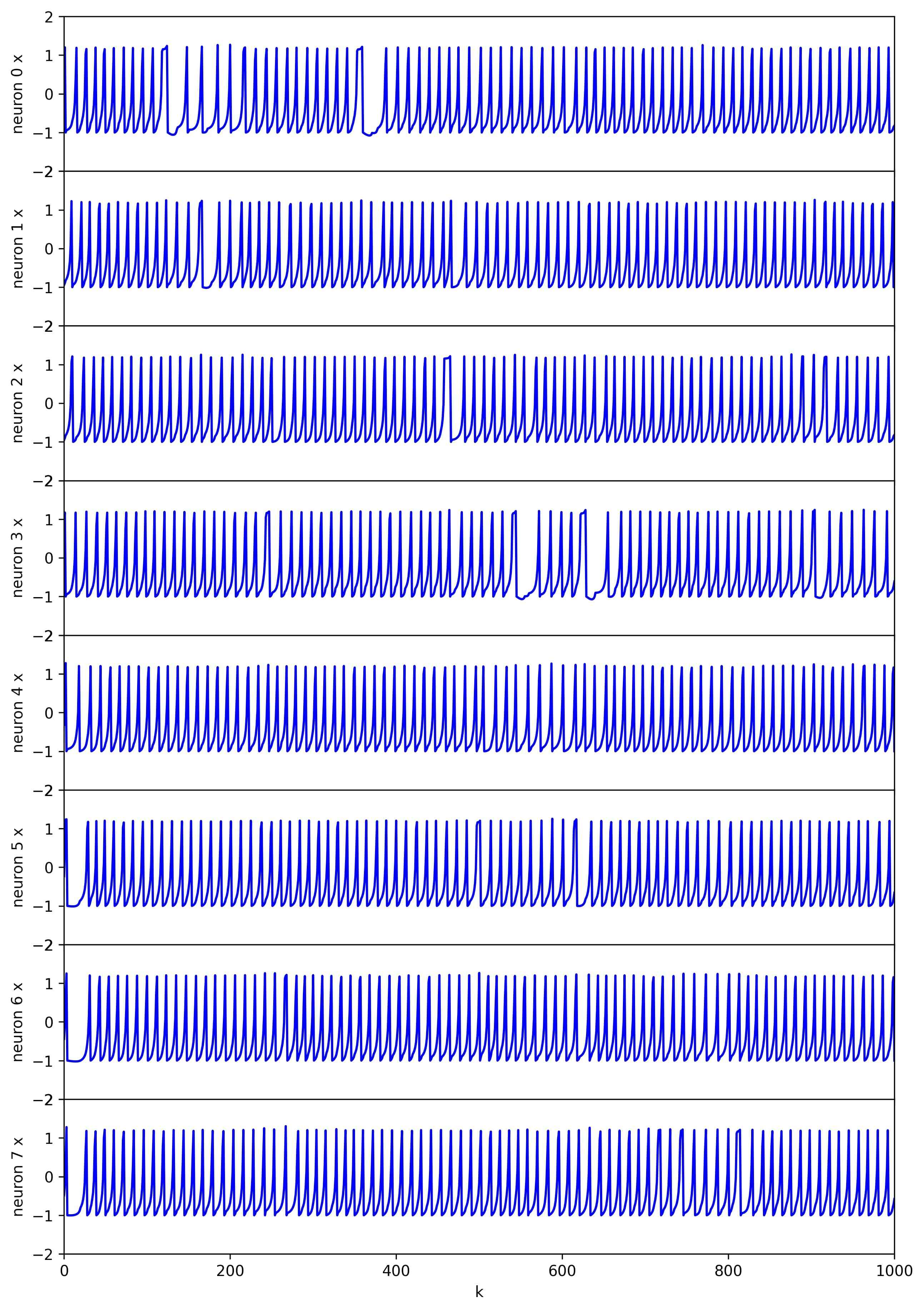}
        \vspace{-0.2cm}
        \caption{$g^e=0.05$, $\lambda_1\approx 0.0491$}
        \label{fig:random_x_ge0.05}
        \vspace{0.2cm}
    \end{subfigure}
    \begin{subfigure}{0.475\textwidth}
        \centering
        \includegraphics[scale=0.08]{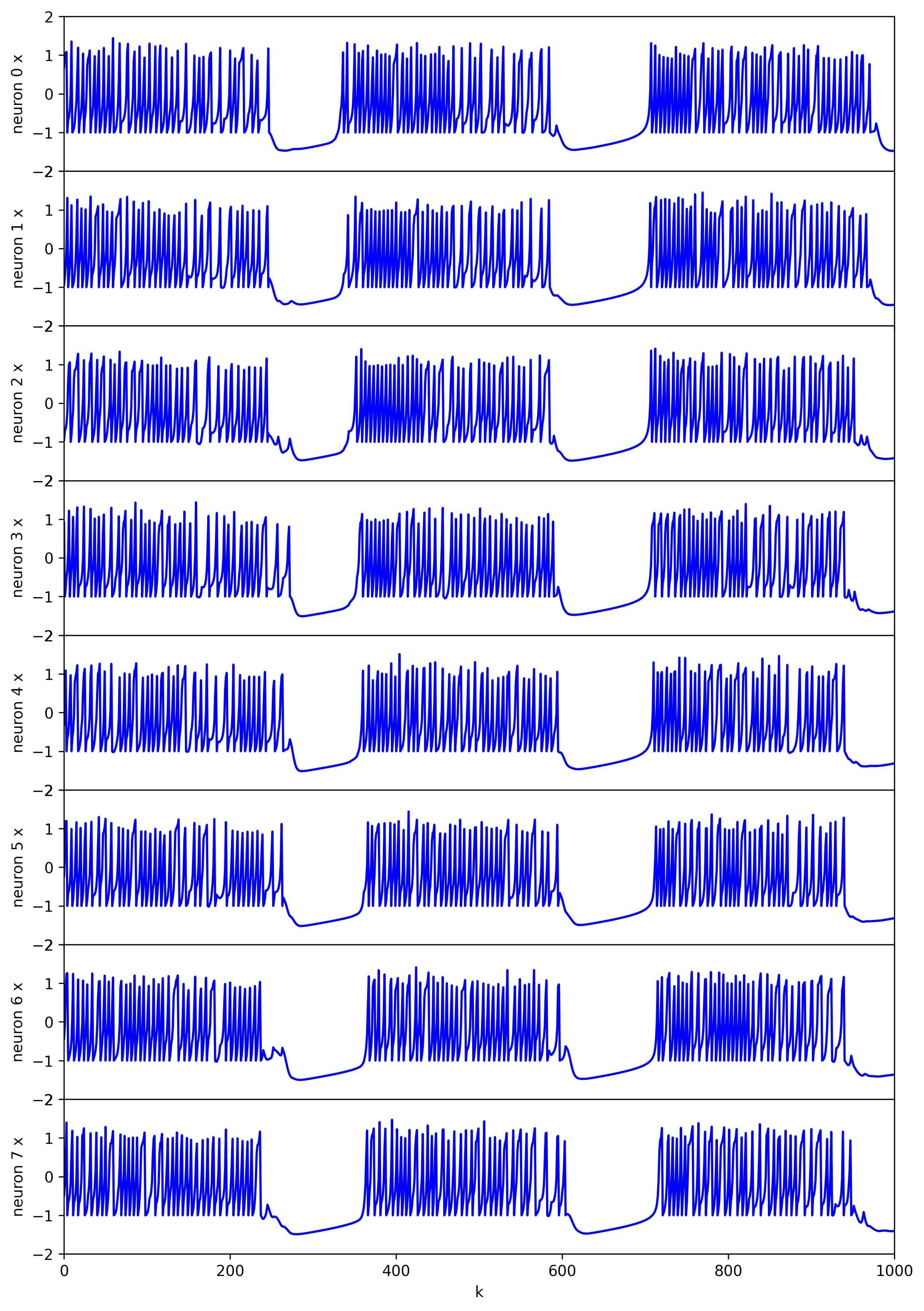}
        \vspace{-0.2cm}
        \caption{$g^e=0.25$, $\lambda_1\approx 0.0595$}
        \label{fig:random_x_ge0.25}
    \end{subfigure}
    \begin{subfigure}{0.475\textwidth}
        \centering
        \includegraphics[scale=0.08]{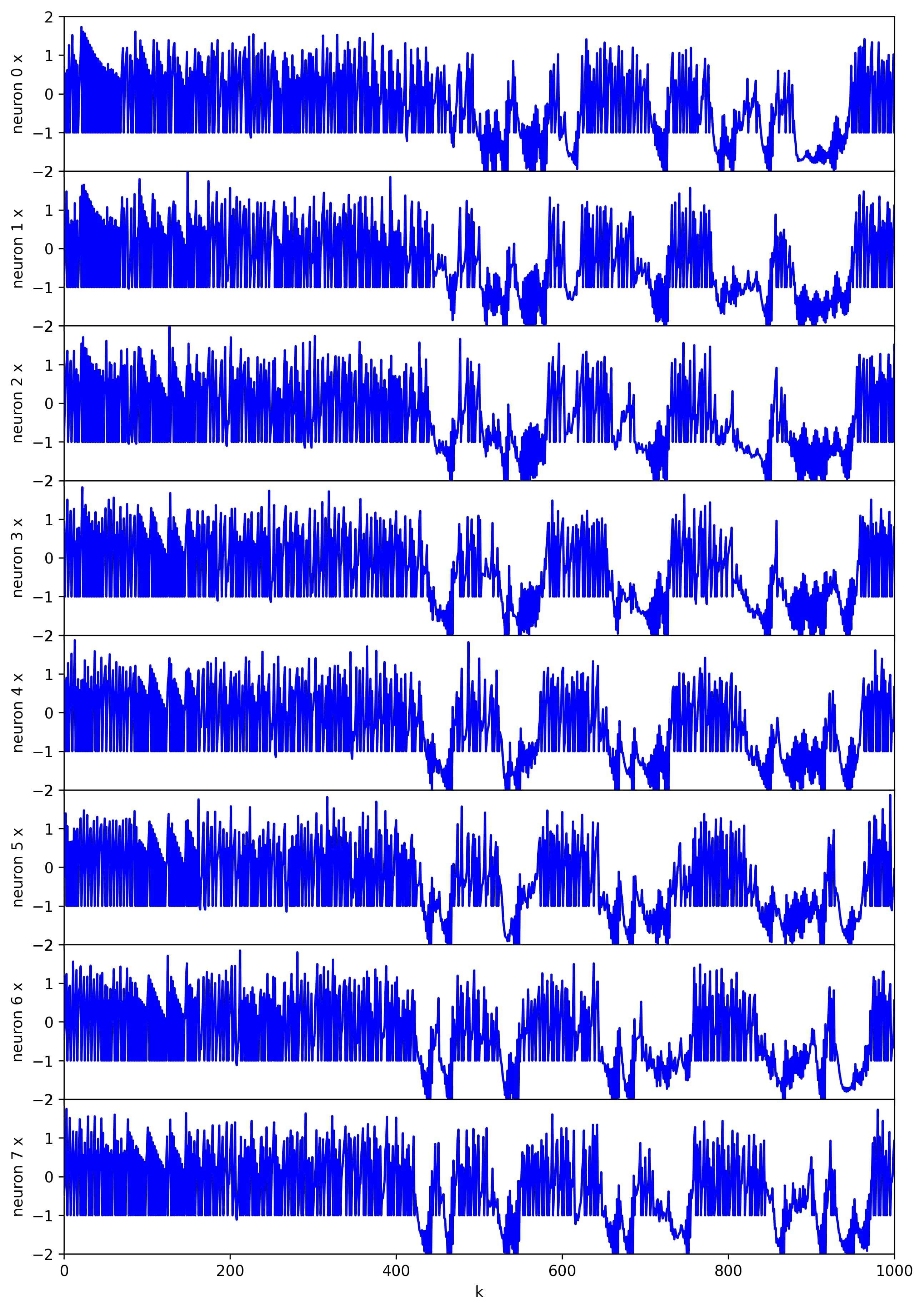}
        \vspace{-0.2cm}
        \caption{$g^e=1$, $\lambda_1\approx 0.1694$}
        \label{fig:random_x_ge1}
    \end{subfigure}
    \caption{Graphs of $x_{i,\,k}$ for eight neurons in a ring of $\zeta=30$ electrically coupled Rulkov 1 neurons, with $x_{i,\,0}\in (-1,\,1)$, $y_{i,\,0}=-3.25$, $\boldsymbol{\sigma} = -0.5\cdot\mathbf{1}$, and $\boldsymbol{\alpha} = 4.5\cdot\mathbf{1}$, visualized and $\lambda_1$ values calculated using the code in Appendix \ref{ring-lattice-code}}
    \label{fig:random_x_graphs}
\end{figure*}

\begin{figure*}[htb!]
    \centering
    \includegraphics[scale=0.21]{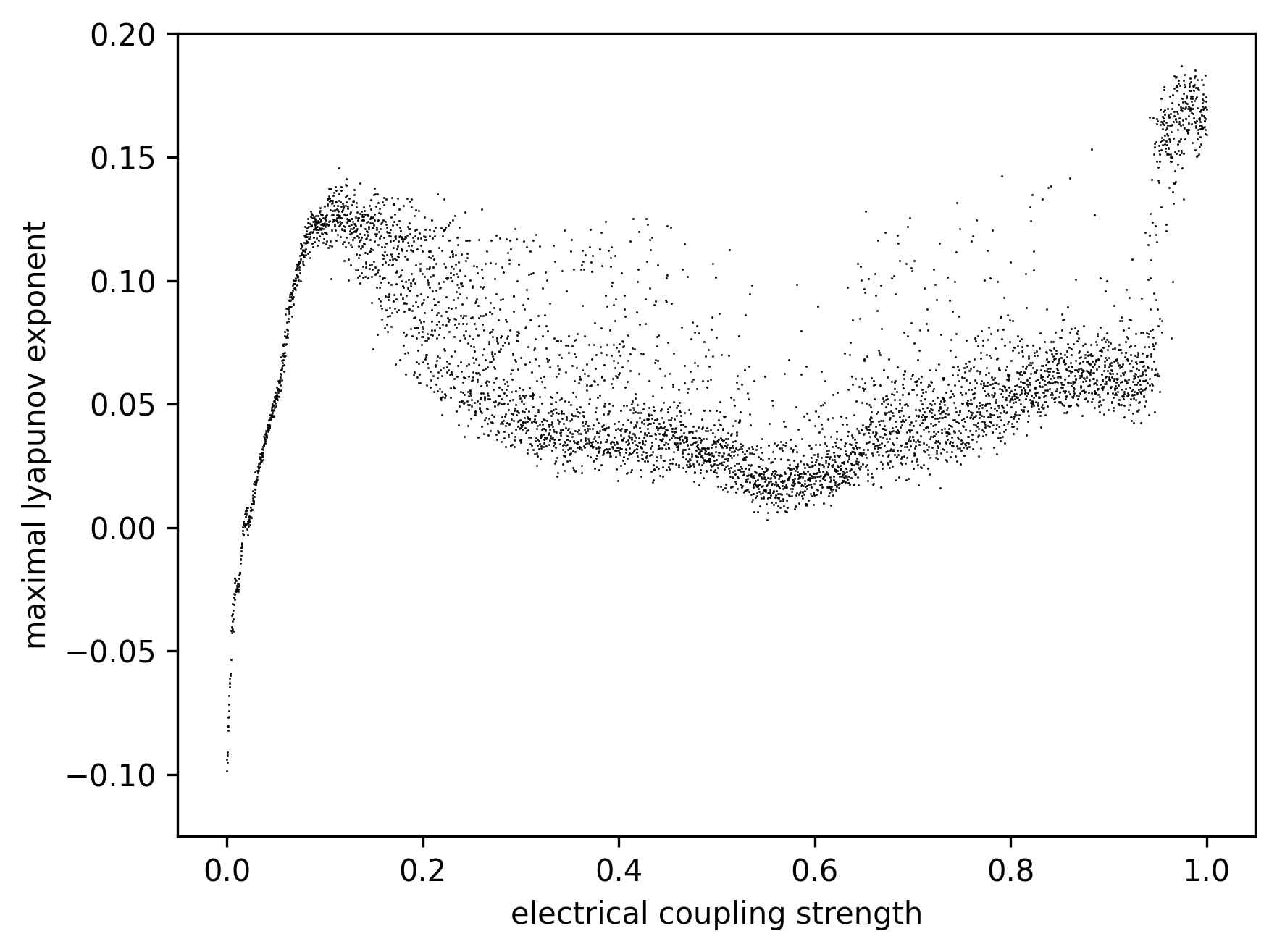}
    \caption{Graph of the maximal Lyapunov exponent $\lambda_1$ against the electrical coupling strength $g^e$ for a ring of $\zeta=30$ electrically coupled Rulkov 1 neurons, with $x_{i,\,0}\in (-1,\,1)$, $y_{i,\,0}=-3.25$, $\boldsymbol{\sigma} = -0.5\cdot\mathbf{1}$, and $\boldsymbol{\alpha} = 4.5\cdot\mathbf{1}$, visualized using the code in Appendix \ref{lyap-exp-and-dim-graphs-code}}
    \label{fig:max_lyap_exp_graph_random_x}
\end{figure*}

Now that all of this is established, let us begin exploring our first system. This system has $\zeta=30$ neurons with parameters $\boldsymbol{\sigma} = -0.5\cdot\mathbf{1}$ and $\boldsymbol{\alpha} = 4.5\cdot\mathbf{1}$. In other words, all the $\sigma_i$ values are $-0.5$ and all the $\alpha_i$ values are $4.5$. Additionally, we set the initial slow variable values for all of the neurons to be $y_{i,\,0} = -3.25$. However, setting the initial fast variable values for all of the neurons to be equal would be pointless because then the neurons would all be the same, resulting in no current flow between them. Instead, we will choose $x_{i,\,0}$ variables randomly from the interval $(-1,\,1)$. For the sake of reproducibility, we will use the initial state
\begin{equation}
    \begin{split}
        \mathbf{X}_0 &= \langle 0.68921784,\, -3.25,\, -0.94561073,\, -3.25,\, \\ 
         &\mathrel{\phantom{=}} -0.95674631,\, -3.25,\,  0.91870134,\, -3.25,\, \\
         &\mathrel{\phantom{=}} -0.32012381,\, -3.25,\, -0.23746836,\, -3.25,\, \\
         &\mathrel{\phantom{=}} -0.43906743,\, -3.25,\, -0.48671017,\, -3.25,\, \\
         &\mathrel{\phantom{=}} -0.37578533,\, -3.25,\, -0.00613823,\, -3.25,\, \\
         &\mathrel{\phantom{=}} 0.25990663,\, -3.25,\, -0.54103868,\, -3.25,\, \\
         &\mathrel{\phantom{=}} 0.12110471,\, -3.25,\,  0.71202085,\, -3.25,\, \\
         &\mathrel{\phantom{=}} 0.689336,\, -3.25,\,   -0.03260047,\, -3.25,\, \\
         &\mathrel{\phantom{=}} -0.90907325,\, -3.25,\,  0.93270227,\, -3.25,\, \\
         &\mathrel{\phantom{=}} 0.51953315,\, -3.25,\, -0.46783677,\, -3.25,\, \\
         &\mathrel{\phantom{=}} -0.96738424,\, -3.25,\, -0.50828432,\, -3.25,\, \\
         &\mathrel{\phantom{=}} -0.60388469,\, -3.25,\, -0.56644705,\, -3.25,\, \\
         &\mathrel{\phantom{=}} -0.42772621,\, -3.25,\,  0.7716625,\, -3.25,\, \\
         &\mathrel{\phantom{=}} -0.60336517,\, -3.25,\,  0.88158364,\, -3.25,\, \\
         &\mathrel{\phantom{=}} 0.0269842,\, -3.25,\,   0.42512831,\, -3.25 \rangle
    \end{split}
    \label{eq:big-initial-state}
\end{equation}
for all our analysis of this particular system. Using the code in Appendix \ref{ring-lattice-code}, we can generate an orbit of this initial state $O(\mathbf{X}_0)$ and approximate the system's Lyapunov spectrum using this orbit. In Figure \ref{fig:random_x_graphs}, we graph the first thousand iterations of the fast variable orbits of the first eight Rulkov 1 neurons in the ring. We start with uncoupled neurons $g^e=0$ in Figure \ref{fig:random_x_ge0}, where we can see these neurons with identical parameters are all out of phase in the non-chaotic spiking domain. As expected, because there is no current flow and all of the individual Rulkov neurons are spiking regularly, the maximal Lyapunov exponent $\lambda_1$ is negative. When we up the electrical coupling strength to $g^e=0.05$, in Figure \ref{fig:random_x_ge0.05}, the neurons still spike relatively periodically, although there are some irregularities when one voltage happens to catch onto another. This small amount of coupling conductance is enough to bump the system into chaos with a positive maximal Lyapunov exponent. Next, we bring the coupling strength up significantly to $g^e=0.25$, where we can see in Figure \ref{fig:random_x_ge0.25} that the ring system now has synchronized chaotic bursts. This is reminiscent of the positive coupling of two symmetrically coupled chaotic bursting Rulkov 1 neurons in Section \ref{two-electrically-coupled-rulkov-1-neurons} (Figure \ref{fig:sym_coup_rulkov_1_ge0.05}), where the bursts happen (mostly) in sync with each other but the spikes within the bursts are still chaotic and unsynchronized. Finally, we take the coupling strength to the extreme, with $g^e=1$ in Figure \ref{fig:random_x_ge1}, where complete chaos ensues due to each Rulkov neuron having a tremendous influence over its nearest neighbors. Here, the maximal Lyapunov exponent jumps up to $\lambda_1=0.1694$, the highest Lyapunov exponent we have seen in this paper thus far.

A natural question to ask is how the maximal Lyapunov exponent varies as we vary $g^e$. This can be answered using the code in Appendix \ref{lyap-exp-and-dim-graphs-code}, which creates a graph of the maximal Lyapunov exponent of a ring lattice system against its electrical coupling strength. For this first system, the results of this code are displayed in Figure \ref{fig:max_lyap_exp_graph_random_x}. Our first observation is that the maximal Lyapunov exponents are rather erratic for higher values of $g^e$, covering a wide range of values for similar values of $g^e$. However, there are some trends we can notice. Because the individual neurons of this system are non-chaotic, $\lambda_1$ values initially start below zero. As the neurons start to affect each other, we reach the range of chaotic spiking, where the $\lambda_1$ values quickly become positive and reach a maximum. Then, as we enter the synchronized bursting regime, the $\lambda_1$ values become much more varied and start a generally downward trend, which can be attributed to the silence between chaotic bursts of spikes being non-chaotic and lowering the Lyapunov exponents. As we reach the extreme values of $g^e$ towards the right side of the graph, the $\lambda_1$ values shoot up to extremely high and chaotic values.

\begin{figure*}[hp!]
    \centering
    \begin{subfigure}{0.475\textwidth}
        \centering
        \includegraphics[scale=0.08]{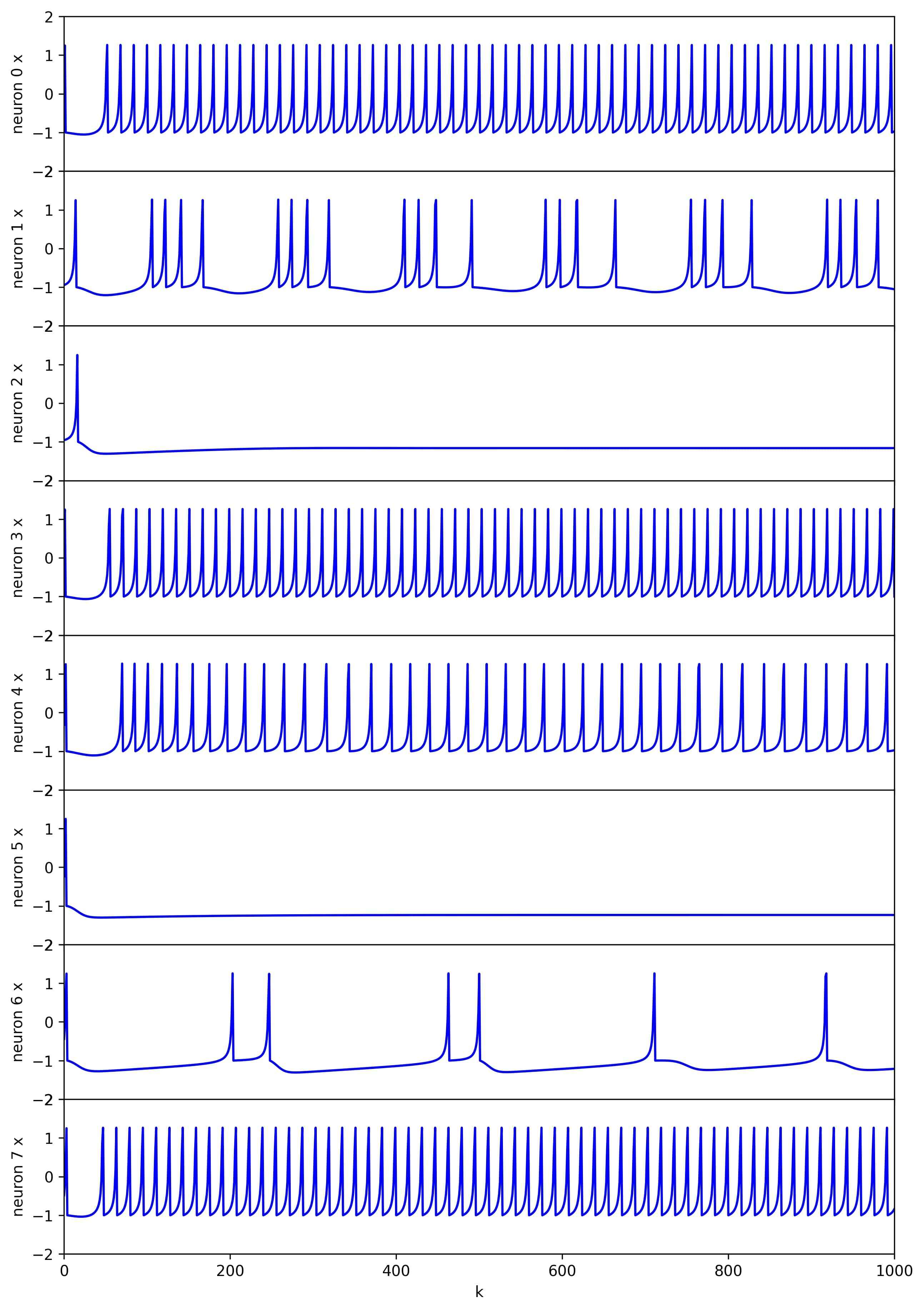}
        \vspace{-0.2cm}
        \caption{$g^e=0$, $\lambda_1\approx 0.0644$}
        \label{fig:random_sigma_ge0}
        \vspace{0.2cm}
    \end{subfigure}
    \begin{subfigure}{0.475\textwidth}
        \centering
        \includegraphics[scale=0.08]{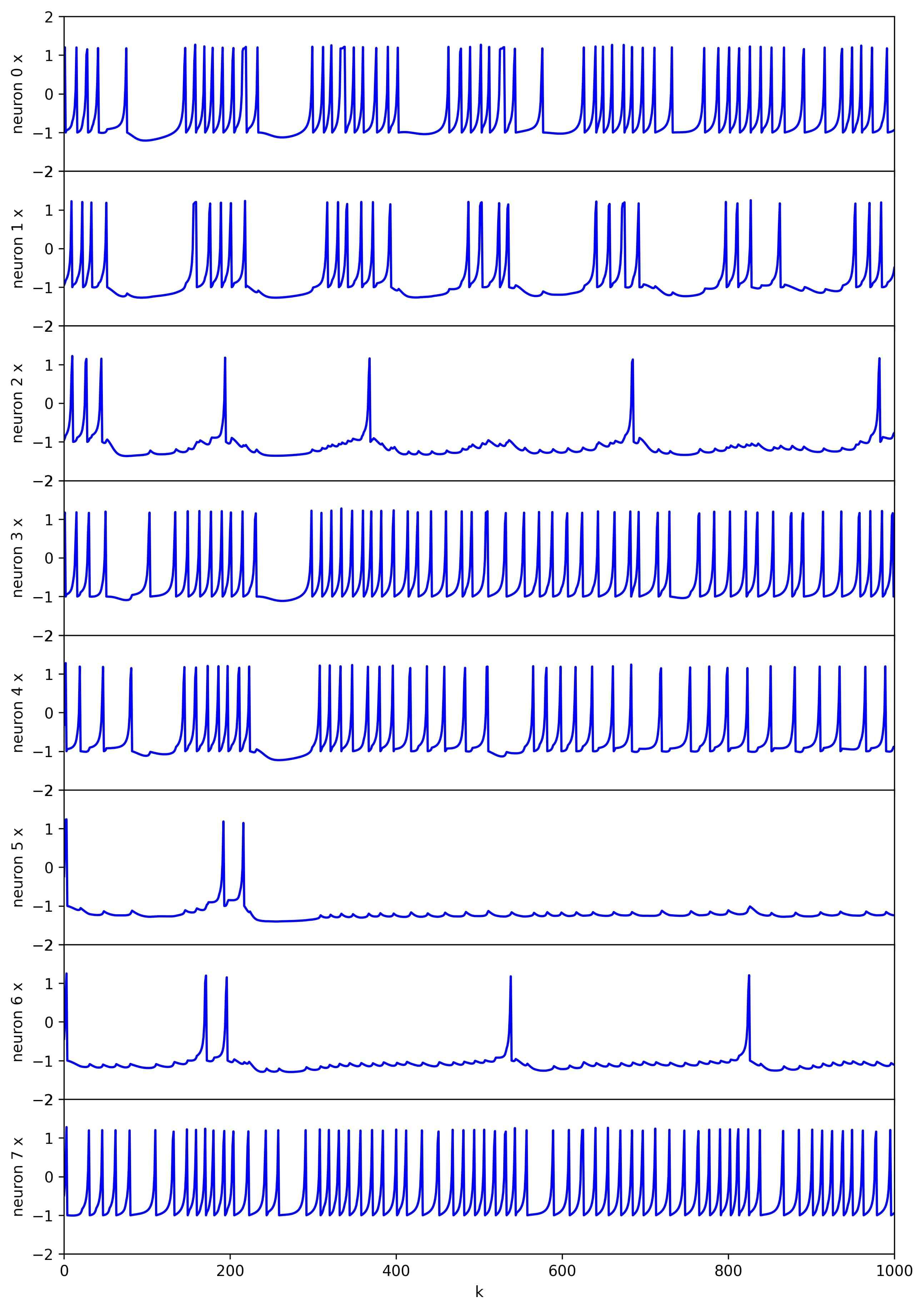}
        \vspace{-0.2cm}
        \caption{$g^e=0.05$, $\lambda_1\approx 0.0686$}
        \label{fig:random_sigma_ge0.05}
        \vspace{0.2cm}
    \end{subfigure}
    \begin{subfigure}{0.475\textwidth}
        \centering
        \includegraphics[scale=0.08]{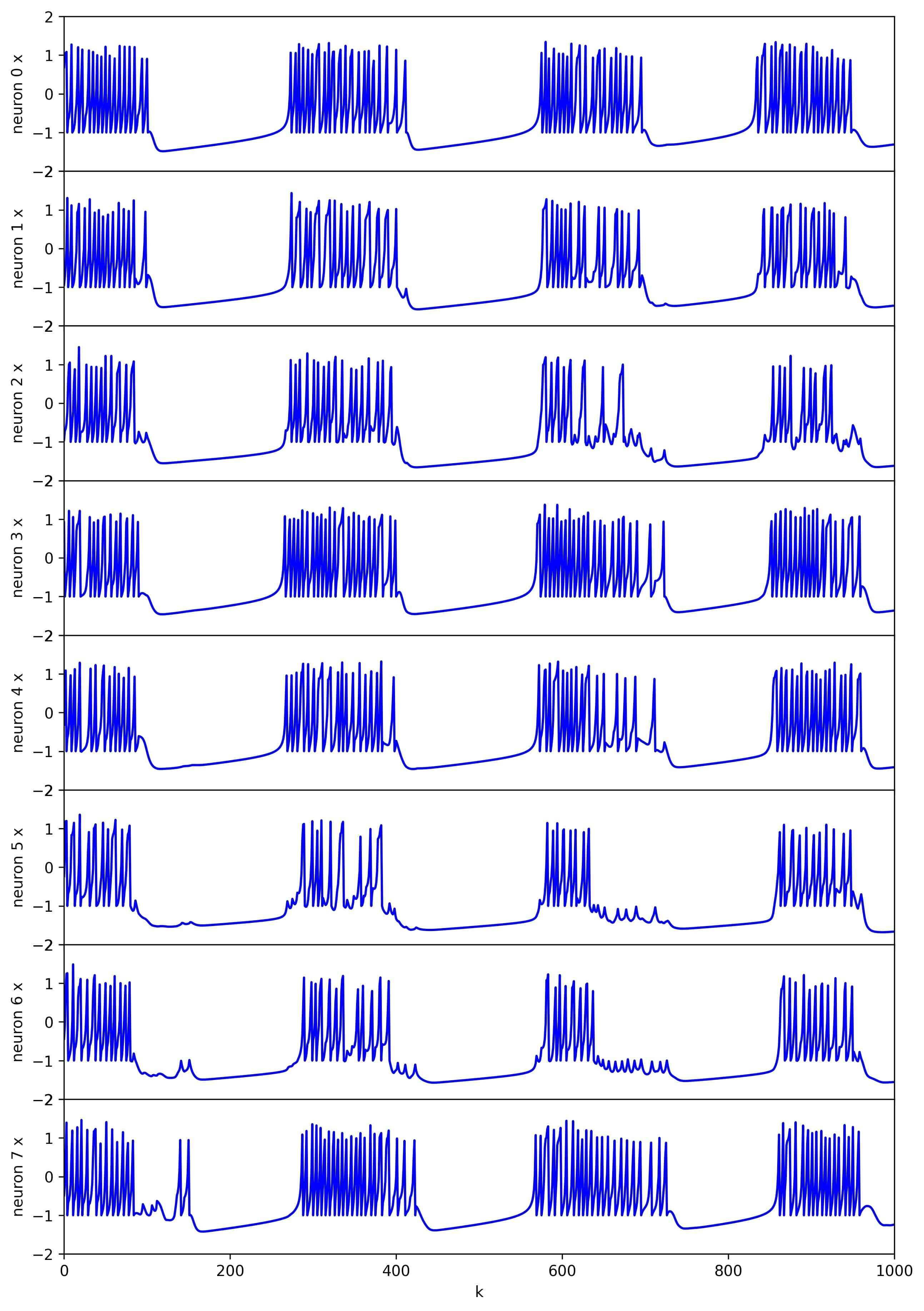}
        \vspace{-0.2cm}
        \caption{$g^e=0.25$, $\lambda_1\approx 0.0663$}
        \label{fig:random_sigma_ge0.25}
    \end{subfigure}
    \begin{subfigure}{0.475\textwidth}
        \centering
        \includegraphics[scale=0.08]{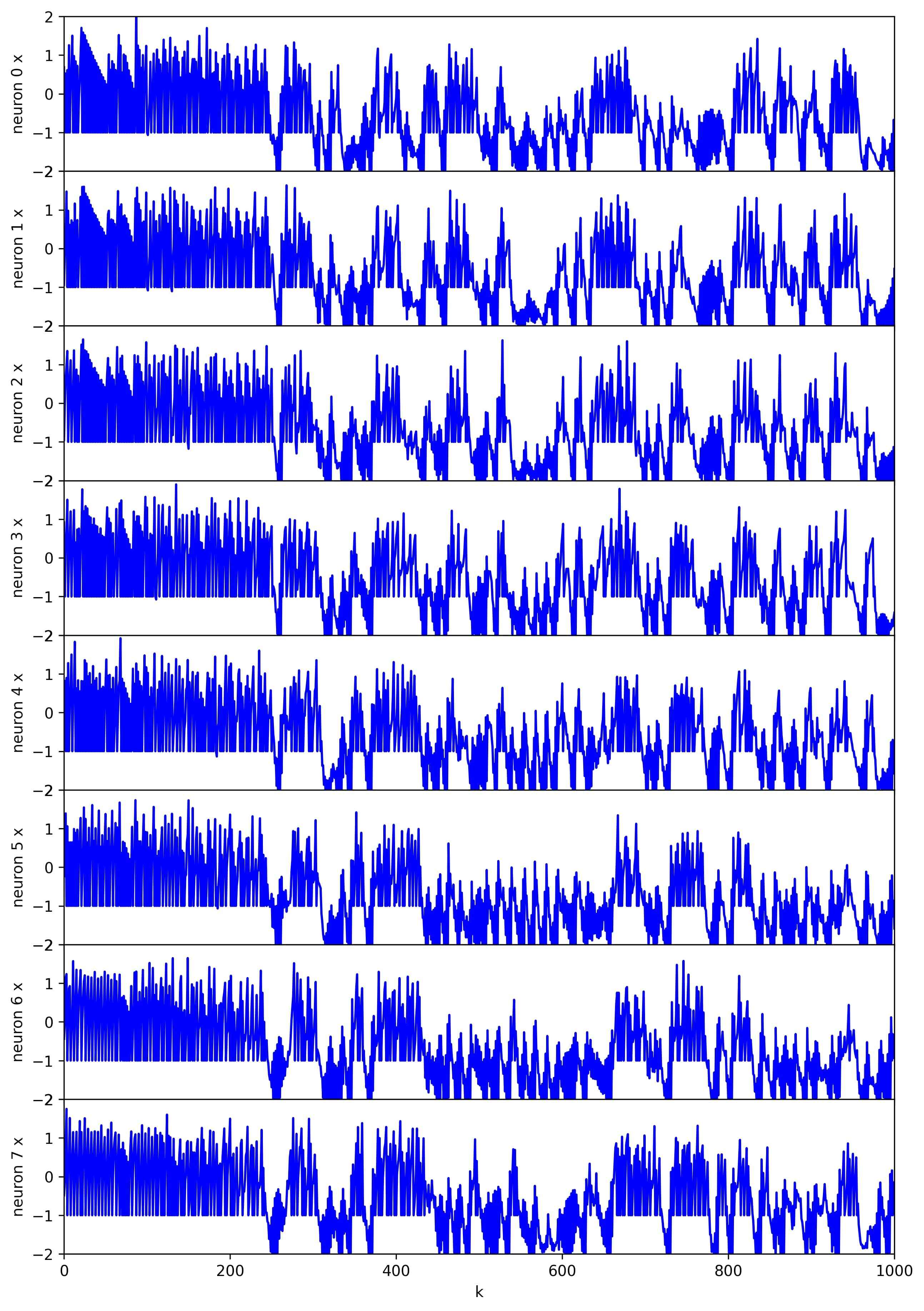}
        \vspace{-0.2cm}
        \caption{$g^e=1$, $\lambda_1\approx 0.2003$}
        \label{fig:random_sigma_ge1}
    \end{subfigure}
    \caption{Graphs of $x_{i,\,k}$ for eight neurons in a ring of $\zeta=30$ electrically coupled Rulkov 1 neurons, with $x_{i,\,0}\in (-1,\,1)$, $y_{i,\,0}=-3.25$, $\sigma_i\in(-1.5,\,-0.5)$, and $\boldsymbol{\alpha} = 4.5\cdot\mathbf{1}$, visualized and $\lambda_1$ values calculated using the code in Appendix \ref{ring-lattice-code}}
    \label{fig:random_sigma_graphs}
\end{figure*}

\begin{figure*}[hp!]
    \centering
    \begin{subfigure}{0.475\textwidth}
        \centering
        \includegraphics[scale=0.08]{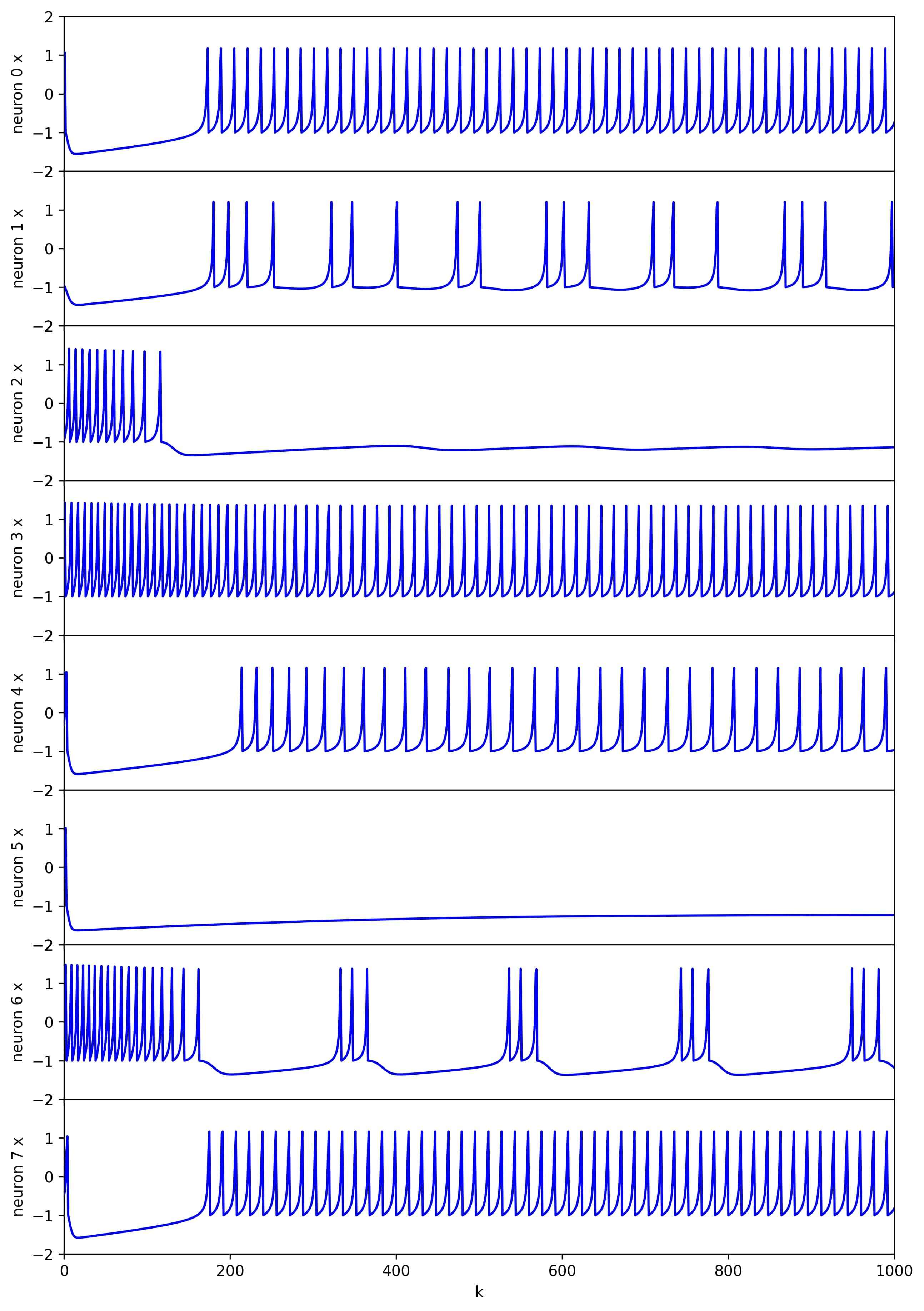}
        \vspace{-0.2cm}
        \caption{$g^e=0$, $\lambda_1\approx 0.0469$}
        \label{fig:random_alpha_ge0}
        \vspace{0.2cm}
    \end{subfigure}
    \begin{subfigure}{0.475\textwidth}
        \centering
        \includegraphics[scale=0.08]{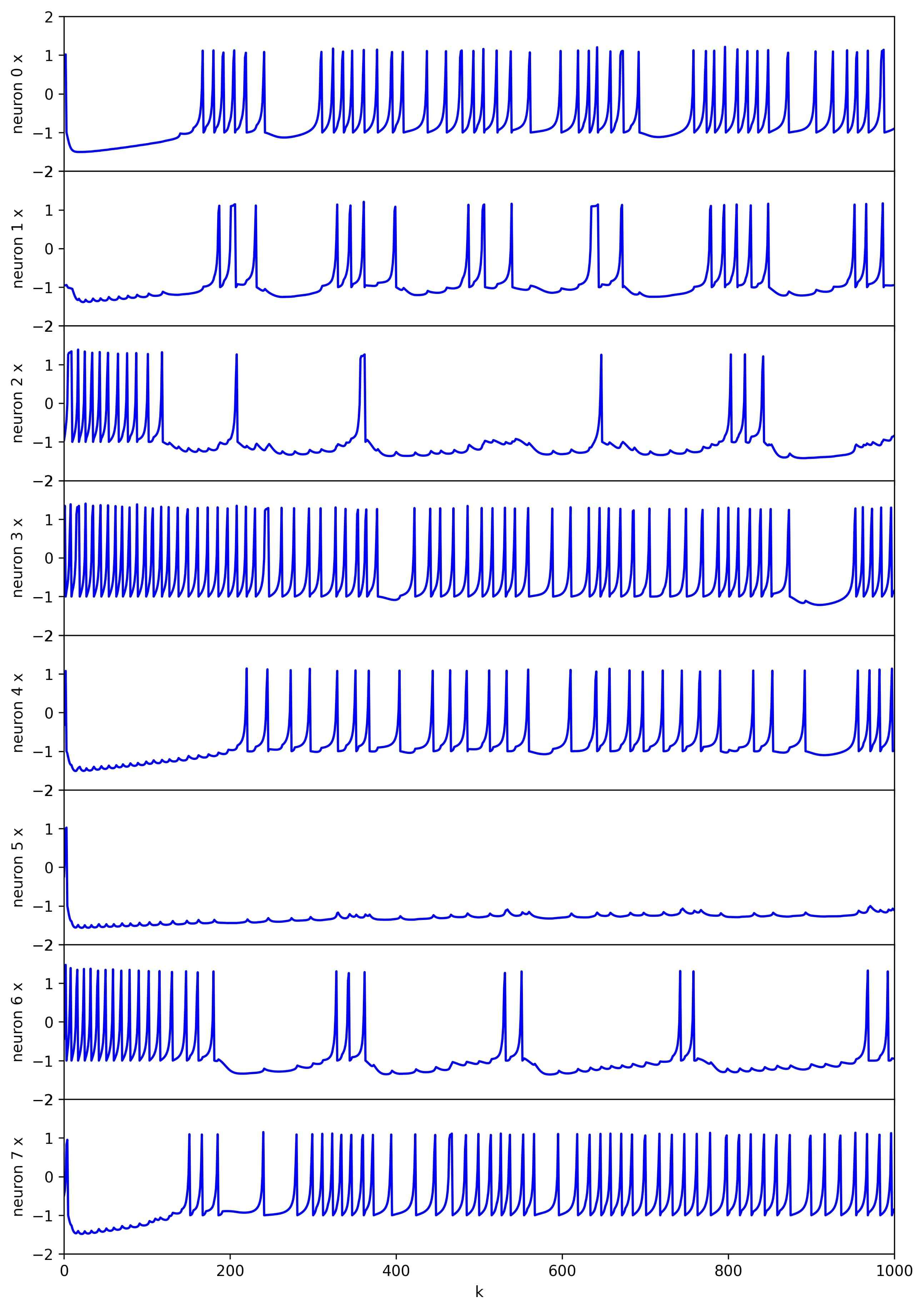}
        \vspace{-0.2cm}
        \caption{$g^e=0.05$, $\lambda_1\approx 0.0563$}
        \label{fig:random_alpha_ge0.05}
        \vspace{0.2cm}
    \end{subfigure}
    \begin{subfigure}{0.475\textwidth}
        \centering
        \includegraphics[scale=0.08]{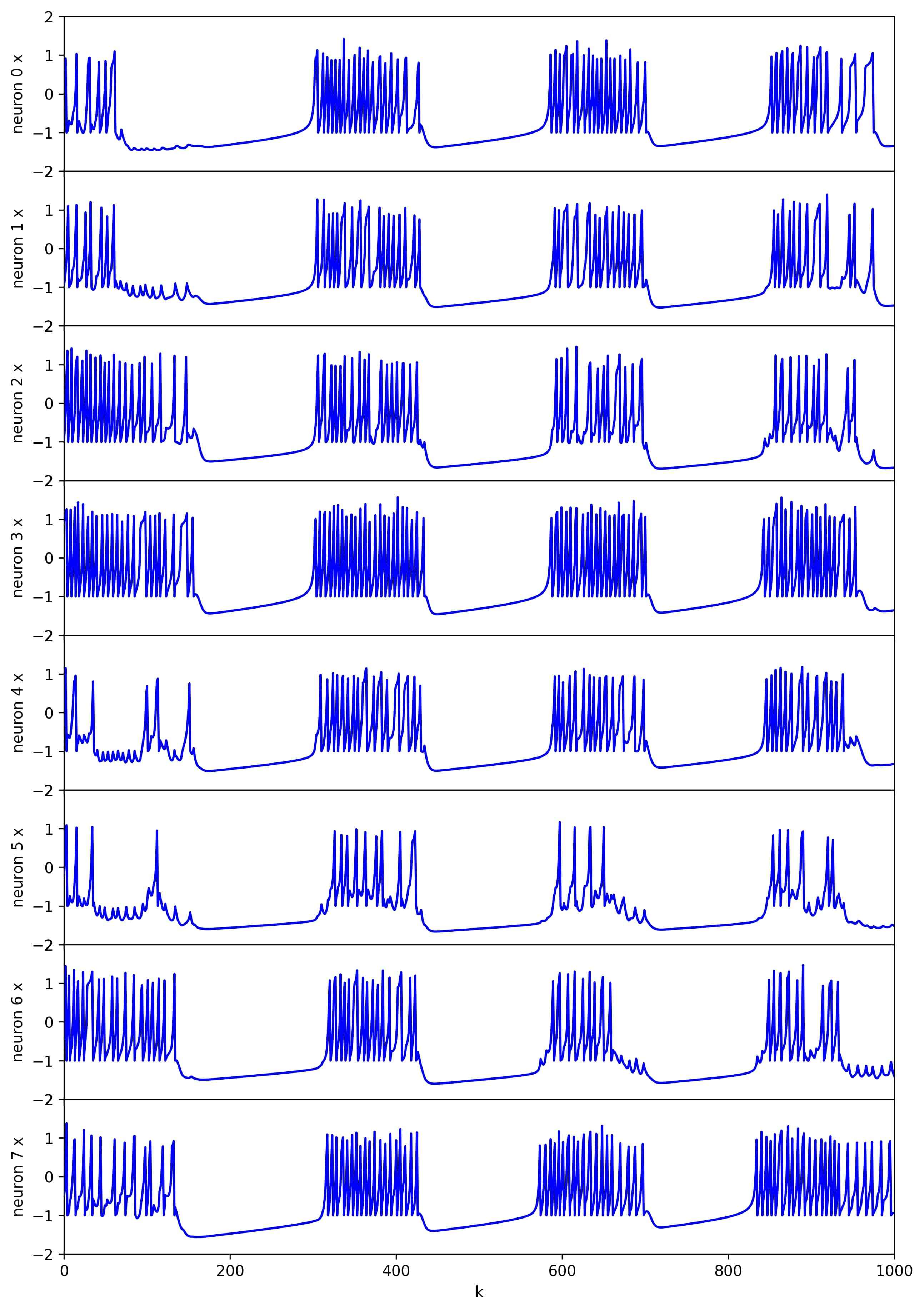}
        \vspace{-0.2cm}
        \caption{$g^e=0.25$, $\lambda_1\approx 0.0633$}
        \label{fig:random_alpha_ge0.25}
    \end{subfigure}
    \begin{subfigure}{0.475\textwidth}
        \centering
        \includegraphics[scale=0.08]{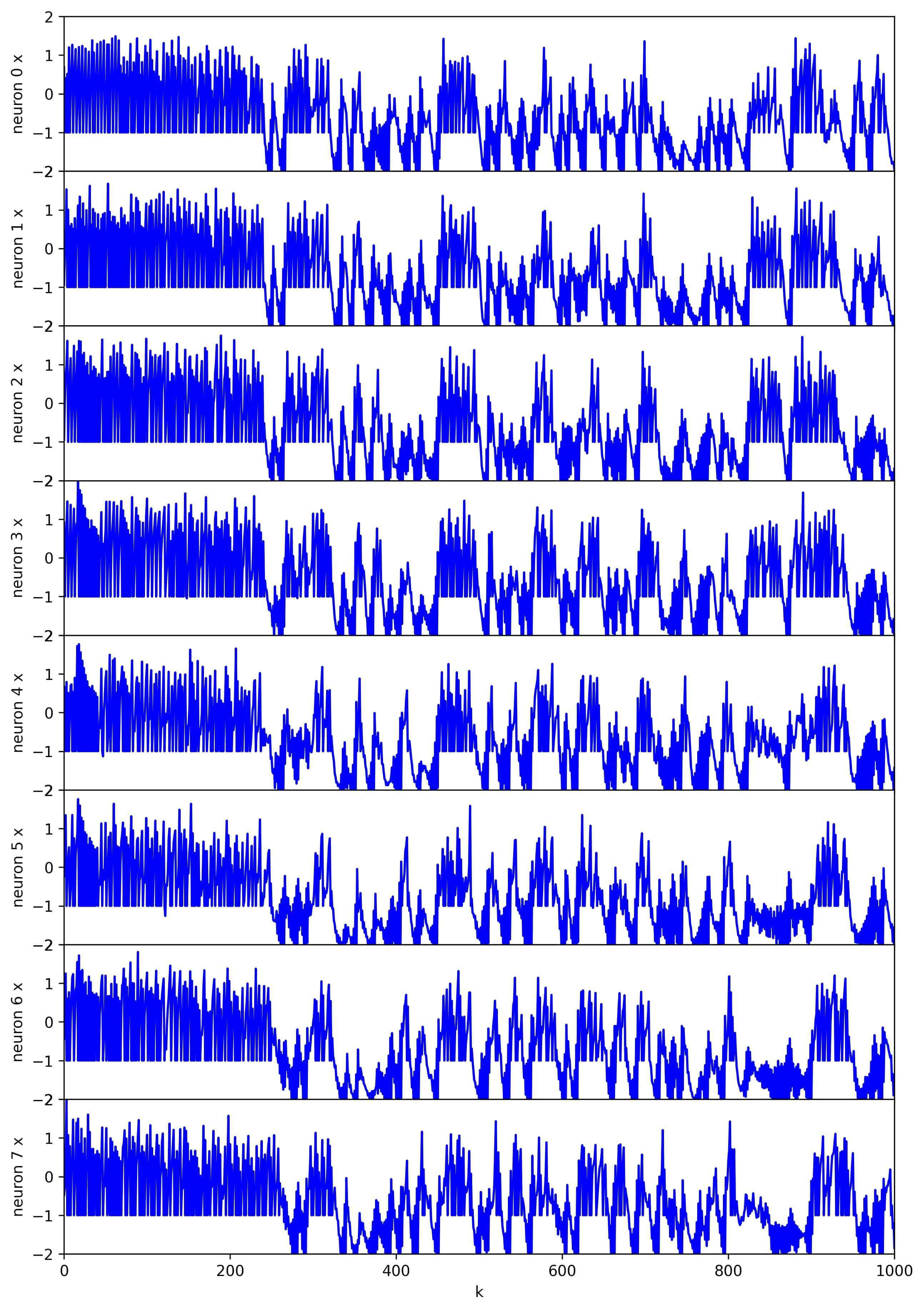}
        \vspace{-0.2cm}
        \caption{$g^e=1$, $\lambda_1\approx 0.2053$}
        \label{fig:random_alpha_ge1}
    \end{subfigure}
    \caption{Graphs of $x_{i,\,k}$ for eight neurons in a ring of $\zeta=30$ electrically coupled Rulkov 1 neurons, with $x_{i,\,0}\in (-1,\,1)$, $y_{i,\,0}=-3.25$, $\sigma_i\in(-1.5,\,-0.5)$, and $\alpha_i\in (4.25,\,4.75)$, visualized and $\lambda_1$ values calculated using the code in Appendix \ref{ring-lattice-code}}
    \label{fig:random_alpha_graphs}
\end{figure*}

We will now examine our second and third Rulkov 1 neuron ring lattice systems, where we have different parameters between different neurons in the ring. The second system we will examine keeps the same randomly distributed $x_{i,\,0}$ values (Equation \ref{eq:big-initial-state}), the same $y_{i,\,0}=-3.25$ values, and the same $\boldsymbol{\alpha} = 4.5\cdot\mathbf{1}$ vector, but we randomly choose $\sigma_i$ values from the interval $(-1.5,\,-0.5)$. Referring back to the bifurcation diagram of Rulkov map 1 in Figure \ref{fig:rulkov-1-bifurc-diag-param-space} from Section \ref{bifurcation-analysis-rulkov-map-1}, the set $\sigma\in(-1.5,\,-0.5)\:\cap\: \alpha=4.5$ gives us individual neurons in silence, bursting, and spiking domains, which can be seen in the uncoupled neuron system (Figure \ref{fig:random_sigma_ge0}). The specific random $\boldsymbol{\sigma}$ vector that we use in this paper is
\begin{equation}
    \begin{split}
        \boldsymbol{\sigma} &= \langle -0.63903048,\, -0.87244087,\, -1.16110093,\, \\
        &\mathrel{\phantom{=}} -0.63908737,\, -0.73103576,\, -1.23516699,\, \\
        &\mathrel{\phantom{=}} -1.09564519,\, -0.57564289,\, -0.75055299,\, \\
        &\mathrel{\phantom{=}} -1.01278976,\, -0.61265545,\, -0.75514189,\, \\
        &\mathrel{\phantom{=}} -0.89922568,\, -1.24012127,\, -0.87605023,\, \\
        &\mathrel{\phantom{=}} -0.94846269,\, -0.78963971,\, -0.94874874,\, \\
        &\mathrel{\phantom{=}} -1.31858036,\, -1.34727902,\, -0.7076453,\, \\
        &\mathrel{\phantom{=}} -1.10631486,\, -1.33635792,\, -1.48435264,\, \\
        &\mathrel{\phantom{=}} -0.76176103,\, -1.17618267,\, -1.10236959,\, \\
        &\mathrel{\phantom{=}} -0.66159308,\, -1.27849639,\, -0.9145025 \rangle
    \end{split}
    \label{eq:big-sigma-vector}
\end{equation}
Finally, the third system we simulate is one where we keep the randomly distributed $x_{i,\,0}$ and $\sigma_i$ values, keep $y_{i,\,0}=-3.25$, but randomly choose $\alpha_i$ values from the interval $(4.25,\,4.75)$. This even further varies the distribution of different possible behaviors between different neurons in the system. This can be seen in the uncoupled neuron system (Figure \ref{fig:random_alpha_ge0}), with some neurons exhibiting very rapid spiking. The specific random $\boldsymbol{\alpha}$ vector we use in this paper is
\begin{equation}
    \begin{split}
        \boldsymbol{\alpha} &= \langle 4.31338267,\, 4.3882788,\,  4.6578449,\, \\
        &\mathrel{\phantom{=}} 4.67308374,\, 4.28873181,\, 4.26278301,\, \\
        &\mathrel{\phantom{=}} 4.73065817,\, 4.29330435,\, 4.44416548,\, \\
        &\mathrel{\phantom{=}} 4.66625973,\, 4.26243104,\, 4.65881579,\, \\
        &\mathrel{\phantom{=}} 4.68086764,\, 4.44092086,\, 4.49639124,\, \\
        &\mathrel{\phantom{=}} 4.55500032,\, 4.33389054,\, 4.38869161,\, \\
        &\mathrel{\phantom{=}} 4.57278526,\, 4.62717616,\, 4.62025928,\, \\
        &\mathrel{\phantom{=}} 4.49780551,\, 4.46750298,\, 4.49561326,\, \\
        &\mathrel{\phantom{=}} 4.66902393,\, 4.60858869,\, 4.6027906,\, \\
        &\mathrel{\phantom{=}} 4.40563641,\, 4.54198743,\, 4.49388045 \rangle
    \end{split}
    \label{eq:big-alpha-vector}
\end{equation}
In Figures \ref{fig:random_sigma_graphs} and \ref{fig:random_alpha_graphs}, we graph the first thousand iterations of the fast variable orbits of the first eight neurons in the ring for the same electrical coupling strength values as the first system: $g^e=0$, 0.05, 0.25, and 1. Comparing both of these systems to the first system, we can see similar patterns emerging. For $g^e=0.05$, the adjacent neurons start to have some effect on each other, but the overall picture of the dynamics stays the same. Raising the electrical coupling strength up to $g^e=0.25$, the neurons all undergo synchronized chaotic bursting, and going to the extreme $g^e=1$, complete chaos ensues. One of the interesting things that is even clearer in these visualizations is the direct effect that neurons have on their adjacent partners. Specifically, we can see that spiking in one neuron is reflected in adjacent neurons through smaller spikes during a time of silence. 

\begin{figure*}[hp!]
    \centering
    \begin{subfigure}{0.9\textwidth}
        \centering
        \includegraphics[scale=0.21]{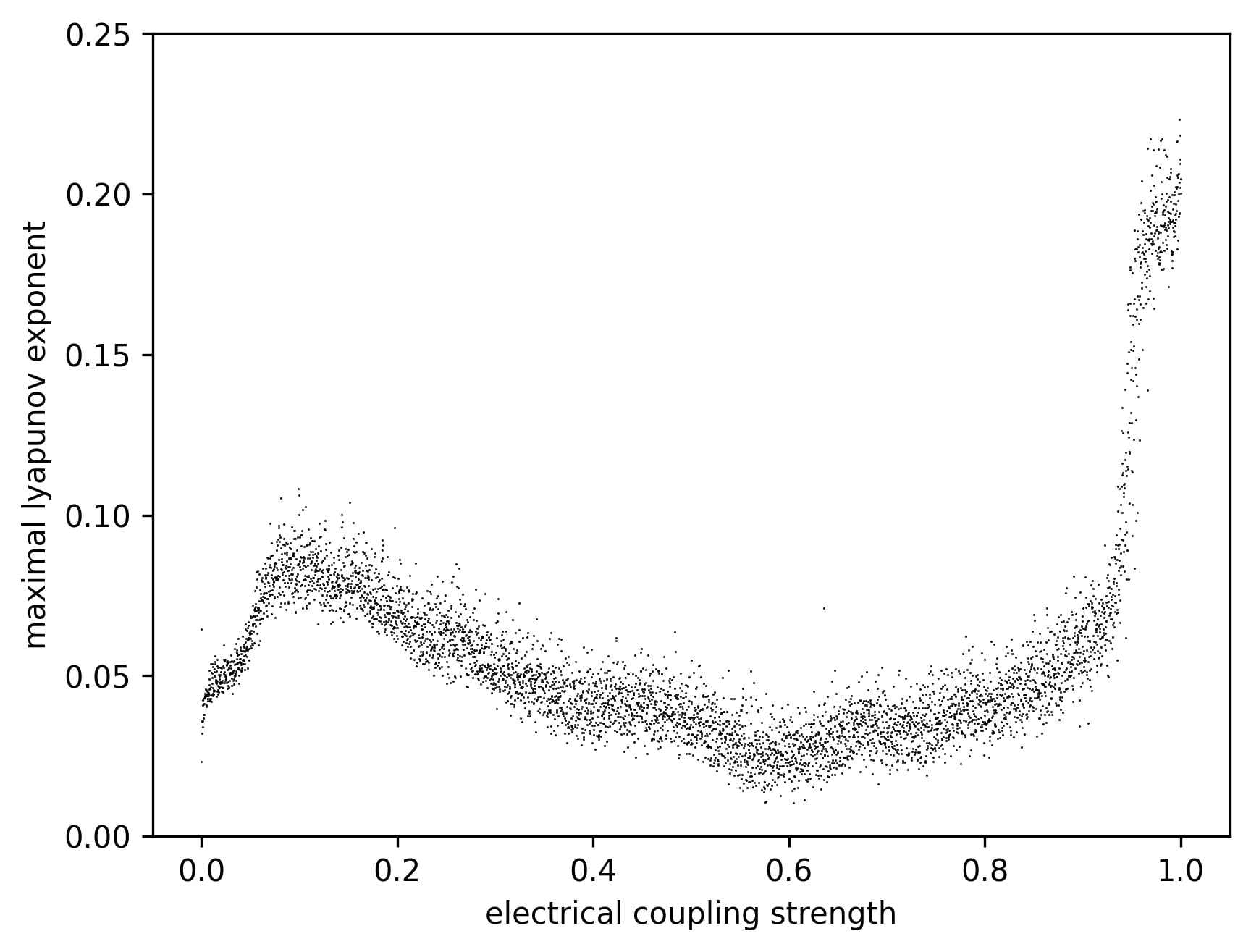}
        \vspace{-0.25cm}
        \caption{$x_{i,\,0}\in (-1,\,1)$, $y_{i,\,0}=-3.25$, $\sigma_i\in(-1.5,\,-0.5)$, $\boldsymbol{\alpha} = 4.5\cdot\mathbf{1}$}
        \label{fig:ring-max-lyap-random-sigma}
        \vspace{0.5cm}
    \end{subfigure}
    \begin{subfigure}{0.9\textwidth}
        \centering
        \includegraphics[scale=0.21]{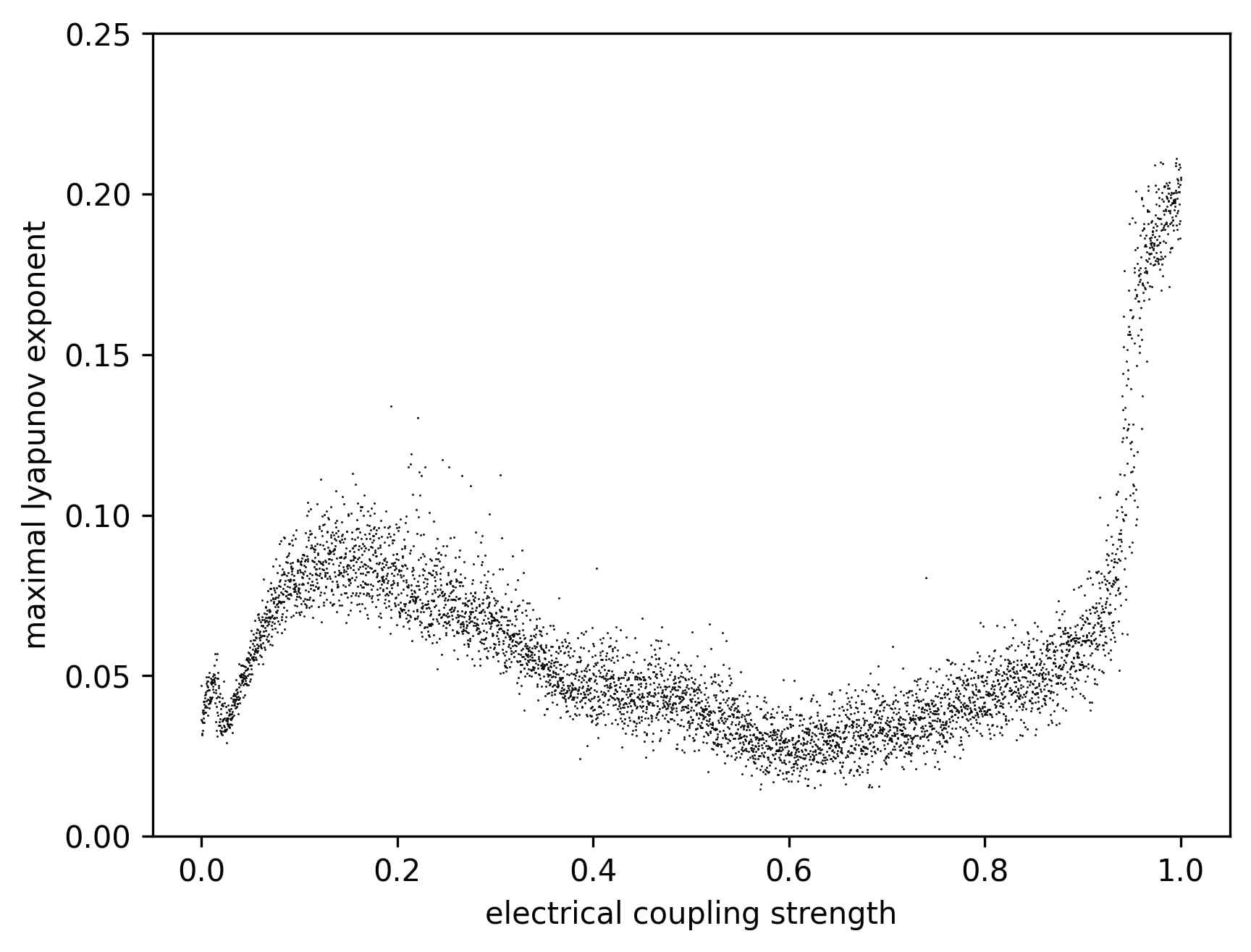}
        \vspace{-0.25cm}
        \caption{$x_{i,\,0}\in (-1,\,1)$, $y_{i,\,0}=-3.25$, $\sigma_i\in(-1.5,\,-0.5)$, $\alpha_i\in(4.25,\,4.75)$}
        \label{fig:ring-max-lyap-random-alpha}
        \vspace{0.5cm}
    \end{subfigure}
    \caption{Graphs of the maximal Lyapunov exponent $\lambda_1$ against the electrical coupling strength $g^e$ for our second and third ring lattice systems, visualized using the code in Appendix \ref{lyap-exp-and-dim-graphs-code}}
    \label{fig:ring-max-lyap-graphs-rand-sigandalph}
\end{figure*}

In Figure \ref{fig:ring-max-lyap-graphs-rand-sigandalph}, we generalize and visualize the maximal Lyapunov exponents of these two systems for many values of $g^e$. An immediate difference we notice when comparing these graphs to the graph in Figure \ref{fig:max_lyap_exp_graph_random_x} is that all of $\lambda_1$ values remain above 0 since some of the individual neurons in these varied parameter systems are chaotic. However, the graphs of maximal Lyapunov exponents of all three of our systems have similar shapes, the major differences being in the beginning where the neurons are weakly coupled and operating under their own parameters. Past this weak coupling domain, all three graphs in Figures \ref{fig:max_lyap_exp_graph_random_x} and \ref{fig:ring-max-lyap-graphs-rand-sigandalph} follow the increase up to chaotic spiking, the swoop down as synchronized chaotic bursts occurring, and the shoot up as we approach the extreme values of $g^e$. This shows that despite making individual neurons more and more different from their neighbors, coupling all of them together makes the systems exhibit similar behaviors and chaotic dynamics.

Although these systems do exhibit multistability in the sense that some $\mathbf{X}_0$ states get attracted to a non-chaotic attractor and some get attracted to a chaotic attractor (like what we saw in the asymmetrically coupled system in the previous section), the multistability exhibited here isn't interesting because, practically, the only states that get attracted to a non-chaotic attractor are ones where all the neurons in the system have the same parameters and the same (or very close to the same) initial states. This is because, with $\zeta=30$ neurons, it is nearly impossible for all the neurons to sync up by chance like it does when the system has only two neurons. For this reason, in Section \ref{ring-lattice-geometry}, we will not be examining complex geometries that come out of multistability, namely, basins and fractal basin boundaries; instead, we will be examining the fractal structure of these systems' chaotic attractors.

\section{Geometrical Analysis of Rulkov Neuron Systems}
\label{geometrical-analysis-of-rulkov-neuron-systems}

In this section, we combine all of the research outlined in this paper, applying our theoretical and computational methods for quantifying the chaos emerging from the complex geometrical structure of attractors and basins (Section \ref{geometry_dynamics_chaos}) to the chaotic neuron systems from our discussion of the Rulkov maps (Sections \ref{rulkov-maps}, \ref{injection-of-current}, \ref{coupling-of-rulkov-neurons}). We will begin by analyzing the fractal geometry of chaotic spiking and chaotic bursting attractors of a single Rulkov 2 neuron. Next, we will perform an in-depth analysis of the attractors, basins, and basin boundaries of a system composed of two asymmetrically electrically coupled Rulkov 1 neurons, which we discover exhibits some very interesting and unexpected geometrical properties. In doing so, we define the concept of a chaotic pseudo-attractor, develop visualizations of higher-dimensional basins, and analyze lower-dimensional slices of basins and basin boundaries. Last, we will conclude by calculating and analyzing the Lyapunov dimensions of the chaotic attractors generated by three systems of $\zeta=30$ electrically coupled Rulkov 1 neurons in a ring lattice. By the Kaplan-Yorke conjecture, these Lyapunov dimensions give a rough approximation of the fractal dimensions of these attractors that live in $60$-dimensional space.

\begin{figure*}[htp!]
    \centering
    \begin{subfigure}{0.9\textwidth}
        \centering
        \hspace{-1.5cm}
        \includegraphics[scale=0.21]{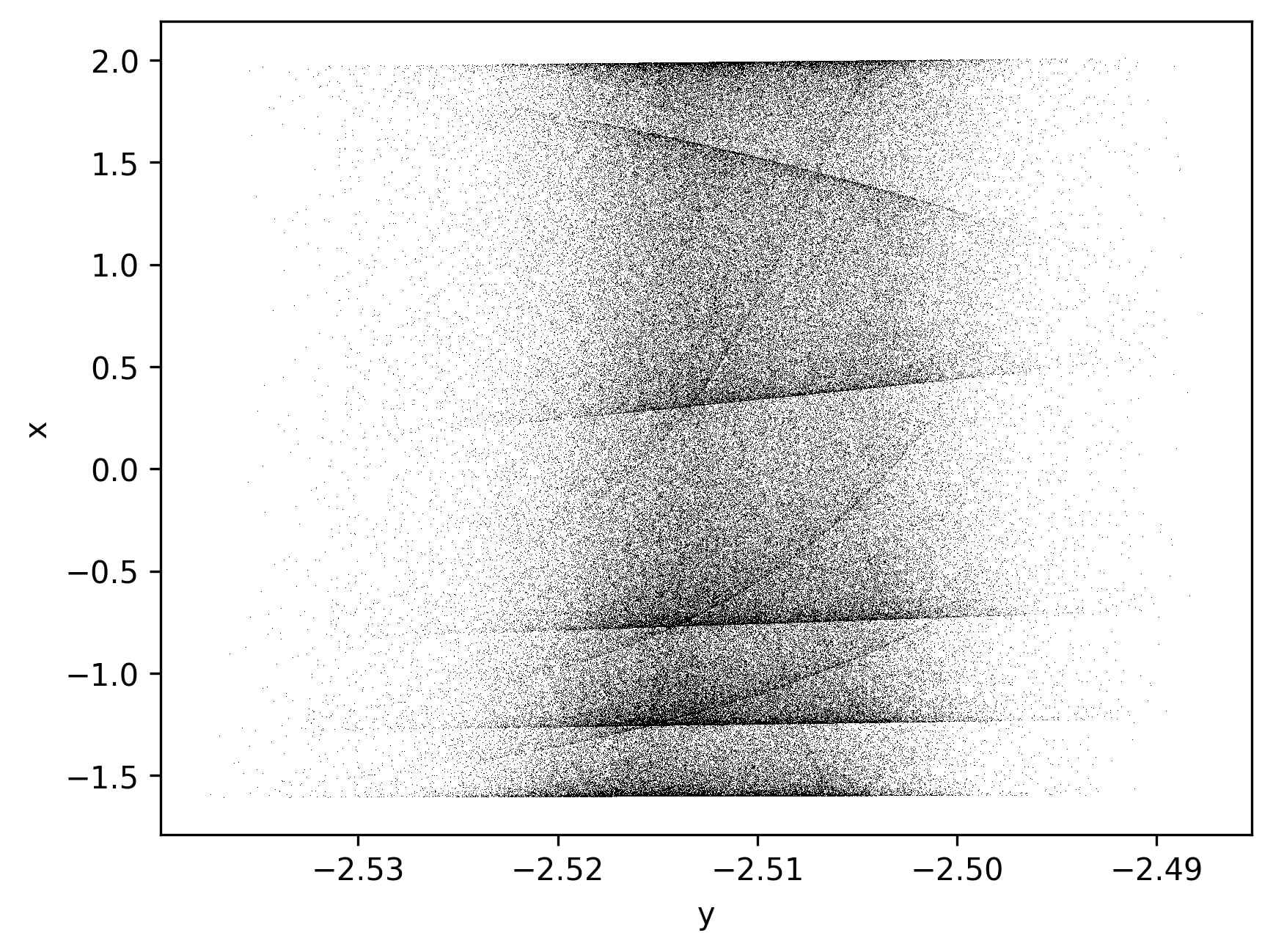}
        \caption{$\alpha=4.5$, $\sigma=0$}
        \label{fig:rulkov-2-spiking-attractor}
        \vspace{0.5cm}
    \end{subfigure}
    \begin{subfigure}{0.9\textwidth}
        \centering
        \hspace{-1.5cm}
        \includegraphics[scale=0.21]{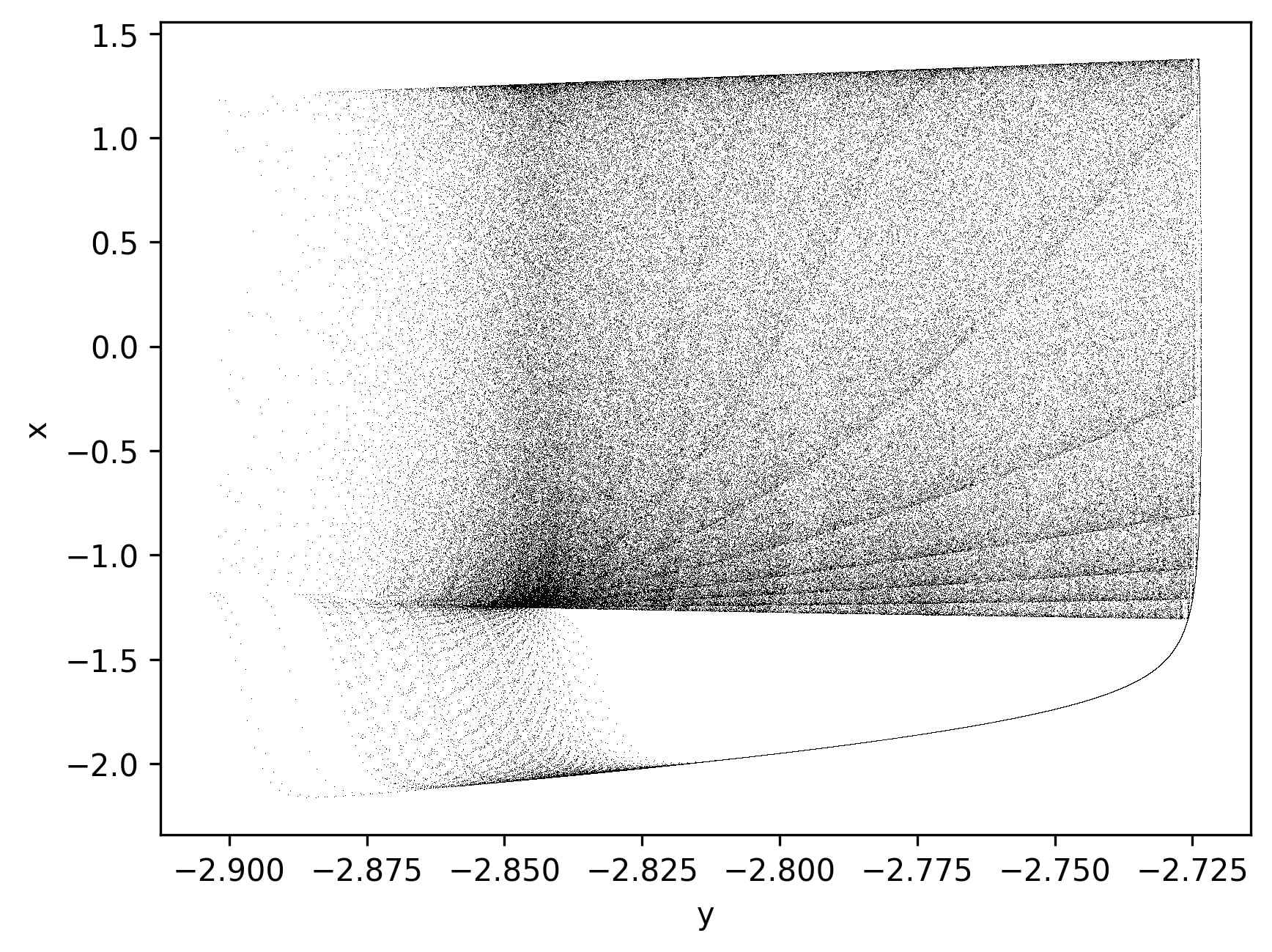}
        \caption{$\alpha=4.1$, $\sigma=-0.5$}
        \label{fig:rulkov-2-bursting-attractor}
    \end{subfigure}
    \caption{Spiking and bursting attractors of systems of one Rulkov 2 neuron in two-dimensional state space $\langle y,\,x\rangle$, visualized using the code in Appendix \ref{rulkov_2_graphs_and_cobweb_code}}
    \label{fig:rulkov-2-attractor-graphs}
\end{figure*}

\subsection{One Rulkov 2 Neuron}
\label{one-rulkov-2-neuron}

Recall from Section \ref{chaotic-dynamics-rulkov-map-1} that the attractor of a single Rulkov 1 neuron is not fractal because the resetting mechanism of the map brings states that deviate from each other back to the same value. However, the chaotic Rulkov map 2 doesn't have a built-in resetting mechanism; it isn't piecewise. Therefore, we suspect that the chaotic attractors of Rulkov map 2 are indeed fractal. In this section, we will examine a Rulkov 2 neuron that exhibits chaotic spiking ($\alpha=4.5$ and $\sigma=0$, shown in Figure \ref{fig:rulkov_2_spiking}) and a Rulkov 2 neuron that exhibits chaotic bursting ($\alpha=4.1$ and $\sigma=-0.5$, shown in Figure \ref{fig:rulkov_2_bursting}). The attractors generated by these systems in two-dimensional state space $\langle y,\,x\rangle$ are visualized in Figure \ref{fig:rulkov-2-attractor-graphs}, where they certainly look complex and fractal-like. Before we analyze the attractors though, one interesting thing to note about the bursting attractor in Figure \ref{fig:rulkov-2-bursting-attractor} is its similarity to the bursting attractor of Rulkov map 1. Comparing the chaotic bursting attractor in Figure \ref{fig:rulkov-2-bursting-attractor} to the diagram of the state space of Rulkov map 1 in Figure \ref{fig:rulkov_1_state_space_diagram_alpha6}, we can see the high $x$ value spiking in both, chaotic in Rulkov map 2 and non-chaotic in Rulkov map 1, followed by a sudden attraction to the stable branch $B_{\text{stable}}$, a slow movement up $B_{\text{stable}}$, then a jump back up to the spiking branch $B_{\text{spikes}}$.

Now, to examine the possible fractal geometry of these Rulkov map 2 spiking and bursting attractors, we will first consider the Kaplan-Yorke conjecture. In Section \ref{dynamics-rulkov-map-2}, we calculated that the Lyapunov spectrum of the spiking attractor generated by the system with parameters $\alpha=4.5$ and $\sigma=0$ is $\lambda \approx \{0.5449,\, -2.070\times 10^{-4}\}$, and the Lyapunov spectrum of the bursting attractor generated by the system with parameters $\alpha=4.1$ and $\sigma=-0.5$ is $\lambda \approx \{0.5025,\, -0.03376\}$. Recall from the end of Section \ref{strangeattractors} that the Kaplan-Yorke conjecture relies on finding an index $\kappa$ such that 
\begin{equation}
    \sum_{i=1}^{\kappa+1} \lambda_i < 0
\end{equation}
However, for both the spiking and bursting attractor, summing both Lyapunov exponents will give us a value greater than 0. Therefore, the Kaplan-Yorke conjecture will be of no use to us here; to calculate the dimension of these attractors, we must return to the method of box-counting.

\begin{table}[t]
    \centering
    \begin{tabular}{c|c}
        \multicolumn{2}{c}{Spiking Attractor} \vspace{3px} \\
        $\epsilon$ & $N(\epsilon)$ \\
        \hline \\ [-10px]
        $1/100$ & 1696 \\
        $1/200$ & 5577 \\
        $1/400$ & 17720 \\
    \end{tabular}
    \hspace{0.5cm}
    \begin{tabular}{c|c}
        \multicolumn{2}{c}{Bursting Attractor} \vspace{3px} \\
        $\epsilon$ & $N(\epsilon)$ \\
        \hline \\ [-10px]
        $1/50$ & 1375 \\
        $1/100$ & 4982 \\
        $1/200$ & 17968 \\
    \end{tabular}
    \caption{Some $N(\epsilon)$ values for a spiking attractor of Rulkov map 2 generated by the system with parameters $\alpha=4.5$ and $\sigma=0$ and a bursting attractor of Rulkov map 2 generated by the system with parameters $\alpha=4.1$ and $\sigma=-0.5$, both calculated approximately on the ranges shown in Figure \ref{fig:rulkov-2-attractor-graphs} using the code in Appendices \ref{rulkov_2_graphs_and_cobweb_code} and \ref{henon-box-counting-code}}
    \label{tab:spiking-attractor-box-counting}
\end{table}

In Section \ref{strangeattractors}, we used the code in Appendix \ref{henon-box-counting-code} to count boxes on the Hénon attractor. However, this box-counting code works for box-counting on any attractor, so we can also use Appendix \ref{henon-box-counting-code} to determine $N(\epsilon)$ values on the spiking and bursting attractors of Rulkov map 2. The results from box-counting on the Rulkov 2 attractors in Figure \ref{fig:rulkov-2-attractor-graphs} are displayed in Table \ref{tab:spiking-attractor-box-counting}. Recalling from Section \ref{strangeattractors} that we expect $N(\epsilon)$ to be proportional to $1/\epsilon^d$, we can take logarithms of the data in the table $N(\epsilon)$ vs. $1/\epsilon$ and perform a linear regression to extract the fractal dimension $d$, which produces stunningly good results. Considering the spiking attractor first, we get that
\begin{equation}
    \log_2 N(\epsilon) = 1.693\log_2\epsilon - 0.509
\end{equation}
with an $R^2$ value of 0.9999. This indicates that the dimension of the spiking attractor is $d\approx 1.693$, which is not an integer, indicating that the spiking attractor shown in Figure \ref{fig:rulkov-2-spiking-attractor} is indeed a fractal. It is certainly believable that the attractor shown in Figure \ref{fig:rulkov-2-spiking-attractor} is closer to being two-dimensional than one-dimensional, as the points that make up the attractor seem to fill up a lot of two-dimensional state space. Performing a similar linear regression on the bursting attractor values in Table \ref{tab:spiking-attractor-box-counting} yields
\begin{equation}
    \log_2 N(\epsilon) = 1.854\log_2\epsilon - 0.037
\end{equation}
with an $R^2$ value of 0.999999. This indicates that the bursting attractor is also fractal with a dimension of $d\approx 1.854$. This higher fractal dimension indicates that the bursting attractor in Figure \ref{fig:rulkov-2-bursting-attractor} is more ``rough'' than the spiking attractor in Figure \ref{fig:rulkov-2-spiking-attractor}, which is not immediately obvious. However, the main result from this section is that an isolated Rulkov neuron can indeed form a true strange attractor in two-dimensional state space, both for chaotic spiking and chaotic bursting systems.

\subsection{Asymmetrical Electrical Coupling of Two Rulkov 1 Neurons}
\label{asym-elec-coup-two-rulkov-1-neurons-geometry}

As we discovered in Section \ref{two-electrically-coupled-rulkov-1-neurons}, multistability and chaotic dynamics appear to exist in a system of two asymmetrically electrically coupled Rulkov 1 neurons. In this section, we will analyze the many complex and interesting geometries of state space that appear in the system from Section \ref{two-electrically-coupled-rulkov-1-neurons}, the asymmetrically coupled neuron system with $\beta^c_1=\sigma^c_1=\beta^c_2=\sigma^c_2=1$, $\sigma_1=\sigma_2=-0.5$, $\alpha_1=\alpha_2=4.5$, $g^e_1=0.05$, and $g^e_2=0.25$. 

\subsubsection{Attractor Analysis}

\begin{figure*}[hp!]
    \centering
    \begin{subfigure}[t]{0.495\textwidth}
        \centering
        \includegraphics[scale=0.093]{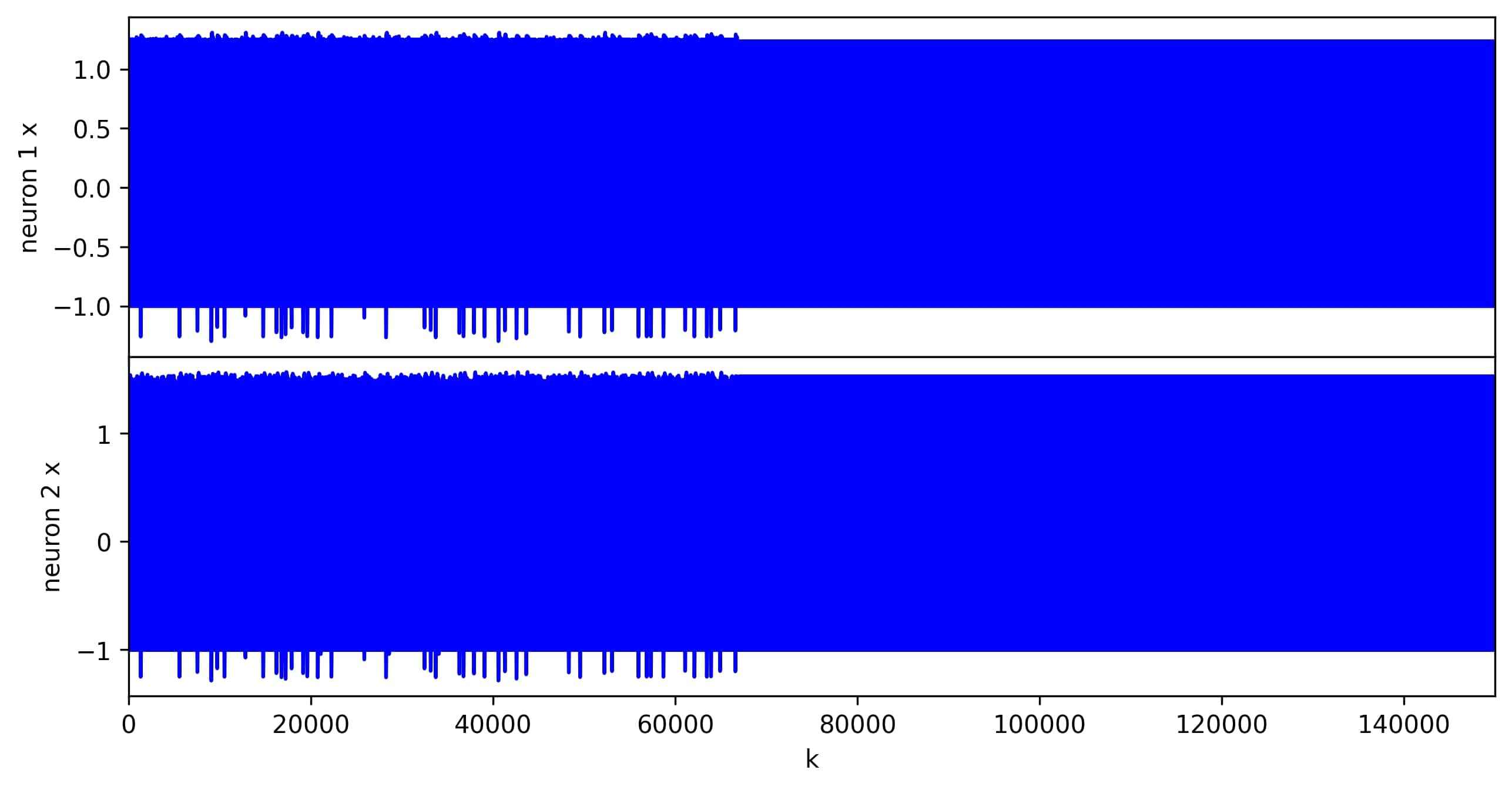}
        \caption{$\mathbf{x}_{1,\,0} = \langle -0.56,\, -3.25 \rangle$, $\mathbf{x}_{2,\,0} = \langle -1,\, -3.25 \rangle$}
        \label{fig:asym_coup_rulkov_1_long_1}
        \vspace{8px}
    \end{subfigure}
    \begin{subfigure}[t]{0.495\textwidth}
        \centering
        \includegraphics[scale=0.093]{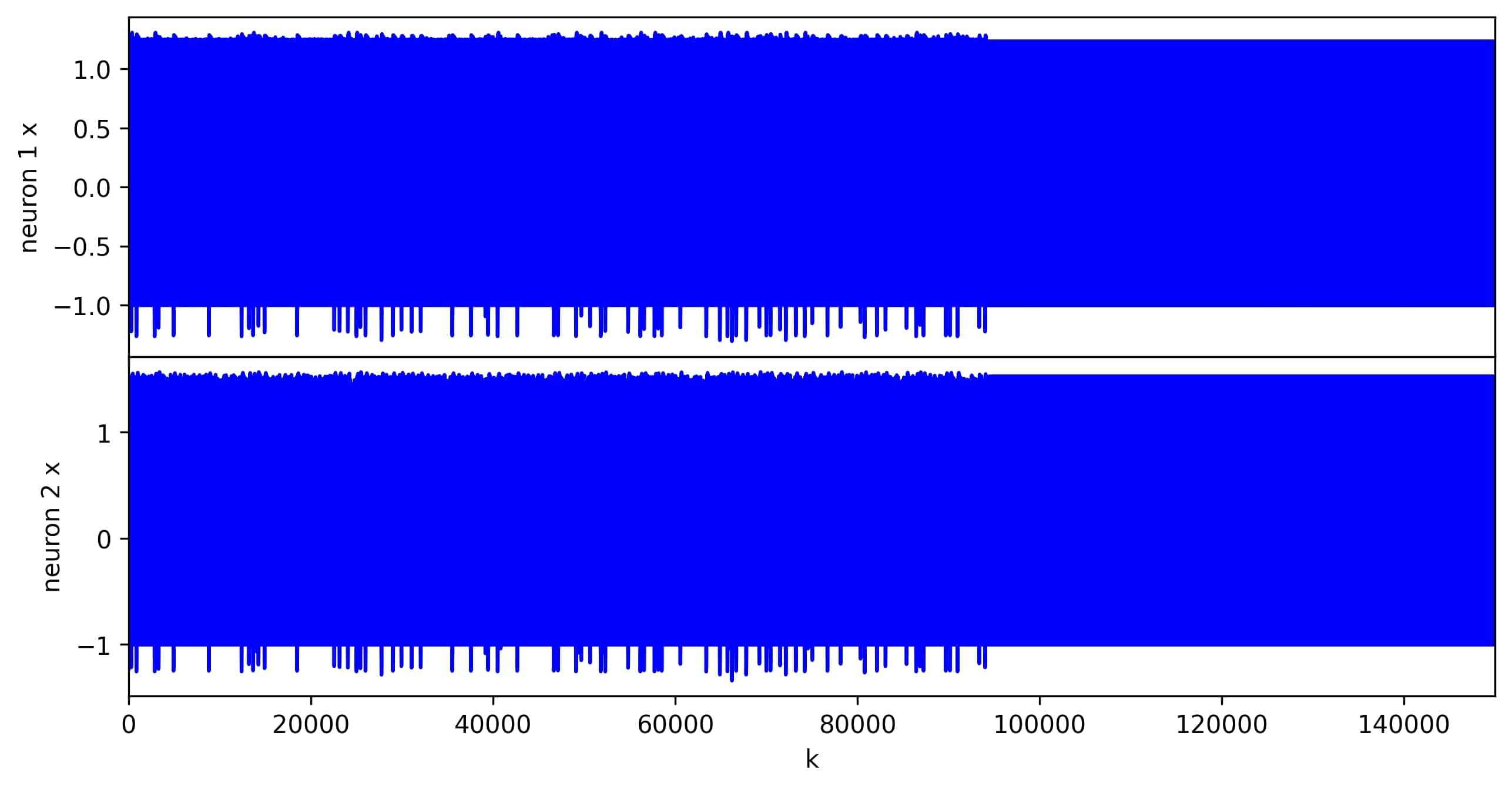}
        \caption{$\mathbf{x}_{1,\,0} = \langle 0.35,\, -3.25 \rangle$, $\mathbf{x}_{2,\,0} = \langle -1.23,\, -3.25 \rangle$}
        \label{fig:asym_coup_rulkov_1_long_2}
        \vspace{8px}
    \end{subfigure}
    \begin{subfigure}[t]{0.495\textwidth}
        \centering
        \includegraphics[scale=0.093]{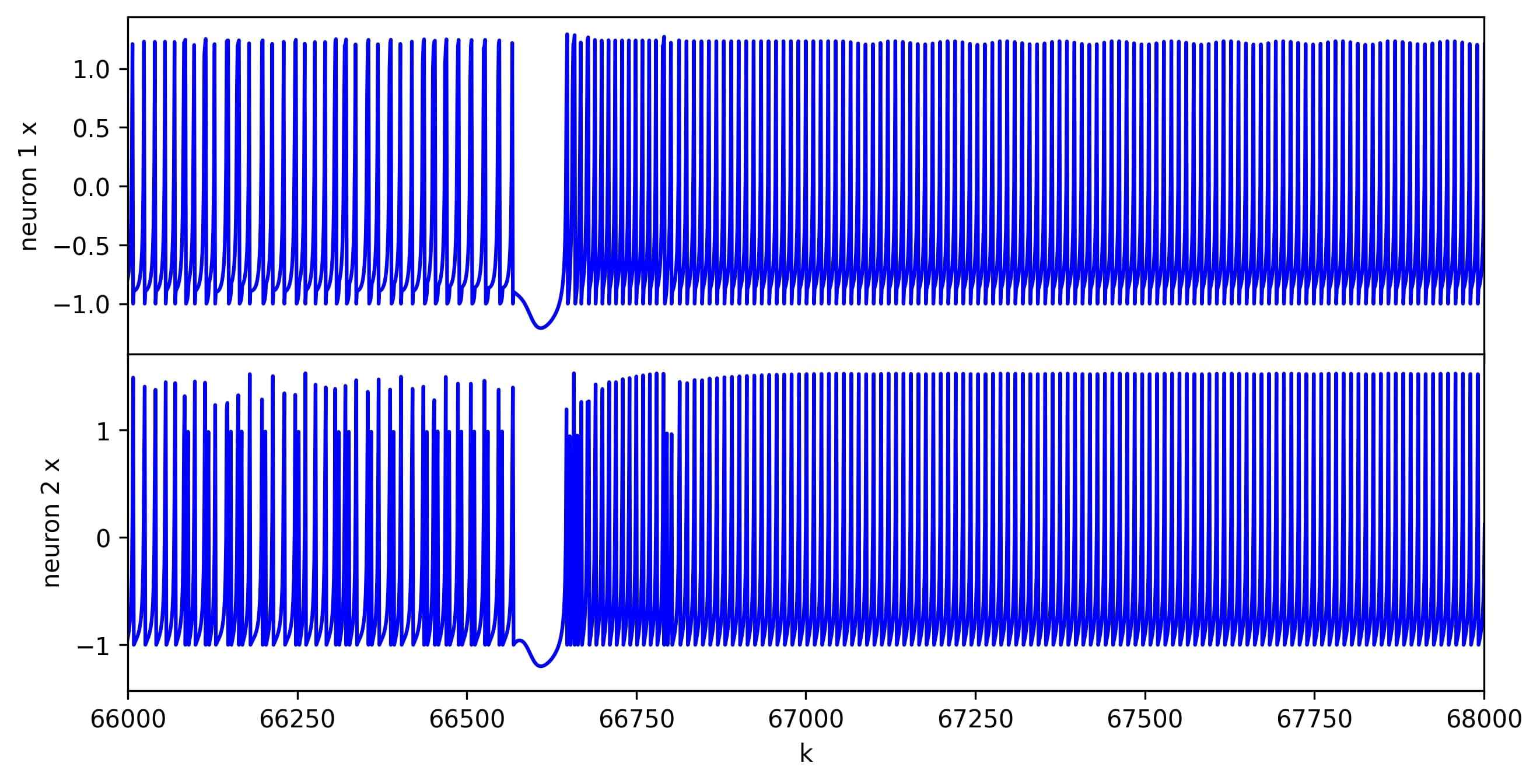}
        \caption{(Zoom) $\mathbf{x}_{1,\,0} = \langle -0.56,\, -3.25 \rangle$, $\mathbf{x}_{2,\,0} = \langle -1,\, -3.25 \rangle$}
        \label{fig:asym_coup_rulkov_1_long_1_zoom}
        \vspace{8px}
    \end{subfigure}
    \begin{subfigure}[t]{0.495\textwidth}
        \centering
        \includegraphics[scale=0.093]{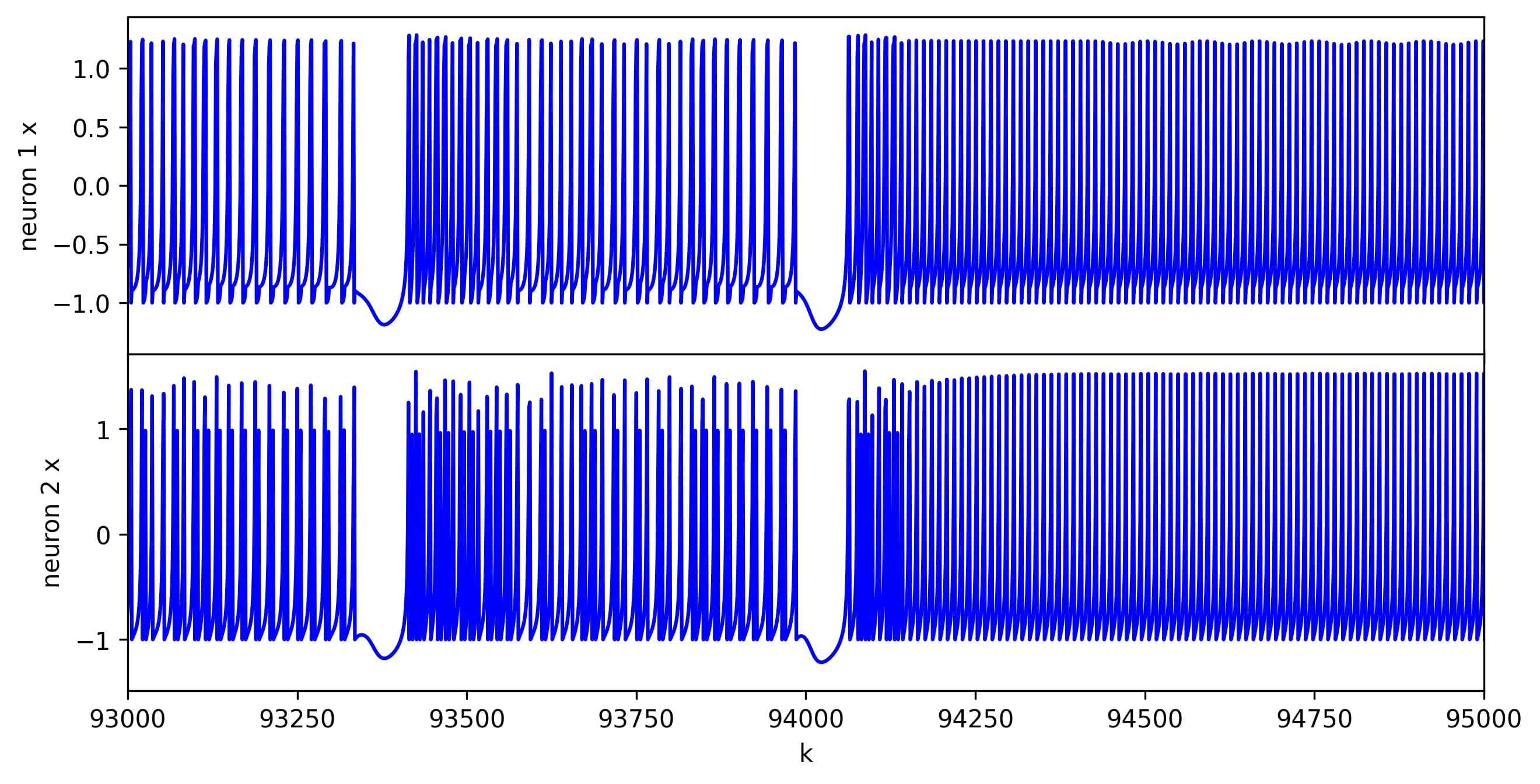}
        \caption{(Zoom) $\mathbf{x}_{1,\,0} = \langle 0.35,\, -3.25 \rangle$, $\mathbf{x}_{2,\,0} = \langle -1.23,\, -3.25 \rangle$}
        \label{fig:asym_coup_rulkov_1_long_2_zoom}
        \vspace{8px}
    \end{subfigure}
    \caption{Graphs showing the eventual attraction of two orbits of the system of two asymmetrically electrically coupled Rulkov 1 neurons to the system's non-chaotic spiking attractor, graphed using the code in Appendix \ref{asym-elec-coup-rulkov-1-neurons-code}}
    \label{fig:asym_coup_rulkov_1_long_graphs}
\end{figure*}

In Figure \ref{fig:asym_coup_rulkov_1_graphs}, we can see that this system appears to have two distinct attractors: a non-chaotic spiking attractor and a chaotic attractor. However, upon further analysis of this system, we discover that this is not completely accurate. To see this, let us examine two seemingly chaotic orbits of this system. First, we will look at the system from Figure \ref{fig:asym_coup_rulkov_1_chaos}, which has initial states $\mathbf{x}_{1,\,0}=\langle -0.56,\, -3.25\rangle$ and $\mathbf{x}_{2,\,0}=\langle -1,\, -3.25\rangle$. Second, we will look at the system with initial states $\mathbf{x}_{1,\,0} = \langle 0.35,\, -3.25 \rangle$ and $\mathbf{x}_{2,\,0} = \langle -1.23,\, -3.25 \rangle$. Using the code in Appendix \ref{asym-elec-coup-rulkov-1-neurons-code}, we graph the fast variable orbits of these systems up to a large value of $k$ ($O(\mathbf{X_0})=\{\mathbf{X}_0,\,\mathbf{X}_1,\,\hdots\,,\,\mathbf{X}_{150000}\}$) in Figures \ref{fig:asym_coup_rulkov_1_long_1} and \ref{fig:asym_coup_rulkov_1_long_2}. As we can see, while the voltage orbits begin chaotic, they eventually fall into the non-chaotic spiking attractor. This is made especially clear in Figures \ref{fig:asym_coup_rulkov_1_long_1_zoom} and \ref{fig:asym_coup_rulkov_1_long_2_zoom}, where we zoom into the region where the orbits transition from chaotic to non-chaotic. We conjecture that these dynamics occur because while the neurons are exhibiting chaotic bursting, their voltages happen to line up and latch onto each other's spiking by chance, which propels the system's orbit to the non-chaotic spiking attractor. If this is true, then a true chaotic attractor of this system doesn't exist because everything will eventually, by chance, get attracted to the non-chaotic spiking attractor. For this reason, true multistability as we have defined it in Section \ref{basins-of-attraction} doesn't exist here. However, we are still interested in analyzing this system because it exhibits sensitive dependence on initial conditions; although all orbits end up in the spiking attractor after enough time, initial conditions still have an effect on the qualitative dynamics of the system in the short term. For this reason, we will treat the short-term chaotic behavior exhibited by this system as being on a chaotic ``pseudo-attractor,'' which is where some orbits end up in before eventually being attracted to the non-chaotic spiking attractor.

\begin{figure*}[hp!]
    \centering
    \hfill
    \begin{subfigure}[t]{0.475\textwidth}
        \centering
        \includegraphics[scale=0.13]{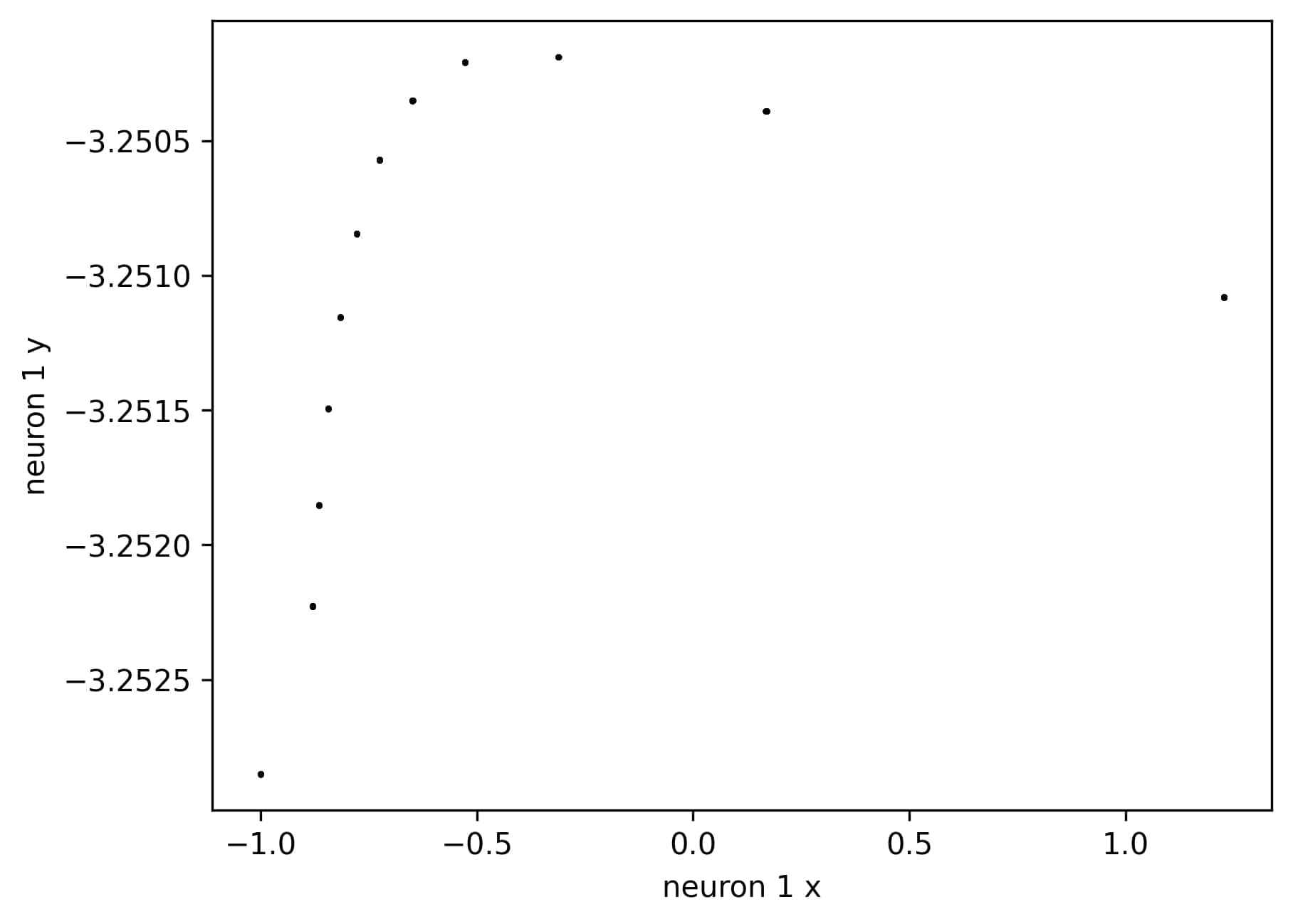}
        \caption{Four-dimensional state space $\langle x_1,\, y_1,\, x_2,\, y_2
        \rangle$ projected onto two-dimensions $\langle x_1,\, y_1
        \rangle$}
        \label{fig:nonchaotic_rulkov1_attractor_neuron1}
    \end{subfigure}
    \hfill
    \begin{subfigure}[t]{0.475\textwidth}
        \centering
        \includegraphics[scale=0.13]{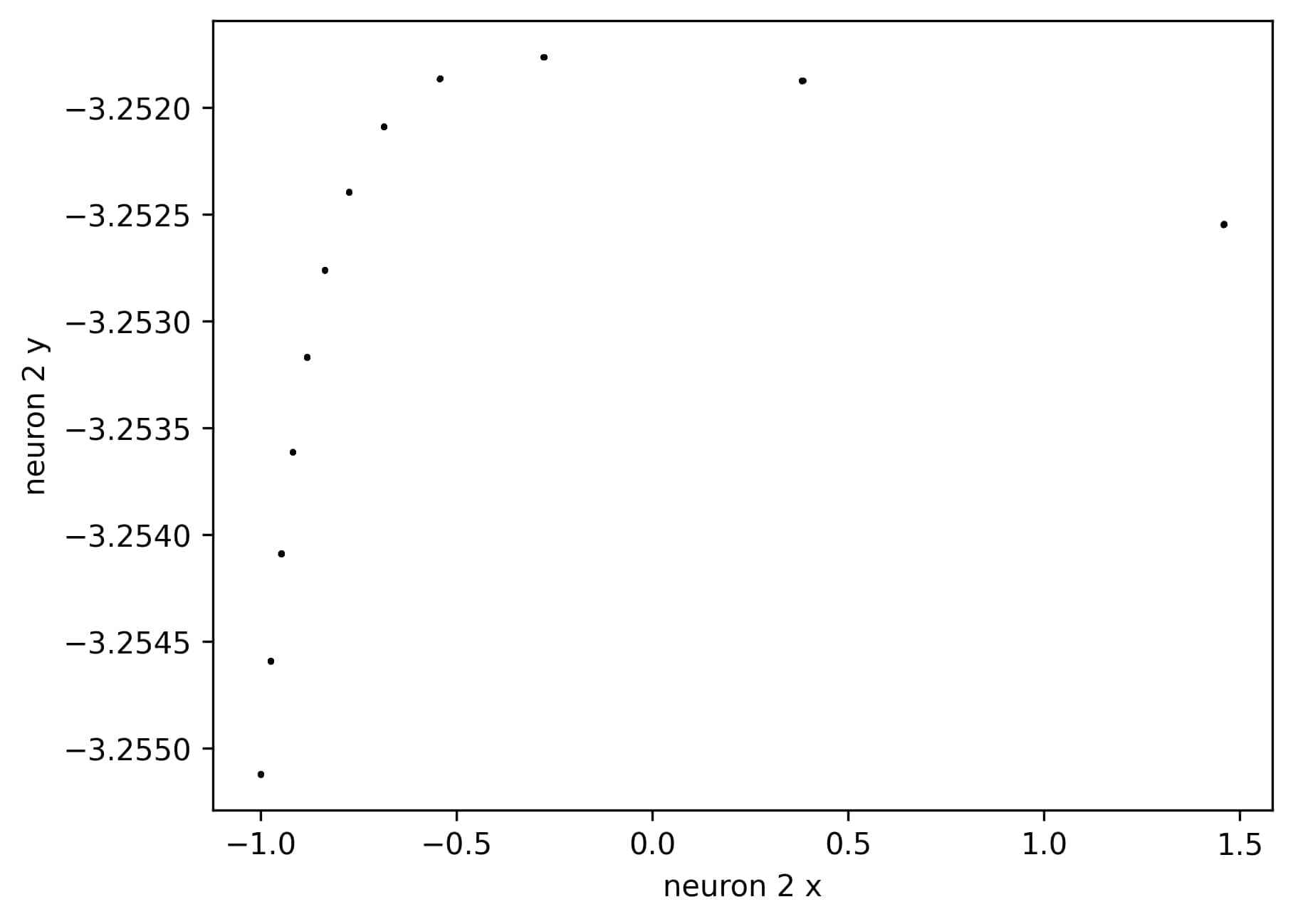}
        \caption{Four-dimensional state space $\langle x_1,\, y_1,\, x_2,\, y_2
        \rangle$ projected onto two-dimensions $\langle x_2,\, y_2
        \rangle$}
        \label{fig:nonchaotic_rulkov1_attractor_neuron2}
    \end{subfigure}
    \hfill
    \caption{Projections of the four-dimensional non-chaotic attractor generated by the system of two asymmetrically coupled Rulkov 1 neurons with initial states $\mathbf{x}_{1,\,0}=\langle -0.54,\, -3.25 \rangle$ and $\mathbf{x}_{2,\,0}=\langle -1,\, -3.25 \rangle$ and Lyapunov spectrum $\lambda\approx\{-0.0057,\,-0.0125,\,-0.0126,\,-\infty\}$, graphed with the code in Appendix \ref{asym-elec-coup-rulkov-1-neurons-code}}
    \label{fig:nonchaotic_rulkov1_attractor}
    \vspace{4px}
\end{figure*}

\begin{figure*}[ht!]
    \centering
    \hfill
    \begin{subfigure}[t]{0.475\textwidth}
        \centering
        \includegraphics[scale=0.13]{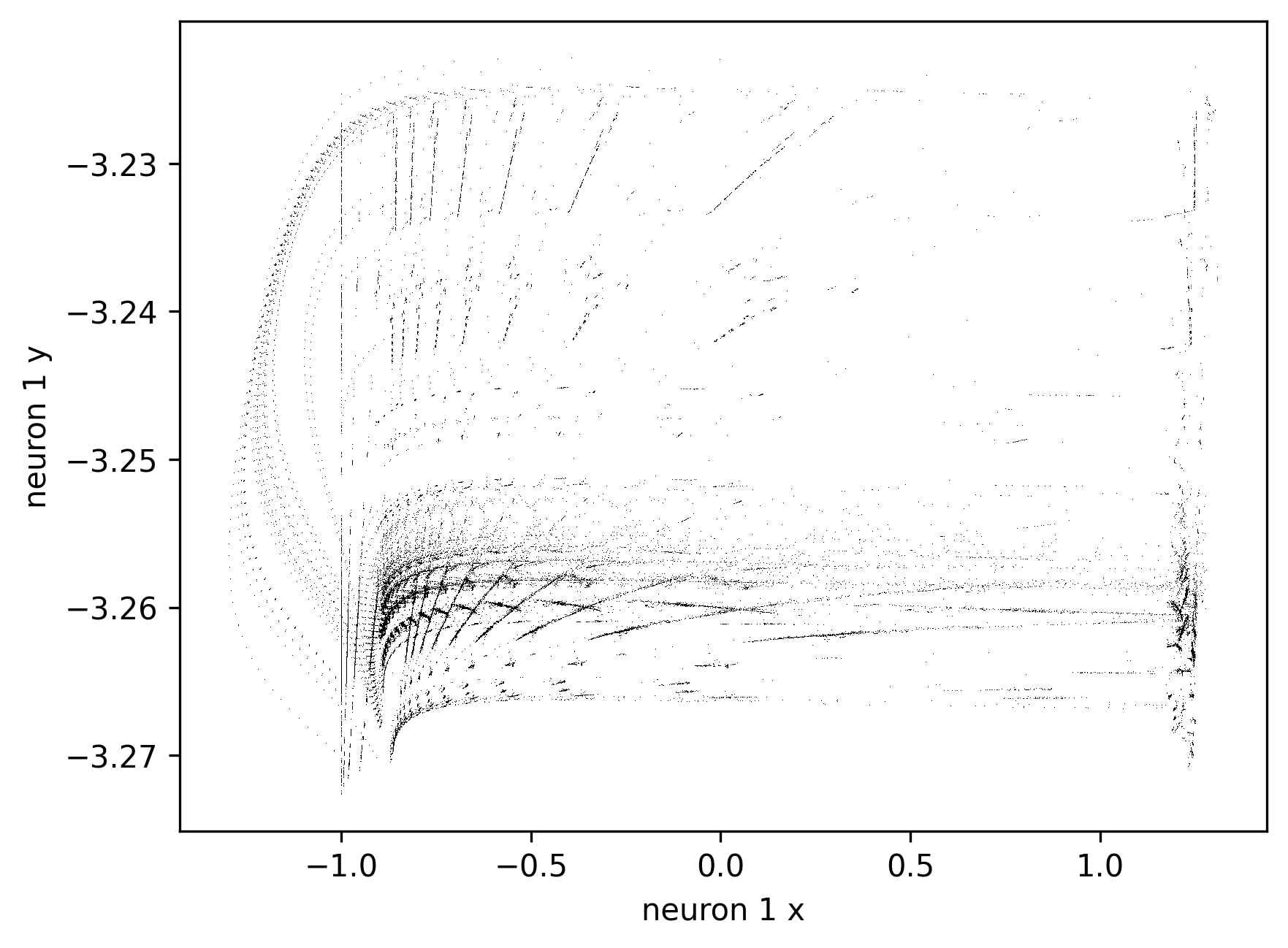}
        \caption{Four-dimensional state space $\langle x_1,\, y_1,\, x_2,\, y_2
        \rangle$ projected onto two-dimensions $\langle x_1,\, y_1
        \rangle$}
        \label{fig:chaotic_rulkov1_attractor_neuron1}
    \end{subfigure}
    \hfill
    \begin{subfigure}[t]{0.475\textwidth}
        \centering
        \includegraphics[scale=0.13]{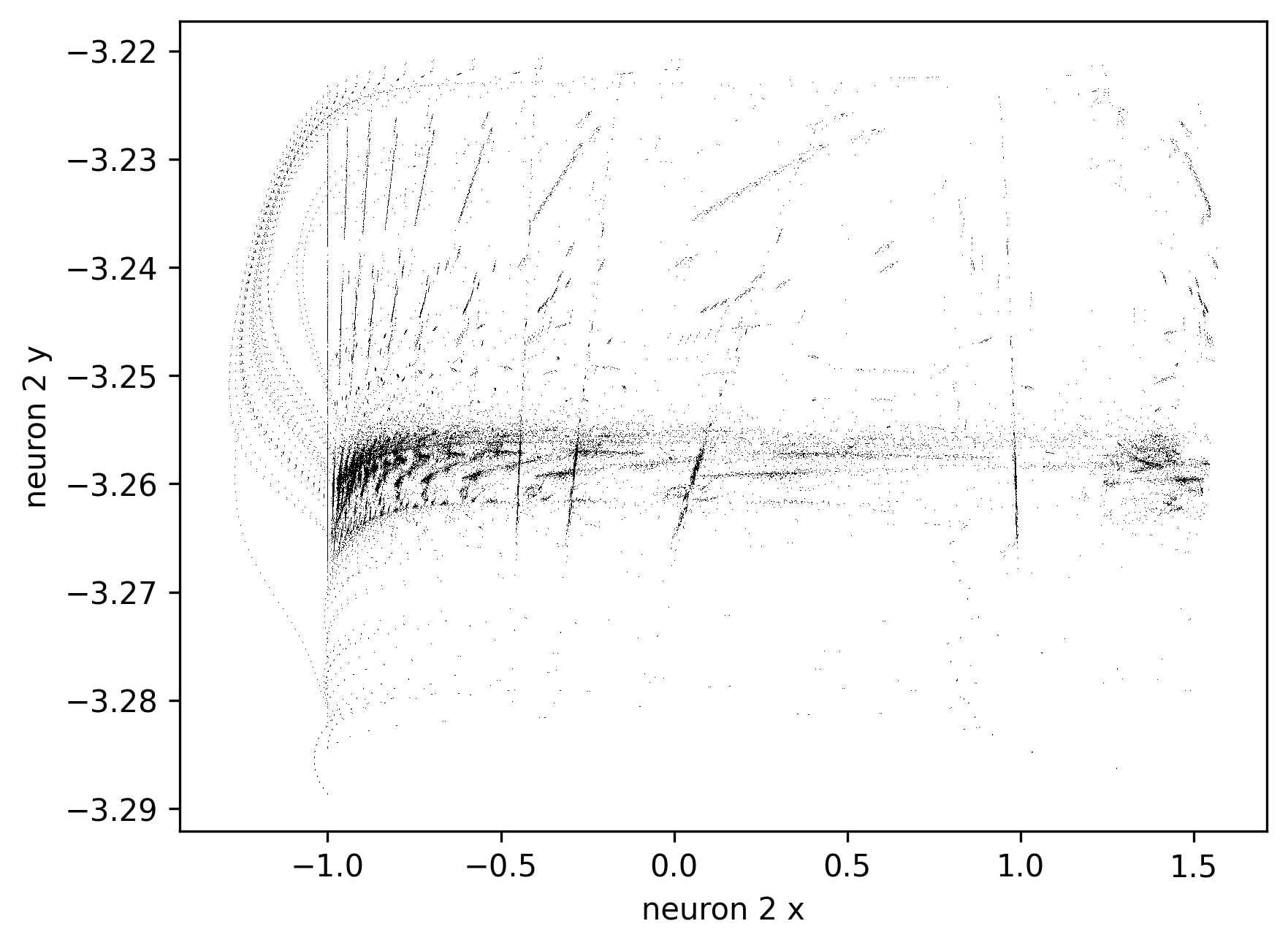}
        \caption{Four-dimensional state space $\langle x_1,\, y_1,\, x_2,\, y_2
        \rangle$ projected onto two-dimensions $\langle x_2,\, y_2
        \rangle$}
        \label{fig:chaotic_rulkov1_attractor_neuron2}
    \end{subfigure}
    \hfill
    \caption{Projections of the four-dimensional strange pseudo-attractor generated by two asymmetrically coupled Rulkov 1 neurons with parameters $\sigma_1=\sigma_2=-0.5$, $\alpha_1=\alpha_2=4.5$, coupling strengths $g^e_1=0.05$, $g^e_2=0.25$, initial states $\mathbf{x}_{1,\,0}=\langle -0.56,\, -3.25 \rangle$, $\mathbf{x}_{2,\,0}=\langle -1,\, -3.25 \rangle$, and Lyapunov spectrum $\lambda\approx\{0.0165,\,-0.0105,\,-0.0810,\,-\infty\}$, graphed with the code in Appendix \ref{asym-elec-coup-rulkov-1-neurons-code}}
    \label{fig:chaotic_rulkov1_attractor}
    \vspace{4px}
\end{figure*}

We are now interested in visualizing the non-chaotic spiking attractor and chaotic pseudo-attractor. Immediately, we run into the problem of the state space of this system being four-dimensional, which we cannot make a graph of. Despite this, as an attempt to visualize the attractors, we will project four-dimensional state space $\langle x_1,\, y_1,\, x_2,\, y_2\rangle$ onto two different two-dimensional surfaces, $\langle x_1,\, y_1\rangle$ and $\langle x_2,\, y_2\rangle$, by plotting only the $\mathbf{x}_1$ or $\mathbf{x}_2$ components of each state vector $\mathbf{X}$, respectively.\footnote{It is important to note that these projections do not capture the true geometry of these attractors in four-dimensional state space, as one of these projections is dimensionally analogous to compacting an entire region of three-dimensional state space onto a line segment. However, because we don't live in four spatial dimensions (at least, macroscopically), this is one of the best ways we have to visualize these attractors.} This is accomplished in Figures \ref{fig:nonchaotic_rulkov1_attractor} and \ref{fig:chaotic_rulkov1_attractor} for the initial conditions $\mathbf{X}_{0}=\langle -0.54,\, -3.25,\, -1,\, -3.25 \rangle$ (associated with Figure \ref{fig:asym_coup_rulkov_1_spiking}) and $\mathbf{X}_{0}=\langle -0.56,\, -3.25,\, -1,\, -3.25 \rangle$ (associated with Figure \ref{fig:asym_coup_rulkov_1_chaos}), respectively. It is immediately clear that the attractor in Figure \ref{fig:nonchaotic_rulkov1_attractor} is non-chaotic and the pseudo-attractor in Figure \ref{fig:chaotic_rulkov1_attractor} is chaotic. 

However, the question now is whether or not the chaotic pseudo-attractor is strange. We know from Section \ref{chaotic-dynamics-rulkov-map-1} that there is a possibility that it isn't because of the resetting mechanism of Rulkov map 1, which is what makes the attractor of a chaotic uncoupled Rulkov 1 neuron not fractal. We could try to use the Kaplan-Yorke conjecture again, but we should be cautious in this case because we are dealing with a pseudo-attractor, not a normal attractor. As a result, the Lyapunov exponents are only approximations inside this pseudo-attractor;\footnote{As shown in the code in Appendix \ref{asym-elec-coup-rulkov-1-neurons-code}, we calculate these Lyapunov exponents up to 65,000 iterations only, which is approximately when the orbit goes to the spiking attractor (see Figure \ref{fig:asym_coup_rulkov_1_long_1_zoom}).} calculating the spectrum for a larger number of iterations will contain information about the orbit on the non-chaotic spiking attractor. For this reason, it would be good to return to our good old-fashioned box-counting method. Using the code in Appendix \ref{box-counting-on-coupled-rulkov-neuron-attractor} for four-dimensional boxes of size $\epsilon=1/20,\,1/30,\,1/40$,\footnote{We calculate $N(\epsilon)$ for a small number of $\epsilon$ values because, since we are dealing with box-counting in four-dimensional state space, the code in Appendix \ref{box-counting-on-coupled-rulkov-neuron-attractor} has time complexity $O(\epsilon^4)$.} we get that the Minkowski–Bouligand dimension of the chaotic pseudo-attractor is $d\approx 1.84$ with an $R^2$ value of 0.9999. This is not an integer, so the chaotic pseudo-attractor is indeed fractal and strange. 

Even though we suspect that the Kaplan-Yorke conjecture will not hold in this case, we can still calculate it for fun since it should still tell us that the pseudo-attractor is fractal even if it doesn't give us the correct dimension. The Lyapunov spectrum of the four-dimensional strange pseudo-attractor that $\mathbf{X}_{0}=\langle -0.56,\, -3.25,\, -1,\, -3.25 \rangle$ goes to is $\lambda\approx\{0.0165,\,-0.0105,\,-0.0810,\,-\infty\}$, which is calculated using the code in Appendix \ref{asym-elec-coup-rulkov-1-neurons-code}. Interestingly, in this case, the $\kappa$ index relevant to calculating the Lyapunov dimension $d_l$ is 2 since
\begin{equation}
    \sum_{i=1}^2 \lambda_i \approx 0.006 > 0
\end{equation}
Therefore, the Lyapunov dimension of the strange pseudo-attractor is 
\begin{equation}
    \begin{split}
        d_l &= \kappa + \frac{1}{|\lambda_{\kappa+1}|}\sum_{i=1}^{\kappa}\lambda_i \\
        &= 2 + \frac{1}{|\lambda_{3}|}\sum_{i=1}^{2}\lambda_i \\
        &\approx 2.07
    \end{split}
\end{equation}
Although this is different from the true fractal dimension of the pseudo-attractor, the Kaplan-Yorke conjecture still indicates that the pseudo-attractor is fractal even though it indicated that the chaotic attractor of an isolated Rulkov 1 neuron wasn't (see Section \ref{chaotic-dynamics-rulkov-map-1}). 

\subsubsection{Basin Analysis}

Due to the short-term multistability of this asymmetrically coupled system, we are now interested in analyzing the basins of the spiking attractor and strange pseudo-attractor for interesting geometrical properties. Because the state space of our system is four-dimensional, these basins are four-dimensional sets. However, because of the difficulty in visualizing four-dimensional objects, as well as the fact that our coupled Rulkov 1 neurons tend to settle down in a small interval of $y$ values (see Figures \ref{fig:nonchaotic_rulkov1_attractor} and \ref{fig:chaotic_rulkov1_attractor}), we will first consider a two-dimensional slice of the basins, namely, the intersection of the four-dimensional basins with the set of state space 
\begin{multline}
    S_2'= \{\mathbf{X} = \langle x_1,\,y_1,\,x_2,\,y_2\rangle: \\-2<x_1<2,\,y_1=-3.25, \\ -2<x_2<2,\,y_2=-3.25\}
    \label{eq:s2'}
\end{multline}
To determine which attractor a given state in $S_2'$ is initially attracted to, we will need to define a specific way of making the distinction between immediately going to the spiking attractor and spending some time in the chaotic pseudo-attractor first. In Figure \ref{fig:asym_coup_rulkov_1_long_graphs}, we see that states that are attracted to the chaotic pseudo-attractor spend on the order of $10^4$ iterations in the pseudo-attractor before going to the spiking attractor. Testing other initial states using the code in Appendix \ref{asym-elec-coup-rulkov-1-neurons-code} gives similar results. We will therefore say that an initial state that has a positive maximal Lyapunov exponent after 5,000 iterations was attracted to the chaotic pseudo-attractor, while an initial state that has a negative $\lambda_1$ after the same amount of time was attracted to the spiking attractor. It is worthwhile to emphasize that this way of detecting which attractor an initial state goes to is very different from the method we used before in Section \ref{geometry_dynamics_chaos}, which was directly testing whether a state was in the vicinity of a given attractor's points. Instead, we are indirectly testing what attractor an initial state goes to using the power of Lyapunov exponents. 

\begin{figure*}[htp!]
    \centering
    \hspace{2.35cm}
    \begin{subfigure}{0.7\textwidth}
        \centering
        \includegraphics[scale=0.035]{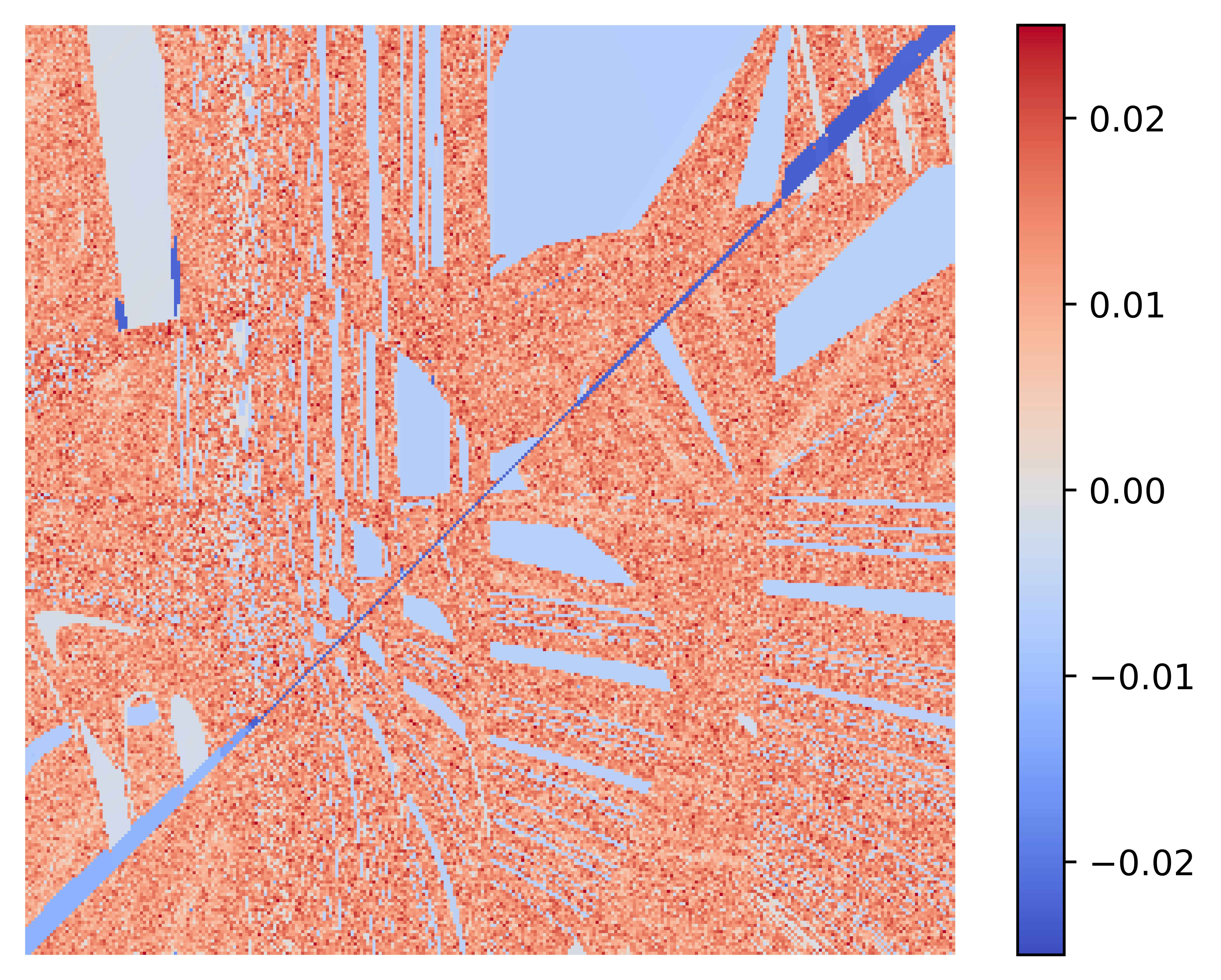}
        \caption{Maximal Lyapunov exponent color map, with red points indicating chaotic \\ dynamics and blue points indicating non-chaotic dynamics}
        \label{fig:asym_basin_color}
    \end{subfigure}
    \begin{subfigure}{0.7\textwidth}
        \centering
        \includegraphics[scale=0.035]{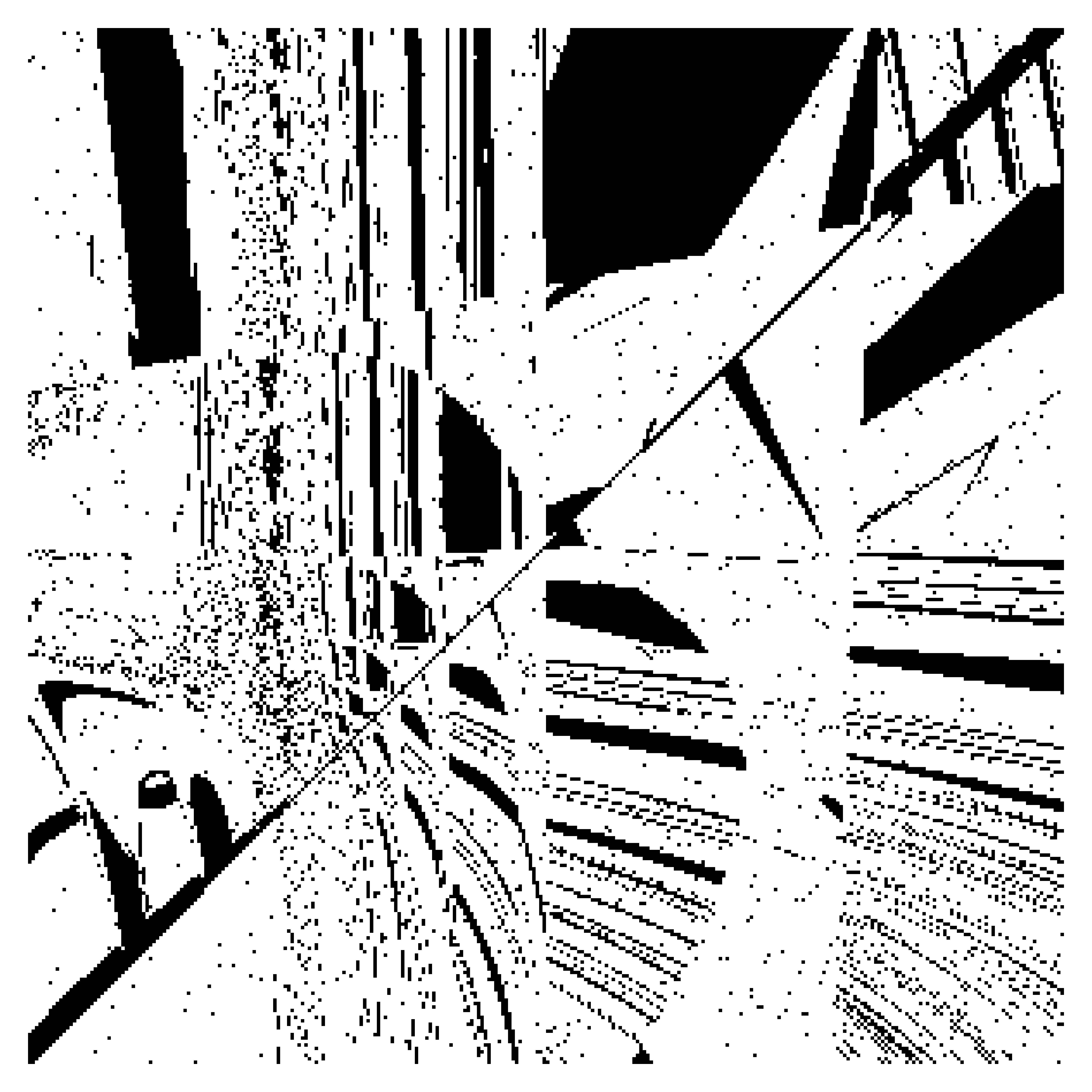}
        \caption{Basin slices of the chaotic (white) and non-chaotic spiking (black) attractors}
        \label{fig:asym_basin_bw}
    \end{subfigure}
    \caption{A slice of the basins of attraction of two asymmetrically electrically coupled Rulkov 1 neurons with $-2\leq x_{1,\,0}\leq 2$ on the horizontal axis and $-2\leq x_{2,\,0}\leq 2$ on the vertical axis, parameters $\sigma_1=\sigma_2=-0.5$, $\alpha_1=\alpha_2=4.5$, coupling strengths $g^e_1=0.05$, $g^e_2=0.25$, and fixed initial $y$ values $y_{1,\,0}=y_{2,\,0}=-3.25$, visualized using the code in Appendix \ref{visualize-basins-asym-coup-rulkov-1-code}}
    \label{fig:asym_basin_graphs}
\end{figure*}

Using the code in Appendix \ref{visualize-basins-asym-coup-rulkov-1-code}, we test a large number of initial states in $S_2'$ using this method and present the results of these Lyapunov exponent calculations using a color map in Figure \ref{fig:asym_basin_color}. An immediate observation is the beautiful complexity of the different maximal Lyapunov exponents in state space. This is further emphasized when we color the basins properly in Figure \ref{fig:asym_basin_bw}, with the basin of the chaotic pseudo-attractor (positive $\lambda_1$) shown in white and the basin of the non-chaotic attractor (negative $\lambda_1$) shown in black. Here, we can see that most of $S_2'$ is taken up by the basin of the white chaotic pseudo-attractor, with some black regions, curves, and scattered points of stability scattered throughout. The most obvious of these features is the black line that goes across the diagonal of Figure \ref{fig:asym_basin_bw} from the bottom left to the top right. This line and the areas surrounding it are places where the initial states of the two neurons are the same or very close to being the same. In this case, it makes sense that they will immediately be synchronized, spiking together in the non-chaotic attractor. 

Now that we have visualized the basins in $S_2'$, we will now classify (using the method detailed in Section \ref{basins-of-attraction}) the basin slices in the infinite two-dimensional plane containing $S_2'$. In other words, we are classifying the intersection of the four-dimensional basins with the infinite plane that lies parallel to and contains the square $S_2'$, which we will denote as $S_2$:
\begin{multline}
    S_2 = \{\mathbf{X} = \langle x_1,\,y_1,\,x_2,\,y_2\rangle: \\x_1\in\mathbb{R},\,y_1=-3.25, \\ x_2\in\mathbb{R},\,y_2=-3.25\}
\end{multline}
To account for $S_2$ being two-dimensional, we will also need to specify what is meant by the mean and standard deviation of the four-dimensional attractors within the two-dimensional plane $S_2$. Let us denote a generic four-dimensional attractor as $A_4$ and its associated basin as $\hat{A}_4$. In this case, $A_4$ can represent either the spiking attractor or the chaotic pseudo-attractor. From Equation \ref{eq:mean}, if 
\begin{equation}
    A_4=\{\mathbf{a}_1,\,\mathbf{a}_2,\,\hdots\}=\left\{\begin{pmatrix}
        a_1\e{1} \\[4pt]
        a_1\e{2} \\[4pt]
        a_1\e{3} \\[4pt]
        a_1\e{4}
    \end{pmatrix},\,
    \begin{pmatrix}
        a_2\e{1} \\[4pt]
        a_2\e{2} \\[4pt]
        a_2\e{3} \\[4pt]
        a_2\e{4}
    \end{pmatrix},\,\hdots
    \right\}
\end{equation}
then the true mean of $A_4$, which we will denote as $\langle A_4\rangle$, is
\begin{equation}
    \langle A_4\rangle = \lim_{j\to\infty}\frac{1}{j}\sum_{i=1}^j\mathbf{a}_i = \begin{pmatrix}
        \langle x_1 \rangle \\[4pt]
        \langle y_1 \rangle \\[4pt]
        \langle x_2 \rangle \\[4pt]
        \langle y_2 \rangle 
    \end{pmatrix}
\end{equation}
We introduce $\langle A_2\rangle$ to be the ``effective mean'' of the four-dimensional attractor $A_4$ that we will use when classifying the two-dimensional slice of a basin, or $\hat{A}_4\cap S_2$. We define it simply as a two-dimensional vector composed of the first and third entries of $\langle A_4\rangle$:
\begin{equation}
    \langle A_2\rangle = \lim_{j\to\infty}\frac{1}{j}\sum_{i=1}^j\begin{pmatrix}
        a_i\e{1} \\[4pt]
        a_i\e{3} 
    \end{pmatrix} = \begin{pmatrix}
        \langle x_1 \rangle \\[4pt]
        \langle x_2 \rangle 
    \end{pmatrix}
\end{equation}
Similarly, the true standard deviation of $A_4$ in four-dimensional state space is, from Equation \ref{eq:standard-dev},
\begin{equation}
    \sigma_{A4} = \sqrt{\lim_{j\to\infty}\frac{1}{j}\sum_{i=1}^j|\mathbf{a}_i-\langle A_4\rangle|^2}
\end{equation}
It follows that the ``effective standard deviation'' $\sigma_{A2}$ of $A_4$ that we will use when classifying $\hat{A}_4\cap S_2$ is
\begin{equation}
    \begin{split}
        \sigma_{A2} &= \sqrt{\lim_{j\to\infty}\frac{1}{j}\sum_{i=1}^j\left|\begin{pmatrix}
            a_i\e{1} \\[4pt]
            a_i\e{3} 
        \end{pmatrix}-\langle A_2\rangle\right|^2} \\
        &= \sqrt{\lim_{j\to\infty}\frac{1}{j}\sum_{i=1}^j\left|\begin{pmatrix}
            a_i\e{1}-\langle x_1 \rangle \\[4pt]
            a_i\e{3}-\langle x_2 \rangle
        \end{pmatrix}\right|^2}
    \end{split}
\end{equation}
For our classification of the basin slices $\hat{A}_4\cap S_2$, we will consider the normalized two-dimensional distance $\xi_2$ from $\langle A_2\rangle$ of some
\begin{equation}
    \mathbf{X} = \begin{pmatrix}
        x_1 \\
        -3.25 \\
        x_2 \\
        -3.25
    \end{pmatrix} \in S_2
\end{equation}
to be
\begin{equation}
    \xi_2 = \frac{1}{\sigma_{A2}}\left|\begin{pmatrix}
        x_1 \\ x_2
    \end{pmatrix} - \langle A_2\rangle\right|
\end{equation}

Recall from Section \ref{basins-of-attraction} that the goal of our basin classification method is to find a function $P(\xi)$ that represents the probability that a randomly selected initial state from an $n$-dimensional ball with radius $\xi$ centered at $\langle A\rangle$ is in a given basin. We know from Equation \ref{eq:probability-function-power-law} that in the limit $\xi\to\infty$, this function follows the power law
\begin{equation}
    P(\xi) = \frac{P_0}{\xi^{\gamma}}
    \label{eq:probability-function-power-law-2}
\end{equation}
where $P_0$ and $\gamma$ are parameters that we use to classify a basin. Our first goal for basin classification is then to find two functions $P_w(\xi_2)$ and $P_b(\xi_2)$, which are the probability functions associated with the white (chaotic) and black (non-chaotic) basins, respectively. In Appendix \ref{classifying-asym-basin-code}, we implement the Monte Carlo algorithm outlined in Section \ref{basins-of-attraction} for classifying basins by altering the code in Appendix \ref{classifying-henon-basin-code}, which we used to classify the basin of the Hénon map. The first alteration we made was changing the way we detect which basin a given state goes to by using Lyapunov exponents rather than a direct location check. Another alteration we made was that because we are testing two basins, we added a boolean to be able to choose which basin we want to be testing for. One important numerical note is that, given we are testing values of $\xi_2=2^k$, we must choose the maximum value of $k$ carefully. This is because it must be large enough that we capture the nature of the basins far away from the attractors since we are interested in the limit $\xi\to\infty$, but it also must be small enough that states get attracted to the attractor fast enough for an accurate Lyapunov exponent calculation on the attractor. By experiment, we determine that a good maximum value of $k$ is 11.

\begin{table}[t]
    \centering
    \begin{tabular}{c|c|c}
        $\xi_2$ & $P_w(\xi_2)$ & $P_b(\xi_2)$ \\
        \hline \\ [-10px]
        $2^6=64$ & 0.8365 & 0.1530 \\
        $2^7=128$ & 0.8736 & 0.1237 \\
        $2^8=256$ & 0.8919 & 0.1010 \\
        $2^9=512$ & 0.9148 & 0.0839 \\
        $2^{10}=1024$ & 0.9419 & 0.0675 \\
        $2^{11}=2048$ & 0.9360 & 0.0544 \\
    \end{tabular}
    \caption{Some approximate $P(\xi_2)$ values of the white and black basins of the asymmetrically coupled Rulkov 1 neuron system, both directly calculated using the code in Appendix \ref{classifying-asym-basin-code}}
    \label{tab:p_function_asym_2_values}
\end{table}

In Table \ref{tab:p_function_asym_2_values}, we show the results from running the code in Appendix \ref{classifying-asym-basin-code}, where we neglect values of $k$ from 0 to 5 as we are interested in large values of $\xi_2$. Our first observation is that $P_w(\xi_2)$ is behaving a bit oddly, as Equation \ref{eq:probability-function-power-law-2} indicates to us that $P(\xi)$ functions should either decrease or stay the same. For this reason, we turn our attention to $P_b(\xi_2)$, which is behaving as we expect. Running a linear regression calculation for $\log_2 P_b(\xi_2)$ against $\log_2\xi_2$, we get
\begin{equation}
    \log_2 P_b(\xi_2) = -0.2957\log_2\xi_2 - 0.9360
    \label{eq:asym-basin-2-lin-reg}
\end{equation}
Since $\gamma\approx0.2957$ is between 0 and $n=2$, we can conclude from Section \ref{basins-of-attraction} that the basin slice of the non-chaotic spiking attractor is Class 3, meaning it extends to infinity in some directions but takes up an increasingly small fraction of state space. This makes sense because the farther away we start from the steady spiking attractor, the less chance there will be for the neurons to immediately synchronize with each other. In the limit $\xi_2\to\infty$, we expect that the only initial conditions that get attracted to the spiking attractor immediately are the ones where $x_{1,\,0}=x_{2,\,0}$, which is the diagonal shown in Figure \ref{fig:asym_basin_graphs}. For this reason, we can conclude that the black basin slice in $S_2$ has finite measure since an infinitely long line contributes nothing to the measure of the two-dimensional black basin slice.\footnote{See Property 4 of measures (Equation \ref{eq:measure-property-4}) in Section \ref{strangeattractors}.} To classify the white basin, we must make the realization that 
\begin{equation}
    P_w(\xi_2) + P_b(\xi_2) = 1
    \label{eq:sum-the-basins}
\end{equation}
which is the case because of the way we defined the spiking attractor and chaotic pseudo-attractor: if an initial state does not immediately get attracted to a spiking orbit with a negative $\lambda_1$, then it goes to a chaotic orbit with a positive $\lambda_1$. Therefore, there is no other basin that would cause the probability of being in either the white or black basin to not be 100\%. This fact can be confirmed by adding the Monte Carlo data in Table \ref{tab:p_function_asym_2_values}, which gives us values very close to 1. Therefore, even though $P_w(2^{11})$ goes down from $P_w(2^{10})$ due to the aforementioned difficulty of calculating Lyapunov exponents for orbits starting from initial states far away from the attractors, the values of $P_w(\xi_2)$ are indeed approaching 1 because from exponentiating both sides of Equation \ref{eq:asym-basin-2-lin-reg},
\begin{equation}
    P_b(\xi_2) \approx \frac{0.5227}{\xi_2^{0.2957}}
\end{equation}
which goes to 0 as $\xi_2\to\infty$. Therefore, because we are interested in this very limit, the basin slice of the chaotic pseudo-attractor must be Class 1. Although it does not take up all of $S_2$, it does take up all of $S_2$ barring a set of finite measure, which is the black basin. If we were to choose a random initial state from the entirety of $S_2$, there would indeed be a 100\% probability that the state gets attracted to the chaotic pseudo-attractor because the black basin has an infinite measure in $S_2$, whereas the white basin has only a finite measure.

Now that we have classified the basin slices that lie in $S_2$, we are now interested in the entire basins living in all of four-dimensional space $S_4$:
\begin{equation}
    S_4 = \{\mathbf{X}=\langle x_1,\,y_1,\,x_2,\,y_2\rangle:\:\mathbf{X}\in\mathbb{R}^4\}
\end{equation}
It is evident that we should define normalized distance in this set as 
\begin{equation}
    \xi_4 = \frac{|\mathbf{X}-\langle A_4\rangle|}{\sigma_{A4}}
\end{equation}
The tricky part now is how to choose a random state in a given four-dimensional ball with radius $\xi_4$ centered at $\langle A_4\rangle$. In two dimensions, this is easy; we simply pick a random number between $0$ and $\xi$ that we say is the distance $r$ away from $\langle A\rangle$ and pick another random number between $0$ and $2\pi$ that we say is the angle $\phi$ from the positive $x_1$ axis. Then, our random initial state $\mathbf{x}_0$ from the two-dimensional disk $|\mathbf{x}-\langle A\rangle|<\xi$ is
\begin{equation}
    \mathbf{x}_0 = \langle A\rangle + 
    \begin{pmatrix}
        r\cos\phi \\
        r\sin\phi
    \end{pmatrix}
\end{equation}
This is the method we implement in Appendices \ref{classifying-henon-basin-code} and \ref{classifying-asym-basin-code} to choose a random state from a given two-dimensional ball. However, the way to do this in four dimensions is not at all obvious, so this challenge in computing $P(\xi_4)$ is worth discussion. If we temporarily take things down a dimension, the analogous way to pick a random state $\mathbf{x}_0$ in three dimensions is to use spherical coordinates. Using physicists' notation, any point in a three-dimensional ball with radius $\xi$ centered at some point $\langle A\rangle\in\mathbb{R}^3$ can be described in spherical coordinates using three numbers
\begin{equation}
    \begin{cases}
        r\in[0,\,\xi) \\
        \theta\in[0,\,\pi) \\
        \phi\in[0,\,2\pi) 
    \end{cases}
\end{equation}
Then, selecting random values of $r$, $\phi$, and $\theta$ in the intervals gives us a random initial state $\mathbf{x}_0$ that can be calculated with
\begin{equation}
    \mathbf{x}_0 = \langle A\rangle + \begin{pmatrix}
        r\sin\theta\cos\phi \\
        r\sin\theta\sin\phi \\
        r\cos\theta
    \end{pmatrix}
\end{equation}
We now have a way of solving our original problem of choosing a random state from a four-dimensional ball with radius $\xi_4$ centered at $\langle A\rangle$ if we have a four-dimensional analogy to spherical coordinates. In a paper by Blumenson \cite{blumenson}, the $n$-dimensional coordinate system analogous to spherical coordinates is derived using linear algebra, so we can utilize these results for our four-dimensional case. The four coordinates used are
\begin{equation}
    \begin{cases}
        r\in[0,\,\xi_4) \\
        \theta_1\in[0,\,\pi) \\
        \theta_2\in[0,\,\pi) \\
        \phi\in[0,\,2\pi) 
    \end{cases}
\end{equation}
Then, choosing these numbers randomly from the intervals, our random initial state $\mathbf{X}_0\in S_4$ is given by
\begin{equation}
    \mathbf{X}_0 = \langle A_4\rangle + \begin{pmatrix}
        r\sin\theta_1\sin\theta_2\cos\phi \\
        r\sin\theta_1\sin\theta_2\sin\phi \\
        r\sin\theta_1\cos\theta_2 \\
        r\cos\theta_1
    \end{pmatrix}
\end{equation}

Now that we know how to pick random states in $S_4$ to test, our Monte Carlo algorithm for classifying the full basins in four-dimensional space is a natural extension from our two-dimensional version, and we display the code used to do it in Appendix \ref{asym-elec-coup-rulkov-1-neurons-code}. However, we wish to make a few numerical notes before we detail our analysis. First, because $y$ is a slow variable, starting far away from where it eventually ends up (around $y=-3.25$) will inevitably result in the state taking even longer to reach its attractor. For this reason, we up the number of times we iterate test points for Lyapunov exponent calculation from 5000 to 20000, which raises computation time significantly. Additionally, recall from Section \ref{basins-of-attraction} that we use a shell method for calculating higher values of $P(\xi)$. As a quick reminder, our shell method says that given we know $P(2^k)$ and $\Delta P(2^k)$, which is the probability that an initial state chosen from the $n$-dimensional shell with inner radius $2^k$ and outer radius $2^{k+1}$ centered at $\langle A\rangle$ is in the basin, we can calculate $P(2^{k+1})$ using Equation \ref{eq:iteration-of-shell-method}, which says that
\begin{equation}
    P(2^{k+1}) = \frac{P(2^k)}{2^n} + \left(1-\frac{1}{2^n}\right)\Delta P(2^k)
\end{equation}
This depends on the dimension of the state space $n$ that the basin lives in because of the way objects scale differently based on their dimension.\footnote{See Section \ref{strangeattractors} for a reminder.} Therefore, we must alter our shell method calculations because we are dealing with four-dimensional state space as opposed to a two-dimensional one.

In Appendix \ref{classifying-asym-basin-code}, we implement our spherical coordinate method of choosing random states in four-dimensional balls and the discussed numerical considerations to calculate $P(\xi_4)$ values for the white and black basins in $S_4$. The results of this code for the white basin $P_w(\xi_4)$ are shown in Table \ref{tab:p_function_asym_4_values}. In the table, $P_b(\xi_4)$ values are indirectly calculated by using the fact that $P_w(\xi_4)+P_b(\xi_4)=1$, which follows directly from Equation \ref{eq:sum-the-basins}. Observing the values in Table \ref{tab:p_function_asym_4_values}, it appears that $P_w(\xi_4)$ and $P_b(\xi_4)$ stay relatively constant as we vary $\xi_4$. Given our analysis of the two-dimensional basin slices, this is a rather unexpected result, so we will proceed carefully with our analysis. 

\begin{table}[t]
    \centering
    \begin{tabular}{c|c|c}
        $\xi_4$ & $P_w(\xi_4)$ & $P_b(\xi_4) = 1-P_w(\xi_4)$ \\
        \hline \\ [-10px]
        $2^0=1$ & 0.8480 & 0.1520 \\
        $2^1=2$ & 0.8649 & 0.1351 \\
        $2^2=4$ & 0.8491 & 0.1509 \\
        $2^3=8$ & 0.8537 & 0.1463 \\
        $2^4=16$ & 0.8690 & 0.1310 \\
        $2^5=32$ & 0.8549 & 0.1451 \\
        $2^6=64$ & 0.8766 & 0.1234 \\
        $2^7=128$ & 0.8760 & 0.1240 \\
        $2^8=256$ & 0.8629 & 0.1371 \\
    \end{tabular}
    \caption{Some approximate $P(\xi_4)$ values of the white and black basins of the asymmetrically coupled Rulkov 1 neuron system, $P_w(\xi_4)$ values directly calculated using the code in Appendix \ref{classifying-asym-basin-code}, $P_b(\xi_4)$ indirectly calculated using the $P_w(\xi_4)$ values}
    \label{tab:p_function_asym_4_values}
\end{table}

Running linear regressions for the log-log plots of both $P_w(\xi_4)$ and $P_b(\xi_4)$, we get that 
\begin{equation}
    P_w(\xi_4) = 0.0042\xi_4 - 0.2315
    \label{eq:asym-white-basin-4-lin-reg}
\end{equation}
with $R^2=0.392$, and
\begin{equation}
    P_b(\xi_4) = -0.0260\xi_4 - 2.7540
    \label{eq:asym-black-basin-4-lin-reg}
\end{equation}
with $R^2=0.386$. These low $R^2$ values confirm our suspicion that there is no upward or downward trend as we change $\xi_4$. This means that within the numerical variability of the Monte Carlo algorithm, the value of the basin classification parameter $\gamma$ is effectively 0. The value of the other basin classification parameter $P_0$ can be found by exponentiating both sides of Equations \ref{eq:asym-white-basin-4-lin-reg} and \ref{eq:asym-black-basin-4-lin-reg}, which yields that for the white basin, $P_0=0.8517$, and for the black basin, $P_0=0.1482$. These are both between 0 and 1, so by the classification method established in Section \ref{basins-of-attraction}, both the white and black basins are Class 2, meaning they both occupy a fixed fraction of four-dimensional state space. This makes sense considering the data in Table \ref{tab:p_function_asym_4_values}, but again, it is unexpected. Our explanation for why the white and black basins are Class 2 is that since Rulkov map 1 is a slow-fast system, $x_1$ and $x_2$ adjust much faster than $y_1$ and $y_2$. Therefore considering some initial condition $\mathbf{X}_0=\langle x_{1,\,0},\,y_{1,\,0},\,x_{2,\,0},\,y_{2,\,0} \rangle$ some distance $\xi_4$ away from either attractor $A_4$, $y_1$ and $y_2$ will slowly drift towards $\langle y_1\rangle$ and $\langle y_2\rangle$ while $x_1$ and $x_2$ easily keep up with their slow attraction. By the time $y_1$ and $y_2$ find their way to $\langle y_1\rangle$ and $\langle y_2\rangle$, $x_1$ and $x_2$ will also be near $\langle x_1\rangle$ and $\langle x_2\rangle$ regardless of how much farther away they started compared to $y_1$ and $y_2$ since they evolve so much more quickly. Thus, once $y_1$ and $y_2$ get close to the attractors, we are effectively starting in $S_2'$ (or a square of similar size close and parallel to it). Indeed, comparing the fractions of state space the white ($P_0=0.8517$) and black ($P_0=0.1482$) basins take up to the $P_w(\xi_2)$ and $P_b(\xi_2)$ fractions for the small values of $\xi_2$ in Table \ref{tab:p_function_asym_2_values}, we find that these align quite well.

\begin{figure*}[htp!]
    \centering
    \hspace{2.35cm}
    \begin{subfigure}{0.7\textwidth}
        \centering
        \includegraphics[scale=0.23]{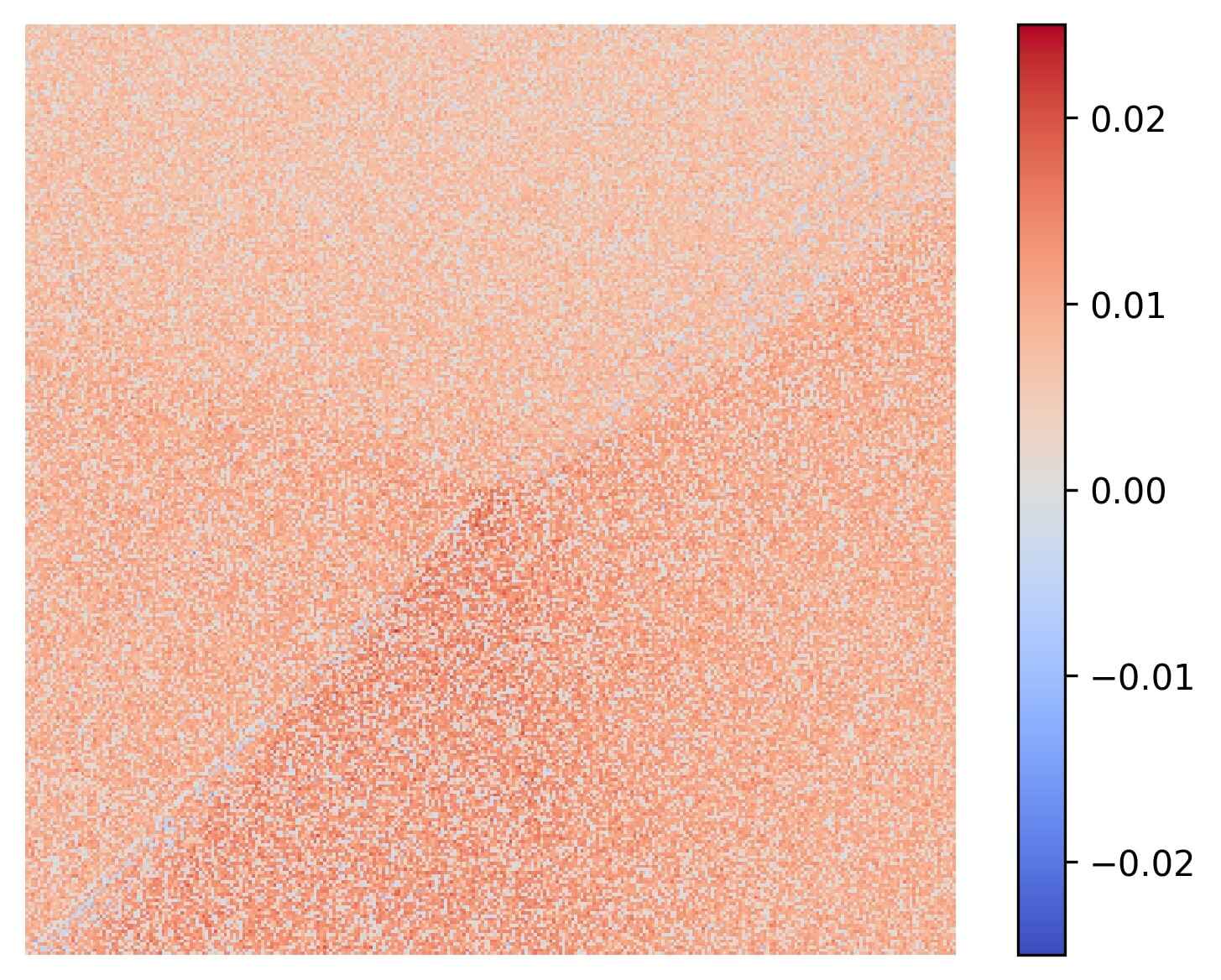}
        \caption{Maximal Lyapunov exponent color map, with red points indicating chaotic \\ dynamics and blue points indicating non-chaotic dynamics}
        \label{fig:asym_basin_color_y}
    \end{subfigure}
    \begin{subfigure}{0.7\textwidth}
        \centering
        \includegraphics[scale=0.23]{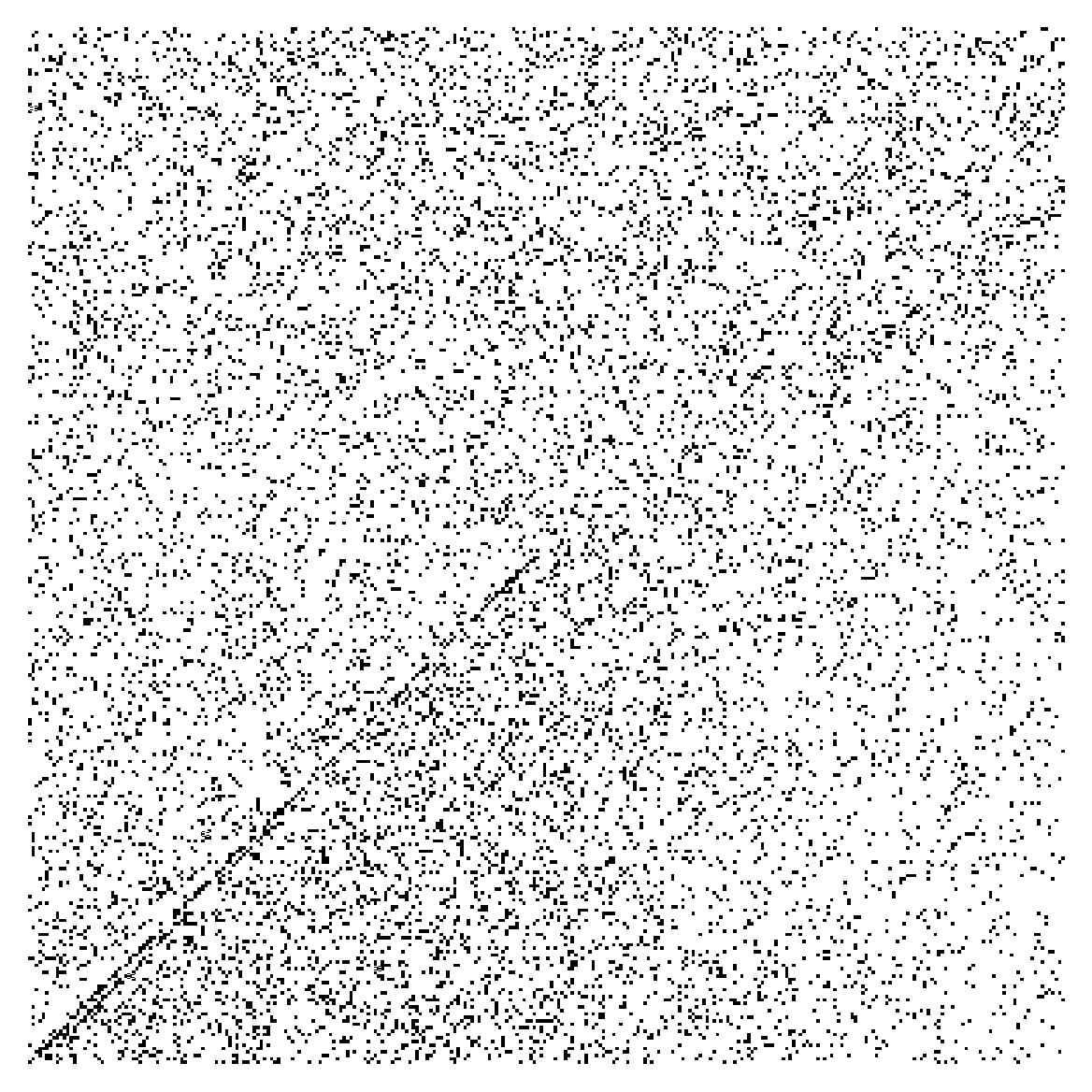}
        \caption{Basin slices of the chaotic (white) and non-chaotic spiking (black) attractors}
        \label{fig:asym_basin_bw_y}
    \end{subfigure}
    \caption{A slice of the basins of attraction of two asymmetrically electrically coupled Rulkov 1 neurons with $-256\leq y_{1,\,0}\leq 256$ on the horizontal axis and $-256\leq y_{2,\,0}\leq 256$ on the vertical axis, parameters $\sigma_1=\sigma_2=-0.5$, $\alpha_1=\alpha_2=4.5$, coupling strengths $g^e_1=0.05$, $g^e_2=0.25$, and fixed initial $x$ values $x_{1,\,0}=-1$ and $x_{2,\,0}=1$, visualized using the code in Appendix \ref{visualize-basins-asym-coup-rulkov-1-code}}
    \label{fig:asym_basin_graphs_y}
\end{figure*}

Although we cannot truly visualize these four-dimensional basins, we can get a grasp of this Class 2 basin behavior by graphing a different two-dimensional slice of the basins, namely, one where we vary $y_1$ and $y_2$ but keep $x_1$ and $x_2$ fixed. In Figure \ref{fig:asym_basin_graphs_y}, we graph a large range of initial $y$ values $-256\leq y_{1,\,0}\leq 256$ and $-256\leq y_{2,\,0}\leq 256$ while keeping the initial $x$ values fixed at $x_{1,\,0}=-1$ and $x_{2,\,0}=1$. In Figure \ref{fig:asym_basin_bw_y}, we can see a dispersed, seemingly random distribution of white and black basin points with no definite pattern in their arrangement. Similarly in Figure \ref{fig:asym_basin_color_y}, we can see that the maximal Lyapunov exponent also appears random, perhaps with the exception of a region of darker red points. What is important to notice from these graphs is that the distribution of white and black points in Figure \ref{fig:asym_basin_bw_y} appears to remain the same no matter where we are on the graph even though we are spanning a huge range of different $y$ values. This is characteristic of Class 2 basins, as the fraction of state space they take up doesn't depend on the distance from their attractors. Something else that we notice in Figure \ref{fig:asym_basin_bw_y} is a lack of the black diagonal that appears in Figure \ref{fig:asym_basin_bw}. This line of stability doesn't appear in these graphs because, as previously mentioned, it is defined by the equation 
\begin{equation}
    x_{1,\,0} = y_{1,\,0} = x_{2,\,0} = y_{2,\,0}
\end{equation}
which the slice in Figure \ref{fig:asym_basin_graphs_y} doesn't intersect since the initial states shown in it have different $x$ values.

\subsubsection{Basin Boundary Analysis}

From Figures \ref{fig:asym_basin_graphs} and \ref{fig:asym_basin_graphs_y}, it is clear that the basins of the non-chaotic spiking attractor and strange pseudo-attractor are not divided by a clear boundary. This naturally leads us to suspect that the basin boundary $\Sigma$ between the white and black basins might be fractal, which we know from Section \ref{basins-of-attraction} leads to geometrical sensitivity to initial conditions with an uncertainty exponent $\mathfrak{u}<1$. Similar to how we classified the intersection of the system's basins with $S_2$ and $S_4$, we are interested in examining the uncertainty exponents and the fractalization of both the intersection of $\Sigma$ with $S_2'$ and the intersection of $\Sigma$ with $S_4'$. We have already defined $S_2'$ as a specific square-shaped subset of $S_2$ (Equation \ref{eq:s2'}), so we will now define $S_4'$ as the subset of $S_4$ 
\begin{multline}
    S_4'= \{\mathbf{X} = \langle x_1,\,y_1,\,x_2,\,y_2\rangle: \\-2<x_1<2,\,-1<y_1<-5, \\ -2<x_2<2,\,-1<y_2<-5\}
    \label{eq:s4'}
\end{multline}
which is a four-dimensional cube, or tesseract.

First, let us examine the set $\Sigma\cap S_2'$, displayed in Figure \ref{fig:asym_basin_bw} as the boundary between the white and black basins. We will denote the uncertainty exponent of this basin boundary set as $\mathfrak{u}_2$ and its associated probability function as $\varrho_2(\epsilon)$. As a reminder, $\varrho_2(\epsilon)$ is the probability that a randomly selected state in $S_2'$ is uncertain, and we expect it to be proportional to $\epsilon^{\mathfrak{u}_2}$. Calculating $\varrho_2(\epsilon)$ for our coupled neuron system mirrors the Monte Carlo method we used for the Hénon map in Section \ref{basins-of-attraction}: we pick a random initial state $\mathbf{X}_0\in S_2'$ and test four specific perturbed states
\begin{equation}
    \begin{pmatrix}
        x_{1,\,0}+\epsilon \\
        -3.25 \\
        x_{2,\,0} \\
        -3.25
    \end{pmatrix},\,
    \begin{pmatrix}
        x_{1,\,0}-\epsilon \\
        -3.25 \\
        x_{2,\,0} \\
        -3.25
    \end{pmatrix},\,
    \begin{pmatrix}
        x_{1,\,0} \\
        -3.25 \\
        x_{2,\,0}+\epsilon \\
        -3.25
    \end{pmatrix},\,
    \begin{pmatrix}
        x_{1,\,0} \\
        -3.25 \\
        x_{2,\,0}-\epsilon \\
        -3.25
    \end{pmatrix}
\end{equation}
to see if any of them end up in a different basin than $\mathbf{X}_0$ does. The only major difference in calculating $\mathfrak{u}_2$ for this system as compared to the Hénon map is that we detect which basin an initial state belongs to using maximal Lyapunov exponents, a distinction that we have already discussed at length.

Although we cannot visualize it, calculating the uncertainty exponent of the basin boundary living in four-dimensional state space $\Sigma\cap S_4'$ is a natural extension from calculating $\mathfrak{u}_2$. Similar to before, let us denote the uncertainty exponent of this four-dimensional basin boundary set as $\mathfrak{u}_4$ and its associated probability function as $\varrho_4(\epsilon)$. The step of picking a random initial state $\mathbf{X}_0\in S_4'$ is much easier than picking a random initial state from a four-dimensional ball (which we did for basin classification) since we chose our region of analysis $S_4'$ to be a simple tesseract. Instead of testing four perturbed initial states, however, we need eight of them because we are now considering a four-dimensional region. Explicitly, these perturbed states are 
\begin{equation}
    \begin{gathered}
        \begin{pmatrix}
            x_{1,\,0}+\epsilon \\
            y_{1,\,0} \\
            x_{2,\,0} \\
            y_{2,\,0}
        \end{pmatrix},\,
        \begin{pmatrix}
            x_{1,\,0}-\epsilon \\
            y_{1,\,0} \\
            x_{2,\,0} \\
            y_{2,\,0}
        \end{pmatrix},\,
        \begin{pmatrix}
            x_{1,\,0} \\
            y_{1,\,0}+\epsilon \\
            x_{2,\,0} \\
            y_{2,\,0}
        \end{pmatrix},\,
        \begin{pmatrix}
            x_{1,\,0} \\
            y_{1,\,0}-\epsilon \\
            x_{2,\,0} \\
            y_{2,\,0}
        \end{pmatrix}, \\
        \begin{pmatrix}
            x_{1,\,0} \\
            y_{1,\,0} \\
            x_{2,\,0}+\epsilon \\
            y_{2,\,0}
        \end{pmatrix},\,
        \begin{pmatrix}
            x_{1,\,0} \\
            y_{1,\,0} \\
            x_{2,\,0}-\epsilon \\
            y_{2,\,0}
        \end{pmatrix},\,
        \begin{pmatrix}
            x_{1,\,0} \\
            y_{1,\,0} \\
            x_{2,\,0} \\
            y_{2,\,0}+\epsilon
        \end{pmatrix},\,
        \begin{pmatrix}
            x_{1,\,0} \\
            y_{1,\,0} \\
            x_{2,\,0} \\
            y_{2,\,0}-\epsilon
        \end{pmatrix}\phantom{,}
    \end{gathered}
\end{equation}

\begin{table}[t]
    \centering
    \begin{tabular}{c|c|c}
        $\epsilon$ & $\varrho_2(\epsilon)$ & $\varrho_4(\epsilon)$ \\
        \hline \\ [-10px]
        $2^0=1$ & 0.710 & 0.404 \\
        $2^{-1}=1/2$ & 0.714 & 0.416 \\
        $2^{-2}=1/4$ & 0.658 & 0.394 \\
        $2^{-3}=1/8$ & 0.571 & 0.392 \\
        $2^{-4}=1/16$ & 0.516 & 0.339 \\
        $2^{-5}=1/32$ & 0.409 & 0.349 \\
        $2^{-6}=1/64$ & 0.340 & 0.341 \\
        $2^{-7}=1/128$ & 0.255 & 0.328 \\
        $2^{-8}=1/256$ & 0.219 & 0.323 \\
        $2^{-9}=1/512$ & 0.172 & 0.308 \\
        $2^{-10}=1/1024$ & 0.126 & 0.306 \\
        $2^{-11}=1/2048$ & 0.092 & 0.302 \\
    \end{tabular}
    \caption{Some approximate $\varrho(\epsilon)$ values for the basin boundary sets $\Sigma\cap S_2'$ and $\Sigma\cap S_4'$ of the asymmetrically coupled Rulkov 1 neuron system, calculated using the code in Appendix \ref{uncert-exp-asym-coup-code}}
    \label{tab:uncertainty_exp_asym_coup_values}
\end{table}

Now that we have discussed the theory behind calculating $\mathfrak{u}_2$ and $\mathfrak{u}_4$, let us begin our analysis. Because $S_2'$ and $S_4'$ are relatively small and close to the attractors, we can safely determine which basin an initial state in either set belongs to by calculating its Lyapunov spectrum using a 5000 iteration long orbit. In Appendix \ref{uncert-exp-asym-coup-code}, we implement the theory and numerical considerations we discussed to calculate the values of the probability functions $\varrho_2(\epsilon)$ and $\varrho_4(\epsilon)$, the results of which are displayed in Table \ref{tab:uncertainty_exp_asym_coup_values}. Taking a linear regression of the log-log plot of $\varrho_2(\epsilon)$ first (neglecting the first two points because we are interested in the limit $\epsilon\to 0$), we get that 
\begin{equation}
    \log_2\varrho_2(\epsilon) = 0.314\log_2\epsilon + 0.196
\end{equation}
with an $R^2$ value of 0.986. This indicates that $\mathfrak{u}_2\approx 0.314$, and since this is less than one, some amount of geometrical sensitivity to initial conditions does indeed exist in $S_2'$. Recalling from Equation \ref{eq:uncertainty-exp-frac-dim} that an uncertainty exponent is related to the fractal dimension of its associated basin boundary, we can say that the fractal dimension $d$ of $\Sigma\cap S_2'$ is
\begin{equation}
    \begin{split}
        d &= n - \mathfrak{u}_2 \\
         &\approx 2 - 0.314 = 1.686
    \end{split}
\end{equation}
Observing the visualization of $S_2'$ in Figure \ref{fig:asym_basin_bw}, we can see that there are some black regions that have boundaries that appear to be smooth, meaning that these particular subsets of the basin boundary have fractal dimensions close to 1. This accounts for the fractal dimension of $\Sigma\cap S_2'$ not being exceptionally close to 2: these boundaries that appear smooth make the overall basin boundary in this slice less ``rough.'' 

Now, let us take a look at the function $\varrho_4(\epsilon)$. Running a linear regression on the values $\log_2\varrho_4(\epsilon)$ vs. $\log_2\epsilon$ from Table \ref{tab:uncertainty_exp_asym_coup_values} gives us
\begin{equation}
    \log_2\varrho_4(\epsilon) = 0.037\log_2\epsilon - 1.341
\end{equation}
with an $R^2$ value of 0.967. This indicates $\mathfrak{u}_4\approx 0.037$, which is exceedingly small and the most extreme example of unpredictability emerging from basin geometry that we have encountered so far in this paper. To put this in perspective, the fact that $\varrho_4(\epsilon)$ is proportional to $\epsilon^{\mathfrak{u}_4}$ means that to reduce the uncertainty in which attractor an initial state in $S_4'$ will end up in by a factor of 10, we will need to reduce our initial uncertainty $\epsilon$ by a factor on the order of $10^{27}$. Using Equation \ref{eq:uncertainty-exp-frac-dim} again, we can calculate that the fractal dimension of the basin boundary $\Sigma\cap S_4'$ is
\begin{equation}
    \begin{split}
        d &= n - \mathfrak{u}_4 \\
         &\approx 4 - 0.037 = 3.963
    \end{split}
\end{equation}
which means that this basin boundary is extremely ``rough,'' behaving similarly to a four-dimensional object even though it divides four-dimensional space. This amount of roughness is analogous to the surface of a human lung ($d\approx 2.97$), just raised one dimension \cite{lung}. We can see this high fractal dimension justified in Figure \ref{fig:asym_basin_bw_y}. Even though Figure \ref{fig:asym_basin_bw_y} doesn't show the four-dimensional set $S_4'$, it is clear that the boundary dividing the seemingly random distribution of white and black points must be close to the dimension of the state space the basins live in. One numerical note is that the $R^2$ value for the linear regression associated with $\varrho_4(\epsilon)$, while certainly telling of a correlation, is less than the $R^2$ value for the regression associated with $\varrho_2(\epsilon)$. We conjecture that this is the case because of the significantly higher uncertainty associated with $\varrho_4(\epsilon)$.

We will conclude this section with a short discussion on how the uncertainty exponents $\mathfrak{u}_2$ and $\mathfrak{u}_4$ change as we expand out from $S_2'$ and $S_4'$. Although the basin classification method we used to classify the white and black basins of this asymmetrically coupled Rulkov neuron system doesn't take into account basin boundaries, we conjecture that $\mathfrak{u}_2$ increases, approaching 1, as we expand the bounds of $x_1$ and $x_2$ away from the set $S_2'$ because the white basin dominates the farther away we go from the attractors. Additionally, we conjecture that $\mathfrak{u}_4$ stays relatively constant as we expand the bounds of $S_4'$ because the white and black basins are both Class 2, so we suspect that they stay similarly overlapped with each other. These conjectures are supported by running the code in Appendix \ref{uncert-exp-asym-coup-code} for different ranges of $x$ and $y$ values.

\subsection{Ring Lattice of \texorpdfstring{$\zeta=30$}{TEXT} Rulkov 1 Neurons}
\label{ring-lattice-geometry}

In Section \ref{neuron-ring-lattice}, we examined three systems of $\zeta=30$ electrically coupled Rulkov 1 neurons in a ring lattice. As a reminder, they are as follows:
\begin{enumerate}
    \item Random fast variable initial states $x_{i,\,0}\in (-1,\,1)$, identical slow variable initial states $y_{i,\,0} = -3.25$, identical $\sigma_i$ values $\boldsymbol{\sigma} = -0.5\cdot\mathbf{1}$, and identical $\alpha_i$ values $\boldsymbol{\alpha} = 4.5\cdot\mathbf{1}$
    \item Random fast variable initial states $x_{i,\,0}\in (-1,\,1)$, identical slow variable initial states $y_{i,\,0} = -3.25$, random $\sigma_i$ values $\sigma\in(-1.5,\,-0.5)$, and identical $\alpha_i$ values $\boldsymbol{\alpha} = 4.5\cdot\mathbf{1}$
    \item Random fast variable initial states $x_{i,\,0}\in (-1,\,1)$, identical slow variable initial states $y_{i,\,0} = -3.25$, random $\sigma_i$ values $\sigma\in(-1.5,\,-0.5)$, and identical $\alpha_i$ values $\alpha_i\in(4.25,\,4.75)$
\end{enumerate}
For the specific initial state $\mathbf{X}_0$, $\boldsymbol{\sigma}$ vector, and $\boldsymbol{\alpha}$ vector we use in this paper, see Equations \ref{eq:big-initial-state}, \ref{eq:big-sigma-vector}, and \ref{eq:big-alpha-vector}, respectively. As we discovered in Section \ref{neuron-ring-lattice}, these systems nearly always exhibit chaotic behavior with positive Lyapunov exponents. Therefore, we can conclude that these systems get attracted to some chaotic attractor in 60-dimensional state space. A 60-dimensional space is so far beyond human comprehension that we will not even attempt to present visualizations of these attractors by projecting them onto two-dimensional surfaces. It would certainly be interesting to analyze the geometry of these 60-dimensional attractors by calculating their fractal dimension $d$, but unfortunately, we do not have the computing power to calculate $d$ in 60-dimensional space for a large number of $g^e$ values. Instead, in this section, we will analyze the geometry of these attractors by approximating their Lyapunov dimensions $d_l$, which we know from the Kaplan-Yorke conjecture at the end of Section \ref{strangeattractors} gives a ballpark approximation (since we are calculating the Lyapunov spectrum for only a thousand iterations) for the true fractal dimensions of these attractors.

\begin{figure*}[hp!]
    \centering
    \begin{subfigure}{0.9\textwidth}
        \centering
        \includegraphics[scale=0.2]{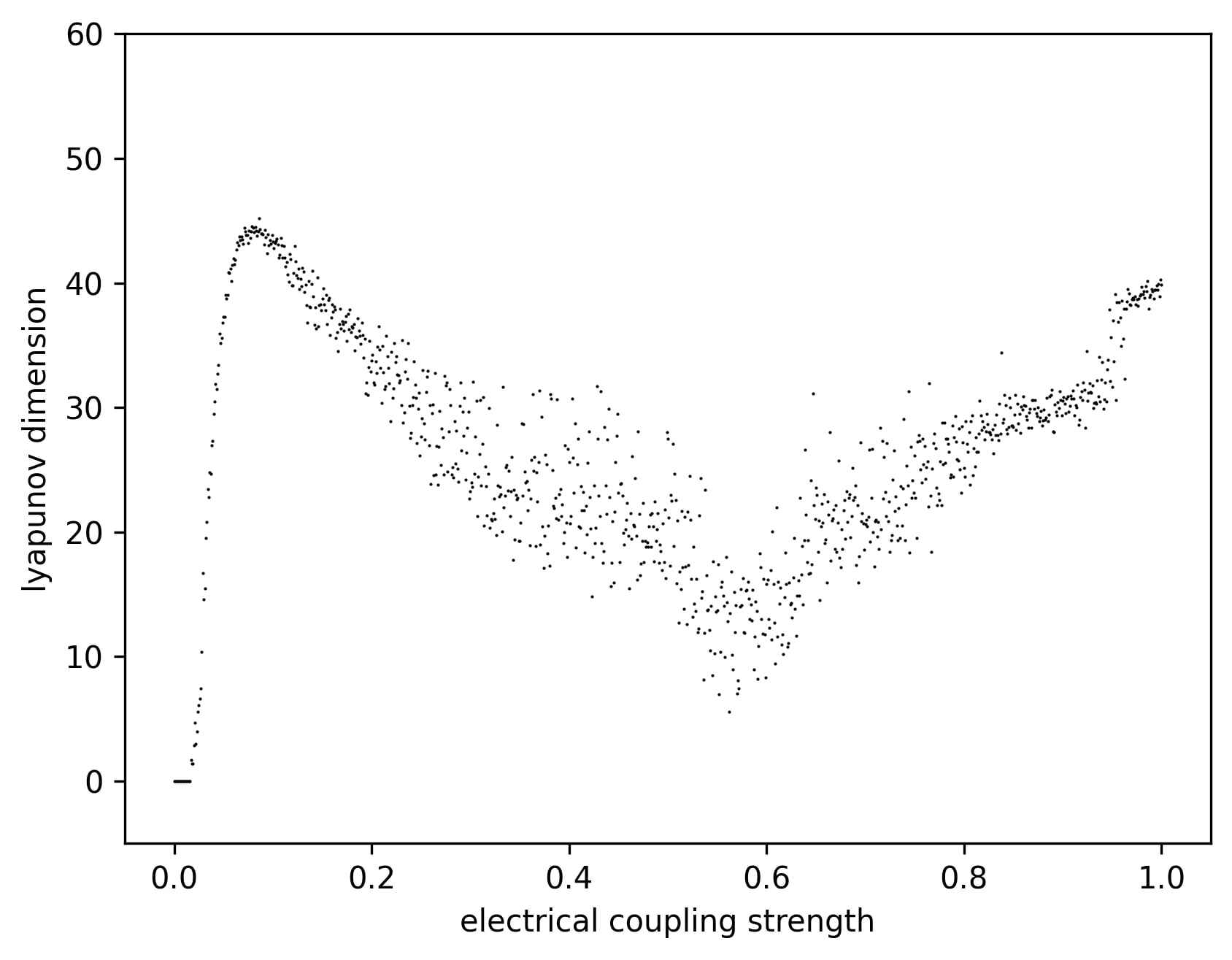}
        \caption{First ring lattice system}
        \label{fig:dl-random-x}
        \vspace{0.75cm}
    \end{subfigure}
    \begin{subfigure}{0.475\textwidth}
        \centering
        \includegraphics[scale=0.1375]{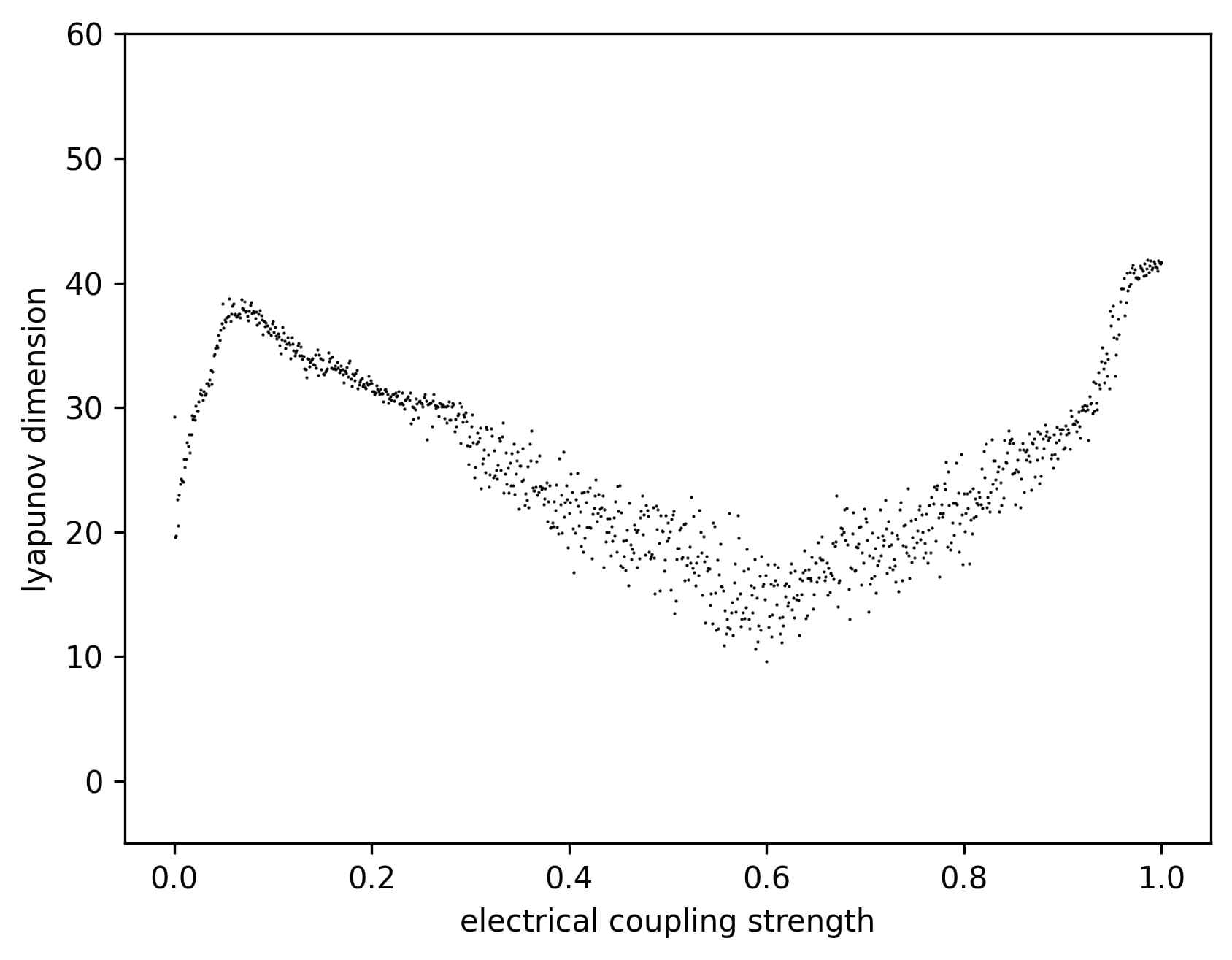}
        \caption{Second ring lattice system}
        \label{fig:dl-random-sigma}
        \vspace{0.5cm}
    \end{subfigure}
    \begin{subfigure}{0.475\textwidth}
        \centering
        \includegraphics[scale=0.1375]{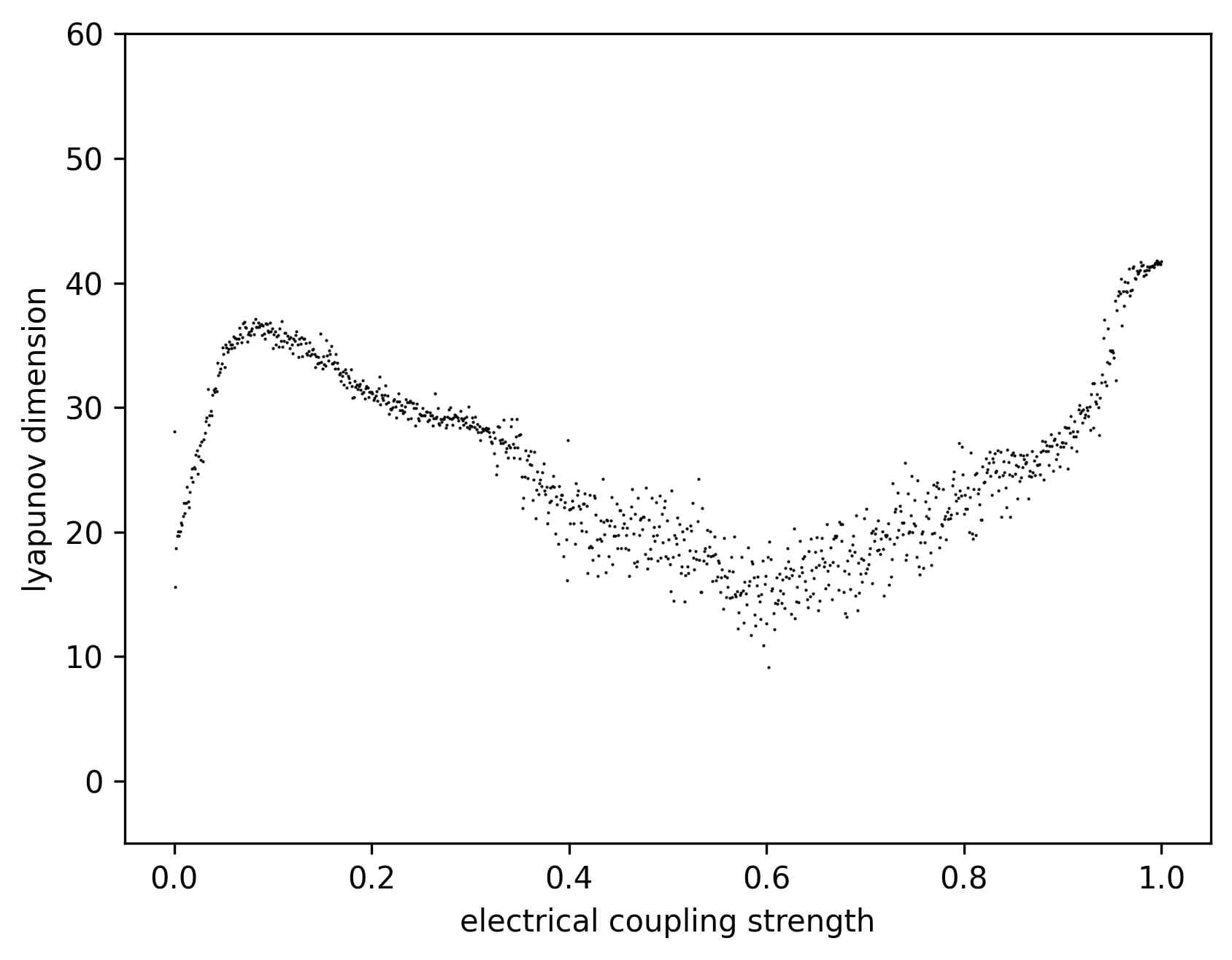}
        \caption{Third ring lattice system}
        \label{fig:dl-random-alpha}
        \vspace{0.5cm}
    \end{subfigure}
    \caption{Graphs of the Lyapunov dimension $d_l$ against the electrical coupling strength $g^e$ for a ring of $\zeta=30$ electrically coupled Rulkov 1 neurons, visualized using the code in Appendix \ref{lyap-exp-and-dim-graphs-code}}
    \label{fig:dl-graphs}
\end{figure*}

Since we are dealing with 60 Lyapunov exponents, it will be more convenient to calculate Lyapunov dimensions by implementing our method of calculating them from Section \ref{strangeattractors} (Equations \ref{eq:kappa} and \ref{eq:dl}) into code, which we do in Appendix \ref{lyap-exp-and-dim-graphs-code}.\footnote{The function that accomplishes this in the appendix looks a little strange compared to Equations \ref{eq:kappa} and \ref{eq:dl} because of the fact that Python indexes from 0 and we index our Lyapunov exponents from 1.} Using this code, we can make graphs similar to the ones in Figures \ref{fig:max_lyap_exp_graph_random_x} and \ref{fig:ring-max-lyap-graphs-rand-sigandalph}, plotting the values of the Lyapunov dimension $d_l$ for many different values of $g^e$, which we do in Figure \ref{fig:dl-graphs}. 

There are many interesting observations we can make about the graphs shown in Figure \ref{fig:dl-graphs}. First, we can assume that all the chaotic attractors of these three systems for different values of $g^e$ are fractal since our approximations of the Lyapunov dimensions are spread out, not sticking to any defined integers. The only true integer dimensions in these graphs are at the very left of Figure \ref{fig:dl-random-x}, where we can see a small number of attractors have dimension 0. This is associated with the non-chaotic attractors we observed in Figure \ref{fig:max_lyap_exp_graph_random_x}, which consist of a finite number of zero-dimensional points. Another notable observation is the sizable amount of state space that these attractors take up. Because the state space of this system is so large, we might expect the attractors to take up only a small fraction of state space, but instead, we see that the strange attractors take up a sizeable portion of it for many values of $g^e$, with some of the ``roughest'' of these attractors taking up close to 45 of the 60 total dimensions. 

Comparing Figure \ref{fig:dl-graphs} to the graphs of $\lambda_1$ vs. $g^e$ in Figures \ref{fig:max_lyap_exp_graph_random_x} and \ref{fig:ring-max-lyap-graphs-rand-sigandalph}, we can see that the Lyapunov dimension $d_l$ follows a similar pattern of increasing as we enter the chaotic spiking domain, decreasing as the neurons start to burst in sync with each other, then increasing again as we enter complete chaos. This is to be expected because the Lyapunov dimension is calculated directly from the Lyapunov spectrum $\lambda$. We can also see a similarity in how the $d_l$ and $\lambda_1$ values are distributed across the different systems. Specifically, the $\lambda_1$ values are more erratic and spread out in the first system than they are in the second and third systems, which is also reflected in the $d_l$ values to some degree, observing that the values of $d_l$ in Figure \ref{fig:dl-random-x} are more spread out vertically in the synchronized bursting region. However, there are some very clear differences between the trends of the maximal Lyapunov exponent $\lambda_1$ and the Lyapunov dimension $d_l$ as we vary $g^e$. The most obvious difference is in comparing the peaks of the $\lambda_1$ vs. $g^e$ graphs and the $d_l$ vs. $g^e$ graphs, with both peaks in both graphs being associated with chaotic spiking around $g^e=0.1$ and complete chaos around $g^e=1$. In the $\lambda_1$ vs. $g^e$ graphs, the peak in the region of complete chaos is always higher than the peak in the region of chaotic spiking, a fact that is especially obvious in Figure \ref{fig:ring-max-lyap-graphs-rand-sigandalph} (the second and third ring systems), where the peaks on the right dwarf the peaks on the left. However, in the graphs of $d_l$ vs. $g^e$, the peaks are similar in height, and in Figure \ref{fig:dl-random-x} (the first ring system), the left peak is actually higher than the right peak. This means that, for this system, the chaotic spiking attractor for the systems with a relatively small electrical coupling strength takes up more of state space and is even ``rougher'' than the attractor for the system of complete chaos $g^e=1$, which is a somewhat surprising result. The main thing to notice from making this comparison, however, is that the maximal Lyapunov exponent $\lambda_1$, the number we use to quantify how chaotic a system is, does not directly correlate to how ``big'' or fractal an attractor is. For that, as the Kaplan-Yorke conjecture indicates, we need the entire Lyapunov spectrum.

\section{Discussion and Conclusion}
\label{discussion-and-conclusion}

We begin by reviewing the original aims of our research and our major results. Then, we discuss our suggestions for future research and the possible implications of our research in different fields.

\subsection{Summary of Results}

The first aim of our research was to analyze and quantify the chaotic dynamics of uncoupled Rulkov neurons. In Rulkov map 1, we accomplished this by expanding on the bifurcation diagram showing the distribution of silence, spiking, and bursting behavior originally presented by Rulkov \cite{rulkov} with an approximation of the bifurcation curve $C_{\text{bs}}$ (see Figure \ref{fig:rulkov-1-bifurc-diag-param-space}). We extended this bifurcation diagram to give information about the chaotic spiking-bursting that occurs for some parameter values of Rulkov map 1 and to show a quantification of how chaotic uncoupled Rulkov 1 neurons are in the region of parameter space $-2\leq\sigma\leq 0$ and $2\leq\alpha\leq 8$ by utilizing the maximal Lyapunov exponent $\lambda_1$ (see Figure \ref{fig:rulkov_1_lyapunov_exponents_visualizations}). We also established a general algorithm for calculating the Lyapunov exponents for any Rulkov 2 neuron and used this to demonstrate the chaotic nature of the spiking and bursting orbits of Rulkov map 2.

The second aim of our research was to model the behavior and analyze the dynamics of complex Rulkov neuron systems. Expanding on the original work by Rulkov \cite{rulkov} involving the symmetrical coupling of two Rulkov 1 neurons, we explored a natural extension of this system, namely, a system of two identical spiking Rulkov 1 neurons with an asymmetrical coupling. The complex neuron systems we chose to perform an in-depth analysis of were based around a structure composed of $\zeta$ Rulkov 1 neurons arranged in a ring lattice with current flow in each neighboring connection. We established a method to calculate the $2\zeta$ Lyapunov exponents in a ring lattice system's Lyapunov spectrum through a detailed derivation of the system's $2\zeta\times 2\zeta$ Jacobian matrix, then we applied this method to analyze the dynamics of three different systems composed of $\zeta=30$ Rulkov 1 neurons arranged in a ring. As we varied the electrical coupling strength between the neurons, we discovered patterns in the dynamics of all three of these systems, including the gradual change from chaotic spiking to synchronized bursting to complete chaos.

The third aim of our research was to explore the possibility of the existence of multistability and fractal geometry in the Rulkov maps. This was a success, as we found at least one of these geometrical properties exhibited in three distinct Rulkov neuron systems. The first system where this arose was an uncoupled Rulkov 2 neuron, which we suspected had a chaotic attractor that exhibited fractal geometry because it lacked a resetting mechanism. The second system we found exhibiting these geometrical properties was the system of two asymmetrically electrically coupled Rulkov 1 neurons. From our analysis, we discovered this system could exhibit both non-chaotic and chaotic dynamics depending on the initial conditions of the two neurons. We also suspected that the system's four-dimensional chaotic attractors might be fractal despite the resetting mechanism in both neurons. Finally, the third distinct system that we suspected exhibited fractal geometry was the ring lattice of $\zeta=30$ Rulkov 1 neurons, whose chaotic dynamics we believed likely resulted in complex fractal attractors in 60-dimensional state space. With the three ring lattice systems we analyzed, we were interested in connecting our work with ``temporal'' chaotic dynamics to the ``geometrical,'' fractal form of chaos by means of the Kaplan-Yorke conjecture.

The fourth and final aim of our research was to detect, classify, and quantify the geometrical properties associated with chaos and unpredictability in Rulkov map systems. Using the detection methods we established, we first confirmed that multistability, fractal geometry, or both were indeed present in all three distinct systems that we suspected had these properties. 

For our first system, the uncoupled Rulkov 2 neuron, we used the computer model from our previous analysis of the neuron's dynamics and the code we used to count boxes on the Hénon attractor to calculate the Minkowski-Bouligand dimensions of attractors generated by a single Rulkov 2 neuron. Our analysis yielded that both chaotic spiking and bursting attractors of Rulkov map 2 are fractal and between one-dimensional and two-dimensional, with the chaotic bursting attractor we analyzed having a higher dimension than the chaotic spiking attractor. 

For our second system, the two asymmetrically electrically coupled Rulkov 1 neurons, we found that true multistability didn't exist in the system because after enough time, the neurons eventually synchronized, but a type of quasi-multistability associated with a non-chaotic spiking attractor and chaotic pseudo-attractor did exist. Analyzing the four-dimensional chaotic pseudo-attractor generated by the initial state $\mathbf{X}_0 = \langle-0.56,\,-3.25,\,-1,\,-3.25\rangle$, we found that the pseudo-attractor was indeed fractal with a Minkowski-Bouligand dimension of $d\approx 1.84$. Then, we classified the two basins of the system, looking first at a specific two-dimensional slice of the basins $S_2$, then all of four-dimensional state space $S_4$. We discovered that the two-dimensional basin slice associated with the non-chaotic spiking attractor was Class 3, while the two-dimensional basin slice associated with the chaotic pseudo-attractor was Class 1. Additionally, we discovered that both full four-dimensional basins were Class 2. Finally, we analyzed the basin boundary of the system, considering the intersection of the basin boundary $\Sigma$ with a subset of the two-dimensional slice $S_2'$ and the intersection of $\Sigma$ with a subset of all four-dimensional space $S_4'$. We discovered that the fractal dimension of the basin boundary in $S_2'$ was $d\approx 1.686$ and the fractal dimension of the basin boundary in $S_4'$ was $d\approx 3.963$. This fractal dimension is extremely high, taking up nearly all of four-dimensional state space and indicating high unpredictability regarding which attractor an initial state will end up in. Finally, we made some computer-experiment-based conjectures on the nature of the uncertainty exponents of the system's basin boundary for different subsets of state space.

For our final distinct system, the ring lattice of $\zeta=30$ neurons, we used our Lyapunov spectrum calculation method to analyze the Lyapunov dimensions of the chaotic attractors of our three ring lattice systems. According to the Kaplan-Yorke conjecture, these Lyapunov dimensions should approximate the attractors' true fractal dimensions. We found that all the chaotic attractors of the three systems were fractal and that for some electrical coupling strength values, the attractors took up significant portions of 60-dimensional state space. Comparing the Lyapunov dimensions of the ring lattice systems to their maximal Lyapunov exponents, we also found that while the Lyapunov dimensions followed a similar pattern of increasing and decreasing as we varied the electrical coupling strength, the two quantities were not directly associated with each other. This indicated that the ``temporal'' chaos measurement of the maximal Lyapunov exponent did not directly relate to the geometrical structure of the attractors.

\subsection{Future Research and Possible Implications}

The research we detail in this paper lends itself to a significant amount of future research that can improve upon and extend our analysis. To start, relating to Section \ref{chaotic-dynamics-rulkov-map-1}, a more rigorous analysis can be done on the distribution of non-chaotic and chaotic dynamics in the parameter space of Rulkov map 1. Specifically, we suggest further analysis be done in dividing parameter space into more than non-chaotic and chaotic regions, perhaps finding a rigorous way to differentiate between the different regimes of behavior: silence, non-chaotic spiking, non-chaotic bursting, chaotic spiking, and chaotic bursting. We also suggest our research in the dynamics of Rulkov map 1 be applied to Rulkov map 2, especially concerning the distribution of non-chaotic and chaotic dynamics in parameter space.

Our research in ring lattice systems of Rulkov 1 neurons sets a precedent for how the chaotic dynamics of different Rulkov neuron lattices may be analyzed and quantified. Specifically, our detailed calculation of the Jacobian matrix of a ring lattice system can be naturally extended to more complex lattices of Rulkov 1 neurons, such as a mesh, torus, or sphere. With more current connections in these two-dimensional lattices, we suspect that more interesting dynamics may appear. A further extension to this is a system where every neuron has an electrical coupling with every other neuron in the system. Although all-to-all couplings have been studied in the context of a mean field of Rulkov 2 neurons,\footnote{See the review by Ibarz, Casado, and Sanjuán \cite{ibarz}.} we believe that this has never been done with the more experimentally-applicable electrical coupling of Rulkov neurons. This amount of coupling connections will likely produce even more complex dynamics and hyperchaotic behavior. In summary, our research regarding ring lattice systems detailed in Sections \ref{neuron-ring-lattice} and \ref{ring-lattice-geometry} provides another step towards modeling a biological neural network.

Additionally, our research in Rulkov ring lattice systems has possible implications for physics, most evidently in statistical mechanics and condensed matter theory. The chaotic dynamics exhibited in our Rulkov lattice models may be applicable to spin models in statistical mechanics, such as the Ising or XY models, as well as crystal lattice structures in condensed matter theory. Chaos in these neuron lattice models may provide a new perspective on the emergence of chaos in lattice models from these branches of physics.

There is also a significant amount of research within dynamical systems theory that could build on our geometrical analysis of Rulkov map systems. For example, research can be done to support or disprove our conjectures about the uncertainty exponents of our asymmetrically electrically coupled Rulkov 1 neuron system in larger regions of space. This may provide more information on how the basin classification method detailed in Section \ref{basins-of-attraction} may or may not be correlated with the nature of the basin boundaries of a system. Additionally, since we used Lyapunov exponents to calculate the Lyapunov dimensions of our ring lattice systems' attractors, researchers with access to the computing power to box count in 60-dimensional space may be interested in confirming or rejecting our use of the Kaplan-Yorke conjecture in these systems. Finally, we recommend future research be done on the possibility of applying the somewhat recent development by Daza \textit{et al.} \cite{basin-entropy} of using basin entropy to measure geometrical sensitivity to initial conditions in these Rulkov systems. To our knowledge, neither basin entropy nor the uncertainty exponents we used in this paper have been used to explore multistability and geometrical sensitivity in these neuron systems before, so we recommend future research be done on how these compare with each other.

Overall, our research provides new insight into how fractal geometry, multistability, and geometrical sensitivity to initial conditions can appear in chaotic neuron systems. This research has large implications for both theoretical and experimental research in the behavior of neuronal models and biological neurons by demonstrating that neuron systems can be sensitive to initial conditions in both the temporal and geometrical sense. We recommend experiments be done with real biological neurons to test our theoretical results of slightly different initial neuron states leading to vastly different eventual behaviors.

\section*{Acknowledgements}

We would first like to thank our lab director, Mr. Mark Hannum, for providing beneficial discussions and guidance throughout the research process. We would also like to acknowledge Dr. Paul So, who gave us the idea for this research project and provided useful information that kickstarted our research. As for our fellow students, we are extremely grateful to Arthur Prudius for his extensive review and practical suggestions in all aspects of our research. We would also like to thank Henry Stievater, Isaac Park, and Kohlen Farah for their peer review, as well as Leonard Schrag for his coding assistance. Finally, we would like to thank you, the reader, for taking the time to read our paper.

\appendix

\newpage

\section{Notation Guide}

This appendix contains some of the notation used in this paper for reference. Notation is presented approximately in order of appearance.
\begin{center}
    \begin{tabular}{cc}
        \multicolumn{2}{c}{\textbf{Basic General Notation}} \vspace{3px} \\
        \hline \\ [-10px]
        $t$ & time \vspace{3px} \\
        $n$ & a number of dimensions \vspace{3px} \\
        $m$ & a dimension \\
         & or dimension index \vspace{3px} \\
        $i,\,j,\,k,\,\kappa,\,p$ & indices \vspace{3px} \\
        $r,\,a,\,b,\,\alpha$ & basic parameters \vspace{3px} \\
        $\sigma$ & a scaling factor \vspace{3px} \\
        $d$ & the box-counting dimension \\
         & of a geometric object \vspace{3px} \\
        \hline
        \vspace{0.1cm}
    \end{tabular}
    
    \begin{tabular}{cc}
        \multicolumn{2}{c}{\textbf{Dynamical Systems Fundamentals}} \vspace{3px} \\
        \hline \\ [-10px]
        $\mathbf{x}$ & a generic state vector \vspace{3px} \\
        $x$ & a state of a one-dimensional system \\
         & or the component of $\mathbf{x}$ \\
         & in the 1st dimension \vspace{3px} \\
        $y$ & the component of $\mathbf{x}$ \\
         & in the 2nd dimension \vspace{3px} \\
        $\mathbf{x}_0,\,x_0$ & the initial state of a system \vspace{3px} \\
        $\mathbf{x}_t,\,x_t$& the state of a system at time $t$ \vspace{3px} \\
        $\mathbf{x}_s,\,x_s$ & a stationary state/fixed point \vspace{3px} \\
        $\mathbf{x}_p,\,x_p$ & a periodic point \vspace{3px} \\
        $q$ & the period of an orbit \vspace{3px} \\
        $\mathbf{g},\,g$ & a differential equation function \vspace{3px} \\
        $\mathbf{f},\,f$ & an iteration function \vspace{3px} \\
        $\mathbf{f}^t,\,f^t$ & $t$ iterations of an iteration function \vspace{3px} \\
        $O(\mathbf{x})$ & an orbit of $\mathbf{x}$ \vspace{3px} \\
        $O^q(\mathbf{x}_p)$ & the smallest periodic orbit of $\mathbf{x}_p$ \vspace{3px} \\
        $O^+(\mathbf{x}_0)$ & the forward orbit of $\mathbf{x}_0$ \vspace{3px} \\
        $x\e{m}$ & the component of $\mathbf{x}$ \\
         & in the $m$th dimension \vspace{3px} \\
        $f\e{m}$ & the iteration function in \\
         & the $m$th dimension of $\mathbf{f}$ \vspace{3px} \\
        $T,\,T'$ & transformations \vspace{3px} \\
        $x',\,y'$ & the result of transforming \\
         & $x$ and $y$ with $T'$ \vspace{3px} \\
        $\delta x_0$ & a perturbation of $x_0$ \vspace{3px} \\
        $\delta x_t$ & the evolution of $\delta x_0$ after $t$ steps \vspace{3px} \\
    \end{tabular}
    
    \begin{tabular}{cc}
        \multicolumn{2}{c}{\textbf{Dynamical Systems Fundamentals (continued)}} \vspace{3px} \\
        \hline \\ [-10px]
        $\delta \mathbf{x}_0$ & a perturbation of $\mathbf{x}_0$ \vspace{3px} \\
        $\delta \mathbf{x}\e{m}_0$ & the component of $\delta\mathbf{x}_0$ \\
         & in the $m$th dimension \vspace{3px} \\
        $\delta \mathbf{x}_t$ & the evolution of $\delta \mathbf{x}_0$ after $t$ steps \vspace{3px} \\
        $d\mathbf{x}_0$ & an infinitesimal perturbation of $\mathbf{x}_0$ \vspace{3px} \\
        $d\mathbf{x}_t$ & the evolution of $d\mathbf{x}_0$ after $t$ steps \vspace{3px} \\
        $\mathbf{u}_0$ & a unit vector in the direction of $d\mathbf{x}_0$ \vspace{3px} \\
        \hline
    \end{tabular}
    
    \begin{tabular}{cc}
        \multicolumn{2}{c}{\textbf{Lyapunov Exponents and Related}} \vspace{3px} \\
        \hline \\ [-10px]
        $\lambda$ & the Lyapunov exponent \\
         & in a one-dimensional system \\
         & or the Lyapunov spectrum \\
         & in a multi-dimensional system \vspace{3px} \\
        $\lambda_i$ & the $i$th Lyapunov exponent \vspace{3px} \\
        $J(\mathbf{x})$ & the Jacobian matrix of a system \vspace{3px} \\
        $J(\mathbf{x}_t)$ & the Jacobian matrix of a system at $\mathbf{x}_t$ \vspace{3px} \\
        $J_{mi}(\mathbf{x}_k)$ & the $m$th row and $i$th \\
         & column of the matrix $J(\mathbf{x}_k)$ \vspace{3px} \\
        $J^t$ & the matrix product \\
         & $J(\mathbf{x}_{t-1})J(\mathbf{x}_{t-2})\hdots J(\mathbf{x}_0)$ \vspace{3px} \\
        $p_i(t)$ & the $i$th principal axis \\
         & of an $n$-dimensional ellipsoid \\
         & of perturbations at time $t$ \vspace{3px} \\
        $\mu_i$ & the $i$th eigenvalue of the matrix $J^{t\intercal}J^t$ \vspace{3px} \\
        $\mathbf{w}_i$ & the eigenvector associated with $\mu_i$ \vspace{3px} \\
        $\nu_i$ & the $i$th eigenvalue of the matrix $J(\mathbf{x}_s)$ \vspace{3px} \\
        $Q^{(1)}R^{(1)}$ & the QR factorization matrices of $J(\mathbf{x}_0)$ \vspace{3px} \\
        $J^*_k$ & the matrix product $J(\mathbf{x}_{k-1})Q^{k-1}$, \\
         & defined recursively from $Q^{(1)}$ \vspace{3px} \\
        $Q^{(k)}R^{(k)}$ & the QR factorization matrices of $J^*_k$ \vspace{3px} \\
        $\Psi(t)$ & the matrix product \\
         & $R^{(t)}R^{(t-1)}\hdots R^{(1)}$ \vspace{3px} \\
        $\psi_{ij}(t)$ & the entry in the $i$th row \\
         & and $j$th column of $\Psi(t)$ \vspace{3px} \\
        $r^{(k)}_{ij}$ & the entry in the $i$th row \\
         & and $j$th column of $R^{(k)}$ \vspace{3px} \\
        $d_l$ & the Lyapunov dimension \vspace{3px} \\
        \hline
    \end{tabular}

    \begin{tabular}{cc}
        \multicolumn{2}{c}{\textbf{Attractors and Basins}} \vspace{3px} \\
        \hline \\ [-10px]
        $A,\,C$ & attractors \vspace{3px} \\
        $\mathcal{A},\,\mathcal{C}$ & Milnor attractors \vspace{3px} \\
        $\hat{A}$ & the basin of attraction of $A$ \vspace{3px} \\
    \end{tabular}
    
    \begin{tabular}{cc}
        \multicolumn{2}{c}{\textbf{Attractors and Basins (continued)}} \vspace{3px} \\
        \hline \\ [-10px]
        $\hat{\mathcal{A}}$ & the basin of attraction of $\mathcal{A}$, \\
         & which could be riddled \vspace{3px} \\
        $A'$ & a proper subset of $A$ \vspace{3px} \\
        $\mathcal{A}'$ & a proper subset of $\mathcal{A}$ \vspace{3px} \\
        $U$ & an open set of states that $A$ attracts \vspace{3px} \\
        $\sigma_A$ & the standard deviation of $A$ \vspace{3px} \\
        $\xi$ & normalized distance from an attractor \vspace{3px} \\
        $S(\xi)$ & the set of all $\mathbf{x}$ \\
         & that lie in an $n$-dimensional ball \\
         & of radius $\xi$ centered at $\langle A\rangle$ \vspace{3px} \\
        $\hat{A}(\xi)$ & the intersection of $\hat{A}$ and $S(\xi)$ \vspace{3px} \\
        $P(\xi)$ & the probability that an initial state \\
         & $\mathbf{x}_0\in S(\xi)$ is also in $\hat{A}$ \vspace{3px} \\
        $\Delta S(2^k)$ & the set of all $\mathbf{x}$ that lie in an \\
         & $n$-dimensional shell with inner radius \\
         & $\xi=2^k$ and outer radius $\xi=2^{k+1}$ \\
         & centered at $\langle A\rangle$ \vspace{3px} \\
        $\Delta \hat{A}(2^k)$ & the intersection of $\hat{A}$ and $\Delta S(2^k)$ \vspace{3px} \\
        $\Delta P(2^k)$ & the probability that an initial state \\
         & $\mathbf{x}_0\in\Delta S(2^k)$ is also in $\hat{A}$ \vspace{3px} \\
        $P_0,\,\gamma$ & parameters for \\
         & basin classification \vspace{3px} \\
        $\xi_0$ & the linear, normalized size \\
         & of a Class 4 basin \vspace{3px} \\
        $\Sigma$ & a basin boundary \vspace{3px} \\
        $\varrho(\epsilon)$ & the fraction of states \\
         & in a given region of state space \\
         & that lie at most a distance $\epsilon$ \\
         & from a basin boundary \vspace{3px} \\
        $\mathfrak{u}$ & the uncertainty exponent \vspace{3px} \\
        $w$ & the number of attractors \\
         & in a system with more \\
         & than 3 attractors \vspace{3px} \\
        $G$ & a grid, or set of boxes \vspace{3px} \\
        $s$ & the number of boxes \\
         & on one side of a grid \vspace{3px} \\
        $\Box_i$ & a box, or an element of a grid $G$ \vspace{3px} \\
        $C(\mathbf{x})$ & the basin $\mathbf{x}$ belongs to \vspace{3px} \\
        $C(\Box_i)$ & the basin that the \\
         & center point of $\Box_i$ belongs to \vspace{3px} \\
        $b(\Box_i)$ & the set of boxes consisting of $\Box_i$ \\
         & and all the boxes sharing at least \\
         & one boundary point with $\Box_i$ \vspace{3px} \\
        $K(\Box_i)$ & the number of distinct \\
         & colors $C(\Box)$ in $b(\Box_i)$ \vspace{3px} \\
    \end{tabular}
    
    \begin{tabular}{cc}
        \multicolumn{2}{c}{\textbf{Attractors and Basins (continued)}} \vspace{3px} \\
        \hline \\ [-10px]
        $G_k^p$ & the set of all $\Box_i$ \\
         & for which $K(\Box_i)=k$ \\
         & immediately after step $p$ \vspace{3px} \\
        $W_k$ & the fraction of boxes on or \\
         & sufficiently close to a boundary \\
         & shared by $k$ basins \vspace{3px} \\
        $S_2$ & a specific, infinite, \\
         & two-dimensional slice \\
         & of four-dimensional state space \vspace{3px} \\
        $S_2'$ & a specific subset of $S_2$ \vspace{3px} \\
        $S_4$ & all of four-dimensional state space \vspace{3px} \\
        $S_4'$ & a specific subset of $S_4$ \vspace{3px} \\
        $A_4$ & a four-dimensional attractor \vspace{3px} \\
        $\hat{A}_4$ & the basin of $A_4$ \vspace{3px} \\
        $\langle A_4\rangle$ & the mean of $A_4$ \vspace{3px} \\
        $\langle A_2\rangle$ & the effective mean of $A_4$ \\
         & used when classifying the \\
         & intersection of $\hat{A}_4$ and $S_2$ \vspace{3px} \\
        $\langle x_1\rangle,\,\langle y_1\rangle,$ & slow and fast variable \\
        $\langle x_2\rangle,\,\langle y_2\rangle$ & mean values on $A_4$ \vspace{3px} \\
        $\sigma_{A4}$ & the standard deviation of $A_4$ \vspace{3px} \\
        $\sigma_{A2}$ & the effective standard \\
         & deviation of $A_4$ \\
         & used when classifying the \\
         & intersection of $\hat{A}_4$ and $S_2$ \vspace{3px} \\
        $\xi_4$ & normalized four-dimensional \\
         & distance from $\langle A_4\rangle$ \vspace{3px} \\
        $\xi_2$ & normalized two-dimensional \\
         & distance from $\langle A_2\rangle$ \vspace{3px} \\
        $P_w(\xi)$ & white basin classification \\
         & probability function \vspace{3px} \\
        $P_b(\xi)$ & black basin classification \\
         & probability function \vspace{3px} \\
        $\mathfrak{u}_2$ & the uncertainty exponent of the \\
         & intersection of $\Sigma$ and $S_2'$ \vspace{3px} \\
        $\varrho_2(\epsilon)$ & the probability function \\
         & associated with $\mathfrak{u}_2$ \vspace{3px} \\
        $\mathfrak{u}_4$ & the uncertainty exponent of the \\
         & intersection of $\Sigma$ and $S_4'$ \vspace{3px} \\
        $\varrho_4(\epsilon)$ & the probability function \\
         & associated with $\mathfrak{u}_4$ \vspace{3px} \\
        \hline
    \end{tabular}

    \begin{tabular}{cc}
        \multicolumn{2}{c}{\textbf{Slow-Fast Systems and Neuron Basics}} \vspace{3px} \\
        \hline \\ [-10px]
        $\eta$ & a very small parameter \vspace{3px} \\
        $\chi,\,\omega$ & continuous slow-fast functions \vspace{3px} \\
        $f,\,g$ & discrete slow-fast functions \vspace{3px} \\
        $E_{\text{ion}}$ & the equilibrium \\
         & potential of an ion \vspace{3px} \\
        $I_{\text{ion}}$ & the current of an ion \vspace{3px} \\
        $g_{\text{ion}}$ & the conductance of an ion \vspace{3px} \\
        $\overline{g}$ & the maximum \\
         & conductance of an ion \vspace{3px} \\
        $C$ & the capacitance of an axon \vspace{3px} \\
        $V,\,n,\,m,\,h$ & variables of the \\
         & Hodgkin-Huxley model \vspace{3px} \\
        $\alpha_n,\,\beta_n,\,\alpha_m,$ & functions of the \\
        $\beta_m,\,\alpha_h,\,\beta_h$ & Hodgkin-Huxley model \vspace{3px} \\
        $v$ & fast voltage variable \\
         & of the Izhikevich model \vspace{3px} \\
        $u$ & slow recovery variable \\
         & of the Izhikevich model \vspace{3px} \\
        $a,\,b,\,c,\,d$ & parameters of the \\
         & Izhikevich model \vspace{3px} \\
        \hline
    \end{tabular}

    \begin{tabular}{cc}
        \multicolumn{2}{c}{\textbf{Rulkov Basics and Current Injection}} \vspace{3px} \\
        \hline \\ [-10px]
        $x$ & the fast or voltage variable \\
         & of the Rulkov maps \vspace{3px} \\
        $y$ & the slow variable \\
         & of the Rulkov maps \vspace{3px} \\
        $\alpha,\,\sigma,\,\eta$ & parameters of the Rulkov maps \vspace{3px} \\
        $f_1$ & the fast variable iteration \\
         & function of Rulkov map 1 \vspace{3px} \\
        $\mathbf{f}_1$ & the iteration function \\
         & of Rulkov map 1 \vspace{3px} \\
        $f_2$ & the fast variable iteration \\
         & function of Rulkov map 2 \vspace{3px} \\
        $\mathbf{f}_2$ & the iteration function \\
         & of Rulkov map 2 \vspace{3px} \\
        $x_{s,\,\text{slow}}$ & the value of $x$ \\
         & that leaves $y$ fixed \vspace{3px} \\
        $x_{s,\,\text{fast}}$ & the fixed points of the \\
         & fast map of Rulkov map 1 \vspace{3px} \\
        $x_{s,\,\text{fast, stable}}$ & the stable fixed point of the \\
         & fast map of Rulkov map 1 \vspace{3px} \\
        $x_{s,\,\text{fast, unstable}}$ & the unstable fixed point of the \\
         & fast map of Rulkov map 1 \vspace{3px} \\
    \end{tabular}

    \begin{tabular}{cc}
        \multicolumn{2}{c}{\textbf{Rulkov Basics and Current Injection (continued)}} \vspace{3px} \\
        \hline \\ [-10px]
        $B_{\text{stable}}$ & the stable branch \\
         & of the Rulkov maps \vspace{3px} \\
        $B_{\text{unstable}}$ & the unstable branch \\
         & of the Rulkov maps \vspace{3px} \\
        $B_{\text{spikes}}$ & the spiking branch \\
         & of the Rulkov maps \vspace{3px} \\
        $\sigma_{\mathrm{th}}$ & the threshold of excitation \\
         & of Rulkov map 1 \vspace{3px} \\
        $C_{\text{bs}}$ & the bifurcation curve between \\
         & spiking and bursting behavior \\
         & in Rulkov map 1 \vspace{3px} \\
        $\sigma_{\text{n-s bif}}$ & the Neimark-Sacker \\
         & bifurcation curve \\
         & of Rulkov map 1 \vspace{3px} \\
        $x_{s,\,1},\,x_{s,\,2},\,x_{s,\,3}$ & the fixed points of the \\
         & fast map of Rulkov map 2 \vspace{3px} \\
        $I_k$ & time-varying injected \\
         & direct current \vspace{3px} \\
        $\beta_k,\,\sigma_k$ & parameters for modeling \\
         & a time-varying current \vspace{3px} \\
        $\beta^c,\,\sigma^c$ & coefficients for DC response \vspace{3px} \\
        \hline
    \end{tabular}

    \begin{tabular}{cc}
        \multicolumn{2}{c}{\textbf{Neuron Couplings}} \vspace{3px} \\
        \hline \\ [-10px]
        $\mathbf{x}_i$ & the $i$th coupled neuron \vspace{3px} \\
        $\mathbf{x}_{i,\,k}$ & the state of the $i$th \\
         & coupled neuron at time $t=k$ \vspace{3px} \\
        $x_i$ & the fast variable of \\
         & the $ith$ coupled neuron \vspace{3px} \\
        $y_i$ & the slow variable of \\
         & the $ith$ coupled neuron \vspace{3px} \\
        $x_{i,\,k}$ & the fast variable state of the \\
         & $i$th couple neuron at time $t=k$ \vspace{3px} \\
        $y_{i,\,k}$ & the slow variable state of the \\
         & $i$th coupled neuron at time $t=k$ \vspace{3px} \\
        $\mathbf{X}$ & the state vector of all the \\
         & coupled neurons in a system \vspace{3px} \\
        $\mathbf{X}_k$ & the state of a coupled \\
         & neuron system at time $t=k$ \vspace{3px} \\
        $\mathbf{F}$ & the iteration function of \\
         & a coupled neuron system \vspace{3px} \\
        $X\e{j}$ & the $j$th dimension of $\mathbf{X}$ \vspace{3px} \\
        $F\e{m}$ & the $m$th dimension of $\mathbf{F}$ \vspace{3px} \\
        $\mathfrak{C}_{i,\,x},\,\mathfrak{C}_{i,\,y}$ & the coupling parameters of \\
         & the $i$th coupled neuron \vspace{3px} \\
    \end{tabular}
    
    \begin{tabular}{cc}
        \multicolumn{2}{c}{\textbf{Neuron Couplings (continued)}} \vspace{3px} \\
        \hline \\ [-10px]
        $\alpha_i,\,\sigma_i$ & the Rulkov map parameters of \\
         & the $ith$ coupled neuron \vspace{3px} \\
        $\beta^c_i,\,\sigma^c_i$ & the coefficients for \\
         & DC response of \\
         & the $i$th coupled neuron \vspace{3px} \\
        $g^e$ & electrical coupling strength \\
         & or coupling conductance \vspace{3px} \\
        $g^e_{i}$ & the electrical coupling strength \\
         & of neuron $\mathbf{x}_i$ \vspace{3px} \\
        $g^e_{ji}$ & the electrical coupling strength \\
         & from neuron $\mathbf{x}_j$ to neuron $\mathbf{x}_i$ \vspace{3px} \\
        $\mathcal{N}_i$ & the set of neurons \\
         & adjacent to $\mathbf{x}_i$ \vspace{3px} \\
        $\zeta$ & the number of neurons \\
         & in a coupled neuron lattice system \vspace{3px} \\
        $J(\mathbf{X})$ & the Jacobian matrix of a \\
         & coupled neuron system \vspace{3px} \\
        $J_{mj}(\mathbf{X})$ & the $m$th row and $j$th column \\
         & of the matrix $J(\mathbf{X})$ \vspace{3px} \\
        $J_{\text{dg},\,a},\,J_{\text{odg},\,b},$ & partition matrices of the \\
        $J_{\text{odg},\,c},\,J_{\text{dg},\,d}$ & Jacobian matrix for two \\
         & coupled Rulkov 1 neurons \vspace{3px} \\
        $\boldsymbol{\sigma}$ & the vector of all $\sigma_i$ values \vspace{3px} \\
        $\boldsymbol{\alpha}$ & the vector of all $\alpha_i$ values \vspace{3px} \\
        \hline
    \end{tabular}

    \begin{tabular}{cc}
        \multicolumn{2}{c}{\textbf{Basic Mathematical Notation}} \vspace{3px} \\
        \hline \\ [-10px]
        $|x|$ & the absolute value of $x$ \vspace{3px} \\
        $z$ & a complex number \vspace{3px} \\
        $|z|,\,r$ & the modulus of $z$ \vspace{3px} \\
        $\Arg(z),\,\varphi$ & the argument of $z$ \vspace{3px} \\
        $z^*$ & the complex conjugate of $z$ \vspace{3px} \\
        $|\mathbf{x}|$ & the magnitude of $\mathbf{x}$ \vspace{3px} \\
        $\phi,\,\theta,\,\theta_i$ & angles describing \\
         & spherical coordinates in $\mathbb{R}^n$ \vspace{3px} \\
        $M$ & a matrix \vspace{3px} \\
        $M\transpose$ & the transpose of $M$ \vspace{3px} \\
        $I$ & the identity matrix \vspace{3px} \\
        $Q$ & an orthogonal matrix \vspace{3px} \\
        $R$ & an upper triangular matrix \vspace{3px} \\
        $\det M,\,\Delta$ & the determinant of $M$ \vspace{3px} \\
        $\tr M,\,\tau$ & the trace of $M$ \vspace{3px} \\
        $\frac{df}{dx}\big|_{x=x_t}\,,\,f'(x_t)$ & the derivative of $f$ with \\
         & respect to $x$ evaluated at $x_t$ \vspace{3px} \\
    \end{tabular}
    
    \begin{tabular}{cc}
        \multicolumn{2}{c}{\textbf{Basic Mathematical Notation (continued)}} \vspace{3px} \\
        \hline \\ [-10px]
        $\dot{\mathbf{x}}$ & the derivative of $\mathbf{x}$ \\
         & with respect to $t$ \vspace{3px} \\
        $\frac{\partial f\e{m}}{\partial x\e{i}}$ & the partial derivative \\
         & of $f\e{m}$ with respect to $x\e{i}$ \vspace{3px} \\
        $\sum_{i=a}^b$ & the sum from $i=a$ to $i=b$ \vspace{3px} \\
        $\prod_{i=a}^b$ & the product from $i=a$ to $i=b$ \vspace{3px} \\
        $\sum_{i\in S}$ & the sum over all indices $i$ in $S$ \vspace{3px} \\
        $\sim$ & is proportional to \vspace{3px} \\
        $\mathbf{1}$ & a vector of all ones \vspace{3px} \\
        \hline
        \vspace{0.1cm}
    \end{tabular}
    
    \begin{tabular}{cc}
        \multicolumn{2}{c}{\textbf{Set Notation}} \vspace{3px} \\
        \hline \\ [-10px]
        $S$ & a set \vspace{3px} \\
        $\{x_0,\,x_1,\,\hdots\,,\,x_t\}$ & the set with elements $x_i$ \\
         & for $0\leq i\leq t$ \vspace{3px} \\
        $(a,\,b)$ & the open interval from $a$ to $b$ \\
         & or an ordered pair \\
         & representing a point \vspace{3px} \\
        $\emptyset$ & the empty set \vspace{3px} \\
        $\mathbb{R}$ & the set of real numbers \vspace{3px} \\
        $\mathbb{N}$ & the set of natural numbers \vspace{3px} \\
        $\mathbb{C}$ & the set of complex numbers \vspace{3px} \\
        $\mathbb{R}^n$ & the set of all vectors of real \\
         & numbers that have length $n$ \vspace{3px} \\
        $\in$ & is an element of \vspace{3px} \\
        $\subseteq$ & is a subset of \vspace{3px} \\
        $\subset$ & is a proper subset of \vspace{3px} \\
        $:$ & such that \vspace{3px} \\
        $S_1\cup S_2$ & the union of $S_1$ and $S_2$ \vspace{3px} \\
        $\bigcup_{i=a}^b S_i$ & the union set \\
         & $S_a\cup S_{a+1}\cup\hdots\cup S_b$ \vspace{3px} \\
        $S_1\cap S_2$ & the intersection of $S_1$ and $S_2$ \vspace{3px} \\
        $S_1\times S_2$ & the Cartesian \\
         & product of $S_1$ and $S_2$ \vspace{3px} \\
        $\mu(S)$ & the measure of $S$ \vspace{3px} \\
        $\mu^n(S)$ & the measure of $S\subset\mathbb{R}^n$ \vspace{3px} \\
        $\langle S\rangle$ & the average of \\
         & all the elements of $S$ \vspace{3px} \\
        $|S|$ & the cardinality of $S$ \vspace{3px} \\
        $N(\epsilon)$ & the number of $n$-dimensional \\
         & boxes of side length $\epsilon$ \\
         & needed to cover a set $S\subset\mathbb{R}^n$ \vspace{3px} \\
        \hline
    \end{tabular}
\end{center}
    
\newpage

\section{Derivations and Proofs}

This appendix contains derivations, proofs, and details of some equations, criteria, and methods utilized in this paper.

\subsection{The One-Dimensional Lyapunov Exponent}
\label{lyap1d-deriv}

From Equation \ref{eq:lyapunovunsimp},
\begin{equation}
    \begin{split}
        \lambda &= \lim_{t\to\infty}\lim_{\delta x_0\to 0}\frac{1}{t}\ln\left|\frac{f^t(x_0+\delta x_0)-f^t(x_0)}{\delta x_0}\right| \\
        &= \lim_{t\to\infty}\frac{1}{t}\ln\left|\frac{df^t(x)}{dx}\bigg|_{x=x_0}\right| \\ 
    \end{split}
\end{equation}
by the limit definition of a derivative. Using the chain rule, this can be rewritten as
\begin{equation}
    \begin{split}
        \lambda &= \lim_{t\to\infty}\frac{1}{t}\ln\Bigg|\frac{df^t(x)}{df^{t-1}(x)}\bigg|_{f^{t-1}(x)=x_{t-1}} \\
        &\mathrel{\phantom{=}} \frac{df^{t-1}(x)}{df^{t-2}(x)}\bigg|_{f^{t-2}(x)=x_{t-2}}\hdots \\
        &\mathrel{\phantom{=}} \frac{df^2(x)}{df(x)}\bigg|_{f(x)=x_1}\frac{df(x)}{dx}\bigg|_{x=x_0}\Bigg|
    \end{split}
    \label{eq:chain_rule_step_1dlyap_deriv}
\end{equation}
Since $f$ is iterative, we can say
\begin{equation}
    \begin{split}
        \lambda &= \lim_{t\to\infty}\frac{1}{t}\ln\Bigg|\frac{df(x)}{dx}\bigg|_{x=x_{t-1}}\frac{df(x)}{dx}\bigg|_{x=x_{t-2}} \\
        &\mathrel{\phantom{=}}\hdots\frac{df(x)}{dx}\bigg|_{x=x_1}\frac{df(x)}{dx}\bigg|_{x=x_0}\Bigg|
    \end{split}
\end{equation}
Condensing and using the product property of logarithms,
\begin{equation}
    \begin{split}
        \lambda &= \lim_{t\to\infty}\frac{1}{t}\ln\left|\,\prod_{i=0}^{t-1}\frac{df(x)}{dx}\bigg|_{x=x_i}\right| \\
        &= \lim_{t\to\infty}\frac{1}{t}\ln\left(\prod_{i=0}^{t-1}\left|\frac{df(x)}{dx}\bigg|_{x=x_i}\right|\right) \\
        &= \lim_{t\to\infty}\frac{1}{t}\sum_{i=0}^{t-1}\ln\left|\frac{df(x)}{dx}\bigg|_{x=x_i}\right| \\
        &= \lim_{t\to\infty}\frac{1}{t}\sum_{i=0}^{t-1}\ln|f'(x_i)|
    \end{split}
\end{equation}
This completes the derivation of Equation \ref{eq:1dlyap}.

\subsection{The Lyapunov Spectrum}
\label{lyapspec-deriv}

Rewriting Equation \ref{eq:perturbation-gendirec} for an infinitesimal perturbation at a general time step $k$,
\begin{equation}
    d\mathbf{x}_k = \begin{pmatrix}
        dx\e{1}_k \\[4pt]
        dx\e{2}_k \\[1pt]
        \vdots \\[1pt]
        dx\e{n}_k
    \end{pmatrix}
\end{equation}
For a general dimension $m$, Equation \ref{eq:expandediteration} reads
\begin{equation}
    x\e{m}_{k+1} = f\e{m}(x\e{1}_k,\, x\e{2}_k,\,\hdots\,,\, x\e{n}_k)
\end{equation}
From calculus \cite[p. 944]{multi}, we know that
\begin{equation}
    dx\e{m}_{k+1} = \frac{\partial f\e{m}}{\partial x\e{1}_k}dx\e{1}_k + \frac{\partial f\e{m}}{\partial x\e{2}_k}dx\e{2}_k + \hdots + \frac{\partial f\e{m}}{\partial x\e{n}_k}dx\e{n}_k 
\end{equation}
Condensing,
\begin{equation}
    dx\e{m}_{k+1} = \sum_{i=1}^n\frac{\partial f\e{m}}{\partial x\e{i}_k}dx\e{i}_k
    \label{eq:appendixmulti}
\end{equation}
We can rewrite Equation \ref{eq:jacobian} for $J(\mathbf{x}_k)$ concisely as
\begin{equation}
    J_{mi}(\mathbf{x}_k) = \frac{\partial f\e{m}}{\partial x\e{i}_k}
\end{equation}
Substituting into Equation \ref{eq:appendixmulti},
\begin{equation}
    dx\e{m}_{k+1} = \sum_{i=1}^nJ_{mi}(\mathbf{x}_k)dx\e{i}_k
\end{equation}
Expanding into the full vector $d\mathbf{x}_{k+1}$,
\begin{align}
    \begin{split}
        \begin{pmatrix}
            dx\e{1}_{k+1} \\[4pt]
            dx\e{2}_{k+1} \\[1pt]
            \vdots \\[1pt]
            dx\e{n}_{k+1}
        \end{pmatrix}
        &= \begin{pmatrix}
            \sum_{i=1}^nJ_{1i}(\mathbf{x}_k)dx\e{i}_k \\[4pt]
            \sum_{i=1}^nJ_{2i}(\mathbf{x}_k)dx\e{i}_k \\[1pt]
            \vdots \\[1pt]
            \sum_{i=1}^nJ_{ni}(\mathbf{x}_k)dx\e{i}_k
        \end{pmatrix} \\
        &= J(\mathbf{x}_k)\begin{pmatrix}
            dx\e{1}_{k} \\[4pt]
            dx\e{2}_{k} \\[1pt]
            \vdots \\[1pt]
            dx\e{n}_{k}
        \end{pmatrix}
    \end{split}
\end{align}
by the nature of matrix multiplication. Concisely, this equation yields Equation \ref{eq:jacobian-iteration}.

Following Equation \ref{eq:lyapunitvector}, consider a generic vector $\mathbf{v}$. It is easy to see that
\begin{align}
    \begin{split}
        \mathbf{v}\transpose\mathbf{v} &= 
        \begin{pmatrix}
            v_1 & v_2 & ... & v_n
        \end{pmatrix}
        \begin{pmatrix}
            v_1 \\
            v_2 \\
            \vdots \\
            v_n
        \end{pmatrix} \\
        &= v_1^2 + v_2^2 + \hdots + v_n^2 = |\mathbf{v}|^2
    \end{split}
    \label{eq:vtransposev}
\end{align}
where $M\transpose$ is the transpose of $M$. Using this,
\begin{align}
    \begin{split}    
        |J^t\mathbf{u}_0|^2 &= (J^t\mathbf{u}_0)\transpose J^t\mathbf{u}_0 \\
        &= \mathbf{u}_0\transpose (J^{t\intercal}J^t)\mathbf{u}_0
    \end{split}
\end{align}
because $(MN)\transpose = N\transpose M\transpose$ \cite[p. 101]{linear}. We can now use this to rewrite Equation \ref{eq:lyapunitvector}:
\begin{align}
    \begin{split}
        \lambda &= \lim_{t\to\infty}\frac{1}{t}\ln\left|J^t\mathbf{u}_0\right| \\
        &= \lim_{t\to\infty}\frac{1}{t}\ln\left[\mathbf{u}_0\transpose (J^{t\intercal}J^t)\mathbf{u}_0\right]^{1/2} \\
        &= \lim_{t\to\infty}\frac{1}{2t}\ln\left[\mathbf{u}_0\transpose (J^{t\intercal}J^t)\mathbf{u}_0\right]
    \end{split}
    \label{eq:appendix-derivation-almost-done}
\end{align}
The matrix $J^{t\intercal}J^t$ is clearly symmetric:\footnote{This is why we took the time to transform Equation \ref{eq:lyapunitvector}. In general, $J^t$ is neither symmetric nor diagonalizable, so it is much more convenient to work with $J^{t\intercal}J^t$, which, in addition to being symmetric and diagonalizable, has a set of $n$ orthonormal eigenvectors corresponding to real and positive eigenvalues.}
\begin{equation}
    (J^{t\intercal}J^t)\transpose = J^{t\intercal}(J^{t\intercal})\transpose = J^{t\intercal}J^t
\end{equation}
Therefore, from linear algebra, we know that $J^{t\intercal}J^t$ has $n$ real eigenvalues (accounting for multiplicities) \cite[p. 397]{linear}. Labelling these eigenvalues $\mu_i$ and their associated normalized eigenvectors $\mathbf{w}_i$ with $i = 1,\,2,\,\hdots\,,\,n$, we can take $\mathbf{u}_0$ to be a normalized eigenvector $\mathbf{w}_i$. Then, it follows from Equation \ref{eq:appendix-derivation-almost-done} that
\begin{align}
    \begin{split}
        \lambda_i &= \lim_{t\to\infty}\frac{1}{2t}\ln\left[\mathbf{w}_i\transpose (J^{t\intercal}J^t)\mathbf{w}_i\right] \\
        &= \lim_{t\to\infty}\frac{1}{2t}\ln\left[\mathbf{w}_i\transpose \mu_i\mathbf{w}_i\right] \\
        &= \lim_{t\to\infty}\frac{1}{2t}\ln\left[\mu_i\mathbf{w}_i\transpose\mathbf{w}_i\right] \\
        &= \lim_{t\to\infty}\frac{1}{2t}\ln\mu_i
    \end{split} 
\end{align}
by the definition of eigenvalues and eigenvectors. By labeling the eigenvalues $\mu_i$ and the normalized eigenvectors $\mathbf{w}_i$ with the appropriate indices to satisfy $\lambda_1\geq\lambda_2\geq\hdots\geq\lambda_n$, this completes the derivation of Equation \ref{eq:lyapunov-eigenvalue}.

\subsection{The Criteria for the Attractiveness of Fixed Points}
\label{fixedpointattractor-criteria}

For a possible fixed point attractor $A=\{\mathbf{x}_s\}$, Property 1 of attractors (see Section \ref{nonchaoticattractors}) is automatically satisfied by Equation \ref{eq:stationary}. Furthermore, if Properties 1 and 2 are satisfied, Property 3 is satisfied because there is no proper subset of $A$ besides the empty set. Therefore, to determine whether or not a fixed point is an attractor, Property 2 is the only property we need to consider.

First, let us consider a very small neighborhood of $A$ as our open set of initial conditions $U$ that $A$ attracts. For some $\mathbf{x}_0\in U$, Property 2 of attractors says that if $A$ is an attractor, the distance from $\mathbf{f}^t(\mathbf{x}_0)=\mathbf{x}_t$ to $\mathbf{x}_s$ goes to 0 as $t$ goes to infinity, or $\lim_{t\to\infty}|\mathbf{x}_t-\mathbf{x}_s|=0$. Since $U$ is very small, the perturbation $\mathbf{x}_t-\mathbf{x}_s$ follows Equation \ref{eq:jacobian-alltheway}:
\begin{equation}
    \mathbf{x}_t-\mathbf{x}_s = J^td\mathbf{x}_s
    \label{eq:jacobian-alltheway-fixed}
\end{equation}
where $d\mathbf{x}_s = \mathbf{x}_0-\mathbf{x}_s$. By Equation \ref{eq:jacobian-alltheway},
\begin{equation}
    J^t = J\left(\mathbf{f}^{t-1}(\mathbf{x}_{s})\right)\,J\left(\mathbf{f}^{t-2}(\mathbf{x}_{s})\right)\,\hdots\, J\left(\mathbf{x}_s\right)
\end{equation}
By Equation \ref{eq:stationary}, this collapses to
\begin{equation}
    J^t = [J(\mathbf{x}_s)]^t
    \label{eq:jtfixed}
\end{equation}
Now, we consider just the first iteration of $d\mathbf{x}_s$:
\begin{equation}
    \mathbf{x}_1-\mathbf{x}_s = J(\mathbf{x}_s)d\mathbf{x}_s
\end{equation}
If we choose $d\mathbf{x}_s$ to be in the direction of an eigenvector of $J(\mathbf{x}_s)$ and $\nu_i$ to be its associated eigenvalue,\footnote{Because the Jacobian matrix $J$ doesn't have the nice properties of $J^{t\intercal}J^t$, which is symmetric and diagonalizable, the eigenvalues of $J(\mathbf{x}_s)$ might be complex. Therefore, in order to choose $d\mathbf{x}_s$ to be in the direction of an eigenvector of $J(\mathbf{x}_s)$, the entries of $d\mathbf{x}_s$ may need to be complex. This is a departure from our established intuition of the state space of dynamical systems, but it is still mathematically sound. For complex entries of $d\mathbf{x}_s$ and complex values of $\nu_i$, the absolute value signs should be interpreted as modulus functions.} this reads
\begin{equation}
    \mathbf{x}_1-\mathbf{x}_s = \nu_i\,d\mathbf{x}_s
\end{equation}
By Equation \ref{eq:jtfixed}, we can now rewrite Equation \ref{eq:jacobian-alltheway-fixed} as
\begin{equation}
    \mathbf{x}_t-\mathbf{x}_s = \nu_i^t\,d\mathbf{x}_s
\end{equation}
Taking the limit of the magnitude on both sides,
\begin{equation}
    \lim_{t\to\infty}|\mathbf{x}_t-\mathbf{x}_s| = \lim_{t\to\infty}|\nu_i^t\,d\mathbf{x}_s| = |d\mathbf{x}_s|\lim_{t\to\infty}|\nu_i|^t
\end{equation}
For $|\nu_i|<1$, 
\begin{equation}
    \lim_{t\to\infty}|\mathbf{x}_t-\mathbf{x}_s| = |d\mathbf{x}_s|\lim_{t\to\infty}|\nu_i|^t = 0
    \label{eq:eigenvalue-attractor}
\end{equation}
For $|\nu_i|>1$,
\begin{equation}
    \lim_{t\to\infty}|\mathbf{x}_t-\mathbf{x}_s| = |d\mathbf{x}_s|\lim_{t\to\infty}|\nu_i|^t\to\infty
    \label{eq:eigenvalue-repeller}
\end{equation}

Now, let us consider the matrix $J^{t\intercal}J^t$ at $\mathbf{x}_s$. Then,
\begin{equation}
    J^{t\intercal}J^t = [J(\mathbf{x}_s)\transpose]^t[J(\mathbf{x}_s)]^t
\end{equation}
The eigenvalues $\mu_i$ of $J^{t\intercal}J^t$ are therefore
\begin{equation}
    \mu_i = \nu_i^{2t}
\end{equation}
since $J$ and $J\transpose$ share the same characteristic polynomial. Equation \ref{eq:lyapunov-eigenvalue} then reads
\begin{equation}
    \begin{split}
        \lambda_i &= \lim_{t\to\infty}\frac{1}{2t}\ln\mu_i \\
        &= \lim_{t\to\infty}\frac{1}{2t}\ln\nu_i^{2t} \\
        &= \lim_{t\to\infty}\ln\nu_i
    \end{split}
    \label{eq:lyapunov-jacobian-eigenvalues}
\end{equation}
Thus, the eigenvalues of the Jacobian at $\mathbf{x}_s$ are directly related to the system's Lyapunov exponents. We know from Section \ref{quantification} that Lyapunov exponents determine the maximum and minimum growth of an infinitesimal $(n-1)$-dimensional sphere of perturbations from an initial condition, so $\nu_1$ and $\nu_n$ provide an upper and lower bound on the growth of $d\mathbf{x}_s$ in any direction.

Therefore, if $|\nu_i|<1$ for all $i = 1,\,2,\,\hdots\,,\,n$, Equation \ref{eq:eigenvalue-attractor} says that $\lim_{t\to\infty}|\mathbf{x}_t-\mathbf{x}_s|=0$ for all $d\mathbf{x}_s$, meaning Property 2 is satisfied and $\mathbf{x}_s$ is an attractor. Similarly, if every $|\nu_i|>1$, then Equation \ref{eq:eigenvalue-repeller} says that $\lim_{t\to\infty}|\mathbf{x}_t-\mathbf{x}_s|\to\infty$ for all $d\mathbf{x}_s$, meaning $\mathbf{x}_s$ is a repeller. If at least one $|\nu_i|<1$ and at least one $|\nu_i|>1$, then $\lim_{t\to\infty}|\mathbf{x}_t-\mathbf{x}_s|=0$ for some $d\mathbf{x}_s$ and $\lim_{t\to\infty}|\mathbf{x}_t-\mathbf{x}_s|\to\infty$ for others, meaning $\mathbf{x}_s$ is a saddle point. This completes the proof of the criteria for the attractiveness of fixed points. It is worthwhile to note that these criteria imply that if $\mathbf{x}_s$ is an attractor, all of the Lyapunov exponents in the Lyapunov spectrum of the forward orbit $O^+(\mathbf{x}_0)$ are negative by Equation \ref{eq:lyapunov-jacobian-eigenvalues}. Similarly, if $\mathbf{x}_s$ is a repeller, all of the Lyapunov exponents of $O^+(\mathbf{x}_0)$ are positive. Finally, if $\mathbf{x}_s$ is a saddle points, some $\lambda_i$ are positive and some are negative.

For a one-dimensional system $x_{k+1} = f(x_k)$ with a fixed point $x_s$, the Jacobian matrix $J(x_s)$ is
\begin{equation}
    J(x_s) = \left(\frac{df}{dx}\bigg|_{x=x_s}\right) = \big(f'(x_s)\big)
\end{equation}
The eigenvalue $\nu_1$ of $\big(f'(x_s)\big)$ is obviously $f'(x_s)$, so the criteria for the attractiveness of fixed points say that if $|f'(x_s)|<1$, $x_s$ is an attractor, and if $|f'(x_s)|>1$, $x_s$ is a repeller. This completes the proof of the criteria for the attractiveness of fixed points in one dimension.

\subsection{A QR Factorization Method of Lyapunov Spectrum Calculation}
\label{qr-meth-lyap-spec-calc}

QR factorization is a method of decomposing some matrix $M$ into the of an orthogonal matrix $Q$ and an upper triangular matrix $R$: $M=QR$ \cite[p. 359]{linear}.\footnote{An orthogonal matrix is a square matrix whose column vectors are orthonormal (orthogonal and normalized). An upper triangular matrix is a matrix whose entries are zero below its main diagonal.} In this appendix, we present the algorithm of Eckmann and Ruelle \cite{eckmann} for calculating Lyapunov spectrums by taking advantage of QR factorization.\footnote{Eckmann and Ruelle \cite{eckmann} were the first to propose this method, but we loosely model our notation after the way it is presented in the paper by Sandri \cite{sandri}.}

Recall from Section \ref{quantification} that when calculating the Lyapunov spectrum, we are concerned with the matrix
\begin{equation}
    J^t = J(\mathbf{x}_{t-1})J(\mathbf{x}_{t-2})\hdots J(\mathbf{x}_0)
    \label{eq:j^t_expansion}
\end{equation}
Let us first decompose the matrix $J(\mathbf{x}_0)$ using the QR factorization method, defining\footnote{In this case, the superscripts of $Q$ and $R$ should be thought of as an index. This is so that we can reference a given entry of one of these matrices by writing its row and column number in the subscript.}
\begin{equation}
    J(\mathbf{x}_0) = Q^{(1)} R^{(1)}
\end{equation}
Now, for $k=2,\,3,\,\hdots\,,\,t$, we can recursively define 
\begin{equation}
    J(\mathbf{x}_{k-1})Q^{(k-1)} = J_k^*
    \label{eq:recursive_jacobian}
\end{equation}
then decompose this matrix:
\begin{equation}
    J_k^* = Q^{(k)} R^{(k)}
    \label{eq:recursive_jacobian_qr}
\end{equation}
Since an orthogonal matrix $Q$ has the property that $Q\transpose Q = QQ\transpose = I$ \cite[p. 345]{linear},\footnote{$I$ is the identity matrix.} we can equate Equations \ref{eq:recursive_jacobian} and \ref{eq:recursive_jacobian_qr} and rearrange:
\begin{align}
    J(\mathbf{x}_{k-1})Q^{(k-1)} = Q^{(k)} R^{(k)} \\
    J(\mathbf{x}_{k-1}) = Q^{(k)} R^{(k)} \left(Q^{(k-1)}\right)\transpose
\end{align}
Substituting into Equation \ref{eq:j^t_expansion},
\begin{equation}
    \begin{split}
        J^t &= Q^{(t)} R^{(t)} \left(Q^{(t-1)}\right)\transpose \\ 
         &\mathrel{\phantom{=}} Q^{(t-1)} R^{(t-1)} \left(Q^{(t-2)}\right)\transpose \\
         &\mathrel{\phantom{=}} \hdots Q^{(1)} R^{(1)}
    \end{split}
\end{equation}
The adjacent $Q$ matrices cancel:
\begin{equation}
    J^t = Q^{(t)} R^{(t)} R^{(t-1)} \hdots R^{(1)}
    \label{eq:j^t_qr_factorized}
\end{equation}
Since the product of upper triangular matrices is an upper triangular matrix,\footnote{This can be seen explicitly in Equation \ref{eq:multiplying_r_matrices}.} we can define
\begin{equation}
    \Psi(t) = \prod_{k=t}^1 R^{(k)} = \begin{pmatrix}
        \psi_{11}(t) & \psi_{12}(t) & \hdots & \psi_{1n}(t) \\
        0 & \psi_{22}(t) & \hdots & \psi_{2n}(t) \\
        \vdots & \vdots & \ddots & \vdots \\
        0 & 0 & \hdots & \psi_{nn}(t)
    \end{pmatrix}
    \label{eq:big_upsilon}
\end{equation}
where $\psi_{ij}$ is the entry in the $i$th row and the $j$th column of $\Psi(t)$. Substituting this into Equation \ref{eq:j^t_qr_factorized},
\begin{equation}
    J^t = Q^{(t)}\,\Psi(t)
\end{equation}
Substituting this representation of $J^t$ into Equation \ref{eq:lyapunitvector},
\begin{equation}
    \lambda = \lim_{t\to\infty}\frac{1}{t}\ln|Q^{(t)}\,\Psi(t)\mathbf{u}_0|
\end{equation}
Since the eigenvalues of a triangular matrix are on its diagonal \cite[p. 271]{linear}, taking $\mathbf{u}_0$ to be a normalized eigenvector of $\Psi(t)$ yields
\begin{equation}
    \lambda_i = \lim_{t\to\infty}\frac{1}{t}\ln|Q^{(t)}\,\psi_{ii}(t)\mathbf{u}_0|
\end{equation}
for $i=1,\,2,\,\hdots\,,\,n$.\footnote{To match the indices, we arrange the column vectors of the $Q$ matrices so that the diagonal entries in their associated $R$ matrices are ordered from greatest to least.} Since $\psi_{ii}(t)$ is a constant within the limit, we can pull it out from between $Q^{(t)}$ and $\mathbf{u}_0$:
\begin{equation}
    \lambda_i = \lim_{t\to\infty}\frac{1}{t}\ln\left(|\psi_{ii}(t)||Q^{(t)}\mathbf{u}_0|\right)
    \label{eq:qr_factorization_step}
\end{equation}
Now, if we consider just the $|Q^{(t)}\mathbf{u}_0|$ factor, using Equation \ref{eq:vtransposev} yields
\begin{equation}
    \begin{split}
        |Q^{(t)}\mathbf{u}_0|^2 &= \left(Q^{(t)}\mathbf{u}_0\right)\transpose Q^{(t)}\mathbf{u}_0 \\
        &= \mathbf{u}_0\transpose \left(Q^{(t)}\right)\transpose Q^{(t)}\mathbf{u}_0 \\
        &= \mathbf{u}_0\transpose\mathbf{u}_0 \\
        &= 1
    \end{split}
\end{equation}
because $Q^{(t)}$ is an orthogonal matrix and $\mathbf{u}_0$ is a unit vector. This shows that $|Q^{(t)}\mathbf{u}_0|=1$, so we can rewrite Equation \ref{eq:qr_factorization_step} as
\begin{equation}
    \lambda_i = \lim_{t\to\infty}\frac{1}{t}\ln |\psi_{ii}(t)|
    \label{eq:qr-factorization-method-of-lyap-spec-calc}
\end{equation}
This completes the derivation of the QR factorization method of calculating a Lyapunov spectrum. 

In practice, when approximating a Lyapunov spectrum computationally using a large value of $t$, the entries of $\Psi(t)$ usually overflow because we are multiplying many $R$ matrices. To counter this, let us consider a simple case, namely, stopping at $t=2$. Then, Equation \ref{eq:big_upsilon} reads
\begin{equation}
    \Psi(2) = \prod_{k=2}^1 R^{(k)} = R^{(2)}R^{(1)}
    \label{eq:upsilon2}
\end{equation}
If we say the entry in the $i$th row and the $j$th column of the matrix $R^k$ is $r^{(k)}_{ij}$, expanding Equation \ref{eq:upsilon2} yields
\begin{equation}
    \begin{gathered}
        \begin{pmatrix}
            \psi_{11}(2) & \psi_{12}(2) & \hdots & \psi_{1n}(2) \\
            0 & \psi_{22}(2) & \hdots & \psi_{2n}(2) \\
            \vdots & \vdots & \ddots & \vdots \\
            0 & 0 & \hdots & \psi_{nn}(2)
        \end{pmatrix} \\
        \begin{aligned}
            &= \begin{pmatrix}
                r^{(2)}_{11} & r^{(2)}_{12} & \hdots & r^{(2)}_{1n} \\[4pt]
                0 & r^{(2)}_{22} & \hdots & r^{(2)}_{2n} \\[1pt]
                \vdots & \vdots & \ddots & \vdots \\[1pt]
                0 & 0 & \hdots & r^{(2)}_{nn}
            \end{pmatrix}
            \begin{pmatrix}
                r^{(1)}_{11} & r^{(1)}_{12} & \hdots & r^{(1)}_{1n} \\[4pt]
                0 & r^{(1)}_{22} & \hdots & r^{(1)}_{2n} \\[1pt]
                \vdots & \vdots & \ddots & \vdots \\[1pt]
                0 & 0 & \hdots & r^{(1)}_{nn}
            \end{pmatrix} \\
            &=
            \begin{pmatrix}
                r^{(2)}_{11}r^{(1)}_{11} & \sum_{k=1}^2r^{(2)}_{1k}r^{(1)}_{k2} & \hdots & \sum_{k=1}^nr^{(2)}_{1k}r^{(1)}_{kn} \\[4pt]
                0 & r^{(2)}_{22}r^{(1)}_{22} & \hdots & \sum_{k=2}^nr^{(2)}_{2k}r^{(1)}_{kn} \\[1pt]
                \vdots & \vdots & \ddots & \vdots \\[1pt]
                0 & 0 & \hdots & r^{(2)}_{nn}r^{(1)}_{nn}
            \end{pmatrix}
        \end{aligned}
    \end{gathered}
    \label{eq:multiplying_r_matrices}
\end{equation}
Notice that the diagonal entries of $\Psi(2)$ are
\begin{equation}
    \psi_{ii}(2) = r^{(2)}_{ii}r^{(1)}_{ii} = r^{(1)}_{ii}r^{(2)}_{ii}
\end{equation}
which are the entries we care about for calculating a Lyapunov spectrum. Generalizing,
\begin{equation}
    \psi_{ii}(t) = \prod_{j=1}^t r^{(j)}_{ii}
\end{equation}
Substituting into Equation \ref{eq:qr-factorization-method-of-lyap-spec-calc},
\begin{equation}
    \lambda_i = \lim_{t\to\infty}\frac{1}{t}\ln \left|\prod_{j=1}^t r^{(j)}_{ii}\right|
\end{equation}
Finally, using the product property of logarithms,
\begin{equation}
    \lambda_i = \lim_{t\to\infty}\frac{1}{t}\sum_{j=1}^t \ln \left|r^{(j)}_{ii}\right|
    \label{eq:qr-comp-friendly}
\end{equation}
This completes the derivation of the computationally friendly algorithm for calculating a Lyapunov spectrum using QR factorization. We use this algorithm to numerically calculate the Lyapunov spectrum of numerous multi-dimensional systems throughout this paper.

\subsection{An Iterative Shell Method for the Numerical Classification of Basins}
\label{iterative-shell-method-derivation}

In our shell method, we are interested in finding some probability value $P(2^{k+1})$ by using the value of $P(2^k)$. Of course, scaling a ball of radius $2^k$ centered at $\langle A\rangle$ by 2 gives us the ball of radius $2^{k+1}$ centered at $\langle A\rangle$. We know from Section \ref{strangeattractors} that scaling an $n$-dimensional ball by 2 scales its measure up by $2^n$, so by Equation \ref{eq:scaling-factor-measure-nonfractal},
\begin{equation}
    \mu(S(2^{k+1}))=\mu(2S(2^k))=2^n\mu(S(2^k))
    \label{eq:scale-a-2k-ball-up}
\end{equation}
Our shell method also involves examining the shell centered at $\langle A\rangle$ with inner radius $\xi=2^k$ and outer radius $\xi=2^{k+1}$, which we will denote as $\Delta S(2^k)$. In other words, it is the $n$-dimensional region between the two $(n-1)$-dimensional spheres centered at $\langle A\rangle$ with radii $\xi=2^k$ and $\xi=2^{k+1}$. Mathematically, this means
\begin{equation}
    \Delta S(2^k) = \{\mathbf{x}:\:2^k\leq|\mathbf{x}-\langle A\rangle|< 2^{k+1}\}
\end{equation}
We can similarly define $\Delta\hat{A}(2^k)$ as being the subset of the basin $\hat{A}$ that lies in the shell between $\xi=2^k$ and $\xi=2^{k+1}$, or mirroring Equation \ref{eq:a-hat-xi-def},
\begin{equation}
    \Delta\hat{A}(2^k) = \hat{A}\cap\Delta S(2^k)
\end{equation}
Mirroring Equation \ref{eq:p-xi-def}, it follows that we can define the probability that a state in the shell $\Delta S(2^k)$ is also in the basin of attraction $\hat{A}$ to be
\begin{equation}
    \Delta P(2^k) = \frac{\mu(\Delta \hat{A}(2^k))}{\mu(\Delta S(2^k))}
    \label{eq:delta-p-2k-def}
\end{equation}

Geometrically, it is clear that the $n$-dimensional ball with radius $\xi=2^{k+1}$ centered at $\langle A\rangle$ is composed of the ball with radius $\xi=2^k$ centered at $\langle A\rangle$ and the shell with inner radius $\xi=2^k$ and outer radius $\xi=2^{k+1}$ centered at $\langle A\rangle$. Of course, this must also be true for the measures of the sets associated with these geometric objects:
\begin{gather}
    \mu(S(2^{k+1})) = \mu(S(2^k)) + \mu(\Delta S(2^k))
    \label{eq:sum-the-measures-of-a-ball-and-shell} \\
    \mu(\hat{A}(2^{k+1})) = \mu(\hat{A}(2^k)) + \mu(\Delta \hat{A}(2^k))
    \label{eq:sum-the-measures-of-basin-ball-and-shell}
\end{gather}
However, we know from Equation \ref{eq:scale-a-2k-ball-up} that the measure of $S(2^{k+1})$ is also equal to the measure of $S(2^k)$ scaled by $2^n$, so equating Equations \ref{eq:scale-a-2k-ball-up} and \ref{eq:sum-the-measures-of-a-ball-and-shell} then solving for $\mu(\Delta S(2^k))$ gives
\begin{equation}
    \mu(\Delta S(2^k)) = (2^n-1)\mu(S(2^k))
    \label{eq:measure-of-shell}
\end{equation}
which means that the shell with inner radius $\xi=2^k$ and outer radius $\xi=2^{k+1}$ centered at $\langle A\rangle$ has $2^n-1$ times more measure than the ball with radius $\xi=2^k$ centered at $\langle A\rangle$.

Now that we have relations between the small ball $S(2^k)$, the shell $\Delta S(2^k)$, and the big ball $S(2^{k+1})$, our goal with this shell method is to find 
\begin{equation}
    P(2^{k+1}) = \frac{\mu(\hat{A}(2^{k+1}))}{\mu(S(2^{k+1}))}
    \label{eq:p-2k1-unsuggestive}
\end{equation}
in terms of $P(2^k)$ and $\Delta P(2^k)$. By definition,
\begin{equation}
    P(2^k) = \frac{\mu(\hat{A}(2^k))}{\mu(S(2^k))}
    \label{eq:p-2k-def}
\end{equation}
Substituting Equation \ref{eq:measure-of-shell} into Equation \ref{eq:delta-p-2k-def} yields
\begin{equation}
    \Delta P(2^k) = \frac{\mu(\Delta \hat{A}(2^k))}{(2^n-1)\mu(S(2^k))}
    \label{eq:delta-p-2k-sub}
\end{equation}
We can now write $P(2^{k+1})$ in terms of the measures of the sets associated with the small ball $S(2^k)$ and shell $\Delta S(2^k)$ by substituting Equations \ref{eq:scale-a-2k-ball-up} and \ref{eq:sum-the-measures-of-basin-ball-and-shell} into Equation \ref{eq:p-2k1-unsuggestive}:
\begin{equation}
    P(2^{k+1}) = \frac{\mu(\hat{A}(2^k)) + \mu(\Delta \hat{A}(2^k))}{2^n\mu(S(2^k))}
\end{equation}
Separating,
\begin{equation}
    P(2^{k+1}) = \frac{\mu(\hat{A}(2^k))}{2^n\mu(S(2^k))} + \frac{\mu(\Delta \hat{A}(2^k))}{2^n\mu(S(2^k))}
\end{equation}
Now, we can write these terms in a suggestive way that will allow us to substitute in $P(2^k)$ and $\Delta P(2^k)$:
\begin{equation}
    \begin{split}
        P(2^{k+1}) &= \frac{1}{2^n}\left(\frac{\mu(\hat{A}(2^k))}{\mu(S(2^k))}\right) \\
        &\mathrel{\phantom{=}} + \frac{2^n-1}{2^n}\left(\frac{\mu(\Delta \hat{A}(2^k))}{(2^n-1)\mu(S(2^k))}\right) \\
        &= \frac{1}{2^n}\left(\frac{\mu(\hat{A}(2^k))}{\mu(S(2^k))}\right) \\
        &\mathrel{\phantom{=}} + \left(1-\frac{1}{2^n}\right)\left(\frac{\mu(\Delta \hat{A}(2^k))}{(2^n-1)\mu(S(2^k))}\right)
    \end{split}
\end{equation}
Finally, substituting in Equations \ref{eq:p-2k-def} and \ref{eq:delta-p-2k-sub} gives us
\begin{equation}
    P(2^{k+1}) = \frac{P(2^k)}{2^n} + \left(1-\frac{1}{2^n}\right)\Delta P(2^k)
\end{equation}
This completes the derivation of Equation \ref{eq:iteration-of-shell-method}.

\subsection{Determinant and Trace of the Jacobian Matrix of a Fixed Point at a Neimark-Sacker Bifurcation}
\label{det-and-trace-ns}

Let us say that the map in Equation \ref{eq:alpha-parameter-2d-map} has a fixed point $\mathbf{x}_s(\alpha)$ that undergoes a Neimark-Sacker bifurcation at $\alpha=\alpha_0$. The Jacobian matrix of this fixed point at $\alpha_0$ is
\begin{equation}
    J(\mathbf{x}_s(\alpha_0)) = \begin{pmatrix}
        \frac{\partial f^{[1]}}{\partial x}\Big|_{\alpha=\alpha_0} & \frac{\partial f^{[1]}}{\partial y}\Big|_{\alpha=\alpha_0} \\[4pt]
        \frac{\partial f^{[2]}}{\partial x}\Big|_{\alpha=\alpha_0} & \frac{\partial f^{[2]}}{\partial y}\Big|_{\alpha=\alpha_0}
    \end{pmatrix}
    = \begin{pmatrix}
        j_{11} & j_{12} \\
        j_{21} & j_{22}
    \end{pmatrix}
\end{equation}
Using this notation, the determinant of the matrix is
\begin{equation}
    \det J(\mathbf{x}_s(\alpha_0)) = j_{11}j_{22} - j_{12}j_{21} = \Delta
\end{equation}
where we assign the determinant to the variable $\Delta$ for ease of notation. Similarly, the trace is
\begin{equation}
    \tr J(\mathbf{x}_s(\alpha_0)) = j_{11} + j_{22} = \tau
\end{equation}
where we assign the trace to the variable $\tau$. 

Recall from linear algebra that we can find the eigenvalues of $J(\mathbf{x}_s(\alpha_0))$, which we denote as $\nu$, using
\begin{equation}
    \det(J(\mathbf{x}_s(\alpha_0)) - \nu I) = 0
\end{equation}
Simplifying this, we get the characteristic equation
\begin{equation}
    \begin{split}
        \det(J(\mathbf{x}_s(\alpha_0)) &= \begin{pmatrix}
            j_{11}-\nu & j_{12} \\
            j_{21} & j_{22}-\nu
        \end{pmatrix} \\
        &= (j_{11}-\nu)(j_{22}-\nu) - j_{12}j_{21} \\
        &= \nu^2 - (j_{11} + j_{22})\nu + j_{11}j_{22} - j_{12}j_{21} \\
        &= \nu^2 - \tau\nu + \Delta = 0
    \end{split}
\end{equation}
Then, we can use the quadratic formula to get
\begin{equation}
    \nu_{1,\,2}(\alpha_0) = \frac{\tau\pm\sqrt{\tau^2-4\Delta}}{2}
\end{equation}
However, since a Neimark-Sacker bifurcation occurs at $\mathbf{x}_s(\alpha_0)$, Equation \ref{eq:eigenvalues-at-ns-bifurcation} tells us that
\begin{equation}
    \nu_{1,\,2}(\alpha_0) = e^{\pm i\varphi}
\end{equation}
This gives us a system of equations
\begin{align}
    \cos\varphi + i\sin\varphi &= \frac{\tau + \sqrt{\tau^2 - 4\Delta}}{2} \label{eq:trace-and-det-system-1}\\
    \cos\varphi - i\sin\varphi &= \frac{\tau - \sqrt{\tau^2 - 4\Delta}}{2}
\end{align}
Adding these two equations together, we get that
\begin{equation}
    \tau = 2\cos\varphi
\end{equation}
Substituting this into Equation \ref{eq:trace-and-det-system-1}, we get
\begin{equation}
    \cos\varphi + i\sin\varphi = \frac{2\cos\varphi + \sqrt{4\cos^2\varphi - 4\Delta}}{2}
\end{equation}
Solving this equation for $\Delta$ yields
\begin{equation}
    \Delta = 1
\end{equation}
This completes the derivation of Equations \ref{eq:det-jacobian-ns-bif} and \ref{eq:trace-jacobian-ns-bif}.

\subsection{Partitioning the Jacobian Matrix of a System of Two Electrically Coupled Rulkov 1 Neurons}
\label{partioning-jacobian-matrix-for-two-coup-rulkov-1-neurons}

Partitioning the $4\times 4$ matrices in Equation \ref{eq:rulkov_1_extremely_big_jacobian_matrix} into four $2\times 2$ submatrices yields five distinct forms, three of which appear on the diagonals and two of which appear on the off-diagonals. We will notate these $2\times 2$ matrices as follows:
\begin{align}
    J_{\text{dg},\,1}(x,\,\alpha,\,g;\,\eta) &= \begin{pmatrix}
        \frac{\alpha}{(1-x)^2}-g & 1 \\
        -\eta(1+g) & 1
    \end{pmatrix} \\
    J_{\text{dg},\,2}(x,\,\alpha,\,g;\,\eta) &= \begin{pmatrix}
        -g & 1 \\
        -\eta(1+g) & 1
    \end{pmatrix} \\
    J_{\text{dg},\,3}(x,\,\alpha,\,g;\,\eta) &= \begin{pmatrix}
        0 & 0 \\
        -\eta(1+g) & 1
    \end{pmatrix} \\ 
    J_{\text{odg},\,1}(g;\,\eta) &= \begin{pmatrix}
        g & 0 \\
        \eta g & 0
    \end{pmatrix} \\
    J_{\text{odg},\,2}(g;\,\eta) &= \begin{pmatrix}
        0 & 0 \\
        \eta g & 0
    \end{pmatrix}
\end{align}
Now, let us assign numbers to some variables $a$, $b$, $c$, and $d$ based on the intervals $x_{1}$ and $x_{2}$ lie on their respective Rulkov 1 fast map piecewise function inputs:
\begin{equation}
    \begin{cases}
        x_{1}\leq 0 & \implies a=1,\,b=1 \\
        0<x_{1}<\alpha_1+y_{1}+\mathfrak{C}_1 & \implies a=2,\,b=1 \\
        x_{1}\geq\alpha_1+y_{1}+\mathfrak{C}_1 & \implies a=3,\,b=2
    \end{cases}
    \label{eq:jacobian-abcd-1}
\end{equation}
\begin{equation}
    \begin{cases}
        x_{2}\leq 0 & \implies d=1,\,c=1 \\
        0<x_{2}<\alpha_2+y_{2}+\mathfrak{C}_2 & \implies d=2,\,c=1 \\
        x_{2}\geq\alpha_2+y_{2}+\mathfrak{C}_2 & \implies d=3,\,c=2
    \end{cases}
    \label{eq:jacobian-abcd-2}
\end{equation}
Then, we can compactify the entirety of the Jacobian matrix in Equation \ref{eq:rulkov_1_extremely_big_jacobian_matrix} into
\begin{equation}
    J(\mathbf{X}) = \begin{pmatrix}
        J_{\text{dg},\,a}(x_{1},\,\alpha_1,\,g^e) & J_{\text{odg},\,b}(g^e) \\
        J_{\text{odg},\,c}(g^e) & J_{\text{dg},\,d}(x_{2},\,\alpha_2,\,g^e)
    \end{pmatrix}
\end{equation}

\newpage

\section{Code}

This appendix contains some of the Python code utilized in this paper.

\subsection{Orbits of the Logistic Map}
\label{orbits_logistic_code}

This code is used to produce the graphs shown in Figure \ref{fig:logistic_orbits}. Figures \ref{fig:logistic_3.25}, \ref{fig:logistic_3.5}, and \ref{fig:logistic_3.75} are produced by changing the value of \verb|r| in Line 24.
\begin{lstlisting}[language=python]
import numpy as np
import matplotlib.pyplot as plt

def logistic_map(r, x):
    return r * x * (1 - x)

def generate_orbit(r, x0, num_iterations):
    orbit = [x0]
    x = x0

    for _ in range(num_iterations):
        x = logistic_map(r, x)
        orbit.append(x)
    
    return orbit

def graph_orbit(num_iterations, orbit):
    plt.plot(orbit, color='black')
    plt.xlabel('t')
    plt.ylabel('x')
    plt.ylim(0, 1)
    plt.show

r = 2.75
x0 = 0.1
num_iterations = 50

orbit = generate_orbit(r, x0, num_iterations)
graph_orbit(num_iterations, orbit)
\end{lstlisting}

\subsection{Bifurcation Diagram of the Logistic Map}
\label{bifurcation_diagram_logistic_code}

This code is used to produce the graph shown in Figure \ref{fig:logistic_bifurcation}. It uses the functions \verb|logistic_map| and \verb|generate_orbit| from Appendix \ref{orbits_logistic_code}.

\begin{lstlisting}[language=python]
import numpy as np
import matplotlib.pyplot as plt

def bifurcation_diagram(
        r_values, x0, 
        num_iterations, num_transient):
    results = []

    for r in r_values:
        x = x0
        for _ in range(num_transient):
            x = logistic_map(r, x)

        attractor_orbit \ 
            = generate_orbit(
                r, x, num_iterations)

        results.append(attractor_orbit)

    for i in range(len(r_values)):
        r = r_values[i]
        plt.plot(
            np.repeat(r, num_iterations + 1), 
            results[i], ',', color='black')

    plt.xlabel('r')
    plt.ylabel('x')
    plt.show()

start = 2
end = 4
r_step = 0.0002
r_values = np.arange(start, end, r_step)
x0 = 0.5
num_iterations = 100
num_transient = 100

bifurcation_diagram(
    r_values, x0, 
    num_iterations, num_transient)
\end{lstlisting}

\subsection{Lyapunov Exponents of the Logistic Map}
\label{logistic_lyapunov_code}

This code is used to produce the graph shown in Figure \ref{fig:logistic_lyapunov}. It uses the functions \verb|logistic_map| and \verb|generate_orbit| from Appendix \ref{bifurcation_diagram_logistic_code}.

\begin{lstlisting}[language=python]
import numpy as np
import matplotlib.pyplot as plt

def lyapunov_calculate(
        r, x0, num_iterations, delta_x0):
    orbit = generate_orbit(
        r, x0, num_iterations)
    lyapunov_sum = 0

    for k in range(num_iterations):
        diff = np.absolute(
            logistic_map(r, 
                orbit[k] + delta_x0) 
            - orbit[k+1])
        lyapunov_sum += np.log(diff / delta_x0)
    
    lyapunov_exp \
        = lyapunov_sum / num_iterations

    return lyapunov_exp
    
def lyapunov_graph(
        r_values, x0, 
        num_iterations, delta_x0):
    lyapunov_exp_values = []

    for r in r_values:
        lyapunov_exp = lyapunov_calculate(
            r, x0, 
            num_iterations, delta_x0)
        lyapunov_exp_values.append(
            lyapunov_exp)
    
    plt.plot(
        r_values, 
        lyapunov_exp_values, 
        ',', color='black', 
        markersize=0.25)

    plt.xlabel('r')
    plt.ylabel('Lyapunov exponent')
    plt.ylim(-3.75, 0.75)
    plt.show()

start = 2.01
end = 4
r_step = 0.00001
r_values = np.arange(start, end, r_step)
x0 = 0.5
num_iterations = 200
delta_x0 = 0.00001

lyapunov_graph(
    r_values, x0, 
    num_iterations, delta_x0)
\end{lstlisting}

\subsection{Hénon Attractor}
\label{henon_attractor_code}

This code is used to produce the graph shown in Figure \ref{fig:henon-attractor}. The graphs in Figure \ref{fig:henon-zooms} are produced by changing the parameters of \verb|plt.xlim| and \verb|plt.ylim| in Lines 40 and 41, respectively, as well as altering the value of \verb|num_iterations| in Line 46.

\begin{lstlisting}[language=python]
import numpy as np
import matplotlib.pyplot as plt

def henon_map(a, b, x, y):
    x_iter = 1 + y - a * x * x 
    y_iter = b * x
    return np.array([x_iter, y_iter])

def generate_orbit_henon(
        a, b, initial_state, 
        num_iterations):
    orbit = [initial_state]
    state = initial_state

    for _ in range(num_iterations):
        state = henon_map(
            a, b, state[0], state[1])
        orbit.append(state)
    
    return orbit

def generate_attractor_henon(
        a, b, initial_state, 
        num_iterations, num_transients):
    state = initial_state
    for _ in range(num_transients):
        state = henon_map(
            a, b, state[0], state[1])
    
    attractor_orbit = generate_orbit_henon(
        a, b, state, num_iterations)

    for k in range(num_iterations):
        plt.plot(
            attractor_orbit[k][0], 
            attractor_orbit[k][1], 
            ',', color='black')
    plt.xlabel('x')
    plt.ylabel('y')
    plt.xlim(-1.5, 1.5)
    plt.ylim(-0.5, 0.5)
    plt.show()

a = 1.4
b = 0.3
num_iterations = 250000
num_transients = 10
initial_state = np.array([0, 0])

generate_attractor_henon(
    a, b, 
    initial_state, 
    num_iterations, num_transients)
\end{lstlisting}

\subsection{Lyapunov Spectrum of the Hénon Map}
\label{lyap_spec_henon_code}

This code implements the method in Appendix \ref{qr-meth-lyap-spec-calc} to calculate the Lyapunov spectrum of the Hénon map on its attractor. It uses the function \verb|generate_orbit_henon| from Appendix \ref{henon_attractor_code}.

\begin{lstlisting}[language=python]
import numpy as np

def generate_jacobians(a, b, orbit):
    J = []
    for k in range(len(orbit)):
        J.append(np.array(
            [[-2 * a * orbit[k][0], 1], 
            [b, 0]]))
    return J

def qr_lyapunov_spectrum_calculate(J):
    QR = [[np.array([[0, 0], [0, 0]]), 
        np.array([[0, 0], [0, 0]])]]
    QR.append(np.linalg.qr(J[0]))
    for k in range(2, len(J)):
        J_star = np.matmul(J[k-1], QR[k-1][0])
        QR.append(np.linalg.qr(J_star))
    lyapunov_sum1 = 0
    lyapunov_sum2 = 0
    for k in range(1, len(QR)):
        lyapunov_sum1 = lyapunov_sum1 \
            + np.log(np.absolute(
                QR[k][1][0][0]))
        lyapunov_sum2 = lyapunov_sum2 \
            + np.log(np.absolute(
                QR[k][1][1][1]))
    lyapunov_spectrum = np.array(
        [lyapunov_sum1 / (len(QR) - 1), 
        lyapunov_sum2 / (len(QR) - 1)])
    return lyapunov_spectrum

a = 1.4
b = 0.3
num_iterations = 10000
initial_state = np.array([0, 0])

orbit = generate_orbit_henon(
    a, b, initial_state, num_iterations)
J = generate_jacobians(a, b, orbit)
lyapunov_spectrum \
    = qr_lyapunov_spectrum_calculate(J)
print(lyapunov_spectrum)
\end{lstlisting}

\subsection{Box-Counting on the Hénon Attractor}
\label{henon-box-counting-code}

This code counts the number of boxes touched by the Hénon attractor for a given box size, and its results are shown in Table \ref{tab:box_henon_values}. It uses the function \verb|henon_map| from Appendix \ref{henon_attractor_code}.

\begin{lstlisting}[language=python]
import numpy as np

def generate_attractor_orbit_henon(
        a, b, initial_state, 
        num_iterations, num_transients):
    state = initial_state
    for _ in range(num_transients):
        state = henon_map(
            a, b, state[0], state[1])
    
    attractor_orbit = generate_orbit_henon(
        a, b, state, num_iterations)

    return attractor_orbit

def box_counting(
        attractor_orbit, 
        xmin, xmax, 
        ymin, ymax, 
        epsilon):
    num_boxes_x = int((xmax - xmin) / epsilon)
    num_boxes_y = int((ymax - ymin) / epsilon)
    running_box_count = 0

    for i in range(num_boxes_x):
        for j in range(num_boxes_y):
            for k in range(num_iterations):
                if ((xmin + epsilon * i) 
                        < attractor_orbit[k][0] 
                        < (xmin + 
                            epsilon * (i + 1))
                        and (ymin 
                            + epsilon * j) 
                        < attractor_orbit[k][1] 
                        < (ymin + 
                            epsilon * (j+1))):
                    running_box_count += 1
                    break

    return running_box_count

a = 1.4
b = 0.3
num_iterations = 250000
num_transients = 10
initial_state = np.array([0, 0])
xmin = -1.5
xmax = 1.5
ymin = -0.5
ymax = 0.5
epsilon = 1/4

attractor_orbit \
    = generate_attractor_orbit_henon(
        a, b, 
        initial_state, 
        num_iterations, 
        num_transients)
boxes = box_counting(
    attractor_orbit, 
    xmin, xmax, 
    ymin, ymax, 
    epsilon)
print(boxes)
\end{lstlisting}

\subsection{Classifying the Basin of the Hénon Attractor}
\label{classifying-henon-basin-code}

This code approximates $P(\xi)$ values for $\xi = 2^k$ using a Monte Carlo method, and its results are displayed in Table \ref{tab:p_function_henon_values}. These values are used to classify the Hénon attractor's basin, which we do in Section \ref{basins-of-attraction}. The code uses the functions \verb|henon_map| and \verb|generate_attractor_orbit_henon| from Appendices \ref{henon_attractor_code} and \ref{henon-box-counting-code}, respectively.

\begin{lstlisting}[language=python]
import numpy as np

def attractor_mean(attractor_orbit):
    mean = (sum(attractor_orbit) 
        / len(attractor_orbit))
    return mean

def attractor_standard_dev(
        attractor_orbit, mean):
    running_sum = 0
    diff = attractor_orbit - mean
    for k in range(len(attractor_orbit)):
        running_sum = running_sum + np.dot(
            diff[k], diff[k])
    standard_dev = np.sqrt(
        running_sum / len(attractor_orbit))
    return standard_dev

def p_function_values(
        a, b, mean, 
        sigma, num_xi_values, 
        num_test_points, 
        num_test_point_iterations):
    count = 0
    distance = np.random.uniform(
        0, sigma, num_test_points)
    angle = np.random.uniform(
        0, 2 * np.pi, num_test_points)
    test_val = 100000 * sigma
    for i in range(num_test_points):
        x_val = (mean[0] 
            + distance[i] * np.cos(angle[i]))
        y_val = (mean[1] 
            + distance[i] * np.sin(angle[i]))
        point = np.array([x_val, y_val])
        flag = 0
        for _ in range(
                num_test_point_iterations):
            point = henon_map(
                a, b, 
                point[0], point[1])
            if np.dot(point, point) > test_val:
                flag = 1
                break
        if flag == 0:
            count += 1
    p_values = [count / num_test_points]

    for k in range(num_xi_values - 1):
        count = 0
        distance = np.random.uniform(
            sigma * 2**k, 
            sigma * 2**(k+1), 
            num_test_points)
        angle = np.random.uniform(
            0, 2 * np.pi, 
            num_test_points)
        test_val = 100000 * sigma * 2**(k+1)
        for i in range(num_test_points):
            x_val = (mean[0] 
                + distance[i] 
                * np.cos(angle[i]))
            y_val = (mean[1] 
                + distance[i] 
                * np.sin(angle[i]))
            point = np.array([x_val, y_val])
            flag = 0
            for _ in range(
                    num_test_point_iterations):
                point = henon_map(
                    a, b, 
                    point[0], point[1])
                if (np.dot(point, point) 
                        > test_val):
                    flag = 1
                    break
            if flag == 0:
                count += 1
        shell_p_value = count / num_test_points
        p_value = (p_values[k] / 4 
            + 3 * shell_p_value / 4)
        p_values.append(p_value)
    
    return p_values
    

a = 1.4
b = 0.3
num_iterations = 1000000
num_transients = 10
num_xi_values = 14
num_test_points = 1000000
num_test_point_iterations = 1000

attractor_orbit \
    = generate_attractor_orbit_henon(
        a, b, 
        initial_state, 
        num_iterations, 
        num_transients)
mean = attractor_mean(attractor_orbit)
sigma = attractor_standard_dev(
    attractor_orbit, mean)
p_values = p_function_values(
    a, b, 
    mean, sigma, 
    num_xi_values, 
    num_test_points, 
    num_test_point_iterations)
print(p_values)
\end{lstlisting}

\subsection{Visualizing the Basin of the Hénon Basin Boundary Attractor}
\label{visualizing-henon-basin-code}

This code is used to produce the graph shown in Figure \ref{fig:henon-basin-nonfractal}. It uses the function \verb|generate_orbit_henon| from Appendix \ref{henon_attractor_code}. 

\begin{lstlisting}[language=python]
import numpy as np
import matplotlib.pyplot as plt

def generate_basin(
        xmin, xmax, 
        ymin, ymax, 
        stepsize, 
        a, b, 
        num_iterations):
    xnumpix = int((xmax - xmin) / stepsize)
    ynumpix = int((ymax - ymin) / stepsize)
    xvals = np.linspace(
        xmin, xmax, 
        xnumpix, 
        endpoint = False)
    yvals = np.linspace(
        ymin, ymax, 
        ynumpix, 
        endpoint = False)
    basin = np.zeros((ynumpix, xnumpix))

    for i in range(xnumpix):
        for j in range(ynumpix):
            state = np.array(
                [xvals[i], yvals[j]])
            orbit = generate_orbit_henon(
                a, b, 
                state, num_iterations)
            if (np.abs(
                    orbit[num_iterations - 1]
                    [0]) > 10000):
                continue
            else:
                basin[ynumpix - 1 - j][i] = 1
    
    plt.imshow(
        basin, cmap='gray', vmin=0, vmax=1)
    plt.axis("off")

xmin = -5
xmax = 5
ymin = -5
ymax = 5
stepsize = 0.001
a = 1.4
b = 0.3
num_iterations = 20

generate_basin(
    xmin, xmax, 
    ymin, ymax, 
    stepsize, 
    a, b, 
    num_iterations)
\end{lstlisting}

\subsection{Uncertainty Exponent of the Hénon Basin Boundary}
\label{uncertainty_exponent_henon_basin_boundary_code}

This code approximates $\varrho(\epsilon)$ values for $\epsilon = 2^{-k}$ using a Monte Carlo method. These values are used to calculate the uncertainty exponent of the Hénon basin boundary, which we do in Section \ref{basins-of-attraction}. The code uses the function \verb|henon_map| from Appendix \ref{henon_attractor_code}.

\begin{lstlisting}[language=python]
import numpy as np

def generate_uncertainty_function_values(
        a, b, 
        num_points, 
        num_iterations, 
        xmin, xmax, 
        ymin, ymax, 
        num_epsilon_values):
    uncertainty_function_values = []
    for k in range(num_epsilon_values):
        epsilon = np.power(2.0, -k)
        count = 0
        for _ in range(num_points):
            basin_flag = 0
            test_point_flag = 0
            xval = (np.random.random() 
                * (xmax - xmin) + xmin)
            yval = (np.random.random() 
                * (ymax - ymin) + ymin)
            point = np.array([xval, yval])
            for _ in range(num_iterations):
                point = henon_map(
                    a, b, 
                    point[0], point[1])
                if (np.dot(point, point) 
                        > 10000):
                    basin_flag = 1
                    break
            
            for i in range(4):
                if i == 0:
                    test_point = np.array(
                        [xval + epsilon, yval])
                elif i == 1:
                    test_point = np.array(
                        [xval - epsilon, yval])
                elif i == 2:
                    test_point = np.array(
                        [xval, yval + epsilon])
                elif i == 3:
                    test_point = np.array(
                        [xval, yval - epsilon])
                
                test_point_flag = 0
                for _ in range(num_iterations):
                    test_point = henon_map(
                        a, b, 
                        test_point[0], 
                        test_point[1])
                    if (np.dot(
                            test_point, 
                            test_point) 
                            > 10000):
                        test_point_flag = 1
                        break
                if (test_point_flag 
                        != basin_flag):
                    count += 1
                    break
        
        uncertainty_value \
            = count / num_points
        uncertainty_function_values.append(
            uncertainty_value)

    return uncertainty_function_values

a = 1.45
b = 0.3
num_points = 500000
num_iterations = 50
xmin = -5
xmax = 5
ymin = -5
ymax = 5
num_epsilon_values = 25

uncertainty_function_values \
    = generate_uncertainty_function_values(
        a, b, 
        num_points, 
        num_iterations, 
        xmin, xmax, 
        ymin, ymax, 
        num_epsilon_values)
print(uncertainty_function_values)
\end{lstlisting}

\subsection{Attractors of the Alternate Hénon Map}
\label{alternate-henon-map-attractors-code}

This code is used to produce the graph shown in Figure \ref{fig:alternate-henon-attractors}.

\begin{lstlisting}[language=python]
import numpy as np
import matplotlib.pyplot as plt

def alternate_henon_map(a, b, x, y):
    x_iter = a - x * x + b * y
    y_iter = x
    return np.array([x_iter, y_iter])

def generate_orbit_alternate_henon(
        a, b, initial_state, num_iterations):
    orbit = [initial_state]
    state = initial_state

    for _ in range(num_iterations):
        state = alternate_henon_map(
            a, b, 
            state[0], state[1])
        orbit.append(state)
    
    return orbit

def generate_attractor_plot_alternate_henon(
        a, b, initial_state_1, initial_state_2,
        num_iterations, num_transients):
    state_1 = initial_state_1
    state_2 = initial_state_2
    for _ in range(num_transients):
        state_1 = alternate_henon_map(
            a, b, state_1[0], state_1[1])
        state_2 = alternate_henon_map(
            a, b, state_2[0], state_2[1])
    
    attractor_orbit_1 \
        = generate_orbit_alternate_henon(
            a, b, state_1, num_iterations)
    attractor_orbit_2 \
        = generate_orbit_alternate_henon(
            a, b, state_2, num_iterations)

    for k in range(num_iterations):
        plt.plot(
            attractor_orbit_1[k][0], 
            attractor_orbit_1[k][1], 
            'o', markersize=5, 
            color=(0, 0.5, 0))
        plt.plot(
            attractor_orbit_2[k][0], 
            attractor_orbit_2[k][1], 
            'o', markersize=5, 
            color=(0, 0, 1))
    plt.xlabel('x')
    plt.ylabel('y')
    plt.show()

a = 0.71
b = 0.9
num_iterations = 100
num_transients = 250
initial_state_1 = np.array([0.75, -0.75])
initial_state_2 = np.array([0.75, -0.3])

generate_attractor_plot_alternate_henon(
    a, b, initial_state_1, initial_state_2, 
    num_iterations, num_transients)
\end{lstlisting}

\subsection{Visualizing the Basins of the Alternate Hénon Map}
\label{visualizing-wada-henon-code}

This code is used to produce the graphs shown in Figure \ref{fig:henon-wada-basins}. Figure \ref{fig:henon-wada-basins-2} is produced by changing the values of \verb|xmin|, \verb|xmax|, \verb|ymin|, \verb|ymax| in Lines 71, 72, 73, and 74, respectively. The code uses the function \verb|alternate_henon_map| from Appendix \ref{alternate-henon-map-attractors-code}. 

\begin{lstlisting}[language=python]
import numpy as np
import matplotlib.pyplot as plt

def generate_henon_wada_basin(
        xmin, xmax, 
        ymin, ymax,
        stepsize, 
        a, b, 
        num_transients,
        num_iterations, 
        error):
    xnumpix = int((xmax - xmin) / stepsize)
    ynumpix = int((ymax - ymin) / stepsize)
    xvals = np.linspace(
        xmin, xmax, xnumpix, 
        endpoint = False)
    yvals = np.linspace(
        ymin, ymax, ynumpix, 
        endpoint = False)
    basin = np.zeros((ynumpix, xnumpix, 3))

    two_cycle_state = np.array([0.75, -0.3])
    for _ in range(num_transients):
        two_cycle_state = alternate_henon_map(
            a, b, 
            two_cycle_state[0], 
            two_cycle_state[1])
    six_cycle_state = np.array([0.75, -0.75])
    for _ in range(num_transients):
        six_cycle_state = alternate_henon_map(
            a, b, 
            six_cycle_state[0], 
            six_cycle_state[1])

    for i in range(xnumpix):
        for j in range(ynumpix):
            flag = 0
            state = np.array(
                [xvals[i], yvals[j]])
            for _ in range(num_transients):
                state = alternate_henon_map(
                    a, b, state[0], state[1])
                if np.dot(
                        state, state) > 10000:
                    flag = 1
                    break
            if flag == 1:
                basin[ynumpix - 1 - j][i] \
                    = [0.75, 0, 0]
                continue
            for _ in range(num_iterations):
                state = alternate_henon_map(
                    a, b, state[0], state[1])
                if (two_cycle_state[0] 
                        - error < state[0] 
                        < two_cycle_state[0] 
                        + error):
                    basin[ynumpix - 1 - j][i] \
                        = [0, 0, 1]
                    break
                elif (six_cycle_state[0] 
                        - error < state[0] 
                        < six_cycle_state[0] 
                        + error):
                    basin[ynumpix - 1 - j][i] \
                        = [0, 0.5, 0]
                    break
    plt.imshow(basin)
    plt.axis("off")
    return basin

xmin = -2.5
xmax = 2.5
ymin = -2.5
ymax = 2.5
stepsize = 0.01
a = 0.71
b = 0.9
num_iterations = 6
num_transients = 1000
error = 0.05

basin = generate_henon_wada_basin(
    xmin, xmax, 
    ymin, ymax, 
    stepsize, 
    a, b, 
    num_iterations, 
    error)
\end{lstlisting}

\subsection{Detecting Basins of Wada in the Alternate Hénon Map}
\label{detecting-wada-alternate-henon-code}

This code implements the method in Section \ref{basins-of-attraction} to calculate $G_k^p$ values for the region shown in Figure \ref{fig:henon-wada-basins-1}. The results are displayed in Table \ref{tab:G_k^p-values-alternate-henon} and graphed in Figure \ref{fig:G_k^p-graph-alternate-henon}, and they show that the map in Equation \ref{eq:alternative-henon-map} with parameters $a=0.71$ and $b=0.9$ exhibits the Wada property, which we discuss in Section \ref{basins-of-attraction}. The code uses the function \verb|alternate_henon_map| from Appendix \ref{alternate-henon-map-attractors-code}.

\begin{lstlisting}[language=python]
import numpy as np

def two_cycle_state(a, b, num_transients):
  two_cycle_state = np.array([0.75, -0.3])
  for _ in range(num_transients):
    two_cycle_state = alternate_henon_map(
      a, b, 
      two_cycle_state[0], 
      two_cycle_state[1])
  return two_cycle_state

def six_cycle_state(a, b, num_transients):
  six_cycle_state = np.array([0.75, -0.75])
  for _ in range(num_transients):
    six_cycle_state = alternate_henon_map(
      a, b, 
      six_cycle_state[0], 
      six_cycle_state[1])
  return six_cycle_state

def generate_grid_centers(
    xmin, xmax, 
    ymin, ymax, s):
  box_side_length = (xmax - xmin) / s
  grid = np.zeros((s, s, 3))
  for i in range(s):
    for j in range(s):
      grid[s - 1 - j][i][0] \
        = (xmin + box_side_length 
          / 2 + i * box_side_length)
      grid[s - 1 - j][i][1] \
        = (ymin + box_side_length 
          / 2 + j * box_side_length)
  return grid

def color_grid(
    s, grid, 
    a, b, 
    two_cycle_state, 
    six_cycle_state, 
    num_transients, 
    num_iterations, 
    error):
  for i in range(s):
    for j in range(s):
      flag = 0
      state = np.array(
        [grid[j][i][0], 
        grid[j][i][1]])
      for _ in range(num_transients):
        state = alternate_henon_map(
          a, b, state[0], state[1])
        if np.dot(state, state) > 10000:
          flag = 1
          break
      if flag == 1:
        grid[j][i][2] = 1
        continue
      for _ in range(num_iterations):
        state = alternate_henon_map(
          a, b, state[0], state[1])
        if (two_cycle_state[0] - error 
            < state[0] 
            < two_cycle_state[0] + error):
          grid[j][i][2] = 2
          break
        elif (six_cycle_state[0] - error 
            < state[0] 
            < six_cycle_state[0] + error):
          grid[j][i][2] = 3
          break
  return grid

def calculate_K_grid(s, colored_grid):
  K_grid = np.zeros((s, s, 3))
  for k in range(s):
    K_grid[0][k] = [1, 0, 0]
    K_grid[s - 1][k] = [1, 0, 0]
    K_grid[k][0] = [1, 0, 0]
    K_grid[k][s - 1] = [1, 0, 0]
  for i in range(s - 2):
    for j in range(s - 2):
      for m in [-1, 0, 1]:
        for n in [-1, 0, 1]:
          color \
            = (colored_grid
                [j + 1 + n][i + 1 + m][2])
          (K_grid
            [j + 1][i + 1][int(color - 1)]) \
                = 1
  return K_grid

def wada_algorithm(
    s, colored_grid, 
    K_grid, a, b, 
    two_cycle_state, 
    six_cycle_state, 
    num_test_steps):
  for i in range(s):
    for j in range(s):
      if sum(K_grid[j][i]) == 2:
        flag = 0
        test_box = colored_grid[j][i]
        for m in [-1, 0, 1]:
          for n in [-1, 0, 1]:
            if (colored_grid[j + n][i + m][2] 
                != test_box[2]):
              compare_box \
                = colored_grid[j + n][i + m]
              flag = 1
              break
          if flag == 1:
            break
              
        for k in np.linspace(
            1, int(num_test_steps), 
            int(num_test_steps)):
          wada = False
          test_step_size \
            = [(test_box[0] - compare_box[0]) 
              / 2**k, 
              (test_box[0] - compare_box[0]) 
              / 2**k]
          for l in range(int(2**(k-1))):
            flag = 0
            test_point \
              = [compare_box[0] 
                + (2 * l + 1) 
                * test_step_size[0], 
                compare_box[1] 
                + (2 * l + 1) 
                * test_step_size[1]]
            test_point_color = 0
            for _ in range(num_transients):
              test_point = alternate_henon_map(
                a, b, 
                test_point[0], 
                test_point[1])
              if np.dot(
                  test_point, 
                  test_point) > 10000:
                flag = 1
                break
            if flag == 1: 
              test_point_color = 1
            else:
              for _ in range(num_iterations):
                state \
                  = alternate_henon_map(
                    a, b, 
                    test_point[0], 
                    test_point[1])
                if (two_cycle_state[0] 
                    - error 
                    < test_point[0] 
                    < two_cycle_state[0] 
                    + error):
                  test_point_color = 2
                  break
                elif (six_cycle_state[0] 
                    - error 
                    < test_point[0] 
                    < six_cycle_state[0] 
                    + error):
                  test_point_color = 3
                  break
            (K_grid[j][i]
              [test_point_color - 1]) = 1
            if sum(K_grid[j][i]) == 3:
              wada = True
              break
          if wada: break
  G_values = [0, 0, 0]
  for i in range(s):
    for j in range(s):
      K_value = sum(K_grid[j][i])
      G_values[int(K_value - 1)] += 1
  return G_values

def steps_G_values(
    s, colored_grid, 
    K_grid, a, b, 
    two_cycle_state, 
    six_cycle_state, 
    max_test_steps):
  steps_G_values = []
  for i in np.linspace(
      1, max_test_steps, 
      max_test_steps):
    G_values = wada_algorithm(
      s, colored_grid, 
      K_grid, a, b, 
      two_cycle_state, 
      six_cycle_state, 
      i)
    steps_G_values.append(G_values)
  return steps_G_values


a = 0.71
b = 0.9
xmin = -2.5
xmax = 2.5
ymin = -2.5
ymax = 2.5
s = 100
num_transients = 250
num_iterations = 6
error = 0.05
max_test_steps = 16

two_cycle_state = two_cycle_state(
  a, b, num_transients)
six_cycle_state = six_cycle_state(
  a, b, num_transients)
grid = generate_grid_centers(
  xmin, xmax, ymin, ymax, s)
colored_grid = color_grid(
  s, grid, a, b,
  two_cycle_state, 
  six_cycle_state, 
  num_transients, 
  num_iterations, error)
K_grid = calculate_K_grid(s, colored_grid)
steps_G_values = steps_G_values(
  s, colored_grid, 
  K_grid, a, b, 
  two_cycle_state, 
  six_cycle_state, 
  max_test_steps)
print(steps_G_values)
\end{lstlisting}

\subsection{Rulkov Map 1 and Cobweb Orbit}
\label{rulkov_1_and_cobweb_code}

This code contains iteration functions for the fast mapof Rulkov map 1 and for producing cobweb orbits of both $q$-cycles and fixed point attractors. It is used to produce the cobweb orbits shown in some of the figures of the fast map of Rulkov map 1 in Section \ref{individual-dynamics-of-rulkov-map-1}.

\begin{lstlisting}[language=python]
import numpy as np

def fast_rulkov_map_1(x, y, alpha):
    if x <= 0:
        x_iter = alpha / (1 - x) + y
    elif 0 < x < (alpha + y):
        x_iter = alpha + y
    elif x >= (alpha + y):
        x_iter = -1
    return x_iter

def generate_periodic_orbit_cobweb(alpha, y):
    x = alpha + y
    x_iter = -1
    cobweb = [[x, x_iter]]
    x = x_iter
    x_iter = fast_rulkov_map_1(x, y, alpha)
    while x_iter != -1:
        cobweb.append([x, x])
        cobweb.append([x, x_iter])
        x = x_iter
        x_iter = fast_rulkov_map_1(x, y, alpha)
    cobweb.append([x, x])
    cobweb.append([x, x_iter])
    return cobweb

def generate_attraction_cobweb(alpha, y):
    attractor = (1 + y - np.sqrt(
        (y - 1) ** 2 - 4 * alpha)) / 2
    x = alpha + y
    x_iter = -1
    cobweb = [[x, x_iter]]
    x = x_iter
    x_iter = fast_rulkov_map_1(x, y, alpha)
    while abs(attractor - x_iter) > 0.001:
        cobweb.append([x, x])
        cobweb.append([x, x_iter])
        x = x_iter
        x_iter = fast_rulkov_map_1(x, y, alpha)
    return cobweb


alpha = 6
y = -3.93

periodic_orbit_cobweb \
    = generate_periodic_orbit_cobweb(alpha, y)
\end{lstlisting}

\subsection{Spiking Branch of Rulkov Map 1}
\label{spiking_branch_of_rulkov_1_code}

This code is used to approximate the location of the spiking branch of Rulkov map 1 for various values of $\alpha$ using the method presented in Section \ref{individual-dynamics-of-rulkov-map-1}. It is used to produce the spiking branch visualizations in Figures \ref{fig:rulkov_1_state_space_diagram_alpha4} and \ref{fig:rulkov_1_state_space_diagram_alpha6}, and it uses the function \verb|fast_rulkov_map_1| from Appendix \ref{rulkov_1_and_cobweb_code}.

\begin{lstlisting}[language=python]
import numpy as np

def generate_spiking_branch(
        alpha, eta, num_points, y_max):
    if alpha <= 4:
        y_min = 1 - 2 * np.sqrt(alpha)
    else:
        y_min = -1 - alpha / 2
    
    spiking_branch = []
    y_vals = np.linspace(
        y_min, y_max, 
        num_points, endpoint=False)
    for i in range(num_points):
        y = y_vals[i]
        x = fast_rulkov_map_1(-1, y, alpha)
        x_vals = [x]
        while x != -1:
            x = fast_rulkov_map_1(x, y, alpha)
            x_vals.append(x)
        x_mean = sum(x_vals) / len(x_vals)
        
        spiking_branch.append([x_mean, y])
    
    return spiking_branch

alpha = 4
eta = 0.001
num_points = 1000
y_max = -2.5

spiking_branch = generate_spiking_branch(
    alpha, eta, num_points, y_max)
\end{lstlisting}

\subsection{Fast Variable Orbits of Rulkov Map 1}
\label{fast_var_orbs_of_rulkov_1_code}

This code graphs the orbits of the fast variable of Rulkov map 1 $x$ against the discrete-time variable $k$ and is used to produce some of the graphs in Section \ref{individual-dynamics-of-rulkov-map-1}. It uses the function \verb|fast_rulkov_map_1| from Appendix \ref{rulkov_1_and_cobweb_code}

\begin{lstlisting}[language=python]
import numpy as np
import matplotlib.pyplot as plt

def fast_rulkov_map_1(x, y, alpha):
    if x <= 0:
        x_iter = alpha / (1 - x) + y
    elif 0 < x < (alpha + y):
        x_iter = alpha + y
    elif x >= (alpha + y):
        x_iter = -1
    return x_iter

def rulkov_map_1(x, y, sigma, alpha, eta):
    x_iter = fast_rulkov_map_1(x, y, alpha)
    y_iter = y - eta * (x - sigma)
    return [x_iter, y_iter]

def generate_orbit_rulkov_1(
        sigma, alpha, eta, 
        initial_state, num_iterations):
    orbit = [initial_state]
    state = initial_state

    for _ in range(num_iterations):
        state = rulkov_map_1(
            state[0], state[1], 
            sigma, alpha, eta)
        orbit.append(state)
    
    orbit = np.asarray(orbit)
    return orbit

def graph_rulkov_neuron_behavior(orbit):
    x, y = orbit.T
    plt.figure(figsize=(7.5, 6))
    plt.plot(x, color='blue')
    plt.xlabel('k')
    plt.ylabel('x')
    plt.xlim(0, 2000)

alpha = 6
eta = 0.001
sigma = -1.25
initial_state = [-1, -3.93]
num_iterations = 10000

orbit = generate_orbit_rulkov_1(
    sigma, alpha, eta, 
    initial_state, num_iterations)
graph_rulkov_neuron_behavior(orbit)
\end{lstlisting}

\subsection{Lyapunov Spectrum of Rulkov Map 1 and Visualizations of the Maximal Lyapunov Exponent in Parameter Space}
\label{lyapunov_spectrum_rulkov_map_1_code}

This code implements the method in Appendix \ref{qr-meth-lyap-spec-calc} using the Jacobian matrix in Section \ref{chaotic-dynamics-rulkov-map-1} to calculate the Lyapunov spectrum of Rulkov map 1 for some given parameters $\alpha$ and $\sigma$. It also graphs visualizations of the maximal Lyapunov exponents of Rulkov map 1 in parameter space, which is shown in Figure \ref{fig:rulkov_1_lyapunov_exponents_visualizations}. The code uses the function \verb|rulkov_map_1| from Appendix \ref{fast_var_orbs_of_rulkov_1_code}.

\begin{lstlisting}[language=python]
import numpy as np
import matplotlib.pyplot as plt

def generate_jacobians_rulkov_1(
        alpha, eta, orbit):
    J = []
    for k in range(len(orbit)):
        x = orbit[k][0]
        y = orbit[k][1]
        if x <= 0:
            J.append(np.array(
                [[alpha / (1 - x)**2, 1], 
                [-eta, 1]]))
        if 0 < x < alpha + y:
            J.append(np.array(
                [[0, 1], 
                [-eta, 1]]))
        if x >= alpha + y:
            J.append(np.array(
                [[0, 0], 
                [-eta, 1]]))
    return J

def qr_lyap_spec_calc_rulkov(J):
    QR = [[np.array([[0, 0], [0, 0]]), 
        np.array([[0, 0], [0, 0]])]]
    QR.append(np.linalg.qr(J[0]))
    for k in range(2, len(J)):
        J_star = np.matmul(J[k-1], QR[k-1][0])
        QR.append(np.linalg.qr(J_star))
    lyapunov_sum1 = 0
    lyapunov_sum2 = 0
    for k in range(1, len(QR)):
        lyapunov_sum1 = lyapunov_sum1 \
            + np.log(np.absolute(
                QR[k][1][0][0]))
        lyapunov_sum2 = lyapunov_sum2 \
            + np.log(np.absolute(
                QR[k][1][1][1]))
    lyapunov_spectrum = np.array(
        [lyapunov_sum1 / (len(QR) - 1), 
        lyapunov_sum2 / (len(QR) - 1)])
    return lyapunov_spectrum

def generate_param_space_lyap_color_diag(
        sigma_min, sigma_max, 
        alpha_min, alpha_max, 
        edge_numpix, num_iterations):
    sigma_vals = np.linspace(
        sigma_min, sigma_max, 
        edge_numpix, endpoint = False)
    alpha_vals = np.linspace(
        alpha_min, alpha_max, 
        edge_numpix, endpoint = False)
    lyapunov_colors \
        = np.zeros((edge_numpix, edge_numpix))

    for i in range(edge_numpix):
        for j in range(edge_numpix):
            sigma = sigma_vals[i]
            alpha = alpha_vals[j]
            initial_state \
                = np.array([-1, -3.5])
            orbit = generate_orbit_rulkov_1(
                sigma, alpha, eta, 
                initial_state, num_iterations)
            J = generate_jacobians_rulkov_1(
                alpha, eta, orbit)
            lyapunov_spectrum \
                = qr_lyap_spec_calc_rulkov(J)
            lyapunov_colors[
                edge_numpix - 1 - j][i] \
                    = lyapunov_spectrum[0]
    
    plt.imshow(
        lyapunov_colors, cmap='coolwarm')
    plt.axis("off")
    plt.colorbar()
    return lyapunov_colors

def generate_3d_lyap_graph(
        sigma_min, sigma_max, 
        alpha_min, alpha_max, 
        edge_numpix, lyapunov_colors):
    fig = plt.figure()
    ax = fig.add_subplot(projection='3d')
    sigma_values = np.linspace(
        sigma_min, sigma_max, 
        edge_numpix, endpoint=False)
    alpha_values = np.linspace(
        alpha_min, sigma_max, 
        edge_numpix, endpoint=False)
    for i in range(edge_numpix):
        for j in range(edge_numpix):
            ax.scatter(
                sigma_values[i], 
                alpha_values[j], 
                lyapunov_colors[
                    edge_numpix - 1 - j][i], 
                c=lyapunov_colors[
                    edge_numpix - 1 - j][i], 
                cmap='coolwarm', 
                vmin=-0.2, vmax=0.2, 
                marker='o')
    ax.set_xlabel('sigma')
    ax.set_ylabel('alpha')
    ax.set_zlabel('maximum Lyapunov exponent')
    plt.show()

eta = 0.001
sigma_min = -2
sigma_max = 0
alpha_min = 2
alpha_max = 8
edge_numpix = 100
num_iterations = 20000

lyapunov_colors \
    = generate_param_space_lyap_color_diag(
        sigma_min, sigma_max, 
        alpha_min, alpha_max, 
        edge_numpix, num_iterations)
\end{lstlisting}

\subsection{Rulkov Map 2 Cobweb Orbit, Graphs, and Lyapunov Exponents}
\label{rulkov_2_graphs_and_cobweb_code}

This code contains many functions related to Rulkov map 2, including code to make a cobweb orbit, to graph $x_k$, $y_k$, and two-dimensional state space $\langle y,\,x\rangle$, and to calculate the Lyapunov spectrum of the map. It is used to produce many of the figures in Sections \ref{dynamics-rulkov-map-2} and \ref{one-rulkov-2-neuron}. The code uses the function \verb|qr_lyapunov_spectrum_calculate_rulkov| from Appendix \ref{lyapunov_spectrum_rulkov_map_1_code}.

\begin{lstlisting}[language=python]
import numpy as np
import matplotlib.pyplot as plt

def fast_rulkov_map_2(x, y, alpha):
    x_iter = alpha / (1 + x*x) + y
    return x_iter

def generate_rulkov2_cobweb(
        alpha, y, num_iterations):
    x = 0
    x_iter = fast_rulkov_map_2(x, y, alpha)
    cobweb = [[x, x_iter]]
    x = x_iter
    x_iter = fast_rulkov_map_2(x, y, alpha)
    for _ in range(num_iterations):
        cobweb.append([x, x])
        cobweb.append([x, x_iter])
        x = x_iter
        x_iter = fast_rulkov_map_2(x, y, alpha)
    cobweb.append([x, x])
    cobweb.append([x, x_iter])
    return cobweb

def rulkov_map_2(x, y, sigma, alpha, eta):
    x_iter = alpha / (1 + x*x) + y
    y_iter = y - eta * (x - sigma)
    return [x_iter, y_iter]

def generate_orbit_rulkov_2(
        sigma, alpha, eta, 
        initial_state, 
        num_iterations):
    orbit = [initial_state]
    state = initial_state

    for _ in range(num_iterations):
        state = rulkov_map_2(
            state[0], state[1], 
            sigma, alpha, eta)
        orbit.append(state)
    
    orbit = np.asarray(orbit)
    return orbit

def graph_rulkov_neuron_behavior(orbit):
    x, y = orbit.T
    plt.figure(figsize=(7.5, 3))
    plt.plot(x, color='blue')
    plt.xlabel('k')
    plt.ylabel('x')
    plt.xlim(0, 5000)
    plt.show()

    plt.figure(figsize=(7.5, 3))
    plt.plot(y, color='red')
    plt.xlabel('k')
    plt.ylabel('y')
    plt.xlim(0, 5000)
    plt.show()

    for i in range(num_iterations - 1000):
        x[i] = x[i + 1000]
        y[i] = y[i + 1000]
    plt.plot(y, x, ',', color='black')
    plt.xlabel('y')
    plt.ylabel('x')

def generate_jacobians_rulkov_2(
        alpha, eta, orbit):
    J = []
    for i in range(len(orbit)):
        x = orbit[i][0]
        y = orbit[i][1]
        J.append(np.array(
            [[(-2*alpha*x) / ((1+x*x)**2), 1],
            [-eta, 1]]))
    return J

alpha = 4.1
eta = 0.001
sigma = -2
initial_state = [0, -2.52]
num_iterations = 250000

orbit = generate_orbit_rulkov_2(
    sigma, alpha, eta, 
    initial_state, num_iterations)
graph_rulkov_neuron_behavior(orbit)
J = generate_jacobians_rulkov_2(
    alpha, eta, orbit)
lyapunov_spectrum \
    = qr_lyapunov_spectrum_calculate_rulkov(J)
print(lyapunov_spectrum)
\end{lstlisting}

\subsection{Bifurcation Diagram of the Fast Variable of Rulkov Map 2}
\label{bifurcation_diagram_rulkov_2_code}

This code is used to produce the graph shown in Figure \ref{fig:bifurc-rulkov-2}. It uses the function \verb|fast_rulkov_map_2| from Appendix \ref{rulkov_2_graphs_and_cobweb_code}.

\begin{lstlisting}[language=python]
import numpy as np
import matplotlib.pyplot as plt

def bifurcation_diagram_rulkov_2(
        x_min, x_max, y_values, alpha, 
        num_iterations, num_transients):
    results = []

    for y in y_values:
        x_val = np.random.random() \
            * (x_max - x_min) + x_min

        for _ in range(num_transients):
            x_val = fast_rulkov_map_2(
                x_val, y, alpha)
        
        attractor_orbit = [x_val]
        for _ in range(num_iterations):
            x_val = fast_rulkov_map_2(
                x_val, y, alpha)
            attractor_orbit.append(x_val)
        attractor_orbit \
            = np.asarray(attractor_orbit)

        results.append(attractor_orbit)

    for i in range(len(y_values)):
        y = y_values[i]
        plt.plot(
            np.repeat(y, num_iterations + 1), 
            results[i], ',', color='black')

    plt.xlabel('y')
    plt.ylabel('x')
    plt.show()

x_min = -2
x_max = 0
y_min = -4.5
y_max = -2
y_step = 0.0001
y_values = np.arange(y_min, y_max, y_step)
num_iterations = 25
num_transients = 500
alpha = 4.1

bifurcation_diagram_rulkov_2(
    x_min, x_max, y_values, alpha, 
    num_iterations, num_transients)
\end{lstlisting}

\subsection{Orbits of Rulkov Map 1 with a Current Pulse Injection}
\label{orbits_rulkov_1_current_pulse_injection_code}

This code graphs both the fast variable and slow variable orbits of Rulkov map 1, modeling the response to an injection of a current pulse $I_k$. It is used to produce the graphs in Figures \ref{fig:inject_current_rulkov_sc1_bc0_graphs} and \ref{fig:inject_current_rulkov_sc1_bc1_graphs}, and it uses the function \verb|fast_rulkov_map_1| from Appendix \ref{rulkov_1_and_cobweb_code}. 

\begin{lstlisting}[language=python]
import numpy as np
import matplotlib.pyplot as plt

def rulkov_map_1_current(
        x, y, k, 
        alpha, eta, 
        sigma_val, beta_val):
    x_iter = fast_rulkov_map_1(
        x, y + beta_val, alpha)
    y_iter = y - eta * (x - sigma_val)
    return [x_iter, y_iter]

def current_pulse(
        pulse_start, pulse_duration, 
        pulse_magnitude_and_direction, 
        num_iterations):
    I = []
    for k in range(pulse_start):
        I.append(0)
    for k in range(
            pulse_start, 
            pulse_start + pulse_duration):
        I.append(pulse_magnitude_and_direction)
    for k in range(
            pulse_start + pulse_duration, 
            num_iterations):
        I.append(0)
    return I

def generate_sigma_vals(
        sigma, sigma_coef, 
        I, num_iterations):
    sigma_vals = []
    for k in range(num_iterations):
        sigma_vals.append(
            sigma + sigma_coef * I[k])
    return sigma_vals

def generate_beta_vals(
        beta_coef, I, 
        num_iterations):
    beta_vals = []
    for k in range(num_iterations):
        beta_vals.append(beta_coef * I[k])
    return beta_vals

def generate_orbit_rulkov_1_current(
        alpha, eta, 
        sigma_vals, beta_vals, 
        I, initial_state, 
        num_iterations):
    orbit = [initial_state]
    state = initial_state

    for k in range(num_iterations):
        state = rulkov_map_1_current(
            state[0], state[1], 
            k, alpha, eta, 
            sigma_vals[k], beta_vals[k])
        orbit.append(state)
    
    orbit = np.asarray(orbit)
    return orbit

def graph_current_response(I, orbit):    
    fig, axs = plt.subplots(
        3, sharex=True, figsize=(8, 6))

    axs[0].plot(I, color='gold')
    axs[0].set(ylabel='I')
    axs[0].set_ylim(-1.25, 1.25)
    axs[0]

    x, y = orbit.T
    axs[1].plot(x, color='blue')
    axs[1].set(ylabel='x')
    axs[1].set_ylim(-3.5, 3.5)

    axs[2].plot(y, color='red')
    axs[2].set(ylabel='y')
    axs[2].set_ylim(-3.65, -3.25)
    plt.xlim(0, 1000)
    plt.xlabel('k')

    plt.subplots_adjust(hspace=0)
    plt.show()

alpha = 5
eta = 0.001
sigma = -0.6
pulse_start = 100
pulse_duration = 100
pulse_magnitude_and_direction = 1
sigma_coef = 1
beta_coef = 0
initial_state = [-1, -3.5]
num_iterations = 1000

I = current_pulse(
    pulse_start, pulse_duration, 
    pulse_magnitude_and_direction, 
    num_iterations)
sigma_vals = generate_sigma_vals(
    sigma, sigma_coef, 
    I, num_iterations)
beta_vals = generate_beta_vals(
    beta_coef, I, num_iterations)
orbit = generate_orbit_rulkov_1_current(
    alpha, eta, 
    sigma_vals, beta_vals, 
    I, initial_state, 
    num_iterations)
graph_current_response(I, orbit)
\end{lstlisting}

\subsection{Two Electrically Coupled Rulkov 1 Neurons: Orbits, Fast Variable Orbit Visualization, and Lyapunov Spectrum Calculation of a Symmetrical Coupling}
\label{sym-elec-coup-rulkov-1-neurons-code}

This code contains many functions relating to the electric coupling of two Rulkov 1 neurons, including functions that calculate the coupling parameters and generate an orbit for this system, visualize the fast variable orbits of this system, and calculate the Lyapunov spectrum of this system using the matrix partitioning method outlined in Appendix \ref{partioning-jacobian-matrix-for-two-coup-rulkov-1-neurons}. The code produces the graphs in Figure \ref{fig:sym_coup_rulkov_1_graphs} and uses the function \verb|fast_rulkov_map_1| from Appendix \ref{rulkov_1_and_cobweb_code}.

\begin{lstlisting}[language=python]
import numpy as np
import matplotlib.pyplot as plt

def rulkov_map_1_coupling(
        neuron, 
        coupling_param_x, 
        coupling_param_y, 
        sigma, alpha, eta):
    x = neuron[0]
    y = neuron[1]
    x_iter = fast_rulkov_map_1(
        x, y + coupling_param_x, alpha)
    y_iter = y - eta * x \
        + eta * (sigma + coupling_param_y)
    return [x_iter, y_iter]

def two_rulkov_1_coupling_params(
        neuron_1, neuron_2, ge_1, ge_2):
    coupling_param_1 \
        = ge_1 * (neuron_2[0] - neuron_1[0])
    coupling_param_2 \
        = ge_2 * (neuron_1[0] - neuron_2[0])
    return coupling_param_1, coupling_param_2

def two_coupled_rulkov_1(
        neuron_1, neuron_2, 
        ge_1, ge_2, 
        sigma_1, alpha_1, 
        sigma_2, alpha_2, eta):
    coupling_param_1, coupling_param_2 \
        = two_rulkov_1_coupling_params(
            neuron_1, neuron_2, ge_1, ge_2)
    neuron_1_iter = rulkov_map_1_coupling(
        neuron_1, 
        coupling_param_1, 
        coupling_param_1, 
        sigma_1, alpha_1, eta)
    neuron_2_iter = rulkov_map_1_coupling(
        neuron_2, 
        coupling_param_2, 
        coupling_param_2, 
        sigma_2, alpha_2, eta)
    return neuron_1_iter, neuron_2_iter

def two_coupled_rulkov_1_orbit(
        initial_state_1, initial_state_2, 
        ge_1, ge_2, 
        sigma_1, alpha_1, 
        sigma_2, alpha_2, 
        eta, num_iterations):
    neuron_1_orbit = [initial_state_1]
    neuron_2_orbit = [initial_state_2]
    neuron_1_state = initial_state_1
    neuron_2_state = initial_state_2

    for _ in range(num_iterations):
        neuron_1_state, neuron_2_state \
            = two_coupled_rulkov_1(
                neuron_1_state, neuron_2_state, 
                ge_1, ge_2, 
                sigma_1, alpha_1, 
                sigma_2, alpha_2, eta)
        neuron_1_orbit.append(neuron_1_state)
        neuron_2_orbit.append(neuron_2_state)
    
    neuron_1_orbit = np.asarray(neuron_1_orbit)
    neuron_2_orbit = np.asarray(neuron_2_orbit)
    return neuron_1_orbit, neuron_2_orbit

def graph_rulkov_1_two_coupled(
        neuron_1_orbit, neuron_2_orbit):
    neuron_1_orbit_x = neuron_1_orbit.T[0]
    neuron_2_orbit_x = neuron_2_orbit.T[0]
    
    fig, axs = plt.subplots(
        2, sharex=True, figsize=(10, 5))
    axs[0].plot(neuron_1_orbit_x, color='blue')
    axs[0].set(ylabel='neuron 1 x')
    axs[0].set_xlim(0, 2000)
    axs[1].plot(neuron_2_orbit_x, color='blue')
    axs[1].set(ylabel='neuron 2 x')
    axs[1].set_xlim(0, 2000)
    plt.xlabel('k')
    plt.subplots_adjust(hspace=0)
    plt.show()

def jacobian_submatrices(
        diagonal, index, 
        x, alpha, g, eta):
    if diagonal:
        if index == 1:
            J_sub = np.array(
                [[alpha / (1 - x)**2 - g, 1], 
                [-eta * (1 + g), 1]])
        elif index == 2:
            J_sub = np.array(
                [[-g, 1], 
                [-eta * (1 + g), 1]])
        else:
            J_sub = np.array(
                [[0, 0], 
                [-eta * (1 + g), 1]])
    else:
        if index == 1:
            J_sub = np.array(
                [[g, 0], 
                [eta * g, 0]])
        else:
            J_sub = np.array(
                [[0, 0], 
                [eta * g, 0]])
    return J_sub

def jacobian_submatrices_indices(
        neuron_1, neuron_2, 
        ge_1, ge_2, 
        alpha_1, alpha_2):
    coupling_param_1, coupling_param_2 = \
        two_rulkov_1_coupling_params(
            neuron_1, neuron_2, ge_1, ge_2)
    x_1 = neuron_1[0]
    y_1 = neuron_1[1]
    x_2 = neuron_2[0]
    y_2 = neuron_2[1]
    
    if x_1 <= 0:
        a, b = 1, 1
    elif (0 < x_1 < alpha_1 
            + y_1 + coupling_param_1):
        a, b = 2, 1
    else:
        a, b = 3, 2
    if x_2 <= 0:
        d, c = 1, 1
    elif (0 < x_2 < alpha_2 
            + y_2 + coupling_param_2):
        d, c = 2, 1
    else:
        d, c = 3, 2

    return a, b, c, d

def generate_rulkov_1_coup_jacobians(
        alpha_1, alpha_2, 
        sigma_1, sigma_2, eta, 
        ge_1, ge_2, 
        neuron_1_orbit, neuron_2_orbit):
    J = []
    for k in range(len(neuron_1_orbit)):
        x_1 = neuron_1_orbit[k][0]
        y_1 = neuron_1_orbit[k][1]
        x_2 = neuron_2_orbit[k][0]
        y_2 = neuron_2_orbit[k][1]
        a, b, c, d \
            = jacobian_submatrices_indices(
                neuron_1_orbit[k], 
                neuron_2_orbit[k], 
                ge_1, ge_2, 
                alpha_1, alpha_2)
        J_a = jacobian_submatrices(
            True, a, x_1, alpha_1, ge_1, eta)
        J_b = jacobian_submatrices(
            False, b, x_1, alpha_1, ge_1, eta)
        J_c = jacobian_submatrices(
            False, c, x_2, alpha_2, ge_2, eta)
        J_d = jacobian_submatrices(
            True, d, x_2, alpha_2, ge_2, eta)
        J.append(np.vstack((
            np.hstack((J_a, J_b)), 
            np.hstack((J_c, J_d)))))
    return J

def qr_lyap_rulkov_two_coup(J):
    QR = [[np.array([
            [0, 0, 0, 0], 
            [0, 0, 0, 0], 
            [0, 0, 0, 0], 
            [0, 0, 0, 0]]), 
        np.array([
            [0, 0, 0, 0], 
            [0, 0, 0, 0], 
            [0, 0, 0, 0], 
            [0, 0, 0, 0]])]]
    QR.append(np.linalg.qr(J[0]))
    for k in range(2, len(J)):
        J_star = np.matmul(J[k-1], QR[k-1][0])
        QR.append(np.linalg.qr(J_star))
    lyapunov_sum1 \
        = lyapunov_sum2 \
        = lyapunov_sum3 \
        = lyapunov_sum4 = 0
    for k in range(1, len(QR)):
        lyapunov_sum1 = lyapunov_sum1 \
            + np.log(np.absolute(
                QR[k][1][0][0]))
        lyapunov_sum2 = lyapunov_sum2 \
            + np.log(np.absolute(
                QR[k][1][1][1]))
        lyapunov_sum3 = lyapunov_sum3 \
            + np.log(np.absolute(
                QR[k][1][2][2]))
        lyapunov_sum4 = lyapunov_sum4 \
            + np.log(np.absolute(
                QR[k][1][3][3]))
    lyapunov_spectrum = np.array(
        [lyapunov_sum1 / (len(QR) - 1), 
        lyapunov_sum2 / (len(QR) - 1),
        lyapunov_sum3 / (len(QR) - 1),
        lyapunov_sum4 / (len(QR) - 1)])
    return lyapunov_spectrum

initial_state_1 = [-1, -3.5]
initial_state_2 = [-1, -3.5]
ge = -0.05
sigma_1 = -0.75
sigma_2 = -0.76
alpha_1 = 4.9
alpha_2 = 5.0
eta = 0.001
num_iterations = 100000

neuron_1_orbit, neuron_2_orbit \
    = two_coupled_rulkov_1_orbit(
        initial_state_1, initial_state_2, 
        ge, ge, 
        sigma_1, alpha_1, 
        sigma_2, alpha_2, 
        eta, num_iterations)
graph_rulkov_1_two_coupled(
    neuron_1_orbit, neuron_2_orbit)
J = generate_rulkov_1_coup_jacobians(
    alpha_1, alpha_2, 
    sigma_1, sigma_2, eta, 
    ge, ge, 
    neuron_1_orbit, neuron_2_orbit)
lyapunov_spectrum \
    = qr_lyap_rulkov_two_coup(J)
print(lyapunov_spectrum)
\end{lstlisting}

\subsection{Two Asymmetrically Electrically Coupled Rulkov 1 Neurons}
\label{asym-elec-coup-rulkov-1-neurons-code}

This code uses the functions in Appendix \ref{sym-elec-coup-rulkov-1-neurons-code} to accomplish the same things with an asymmetrical electric coupling of two Rulkov 1 neurons.

\begin{lstlisting}[language=python]
initial_state_1 = [-0.54, -3.25]
initial_state_2 = [-1, -3.25]
ge_1 = 0.05
ge_2 = 0.25
sigma_1 = -0.5
sigma_2 = -0.5
alpha_1 = 4.5
alpha_2 = 4.5
eta = 0.001
num_iterations = 65000

neuron_1_orbit, neuron_2_orbit \
    = two_coupled_rulkov_1_orbit(
        initial_state_1, initial_state_2, 
        ge_1, ge_2, 
        sigma_1, alpha_1, 
        sigma_2, alpha_2, 
        eta, num_iterations)
graph_rulkov_1_two_coupled(
    neuron_1_orbit, neuron_2_orbit)
J = generate_rulkov_1_coup_jacobians(
    alpha_1, alpha_2, 
    sigma_1, sigma_2, eta, 
    ge_1, ge_2, 
    neuron_1_orbit, neuron_2_orbit)
lyapunov_spectrum \
    = qr_lyap_rulkov_two_coup(J)
print(lyapunov_spectrum)

neuron_1_orbit_x = neuron_1_orbit.T[0]
neuron_1_orbit_y = neuron_1_orbit.T[1]
neuron_2_orbit_x = neuron_2_orbit.T[0]
neuron_2_orbit_y = neuron_2_orbit.T[1]
for k in range(1000, num_iterations):
    plt.plot(
        neuron_1_orbit_x[k], 
        neuron_1_orbit_y[k], 
        'o', color='black', 
        markersize=1)
plt.xlabel('neuron 1 x')
plt.ylabel('neuron 1 y')
plt.figure(figsize=(8, 6))
plt.show()
\end{lstlisting}

\subsection{Ring Lattice System: Orbits, Graphs, and Lyapunov Spectrum}
\label{ring-lattice-code}

This code implements the theory from Section \ref{neuron-ring-lattice} to model the behavior of a ring lattice of $\zeta$ electrically coupled Rulkov 1 neurons. It contains many functions, including one to calculate the orbit of the $2\zeta$-dimensional system, one to graph some of the fast variable orbits of the system, and one to calculate the Lyapunov spectrum of the system. It uses the function \verb|rulkov_map_1_coupling| from Appendix \ref{sym-elec-coup-rulkov-1-neurons-code}.

\begin{lstlisting}[language=python]
import numpy as np
import matplotlib.pyplot as plt

def ring_coup_params(system_state, zeta, ge):
  coup_params \
    = [ge/2 * (system_state[zeta-1][0] 
      + system_state[1][0] 
      - 2*system_state[0][0])]
  for i in range(1, zeta-1):
    coup_params.append(ge/2 
      * (system_state[i-1][0] 
        + system_state[i+1][0] 
        - 2*system_state[i][0]))
  coup_params.append(ge/2 
    * (system_state[zeta-2][0] 
      + system_state[0][0] 
      - 2*system_state[zeta-1][0]))
  return coup_params

def ring_coup_rulkov_1(
    system_state, zeta, 
    ge, sigma_vals, alpha_vals, eta):
  coup_params = ring_coup_params(
    system_state, zeta, ge)
  system_state_iter = []
  for i in range(0, zeta):
    neuron_iter = rulkov_map_1_coupling(
      system_state[i], 
      coup_params[i], coup_params[i], 
      sigma_vals[i], alpha_vals[i], eta)
    system_state_iter.append(neuron_iter)
  return system_state_iter

def generate_ring_orbit(
    initial_system_state, zeta, 
    ge, sigma_vals, alpha_vals, eta, 
    num_iterations):
  ring_orbit = [initial_system_state]
  system_state = initial_system_state
  for _ in range(num_iterations):
    system_state = ring_coup_rulkov_1(
      system_state, zeta, 
      ge, sigma_vals, alpha_vals, eta)
    ring_orbit.append(system_state)
  ring_orbit = np.asarray(ring_orbit)
  return ring_orbit

def graph_some_ring_neuron_orbits(ring_orbit):
  neuron_0_orbit_x = ring_orbit.T[0][0]
  neuron_1_orbit_x = ring_orbit.T[0][1]
  neuron_2_orbit_x = ring_orbit.T[0][2]
  neuron_3_orbit_x = ring_orbit.T[0][3]
  neuron_4_orbit_x = ring_orbit.T[0][4]
  neuron_5_orbit_x = ring_orbit.T[0][5]
  neuron_6_orbit_x = ring_orbit.T[0][6]
  neuron_7_orbit_x = ring_orbit.T[0][7]
  fig, axs = plt.subplots(
    8, sharex=True, figsize=(10, 15))
  axs[0].plot(neuron_0_orbit_x, color='blue')
  axs[0].set(ylabel='neuron 0 x')
  axs[0].set_xlim(0, 1000)
  axs[0].set_ylim(-2, 2)
  axs[1].plot(neuron_1_orbit_x, color='blue')
  axs[1].set(ylabel='neuron 1 x')
  axs[1].set_xlim(0, 1000)
  axs[1].set_ylim(-2, 2)
  axs[2].plot(neuron_2_orbit_x, color='blue')
  axs[2].set(ylabel='neuron 2 x')
  axs[2].set_xlim(0, 1000)
  axs[2].set_ylim(-2, 2)
  axs[3].plot(neuron_3_orbit_x, color='blue')
  axs[3].set(ylabel='neuron 3 x')
  axs[3].set_xlim(0, 1000)
  axs[3].set_ylim(-2, 2)
  axs[4].plot(neuron_4_orbit_x, color='blue')
  axs[4].set(ylabel='neuron 4 x')
  axs[4].set_xlim(0, 1000)
  axs[4].set_ylim(-2, 2)
  axs[5].plot(neuron_5_orbit_x, color='blue')
  axs[5].set(ylabel='neuron 5 x')
  axs[5].set_xlim(0, 1000)
  axs[5].set_ylim(-2, 2)
  axs[6].plot(neuron_6_orbit_x, color='blue')
  axs[6].set(ylabel='neuron 6 x')
  axs[6].set_xlim(0, 1000)
  axs[6].set_ylim(-2, 2)
  axs[7].plot(neuron_7_orbit_x, color='blue')
  axs[7].set(ylabel='neuron 7 x')
  axs[7].set_xlim(0, 1000)
  axs[7].set_ylim(-2, 2)
  plt.xlabel('k')
  plt.subplots_adjust(hspace=0)
  plt.show()

def generate_ring_jacobians(
    ring_orbit, zeta, 
    ge, sigma_vals, alpha_vals, eta):
  J = []
  for k in range(len(ring_orbit)):
    one_jacobian = np.zeros((2*zeta, 2*zeta))
    for a in range(2*zeta):
      for b in range(2*zeta):
        m = a+1
        j = b+1
        X = ring_orbit[k]

        if m%2 == 1:
          i = int((m-1)/2)
          coup_param = (ge / 2 
            * (X[(i-1)%zeta][0] 
              + X[(i+1)%zeta][0] 
              - 2 * X[i][0]))

          if (X[i][0] <= 0):
            if j == m+1:
              one_jacobian[a][b] = 1
            elif j == m:
              one_jacobian[a][b] \
                = (alpha_vals[i] 
                  / (1-X[i][0])**2 - ge)
            elif ((j == m-2 and m != 1)
                or (j == 2*zeta-1 and m == 1)
                or (j == m+2 and m != 2*zeta-1)
                or (j == 1 and m == 2*zeta-1)):
              one_jacobian[a][b] = ge / 2
            else:
              one_jacobian[a][b] = 0
        
          elif (0 < X[i][0] 
              < alpha_vals[i] 
                + X[i][1] + coup_param):
            if j == m+1:
              one_jacobian[a][b] = 1
            elif j == m:
              one_jacobian[a][b] = -ge
            elif ((j == m-2 and m != 1)
                or (j == 2*zeta-1 and m == 1)
                or (j == m+2 and m != 2*zeta-1)
                or (j == 1 and m == 2*zeta-1)):
              one_jacobian[a][b] = ge / 2
            else:
              one_jacobian[a][b] = 0

          elif (X[i][0] >= alpha_vals[i] 
              + X[i][1] + coup_param):
            one_jacobian[a][b] = 0
        
        if m%2 ==0:
          if j == m:
            one_jacobian[a][b] = 1
          elif j == m-1:
            one_jacobian[a][b] = -eta*(1+ge)
          elif ((j == m-3 and m != 2)
              or (j == 2*zeta-1 and m == 2)
              or (j == m+1 and m != 2*zeta)
              or (j == 1 and m == 2*zeta)):
            one_jacobian[a][b] = eta*ge/2
          else:
            one_jacobian[a][b] = 0
    J.append(one_jacobian)

  return J

def qr_lyap_rulkov_two_coup(J):
  QR = [[np.zeros((2*zeta, 2*zeta)), 
    np.zeros((2*zeta, 2*zeta))]]
  QR.append(np.linalg.qr(J[0]))
  for k in range(2, len(J)):
    J_star = np.matmul(J[k-1], QR[k-1][0])
    QR.append(np.linalg.qr(J_star))
  lyapunov_sums = np.zeros(2*zeta)
  for k in range(1, len(QR)):
    for j in range(2*zeta):
      lyapunov_sums[j] = (lyapunov_sums[j] 
        + np.log(np.absolute(QR[k][1][j][j])))
  lyapunov_spectrum \
    = lyapunov_sums / (len(QR) - 1)
  lyapunov_spectrum = list(reversed(
    np.sort(lyapunov_spectrum)))
  return lyapunov_spectrum

zeta = 30
'''initial_x_states \
  = np.random.uniform(-1, 1, zeta)'''
initial_x_states \
  = [0.68921784, -0.94561073, -0.95674631,  
    0.91870134, -0.32012381, -0.23746836,
    -0.43906743, -0.48671017, -0.37578533, 
    -0.00613823,  0.25990663, -0.54103868,
    0.12110471,  0.71202085,  0.689336,   
    -0.03260047, -0.90907325,  0.93270227,
    0.51953315, -0.46783677, -0.96738424, 
    -0.50828432, -0.60388469, -0.56644705,
    -0.42772621,  0.7716625,  -0.60336517,  
    0.88158364,  0.0269842,   0.42512831]
initial_y_states = np.repeat(-3.25, zeta)
initial_system_state \
  = np.column_stack((
    initial_x_states, 
    initial_y_states))
ge = 0.05
sigma_vals = np.repeat(-0.5, zeta)
'''sigma_vals \
  = np.random.uniform(-1.5, -0.5, zeta)'''
'''sigma_vals \
  = [-0.63903048, -0.87244087, -1.16110093, 
    -0.63908737, -0.73103576, -1.23516699,
    -1.09564519, -0.57564289, -0.75055299, 
    -1.01278976, -0.61265545, -0.75514189,
    -0.89922568, -1.24012127, -0.87605023, 
    -0.94846269, -0.78963971, -0.94874874,
    -1.31858036, -1.34727902, -0.7076453,  
    -1.10631486, -1.33635792, -1.48435264,
    -0.76176103, -1.17618267, -1.10236959, 
    -0.66159308, -1.27849639, -0.9145025 ]'''
alpha_vals = np.repeat(4.5, zeta)
'''alpha_vals \
  = np.random.uniform(4.25, 4.75, zeta)'''
'''alpha_vals \
  = [4.31338267, 4.3882788,  4.6578449,  
    4.67308374, 4.28873181, 4.26278301,
    4.73065817, 4.29330435, 4.44416548, 
    4.66625973, 4.26243104, 4.65881579,
    4.68086764, 4.44092086, 4.49639124, 
    4.55500032, 4.33389054, 4.38869161,
    4.57278526, 4.62717616, 4.62025928, 
    4.49780551, 4.46750298, 4.49561326,
    4.66902393, 4.60858869, 4.6027906,  
    4.40563641, 4.54198743, 4.49388045]'''
eta = 0.001
num_iterations = 1000

ring_orbit = generate_ring_orbit(
  initial_system_state, zeta, 
  ge, sigma_vals, alpha_vals, eta, 
  num_iterations)
graph_some_ring_neuron_orbits(ring_orbit)
J = generate_ring_jacobians(
  ring_orbit, zeta, 
  ge, sigma_vals, alpha_vals, eta)
lyapunov_spectrum = qr_lyap_rulkov_two_coup(J)
print(lyapunov_spectrum)
\end{lstlisting}

\subsection{Box-Counting on a Coupled Rulkov Neuron Attractor}
\label{box-counting-on-coupled-rulkov-neuron-attractor}

This code implements the box-counting method established in Section \ref{strangeattractors} to count the number of four-dimensional boxes touched by a coupled Rulkov neuron attractor. It is used to calculate the fractal dimension of the system of two asymmetrically electrically coupled Rulkov 1 neurons in Section \ref{asym-elec-coup-two-rulkov-1-neurons-geometry}, and it uses the function \verb|two_coupled_rulkov_1_orbit| from Appendix \ref{sym-elec-coup-rulkov-1-neurons-code}.

\begin{lstlisting}[language=python]
import numpy as np

def box_counting_neurons(
    neuron_1_orbit, 
    neuron_2_orbit, 
    num_iterations,
    x1_min, x1_max, 
    y1_min, y1_max, 
    x2_min, x2_max, 
    y2_min, y2_max, 
    epsilon):
  neuron_1_orbit_x = neuron_1_orbit.T[0]
  neuron_1_orbit_y = neuron_1_orbit.T[1]
  neuron_2_orbit_x = neuron_2_orbit.T[0]
  neuron_2_orbit_y = neuron_2_orbit.T[1]
  
  num_boxes_x1 = int(
    (x1_max - x1_min) / epsilon)
  num_boxes_y1 = int(
    (y1_max - y1_min) / epsilon)
  num_boxes_x2 = int(
    (x2_max - x2_min) / epsilon)
  num_boxes_y2 = int(
    (y2_max - y2_min) / epsilon)
  running_box_count = 0
  total_box_count \
    = (num_boxes_x1 
      * num_boxes_x2 
      * num_boxes_y1 
      * num_boxes_y2)

  for a in range(num_boxes_x1):
    for b in range(num_boxes_y1):
      for c in range(num_boxes_x2):
        for d in range(num_boxes_y2):
          for k in range(num_iterations):
            if (((x1_min + epsilon * a) 
                < neuron_1_orbit_x[k] 
                < (x1_min + epsilon * (a + 1)))
                and ((y1_min + epsilon * b) 
                < neuron_1_orbit_y[k] 
                < (y1_min + epsilon * (b + 1)))
                and ((x2_min + epsilon * c) 
                < neuron_2_orbit_x[k] 
                < (x2_min + epsilon * (c + 1)))
                and ((y2_min + epsilon * d) 
                < neuron_2_orbit_y[k] 
                < (y2_min + epsilon * (d + 1)))
                ):
              running_box_count += 1
              break

  return running_box_count, total_box_count

initial_state_1 = [-0.54, -3.25]
initial_state_2 = [-1, -3.25]
ge_1 = 0.05
ge_2 = 0.25
sigma_1 = -0.5
sigma_2 = -0.5
alpha_1 = 4.5
alpha_2 = 4.5
eta = 0.001
num_iterations = 65000
x1_min = -1.5
x1_max = 1.5
y1_min = -3.3
y1_max = -3.2
x2_min = -1.5
x2_max = 2
y2_min = -3.3
y2_max = -3.2
epsilon = 1/30

neuron_1_orbit, neuron_2_orbit \
  = two_coupled_rulkov_1_orbit(
    initial_state_1, 
    initial_state_2, 
    ge_1, ge_2, 
    sigma_1, alpha_1, 
    sigma_2, alpha_2, 
    eta, num_iterations)
boxes = box_counting_neurons(
  neuron_1_orbit, 
  neuron_2_orbit, 
  num_iterations,
  x1_min, x1_max, 
  y1_min, y1_max, 
  x2_min, x2_max, 
  y2_min, y2_max, 
  epsilon)
print(boxes)
\end{lstlisting}

\subsection{Visualizing a Slice of the Basins of Attraction of Two Asymmetrically Coupled Rulkov 1 Neurons}
\label{visualize-basins-asym-coup-rulkov-1-code}

This code is used to visualize the basins of the non-chaotic spiking attractor and chaotic pseudo-attractor of two asymmetrically coupled Rulkov 1 neurons. It is used to produce the graphs shown in Figures \ref{fig:asym_basin_graphs} and \ref{fig:asym_basin_graphs_y}, and it uses various functions from Appendix \ref{sym-elec-coup-rulkov-1-neurons-code}. 

\begin{lstlisting}[language=python]
import numpy as np
import matplotlib.pyplot as plt

def generate_asym_rulkov_1_basin_x(
    x1_min, x1_max, 
    x2_min, x2_max, 
    edge_numpix, num_iterations):
  x1_vals = np.linspace(
    x1_min, x1_max, 
    edge_numpix, endpoint = False)
  x2_vals = np.linspace(
    x2_min, x2_max, 
    edge_numpix, endpoint = False)
  basin = np.zeros((
    edge_numpix, edge_numpix))

  for i in range(edge_numpix):
    for j in range(edge_numpix):
      initial_state_1 = np.array(
        [x1_vals[i], -3.25])
      initial_state_2 = np.array(
        [x2_vals[j], -3.25])
      neuron_1_orbit, neuron_2_orbit \
        = two_coupled_rulkov_1_orbit(
          initial_state_1, initial_state_2, 
          ge_1, ge_2, 
          sigma_1, alpha_1, 
          sigma_2, alpha_2, 
          eta, num_iterations)
      J = generate_rulkov_1_coup_jacobians(
        alpha_1, alpha_2, 
        sigma_1, sigma_2, eta, 
        ge_1, ge_2, 
        neuron_1_orbit, neuron_2_orbit)
      lyapunov_spectrum \
        = qr_lyap_rulkov_two_coup(J)
      max_lyapunov_exp = max(lyapunov_spectrum)
      basin[edge_numpix - 1 - j][i] \
        = max_lyapunov_exp
    
  plt.imshow(basin, cmap='coolwarm')
  plt.colorbar()
  plt.axis("off")
  return basin

def generate_asym_rulkov_1_basin_y(
    y1_min, y1_max, 
    y2_min, y2_max, 
    edge_numpix, num_iterations):
  y1_vals = np.linspace(
    y1_min, y1_max, 
    edge_numpix, endpoint = False)
  y2_vals = np.linspace(
    y2_min, y2_max, 
    edge_numpix, endpoint = False)
  basin = np.zeros((
    edge_numpix, edge_numpix))

  for i in range(edge_numpix):
    for j in range(edge_numpix):
      initial_state_1 = np.array(
        [-1, y1_vals[i]])
      initial_state_2 = np.array(
        [1, y2_vals[j]])
      neuron_1_orbit, neuron_2_orbit \
        = two_coupled_rulkov_1_orbit(
          initial_state_1, initial_state_2, 
          ge_1, ge_2, 
          sigma_1, alpha_1, 
          sigma_2, alpha_2, 
          eta, num_iterations)
      J = generate_rulkov_1_coup_jacobians(
        alpha_1, alpha_2, 
        sigma_1, sigma_2, eta, 
        ge_1, ge_2, 
        neuron_1_orbit, neuron_2_orbit)
      lyapunov_spectrum \
        = qr_lyap_rulkov_two_coup(J)
      max_lyapunov_exp = max(lyapunov_spectrum)
      basin[edge_numpix - 1 - j][i] \
        = max_lyapunov_exp
    
  plt.imshow(basin, cmap='coolwarm', 
    vmin=-0.025, vmax=0.025)
  plt.colorbar()
  plt.axis("off")
  return basin

x1_min = -2
x1_max = 2
x2_min = -2
x2_max = 2
edge_numpix = 300
ge_1 = 0.05
ge_2 = 0.25
sigma_1 = -0.5
sigma_2 = -0.5
alpha_1 = 4.5
alpha_2 = 4.5
eta = 0.001
num_iterations = 5000

basin = generate_asym_rulkov_1_basin_x(
  x1_min, x1_max, 
  x2_min, x2_max, 
  edge_numpix, num_iterations)
np.savetxt('aym_coupled_basin_color', basin)
\end{lstlisting}

\subsection{Classifying the Basins of the Asymmetrically Coupled Rulkov 1 Neuron System}
\label{classifying-asym-basin-code}

This code is used to classify the four-dimensional basins and two-dimensional basin slices of the asymmetrically coupled Rulkov 1 neuron system. Results from this code are displayed in Tables \ref{tab:p_function_asym_2_values} and \ref{tab:p_function_asym_4_values}, and it uses various functions from Appendix \ref{sym-elec-coup-rulkov-1-neurons-code}. 

\begin{lstlisting}[language=python]
import numpy as np

def rulkov_asym_attractor_mean_2d(
    neuron_1_orbit, neuron_2_orbit):
  orbit_length = len(neuron_1_orbit)
  neuron_orbit_x = np.column_stack((
    neuron_1_orbit.T[0], neuron_2_orbit.T[0]))
  mean = sum(neuron_orbit_x) / orbit_length
  return mean

def rulkov_asym_attractor_mean_4d(
    neuron_1_orbit, neuron_2_orbit):
  orbit_length = len(neuron_1_orbit)
  neuron_orbit = np.hstack((
    neuron_1_orbit, neuron_2_orbit))
  mean = sum(neuron_orbit) / orbit_length
  return mean

def rulkov_asym_attractor_standard_dev_2d(
    neuron_1_orbit, neuron_2_orbit, mean):
  attractor_orbit = np.column_stack((
    neuron_1_orbit.T[0], neuron_2_orbit.T[0]))
  running_sum = 0
  diff = attractor_orbit - mean
  for k in range(len(attractor_orbit)):
    running_sum = running_sum + np.dot(
      diff[k], diff[k])
  standard_dev = np.sqrt(
    running_sum / len(attractor_orbit))
  return standard_dev

def rulkov_asym_attractor_standard_dev_4d(
    neuron_1_orbit, neuron_2_orbit, mean):
  attractor_orbit = np.hstack((
    neuron_1_orbit, neuron_2_orbit))
  running_sum = 0
  diff = attractor_orbit - mean
  for k in range(len(attractor_orbit)):
    running_sum = running_sum + np.dot(
      diff[k], diff[k])
  standard_dev = np.sqrt(
    running_sum / len(attractor_orbit))
  return standard_dev

def asym_p_function_values_2d(
    ge_1, ge_2, 
    sigma_1, alpha_1, 
    sigma_2, alpha_2, eta, 
    mean, sigma, 
    num_xi_values, 
    num_test_points, 
    num_test_point_iterations, 
    white):
  count = 0
  distance = np.random.uniform(
    0, sigma, num_test_points)
  angle = np.random.uniform(
    0, 2 * np.pi, num_test_points)
  for i in range(num_test_points):
    chaotic = False
    x1_val = (mean[0] 
      + distance[i] * np.cos(angle[i]))
    x2_val = (mean[1] 
      + distance[i] * np.sin(angle[i]))
    initial_state_1 \
      = np.array([x1_val, -3.25])
    initial_state_2 \
      = np.array([x2_val, -3.25])
    neuron_1_orbit, neuron_2_orbit \
      = two_coupled_rulkov_1_orbit(
        initial_state_1, 
        initial_state_2, 
        ge_1, ge_2, 
        sigma_1, alpha_1, 
        sigma_2, alpha_2, eta, 
        num_test_point_iterations)
    J = generate_rulkov_1_coup_jacobians(
      alpha_1, alpha_2, 
      sigma_1, sigma_2, eta, 
      ge_1, ge_2, 
      neuron_1_orbit, neuron_2_orbit)
    lyapunov_spectrum \
      = qr_lyap_rulkov_two_coup(J)
    max_lyapunov_exp = max(lyapunov_spectrum)
    if max_lyapunov_exp > 0:
      chaotic = True
    if chaotic == white:
      count += 1
  p_values = [count / num_test_points]

  for k in range(num_xi_values - 1):
    count = 0
    distance = np.random.uniform(
      sigma * 2**k, sigma * 2**(k+1), 
      num_test_points)
    angle = np.random.uniform(
      0, 2 * np.pi, num_test_points)
    for i in range(num_test_points):
      chaotic = False
      x1_val = (mean[0] 
        + distance[i] * np.cos(angle[i]))
      x2_val = (mean[1] 
        + distance[i] * np.sin(angle[i]))
      initial_state_1 \
        = np.array([x1_val, -3.25])
      initial_state_2 \
        = np.array([x2_val, -3.25])
      neuron_1_orbit, neuron_2_orbit \
        = two_coupled_rulkov_1_orbit(
          initial_state_1, 
          initial_state_2, 
          ge_1, ge_2, 
          sigma_1, alpha_1, 
          sigma_2, alpha_2, eta, 
          num_test_point_iterations)
      J = generate_rulkov_1_coup_jacobians(
        alpha_1, alpha_2, 
        sigma_1, sigma_2, eta, 
        ge_1, ge_2, 
        neuron_1_orbit, 
        neuron_2_orbit)
      lyapunov_spectrum \
        = qr_lyap_rulkov_two_coup(J)
      max_lyapunov_exp = max(lyapunov_spectrum)
      if max_lyapunov_exp > 0:
        chaotic = True
      if chaotic == white:
        count += 1
    shell_p_value = count / num_test_points
    p_value \
      = p_values[k] / 4 + 3 * shell_p_value / 4
    p_values.append(p_value)
    
  return p_values

def asym_p_function_values_4d(
    ge_1, ge_2, 
    sigma_1, alpha_1, 
    sigma_2, alpha_2, eta, 
    mean, sigma, 
    num_xi_values, 
    num_test_points, 
    num_test_point_iterations, 
    white):
  count = 0
  r = np.random.uniform(
    0, sigma, num_test_points)
  theta_1 = np.random.uniform(
    0, np.pi, num_test_points)
  theta_2 = np.random.uniform(
    0, np.pi, num_test_points)
  phi = np.random.uniform(
    0, 2 * np.pi, num_test_points)
  for i in range(num_test_points):
    chaotic = False
    x1_val = (mean[0] 
      + r[i] * np.sin(theta_1[i]) 
      * np.sin(theta_2[i]) * np.cos(phi[i]))
    y_1_val = (mean[1] 
      + r[i] * np.sin(theta_1[i]) 
      * np.sin(theta_2[i]) * np.sin(phi[i]))
    x2_val = (mean[2] 
      + r[i] * np.sin(theta_1[i]) 
      * np.cos(theta_2[i]))
    y_2_val = (mean[3] 
      + r[i] * np.cos(theta_1[i]))
    initial_state_1 \
      = np.array([x1_val, y_1_val])
    initial_state_2 \
      = np.array([x2_val, y_2_val])
    neuron_1_orbit, neuron_2_orbit \
      = two_coupled_rulkov_1_orbit(
        initial_state_1, 
        initial_state_2, 
        ge_1, ge_2, 
        sigma_1, alpha_1, 
        sigma_2, alpha_2, eta, 
        num_test_point_iterations)
    J = generate_rulkov_1_coup_jacobians(
      alpha_1, alpha_2, 
      sigma_1, sigma_2, eta, 
      ge_1, ge_2, 
      neuron_1_orbit, neuron_2_orbit)
    lyapunov_spectrum \
      = qr_lyap_rulkov_two_coup(J)
    max_lyapunov_exp = max(lyapunov_spectrum)
    if max_lyapunov_exp > 0:
      chaotic = True
    if chaotic == white:
      count += 1
  p_values = [count / num_test_points]

  for k in range(num_xi_values - 1):
    count = 0
    r = np.random.uniform(
      sigma * 2**k, sigma * 2**(k+1), 
      num_test_points)
    theta_1 = np.random.uniform(
      0, np.pi, num_test_points)
    theta_2 = np.random.uniform(
      0, np.pi, num_test_points)
    phi = np.random.uniform(
      0, 2 * np.pi, num_test_points)
    for i in range(num_test_points):
      chaotic = False
      x1_val = (mean[0] 
        + r[i] * np.sin(theta_1[i]) 
        * np.sin(theta_2[i]) * np.cos(phi[i]))
      y_1_val = (mean[1] 
        + r[i] * np.sin(theta_1[i]) 
        * np.sin(theta_2[i]) * np.sin(phi[i]))
      x2_val = (mean[2] 
        + r[i] * np.sin(theta_1[i]) 
        * np.cos(theta_2[i]))
      y_2_val = (mean[3] 
        + r[i] * np.cos(theta_1[i]))
      initial_state_1 \
        = np.array([x1_val, y_1_val])
      initial_state_2 \
        = np.array([x2_val, y_2_val])
      neuron_1_orbit, neuron_2_orbit \
        = two_coupled_rulkov_1_orbit(
          initial_state_1, 
          initial_state_2, 
          ge_1, ge_2, 
          sigma_1, alpha_1, 
          sigma_2, alpha_2, eta, 
          num_test_point_iterations)
      J = generate_rulkov_1_coup_jacobians(
        alpha_1, alpha_2, 
        sigma_1, sigma_2, eta, 
        ge_1, ge_2, 
        neuron_1_orbit, 
        neuron_2_orbit)
      lyapunov_spectrum \
        = qr_lyap_rulkov_two_coup(J)
      max_lyapunov_exp = max(lyapunov_spectrum)
      if max_lyapunov_exp > 0:
        chaotic = True
      if chaotic == white:
        count += 1
    shell_p_value = count / num_test_points
    p_value \
      = (p_values[k] / 16 
        + 15 * shell_p_value / 16)
    p_values.append(p_value)
    
  return p_values
    

attractor_initial_state_1 = [-0.56, -3.25]
attractor_initial_state_2 = [-1, -3.25]
ge_1 = 0.05
ge_2 = 0.25
sigma_1 = -0.5
sigma_2 = -0.5
alpha_1 = 4.5
alpha_2 = 4.5
eta = 0.001
num_attractor_iterations = 65000
num_xi_values = 9
num_test_points = 500
num_test_point_iterations = 20000
white = True

(attractor_neuron_1_orbit, 
  attractor_neuron_2_orbit) \
    = two_coupled_rulkov_1_orbit(
      attractor_initial_state_1, 
      attractor_initial_state_2, 
      ge_1, ge_2, 
      sigma_1, alpha_1, 
      sigma_2, alpha_2, eta, 
      num_attractor_iterations)
mean = rulkov_asym_attractor_mean_4d(
  attractor_neuron_1_orbit, 
  attractor_neuron_2_orbit)
sigma = rulkov_asym_attractor_standard_dev_4d(
  attractor_neuron_1_orbit, 
  attractor_neuron_2_orbit, 
  mean)
p_values = asym_p_function_values_4d(
  ge_1, ge_2, 
  sigma_1, alpha_1, 
  sigma_2, alpha_2, eta, 
  mean, sigma, 
  num_xi_values, 
  num_test_points, 
  num_test_point_iterations, 
  white)
print(p_values)
\end{lstlisting}

\subsection{Uncertainty Exponents of the Asymmetrically Coupled Rulkov 1 Neuron System}
\label{uncert-exp-asym-coup-code}

This code is used to calculate the uncertainty exponents of the basin boundary between the four-dimensional white and black basins, as well as the two-dimensional white and black basin slices. Its results are shown in Table \ref{tab:uncertainty_exp_asym_coup_values}, and it uses various functions from Appendix \ref{sym-elec-coup-rulkov-1-neurons-code}.

\begin{lstlisting}[language=python]
import numpy as np
import matplotlib.pyplot as plt

def generate_uncertainty_function_values_2d(
    ge_1, ge_2, 
    sigma_1, alpha_1, 
    sigma_2, alpha_2, eta, 
    num_points, num_iterations, 
    x1_min, x1_max, 
    x2_min, x2_max, 
    num_epsilon_values):
  uncertainty_function_values = []
  for k in range(num_epsilon_values):
    epsilon = np.power(2.0, -k)
    count = 0
    for _ in range(num_points):
      white = False
      x1_val = np.random.random() \
        * (x1_max - x1_min) + x1_min
      x2_val = np.random.random() \
        * (x2_max - x2_min) + x2_min
      initial_state_1 \
        = np.array([x1_val, -3.25])
      initial_state_2 \
        = np.array([x2_val, -3.25])
      neuron_1_orbit, neuron_2_orbit \
        = two_coupled_rulkov_1_orbit(
          initial_state_1, 
          initial_state_2, 
          ge_1, ge_2, 
          sigma_1, alpha_1, 
          sigma_2, alpha_2, eta, 
          num_iterations)
      J = generate_rulkov_1_coup_jacobians(
        alpha_1, alpha_2, 
        sigma_1, sigma_2, eta, 
        ge_1, ge_2, 
        neuron_1_orbit, 
        neuron_2_orbit)
      lyapunov_spectrum \
        = qr_lyap_rulkov_two_coup(J)
      max_lyapunov_exp = max(lyapunov_spectrum)
      if max_lyapunov_exp > 0:
        white = True

      for i in range(4):
        test_white = False
        if i == 0:
          test_state_1 = np.array(
            [x1_val + epsilon, -3.25])
          test_state_2 = np.array(
            [x2_val, -3.25])
        elif i == 1:
          test_state_1 = np.array(
            [x1_val - epsilon, -3.25])
          test_state_2 = np.array(
            [x2_val, -3.25])
        elif i == 2:
          test_state_1 = np.array(
            [x1_val, -3.25])
          test_state_2 = np.array(
            [x2_val + epsilon, -3.25])
        elif i == 3:
          test_state_1 = np.array(
            [x1_val, -3.25])
          test_state_2 = np.array(
            [x2_val - epsilon, -3.25])
                
        test_neuron_1_orbit, \
          test_neuron_2_orbit \
            = two_coupled_rulkov_1_orbit(
              test_state_1, test_state_2, 
              ge_1, ge_2, 
              sigma_1, alpha_1, 
              sigma_2, alpha_2, eta, 
              num_iterations)
        J = generate_rulkov_1_coup_jacobians(
          alpha_1, alpha_2, 
          sigma_1, sigma_2, eta, 
          ge_1, ge_2, 
          test_neuron_1_orbit, 
          test_neuron_2_orbit)
        lyapunov_spectrum \
          = qr_lyap_rulkov_two_coup(J)
        max_lyapunov_exp \
          = max(lyapunov_spectrum)
        if max_lyapunov_exp > 0:
          test_white = True
        if test_white != white:
          count += 1
          break
        
    uncertainty_value = count / num_points
    uncertainty_function_values.append(
      uncertainty_value)

  return uncertainty_function_values

def generate_uncertainty_function_values_4d(
    ge_1, ge_2, 
    sigma_1, alpha_1, 
    sigma_2, alpha_2, eta, 
    num_points, num_iterations, 
    x1_min, x1_max, 
    x2_min, x2_max, 
    y1_min, y1_max, 
    y2_min, y2_max, 
    num_epsilon_values):
  uncertainty_function_values = []
  for k in range(num_epsilon_values):
    epsilon = np.power(2.0, -k)
    count = 0
    for _ in range(num_points):
      white = False
      x1_val = np.random.random() \
        * (x1_max - x1_min) + x1_min
      x2_val = np.random.random() \
        * (x2_max - x2_min) + x2_min
      y1_val = np.random.random() \
        * (y1_max - y1_min) + y1_min
      y2_val = np.random.random() \
        * (y2_max - y2_min) + y2_min
      initial_state_1 \
        = np.array([x1_val, y1_val])
      initial_state_2 \
        = np.array([x2_val, y2_val])
      neuron_1_orbit, neuron_2_orbit \
        = two_coupled_rulkov_1_orbit(
          initial_state_1, 
          initial_state_2, 
          ge_1, ge_2, 
          sigma_1, alpha_1, 
          sigma_2, alpha_2, eta, 
          num_iterations)
      J = generate_rulkov_1_coup_jacobians(
        alpha_1, alpha_2, 
        sigma_1, sigma_2, eta, 
        ge_1, ge_2, 
        neuron_1_orbit, 
        neuron_2_orbit)
      lyapunov_spectrum \
        = qr_lyap_rulkov_two_coup(J)
      max_lyapunov_exp = max(lyapunov_spectrum)
      if max_lyapunov_exp > 0:
        white = True

      for i in range(8):
        test_white = False
        if i == 0:
          test_state_1 = np.array(
            [x1_val + epsilon, y1_val])
          test_state_2 = np.array(
            [x2_val, y2_val])
        elif i == 1:
          test_state_1 = np.array(
            [x1_val - epsilon, y1_val])
          test_state_2 = np.array(
            [x2_val, y2_val])
        elif i == 2:
          test_state_1 = np.array(
            [x1_val, y1_val + epsilon])
          test_state_2 = np.array(
            [x2_val, y2_val])
        elif i == 3:
          test_state_1 = np.array(
            [x1_val, y1_val - epsilon])
          test_state_2 = np.array(
            [x2_val, y2_val])
        elif i == 4:
          test_state_1 = np.array(
            [x1_val, y1_val])
          test_state_2 = np.array(
            [x2_val + epsilon, y2_val])
        elif i == 5:
          test_state_1 = np.array(
            [x1_val, y1_val])
          test_state_2 = np.array(
            [x2_val - epsilon, y2_val])
        elif i == 6:
          test_state_1 = np.array(
            [x1_val, y1_val])
          test_state_2 = np.array(
            [x2_val, y2_val + epsilon])
        elif i == 7:
          test_state_1 = np.array(
            [x1_val, y1_val])
          test_state_2 = np.array(
            [x2_val, y2_val - epsilon])
                
        test_neuron_1_orbit, \
          test_neuron_2_orbit \
            = two_coupled_rulkov_1_orbit(
              test_state_1, test_state_2, 
              ge_1, ge_2, 
              sigma_1, alpha_1, 
              sigma_2, alpha_2, eta, 
              num_iterations)
        J = generate_rulkov_1_coup_jacobians(
          alpha_1, alpha_2, 
          sigma_1, sigma_2, eta, 
          ge_1, ge_2, 
          test_neuron_1_orbit, 
          test_neuron_2_orbit)
        lyapunov_spectrum \
          = qr_lyap_rulkov_two_coup(J)
        max_lyapunov_exp \
          = max(lyapunov_spectrum)
        if max_lyapunov_exp > 0:
          test_white = True
        if test_white != white:
          count += 1
          break
        
      uncertainty_value = count / num_points
      uncertainty_function_values.append(
        uncertainty_value)

  return uncertainty_function_values

ge_1 = 0.05
ge_2 = 0.25
sigma_1 = -0.5
sigma_2 = -0.5
alpha_1 = 4.5
alpha_2 = 4.5
eta = 0.001
num_points = 1000
num_iterations = 5000
x1_min = -2
x1_max = 2
x2_min = -2
x2_max = 2
y1_min = -5
y1_max = -1
y2_min = -5
y2_max = -1
num_epsilon_values = 12

uncertainty_function_values \
  = generate_uncertainty_function_values_4d(
    ge_1, ge_2, 
    sigma_1, alpha_1, 
    sigma_2, alpha_2, eta, 
    num_points, num_iterations, 
    x1_min, x1_max, 
    x2_min, x2_max, 
    y1_min, y1_max, 
    y2_min, y2_max, 
    num_epsilon_values)
print(uncertainty_function_values)
\end{lstlisting}

\subsection{Lyapunov Exponent and Dimension Graphs for the Rulkov 1 Ring Lattice System}
\label{lyap-exp-and-dim-graphs-code}

This code is used to produce the Lyapunov exponent and Lyapunov dimension graphs in Sections \ref{neuron-ring-lattice} and \ref{ring-lattice-geometry}. It uses many of the functions from Appendix \ref{ring-lattice-code}.

\begin{lstlisting}[language=python]
import numpy as np
import matplotlib.pyplot as plt

def max_lyap_graph_ring(
    initial_system_state, zeta, 
    ge_min, ge_max, ge_num_steps, 
    sigma_vals, alpha_vals, eta, 
    num_iterations):
  ge_vals = np.linspace(
    ge_min, ge_max, ge_num_steps)
  max_lyap_vals = []
  for a in range(len(ge_vals)):
    ring_orbit = generate_ring_orbit(
      initial_system_state, zeta, 
      ge_vals[a], sigma_vals, alpha_vals, eta, 
      num_iterations)
    J = generate_ring_jacobians(
      ring_orbit, zeta, 
      ge_vals[a], sigma_vals, alpha_vals, eta)
    lyapunov_spectrum \
      = qr_lyap_rulkov_two_coup(J)
    max_lyap_exp = lyapunov_spectrum[0]
    max_lyap_vals.append(max_lyap_exp)
  plt.scatter(ge_vals, max_lyap_vals, 
    s=0.5, c='black', linewidths=0)
  plt.xlabel('electrical coupling strength')
  plt.ylabel('maximal lyapunov exponent')
  return max_lyap_vals

def calc_lyap_dim(lyap_spec):
  spectrum_sum = 0
  for a in range(len(lyapunov_spectrum)):
    spectrum_sum = spectrum_sum + lyap_spec[a]
    if spectrum_sum <= 0:
      kappa = a
      spectrum_sum \
        = spectrum_sum - lyap_spec[a]
      break
  lyapunov_dim \
    = kappa - spectrum_sum / lyap_spec[kappa]
  return lyapunov_dim
    
def lyap_dim_graph_ring(
    initial_system_state, zeta, 
    ge_min, ge_max, ge_num_steps, 
    sigma_vals, alpha_vals, eta, 
    num_iterations):
  ge_vals = np.linspace(
    ge_min, ge_max, ge_num_steps)
  lyap_dims = []
  for a in range(len(ge_vals)):
    ring_orbit = generate_ring_orbit(
      initial_system_state, zeta, 
      ge_vals[a], sigma_vals, alpha_vals, eta, 
      num_iterations)
    J = generate_ring_jacobians(
      ring_orbit, zeta, 
      ge_vals[a], sigma_vals, alpha_vals, eta)
    lyapunov_spectrum \
      = qr_lyap_rulkov_two_coup(J)
    lyapunov_dimension \
      = calc_lyap_dim(lyapunov_spectrum)
    lyap_dims.append(lyapunov_dimension)
  plt.scatter(ge_vals, lyap_dims, 
    s=0.5, c='black', linewidths=0)
  plt.xlabel('electrical coupling strength')
  plt.ylabel('lyapunov dimension')
  return lyap_dims

zeta = 30
'''initial_x_states \
  = np.random.uniform(-1, 1, zeta)'''
initial_x_states \
  = [0.68921784, -0.94561073, -0.95674631,  
    0.91870134, -0.32012381, -0.23746836,
    -0.43906743, -0.48671017, -0.37578533, 
    -0.00613823,  0.25990663, -0.54103868,
    0.12110471,  0.71202085,  0.689336,   
    -0.03260047, -0.90907325,  0.93270227,
    0.51953315, -0.46783677, -0.96738424, 
    -0.50828432, -0.60388469, -0.56644705,
    -0.42772621,  0.7716625,  -0.60336517,  
    0.88158364,  0.0269842,   0.42512831]
initial_y_states = np.repeat(-3.25, zeta)
initial_system_state \
  = np.column_stack((
    initial_x_states, 
    initial_y_states))
ge_min = 0
ge_max = 1
ge_num_steps = 5001
sigma_vals = np.repeat(-0.5, zeta)
'''sigma_vals \
  = np.random.uniform(-1.5, -0.5, zeta)'''
'''sigma_vals \
  = [-0.63903048, -0.87244087, -1.16110093, 
    -0.63908737, -0.73103576, -1.23516699,
    -1.09564519, -0.57564289, -0.75055299, 
    -1.01278976, -0.61265545, -0.75514189,
    -0.89922568, -1.24012127, -0.87605023, 
    -0.94846269, -0.78963971, -0.94874874,
    -1.31858036, -1.34727902, -0.7076453,  
    -1.10631486, -1.33635792, -1.48435264,
    -0.76176103, -1.17618267, -1.10236959, 
    -0.66159308, -1.27849639, -0.9145025 ]'''
alpha_vals = np.repeat(4.5, zeta)
'''alpha_vals \
  = np.random.uniform(4.25, 4.75, zeta)'''
'''alpha_vals \
  = [4.31338267, 4.3882788,  4.6578449,  
    4.67308374, 4.28873181, 4.26278301,
    4.73065817, 4.29330435, 4.44416548, 
    4.66625973, 4.26243104, 4.65881579,
    4.68086764, 4.44092086, 4.49639124, 
    4.55500032, 4.33389054, 4.38869161,
    4.57278526, 4.62717616, 4.62025928, 
    4.49780551, 4.46750298, 4.49561326,
    4.66902393, 4.60858869, 4.6027906,  
    4.40563641, 4.54198743, 4.49388045]'''
eta = 0.001
num_iterations = 1000

lyapunov_dimensions \
  = lyap_dim_graph_ring(
    initial_system_state, zeta, 
    ge_min, ge_max, ge_num_steps, 
    sigma_vals, alpha_vals, eta, 
    num_iterations)
np.savetxt('lyap_dims_random_x', 
  lyapunov_dimensions)
print(lyapunov_dimensions)
\end{lstlisting}

\section{A Brief Review of Complex Algebra}
\label{complex-algebra}

A complex number $z\in\mathbb{C}$ can be written in the form 
\begin{equation}
    z=a+bi
\end{equation} 
where $a,\,b\in\mathbb{R}$ and $i^2=-1$. Complex numbers live in the complex plane, where the horizontal axis represents the set of the real numbers $\mathbb{R}$ and the vertical axis represents the set of the purely imaginary numbers. Then, the real part of $z$ is $\mathrm{Re}(z)=a$ and the imaginary part of $z$ is $\mathrm{Im}(z)=b$. We call the distance $z$ is from the origin of the complex plane the modulus of $z$, denoted by $|z|$. We call the angle $z$ makes from the positive real axis the argument of $z$, denoted by $\Arg(z)$, where $-\pi<\Arg(z)\leq\pi$. Geometrically, we can see that 
\begin{equation}
    |z|=\sqrt{a^2+b^2} 
\end{equation}
and 
\begin{equation}
    \Arg(z) = \tan^{-1}(b/a)
\end{equation} 
Then, we can represent $z$ in polar form as 
\begin{equation}
    z = r\cos\varphi + ir\sin\varphi
\end{equation}
where $r=|z|$ and $\varphi=\Arg(z)$. 

Euler's formula tells us that 
\begin{equation}
    e^{i\varphi} = \cos\varphi + i\sin\varphi
\end{equation}
which can be easily proved by writing out the Taylor series for $e^{i\varphi}$, $\cos\varphi$, and $\sin\varphi$. Therefore, we can represent any complex number in compact polar form as 
\begin{equation}
    z = re^{i\varphi}
\end{equation}

The complex conjugate of a given complex number $z=a+bi$ is
\begin{equation}
    z^* = a-bi
\end{equation}
which can be thought of as reflecting $z$ over the real axis. Then, in polar form,
\begin{equation}
    z^* = re^{-i\varphi} = r\cos\varphi - ir\sin\varphi
\end{equation}

\bibliography{refs}

\end{document}